\documentclass[12pt]{book}

\usepackage[LGR, T1]{fontenc}
\usepackage[utf8]{inputenc}
\usepackage[greek,main=english]{babel}

\usepackage{lettrine} 
\usepackage[object=vectorian]{pgfornament} 
\usepackage{pgfornament-han} 

\usepackage{pgfplots}
\usepackage{pgfplotstable}
\usepgfplotslibrary{patchplots}
\usepgfplotslibrary{colormaps}
\usepackage{scalerel}
\usepackage{tikz}
\usetikzlibrary{patterns}
\usetikzlibrary{snakes}
\usetikzlibrary{svg.path}
\usetikzlibrary{arrows, calc} 
\usetikzlibrary{decorations.markings}

\pgfmathsetmacro\sprayRadius{0.6pt}
\pgfmathsetmacro\sprayPeriod{0.6cm}

\pgfdeclarepatternformonly{spray}{\pgfpoint{-\sprayRadius}{-\sprayRadius}}{\pgfpoint{1cm + \sprayRadius}{1cm + \sprayRadius}}{\pgfpoint{\sprayPeriod}{\sprayPeriod}}{
    \foreach \x/\y in {2/53,6/52,11/48,23/49,20/47,32/46,41/47,47/51,56/52,46/44,4/43,16/42,33/41,41/37,49/35,55/31,37/35,44/30,28/37,24/36,17/37,7/38,0/31,8/29,18/31,28/30,37/28,30/27,46/24,51/21,24/23,12/24,4/21,18/19,12/16,31/21,38/18,26/16,46/16,56/12,52/10,45/8,51/4,37/12,35/7,24/9,14/9,2/12,8/6,15/4,27/0,34/1,40/1} {
        \pgfpathcircle{\pgfpoint{(\x + random()) / 57 * \sprayPeriod}{\sprayPeriod - (\y + random()) / 55 * \sprayPeriod}}{\sprayRadius}
    }
    \pgfusepath{fill}
}

\usepackage{tocloft}
\usepackage[unicode,hyperfootnotes=false]{hyperref}
\usepackage{float}
\usepackage{titlesec} 
\usepackage{appendix}
\usepackage{graphicx}
\usepackage{graphics}
\graphicspath{{thesis_images/}} 
\usepackage{epigraph} 
\usepackage[rightcaption]{sidecap} 
\usepackage{multirow} 
\usepackage{fancyhdr, ragged2e}
\usepackage{setspace}
\usepackage[multiple]{footmisc}
\usepackage{empheq}
 
\usepackage{color,soul}

\usepackage{enumerate}
\usepackage{mathtools}
\usepackage[figuresright]{rotating} 
\usepackage{cancel}
\usepackage{bigints}
\usepackage{cite}

\usepackage{afterpage}
\newcommand\blankpage{%
    \null
    \thispagestyle{empty}%
    \newpage}   

\usepackage{relsize}
\usepackage{indentfirst}
\usepackage{amssymb}
\usepackage{accents}
\usepackage{fixmath}
\usepackage{subcaption}
\usepackage{sidecap}
\usepackage{bookmark}
\usepackage{amsthm}
\usepackage{wrapfig}
\usepackage{amsmath}

\newcommand{\newc}{\newcommand}
\usepackage{verbatim}  
\usepackage{vector}  
\allowdisplaybreaks[1] 

\usepackage[margin=2.6cm]{geometry}

\titlespacing{\section}{0pt}{1em}{0.8em}
\titlespacing{\subsection}{0pt}{0.8em}{0.6em}

\definecolor{mygray}{rgb}{0.2, 0.3, 0.3}
\definecolor{azure(colorwheel)}{rgb}{0.0, 0.5, 1.0}
\definecolor{chapter(color)}{rgb}{0.01, 0.35, 0.55}  
\definecolor{cerulean}{rgb}{0.0, 0.48, 0.65}
\definecolor{brightcerulean}{rgb}{0.11, 0.67, 0.84}
\definecolor{babyblueeyes}{rgb}{0.63, 0.79, 0.95}
\definecolor{babyblue}{rgb}{0.54, 0.81, 0.94}
\definecolor{nicered}{rgb}{0.7,0.1,0.1}
\definecolor{coral}{rgb}{1.0, 0.5, 0.31}
\definecolor{nicegreen}{rgb}{0.1,0.5,0.1}
\definecolor{DarkBlue}{RGB}{0,0,153}

\newc{\chapc}[1]{\textcolor{chapter(color)}{#1}}
\newc{\teo}[1]{\textcolor{azure(colorwheel)}{#1}}
\newc{\corr}[1]{\textcolor{red}{#1}}

\usepackage{xcolor}
\usepackage{caption}
\usepackage[labelfont={color=chapter(color),bf}]{caption}

\usepackage[Sonny]{fncychap}
\ChNameVar{\fontsize{18}{18}\bfseries\rm}
\ChNumVar{\fontsize{60}{62}\bfseries\rm}
\ChTitleVar{\Huge\bfseries\rm}
\ChRuleWidth{1pt}

\newcommand{\mychapter}[1]{\textcolor{chapter(color)}{\chapter{\textbf{#1}}}}
\newcommand{\mychapterbl}[1]{\textcolor{chapter(color)}{\chapter*{\textbf{#1}}}}
\newcommand{\mysection}[1]{\vspace*{-1em}\textcolor{chapter(color)}{\section{#1}}}
\newcommand{\mysubsection}[1]{\textcolor{chapter(color)}{\subsection{#1}}}

\hypersetup{colorlinks,bookmarksopen,bookmarksnumbered,
citecolor={nicered},
linktoc=all,
linkcolor={chapter(color)},pdfstartview=FitH,
urlcolor={teal}}

\usepackage{libertine}

\newcommand{\sectionlinetwo}[2]{%
  \nointerlineskip \vspace{.5\baselineskip}\hspace{\fill}
  {\color{#1}
    \resizebox{0.5\linewidth}{1.2ex}
    {{%
    {\begin{tikzpicture}
    \node  (C) at (0,0) {};
    \node (D) at (9,0) {};
    \path (C) to [ornament=#2] (D);
    \end{tikzpicture}}}}}%
    \hspace{\fill}
    \par\nointerlineskip \vspace{.5\baselineskip}
  }
  
\def\eq$#1${\begin{equation}#1\end{equation}}
\def\gat$#1${\begin{gather}#1\end{gather}}
\def\bal$#1${\begin{align}#1\end{align}}
\def\eqarr$#1${\begin{eqnarray}#1\end{eqnarray}}
\def\beq{\begin{equation}}
\def\eeq{\end{equation}}
\def\bea{\begin{eqnarray}}
\def\eea{\end{eqnarray}}

\newc{\pa}{\partial}
\newc{\alp}{\alpha}
\newc{\gam}{\gamma}
\newc{\Gam}{\Gamma}
\newc{\del}{\delta}
\newc{\eps}{\epsilon}
\newc{\lam}{\lambda}
\newc{\sig}{\sigma}
\newc{\ups}{\upsilon}
\newc{\ome}{\omega}
\newc{\pphi}{\varphi}
\newc{\nonum}{\nonumber}
\newc{\hami}{\text{\textbf{\lat{H}}}}
\newc{\reals}{\mathbb{R}}
\newc{\gren}{\mathcal{G}}
\newc{\lagr}{\mathcal{L}}
\newc{\timor}{\mathcal{T}}
\newc{\prop}{\mathcal{K}}
\newc{\zcal}{\mathcal{Z}}
\newc{\cinf}{\mathcal{C}_\infty}
\newc{\operx}{\text{\textbf{\lat{x}}}}
\newc{\opera}{\text{\textbf{\lat{a}}}}
\newc{\operp}{\text{\textbf{\lat{p}}}}
\newc{\operl}{\text{\textbf{\lat{L}}}}
\newc{\gfv}{g^{(5)}}
\newc{\kfv}{\kappa_{(5)}}
\newc{\tf}{\tilde{f}}
\newc{\tlam}{\tilde{\Lambda}}
\newc{\tl}{\tilde{\lam}}
\newc{\dist}{\displaystyle}
\newc{\ra}{\rightarrow}
\newc{\Ra}{\Rightarrow}
\newc{\hsp}{\hspace{1em}}
\newc{\wtild}{\widetilde}
\newc{\ssst}{\scriptscriptstyle}
\newc{\sstar}{\text{\fontsize{5}{5}{$\bigstar$}}}
\def\lat#1{\textlatin{#1}}
\def\gr#1{\textgreek{#1}}
\newcommand{\mathsym}[1]{{}}
\newcommand{\unicode}[1]{{}}

\newcommand{\fref}[1]{{\color{chapter(color)}Fig.\hspace{0.3em}\ref{#1}}} 
\newcommand{\chapref}[1]{{\color{chapter(color)}Chap.\hspace{0.3em}\ref{#1}}} 
\newcommand{\secref}[1]{{\color{chapter(color)}Sec.\hspace{0.3em}\ref{#1}}} 
\newcommand{\myref}[2]{{\color{chapter(color)}Fig.\hspace{0.3em}\ref{#1}(\subref{#2})}} 

\newc{\form}[1]{\accentset{\leftarrow}{#1}} 
\newcommand{\trank}{\genfrac{\{}{\}}{0pt}{}} 
\newc{\tensor}[1]{\boldsymbol{\mathcal{#1}}} 
\newc{\vecpr}[2]{\vec{#1}{\, '}_{\hspace{-0.4em}#2}}
\newc{\formpr}[2]{\form{#1}{'}^{\hspace{0em}#2}}

\numberwithin{equation}{chapter}

\renewcommand{\headrulewidth}{1pt}

\renewcommand{\bibname}{\csname \gdef\@bibname{Bla} \endcsname}

\begin{document}

\fontdimen2\font=5pt 


\begin{titlepage}
\centering

\includegraphics[width=2cm]{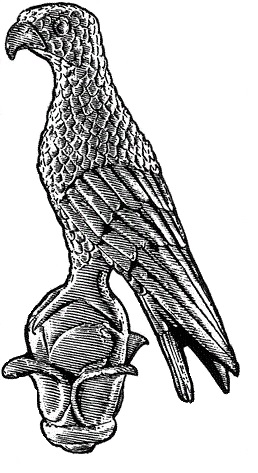}\\ [0.5cm]

\textbf{UNIVERSITY OF IOANNINA\\
SCHOOL OF NATURAL SCIENCES\\
PHYSICS DEPARTMENT} \\[2cm]

{\fontsize{22}{22} \textbf{Five-dimensional gravitational brane-world models: Solutions for black 
strings and\\[1.3mm] black holes}} \\[3.5cm]

{\fontsize{15}{15} \bf Theodoros Nakas } \\[3.5cm]

{\fontsize{15}{15} \textit{Ph.D. Dissertation}}\\[3.5cm]

{\fontsize{15}{15} \textbf{Ioannina 2022}}

\thispagestyle{empty}
\afterpage{\blankpage}

\includegraphics[width=2cm]{logo.jpg}\\ [0.5cm]


\textgreek{\textbf{ΠΑΝΕΠΙΣΤΗΜΙΟ ΙΩΑΝΝΙΝΩΝ\\
ΣΧΟΛΗ ΘΕΤΙΚΩΝ ΕΠΙΣΤΗΜΩΝ\\
ΤΜΗΜΑ ΦΥΣΙΚΗΣ} \\[2cm]

{\fontsize{22}{22}  \textbf{Πενταδιάστατες βαρυτικές θεωρίες με μεμβράνες: Λύσεις μαύρων χορδών 
και μαύρων οπών}} \\[3.5cm]

{\fontsize{15}{15}  \textbf{Θεόδωρος Νάκας}} \\[3.5cm]

{\fontsize{15}{15} \textit{Διδακτορική Διατριβή}}\\[3.5cm]

{\fontsize{15}{15} \textbf {Ιωάννινα 2022}}}

\thispagestyle{empty}
\afterpage{\blankpage}

\end{titlepage}


\thispagestyle{empty}


\thispagestyle{empty}

\vspace*{15.8cm}

\noindent{\gr{\fontdimen2\font=5pt Το έργο συγχρηματοδοτείται από την Ελλάδα και την 
Ευρωπαϊκή Ένωση (Ευρωπαϊκό Κοι\-νω\-νικό Ταμείο) μέσω του Επιχειρησιακού Προγράμματος 
«Ανάπτυξη Ανθρωπίνου Δυναμικού, Εκπαίδευση και Δια Βίου Μάθηση», στο πλαίσιο της Πράξης 
«Ενίσχυση του ανθρώπινου ε\-ρευ\-νη\-τι\-κού δυναμικού μέσω της υλοποίησης διδακτορικής 
έρευνας – 2ος Κύκλος» (\lat{MIS}-5000432), που υλοποιεί το Ίδρυμα Κρατικών 
Υποτροφιών (ΙΚΥ).}}

\vspace*{-0.5em}

\includegraphics[width=0.96\textwidth]{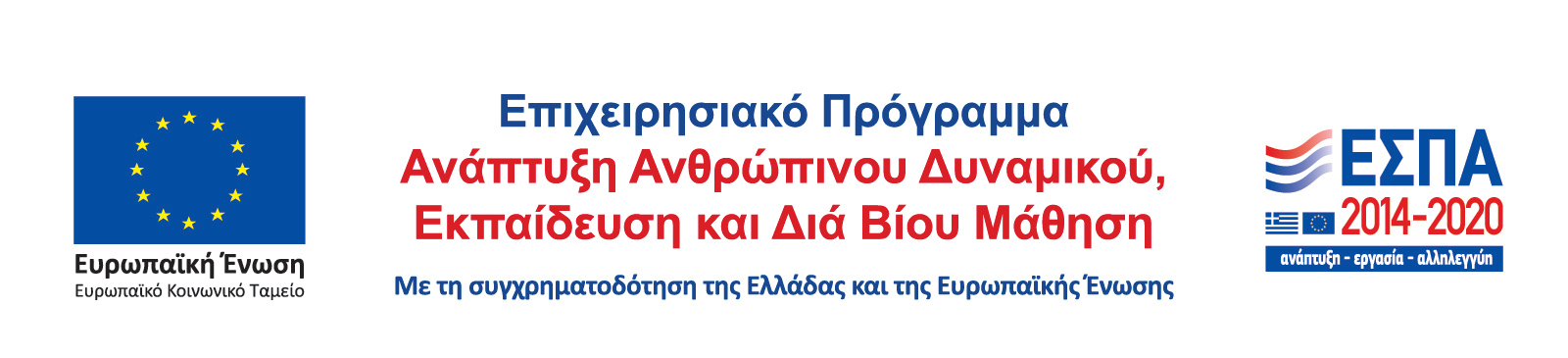}

\newpage
\blankpage

\thispagestyle{empty}

\vspace*{15.8cm}

\justify
This research is co-financed by Greece and the European Union (Eu\-ro\-pean 
So\-cial Fund- ESF) through the Op\-er\-a\-tional Pro\-gramme «Human 
Resources Development, Education and Lifelong Learning» in the context of the 
project “Strengthening Human Resources Research Potential via Doctorate 
Research – 2nd Cycle” (MIS-5000432), implemented by the State Scholarships 
Foundation (IKY).

\vspace*{2em}

\includegraphics[width=0.96\textwidth]{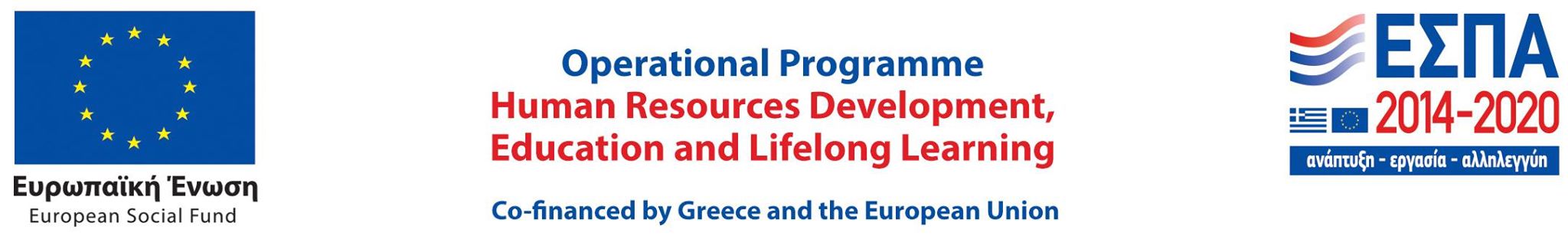}

\newpage
\blankpage


\pagenumbering{roman}


\thispagestyle{empty}

\vspace*{1cm}

\justifying

\begin{center}
\large{\textbf{Three-member advisory committee}}:
\end{center}

\begin{itemize}
\item Panagiota Kanti (advisor) --  Professor, Physics Department, University of Ioannina, Ioannina, Greece.
\item Kyriakos Tamvakis -- Emeritus Professor, Physics Department, University of Ioannina, Ioannina, Greece.
\item George Leontaris -- Emeritus Professor, Physics Department, University of Ioannina, Ioannina, Greece.
\end{itemize}

\vspace*{1cm}

\begin{center}
\large{\textbf{Seven-member PhD examination committee}}:
\end{center}

\begin{itemize}
\item Panagiota Kanti (advisor) --  Professor, Physics Department, University of Ioannina, Ioannina, Greece.
\item Kyriakos Tamvakis -- Emeritus Professor, Physics Department, University of Ioannina, Ioannina, Greece.
\item George Leontaris -- Emeritus Professor, Physics Department, University of Ioannina, Ioannina, Greece.
\item John Rizos -- Professor, Physics Department, University of Ioannina, Ioannina, Greece.
\item Leandros Perivolaropoulos -- Professor, Physics Department, University of Ioannina, Ioannina, Greece.
\item Christos Charmousis -- Professor, Universit\'{e} Paris-Saclay, CNRS/IN2P3, IJCLab, 91405 Orsay, France.
\item George Koutsoumbas -- Professor, School of Applied Mathematical and Physical Sciences, National Technical University of Athens, Athens, Greece.

\end{itemize} 

\newpage
\thispagestyle{empty}


\renewcommand{\headrulewidth}{0pt}
\fancyhead{}
\fancyfoot[CE,CO]{\vspace*{-0.5em}\sectionlinetwo{chapter(color)}{88} \chapc{--\,\thepage\,--}}



\mychapterbl{\begin{center}\vspace*{-2em}List of Publications \vspace*{-1em}\end{center}}
\phantomsection
\addcontentsline{toc}{chapter}{List of Publications}
\setcounter{Item}{1}

\thispagestyle{empty}

The current Ph.D. dissertation is mainly based on the following peer-reviewed articles which are listed below in ascending chronological order:

\begin{enumerate}
	
	\item P. Kanti, T. Nakas, and N. Pappas,
	``\textbf{Antigravitating\, braneworld\, solutions\, for\, a\, de\, Sitter\, brane\, in\, scalar-tensor\, gravity}'',
	\textit{Phys. Rev. D} 98 (2018) 6, 064025,
	\cite{KNP1}.
	
	\item T. Nakas, N. Pappas, and  P. Kanti,
	``\textbf{New\, black-string\, solutions\, for\, an\, anti–de\, Sitter\, brane\, in\, scalar-tensor\, gravity}'',
	\textit{Phys. Rev. D} 99 (2019) 12, 124040,
	\cite{KNP2}.

	\item T. Nakas, P. Kanti, and N. Pappas, 
	``\textbf{Incorporating\, physical\, constraints\, in\, brane\-world\, black-string\, solutions\, for\, a\, Minkowski\, brane\, in\, scalar-tensor\, gravity}'',
	\textit{Phys. Rev. D} 101 (2020) 8, 084056,
	\cite{KNP3}.

	\item T. Nakas and P. Kanti,
	``\textbf{Localized\, brane-world\, black\, hole\, analytically\, connected\, to\, an\, AdS$\bold{_5}$\, boundary}'',
	\textit{Phys. Lett. B} 816 (2021) 136278,
	\cite{NK1}.

	\item T. Nakas and P. Kanti,
	``\textbf{Analytic\, and\, exponentially\, localized\, brane-world\, Reissner-Nordström-AdS\, solution:\, a\, top-down\, approach}'',
    \textit{Phys. Rev. D} 104 (2021) 10, 104037,
	\cite{NK2}. 

\end{enumerate}

Moreover, during my Ph.D. studies, I also worked in parallel on the following two articles:

\begin{enumerate}

	\item T. Nakas and K. Rigatos,
	``\textbf{Fermions\, and\, baryons\, as\, open-string\, states\, from\, brane\, junctions}'',
	\textit{JHEP}  12 (2020) 157,
	\cite{NR}.

	\item A. Bakopoulos and T. Nakas,
	``\textbf{Analytic\, and\, asymptotically\, flat\, hairy\, (ultra-com\-pact)\, black-hole\, solutions\, and\, their\, axial\, perturbations}'',
	\textit{JHEP} 04 (2022) 096,
	\cite{BN}.

\end{enumerate}

\newpage
\blankpage



\thispagestyle{empty}
\vspace*{4cm}

\begin{flushright}

\begin{flushright}

\textit{Dedicated to my father\\ for teaching me to question,\\ and to my mother\\ 
for always being there for me}
\end{flushright}

\end{flushright}

\newpage
\thispagestyle{empty}

\mychapterbl{\begin{center}\vspace*{-2em}Acknowledgements \vspace*{-1em}\end{center}}
\phantomsection
\addcontentsline{toc}{chapter}{Acknowledgements}
\setcounter{Item}{2}

\thispagestyle{empty}

\justifying

The past four years of my Ph.D. studies were not only academically edifying but also very enjoyable 
despite the COVID pandemic. I owe this to a handful of people, thereby I would like to take the 
chance to thank:

-- My advisor Professor Panagiota Kanti for giving me the opportunity to work with her and deepen
my knowledge and understanding of gravitational theories beyond General Relativity. 

-- My friends and collaborators: Athanasios Bakopoulos, Nikolaos Pappas, and Konstantinos Rigatos 
for fruitful discussions both on scientific and philosophical issues.

-- My brother, my friends and my colleagues for all the great moments and ideas that we shared 
over the past years.

-- Georgia, for her love, support, and patience.

In addition, I would like to thank:

--  All seven members of the dissertation evaluation committee for the time they devoted to read 
it and their valuable remarks.

-- The Greek State Scholarships Foundation (IKY) for the financial support it provided me for almost 
the whole duration of my Ph.D. studies.

\newpage
\blankpage



\thispagestyle{empty}
\vspace*{4cm}

\justifying

\begin{center}
\Large{\textbf{Questioning Doubt}}
\vspace*{-1em}
\end{center}

\noindent{\rule{16.5cm}{0.1mm}}
\textit{\fontdimen2\font=5pt
They say ``doubt everything'', but I disagree. Doubt is useful in small amounts, but too much of it leads to apathy 
and confusion. No, don't doubt everything. QUESTION everything. That's the real trick. Doubt is just a lack of 
certainty. If you doubt everything, you'll doubt evolution, science, faith, morality, even reality itself---and you'll 
end up with nothing, because doubt doesn't give anything back. But questions have answers, you see. If you question 
everything, you'll find that a lot of what we believe is untrue… but you might also discover that some things ARE true. 
You might discover what your own beliefs are. And then you'll question them again, and again, eliminating flaws, 
discovering lies, until you get as close to the truth as you can.}

\noindent{\textit{\fontdimen2\font=5pt
Questioning is a lifelong process. That's precisely what makes it so unlike doubt. Questioning engages with reality, 
interrogating all it sees. Questioning leads to a constant assault on the intellectual status quo, where doubt is far 
more likely to lead to resigned acceptance. After all, when the possibility of truth is doubtful, why not simply play 
along with the most convenient lie?}}

\noindent{\textit{\fontdimen2\font=5pt Questioning is progress, but doubt is stagnation. }}
\vspace*{-2mm}\\
\rule{16.5cm}{0.1mm}
\noindent{The Talos Principle \hspace{10cm}Nadya Sarabhai}

\blankpage
\thispagestyle{empty}



\mychapterbl{\gr{Πρόλογος}}
\phantomsection
\addcontentsline{toc}{chapter}{\gr{\textbf{Πρόλογος}}}
\setcounter{Item}{3}

\thispagestyle{empty}

\epigraph{\justify\textit{\gr{<<Στόχος της επιστήμης δεν είναι να ανοίξει την πόρτα σε απέραντη σοφία, αλλά να θέσει ένα όριο στο απέραντο 
σφάλμα.>>}}}{\gr{Μπέρτολτ Μπρεχτ, Η ζωή του Γαλιλαίου}}

{\lettrine[lines=3, lhang=0.03]{\color{chapter(color)}\gr{Ό}}{\hspace*{5.5px}\gr{πως}} \gr{\fontdimen2\font=5pt
αναφέρει ο Νίτσε (1844,\,\lat{R\"{o}cken}--1900,\,Βαϊμάρη) στο βιβλίο του <<Η Γέννηση της Φι\-λο\-σο\-φί\-ας στα χρόνια 
της Ελληνικής τραγωδίας>>, η Ελληνική φιλοσοφία φαίνεται να ξε\-κι\-νά από την παράλογη αντίληψη του Θαλή
(624--547\,π.Κ.Ε.,\,Μί\-λη\-τος)\,\footnote{\gr{Κ.Ε. σημαίνει Κοινή Εποχή, και παρομοίως π.Κ.Ε. σημαίνει προ Κοινής Εποχής.
Οι προηγούμενες συ\-ντο\-μο\-γραφίες αποτελούν εναλλακτικές των μ.Χ. και π.Χ., αντίστοιχα, και είναι χρονολογικά ισοδύναμες με τις τελευταίες.
Οι ορολογίες Κ.Ε. και π.Κ.Ε. φαίνεται ότι χρησιμοποιήθηκαν για πρώτη φορά το 1615 από τον \lat{Johannes Kepler}, ενώ από τα τέλη του
20ού αιώνα και έπειτα χρησιμοποιούνται ευρέως στην ακαδημαϊκή και επιστημονική κοινότητα λόγω της θρησκευτικής τους ουδετερότητας.
Το ίδιο λοιπόν θα ισχύσει και στην παρούσα διατριβή. }} 
ότι το νερό είναι το θεμελιώδες συστατικό των πάντων.  
Αξιολογώντας όμως σοβαρά τη συ\-γκεκριμένη πρόταση, γρήγορα συνειδητοποιεί κανείς 
ότι η υπόθεση του Θαλή αποτέλεσε ένα διανοητικό άλμα σε σχέση με το επίπεδο των φυσικών θεωριών εκείνης της εποχής, κυρίως
διότι δεν είχαν αναπτυχθεί φυσικές θεωρίες μέχρι τότε. Ο Θαλής, 
αντιλαμβανόμενος την κεντρική σημασία που έχει το νερό για την ζωή όλων των ζωντανών οργανισμών, επιχείρησε μία γιγαντιαία 
γενίκευση τοποθετώντας το νερό ως την πρωταρχική αιτία των πάντων. Η σύλληψη του Θαλή έχει ιδιαίτερη αξία διότι στον πυρήνα της 
πραγματεύεται την ιδέα ότι}
<<\gr{εν το παν}>>
\gr{\fontdimen2\font=5pt και ταυτόχρονα αποτελεί μία πρόταση για την καταγωγή των όντων. Κατά τη γνώμη μου όμως, αυτό
που έχει ακόμα μεγαλύτερη σημασία στη σύλληψη του Θαλή, είναι ότι τοποθετήθηκε σε ένα τόσο θεμελιώδες ερώτημα, με έναν 
καθαρά ορθολογικό και φυσικό τρόπο, παρατηρώντας απλώς τον κεντρικό ρόλο που έχει το νερό για την ζωή επάνω στη Γη. Η 
σκέψη του Θαλή είναι πλήρως απαλλαγμένη από τους μύθους, τους θεούς και τις δεισιδαιμονίες της εποχής του, για αυτό αναζητά 
απαντήσεις μέσα στη φύση και όχι έξω από αυτή. Με γνώμονα αυτό το τελευταίο χαρακτηριστικό---δηλαδή την αναζήτηση της αλήθειας
παρατηρώντας τη φύση και αφήνοντας στην άκρη μεταφυσικές ερμηνείες---ο Θαλής μαζί με τους υπόλοιπους προσωκρατικούς 
φιλοσόφους,\,\footnote{Προσωκρατικοί ονομάζονται οι φιλόσοφοι που έζησαν από τον 7ο αιώνα π.Χ. μέχρι και την εποχή του 
Σωκράτη (469\,π.Κ.Ε.,\,Αλωπεκή--399\,π.Κ.Ε.,\,Αθήνα). Κυριότεροι εκπρόσωποι της προσωκρατικής φιλοσοφίας είναι οι Θαλής, 
Αναξίμανδρος, Πυθαγόρας, Ηράκλειτος, Παρμενίδης, Αναξαγόρας, Εμπεδοκλής και Δημόκριτος.} αποτελούν τους πρώτους φυσικούς 
φιλοσόφους.}

}

\gr{\fontdimen2\font=5pt
Μόλις 100 χρόνια μετά τον Θαλή, ο Δημόκριτος (460--370\,π.Κ.Ε.,\,Άβδηρα), επηρε\-α\-σμένος από τις ιδέες των προγενέστερων φιλοσόφων, 
διατύπωσε την έννοια του}
{\gr{\textit{ατόμου}}}\,\footnote{\gr{\fontdimen2\font=5pt Ετυμολογικά, η λέξη άτομο σημαίνει το συστατικό που δεν μπορεί να 
διαιρεθεί περαιτέρω. Να σημειωθεί επίσης, ότι η λέξη άτομο θα γράφεται πάντοτε με πλάγια γράμματα όταν χρησιμοποιείται με την ετυμολογική 
της έννοια, αλλιώς θα εννοείται με την επιστημονική της σημασία.}}
\gr{\fontdimen2\font=5pt 
ως τον θεμέλιο λίθο της φύσης. Υπέθεσε δηλαδή την ύπαρξη ενός απειροελάχιστα μικρού και θεμελιώδους συστατικού στη φύση από το 
οποίο πηγάζουν τα πάντα. 
Η ιδέα του Δημόκριτου ήταν πραγματικά επαναστατική για τα δεδομένα της εποχής. Απόδειξη αποτελεί το γεγονός ότι χρειάστηκε να 
περάσουν περίπου 2200 χρόνια για την εύρεση των πρώτων ενδείξεων της ύπαρξης θεμελιωδών συστατικών στη δομή της ύλης. Ο
\lat{John Dalton} (1766,\,Κά\-μπερ\-λαντ--1844,\,Μάντσεστερ) ήταν αυτός που στις αρχές του 19ου αιώνα ανέπτυξε το <<νόμο των 
πολλαπλών αναλογιών>>, καθώς και αυτός που εισήγαγε τον όρο άτομο με την έννοια που μαθαίνουμε σήμερα στη Φυσική και τη Χημεία. 
Παρόλα αυτά, δεν άργησε πολύ να αποδειχθεί ότι αυτό που αποκάλεσε ο \lat{Dalton} ως άτομο δεν ήταν παρά ένα σύνολο από επιμέρους, 
ακόμα μικρότερα σωμάτια. Αντιλαμβανόμαστε λοιπόν, ότι το άτομο του \lat{Dalton} κατέληξε να διαφέρει από το \textit{άτομο} του 
Δημόκριτου.}

\gr{\fontdimen2\font=5pt 
Σήμερα, περίπου 25 αιώνες μετά τους προσωκρατικούς φιλοσόφους, παρόλο που είμαστε σε θέση να απορρίψουμε την αντίληψη που είχε 
ο Θαλής για το βασικό στοιχείο του σύμπαντος, συνεχίζουμε να ψάχνουμε το \textit{άτομο} του Δημόκριτου. 
Έχοντας τη Γενική Θεωρία της Σχετικότητας (ΓΘΣ) από τη μία, και την Κβαντομηχανική μαζί με το Κα\-θιε\-ρω\-μέ\-νο Πρότυπο Στοιχειωδών 
Σωματιδίων (ΚΠΣΣ) από την άλλη, μπορούμε να ερμηνεύσουμε πλήρως όλα τα φαινόμενα της καθημερινότητaς, ενώ παράλληλα υπάρχει η 
δυνατότητα εύρεσης απαντήσεων ακόμα και σε ερωτήματα που σχετίζονται με την εξέλιξη του ίδιου του σύμπαντος.}

\gr{\fontdimen2\font=5pt 
Η Γενική Θεωρία της Σχετικότητας \cite{einstein1,einstein2,einstein3} (δείτε επίσης \cite{MTW,carroll:2019,Groen:2007zz}) 
θεμελιώθηκε το 1915 από τον \lat{Albert Einstein} (1879,\,Ουλμ--1955,\,Πρίνστον) και αποτελεί μία καθαρά γεωμετρική θεωρία η οποία έχει τις βάσεις 
της στη Ριμάνεια γεωμετρία. Για την ακρίβεια, ο τετραδιάστατος χωρόχρονος της ΓΘΣ καθιστά μία ψευδοριμάνεια πολλαπλότητα. 
Στο πλαίσιο της Γενικής Σχετικότητας, το πεδίο βαρύτητας ταυτίζεται με την κα\-μπυ\-λό\-τη\-τα του τε\-τρα\-δι\-ά\-στα\-του 
χωρόχρονου, ο οποίος, μπορεί να περιγράφει από το βαρυτικό α\-πο\-τύ\-πω\-μα ενός αστέρα ή μίας μελανής οπής μέχρι και το ίδιο 
το σύμπαν ως όλον. Το 2015, ακριβώς 100 χρόνια μετά τη θεμελίωση της ΓΘΣ, μέσω του πειράματος \lat{LIGO}-\lat{Virgo} 
\cite{Abbott:2016blz,LIGOScientific:2018mvr,Abbott:2020uma} ανιχνεύθηκαν για πρώτη φορά βαρυτικά κύματα. Αυτή η παρατήρηση είχε 
ως αποτέλεσμα να επιβεβαιώσει με 
απόλυτη σαφήνεια πως η έννοια του χωρόχρονου δεν αποτελεί απλώς ένα βολικό μαθηματικό κατασκεύασμα, αλλά είναι μία υπαρκτή φυσική οντότητα. 
Επιπροσθέτως, τον Απρίλιο του 2019, δημοσιεύτηκε από τη διεθνή ερευνητική συνεργασία \lat{Event Horizon Telescope (EHT)} η πρώτη 
εικόνα μίας υπερμαζικής μελανής οπής που βρίσκεται στο κέντρο του γειτονικού μας γαλαξία \lat{Messier 87 (M87)} \cite{Akiyama:2019cqa}. 
Η εικόνα της μαύρης οπής \lat{M87} καθώς και οι δεκάδες παρατηρήσεις βαρυτικών κυμάτων από διαφορετικές πηγές τα τελευταία χρόνια, 
αποδεικνύουν περίτρανα ότι η ΓΘΣ αποτελεί πράγματι μία φυσική θεωρία. Πέρα όμως από τη ΓΘΣ, η οποία είναι μία κλασική 
θεωρία,\,\footnote{Με τον όρο κλασική θεωρία εννοείται οποιαδήποτε φυσική θεωρία στερείται κβαντικής περιγραφής.} 
έχουμε στην διάθεσή μας και την Κβαντομηχανική μαζί με το Καθιερωμένο Πρότυπο Στοιχειωδών Σωματιδίων. Η Κβα\-ντο\-μη\-χα\-νική 
περιγράφει με εξαιρετική ακρίβεια όλες τις φυσικές διεργασίες που πραγματοποιούνται σε ατομικό και μοριακό επίπεδο, ενώ το ΚΠΣΣ
φθάνει σε ακόμα μεγαλύτερο βάθος περιγράφοντας τις αλληλεπιδράσεις όλων των μέχρι σήμερα γνωστών στοιχειωδών σωματιδίων σε 
ενέργειας που φθάνουν μέχρι και τις δεκάδες \lat{TeV}.}

\gr{\fontdimen2\font=5pt 
Παρόλα αυτά, ενώ οι παραπάνω θεωρίες εξηγούν πλήρως και σε εντυπωσιακό βάθος όλα τα φαινόμετα που συναντάμε στην 
καθημερινότητά μας, και όχι μόνο, η κατανόησή μας για τους θεμελιώδεις νόμους του σύ\-μπα\-ντος παραμένει ακόμα εξαιρετικά μικρή. 
Ο λόγος είναι ότι με κάθε νέα γνώση, δημιουργούνται και νέα ερωτήματα, που ενώ μερικές φορές απαντώνται εύκολα, τις περισσότερες 
φορές οι απαντήσεις προϋποθέτουν μία ριζοσπαστική σκέψη ικανή να δημιουργήσει έναν νέο και πιο θεμελιακό τρόπο ερμηνείας της φύσης. 
Το ίδιο έγινε με την Γενική Θεωρία της Σχετικότητας το 1915, με την Κβα\-ντο\-μη\-χα\-νι\-κή την δεκαετία του 1920, με την Κβαντική 
Θεωρία Πεδίου την δεκαετία του 1950, καθώς και με το Κα\-θι\-ε\-ρω\-μέ\-νο Πρότυπο Στοιχειωδών Σωματιδίων την δεκαετία του 
1970. Σήμερα, υπάρχει μία πληθώρα πειραματικών μετρήσεων και θεωρητικών συλλογισμών που μας κάνουν να πιστεύουμε ότι η φύση σε 
θεμελιώδες επίπεδο διαφέρει ριζικά από αυτό που αντιλαμβανόμαστε μέχρι στιγμής. Καταρχάς, σύμφωνα με αστροφυσικές παρατηρήσεις 
εκτιμάται ότι η ύλη που περιγράφεται από το Καθιερωμένο Πρότυπο αποτελεί μόνο το 5$\%$ των συστατικών του σύμπαντος, ενώ το 
υπόλοιπο 95$\%$ αναλύεται κατά 27$\%$ σε σκοτεινή ύλη\,\footnote{Η σκοτεινή ύλη είναι μία υποθετική μορφή ύλης, μη βαρυονικής
φύσεως, η οποία δεν αλληλεπιδρά καθόλου με το ηλεκτρομαγνητικό πεδίο. Λόγω του τελευταίου χαρακτηριστικού της ονομάστηκε
<<σκοτεινή>>, βέβαια στην πράξη είναι περισσότερο διαφανής παρά σκοτεινή.} 
και κατά 68$\%$ σε σκοτεινή ενέργεια.\,\footnote{Η σκοτεινή ενέργεια είναι μία άγνωστη μορφή ενέργειας, η ύπαρξη της οποίας είναι 
άρρηκτα συνδεδεμένη με την επιταχυνόμενη διαστολή του σύμπαντος.} Η πραγματική φύση των δύο αυτών συστατικών παραμένει ακόμα 
και σήμερα παντελώς άγνωστη. Όμως, χάρη στη βαρυτική τους επίδραση επάνω στην συνηθισμένη ύλη, αλλά και στον ρόλο που παίζουν 
στην εξέλιξη του σύ\-μπα\-ντος, μπορούμε να διαπιστώσουμε την ύπαρξή τους. Επίσης, ένα από τα σημαντικότερα ερωτήματα που 
παραμένει ακόμα αναπάντητο είναι η κβαντική φύση της βαρύτητας, η οποία περιμένουμε να παίζει καθοριστικό ρόλο στο αρχέγονο 
σύμπαν, όπου η ενέργεια των αλληλεπιδράσεων μεταξύ των σωματιδίων φθάνει την κλίμακα \lat{Planck ($10^{19}$ GeV)}, καθώς
και στην ερμηνεία της ιδιομορφίας που εμφανίζεται σε όλες τις λύσεις μαύρων οπών. Τέλος, έ\-χο\-ντας καταφέρει να περιγράψουμε 
με ενοποιημένο τρόπο τις τρεις εκ των τεσσάρων θεμελιωδών αλληλεπιδράσεων, φαντάζει εξαιρετικά απίθανο να μην υ\-πάρ\-χει μία 
βαθύτερη θεωρία η οποία θα ενοποιεί και τη βαρύτητα μαζί με τις υπόλοιπες τρεις. Προς το παρόν, η θεωρία χορδών, σε συνδυασμό με 
την υπερσυμμετρία, αποτελεί τη μοναδική θεωρία που προσφέρει μια ενοποιημένη αντίληψη όλων των στοιχειωδών σωματιδίων στη φύση 
(μποζονίων και φερμιονίων). Παρόλα αυτά, η έλλειψη πειραματικών ενδείξεων για την ορθότητά της σε συνδυασμό με την αδυναμία της
να εξηγήσει την θετική κοσμολογική σταθερά που παρατηρούμε στο σύμπαν μας, έχει ως αποτελέσμα η θεωρία χορδών να δέχεται κριτική. 
Το μόνο σίγουρο είναι ότι το ταξίδι της γνώσης συνεχίζεται, και για όσο θα υπάρχει ο άνθρωπος, θα υπάρχουν και προβλήματα 
προς επίλυση.  Άλλωστε, η δυσκολία ανακάλυψης της αλήθειας είναι και αυτή που στο τέλος της δίνει αξία. }


\mychapterbl{Prologue}
\phantomsection
\addcontentsline{toc}{chapter}{Prologue}
\setcounter{Item}{4}

\epigraph{\justify\textit{``The aim of science is not to open the door to infinite wisdom, but to set a limit to infinite 
error.''}}{Bertolt Brecht, Life of Galileo}

\thispagestyle{empty}

{\lettrine[lines=3, lhang=0.03]{\color{chapter(color)}A}{\hspace*{5.5px}s} Nietzsche (1844,\,R\"{o}cken--1900,\,Weimar) 
states in his book ``Philosophy in the Tragic Age of the Greeks'', the Greek 
philosophy seems to start by the absurd perception of Thales (624--547\,BCE,\,Miletus)\,\footnote[1]{CE stands for 
``common (or current) era'', and likewise BCE stands for ``before common era''. The previous notations constitute alternatives to
the AD and BC, respectively, while both systems are numerically equivalent. The terms CE and BCE trace back to 1615, when they
were used for the first time by Johannes Kepler, while since the late 20th century they are widely used in academic and scientific
publications due to the fact that they are religiously neutral. Consequently, the aforementioned notation will be used in the current 
dissertation as well.}
that water is the fundamental substance 
of everything. Although it may seem futile, considering this proposition seriously, one shortly realizes that Thales' assumption 
constitutes an intellectual leap compared to the physical theories of his era, mainly because no physical theory had been 
formulated before that time. 
Having realized the central role that water plays for life, Thales attempted a stupendous generalization by putting water as 
the primal cause of everything. This conception of Thales is of particular importance since in its core addresses the idea of 
<<omnia ab uno>> (everything from one), and simultaneously constitutes a proposition for the origin of beings. In my opinion though, 
what is even more important in the conception of Thales, is that he approached such a fundamental question with a purely natural, 
and rational way of thinking, by just observing the key role that water plays for the life on Earth. The thinking of Thales is 
completely unrestrained from the myths, the gods, and the superstitions of his era, and thus he seeks explanations to natural 
phenomena via rational hypotheses which reference natural processes themselves.
Based on this particular characteristic---namely, the pursuit of truth via observations of the natural world, and the rejection of 
any other metaphysical interpretation---Thales, together with the rest of presocratic 
philosophers\,\footnote[2]{\fontdimen2\font=5pt
Presocratic philosophers are the Greek thinkers who lived from the 7th century BCE until the era of Socrates 
(469\,BCE,\,Alopece--399\,BCE,\,Athens).  The main representatives of presocratic philosophy are Thales, Anaximander, Pythagoras, 
Heraclitus, Parmenides, Anaxagoras, Empedocles, and Democritus.} constitute the first natural philosophers.

}

Only 100 years after Thales, Democritus (460--370\,BCE,\,Abdera), influenced by the ideas of his preceding philosophers,
proposed the notion of \textit{atom}\,\footnote[3]{\fontdimen2\font=5pt The word \textit{atom} is derived from the ancient 
Greek word ``\gr{άτομον}'', which denotes the substance which cannot be divided into smaller pieces. In what follows, the word 
\textit{atom} is meant with its etymological meaning when it is written in italics, while its scientific meaning will be implied when 
it is formally written. } as the fundamental building block of the natural world, that is, he assumed the existence of a miniscule 
substance in nature from which everything else originates.  Contemplating how primitive was the scientific knowledge in the era 
of Democritus, it is not hard to imagine how revolutionary this idea was. As a matter of fact, the first indications for the existence 
of elementary constituents in the structure of matter had only been possible about 2200 years after Democritus. It was John Dalton 
(1766,\,Cumberland--1844,\,Manchester) who developed in the early 19th century ``the law of multiple portions'', and also the 
one who addressed the term atom with its current scientific meaning. However, soon after Dalton's proposition, it was shown 
that atoms consist of even smaller individual particles. Therefore, it becomes clear that Dalton's atom ended up differing 
from Democritus'  \textit{atom}.

Today, almost 25 centuries after presocratic philosophers, although we have the knowledge to reject Thales' assumption regarding 
the fundamental building block of nature, we are still searching for Democritus' \textit{atom}. Having General theory of Relativity (GR) 
for the description of gravity, and Quantum Mechanics (QM) together with the Standard Model (SM) of elementary particles for the 
description of the quantum world, we can fully explain the every day phenomena, while we are even capable of answering questions 
regarding the evolution of our universe.

General theory of Relativity \cite{einstein1,einstein2,einstein3} (see also \cite{MTW,carroll:2019,Groen:2007zz}) was formulated
by Albert Einstein (1879,\,Ulm--1955,\,Princeton) back in 1915 and constitutes a purely geometrical theory which has its basis in 
Riemannian geometry. As a matter of fact, the four-dimensional spacetime of GR renders a pseudo-riemannian manifold. In the context 
of GR, the gravitational field is identified with the curvature of the four-dimensional spacetime. Hence, the curvature of the four-dimensional 
spacetime may describe the gravitational imprint of a star, a black hole, or even describe the universe as a whole. In 2015, exactly 
100 years after the formulation of GR, the LIGO-Virgo experiment \cite{Abbott:2016blz,LIGOScientific:2018mvr,Abbott:2020uma} observed 
gravitational waves for the first time. 
This observation has undoubtedly verified that the notion of spacetime is not just a convenient mathematical edifice but rather a real 
physical entity. Furthermore, in April of 2019, the international collaboration Event Horizon Telescope (EHT) published the first image 
of a supermassive black hole which resides in the center of our neighbour galaxy, Messier 87 (M87) \cite{Akiyama:2019cqa}. The image 
of M87 black hole, together with the dozens of more recent gravitational-wave observations from various sources have irrefutably proved 
that GR is indeed a physical theory. In addition to GR, which is a classical theory,\,\footnote[4]{\fontdimen2\font=5pt The term 
<<classical>> denotes every physical theory which is deprived of a quantum description.} we have also at our disposal both Quantum 
Mechanics and the Standard Model of elementary particles. QM describes in extreme detail every physical process which takes place in the atomic
and molecular scale, while the SM goes deeper than that and describes the interactions of all known elementary particles up to the energy scale of
tens of TeV.

Although the aforementioned theories explain in great detail and with excessive precision all the physical phenomena which are met in our 
every-day life and in our laboratories, our understanding of the fundamental laws of the universe remains still very limited. The reason
is that every new knowledge begets new questions, while their answers, most often than not, demand a radical idea, capable of creating a
new and more fundamental way of perceiving nature. The same thing happened with General Relativity in 1915, with Quantum Mechanics
in 1920s, with Quantum Field Theory in 1950s, and with the Standard Model of elementary particles in 1970s. Today, there are a plethora 
of experimental evidence and theoretical calculations which make us believe that nature might differ dramatically in a more fundamental level.
First of all, according to astrophysical measurements it is estimated that the particles of the SM constitute only the $5\%$ of the ingredients 
of our universe, while the rest of $95\%$ is analysed into $27\%$ dark matter,\,\footnote[5]{Dark matter is a hypothetical, non-baryonic form 
of matter which does not interact with the electromagnetic field. Due to this last characteristic it was named as ``dark'', however, in practice, it 
is more transparent than dark.} and $68\%$ dark 
energy.\,\footnote[6]{\fontdimen2\font=5pt Dark energy is an unknown form of energy, the existence of which is inextricably linked with the 
accelerated expansion of our universe.}
The real nature of these two ingredients remain completely unknown at the moment. However, due to their gravitational effect on regular
illuminating matter, and from the pivotal role they play in the evolution of the universe, we can deduce their existence. Moreover, one of the
most important scientific inquiries that still remains unresolved, is the quantum nature of gravity. Quantum gravity is expected to play a crucial 
role in the very early universe, where the energy of the particle interactions reaches the Planck scale ($10^{19}$ GeV), and also is of key 
importance for the interpretation of black-hole singularities. Finally, having accomplished to describe the three out of four fundamental interactions
in a unified way, it seems very unlikely that a deeper theory able to unify all known interactions (including gravity) does not exist.
At the moment, String Theory constitutes the only theory which provides a framework capable of unifying all known particles (bosons and
fermions). However, the lack of experimental evidence for its validity, together with its inability to explain the positive-definite cosmological
constant that we observe in our universe resulted to an intense criticism about its connection to the real world. 
One thing is certain though, that the journey of knowledge will never end. As long as humans exist, there will always exist problems to be solved. 
After all, the difficulty of discovering the truth is what gives truth its value.

\afterpage{\blankpage}



\mychapterbl{\gr{Περίληψη}}
\phantomsection
\addcontentsline{toc}{chapter}{\gr{\textbf{Περίληψη}}}
\setcounter{Item}{5}

\epigraph{\textit{\gr{<<Αν γνωρίζεις επακριβώς τι είναι το τίποτα, τότε γνωρίζεις τα πάντα.>>}}}{\gr{Απόστολος Πιλαφτσής}}

\thispagestyle{empty}

{\lettrine[lines=3, lhang=0.03]{\color{chapter(color)}\gr{Η}}{\hspace*{5.5px}}
\gr{συγκεκριμένη διατριβή είναι αφιερωμένη στη μελέτη λύσεων μαύρων χορδών και μαύρων οπών στο πλαίσιο πενταδιάστατων μοντέλων
βαρύτητας με μεμβράνες. 
Αφετηρία για τη μελέτη τέτοιων θεωριών αποτέλεσε το μοντέλο \lat{Randall-Sundrum} \cite{RS1,RS2} το οποίο διατυπώθηκε από τη \lat{Lisa Randall} και τον \lat{Raman Sundrum} το 1999.
Πιο συγκεκριμένα, στο πλαίσιο της παρούσας διατριβής λαμβάνει χώρα μία λε\-πτο\-μερής ανάλυση των φυσικών χαρακτηριστικών και ιδιοτήτων τόσο των μαύρων χορδών 
όσο και των εντοπισμένων μελανών οπών που εμφανίζονται στα εκά\-στοτε πενταδιάστατα μοντέλα βαρύτητας. 
Επιπροσθέτως, θεωρώντας κατάλληλες θεωρίες πεδίου, προσδιορίζεται πλήρως το ενεργειακό ή υλικό περιεχόμενο το οποίο είναι απαραίτητο για να υπο\-στηρίξει τη γεωμετρία κάθε
μίας από τις προαναφερθείσες λύσεις. Το σχεδιάγραμμα της παρούσας διατριβής περιγράφεται παρακάτω.}

}

\gr{Στο Κεφ.\hspace{0.3em}1 παρουσιάζεται μία σύντομη εισαγωγή στη Γενική Θεωρία της Σχετικότητας (ΓΘΣ), ενώ συμπεριλαμβάνονται και όλες οι βασικές 
μαθηματικές έννοιες που απαιτού\-νται από τη θεωρία.
Επίσης, η γεωμετρία \lat{Schwarzschild} αναλύεται λεπτομερώς, ενώ παρουσιάζο\-νται και άλλες, πιο γενικές, λύσεις μαύρων τρυπών, οι οποίες ενσωματώνουν τόσο την ύπαρξη
κοσμολογικής σταθεράς όσο και την ύπαρξη ηλεκτρικού φορτίου. 
Τέλος, μία εκτεταμένη συζήτηση για την προέλευση και τα κίνητρα της ιδέας για την ύπαρξη επιπλέον χωρικών διαστάσεων---καθώς και των βαρυτικών μοντέλων με 
μεμβράνες---λαμβάνει χώρα πριν την ολοκλήρωση του κεφαλαίου. }

\gr{Στο Κεφ.\hspace{0.3em}2 μελετάμε την ύπαρξη μαύρων χορδών ή μέγιστα συμμετρικών λύσεων, στο πλαίσιο μίας πενταδιάστατης βαθμοτανυστικής θεωρίας στην οποία ένα βαθμωτό 
πεδίο είναι συζευγμένο με μη-τετριμμένο τρόπο με τη βαρύτητα. 
Το στοιχείο μήκους επάνω στη μεμβράνη περιγράφεται από τη γεωμετρία \lat{Schwarzschild (anti-)de Sitter}, ενώ στο πλαίσιο του συγκεκριμένου κεφαλαίου έχουμε
επιλέξει να μελετήσουμε λύσεις θετικής τετραδιά\-στατης κοσμολογικής σταθεράς.
Στο συγκεκριμένο μοντέλο εξετάζονται δύο διαφορετικές μορφές για τη συνάρτηση σύζευξης $f(\Phi)$, μία γραμμική και μία τετρα\-γωνική.
Στην περίπτωση της γραμμικής συνάρτησης σύζευξης βρίσκουμε λύσεις για τις οποίες η θεωρία κοντά στην τετραδιάστατη μεμβράνη μοιάζει να μιμείται  
μια συνηθισμένη βαρυτική θεωρία με τετριμμένη σύζευξη μεταξύ του βαθμωτού πεδίου και της βαρύτητας. 
Η συγκεκριμένη λύση χαρακτηρίζεται από την παρουσία μίας περιοχής εκατέρωθεν της μεμβράνης όπου η βαρύτητα λειτουργεί ελκτικά,---όπως είναι γνωστό 
άλλωστε---ενώ μία περιοχή απωστικής βαρύτητας εμφανίζεται καθώς απομακρυνόμαστε από τη μεμβράνη.
Α\-ντι\-θέτως, στην περίπτωση της τετραγωνικής συνάρτησης σύζευξης, δεν υπάρχει καθόλου η περιοχή της ελκτικής βαρύτητας.
Όπως αναλύουμε και στο τέλος του κεφαλαίου, η ύπαρξη μιας θετικής κοσμολογικής σταθεράς επάνω στη μεμβράνη μας συνοδεύεται πάντοτε από
μία περιοχή απωστικής βαρύτητας κατά μήκος της πέμπτης διάστασης.}

\gr{Στο Κεφ.\hspace{0.3em}3 συνεχίζουμε την ανάλυση του προηγούμενου κεφαλαίου εξετάζοντας την πε\-ρί\-πτω\-ση της αρνητικής τετραδιάστατης κοσμολογικής σταθεράς
επάνω στη μεμβράνη μας. 
Κατά συνέπεια, ο τετραδιάστατος χωρόχρονος επάνω στη μεμβράνη περιγράφεται από τη γε\-ωμετρία \lat{Schwarzschild anti-de Sitter}.
Διαλέγοντας κατάλληλα την έκφραση της συνάρτησης σύζευξης επιλύουμε αναλυτικά τις εξισώσεις πεδίου της θεωρίας και οδηγούμαστε στην εύρεση λύσεων
μαύρων χορδών.
Παράγουμε δύο τέτοιες λύσεις με την κάθε μία από αυτές να χαρακτηρίζεται από ένα ομαλό βαθμωτό πεδίο $\Phi$, έναν εντοπισμένο (κοντά στη μεμβράνη μας) τανυστή 
ενέργειας και ορμής, καθώς και ένα αρνητικά ορι\-σμένο βαθμωτό δυναμικό $V_B(\Phi)$, το οποίο υποστηρίζει από μόνο του τη στρέβλωση του πενταδιάστατου χωρόχρονου
ακόμα και υπό την απουσία της πενταδιάστατης αρνητικής κοσμολογικής σταθεράς στο \lat{bulk}.
Παρόλο που η ιδιομορφία της μαύρης χορδής εκτείνεται μέχρι το άπειρο, η τετραδιάστατη ενεργός θεωρία επάνω στη μεμβράνη είναι ομαλότατη.
Και στις δύο επιμέρους περιπτώσεις που μελετώνται, η τετραδιάστατη κλίμακα της βαρύτητας εκφράζεται αναλυτικά ως συνάρτηση των παραμέτρων της
πενταδιά\-στατης θεμελιώδους θεωρίας.
Αξίζει τέλος να σημειωθεί ότι αν επιλέξουμε να μηδενίσουμε τη μάζα της μαύρης χορδής, τότε η μαύρη χορδή εξαφανίζεται τελείως από το μοντέλο μας αφήνοντας πίσω
ένα σύνηθες πενταδιάστατο μοντέλο όπως αυτό των \lat{Randall-Sundrum} \lat{(RS)}. }

\gr{Στο Κεφ.\hspace{0.3em}4 ολοκληρώνουμε την ανάλυση των προηγούμενων δύο κεφαλαίων μελετώ\-ντας την περίπτωση της \lat{Minkowski} μεμβράνης, δηλαδή την
περίπτωση όπου η κοσμολογική σταθερά επάνω στη μεμβράνη είναι ταυτοτικά μηδέν.
Εξετάζοντας μία πληθώρα από συναρτήσεις σύζευξης, καταφέρνουμε κάθε φορά να επιλύσουμε τις βαρυτικές εξισώσεις με αναλυτικό τρόπο και
να προσδιορίσουμε όλες τις άγνωστες συναρτήσεις του μοντέλου, όπως για παράδειγμα την έκφραση του δυναμικού του βαθμωτού πεδίου $V_B(\Phi)$.
Οι λύσεις και σε αυτή την περίπτωση χαρακτηρίζονται από την ύπαρξη ενός ομαλού βαθμωτού πεδίου, ενός εντοπισμένου τανυστή ενέργειας-ορμής καθώς και
ενός εκθετικά αποσβένον παράγοντα στρέβλωσης ακόμα και κατά το μηδενισμό της πενταδιάστατης κοσμολογικής σταθεράς.
Όπως είναι σαφές από τα παραπάνω η γεωμετρία του χωρόχρονου επάνω στη μεμβράνη δίνεται από την λύση \lat{Schwarzschild} και οδηγεί σε μία μη-ομογενή
μαύρη χορδή στο \lat{bulk}.
Προφανώς, αν μηδενίσουμε την μάζα της μαύρης χορδής, οδηγούμαστε φυσιολογικά στο μοντέλο \lat{Randall-Sundrum}. 
Κανείς μπορεί να κατασκευάσει φυσικά απο\-δε\-κτές λύσεις εάν υποθέσει μία θετικά ορισμένη τετραδιάστατη βαρυτική σταθερά επάνω στη μεμβράνη,
μία θετικά ορισμένη πυκνότητα ενέργειας επάνω στη μεμβράνη, καθώς και την ικανοποίηση των ασθενών ενεργειακών συνθηκών (\lat{weak energy conditions}) στο \lat{bulk}. 
Όπως διαπιστώνουμε μετά από εκτεταμένη ανάλυση, παρόλο που είναι αδύνατον να ικανοποιηθούν ταυτόχρονα και οι τρεις προαναφερθείσες συνθήκες, μπορούμε κάθε φορά
να ικανοποιούμε δύο από τις τρεις.
Ας σημειωθεί τέλος ότι οι πρώτες δύο συνθήκες είναι και οι πιο σημαντικές.}

\gr{Στο Κεφ.\hspace{0.3em}5 στρέφουμε την προσοχή μας σε λύσεις εντοπισμένων μελανών οπών.
Συ\-γκεκριμένα, καταφέρνουμε να παράξουμε από πρώτες αρχές τη γεωμετρία μιας αναλυτικής και εκθετικά εντοπισμένης πενταδιάστατης μαύρης οπής.
Η σημειακή ιδιομορφία της μελανής οπής κείται εξ ολοκλήρου επάνω στην τετραδιάστατη μεμβράνη, ενώ ο ορίζοντας γεγονότων της μελανής οπής φαίνεται να έχει
την τοπολογία μίας πενταδιά\-στατης <<τηγανίτας>> που οφείλεται στον εκθετικό εντοπισμό της μαύρης οπής γύρω από τη μεμβράνη.
Το στοιχείο μήκους της μαύρης οπής επάνω στη μεμβράνη δίνεται από τη γεωμετρία \lat{Schwarzschild}, ενώ ο πενταδιάστατος χωρόχρονος έξω από
τον ορίζοντα γεγονότων της μελανής οπής είναι πρακτικά \lat{AdS$_5$}.
Η γεωμετρία της μελανής οπής υ\-πο\-στηρίζεται από ένα ανισοτροπικό ρευστό το οποίο χαρακτηρίζεται μόνο από δύο ανεξάρτητες συνιστώσες του τανυστή
ενέργειας και ορμής. Οι συνιστώσες αυτές είναι η πυκνότητα ενέργειας $\rho_E$ και η γωνιακή ή αζιμουθιακή πίεση $p_\theta$.
Επιπλέον, αποδεικνύουμε ότι δεν χρειάζεται να εισαχθεί επιπρόσθετη ύλη επάνω στη μεμβράνη για τον αυτοσυνεπή εμβαπτισμό της μέσα στον πενταδιάστατο
χωρόχρονο. }

\gr{Στο Κεφ.\hspace{0.3em}6 γενικεύουμε τη γεωμετρία της εντοπισμένης μελανής οπής του προηγούμενου κεφαλαίου.
Χρησιμοποιώντας την ίδια διαδικασία για τον εντοπισμό της μελανής οπής, καταφέρνουμε να υποστηρίζουμε τη γεωμετρία μίας \lat{Reissner-Nordstr\"{o}m-(A)dS}
μαύ\-ρης τρύπας επάνω στη μεμβράνη.
Εκτελούμε μία προσεκτική κατηγοριοποίηση των οριζόντων γεγονότων της παραπάνω λύσης και δείχνουμε ότι όλοι τους βρίσκονται εκθετικά ε\-ντο\-πισμένοι κοντά στην
τετραδιάστατη μεμβράνη (το σύμπαν μας).
Να σημειωθεί ότι η πενταδιάστατη γεωμετρία του χωρόχρονου είναι παντού ομαλή, ενώ ο χωρόχρονος έξω από τον ορίζοντα γεγονότων της μελανής οπής είναι και σε
αυτή την περίπτωση πρα\-κτι\-κά \lat{AdS$_5$}.
Η γεωμετρία της μελανής οπής υποστηρίζεται και εδώ από την ύπαρξη ενός ανισοτροπικού ρευστού με τις ανεξάρτητες συνιστώσες να είναι η πυκνότητα ενέργειας $\rho_E$ 
και η γωνιακή πίεση $p_\theta$.
Όλες οι ενεργειακές συνθήκες του τανυστή ενέργειας και ορμής ικανοποιούνται επάνω και στη γειτονιά της τετραδιάστατης μεμβράνης, ενώ μία τοπική παραβίαση των συνθηκών
λαμβάνει χώρα στο εσωτερικό του ορίζοντα γεγονότων της μελανής οπής καθώς αυτός εκτείνεται στην πέμπτη διάσταση.
Σε μία προσπάθεια να προσδιορίσουμε επακριβώς την θεωρία πεδίου που είναι ικανή να υποστηρίζει την γεωμετρία της συγκεκριμένης μελανής οπής, υποθέσαμε μία θεωρία
που περιέχει τόσο βαθμωτούς όσο και διανυσματικούς βαθμούς ελευθερίας \lat{(scalar and gauge field theory)}.
Παρόλα αυτά, όπως δείχνουμε στην σχετική ενότητα του κεφαλαίου η συ\-γκεκριμένη θεωρία πεδίου δεν μπορεί να υποστηρίξει ικανοποιητικά την προαναφερθείσα γεωμετρία,
λόγω του ότι τα πεδία αποκτούν αναπόφευκτα μία \lat{phantom-like} συμπεριφορά σε κάποιο σημείο του πενταδιάστατου χωρόχρονου.
Μελετώντας τις συνοριακές συνθήκες επάνω στη μεμβράνη αποδεικνύουμε ότι δεν είναι απαραίτητη η προσθήκη επιπρόσθετης ύλης για τον επιτυχή εμβαπτισμό της 
μέσα στον πενταδιάστατο χωρόχρονο.
Τέλος, αποδεικνύουμε ότι η γεωμετρία της \lat{Reissner-Nordstr\"{o}m-(A)dS} μαύ\-ρης οπής επάνω στη μεμβράνη αποτελεί ένα συνδυαστικό αποτέλεσμα της πενταδιάστατης
γεωμετρίας και πε\-νταδιάστατης ενέργειας/ύλης που βρίσκονται στο \lat{bulk}, ενώ το <<φορτίο>> της μελανής οπής δεν είναι παρά ένα παλιρροιακό φορτίο (\lat{tidal charge}) και όχι φορτίο ενός πεδίου βαθμίδας (\lat{gauge charge}).}

\gr{Τέλος, στο Κεφ.\hspace{0.3em}7 συγκεντρώνουμε τα αποτελέσματά μας και συζητάμε ενδεχόμενες μελλοντικές ερευνητικές προεκτάσεις που πηγάζουν από την παρούσα εργασία.}


\mychapterbl{\textbf{Abstract and Outline}}
\phantomsection
\addcontentsline{toc}{chapter}{Abstract and Outline}
\setcounter{Item}{6}

\epigraph{\textit{``If you know exactly what nothing is, then you know everything.''}}{Apostolos Pilaftsis}

\thispagestyle{empty}

{\lettrine[lines=3, lhang=0.03]{\color{chapter(color)}I}{\hspace*{5.5px}n}
the content of this dissertation, we study the emergence of black-string and black-hole solutions in the
framework of five-dimensional braneworld models. The main motivation for studying such theories stems from the Randall-Sundrum model \cite{RS1,RS2} which was formulated by Lisa Randall and Raman Sundrum  back in 1999. To be more precise, we investigate the physical characteristics of both black-string and localized black-hole solutions which emerge in the corresponding five-dimensional braneworld models, while, by considering appropriate 
field theories, we examine the necessary matter/energy content which is able to support the geometry of each one of the aforementioned solutions.
The outline of the dissertation is described below.

}

In \chapref{Chap: intro} we commence with a prelude to General theory of Relativity (GR) which includes all the basic notions and mathematical tools 
required by the theory. 
We then discuss in great detail the Schwarzschild solution, as well as additional and more general black-hole solutions which
incorporate the existence of both a cosmological constant and electric charge.
Finally, we discuss the origin and motives behind the idea of extra spatial dimensions and braneworld models.

In \chapref{Chap: P1}, we consider a five-dimensional theory with a scalar field nonminimally coupled to
gravity, and we look for solutions that describe novel black-string or maximally symmetric solutions in the bulk. 
The brane line element is found to describe a Schwarzschild (anti-)de
Sitter spacetime, where we choose to study solutions with a positive four-dimensional
cosmological constant.
We consider two different forms of the coupling function of the
scalar field to the bulk scalar curvature, a linear and a quadratic one. In the linear
case, we find solutions where the theory, close to our brane, mimics an ordinary gravitational 
theory with a minimally coupled scalar field giving rise to an exponentially
decreasing warp factor in the absence of a negative bulk cosmological constant. The so-
lution is characterized by the presence of a normal gravity regime around our brane and
an antigravitating regime away from it. In the quadratic case, there is no normal-gravity
regime at all; however, scalar field and energy-momentum tensor components are well de-
fined and an exponentially decreasing warp factor emerges again. We demonstrate that,
in the context of this theory, the emergence of a positive cosmological constant on our
brane is always accompanied by an antigravitating regime in the five-dimensional bulk.

In \chapref{Chap: P2} we continue the study of the previous Chapter and we consider the case of an anti-de Sitter brane.
Hence, the four-dimensional brane spacetime is described by the Schwarzschild anti-de Sitter geometry.
By appropriately choosing the non-minimal coupling function
of the scalar field, we analytically solve the gravitational and scalar-field equations in the bulk to produce
black-string solutions. We produce two complete such solutions that are both characterized by a regular scalar field, a localized-close-to-our
brane energy-momentum tensor and a negative-definite, non-trivial bulk potential that may support by
itself the warping of the spacetime even in the absence of the traditional, negative, bulk cosmological
constant. Despite the infinitely long string singularity in the bulk, the four-dimensional effective theory
on the brane is robust with the effective gravity scale being related to the fundamental one and the
warping scale. It is worth noting that if we set the mass of the black hole on the brane equal to zero,
the black string disappears leaving behind a regular braneworld model with only a true singularity at
the boundary of the fifth dimension. 

In \chapref{Chap: P3} we complete the study of the model, which commenced in \chapref{Chap: P1}
and continued in \chapref{Chap: P2}, by investigating the case of a Minkowski brane, namely $\Lambda=0$.
By assuming a variety of forms for the coupling function, we solve the field equations in the
bulk, and determine in an analytic way the form of the gravitational background and scalar field in
each case. The solutions are always characterized by a regular scalar field, a finite energy-momentum
tensor, and an exponentially decaying warp factor even in the absence of a negative bulk cosmological
constant. The spacetime on the brane is described by the Schwarzschild solution leading to either a
non-homogeneous black-string solution in the bulk, when the mass parameter M is non-zero, or a regular
anti-de Sitter spacetime, when $M = 0$. We construct physically-acceptable solutions by demanding in
addition a positive effective gravitational constant on our brane, a positive total energy-density for our
brane and the validity of the weak energy condition in the bulk. We find that, although the theory does
not allow for all three conditions to be simultaneously satisfied, a plethora of solutions emerge which
satisfy the first two, and most fundamental, conditions.

In \chapref{Chap: P4} we turn to braneworld black-hole solutions, and we manage to construct from first 
principles the geometry of an analytic, exponentially localized five-dimensional braneworld black hole. 
The black-hole singularity lies entirely on the 3-brane, while
the event horizon is shown to have a ``pancake'' shape. The induced line-element on the brane assumes
the form of the Schwarzschild solution while the bulk geometry is effectively AdS$_5$ outside the horizon. 
The derived geometry is supported by an anisotropic fluid in the bulk described only by two
independent components, the energy density and tangential pressure, whereas no matter needs to
be introduced on the brane for its consistent embedding in the bulk.

In \chapref{Chap: P5} we generalize the black-hole solution of the previous Chapter, by constructing a 
five-dimensional spherically-symmetric, charged and asymptotically Anti-de Sitter
black hole with its singularity being point-like and strictly localized on our brane. In addition, the induced
brane geometry is described by a Reissner-Nordstr\"{om}m-(A)dS line-element. We perform a careful classification of
the horizons, and demonstrate that all of them are exponentially localized close to the brane thus exhibiting a
``pancake'' shape. The bulk gravitational background is everywhere regular, and reduces to an AdS$_5$ spacetime right
outside the black-hole event horizon. This geometry is supported again by an anisotropic fluid with only two independent
components, the energy density $\rho_E$ and tangential pressure $p_\theta$. All energy conditions are respected close to and
on our brane, but a local violation takes place within the event horizon regime in the bulk. A tensor-vector-scalar
field-theory model is built in an attempt to realize the necessary bulk matter, however, in order to do so, both
gauge and scalar degrees of freedom need to turn phantom-like at the bulk boundary. The study of the junction
conditions reveals that no additional matter needs to be introduced on the brane for its consistent embedding
in the bulk geometry apart from its constant, positive tension. We finally compute the effective gravitational
equations on the brane, and demonstrate that the Reissner-Nordstr\"{o}m-(A)dS geometry on our brane is caused by
the combined effect of the five-dimensional geometry and bulk matter with its charge being in fact a tidal charge.

Finally, in \chapref{Chap: Concl} we conclude by reviewing our results and discussing future research projects that one might 
undertake based on our work.

\newpage



\renewcommand{\headrulewidth}{1pt}
\fancyhead{}
\fancyhead[LO]{\chapc{\slshape\nouppercase{\rightmark}}}
\fancyhead[LE]{\chapc{\slshape\nouppercase{\leftmark}}}
\fancyfoot[CE,CO]{\vspace*{-0.8em}\sectionlinetwo{chapter(color)}{88} \chapc{--\,\thepage\,--}}


{\pagestyle{empty}
\phantomsection\addcontentsline{toc}{chapter}{Contents}
\hypersetup{linktocpage}
\renewcommand\contentsname{\thispagestyle{empty}
	\chapc{\noindent{\rule{16.5cm}{0.5mm}}\\[0.5em]
	{\fontsize{30}{32} \textbf{Contents}}\\[0.25em]
	\noindent{\rule{16.5cm}{0.5mm}}}} 
\tableofcontents


\mychapterbl{\begin{center}\vspace*{-2em}Basic Notation\vspace*{-1em}\end{center}}
\phantomsection
\addcontentsline{toc}{chapter}{Basic Notation}

\thispagestyle{empty}

\begin{itemize}

\item[\chapc{$\blacktriangleright$}] In the context of this dissertation the signature for the metric tensor is chosen to be $(-,+,+,\ldots,+)$.
Consequently, a flat four-dimensional spacetime has the following line-element:
$$ds^2=-(c\,dt)^2+dx^2+dy^2+dz^2\,.$$

\item[\chapc{$\blacktriangleright$}] In cases where it is not explicitly specified the domain in which tensor indices run, we will follow the subsequent notation.

Upper-case Latin indices $M,N,\ldots$ will denote bulk coordinates. 
Thus, for an $(n+1)$-dimensional spacetime they will take the values $0,1,2,\ldots,n$. 
Greek indices $\mu,\nu,\ldots$ will be used for brane coordinates, hence, they will take the values $0,1,2,3$.
Finally, lower-case Latin indices $a,b,\ldots$ will denote the three spatial coordinates $1,2,3$.

\item[\chapc{$\blacktriangleright$}] In most cases, natural or Planck units will be used, that is $c=\hbar=1$ or $c=\hbar=G_N=1$, respectively. 

\end{itemize}



\mychapterbl{\begin{center}\vspace*{-2em}Abbreviations/\gr{Συντομογραφίες}\vspace*{-1em}\end{center}}
\phantomsection
\addcontentsline{toc}{chapter}{Abbreviations}

\thispagestyle{empty}

\vspace*{-1em}

\gr{Ελληνικές συντομογραφίες σε αλφαβητική σειρά:}\vspace*{1em}

\begin{center}
\begin{tabular}{ l l }
\gr{\hspace{-0.8em}\textbf{ΓΘΣ}} & \hspace{2.7em}\gr{\teo{Γ}ενική \teo{Θ}εωρία \teo{Σ}χετικότητας}\\[2mm]
\gr{\hspace{-0.8em}\textbf{ΚΠΣΣ}} & \hspace{2.7em}\gr{\teo{Κ}αθιερωμένο \teo{Π}ρότυπο \teo{Σ}τοιχειωδών \teo{Σ}ωματιδίων}
\end{tabular}
\end{center}

English abbreviations in alphabetical order:

\begin{center}
\begin{tabular}{ l l }
\textbf{ADD} & \hspace{2em}\teo{A}rkani-Hamed, \teo{D}imopoulos, \teo{D}vali\\[2mm]
\textbf{(A)dS} & \hspace{2em}(\teo{A}nti-)\teo{d}e \teo{S}itter\\[2mm]
\textbf{CHR} & \hspace{2em}\teo{C}hamblin \teo{H}awking \teo{R}eall\\[2mm]
\textbf{EHT} & \hspace{2em}\teo{E}vent \teo{H}orizon \teo{T}elescope\\[2mm]
\textbf{GR} & \hspace{2em}\teo{G}eneral \teo{R}elativity or \teo{G}eneral theory of \teo{R}elativity\\[2mm]
\textbf{KK} & \hspace{2em}\teo{K}aluza-\teo{K}lein\\[2mm]
\textbf{l.h.s.} & \hspace{2em}\teo{l}eft-\teo{h}and \teo{s}ide\\[2mm]
\textbf{M87} & \hspace{2em}\teo{M}essier 87\\[2mm]
\textbf{QM} & \hspace{2em}\teo{Q}uantum \teo{M}echanics\\[2mm]
\textbf{r.h.s.} & \hspace{2em}\teo{r}ight-\teo{h}and \teo{s}ide\\[2mm]
\textbf{RN} & \hspace{2em}\teo{R}eissner-\teo{N}ordstr\"{o}m\\[2mm]
\textbf{RS} & \hspace{2em}\teo{R}andall-\teo{S}undrum\\[2mm]
\textbf{S(a-)dS} & \hspace{2em}\teo{S}chwarzschild (\teo{a}nti-)\teo{d}e \teo{S}itter\\[2mm]
\textbf{SM} & \hspace{2em}\teo{S}tandard \teo{M}odel (of elementary particles)\\[2mm]
\textbf{SR} & \hspace{2em}\teo{S}pecial \teo{R}elativity\\[2mm]
\textbf{SUGRA} & \hspace{2em}\teo{Su}per\teo{gra}vity\\[2mm]
\end{tabular}
\end{center}



\mychapter{Introduction \label{Chap: intro}}
\phantomsection
\setcounter{chapter}{1}

\pagenumbering{arabic}

\epigraph{\textit{``Believe those who are seeking the truth; doubt those who find it.''}}{Andr\'{e} Gide}

\thispagestyle{empty}}

\renewcommand{\headrulewidth}{1pt}
\fancyhead{}
\fancyhead[LO]{\chapc{\slshape\nouppercase{\rightmark}}}
\fancyhead[LE]{\chapc{\slshape\nouppercase{\leftmark}}}
\fancyfoot[CE,CO]{\vspace*{-0.8em}\sectionlinetwo{chapter(color)}{88} \chapc{--\,\thepage\,--}}


{\lettrine[lines=3, lhang=0.03]{\color{chapter(color)}S}{\hspace*{5.5px}ince} the current dissertation is occupied with black-string and 
black-hole solutions in the context of brane-world models, namely gravitational models with more than four spacetime 
dimensions, it seems only reasonable to begin with a prelude to General Relativity and build the remaining concepts. 

}

\mysection{Prelude to General theory of Relativity}

In the late 19th century, the only physical theories which were available to interpret the natural world were Classical 
Mechanics, and Electromagnetism. Classical Mechanics was formulated by Isaac Newton (1643,\,Lincolnshire--1727,\,Kens\-ing\-ton), 
Joseph-Louis Lagrange (1736,\,Turin--1813,\,Paris), and William Rowan Hamilton (1805--1865,\,Dublin), and it provides a theoretical 
framework capable of describing the motion of macroscopic objects either under the influence of external forces like gravity or not. 
Electromagnetism on the other hand is the theory which describes the interaction that emerges between electrically 
charged particles. Since the electromagnetic interaction comprises electric and magnetic fields, which by their turn produce
the electromagnetic radiation, we could offhandedly say that electromagnetism constitutes the theory of light. 
It was the work of James Clerk Maxwell (1831,\,Edinburgh--1879,\,Cambridge), ``\textit{A Treatise on Electricity and 
Magnetism}'',---pub\-lished in 1873---that provided the unified picture of electromagnetism as we learn it today.
In addition to the aforementioned physical theories, the notion of \textit{luminiferous ether}, namely the propagation medium of
light, was also widely accepted at that point in time. Since all waves in Classical Mechanics propagate through a medium, it seemed
very reasonable to the physicists at the time to assume the existence of a medium for the electromagnetic waves as well.

In 1887, soon after the formulation of Electromagnetism, Albert Abraham Michelson (1852,\,Strzelno--1931,\,Pasadena) and
Edward Williams Morley (1838,\,Newark--1923,\,West Hartford)  in an attempt to measure the speed of Earth relative to 
the presumed stationary luminiferous ether, conducted the renowned Michelson-Morley experiment. However, instead of 
confirming the ether hypothesis, they showed that there is no difference in the speed of light at different directions.
Moreover, the incompatibility of the Maxwell's equations with the Galilean transformations led physicists to develop new 
transformations and investigate their physical implications. The transformations which were proven to be compatible with 
Maxwell's equations are known as Lorentz transformations and they are named after the Dutch physicist Hendrik Antoon 
Lorentz (1853,\,Arnhem--1928,\,Haarlem).

Taking into consideration the above results, Albert Einstein (1879,\,Ulm--1955,\,Princeton) published in 1905 his eminent paper
``\textit{On the electrodynamics of moving bodies}'' \cite{EinsteinSR}, where he formulated the Special theory of Relativity 
(SR) by introducing two simple postulates:

\begin{enumerate}

\item[\textbf{(i)}] \textbf{\textit{Principle of Relativity:}} \textit{No experiment can measure the absolute velocity of an
observer, and the physical laws are invariant in all inertial reference frames, i.e. frames of reference with no acceleration.}

\item[\textbf{(ii)}] \textbf{\textit{Principle of Constancy of the speed of light:}} \textit{Every ray of light in vacuum moves 
in the ``stationary coordinate system'' with the same speed $c$. This speed is independent of whether this ray of light is 
emitted by a body at rest or in motion.}

\end{enumerate} 

\noindent{It is evident that the second postulate incorporates the result of the Michelson-Morley experiment, while an 
appropriate utilization of both axioms may lead to the Lorentz transformations. Consequently, from these two very simple 
postulates, Einstein not only explained the results of the Michelson-Morley experiment, but he also gave physical meaning to 
the Lorentz transformations.}

Special Relativity has a huge impact on the way we perceive nature today. Although it corrected the kinematics of Newtonian 
mechanics for velocities comparable with the speed of light, and provided the most famous equation in all physics ($E=mc^2$), 
in its core Special Relativity is a theory of spacetime. It elucidates that in high speeds (or high energies) space 
and time cannot be considered as two independent physical quantities, but they constitute a single unified entity called 
spacetime. It was indeed this particular realization, and its geometrical manifestation, that allowed Einstein to extend Special 
Relativity, a theory which involves flat spacetime geometries, to General Relativity (GR) which constitutes a theory of gravity. 
General Relativity not only permits curved spacetime geometries, but also gives physical meaning to the notion of spacetime by 
identifying its curvature with the gravitational field. 

In the following section we will present the basic mathematical elements of General Relativity. This set of mathematical tools
and notions constitutes the basis of all modern theories of gravity and will accompany us throughout this dissertation.

\mysection{Basic mathematical notions of General Relativity}

As we discussed earlier, General Relativity \cite{einstein1,einstein2,einstein3} is a theory of gravity which was formulated by 
Albert Einstein in 1915. Its mathematical foundations lie entirely on the Riemannian geometry with the only difference being
that GR involves pseudo-Riemannian manifolds, instead of the ordinary Riemannian ones. From a physical point of view though,
this peculiar prefix ``pseudo'' is what creates all the magic in GR; black hole solutions for example would not have been 
possible without this prefix. Let us now clarify the difference between Riemannian and pseudo-Riemannian manifolds.
\footnotetext[1]{In mathematics, an immersion is a differentiable function between differentiable manifolds whose derivative 
is everywhere injective. Strictly speaking, $f : M \ra N$ is an immersion if $D_p f : T_p M \ra T_{f(p)} N$ is an injective (or 
one-to-one) function at every point $p$ on $M$.}

\begin{figure}[H]
\centering
\includegraphics[width=0.6\textwidth]{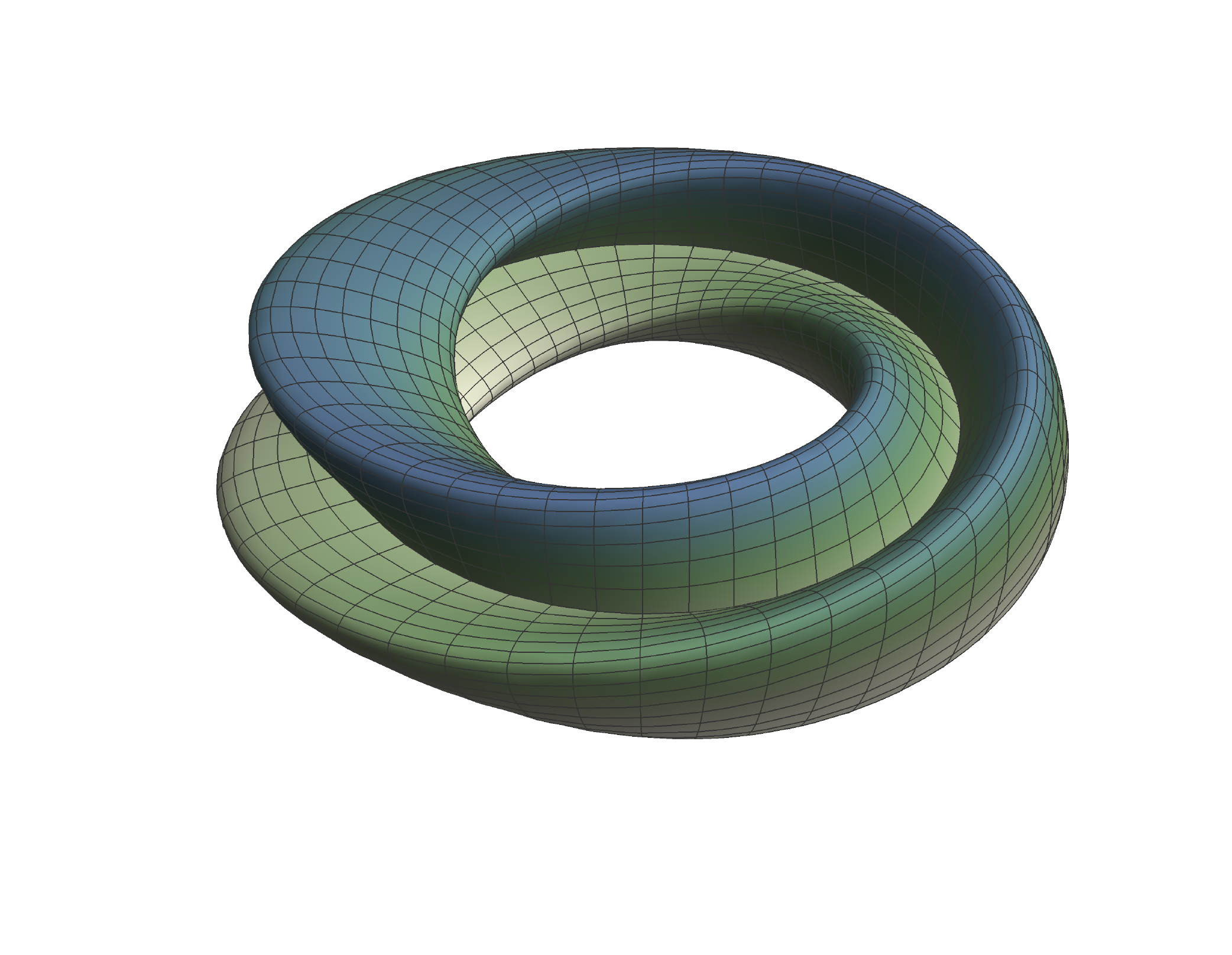}
\vspace*{-1cm}
\caption[]{The ``figure 8'' or ``bagel'' immersion\,\footnotemark \ of the Klein bottle.}
\label{intro-fig: bagel}
\end{figure}

\textit{An $\boldsymbol{n}$\textbf{-dimensional Riemannian manifold} is a topological space with a globally defined differential structure 
supplemented 
with the property that at each one of its points there is a neighborhood which is homeomorphic to an open subset of $n$-dimensional 
Euclidean space, whilst it is also equipped with a positive-definite inner product.} A very interesting example of a $2$-dimensional Riemannian 
manifold (or surface) is depicted in \fref{intro-fig: bagel}. The illustrated ``figure 8'' immersion of the Klein bottle results from a 
particular parametrization of the well-known Klein bottle geometry, while its name follows from the method of its construction. 
One may obtain the surface in \fref{intro-fig: bagel} by glueing the ends of a hollow ``figure-8'' cylinder after performing a half-twist 
(a twist of 180 degrees) on one of its ends. 

Similarly to the preceding definition, \textit{an $\boldsymbol{(n+1)}$\textbf{-dimensional pseudo-Riemannian manifold} ($n$ spatial and one 
temporal dimensions) 
is locally homeomorphic to an $(n+1)$-dimensional Minkowski spacetime, and it is also equipped with a non-degenerate metric tensor.}
The metric tensor (or simply ``metric'') is the most valuable mathematical object when it comes to the study of curved spacetimes.
The metric tensor is symmetric and nondegenerate while its covariant components are symbolized as $g_{\mu\nu}$. In terms of its
components, the two aforementioned properties of the metric tensor are translated as $g_{\mu\nu}=g_{\nu\mu}$ and $\det(g_{\mu\nu})\neq 0$, 
respectively.

Tensors have a pivotal role in General Relativity. We already mentioned the importance of the metric tensor, but even the gravitational field 
equations in the context of GR are given in tensorial form. Therefore, in an attempt to be self-contained, we will 
discuss briefly but with the right amount of mathematical rigor the notions of vectors, one-forms, and tensors. However, before we proceed
it is necessary to introduce and define some preliminary concepts, which will prove essential for the understanding of how a coordinate system 
transformation alters the components of a tensor.

\mysubsection{Coordinate system and coordinate transformation}

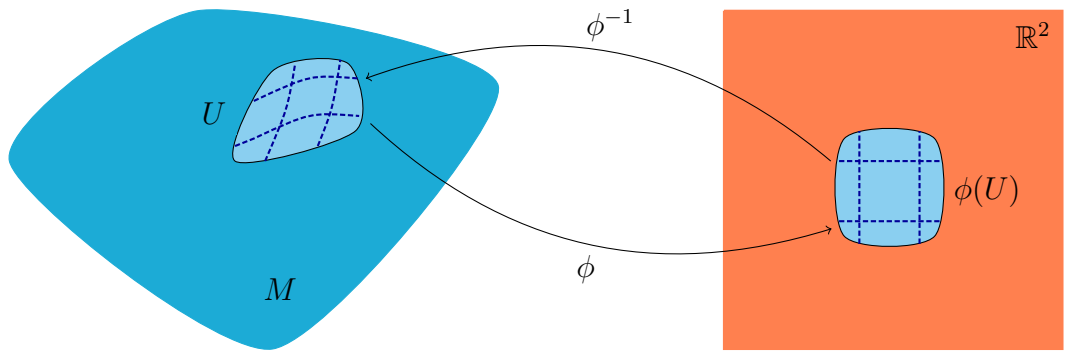
\begin{figure}[b]

\centering

\begin{tikzpicture}

    \filldraw[smooth cycle, tension=0.4, color=brightcerulean] 
    plot coordinates{(2,2) (-0.5,0) (3,-2.5) (6,1)} node at (3.1,-1.7) {$\textcolor{black}{M}$};
	
    \draw[smooth cycle, fill=babyblue]
     plot coordinates {(2.5,0) (3, 1.2) (4,1.3) (4.1, 0.4)} 
     node [label={[label distance=-0.3cm, xshift=-2.08cm, fill=brightcerulean]:$\textcolor{black}{U}$}] at (4.3,0.5)  {};	
	
	\filldraw[thick,color=coral] (9,2) -- ++(4.5,0) node [below left] {$\textcolor{black}{\reals^2}$}--++(0,-4.5) --++(-4.5,0) --++(0,4.5);   
    
    \draw[smooth cycle, fill=babyblue] plot coordinates{(-2+12.6, -4.5+3.5) (-2+12.6, -3.2+3.5) (-0.8+12.6, -3.2+3.5) (-0.8+12.6, -4.5+3.5)}
    node [label={[label distance=0cm, xshift=2.1cm, yshift=-1.4cm]:$\textcolor{black}{\phi(U)}$}] at (10.8-0.4,0.5) (n2) {};
    
    \draw[color=DarkBlue, dash pattern=on 2pt off 1pt, thick] (2.9,0) .. controls (3.2,0.7) .. (3.3,1.3);
    \draw[color=DarkBlue, dash pattern=on 2pt off 1pt, thick] (3.6,0.2) .. controls (3.8,0.7) .. (3.9,1.35);
    \draw[color=DarkBlue, dash pattern=on 2pt off 1pt, thick] (2.75,0.8) .. controls (3.5,1.15) .. (4.13,1.1);
    \draw[color=DarkBlue, dash pattern=on 2pt off 1pt, thick] (2.5,0.2) -- (2.75,0.8-0.5) .. controls (3.5,1.15-0.5) .. (4.15,1.1-0.5);
    
	\draw[color=DarkBlue, dash pattern=on 2pt off 1pt, thick] (10.8,-1.1) -- (10.8,0.4); 
    \draw[color=DarkBlue, dash pattern=on 2pt off 1pt, thick] (11.6,-1.1) -- (11.6,0.4);
    \draw[color=DarkBlue, dash pattern=on 2pt off 1pt, thick] (10.53,-0.8) -- (11.88,-0.8);
    \draw[color=DarkBlue, dash pattern=on 2pt off 1pt, thick] (10.53,-0) -- (11.9,-0.);
    
    \path[->] (4.3,0.5) edge [bend right] node [below, yshift=0mm] {$\phi$} (10.43,-0.9) ;
    \path[->] (10.42,-0) edge [bend right] node [above, yshift=0mm] {$\phi^{-1}$} (4.24,1.1);
	
\end{tikzpicture}

\caption{The mapping of an open region $U$ of a $2$-dimensional manifold $M$ to $\reals^2$.}
\label{intro-fig: map}

\end{figure}

Consider an $n$-dimensional Riemannian manifold $M$. According to the definition we gave earlier, for an arbitrary open region 
$U$ of the manifold, it will exist an one-to-one mapping of the form $\phi: U\ra \reals^n$. A mapping of this kind is called a
\textbf{\textit{coordinate system}}, while $U$ is the coordinate region of $M$. Therefore, a set of maps $\{x^1,\ldots,x^n\}$ constitutes a
coordinate system. The coordinate system represents any point $P$ of $U$ via $n$-tuples $(x^1,\ldots,x^n)$. In \fref{intro-fig: map}
below we present a two-dimensional example of the concepts that we just discussed.


Imagine now that two regions $U$ and $V$ of the manifold $M$ overlap at some point, hence it holds that $U\cap V\neq \emptyset$. 
Assuming that the coordinate systems of $U$ and $V$ are $\{x^\mu\}$ and $\{x'^\mu\}$ ($\mu=1,\ldots,n$), respectively, 
then in the region $U\cap V$ the coordinate systems are related with each other in the following way
\eq$\label{intro-eq: coord-trans}
x'^\mu=x'^\mu(x^\lam)\,.$
This particular relation constitutes a \textbf{\textit{coordinate transformation}}. Note that the functions $x'^\mu$ must be invertible. The condition of
invertibility can be checked using the Jacobian determinant. Hence, a valid coordinate transformation should satisfy the relation
\eq$\label{intro-eq: Jacob}
\det \left(\frac{\pa x'^\mu}{\pa x^\nu}\right)\equiv\left|\begin{array}{ccc}
\frac{\pa x'^1}{\pa x^1} & \cdots & \frac{\pa x'^1}{\pa x^n}\\[1mm]
\vdots & \ddots &\vdots\\[1mm]
\frac{\pa x'^n}{\pa x^1} & \cdots & \frac{\pa x'^n}{\pa x^n}
\end{array}\right|\neq 0\,.$
 If, in addition to the above, the functions $x'^\mu$ are also of class $C^\infty$, then we say that $x'^\mu$ are \textit{smooth functions}. 
 In the same sense, \textit{smooth manifolds} are called the manifolds which possess smooth coordinate mappings.

\mysubsection{Tangent space and holonomic basis
\label{intro-subsec: tang}}

\vspace*{-0.1em}

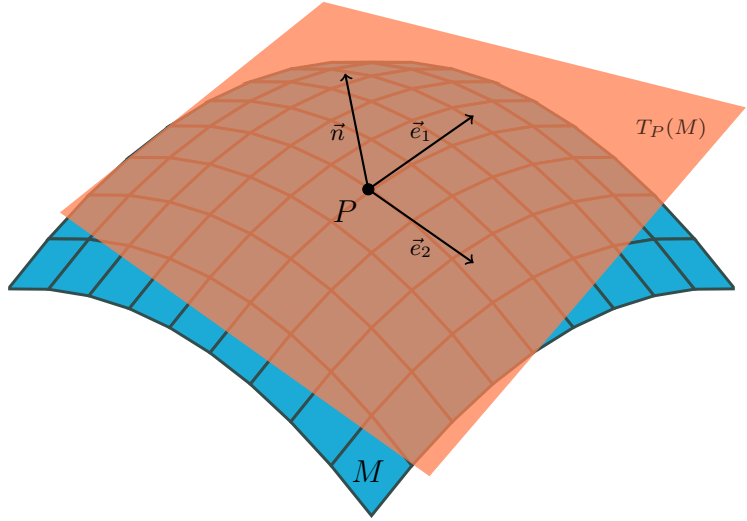
\begin{figure}[t]

\centering

\begin{tikzpicture}[scale=1.4]
	
	\begin{axis}[hide axis,
		colormap={bw}{color=(brightcerulean) color=(brightcerulean)}, 
		view={45}{65}]
		\addplot3[surf,
			faceted color = mygray,
			thick, 
			samples=10] {-x^2-y^2};
	\end{axis}
	
	\filldraw[brightcerulean] (3.42,0.33) circle (1pt)
	node[above] {$\textcolor{black}{M}$};
	
	\filldraw[tension=0.4, color=coral, opacity=0.75] 
    plot coordinates{(2+1,2+3) (-0.5+1,0+3) (3+1,-2.5+3) (6+1,1+3)} 
    node at (3.1+3.2,-1.7+5.5) {\fontsize{8}{8}{$\textcolor{black}{T_P(M)}$}};
    
	\filldraw[black] (3.42,3.22) circle (1.5pt)
	node[below left]{$\textcolor{black}{P}$};
	
	\draw[thick, black,->] (3.42,3.22) -- (4.42,3.92)
	node[above,midway]{\fontsize{9}{9}{$\vec{e}_1$}};
	
	\draw[thick, black,->] (3.42,3.22) -- (4.42,2.52)
	node[below,midway]{\fontsize{9}{9}{$\vec{e}_2$}};
	
	\draw[thick, black,->] (3.42,3.22) -- (3.2,4.32)
	node[left,midway]{\fontsize{9}{9}{$\vec{n}$}};
	

\end{tikzpicture}

\caption{A $2$-dimensional manifold $M$ and its tangent plane $T_P(M)$ at the point $P$.}
\label{intro-fig: tang-pl}

\end{figure}

Given an $n$-dimensional manifold $M$ which is embedded in an $(n+1)$-dimensional Euclidean space, we can define at an arbitrary 
point $P$ of $M$ an $n$-dimensional tangent space\,\footnote{The tangent space generalizes the notion of tangent plane in cases where the
dimensionality of manifolds exceeds $2$.}  $T_P(M)$ by using the linearly independent vectors $\vec{e}_\mu$ which are 
defined as
\eq$\label{intro-eq: hol-vecs}
\vec{e}_\mu=\frac{\pa \vec{r}}{\pa x^\mu}\,, \hspace{1em} \mu\in\{1,\ldots,n\}\,.$
In the above, the set $\{x^\mu\}$ is the coordinate system, the vector $\vec{r}=\vec{r}(x^\mu)$ denotes the \textit{canonical parametric 
representation} of the manifold, while the vectors $\{\vec{e}_\mu\}$ constitute at each point $P$ a \textbf{\textit{holonomic basis (or coordinate
basis)}} on the tangent space. In \fref{intro-fig: tang-pl} (next page) it is depicted a $2$-dimensional Riemannian manifold with its tangent plane 
and its holonomic basis 
$\{\vec{e}_1,\vec{e}_2\}$ at an arbitrary point $P$. Note that the vector $\vec{n}$ is the normal vector of the plane.
The important difference of a holonomic basis compared to any other vector basis on the tangent space is that the vectors of a 
non-holonomic basis cannot be expressed in the form of eq.\,\eqref{intro-eq: hol-vecs}.


Let us now examine how two holonomic bases are related to each other under a coordinate transformation. To this end, we consider the
coordinate systems $\{x^\mu\}$ and $\{x'^\mu\}$ which by their turn lead to the holonomic bases $\{\vec{e}_\mu=\frac{\pa \vec{r}}{\pa 
x^\mu}\}$ and  $\{\vecpr{e}{\mu}=\frac{\pa \vec{r}}{\pa x'^\mu}\}$, respectively. Using eq.\,\eqref{intro-eq: hol-vecs}
and the chain rule for derivatives, its is straightforward to calculate that
\eq$\label{intro-eq: base-trans1}
\vec{e}_\mu=\frac{\pa x'^\nu}{\pa x^\mu}\vecpr{e}{\nu}\,,$
with its inverse being
\eq$\label{intro-eq: base-trans2}
\vecpr{e}{\mu}=\frac{\pa x^\nu}{\pa x'^\mu}\vec{e}_{\nu}\,.$

All the concepts which were considered previously can be applied without any modification to pseudo-Riemannian manifolds as well. 
Eqs.\,\eqref{intro-eq: coord-trans}-\eqref{intro-eq: base-trans2} will continue to hold, with the only difference being the running of the indices.
In the context of GR, the indices run from zero, with the zeroth index denoting the temporal coordinate.

\vspace*{-0.1em}

\mysubsection{Vectors}

\textit{A \textbf{vector} is a quantity that has both magnitude and direction, and is usually denoted by a letter 
with an arrow above it, e.g. $\vec{v}$.} In an $(n+1)$-dimensional spacetime, if $\alpha^\mu$ are real numbers and $\vec{e}_\mu$ are 
vectors, with $\mu=0,1,\ldots, n$, then the expression $\alpha^\mu \vec{e}_\mu$ constitutes a 
linear combination of the $\vec{e}_\mu$ vectors. 

\textit{A \textbf{basis} $B$ of an $(n+1)$-dimensional vector space $V$ over $\reals$  (real numbers) is  a linearly independent subset of 
$V$ that spans $V$.} In other words, a basis $B$ of $V$ satisfies the following conditions:

\begin{itemize}

\item[\chapc{$\blacktriangleright$}] \textit{Linear independence condition:} For every subset $\{\vec{e}_\nu\}$ of $B$, with cardinality 
$|B|\leq n+1$, if 
$\alpha^\nu\,\vec{e}_\nu=0$ for some numbers $\alpha^\nu$ in $\reals$, then it holds that each one of the numbers $\alpha^\nu$ 
should be identically zero; $\alpha^\nu=0,\,\forall\, \nu$.

\item[\chapc{$\blacktriangleright$}] \textit{The spanning condition:} Every vector $\vec{v}$ in $V$ can be represented as a linear combination 
of the basis
vectors $\{\vec{e}_\mu\} $ of $B$, $\mu\in\{0,1,\ldots,n\}$. This means that $\vec{v}= v^\mu\, \vec{e}_\mu$, for some numbers 
$v^\mu \in \reals$. The numbers $v^\mu$ are called \textbf{\textit{components}} of the vector $\vec{v}$ relative to the basis $B$.\\[-1.5cm]

\end{itemize}

\mysubsection{One-forms/dual vectors}

\textit{A \textbf{one-form (or dual vector)} is a linear function which maps vectors into real numbers.} One-forms will be denoted by letters with a 
left arrow above them, e.g. $\accentset{\leftarrow}{\alp}$. In a sense this notation is self-evident, since the action of objects with left 
arrows on objects with right arrows results to objects with no arrows, i.e. real numbers.  Given two one-forms $\form{\alpha}$ and 
$\form{\beta}$ it holds that
\gat$\label{intro-eq: form1}
\form{\alpha}(b\,\vec{u}+c\,\vec{v})=b\,\form{\alp}(\vec{u})+c\,\form{\alp}(\vec{v})\in\reals\,,\\[2mm]
\label{intro-eq: form2}
(\form{\alpha}+\form{\beta})(\vec{v})=\form{\alpha}(\vec{v})+\form{\beta}(\vec{v})\in\reals\,,\\[2mm]
\label{intro-eq: form3}
(b\,\form{\alpha})(\vec{v})=b[\form{\alpha}(\vec{v})]\in\reals\,,$
with $b,c\in\reals$ and $\vec{u},\vec{v}\in V$, where $V$ is a vector space.  

Using the basis vectors $\{\vec{e}_\nu\}$ of the basis $B$, one may also define a basis $\bar{B}$ for the one-form space $\bar{V}$
 via the relation
\eq$\label{intro-eq: form-basis}
\form{\omega}^\mu(\vec{e}_\nu)=\del^{\mu}{}_\nu\,.$
In the above equation, $\{\form{\omega}^\mu\}$ are the one-forms of the basis, while $\del^{\mu}{}_\nu$ is the 
Kronecker-delta symbol. An arbitrary one-form $\form{\alp}$ can be expressed as a linear combination 
of the one-forms $\{\form{\omega}^\mu\}$ of the basis $\bar{B}$ in the following way
\eq$\label{intro-eq: form-comp}
\form{\alp}=\alp_\mu\,\form{\omega}^\mu\,.$
In this case, we say that the numbers $\alp_\mu$ are the \textbf{\textit{components}} of $\form{\alp}$ relative to the basis $\bar{B}$. 

Just like dual vectors, vectors can also be perceived as linear functions of one-forms. Hence, given a one-form $\form{\alp}$ 
and a vector $\vec{v}$, the functions $\form{\alp}(\vec{v})$ and $\vec{v}(\form{\alp})$ should result to the same number.
Using the above properties, we can calculate that
\gat$\label{intro-eq: form-comp2}
\form{\alp}(\vec{e}_\mu)=\alp_\nu\,\form{\omega}^\nu(\vec{e}_\mu)=\alp_\mu\,,\\[2mm]
\label{intro-eq: vec-comp}
\vec{v}(\form{\omega}^\mu)=v^\nu\,\vec{e}_\nu(\form{\omega}^\mu)=v^\mu\,,\\[2mm]
\label{intro-eq: form-contr}
\vec{v}(\form{\alp})=\form{\alp}(\vec{v})=\alp_\mu\,\form{\omega}^\mu(v^\nu\,\vec{e}_\nu)=\alp_\mu v^\nu\,
\form{\omega}^\mu(\vec{e}_\nu)=\alp_\mu v^\mu\,.$
In eqs.\,\,(\ref{intro-eq: form-comp2})\,\,and\,\,\eqref{intro-eq: vec-comp} we see the underlying connection between the components of 
one-forms and vectors with the bases $B$ and $\bar{B}$, respectively. In addition, the operation in eq.\,\eqref{intro-eq: form-contr} is 
called \textbf{\textit{contraction}}. Contraction will prove a very powerful concept later on, since it allows one to create scalar quantities by suitably
combining objects with indices. Scalars constitute useful quantities in physics due to their innate property to remain invariant under 
coordinate transformations.

\newpage
\vspace*{-2cm}

\mysubsection{Tensors}

A generalization of the previous concepts leads us to tensors. \textit{A \textbf{tensor} is a multi-linear function---linear in all its 
arguments---which maps both vectors and one-forms into real numbers.} Obviously, if a tensor maps a single one-form into real
numbers then it is a vector, and vice versa. Given a tensor $\tensor{T}$ which maps a one-form and a vector, the multi-linear
property leads to
\eq$\label{intro-eq: mult-lin}
\tensor{T}(a\form{\eta}+b\form{\theta};c\vec{u}+d\vec{v})=ac\tensor{T}(\form{\eta};\vec{u})+ad\tensor{T}(\form{\eta};\vec{v})
+bc\tensor{T}(\form{\theta};\vec{u})+bd\tensor{T}(\form{\theta};\vec{v})\in\reals\,,$
where $a,b,c,d\in\reals$. Depending on the number of vectors and one-forms
that a tensor takes as arguments we can identify the following categories:
\begin{itemize}

\item[\chapc{$\blacktriangleright$}] \textit{\textbf{Covariant tensors:}} Covariant tensors map only vectors.

\item[\chapc{$\blacktriangleright$}] \textit{\textbf{Contravariant tensors:}} Contravariant tensors map only one-forms.

\item[\chapc{$\blacktriangleright$}] \textit{\textbf{Mixed tensors:}} Mixed tensors map both vectors and one-forms.

\end{itemize}
The \textbf{\textit{rank}} of a tensor denotes explicitly the number of one-forms and vectors that the tensor maps into real numbers.  
We say that a tensor $\tensor{T}$ is of rank $\trank{n}{m}$ if it maps $n$ one-forms and $m$ vectors, namely it holds that
$\tensor{T}(\form{\alp}_1,\ldots,\form{\alp}_n;\vec{v}_1,\ldots,\vec{v}_m)\in\reals$. 

Considering now two tensors $\tensor{T}$ and $\tensor{S}$ of rank $\trank{0}{n}$ and $\trank{0}{m}$, respectively, the 
\textbf{\textit{tensor product}} between them is denoted by $\otimes$ and is defined as
\eq$\label{intro-eq: tprod}
\tensor{T}\otimes\tensor{S}(\vec{v}_1,\ldots,\vec{v}_n,\vec{u}_1,\ldots,\vec{u}_m)=\tensor{T}(\vec{v}_1,\ldots,\vec{v}_n)\,
\tensor{S}(\vec{u}_1,\ldots,\vec{u}_m)\in\reals\,.$
Notice in the previous example that the tensor product creates a new tensor of rank $\trank{0}{n+m}$. In general, the tensor product between
two tensors of rank $\trank{n}{m}$ and $\trank{k}{l}$ results to a tensor of rank $\trank{n+k}{m+l}$. 
It is also important to note that the tensor product is \textit{bilinear}, i.e.
\eq$\label{intro-eq: tprod2}
(a\tensor{S}+b\tensor{T})\otimes(c\,\tensor{U}+d\tensor{V})=ac\,\tensor{S}\otimes\tensor{U}+ad\,\tensor{S}\otimes\tensor{V}
+bc\,\tensor{T}\otimes\tensor{U}+bd\,\tensor{T}\otimes\tensor{V}\,,$
but it is \textit{not commutative}. To elucidate how commutativity breaks down, we may use as a point of reference the tensors in 
eq.\,\eqref{intro-eq: tprod}. Without loss of generality we may also assume that $n>m$, and we obtain
\eq$\label{intro-eq: tprod3}
\tensor{S}\otimes\tensor{T}(\vec{v}_1,\ldots,\vec{v}_n,\vec{u}_1,\ldots,\vec{u}_m)=
\tensor{S}(\vec{v}_1,\ldots,\vec{v}_m)\tensor{T}(\vec{v}_{m+1},\ldots,\vec{v}_n,\vec{u}_1,\ldots,\vec{u}_m)\,.$
Comparing now the right-hand sides (r.h.s.) of eqs.\,\,\eqref{intro-eq: tprod}\,\,and\,\,\eqref{intro-eq: tprod3} it becomes obvious that 
$\tensor{T}\otimes\tensor{S}\neq \tensor{S}\otimes\tensor{T}$.

As we discussed a couple of lines earlier, the tensor product between two tensors increases the rank of the ensuing tensor. 
Consequently, by exploiting this property one may construct a basis for tensors by combining in an appropriate way basis one-forms 
and basis vectors. For example, the basis of covariant tensors of rank $p$ is expected to be composed of a linearly independent set of basis one-forms, i.e. $\{\form{\omega}^{\mu_1},\ldots,
\form{\omega}^{\mu_p}\}$. Thus, a covariant tensor $\tensor{S}$ of rank $p$ can be expressed as follows
\eq$\label{intro-eq: tcomp}
\tensor{S}=S_{\mu_1\ldots\mu_p}\,\form{\omega}^{\mu_1}\otimes\cdots\otimes\form{\omega}^{\mu_p}\,.$
The numbers $S_{\mu_1\ldots\mu_p}$ are the \textbf{\textit{components of the tensor}} relative to the specific basis and they are
defined through the relation
\eq$\label{intro-eq: cov-comp}
S_{\mu_1\ldots\mu_p}\equiv \tensor{S}(\vec{e}_{\mu_1},\ldots,\vec{e}_{\mu_p})\,.$
Following a similar reasoning, it is not hard to deduce that a contravariant tensor $\tensor{W}$ of rank $p$ will be expressed as
\eq$\label{intro-eq: contr-comp}
\tensor{W}=W^{\mu_1\ldots\mu_p}\,\vec{e}_{\mu_1}\otimes\cdots\otimes\vec{e}_{\mu_p}\,,$
while a mixed tensor $\tensor{M}$ of rank $\trank{p}{q}$ will be expressed as
\eq$\label{intro-eq: mix-comp}
\tensor{M}=M^{\mu_1\ldots\mu_p}{}_{\nu_1\ldots\nu_q}\,\vec{e}_{\mu_1}\otimes\cdots\otimes\vec{e}_{\mu_p}\otimes
\form{\omega}^{\nu_1}\otimes\cdots\otimes\form{\omega}^{\nu_q}\,.$
In this case, the components of the mixed tensor $\tensor{M}$ are defined through the relation
\eq$\label{intro-eq: mix-comp2}
M^{\mu_1\ldots\mu_p}{}_{\nu_1\ldots\nu_q}\equiv\tensor{M}(\form{\omega}^{\mu_1},\ldots,\form{\omega}^{\mu_p};
\vec{e}_{\nu_1},\ldots,\vec{e}_{\nu_q})\,.$

\vspace*{-2em}

\mysubsection{Vectors, one-forms and tensors revisited}

In the subsection \ref{intro-subsec: tang} we defined the notion of holonomic basis on the tangent space, and we proved that under a coordinate system transformation 
the holonomic bases are related to each other via eqs.\,\,\eqref{intro-eq: base-trans1}\,\,and\,\,\eqref{intro-eq: base-trans2}. 
It is now reasonable to wonder if vectors, one-forms or tensors, defined on the tangent space, are affected by coordinate system transformations. 

The answer to the preceding question resides exclusively in the definitions of these quantities, and as one can directly deduce, a coordinate
transformation does not have an impact on them. Although coordinate systems help us describe objects like vectors, one-forms and tensors,
the particular choice of a coordinate system does not affect by any means the characteristics of these objects. Therefore, \textit{\textbf{vectors, 
one-forms and tensors remain invariant under coordinate transformations}}. However, since the basis relative to which they are expressed changes 
under a coordinate transformation, their components will change accordingly in order the final result to remain the same.

Let us now examine in more detail the above inferences. Without loss of generality we will assume that the original coordinate system is $\{x^\mu\}$ 
while $\{x'^\mu\}$ is the one after the transformation. Consequently, the holonomic bases which accompany the coordinate systems on the tangent 
space will be 
$\{\vec{e}_\mu\}$ and $\{\vecpr{e}{\mu}\}$, respectively. 

For a vector $\vec{A}$ on the tangent space it holds that
\eq$\label{intro-eq: vec-inv}
\vec{A}=A^\nu\,\vec{e}_\nu=A'^\mu\,\vecpr{e}{\mu}\,.$
With the use of eq.\,\,\eqref{intro-eq: base-trans1} in \eqref{intro-eq: vec-inv} we obtain
\eq$\label{intro-eq: vec-trans1}
A'^\mu=\frac{\pa x'^\mu}{\pa x^\nu}A^\nu\,,$
while the inverse transformation ensues from the combination of eqs.\,\,\eqref{intro-eq: base-trans2} and \eqref{intro-eq: vec-inv},
\eq$\label{intro-eq: vec-trans2}
A^\nu=\frac{\pa x^\nu}{\pa x'^\mu}A'^\mu\,.$
Equations \eqref{intro-eq: vec-trans1} and \eqref{intro-eq: vec-trans2} describe how the components of a vector on the tangent space change
due to a coordinate transformation.

To determine how the components of a one-form $\form{\alp}$ vary under the same transformation we make use of 
eqs.\,\,\eqref{intro-eq: base-trans2},\,\,\eqref{intro-eq: form-comp2}, and we get
\eq$\label{intro-eq: form-trans1}
\alp'_\mu\equiv\form{\alp}(\vecpr{e}{\mu})=\form{\alp}\left(\frac{\pa x^\nu}{\pa x'^\mu}\vec{e}_\nu\right)=
\frac{\pa x^\nu}{\pa x'^\mu}\,\form{\alp}(\vec{e}_\nu)=\frac{\pa x^\nu}{\pa x'^\mu}\,\alp_\nu\,.$
In the same way, by using \eqref{intro-eq: base-trans1} instead of \eqref{intro-eq: base-trans2} one can identify the inverse 
transformation which is given below
\eq$\label{intro-eq: form-trans2}
\alp_\mu=\frac{\pa x'^\nu}{\pa x^\mu}\,\alp'_\nu\,.$
From the above relations and the fact that one-forms are invariant under coordinate transformations, i.e.
\eq$\label{intro-eq: form-inv}
\form{\alp}=\alp_\mu\,\form{\omega}^\mu=\alp'_\nu\,\formpr{\omega}{\nu}\,,$
it is straightforward to obtain the transformation relations between the bases $\{\form{\omega}^\mu\}$ and 
$\{\formpr{\omega}{\mu}\}$. Hence, we are led to
\eq$\label{intro-eq: form-base-trans1}
\formpr{\omega}{\mu}=\frac{\pa x'^\mu}{\pa x^\nu}\,\form{\omega}^\nu\,, \hspace{2em} 
\form{\omega}^{\mu}=\frac{\pa x^\mu}{\pa x'^\nu}\,\formpr{\omega}{\nu}\,.$

Finally, having in our disposal the transformation relations for both vector and one-form bases, we are now able to determine
the transformation relations for the components of an arbitrary tensor. Thus, given a tensor $\tensor{S}$ of rank $\trank{p}{q}$, 
we may use eqs.\,\,\eqref{intro-eq: base-trans2}\,\,and\,\,\eqref{intro-eq: form-trans1} in \eqref{intro-eq: mix-comp2} to obtain
\eq$\label{intro-eq: ten-inv}
S'^{\mu_1\ldots\mu_p}{}_{\nu_1\ldots\nu_q}=\frac{\pa x'^{\mu_1}}{\pa x^{\alp_1}}\cdots\frac{\pa x'^{\mu_p}}{\pa x^{\alp_p}}
\frac{\pa x^{\beta_1}}{\pa x'^{\nu_1}}\cdots\frac{\pa x^{\beta_q}}{\pa x'^{\nu_q}}S^{\alp_1\ldots\alp_p}{}_{\beta_1\ldots\beta_q}
\,.$ 
The inverse transformation can be derived from eq.\,\,\eqref{intro-eq: ten-inv} without additional calculations; one simply needs to transfer 
the primes to the unprimed quantities.

\vspace*{-0.5em}

\mysubsection{First fundamental form and the metric tensor}

\vspace*{-0.5em}

As stated earlier, the importance of the metric tensor is absolutely fundamental when it comes to the study of curved spacetime geometries in 
the context of GR. The metric tensor provides all the necessary information for the measurement of lengths and time intervals on 
the spacetime manifold. It is essential to understand that a metric tensor literally grants structure to the manifold. As we will see in a subsequent
section, the metric tensor defines explicitly the curvature of the manifold.

\textit{The \textbf{metric tensor (or metric), g,} is a covariant tensor of rank 2 with its components being defined via}
\eq$\label{intro-eq: metr}
\text{\textit{\textbf{g}}}(\vec{e}_\mu,\vec{e}_\nu)=g_{\mu\nu}\equiv \vec{e}_\mu\cdot\vec{e}_\nu\,,$
\textit{where $\{\vec{e}_\mu\}$ are the vectors of a holonomic basis.}
In the context of this dissertation, we will use the signature $(-+\ldots+)$ for the components of the metric tensor. 
This means that in the trivial case of a flat spacetime ($g_{\mu\nu}\ra\eta_{\mu\nu}$), one gets $\eta_{\mu\nu}=\text{diag}\{-1,+1,\ldots,+1\}$.
The metric tensor has the following properties:

\begin{itemize}

\item[\chapc{$\blacktriangleright$}] It is \textit{symmetric}. This can be instantly deduced from eq.\,\eqref{intro-eq: metr}.

\item[\chapc{$\blacktriangleright$}] It is \textit{non-degenerate, i.e. $\det(g_{\mu\nu})\neq 0$}. Interpreting the components 
$g_{\mu\nu}$ as elements of a $2\times 2$ matrix, this property implies that the inverse matrix $g^{\mu\nu}$ will also exist.
Consequently, from the covariant components of a metric tensor one may also determine the contravariant components $g^{\alp\beta}$ 
through the relation
\eq$\label{intro-eq: metr-comp}
g_{\nu\lam}g^{\lam\mu}=g^{\mu\lam}g_{\lam\nu}=\del^{\mu}{}_\nu\,.$

\item[\chapc{$\blacktriangleright$}] \textit{The covariant components of the metric tensor ($g_{\mu\nu}$) have the power to map a tensor 
of rank 
$\trank{p}{q}$ to a tensor of rank $\trank{p-1}{q+1}$, while its contravariant components ($g^{\mu\nu}$) can map a tensor of rank 
$\trank{p}{q}$ to a $\trank{p+1}{q-1}$ tensor.} Consider for example the components $S_{\alp\beta}{}^\gam$ of a $\trank{1}{2}$
tensor. According to the preceding assertion we may write
\gat$\label{intro-eq: rais-low1}
S^{\mu}{}_{\beta}{}^\gam=g^{\mu\alp}S_{\alp\beta}{}^\gam\,,\\[1mm]
\label{intro-eq: rais-low2}
S_{\alp}{}^{\nu\gam}=g^{\nu\beta}S_{\alp\beta}{}^\gam\,,\\[1mm]
\label{intro-eq: rais-low3}
S_{\alp\beta\lam}=g_{\lam\gam}S_{\alp\beta}{}^\gam\,.$
Combining now this property with eq.\,\eqref{intro-eq: metr-comp} it is straightforward to infer that $g^{\mu}{}_\nu=\del^{\mu}{}_\nu\,.$

\end{itemize}

Let us now define the first fundamental form (or line-element). To this end, we assume an $(n+1)$-dimensional spacetime 
manifold  with canonical parametric representation 
$\vec{r}=\vec{r}(x^\mu)$ ($\mu=0,1,\ldots,n$). The vector $d\vec{r}$, at an arbitrary point $P$ of the manifold, is given by
\eq$\label{intro-eq: dr}
d\vec{r}=\frac{\pa \vec{r}}{\pa x^\mu}dx^\mu= dx^\mu\,\vec{e}_\mu\,,$ 
and belongs to the tangent $(n+1)$-dimensional Minkowski spacetime at $P$. In the above, the vectors $\{\vec{e}_\mu\}$ 
constitute the holonomic basis on the tangent spacetime.

\textit{The \textbf{first fundamental form (or line-element)} is a scalar quantity which is denoted as $ds^2$ and 
is defined through the relation}
\eq$\label{intro-eq: fff}
ds^2\equiv d\vec{r}\cdot d\vec{r}=(\vec{e}_\mu\cdot\vec{e}_\nu)\,dx^\mu dx^\nu=g_{\mu\nu}\,dx^\mu dx^\nu\,.$
In respect of the line-element, the difference between Riemannian and pseudo-Riemannian manifolds becomes apparent. The line-element
of Riemannian manifolds is everywhere positive-definite, while in pseudo-Riemannian manifolds one may obtain either
$ds^2>0$ or $ds^2\leq 0$. In fact, \textit{according to our convention for the signature of metric tensor, the physically acceptable
line-elements in the context of GR are those with $ds^2\leq 0$.}

\vspace*{-1em}

\mysubsection{Connections, Christoffel symbols and geodesic curves}

Up until this point, in our discussion of manifolds, we have emphasized many times the significance of the metric tensor. However, in
differential geometry\,\footnote{Differential geometry includes both Riemannian and pseudo-Riemannian manifolds.} the affine 
connection (or just connection) is equally important as the metric tensor. In simple terms, \textit{an \textbf{affine connection} is an object which 
decides how nearby tangent spaces of a manifold connect with each other. As a result, the connection allows the differentiation of tangent vector
fields as if they were functions on the manifold.} Although, in general, an affine connection may or may not depend on the metric tensor, 
in the context of GR---which, as far as this dissertation is concerned, is our main interest---the connection is expressed solely in terms of 
the metric tensor and its derivatives. Note additionally that \textit{an affine connection with no torsion, i.e. $\Gam^\sigma_{\mu\nu}=
\Gam^\sigma_{\nu\mu}$, is also called 
Levi-Civita\,\footnote{Tullio Levi-Civita (1873,\,Padua--1941,\,Rome) was an Italian mathematician.} connection, while its structure coefficients,
which are denoted as $\Gam^\sigma_{\mu\nu}$, are called \textbf{Christoffel symbols}.\,\footnote{Elwin Bruno Christoffel 
(1829,\,Monschau--1900,\,Strasbourg) was a German mathematician and physicist.}} It is essential to stress here that although
Christoffel symbols resemble the components of a $\trank{1}{2}$ tensor, they are \underline{not} such. 

Christoffel symbols are also closely related to a special type of curves on a manifold, which are called geodesic curves (or geodesics).
Assuming that $G$ is a curve on a manifold $M$ with parametric equation $\vec{r}=\vec{r}(x^\mu(\xi))$, we say that \textit{$G$ is a 
\textbf{geodesic curve} on $M$ if its tangent vector $\vec{t}$, with components $t^\mu=\dot{x}^\mu\equiv dx^\mu/d\xi$, is parallelly 
transported along it.} In mathematical language this translates to
\eq$\label{intro-eq: geo-def1}
\nabla_{\vec{t}}\,\vec{t}=0\Ra \ddot{x}^\sigma+\Gam^\sigma_{\mu\nu}\,\dot{x}^\mu\dot{x}^\nu=0\,,$
\textit{and it means that geodesic curves represent the shortest path between two points on a manifold}. 
In eq.\,\eqref{intro-eq: geo-def1} we made use of the defining relations of the covariant derivative
\eq$\label{intro-eq: cov-der-vec}
\nabla_{\vec{u}} \vec{A}\equiv u^\nu(\nabla_\nu A^\mu)\,\vec{e}_\mu\,,$
\bal$\label{intro-eq: cov-der-ten}
\nabla_\mu S^{\alp_1\ldots\alp_p}{}_{\beta_1\ldots\beta_q}\equiv\pa_\mu S^{\alp_1\ldots\alp_p}{}_{\beta_1\ldots\beta_q}
&+\Gam^{\alp_1}_{\mu\nu}S^{\nu\ldots\alp_p}{}_{\beta_1\ldots\beta_q}+\ldots+\Gam^{\alp_p}_{\mu\nu}
S^{\alp_1\ldots\nu}{}_{\beta_1\ldots\beta_q}\nonum\\[1mm]
&-\Gam^{\nu}_{\mu\beta_1}S^{\alp_1\ldots\alp_p}{}_{\nu\ldots\beta_q}-\ldots-\Gam^{\nu}_{\mu\beta_q}
S^{\alp_1\ldots\alp_p}{}_{\beta_1\ldots\nu}\,.$ Specifically in the context of GR, 
geodesic curves 
are of significant importance, since they are identified with the world lines of freely falling particles inside a gravitational field.

Using the aforementioned property of geodesic curves, one may determine the Christoffel symbols $\Gam^\sigma_{\mu\nu}$ in terms of the 
metric tensor and its derivatives. To achieve this, we parametrize the geodesic curve in terms of the parameter $\xi$, hence, $x^\mu=
x^\mu(\xi)$. Consequently, the length of the geodesic curve between two points residing at $\xi_1$ and $\xi_2$ satisfies the relation
\eq$\label{intro-eq: geo-def2}
\del \int_{\xi_1}^{\xi_2}ds=0\,.$
Employing equation \eqref{intro-eq: fff} in \eqref{intro-eq: geo-def2} we obtain 
\eq$\label{intro-eq: geo1}
\del \int_{\xi_1}^{\xi_2}\sqrt{g_{\alp\beta}\,\dot{x}^\alp\dot{x}^\beta}\,d\xi=0\,.$
In the above, the dot implies derivative with respect to $\xi$. Performing the variation of \eqref{intro-eq: geo1} with respect to the variables
$\{x^\kappa,\dot{x}^\lam\} $, we are led to the relation
\eq$\label{intro-eq: geo2}
\ddot{x}^\sigma+\frac{1}{2}\,g^{\sigma\rho}(\pa_\mu g_{\rho\nu}+\pa_\nu g_{\rho\mu}-\pa_\rho g_{\mu\nu})\dot{x}^\mu
\dot{x}^\nu=0\,.$
A simple comparison of eqs. \eqref{intro-eq: geo-def1} and \eqref{intro-eq: geo2} is now sufficient to provide us with the expression of 
Christoffel symbols in terms of the metric tensor, that is
\eq$\label{intro-eq: chr-symb}
\Gam^\sigma_{\mu\nu}=\frac{1}{2}\,g^{\sigma\rho}(\pa_\mu g_{\rho\nu}+\pa_\nu g_{\rho\mu}-\pa_\rho g_{\mu\nu})\,.$

\vspace*{-1em}

\mysubsection{The Riemann curvature tensor}

Given a $2$-dimensional surface it is intuitively obvious, although sometimes tricky, to deduce whether this surface has some kind of curvature or not.
However, when we are dealing with higher-dimensional manifolds, intuition cannot be trusted. Hence, a systematic way of measuring the curvature of
a manifold, regardless of its dimensionality, is essential. The mathematical object able to fulfil this purpose is the Riemann curvature tensor with
components of the form $R^{\rho}{}_{\sig\mu\nu}$. 


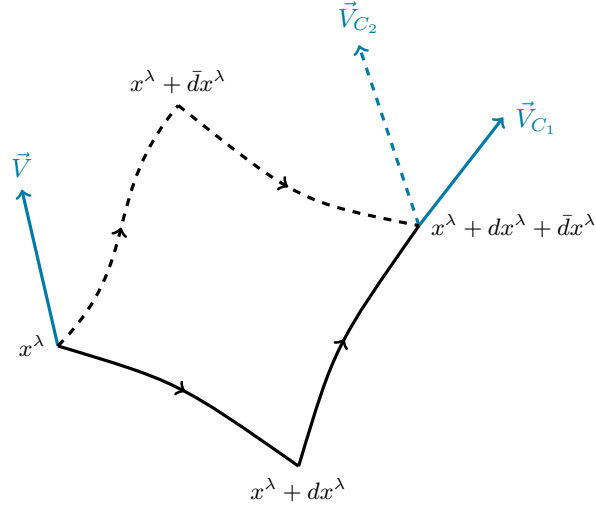
\begin{figure}[t]

\centering

\begin{tikzpicture}[scale=1.6]
	
	\draw[->,color=cerulean,
		very thick
	]
	(0,0) -- (-0.3,1.3) node[above]{\fontsize{10}{10}{\textcolor{cerulean}{$\vec{V}$}}};

	\draw[->,color=cerulean,
		very thick
	]
	(3,1) -- (3.7,1.9) node[above,right]{\fontsize{10}{10}{\textcolor{cerulean}{$\vec{V}_{C_1}$}}};

	\draw[->,dashed,color=cerulean,
		very thick
	]
	(3,1) -- (2.5,2.5) node[above]{\fontsize{10}{10}{\textcolor{cerulean}{$\vec{V}_{C_2}$}}};

	\draw[very thick,
		decoration={markings,
	    	mark=at position 0.5 with {\arrow{>}}
	    },
	    postaction={decorate}	    
	]    	 
	(0,0) node[below,left]{\fontsize{9}{9}{$x^\lam$}} .. controls (1,-0.3) .. (2,-1);

	\draw[very thick,
		decoration={markings,
	    	mark=at position 0.5 with {\arrow{>}}
	    },
	    postaction={decorate}	    
	]    	 
	(2,-1) node[below]{\fontsize{9}{9}{$x^\lam+dx^\lam$}} .. controls (2.3,0) ..  (3,1);

	\draw[dashed,very thick,
		decoration={markings,
	    	mark=at position 0.5 with {\arrow{>}}
	    },
	    postaction={decorate}	    
	]    	 
	(0,0)  .. controls +(45:1cm) and +(230:1cm) .. (1,2) node[above]{\fontsize{9}{9}{$x^\lam+\bar{d}x^\lam$}};

	\draw[dashed,very thick,
		decoration={markings,
	    	mark=at position 0.5 with {\arrow{>}}
	    },
	    postaction={decorate}	    
	]    	 
	(1,2) .. controls (2,1.2) .. (3,1) node[above,right]{\fontsize{9}{9}{$x^\lam+dx^\lam+\bar{d}x^\lam$}};

\end{tikzpicture}

\caption{The parallel transportation of a vector $\vec{V}$ following two different paths $C_1$ and $C_2$ (dashed) of an infinitesimal loop.}
\label{intro-fig: par-trans}

\end{figure}


To understand how the components of the Riemann tensor measure the curvature of a manifold, we
first need to define what we mean by curvature. To this end, we consider the parallel transport of a vector $\vec{V}$ around an infinitesimal loop 
on a manifold. On a flat manifold, where the Christoffel symbols are all zero, such an action would have no effect on the direction of the vector 
$\vec{V}$. However, the same does not apply to a curved manifold. In the latter case the direction of the vector will tilt, and since the action of parallel 
transport is
independent of the coordinate system, there should be a tensor of some rank which quantifies this change. Consequently, \textit{the curvature of a 
manifold is directly related to the change in the direction of a vector under a parallel transport around an infinitesimal loop, while the tensor which 
quantifies this change is the \textbf{Riemann curvature tensor}.} 

The change of the vector $\vec{V}$ under a parallel transport around a loop is given by the vector $\Delta\vec{V}
\equiv\vec{V}_{\circlearrowleft}-\vec{V}$. In \fref{intro-fig: par-trans} we depict the parallel transport of the vector $\vec{V}$ from $x^\lam$ to
$x^\lam+dx^\lam+\bar{d}x^\lam$ following two different paths, $C_1$ and $C_2$ (dashed), which form an infinitesimal loop. 
Since $\vec{V}_{C_1}-\vec{V}_{C_2}=\vec{V}_{\circlearrowleft}-\vec{V}$, we may use the paths $C_1$ and $C_2$ to extract 
the required information for the vector  $\Delta\vec{V}$.\,\footnote{A hand-waving proof of the preceding 
assertion: Assuming that $\vec{V}_{C_1}=\vec{V}+\vec{v}_1$ and $\vec{V}_{C_2}=\vec{V}+\vec{v}_2$, then $\vec{V}_{\circlearrowleft}=\vec{V}
+\vec{v}_1-\vec{v}_2$ and $\Delta\vec{V}\equiv \vec{V}_{\circlearrowleft}-\vec{V}=\vec{V}_{C_1}-\vec{V}_{C_2}=\vec{v}_1-\vec{v}_2$.}

We first consider the path $C_1$. The parallel transport of $\vec{V}$ from $x^\lam$ to $x^\lam+dx^\lam$ leads to
\eq$\label{intro-eq: path-c1-1}
V^\rho(x^\lam+dx^\lam)=V^\rho(x^\lam)+dx^\mu\,\pa_\mu V^\rho(x^\lam)\,,$
where we have kept only the linear term in the expansion. Since the vector is parallelly transported, it also holds that
\gat$
\nabla_{d\vec{x}}\vec{V}=dx^\mu(\nabla_\mu V^\rho)\vec{e}_\rho=0\Ra\nonum\\[1mm]
\label{intro-eq: par-trans}
\pa_\mu V^\rho=-\Gam^\rho_{\mu\sig}V^\sig\,.$
Consequently, it is
\eq$\label{intro-eq: path-c1-2}
V^\rho(x^\lam+dx^\lam)=V^\rho(x^\lam)-dx^\mu\Gam^\rho_{\mu\sig}(x^\lam)V^\sig(x^\lam)\,.$
Henceforth, when a mathematical object is defined at the starting point $x^\lam$ we may simply write the object without denoting the point of reference. 
Transporting now the vector from $x^\lam+dx^\lam$ to $x^\lam+dx^\lam+\bar{d}x^\lam$ we obtain
\eq$\label{intro-eq: path-c1-3}
V^\rho(x^\lam+dx^\lam+\bar{d}x^\lam)=V^\rho(x^\lam+dx^\lam)-\bar{d}x^\mu\Gam^\rho_{\mu\kappa}(x^\lam+dx^\lam)
V^\kappa(x^\lam+dx^\lam)\,.$
Using eq.\,\eqref{intro-eq: path-c1-2} and expanding the Christoffel symbols we get
\bal$\label{intro-eq: path-c1-4}
V^\rho(x^\lam+dx^\lam+\bar{d}x^\lam)&=V^\rho-dx^\mu\Gam^\rho_{\mu\sig}V^\sig-\bar{d}x^\mu\left(\Gam^\rho_{\mu\kappa}+dx^\nu\pa_\nu
\Gam^\rho_{\mu\kappa}\right)(V^\kappa-dx^\alp\Gam^\kappa_{\alp\sig}V^\sig)\nonum\\[1mm]
&=V^\rho-[dx^\mu\Gam^\rho_{\mu\sig}+\bar{d}x^\mu\Gam^\rho_{\mu\sig}-\bar{d}x^\mu dx^\nu(\Gam^\rho_{\mu\kappa}\Gam^\kappa_{\nu\sig}
-\pa_\nu \Gam^\rho_{\mu\sig})]V^\sig\,.$
In the second line we have neglected terms of third order. 

To evaluate the components $V^\rho(x^\lam+\bar{d}x^\lam+dx^\lam)$ following the path $C_2$ we only need to interchange 
$dx^\mu$ with $\bar{d}x^\mu$ in \eqref{intro-eq: path-c1-4}. By doing so we are led to
\bal$\label{intro-eq: path-c2-1}
V^\rho(x^\lam+\bar{d}x^\lam+dx^\lam)&=V^\rho-[\bar{d}x^\mu\Gam^\rho_{\mu\sig}+dx^\mu\Gam^\rho_{\mu\sig}
-dx^\mu \bar{d}x^\nu(\Gam^\rho_{\mu\kappa}\Gam^\kappa_{\nu\sig}-\pa_\nu \Gam^\rho_{\mu\sig})]V^\sig\nonum\\[1mm]
&=V^\rho-[\bar{d}x^\mu\Gam^\rho_{\mu\sig}+dx^\mu\Gam^\rho_{\mu\sig}
-\bar{d}x^\mu dx^\nu(\Gam^\rho_{\nu\kappa}\Gam^\kappa_{\mu\sig}-\pa_\mu \Gam^\rho_{\nu\sig})]V^\sig\,.$

The 	components of the vector $\Delta \vec{V}$ can now be determined using eqs. \eqref{intro-eq: path-c1-4} and \eqref{intro-eq: path-c2-1}. 
Thus, it is
\bal$\label{intro-eq: DV}
\Delta V^\rho&=V^\rho(x^\lam+dx^\lam+\bar{d}x^\lam)-V^\rho(x^\lam+\bar{d}x^\lam+dx^\lam)\nonum\\[1mm]
&=(\pa_\mu \Gam^\rho_{\nu\sig}-\pa_\nu \Gam^\rho_{\mu\sig}+\Gam^\rho_{\mu\kappa}\Gam^\kappa_{\nu\sig}
-\Gam^\rho_{\nu\kappa}\Gam^\kappa_{\mu\sig})V^\sig dx^\nu\bar{d}x^\mu\nonum\\[1mm]
&=R^{\rho}{}_{\sig\mu\nu}V^\sig dx^\nu\bar{d}x^\mu\,.$
From the above relation one can readily deduce the components of the Riemann tensor, namely
\eq$\label{intro-eq: riem-def}
R^{\rho}{}_{\sig\mu\nu}\equiv \pa_\mu \Gam^\rho_{\nu\sig}-\pa_\nu \Gam^\rho_{\mu\sig}+\Gam^\rho_{\mu\kappa}\Gam^\kappa_{\nu\sig}
-\Gam^\rho_{\nu\kappa}\Gam^\kappa_{\mu\sig}\,.$
Below we present the basic symmetries and properties of the Riemann tensor components.

\begin{itemize}

\item[\chapc{$\blacktriangleright$}] Skew-symmetry: $R_{(\alp\beta)\gam\del}=R_{\alp\beta(\gam\del)}=0$.\,\footnote{The parentheses 
in the indices denote symmetrization:
$$R_{\alp\beta(\gam\del)}=\frac{1}{2!}\left(R_{\alp\beta\gam\del}+R_{\alp\beta\del\gam}\right)\,. $$}

\item[\chapc{$\blacktriangleright$}] Symmetry with respect to a pair of indices: $R_{\alp\beta\gam\del}=R_{\gam\del\alp\beta}$.

\item[\chapc{$\blacktriangleright$}] Algebraic Bianchi identity: $R_{\alp[\beta\gam\del]}=R_{\alp\beta\gam\del}+R_{\alp\del\beta\gam}
+R_{\alp\gam\del\beta}=0$.\,\footnote{The square brackets in the indices denote anti-symmetrization:
$$R_{\alp[\beta\gam\del]}=\frac{1}{3!}\left(R_{\alp\beta\gam\del}-R_{\alp\beta\del\gam}-R_{\alp\gam\beta\del}+R_{\alp\gam\del\beta}
+R_{\alp\del\beta\gam}-R_{\alp\del\gam\beta}\right)\,.$$}

\item[\chapc{$\blacktriangleright$}] Differential Bianchi identity: $\nabla_\kappa R^\alp{}{}_{\beta\gam\del}
+\nabla_\del R^\alp{}{}_{\beta\kappa\gam}+\nabla_\gam R^\alp{}{}_{\beta\del\kappa}=0$.

\end{itemize}

\mysection{General Relativity}

The gravitational field equations of General Relativity constitute a relativistic generalization of the Newton's law of universal gravitation. 
Based on the equivalence between the inertial and gravitational mass, Einstein re-conceived gravity in purely geometrical terms. What is
considered as the motion of a particle under the force of gravity in Newtonian gravity, according to GR is the effect of a freely falling
particle following a geodesic curve on a curved spacetime. Similarly, as the Newton's law of gravity explains the connection between the 
force of gravity and the mass of an object, GR describes how matter and energy deform the fabric of spacetime.

Generalizing the stress tensor of Newtonian physics, one obtains the stress-energy tensor (or energy-momentum tensor) which, as its name
denotes, describes 
the density and the flux of energy and momentum in spacetime. \textit{The \textbf{energy-momentum tensor}, with components $T^{\mu\nu}$, is a
symmetric tensor of rank two, which in the context of GR operates as the source of the gravitational field.}\,\footnote{
Note that $T^{00}$ is the energy density, $T^{0i}$ is the momentum density, $T^{i0}$ is the energy flux, and $T^{ij}$ is the momentum flux,
while the indices $i,j$ run from $1$ to $3$.} In terms of the stress-energy 
tensor, the conservation of energy and momentum are expressed in the compact form
\eq$\label{intro-eq: cons-ene}
\nabla_{\mu}T^{\mu\nu}=0\,.$
\textit{A tensor with zero covariant derivative is also called a divergence free tensor.}\,\footnote{In \eqref{intro-eq: cons-ene}, $\nu=0$ leads to the 
conservation of energy, while for $\nu=1,2,3$ one obtains the momentum conservation.} 

Consequently, to obtain the gravitational field equations of General Relativity, we just need to equate the stress-energy tensor with a symmetric 
and divergence free tensor of rank two, which carries information about the curvature of the spacetime. Using the differential Bianchi identity 
and contracting the indices $\alp$ and $\gam$, we get 
\eq$\label{intro-eq: bianchi-2}
\nabla_\kappa R_{\beta\del}-\nabla_\del R_{\beta\kappa}+\nabla_\alp R^\alp{}{}_{\beta\del\kappa}=0\,,$
where \textit{$R_{\mu\nu}\equiv R^{\rho}{}_{\mu\rho\nu}$ are the covariant components of the \textbf{Ricci tensor}}. Contracting now the l.h.s. 
of \eqref{intro-eq: bianchi-2} with $g^{\beta\del}$, and utilizing the fact that $\nabla_{\lam}g^{\mu\nu}=\nabla_{\lam}g_{\mu\nu}=0$, we are led to
\eq$\label{intro-eq: bianchi-3}
\nabla_\lam\left(R^\lam{}_\kappa-\frac{1}{2}\del^\lam{}_\kappa R\right)=0\,,$
where $R\equiv g^{\mu\nu}R_{\mu\nu}$ is the \textit{\textbf{Ricci scalar}}. \textit{The contravariant components of \textbf{Einstein tensor} are defined via}
\eq$\label{intro-eq: eins-ten}
G^{\mu\nu}\equiv R^{\mu\nu}-\frac{1}{2}g^{\mu\nu}R\,,$
and it is straightforward to deduce from \eqref{intro-eq: bianchi-3} that $\nabla_\mu G^{\mu\nu}=0$. In 1972, David Lovelock (born 1938,\,Bromley)
proved that in a differentiable manifold of dimensionality four, the Einstein tensor is the only divergence free tensorial function of the metric tensor 
and their first and second partial derivatives \cite{Lovelock:1972vz}.

From the above discussion, one may readily guess the gravitational field equations of General Relativity, which in covariant form are expressed as
\eq$\label{intro-eq: GR}
G_{\mu\nu}+\Lambda\, g_{\mu\nu}=\kappa^2\, T_{\mu\nu}\,.$ 
In the preceding equation $\kappa^2=8\pi G_N/c^4$, $G_N$ is the Newton's gravitational constant, while we have included the cosmological term as well.

\mysubsection{Lagrangian formulation of General Relativity}

An alternative but arguably more interesting route to the gravitational field equations of General Relativity is through the Einstein-Hilbert action 
and the principle of least action. The idea is to use an appropriate action functional which under its variation with respect to the metric tensor leads to
the field equations of GR. David Hilbert (1862,\,K\"{o}nigsberg--1943,\,G\"{o}ttingen) showed that the simplest action able to produce the
field equations of GR is the following
\gat$\label{intro-eq: EH}
S_{E\text{-}H}=\int_{-\infty}^{+\infty} d^4x\sqrt{-g}\left[\frac{1}{2\kappa^2}(R-2\Lambda)+\lagr_m\right]\,.$
In the above, $R$ is the Ricci scalar, $\kappa^2=8\pi G_N$, while the Lagrangian density $\lagr_m$ incorporates the rest 
of the fields appearing in the theory. Applying the principle of the least action to \eqref{intro-eq: EH} we obtain
\eq$\label{intro-eq: var-EH}
\del S_{E\text{-}H}=0=\int_{-\infty}^{+\infty} d^4x\, \frac{1}{2\kappa^2}\left[\del(\sqrt{-g})\,(R-2\Lambda)+\sqrt{-g}\,\del R\right]
+\int_{-\infty}^{+\infty} d^4x\, \del(\sqrt{-g}\,\lagr_m)\,.$
where
\gat$\label{intro-eq: var-metr}
\del(\sqrt{-g})=-\frac{1}{2}\sqrt{-g}\, g_{\mu\nu}\del g^{\mu\nu}\,,\\[2mm]
\label{intro-eq: var-R}
\del R=R_{\mu\nu}\del g^{\mu\nu}-\nabla_{\mu}(\nabla_{\nu}\del g^{\mu\nu})+\nabla_\mu(g_{\alp\beta} \nabla^\mu \del g^{\alp\beta})\,,\\[2mm]
\label{intro-eq: var-ene}
\del(\sqrt{-g}\, \lagr_m)=-\frac{1}{2}\sqrt{-g}\, T_{\mu\nu}\del g^{\mu\nu}\,.$
For rigorous proofs of the preceding relations see Appendix F of \cite{Nakas}. Using now eqs.\,\eqref{intro-eq: var-metr}-\eqref{intro-eq: var-ene} 
in \eqref{intro-eq: EH} together with the identity 
\eq$\label{intro-eq: nabla-iden}
\nabla_\kappa A^\kappa=\frac{1}{\sqrt{-g}}\pa_\kappa(\sqrt{-g}\, A^\kappa)\,,$
we are led to
\bal$
0&=\int_{-\infty}^{+\infty} d^4x\,\frac{\sqrt{-g}}{2\kappa^2}\left(R_{\mu\nu}-\frac{1}{2}g_{\mu\nu}R+\Lambda\, g_{\mu\nu}-\kappa^2\, T_{\mu\nu}
\right)\del g^{\mu\nu}\nonum\\[1mm]
\label{intro-eq: EH-var}
&\hspace{1em}+\int_{-\infty}^{+\infty} d^4x\, \pa_\mu[\sqrt{-g}\,(g_{\alp\beta} \nabla^\mu \del g^{\alp\beta}-\nabla_\nu\del g^{\mu\nu})]\,.$
The term in the second line of the above equation vanishes at the spacetime boundary since $\del g^{\alp\beta}\xrightarrow{x^\lam\ra\pm\infty} 0$. 
Consequently, eq.\,\eqref{intro-eq: EH-var} reduces to the Einstein's equations \eqref{intro-eq: GR}, while the Einstein-Hilbert action correctly formulates
General Relativity in Lagrangian form. It is important to stress here that the Lagrangian formulation of General Relativity is not only useful for GR itself,
but it is essential for the construction of gravitational models beyond GR as well. Lovelock gravity \cite{Lovelock:1971yv,Padmanabhan:2013xyr}, 
$f(R)$ gravity \cite{Buchdahl:1970ynr,DeFelice:2010aj}, and scalar-tensor theories \cite{Brans:1961sx,Horndeski:1974wa,Kobayashi:2019hrl} are only 
some of the gravitational theories which would not have been developed without the Lagrangian formulation of GR.

\vspace*{0.5em}
\mysection{Black holes in General Relativity}

The gravitational field equations of General Relativity although it may seem simple in their compact tensorial form, in practice they constitute a set of
differential equations for the components of the metric tensor.  In general, these differential equations are very complicated, and in order to have some 
hope of solving them we need to assume a specific matter and energy distribution, as well as a suitable form for the metric tensor.

\vspace*{-1em}
\mysubsection{The Schwarzschild solution}

The first solution to the Einstein's equations came from Karl Schwarzschild (1873,\,Frank\-furt am Main--1916,\,Potsdam)  \cite{Schwarzschild:1916uq} 
only a year after the publication of General Relativity. Schwarz\-schild considered the simplest possible case and examined the geometry of a 
\textbf{static} and \textbf{spherically symmetric} object in \textbf{vacuum}. In a nutshell, when we refer to a \textit{\textbf{static spacetime} we
mean that there is no rotation of any kind in this spacetime, and that if we could photograph the geometry in different moments in time we would get the
same photo over and over again. 
In mathematical language this means that the components of the metric tensor are time-independent, and that the components $g_{0i}$ (with $i=1,2,3$) 
vanish.} Hence, the most general expression for the line-element of a static and spherically symmetric spacetime is the following:
\eq$\label{intro-eq: stat-spher}
ds^2=-e^{F(r)}dt^2+e^{G(r)}dr^2+e^{H(r)}r^2(d\theta^2+\sin^2\theta\,d\varphi^2)\,.$
The exponentials in the above equation are not of physical importance, but they make the forthcoming calculations easier.
Notice also that eq.\,\eqref{intro-eq: stat-spher} implies that the temporal coordinate $t$ has the same
units as the spatial coordinates; this happens because we are working in a \textbf{\textit{geometric unit system}} where it holds that $c=G_N=1$. This 
particular unit system will be used extensively throughout this dissertation, and therefore henceforth its use should be considered self-evident. 
Exploiting now our freedom to change coordinate systems we can further simplify the line-element \eqref{intro-eq: stat-spher}. To this end, we introduce 
a new radial coordinate 
$\tilde{r}$ defined as
\eq$\label{intro-eq: new-rad}
\tilde{r}\equiv e^{H(r)/2} r\,,$
and we obtain
\eq$\label{intro-eq: dr-tild}
d\tilde{r}=e^{H(r)/2}\left(1+\frac{r}{2}\frac{dH}{dr}\right)dr\,.$
Substituting the above into \eqref{intro-eq: stat-spher} we are led to
\eq$\label{intro-eq: stat-spher2}
ds^2=-e^{F(r)}dt^2+e^{G(r)-H(r)}\left(1+\frac{r}{2}\frac{dH}{dr}\right)^{-2}d\tilde{r}^2+\tilde{r}^2\, d\Omega_2^2\,,$
where $d\Omega_2^2=d\theta^2+\sin^2\theta\,d\varphi^2$. Introducing the functions
\gat$
\tilde{F}(\tilde{r})=F(r)\,,\\[1mm]
e^{\tilde{G}(\tilde{r})}=e^{G(r)-H(r)}\left(1+\frac{r}{2}\frac{dH}{dr}\right)^{-2}\,,$
and renaming everything in order to get rid of the tildes, we finally get
\eq$\label{intro-eq: stat-spher3}
ds^2=-e^{F(r)}dt^2+e^{G(r)}dr^2+r^2\, d\Omega_2^2\,.$
As we have just shown the line-elements \eqref{intro-eq: stat-spher} and \eqref{intro-eq: stat-spher3} are equally general. Henceforth we will
use the latter one for our convenience in calculations.

We mentioned earlier that the Schwarzschild solution is a vacuum solution ($\Lambda=0$ and $\lagr_m=0$), thus, the gravitational field equations are of the form $G_{\mu\nu}=
R_{\mu\nu}-g_{\mu\nu}R/2=0$. The functions $F(r)$ and $G(r)$ will be determined via these equations. Contracting $G_{\mu\nu}$ with $g^{\mu\nu}$ 
one finds that $R=0$, and therefore the Einstein's equations in vacuum reduce to
\eq$\label{intro-eq: GR-vac}
R_{\mu\nu}=0\,.$
With the use of eqs. \eqref{intro-eq: chr-symb} and \eqref{intro-eq: riem-def} one can calculate the components of the Ricci tensor. 
By doing so, eq. \eqref{intro-eq: GR-vac} leads to the following three independent differential equations:
\bal$\label{intro-eq: Schw-feq1}
\text{\textit{(tt)-component}}:\hspace{1em}& 2\,\pa_r^2 F + 4\,\frac{\pa_r F}{r} - (\pa_r G)(\pa_r F) + (\pa_r F)^2 = 0\,,\\[3mm]
\label{intro-eq: Schw-feq2}
\text{\textit{(rr)-component}}:\hspace{1em}& \frac{\pa_r G}{r} + \frac{1}{4}\left[(\pa_r G)(\pa_r F) - (\pa_r F)^2 - 2\,\pa_r^2 F\right] = 0\,,\\[3mm]
\label{intro-eq: Schw-feq3}
\text{\textit{(\gr{θθ})-component}}:\hspace{1em}& 1 + e^{-G(r)} \left( \frac{r}{2}\,\pa_r G - \frac{r}{2}\,\pa_r F - 1 \right) = 0\,.$
Combining eqs. \eqref{intro-eq: Schw-feq1} and \eqref{intro-eq: Schw-feq2}, we get $\pa_r F + \pa_r G = 0 $ which results to $F(r)=-G(r)+C$. The constant
$C$ may be absorbed in the time coordinate by performing the transformation $t\ra e^{-C/2}\,t$. Hence, we can set $C=0$ and we are left with
\eq$\label{intro-eq: Schw-feq4}
F(r)=-G(r)\,.$
Turning now our attention to eq. \eqref{intro-eq: Schw-feq3} and making use of \eqref{intro-eq: Schw-feq4}, it is not hard to show that it takes the form
$\pa_r\left(r e^{F(r)}\right)=1$, while its solution is given by the relation
\eq$\label{intro-eq: Schw-sol}
e^{F(r)}=1-\frac{r_s}{r}\,.$
The constant $r_s$ is called \textbf{\textit{Schwarzschild radius}} and it can be determined via the \textit{weak-field approximation}. The weak-field 
limit emerges at large distances away from the massive object ($r\gg r_s$) and it is defined as the region of spacetime in which General Relativity and 
Newtonian gravity can be considered equivalent.

According to Newtonian mechanics, a free test particle outside a spherical object of mass $M$ feels an acceleration

\eq$\label{intro-eq: New-g}
g_{N}=-\frac{G_N M}{r^2}\,,$

\noindent{due} to the gravitational field. However, in the language of General Relativity, the free particle moves towards the massive object following a 
geodesic curve on the curved spacetime. 
The maxim of John Wheeler\,\footnote{John Archibald Wheeler (1911,\,Florida--2008,\,New Jersey) was an American theoretical physicist.} 
\textit{``Matter tells spacetime how to curve; spacetime tells matter how to move''} encapsulates precisely this particular <<interaction>>. 
The equation of the geodesic curve is of the form 
\eq$\label{intro-eq: schw-geo1}
\frac{d^2x^\mu}{d\tau^2}+\Gam^\mu_{\alp\beta}\frac{dx^\alp}{d\tau}\frac{dx^\beta}{d\tau}=0\,,$
where the parameter $\xi$ in \eqref{intro-eq: geo-def1} has been replaced by the proper time $\tau$ of the moving particle.
In the weak-field limit, the test particle is considered to move slowly compared to the speed of light. This means that $dt/d\tau\gg dx^i/d\tau$. 
Also, the proper time $\tau$ can be approximated with $t$, and thus $dx^\mu/d\tau=(1,0,0,0)$. The geodesic equation now simplifies to
\eq$\label{intro-eq: schw-geo2}
\frac{d^2x^\mu}{dt^2}\approx -\Gam^\mu_{tt}\,,$
with
\bal$
\Gam^\mu_{tt}&=\frac{1}{2}g^{\mu\lam}(\pa_t g_{\lam t}+\pa_t g_{\lam t}-\pa_\lam g_{tt})\nonum\\[2mm]
&=-\frac{1}{2}g^{\mu r}\pa_r g_{tt}=-\frac{1}{2}\del^\mu{}_ r\, g^{rr}\pa_r g_{tt}\nonum\\[2mm]
&=\frac{1}{2}\del^\mu{}_ r\left(1-\frac{r_s}{r}\right)\frac{r_s}{r^2}\approx \frac{1}{2}\del^\mu{}_ r\, \frac{r_s}{r^2}\,.$
Comparing with the classical case, we are led to
\eq$ 
g_{N}=\frac{d^2r}{dt^2}=-\Gam^r_{tt}\,.$
Hence, the constant $r_s$ is related to the mass $M$ of the spherical object via the relation $r_s=2 M$ in geometric units, whilst in 
the MKS unit system it becomes $r_s=2G_N M/c^2$.
Consequently, the Schwarz\-schild solution which describes the geometry of a static and spherically symmetric massive object in vacuum, is characterized by
the line-element 
\eq$\label{intro-eq: schaw-line}
ds^2=-\left(1-\frac{2M}{r}\right)dt^2+\left(1-\frac{2M}{r}\right)^{-1}dr^2+r^2\, d\Omega_2^2\,.$

Let us now scrutinize the line-element \eqref{intro-eq: schaw-line} in order to extract more information about the spacetime geometry. 
To begin with, notice that at $r=0$ the metric component $g_{tt}$ becomes infinite, while the metric component $g_{rr}$ exhibits a similar behaviour at 
$r=2M$. 
So, the questions that come naturally to mind are: ``What do these infinities indicate about the geometry of the spacetime?'', ``Do they have physical 
consequences or they are just ill-defined points associated with the particular coordinate system that we are using?''. 
To answer these questions we first need to be able to distinguish unequivocally the real spacetime singularities from the coordinate singularities. 
The latter ones can be remedied by an appropriate coordinate system transformation, while the former ones are true spacetime singularities which
exist independently of the coordinate system.
To identify the true spacetime singularities, we need to utilize the Riemann curvature tensor, since, as we showed in a preceding section the Riemann tensor 
encapsulates the whole information about the curvature of the spacetime.
However, since the components of a tensor are coordinate-dependent quantities, we need to construct scalar quantities from which we can make
our inferences.
The most reliable scalar curvature quantity for this task is the Kretschmann\,\footnote{Erich Justus Kretschmann (1887--1973) was a German physicist.} scalar 
$\mathcal{K}\equiv R^{\mu\nu\kappa\lam}R_{\mu\nu\kappa\lam}$, while the scalar quantities $R\equiv g^{\mu\nu}R_{\mu\nu}$ and $\mathcal{R}\equiv 
R^{\mu\nu}R_{\mu\nu}$ although they provide information about the spacetime curvature, they are not of the same credibility. 
The reason for this comes from the fact that the Ricci tensor incorporates, by construction, only a part of the Riemann tensor, consequently some information 
about the curvature of the spacetime can be missed.
The previous assertion becomes fully evident in the case of the Schwarzschild geometry. 
For the line-element \eqref{intro-eq: schaw-line}, although the components of the Ricci tensor are identically zero, and therefore $R=\mathcal{R}=0$,  
the Kretschmann scalar is evaluated to be
\eq$\label{intro-eq: schw-krets}
\mathcal{K}=\frac{48M^2}{r^6}\,.$
As one may readily deduce from the equation above, the Kretschmann scalar diverges at $r=0$ and the curvature goes to infinity. 
This means that at this particular point in space resides a true singularity. According to the preceding discussion, we can formulate the following criterion:
\textit{``A true singularity constitutes a point in space at which the Kretschmann scalar diverges. The divergence of the other two scalar curvature 
quantities is of secondary importance, since they will either diverge in the same point(s) as the Kretschmann scalar, or they will not diverge at all''.}

Employing the aforementioned criterion, it is now apparent that the point $r=r_s$ constitutes a coordinate singularity. 
However, as we will see below, this particular limit is of special importance only when it comes to the study of black holes. 
But first, it is essential to stress that although the Schwarzschild spacetime has been strongly connected with black-hole solutions (and for a good reason), 
it is also possible to describe the gravitational imprint outside any astrophysical object which can be considered static and spherically symmetric, 
such as a planet (but a very round and smooth one) or a star. 
In these cases, both the singularity and the Schwarzschild radius $r_s$ reside way inside these astrophysical objects, and thus they do not affect their exterior 
spacetime geometry. 
Note, for intuitive purposes, that the Schwarzschild radius for the Earth is evaluated to be 8.9\,mm, while for our Sun it is approximately 2950\,m. 

As a side remark, before we dive into the dark waters of black holes, let us to emphasize that relativistic calculations are not only met in the papers of
physicists, but they are extensively used in our everyday life as well. 
Every time one opens his/her GPS (Global Positioning System) to move from one place to another, a plethora of relativistic calculations are being performed every 
split second. 
Without the relativistic corrections for the Earth's gravity and the relativistic consideration of time, the GPS would fail in its navigational functions 
within about 2 minutes.\,\footnote{For more details about this subject, see the article 
\href{https://physicscentral.com/explore/writers/will.cfm}{https://physicscentral.com/explore/writers/will.cfm}.}

\mysubsection{Schwarzschild black holes}

\begin{figure}[t]

\centering

\begin{tikzpicture}

\draw[->,very thick] (0,0)--(15,0) node[right]{$r$};
\draw[->,very thick] (0,-2)--(0,2) node[above]{$t$};

\draw[dashed,very thick] (2,-1.8) -- (2,0) -- (2,1.8) node[above]{$r=2M$} ;

\filldraw[color = black, fill = brightcerulean, thick] (13,0) -- ({13+tan(45)*1.5},{1.5}) -- ({13-tan(45)*1.5},{1.5}) -- (13,0);
\filldraw[color = black, fill = babyblue, thick] (13,1.5)
	ellipse [ x radius = {tan(45)*1.5}, y radius = {tan(45)*1.5/6}];
\filldraw[color = black, fill = brightcerulean, thick] (13,-1.5)
	ellipse [ x radius = {tan(45)*1.5}, y radius = {tan(45)*1.5/6}];
\filldraw[color = black, fill = brightcerulean, thick] (13,0) -- ({13+tan(45)*1.5},{-1.5}) -- ({13-tan(45)*1.5},{-1.5}) -- (13,0);
\draw[ultra thick, color = brightcerulean] ({13+tan(45)*1.5},{-1.5}) -- ({13-tan(45)*1.5},{-1.5});

\filldraw[color = black, fill = brightcerulean, thick] (9,0) -- ({9+tan(30)*1.5},{1.5}) -- ({9-tan(30)*1.5},{1.5}) -- (9,0);
\filldraw[color = black, fill = babyblue, thick] (9,1.5)
	ellipse [ x radius = {tan(30)*1.5}, y radius = {tan(30)*1.5/6}];
\filldraw[color = black, fill = brightcerulean, thick] (9,-1.5)
	ellipse [ x radius = {tan(30)*1.5}, y radius = {tan(30)*1.5/6}];
\filldraw[color = black, fill = brightcerulean, thick] (9,0) -- ({9+tan(30)*1.5},{-1.5}) -- ({9-tan(30)*1.5},{-1.5}) -- (9,0);
\draw[ultra thick, color = brightcerulean] ({9+tan(30)*1.5},{-1.5}) -- ({9-tan(30)*1.5},{-1.5});

\filldraw[color = black, fill = brightcerulean, thick] (6,0) -- ({6+tan(20)*1.5},{1.5}) -- ({6-tan(20)*1.5},{1.5}) -- (6,0);
\filldraw[color = black, fill = babyblue, thick] (6,1.5)
	ellipse [ x radius = {tan(20)*1.5}, y radius = {tan(20)*1.5/6}];	
\filldraw[color = black, fill = brightcerulean, thick] (6,-1.5)
	ellipse [ x radius = {tan(20)*1.5}, y radius = {tan(20)*1.5/6}];
\filldraw[color = black, fill = brightcerulean, thick] (6,0) -- ({6+tan(20)*1.5},{-1.5}) -- ({6-tan(20)*1.5},{-1.5}) -- (6,0);	
\draw[ultra thick, color = brightcerulean] ({6+tan(20)*1.5},{-1.5}) -- ({6-tan(20)*1.5},{-1.5});

\filldraw[color = black, fill = brightcerulean, thick] (3.5,0) -- ({3.5+tan(5)*1.5},{1.5}) -- ({3.5-tan(5)*1.5},{1.5}) -- (3.5,0);
\filldraw[color = black, fill = babyblue, thick] (3.5,1.5)
	ellipse [ x radius = {tan(5)*1.5}, y radius = {tan(5)*1.5/6}];
\filldraw[color = black, fill = brightcerulean, thick] (3.5,-1.5)
	ellipse [ x radius = {tan(5)*1.5}, y radius = {tan(5)*1.5/6}];	
\filldraw[color = black, fill = brightcerulean, thick] (3.5,0) -- ({3.5+tan(5)*1.5},{-1.5}) -- ({3.5-tan(5)*1.5},{-1.5}) -- (3.5,0);
\draw[ultra thick, color = brightcerulean] ({3.5+tan(5)*1.5},{-1.5}) -- ({3.5-tan(5)*1.5},{-1.5});

\end{tikzpicture}

\caption{The light cones in the Schwarzschild spacetime.}
\label{intro-fig: schw-null-traj}

\end{figure}
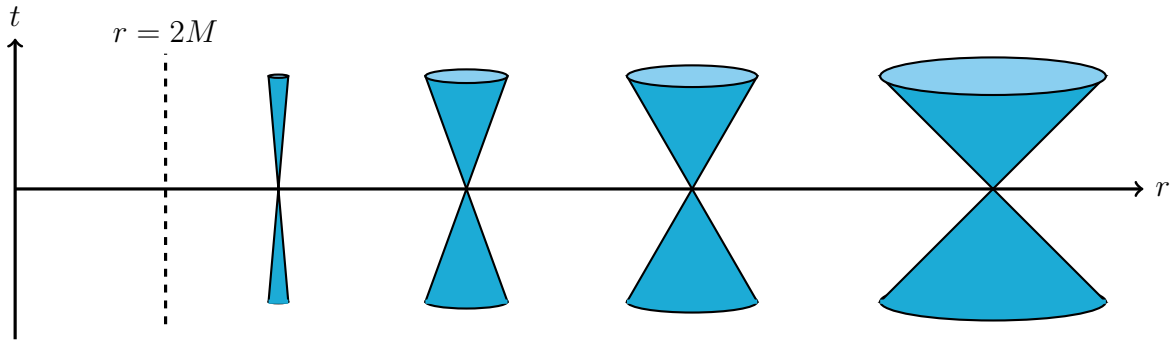

Let us now imagine an object with a thousand times the mass of our Sun, and its mass concentrated in a single point. 
In this scenario, even the points of the spacetime with radial coordinate smaller than the Schwarzschild radius lie on the exterior of the object.
Henceforth, \textit{any static and spherically symmetric object which has its mass $M$ concentrated in an area with volume smaller than the volume of a 
sphere with radius $r_s$ will be called a \textbf{Schwarzschild black hole}.} 
The reasoning behind this name will become apparent from the forthcoming analysis. 
However, it should be obvious from the preceding definition, that when we are dealing with Schwarzschild black holes, the behaviour of the spacetime in the 
region with $r\leq r_s$ is of main importance.
To this end, we need to study the casual structure of the spacetime, as it is defined by the light cones.
Consequently, we consider radial null trajectories, that is radially moving photons, in the Schwarzschild spacetime.
Since we are investigating the movement of photons in the spacetime, the line-element is identically zero ($ds^2=0$), while the radial movement of photons 
indicate that the angular part of the line-element vanishes as well.
Therefore, one obtains\vspace*{0.7em}
\eq$
ds^2=-\left(1-\frac{2M}{r}\right)dt^2+\left(1-\frac{2M}{r}\right)^{-1}dr^2=0\,,$
which leads to
\eq$\label{intro-eq: schw-null-t-r}
\frac{dt}{dr}=\pm \left(1-\frac{2M}{r}\right)^{-1}\,.$
The positive sign corresponds to outward moving photons, while inward moving photons are described by the negative sign. 
It is now straightforward to find that at the spacetime boundary, eq.\,\eqref{intro-eq: schw-null-t-r} simplifies to
\eq$\label{intro-eq: schw-null-t-r-infty}
\lim_{r\ra+\infty}\frac{dt}{dr}=\pm 1\,.$
This simply means that the Schwarzschild geometry is asymptotically flat, something which may also be deduced by the line-element \eqref{intro-eq: schaw-line},
and it is intuitively sensible from a physical point of view. 
However, as one approaches the Schwarzschild radius peculiar things seem to happen.
As it is depicted in \fref{intro-fig: schw-null-traj}, the light cones shrink as we get closer and closer to the black hole, and they finally collapse at $r=2M$. 
At $r=2M$, eq.\,\eqref{intro-eq: schw-null-t-r} results to
\eq$\label{intro-eq: schw-null-t-r-hor}
\lim_{r\ra 2M^+}\frac{dt}{dr}=\pm \infty\,.$
For an observer far away from the black hole, who measures time in the parameter $t$, it appears that nothing, not even light, can reach the distance
$r=2M$. Therefore, an observer far away from the black hole observes a \textit{\textbf{horizon}} at $r=2M$, which cannot be reached even
by a photon moving towards the horizon. 
This inference seems not only counter-intuitive but also quite absurd. 
How can it be that light which is moving towards a massive object cannot reach the object?

As we will shortly see, this ostensible inability to approach the horizon is only an illusion with its roots at the coordinate system that the outside observer uses 
to draw his/her conclusions. 
To be more specific, the Schwarzschild coordinates are ill-defined near the horizon, since they result to $dt/dr\ra \pm \infty$.
Therefore, we need to make an appropriate coordinate transformation to resolve this problem.
To this end, we return to the eq.\,\eqref{intro-eq: schw-null-t-r}, which characterises the radial null trajectories, and we solve it explicitly.
After some calculations, one finds that
\eq$\label{intro-eq: null-sol}
t=\pm r^*+C\,.$
In the above, $C$ is a constant, while $r^*$ is a new radial coordinate, which is often called as \textbf{tortoise coordinate}, and is defined via the
relation
\eq$\label{intro-eq: tortoise}
r^*\equiv r+2M\ln\left(\frac{r}{2M}-1\right)\,.$
The function $r^*(r)$ is injective and therefore every value of $r$ is mapped to a unique value of $r^*$, and vice versa. 

\begin{figure}[t]

\centering

\begin{tikzpicture}

\def\x{12}
\def\y{0}
\def\r{1.4}
\def\u{atan(2)}
\def\xx{7}
\def\uu{atan(10000)}
\def\xxx{2.5}
\def\uuu{123.7}
\draw[->,very thick] (0,0)--(15,0) node[right]{$r$};
\draw[->,very thick] (0,-2)--(0,2) node[above]{$v$};

\draw[dashed,very thick] ({\xx},-1.8) -- ({\xx},2) node[above]{$r=2M$} ;

\filldraw[color = black, fill = brightcerulean, thick] ({\x-\r},\y) -- ({\x},\y) -- ({\x+\r*cos(\u)},{\y+\r*sin(\u)}) -- ({\x-\r},\y);
\filldraw [ color = black, fill = babyblue, thick] ({\x+\r*(cos(\u)-1)/2},{\y+\r*sin(\u)/2})
	ellipse [ rotate = { \u/2 } , x radius = { \r*sqrt( cos(\u)*cos(\u)+1 )/sqrt(2)+0.1 }, y radius = { \r*sqrt( cos(\u)*cos(\u)+1 )/sqrt(2)/8 }];
\filldraw[color = black, fill = brightcerulean, thick] ({\x+\r*(1-cos(\u))/2},{\y-\r*sin(\u)/2})
	ellipse [ rotate = { \u/2 } , x radius = { \r*sqrt( cos(\u)*cos(\u)+1 )/sqrt(2)+0.1 }, y radius = { \r*sqrt( cos(\u)*cos(\u)+1 )/sqrt(2)/8 }];
\filldraw[color = black, fill = brightcerulean, thick] ({\x},\y) -- ({\x+\r},\y) -- ({\x-\r*cos(\u)},{\y-\r*sin(\u)}) -- ({\x},\y);
\draw[ultra thick, color = brightcerulean] ({\x+\r},\y) -- ({\x-\r*cos(\u)},{\y-\r*sin(\u)});

\filldraw[color = black, fill = brightcerulean, thick] ({\xx-\r},\y) -- ({\xx},\y) -- ({\xx+\r*cos(\uu)},{\y+\r*sin(\uu)}) -- ({\xx-\r},\y);
\filldraw [ color = black, fill = babyblue, thick] ({\xx+\r*(cos(\uu)-1)/2},{\y+\r*sin(\uu)/2})
	ellipse [ rotate = { \uu/2 } , x radius = { \r*sqrt( cos(\uu)*cos(\uu)+1 )/sqrt(2) }, y radius = { \r*sqrt( cos(\uu)*cos(\uu)+1 )/sqrt(2)/8 }];
\filldraw[color = black, fill = brightcerulean, thick] ({\xx+\r*(1-cos(\uu))/2},{\y-\r*sin(\uu)/2})
	ellipse [ rotate = { \uu/2 } , x radius = { \r*sqrt( cos(\uu)*cos(\uu)+1 )/sqrt(2) }, y radius = { \r*sqrt( cos(\uu)*cos(\uu)+1 )/sqrt(2)/8 }];
\filldraw[color = black, fill = brightcerulean, thick] ({\xx},\y) -- ({\xx+\r},\y) -- ({\xx-\r*cos(\uu)},{\y-\r*sin(\uu)}) -- ({\xx},\y);
\draw[ultra thick, color = brightcerulean] ({\xx+\r},\y) -- ({\xx-\r*cos(\uu)},{\y-\r*sin(\uu)});

\filldraw[color = black, fill = brightcerulean, thick] ({\xxx-\r},\y) -- ({\xxx},\y) -- ({\xxx+\r*cos(\uuu)},{\y+\r*sin(\uuu)}) -- ({\xxx-\r},\y);
\filldraw [ color = black, fill = babyblue, thick] ({\xxx+\r*(cos(\uuu)-1)/2},{\y+\r*sin(\uuu)/2})
	ellipse [ rotate = { \uuu/2 } , x radius = { \r*sqrt( cos(\uuu)*cos(\uuu)+1 )/sqrt(2)-0.48 }, y radius = { \r*sqrt( cos(\uuu)*cos(\uuu)+1 )/sqrt(2)/8 }];
\filldraw[color = black, fill = brightcerulean, thick] ({\xxx+\r*(1-cos(\uuu))/2},{\y-\r*sin(\uuu)/2})
	ellipse [ rotate = { \uuu/2 } , x radius = { \r*sqrt( cos(\uuu)*cos(\uuu)+1 )/sqrt(2)-0.48 }, y radius = { \r*sqrt( cos(\uuu)*cos(\uuu)+1 )/sqrt(2)/8 }];
\filldraw[color = black, fill = brightcerulean, thick] ({\xxx},\y) -- ({\xxx+\r},\y) -- ({\xxx-\r*cos(\uuu)},{\y-\r*sin(\uuu)}) -- ({\xxx},\y);
\draw[ultra thick, color = brightcerulean] ({\xxx+\r},\y) -- ({\xxx-\r*cos(\uuu)},{\y-\r*sin(\uuu)});

\end{tikzpicture}

\caption{The light cones in the Eddington-Finkelstein coordinates.}
\label{intro-fig: EF-null-traj}

\end{figure}
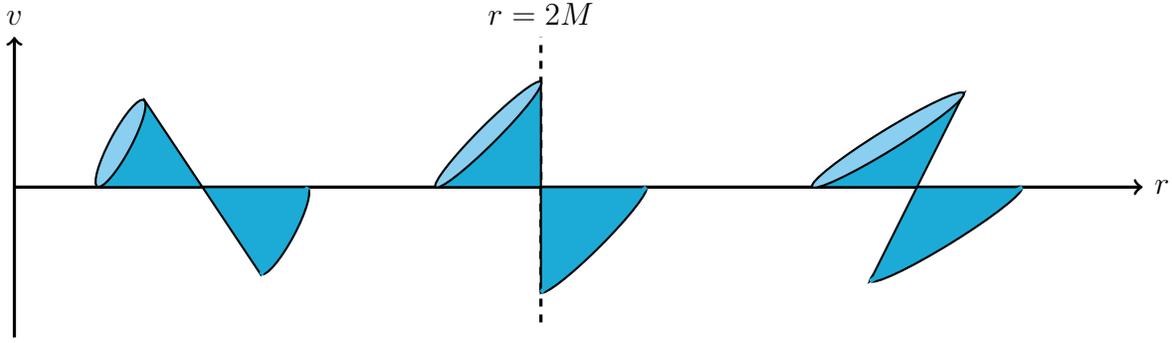

Let us now use the tortoise coordinate $r^*$ to define a new temporal coordinate $v$ as follows
\gat$\label{intro-eq: v-def}
v=t+r^*\,.$
In the coordinate system $\{v,r,\theta,\varphi\}$, which is known as \textbf{Eddington-Finkelstein coordinate system}\,\footnote{Sir Arthur 
Stanley Eddington (1882,\,Kendal--1944,\,Cambridge) was an English astronomer, physicist, and mathematician.}\footnote{David Ritz Finkelstein (1929--2016)
was an American physicist.} \cite{Eddington,Finkelstein,Penrose}, the metric 
\eqref{intro-eq: schaw-line} takes the form
\eq$\label{intro-eq: EF-in}
ds^2=-\left(1-\frac{2M}{r}\right)dv^2+2dv dr+r^2\, d\Omega_2^2\,.$
In terms of these coordinates, the equation for radial null trajectories is given by
\eq$
-\left(1-\frac{2M}{r}\right)\left(\frac{dv}{dr}\right)^2+2\,\frac{dv}{dr}=0\,,$
which has the solutions
\eq$\label{intro-eq: EF-null}
\frac{dv}{dr}=\left\{\begin{array}{cl}
0\,, & \text{(\textit{inward} moving photons)}\\[1mm]
2\left(1-\frac{2M}{r}\right)^{-1}\,, & \text{(\textit{outward} moving photons)}
\end{array}\right\}\,.$
In \fref{intro-fig: EF-null-traj} we depict the light cones as they ensue from eq.\,\eqref{intro-eq: EF-null}.
As one can readily deduce from the figure, the light cones in the Eddington-Finkelstein coordinates do not collapse at the Schwarzschild radius, 
but they gradually tilt towards the singularity. 
Notice that at the Schwarzschild radius $r=2M$, the future-directed worldline of an outward moving photon is described by $dv/dr\ra +\infty$, 
while for $r<2M$ it is $dv/dr<0$. 
This means that all future-directed worldlines are heading towards the singularity at $r=0$. 
This is indeed a very interesting result. 
In a sense, the surface $r=2M$ divides the universe into two regions. 
The <<inside>> region contains all the spacetime events with $r<2M$, while the <<outside>> one contains the events with $r>2M$. 
Although it is possible to move from the <<outside>> to the <<inside>> region, once something enters the latter one there is no turning back. 
\textit{Henceforth, any surface with the aforementioned property will be called \textbf{event horizon}; and since it is impossible, even for light,
to escape the event horizon, the objects which create such a geometry will appear altogether black. Consequently, it seems only reasonable to name
these objects as \textbf{black holes}.}\,\footnote{The Hawking radiation is obviously not considered in this description. If one takes into account
the Hawking radiation, then black holes do not appear black at all.}

It is important to mention at this point that in the $\{v,r,\theta,\varphi\}$ coordinate system the event horizon can \textit{only} be met on 
future-directed worldlines, not on past-directed ones. 
Although this might seem unreasonable, since we started from a time-independent solution, it is well justified by our choice for the coordinate $v$. 
Equation \eqref{intro-eq: v-def} dictates that for $v=\text{constant}$ we obtain the ingoing radial null geodesics, since as the time $t$ increases the 
tortoise coordinate $r^*$ decreases, and therefore $r$ decreases as well.
To obtain the outgoing radial null geodesics we need to define the coordinate $u$ in the following way
\eq$\label{intro-eq: u-def}
u=t-r^*\,.$
Using now the coordinate system $\{u,r,\theta,\varphi\}$, instead of the $\{v,r,\theta,\varphi\}$ one, and following a similar procedure as before, we 
can trace the past-directed worldlines which may cross the event horizon. 
The question that emerges now is: ``Can we describe in a unified way both the ingoing and outgoing radial null trajectories?''. 

Our mathematical instinct tells us that there must be a positive answer to the question, but first we need to find the appropriate coordinate system
to carry out the job. 
So, let us combine both $v$ and $u$ in a single coordinate system, and see where this leads us.
From eqs. \eqref{intro-eq: tortoise}, \eqref{intro-eq: v-def} and \eqref{intro-eq: u-def} one can easily find that
\eq$\label{intro-eq: v-u-r}
\frac{v-u}{2}=r+2M\ln\left(\frac{r}{2M}-1\right)\,,$
and hence
\eq$\label{intro-eq: dr-dv-du}
dr=\left(1-\frac{2M}{r}\right)\frac{dv-du}{2}\,.$
With the use of \eqref{intro-eq: dr-dv-du} in the line-element \eqref{intro-eq: EF-in} it is straightforward to obtain the line-element in the 
$\{v,u,\theta,\varphi\}$ coordinates. By doing so, we get
\eq$\label{intro-eq: v-u-line-el}
ds^2=-\left(1-\frac{2M}{r}\right)dvdu+r^2\, d\Omega_2^2\,,$
where the relation between $r$ and the coordinates $\{v,u\}$ is given in \eqref{intro-eq: v-u-r}. 
However, it is the same relation which manifests the problem with this specific coordinate system. 
As one can instantly deduce, the event horizon at $r=2M$ has moved infinitely far away.  
For $r\ra 2M$, equation \,\eqref{intro-eq: v-u-r} indicates that either $v\ra -\infty$ or $u\ra +\infty$. 
To bring these points at finite distances we introduce two new coordinates $\{\bar{v},\bar{u}\}$ which are defined as
\bal$\label{into-eq: bar-v}
\bar{v}& \equiv e^{v/4M}=\left|\frac{r}{2M}-1\right|^{1/2}\, e^{(r+t)/4M}\,,\\[1mm]
\label{into-eq: bar-u}
\bar{u}& \equiv -e^{-u/4M}=-\left|\frac{r}{2M}-1\right|^{1/2}\, e^{(r-t)/4M}\,.$
In the above equations the absolute value indicates that
\eq$\label{intro-eq: abs-val-r-2M}
\left|\frac{r}{2M}-1\right|=\left\{\begin{array}{cl}
\displaystyle{\frac{r}{2M}-1}\,, & r>2M\\[5mm]
\displaystyle{1-\frac{r}{2M}}\,, & r<2M\end{array}\right\}\,.$
Using eqs. \eqref{intro-eq: v-u-r} and \eqref{intro-eq: v-u-line-el}--\eqref{into-eq: bar-u}, the line-element in the coordinate system 
$\{\bar{v},\bar{u},\theta,\varphi\}$ can be easily evaluated and it is of the form
\eq$\label{intro-eq: bar-v-u-line-el}
ds^2=-\frac{32M^3}{r}\, e^{-r/2M}\, d\bar{v} d\bar{u}+r^2\, d\Omega_2^2\,.	$

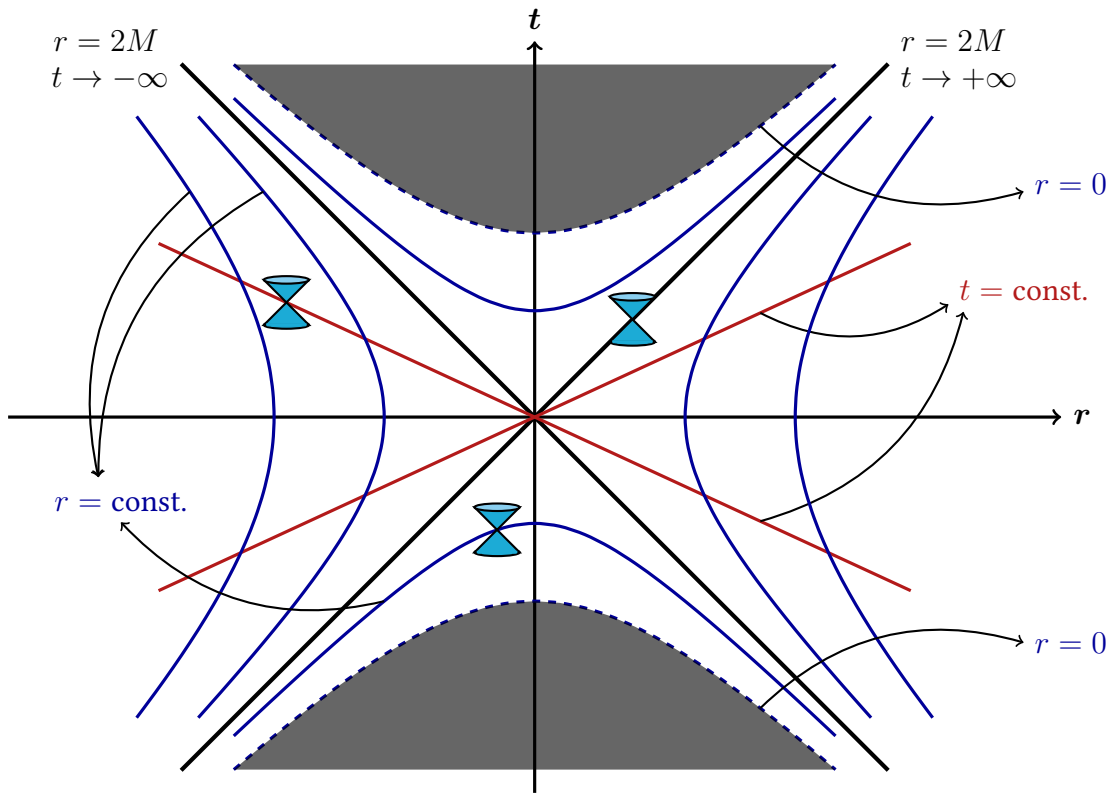
\begin{figure}[t]

\centering

\begin{tikzpicture}

\draw[->,very thick] (-7,0)--(7,0) node[right]{$\boldsymbol{r}$};
\draw[->,very thick] (0,-5)--(0,5) node[above]{$\boldsymbol{t}$};


\draw [ 
	color = black, 
	ultra thick] (-4.7,-4.7) -- (4.7,4.7) node[above right]{$r=2M$} node[above right, yshift=-1.2em] {$t\ra +\infty$};

\draw [ 
	color = black, 
	ultra thick] (4.7,-4.7) -- (-4.7,4.7) node[above left]{$r=2M \hspace{0.3em}$} node[above left, yshift=-1.2em] {$t\ra -\infty$};

\draw [color = nicered, domain = -5:5, smooth, very thick ] plot(\x,{tanh(1/2)*\x});
\path[->, thick] (3,{tanh(1/2)*3}) edge [bend right]  (5.5,1.5);
\node at (5.5,1.5) [above right, yshift=-1mm] {\textcolor{nicered}{$t=\text{const.}$}};

\draw [color = nicered, domain = -5:5, smooth, very thick ] plot(\x,{-tanh(1/2)*\x});
\path[->, thick] (3,{-tanh(1/2)*3}) edge [bend right]  (5.7,1.4);



\draw[ color = DarkBlue, domain = -4:4, smooth, very thick, dashed ] plot (\x, {sqrt((\x)^2+6)});
\fill [ color = black , domain = -4:4, opacity = 0.6 ] plot (\x, {sqrt((\x)^2+6)}) |- (4,{sqrt(22)});

\draw[ color = DarkBlue, domain = -4:4, smooth, very thick, dashed ] plot (\x, {-sqrt((\x)^2+6)});
\fill [ color = black , domain = -4:4, opacity = 0.6 ] plot (\x, {-sqrt((\x)^2+6)}) |- (4,{-sqrt(22)});

\draw [ color = DarkBlue, domain = -4:4, smooth, very thick ] plot (\x, {sqrt((\x)^2+2)});

\draw [ color = DarkBlue, domain = -4:4, smooth, very thick ] plot (\x, {-sqrt((\x)^2+2)});

\draw [ color = DarkBlue, domain = -4:4, smooth, very thick ] plot ( {-sqrt((\x)^2+4)},\x);

\draw [ color = DarkBlue, domain = -4:4, smooth, very thick ] plot ( {-sqrt((\x)^2+12)},\x);

\draw [ color = DarkBlue, domain = -4:4, smooth, very thick ] plot ( {sqrt((\x)^2+4)},\x);

\draw [ color = DarkBlue, domain = -4:4, smooth, very thick ] plot ( {sqrt((\x)^2+12)},\x);

\node at (6.5,3) [right, yshift=1mm] {\textcolor{DarkBlue}{$r=0$}};
\path[->, thick] (3,{sqrt(15)}) edge [bend right]  (6.5,3);
\node at (6.5,-3) [right, yshift=0mm] {\textcolor{DarkBlue}{$r=0$}};
\path[->, thick] (3,{-sqrt(15)}) edge [bend left]  (6.5,-3) ;

\path[->, thick] ({-sqrt((3)^2+4)},3) edge [bend right]  (-5.8,-0.8);
\path[->, thick] ({-sqrt((3)^2+12)},3) edge [bend right]  (-5.8,-0.8);
\path[->, thick] (-2,{-sqrt((-2)^2+2)}) edge [bend left]  (-5.5,-1.4);
\node at (-5.5,-1.4) [above, yshift=0mm] {\textcolor{DarkBlue}{$r=\text{const.}$}};


\def\dx{1.3}
\def\dy{1.3}
\def\rr{0.3}

\filldraw[color = black, fill = brightcerulean, thick] (\dx,\dy) -- ({\dx+tan(45)*\rr},{\dy+\rr}) -- ({\dx-tan(45)*\rr},{\dy+\rr}) -- (\dx,\dy);
\filldraw[color = black, fill = babyblue, thick] (\dx,\dy+\rr)
	ellipse [ x radius = {tan(45)*\rr}, y radius = {tan(45)*\rr/6}];
\filldraw[color = black, fill = brightcerulean, thick] (\dx,\dy-\rr)
	ellipse [ x radius = {tan(45)*\rr}, y radius = {tan(45)*\rr/6}];
\filldraw[color = black, fill = brightcerulean, thick] (\dx,\dy) -- ({\dx+tan(45)*\rr},{\dy-\rr}) -- ({\dx-tan(45)*\rr},{\dy-\rr}) -- (\dx,\dy);
\draw[line width = 0.3mm, color = brightcerulean] ({\dx+tan(45)*\rr-0.03},{\dy-\rr+0.005}) -- ({\dx-tan(45)*\rr+0.03},{\dy-\rr+0.005});
\draw[line width = 0.25mm, color = brightcerulean] ({\dx+tan(45)*\rr-0.1},{\dy-\rr-0.01}) -- ({\dx-tan(45)*\rr+0.1},{\dy-\rr-0.01});
\draw[line width = 0.15mm, color = brightcerulean] ({\dx+tan(45)*\rr-0.04},{\dy-\rr-0.01}) -- ({\dx-tan(45)*\rr+0.04},{\dy-\rr-0.01});

\def\dxx{-0.5}
\def\dyy{-1.5}

\filldraw[color = black, fill = brightcerulean, thick] (\dxx,\dyy) -- ({\dxx+tan(45)*\rr},{\dyy+\rr}) -- ({\dxx-tan(45)*\rr},{\dyy+\rr}) -- (\dxx,\dyy);
\filldraw[color = black, fill = babyblue, thick] (\dxx,\dyy+\rr)
	ellipse [ x radius = {tan(45)*\rr}, y radius = {tan(45)*\rr/6}];
\filldraw[color = black, fill = brightcerulean, thick] (\dxx,\dyy-\rr)
	ellipse [ x radius = {tan(45)*\rr}, y radius = {tan(45)*\rr/6}];
\filldraw[color = black, fill = brightcerulean, thick] (\dxx,\dyy) -- ({\dxx+tan(45)*\rr},{\dyy-\rr}) -- ({\dxx-tan(45)*\rr},{\dyy-\rr}) -- (\dxx,\dyy);
\draw[line width = 0.3mm, color = brightcerulean] ({\dxx+tan(45)*\rr-0.03},{\dyy-\rr+0.005}) -- ({\dxx-tan(45)*\rr+0.03},{\dyy-\rr+0.005});
\draw[line width = 0.25mm, color = brightcerulean] ({\dxx+tan(45)*\rr-0.1},{\dyy-\rr-0.01}) -- ({\dxx-tan(45)*\rr+0.1},{\dyy-\rr-0.01});
\draw[line width = 0.15mm, color = brightcerulean] ({\dxx+tan(45)*\rr-0.04},{\dyy-\rr-0.01}) -- ({\dxx-tan(45)*\rr+0.04},{\dyy-\rr-0.01});

\def\dxxx{-3.3}
\def\dyyy{1.525}

\filldraw[color = black, fill = brightcerulean, thick] (\dxxx,\dyyy) -- ({\dxxx+tan(45)*\rr},{\dyyy+\rr}) -- ({\dxxx-tan(45)*\rr},{\dyyy+\rr}) -- (\dxxx,\dyyy);
\filldraw[color = black, fill = babyblue, thick] (\dxxx,\dyyy+\rr)
	ellipse [ x radius = {tan(45)*\rr}, y radius = {tan(45)*\rr/6}];
\filldraw[color = black, fill = brightcerulean, thick] (\dxxx,\dyyy-\rr)
	ellipse [ x radius = {tan(45)*\rr}, y radius = {tan(45)*\rr/6}];
\filldraw[color = black, fill = brightcerulean, thick] (\dxxx,\dyyy) -- ({\dxxx+tan(45)*\rr},{\dyyy-\rr}) -- ({\dxxx-tan(45)*\rr},{\dyyy-\rr}) -- (\dxxx,\dyyy);
\draw[line width = 0.3mm, color = brightcerulean] ({\dxxx+tan(45)*\rr-0.03},{\dyyy-\rr+0.005}) -- ({\dxxx-tan(45)*\rr+0.03},{\dyyy-\rr+0.005});
\draw[line width = 0.25mm, color = brightcerulean] ({\dxxx+tan(45)*\rr-0.1},{\dyyy-\rr-0.01}) -- ({\dxxx-tan(45)*\rr+0.1},{\dyyy-\rr-0.01});
\draw[line width = 0.15mm, color = brightcerulean] ({\dxxx+tan(45)*\rr-0.04},{\dyyy-\rr-0.01}) -- ({\dxxx-tan(45)*\rr+0.04},{\dyyy-\rr-0.01});

\end{tikzpicture}

\caption{The structure of the spacetime in the Kruskal-Szekeres coordinates.}
\label{intro-fig: KS-spacetime}

\end{figure}

As a final step, we will make one last coordinate transformation in order to establish a coordinate system with one timelike and three spacelike coordinates.
In this way, any information about the geometry of the spacetime will be much easier to be extracted.
To this end, we define the coordinates $\{\boldsymbol{t},\boldsymbol{r}\}$ via the relations
\gat$\label{intro-eq: KS-t-r}
\boldsymbol{t} \equiv \frac{1}{2}(\bar{v}+\bar{u})=\left|\frac{r}{2M}-1\right|^{1/2}\, e^{r/4M}\, \sinh\left(\frac{t}{4M}\right)\,,\\[1mm]
\boldsymbol{r} \equiv \frac{1}{2}(\bar{v}-\bar{u})=\left|\frac{r}{2M}-1\right|^{1/2}\, e^{r/4M}\, \cosh\left(\frac{t}{4M}\right)\,.$
The coordinates $\{\boldsymbol{t},\boldsymbol{r},\theta,\varphi\}$ are known as 
\textbf{Kruskal-Szekeres coordinates}\,\footnote{Martin David Kruskal (1925,\,New York--2006,\,Princeton) was an American mathematician and 
physicist.}\footnote{George Szekeres (1911,\,Budabest--2005,\,Adelaide) was a Hungarian-Australian mathematician.} \cite{Kruskal,Szekeres}, and the 
line-element which ensues from this coordinate system has the form
\eq$\label{intro-eq: KS-t-r-line-ell}
ds^2=\frac{32M^3}{r}\, e^{-r/2M}\left(-d\boldsymbol{t}^2+d\boldsymbol{r}^{2}\right)+r^2\, d\Omega_2^2\,.$
Notice that the coordinate singularity at the Schwarzschild radius has been eliminated altogether in this coordinate system. 
In terms of the new coordinates, the original radial coordinate $r$ is defined via the relation	
\eq$\label{intro-eq: KS-t-r--r}
\boldsymbol{r}^2-\boldsymbol{t}^2=\left(\frac{r}{2M}-1\right)\, e^{r/2M}\,,$
while the initial timelike coordinate $t$ is evaluated through
\eq$\label{intro-eq: KS-t-r--t}
\frac{\boldsymbol{t}}{\boldsymbol{r}}=\tanh\left(\frac{t}{4M}\right)\,.$
As it can be immediately determined from the line-element \eqref{intro-eq: KS-t-r-line-ell}, the radial null trajectories in the Kruskal-Szekeres coordinates 
satisfy the equation $d\boldsymbol{t}/d\boldsymbol{r}=\pm 1$, which by its turn leads to
\eq$\label{intro-eq: KS-null-traj}
\boldsymbol{t}=\pm \boldsymbol{r}+\text{constant}\,.$
Consequently, the light-cone structure remains invariant independently of where we are in the spacetime.
Note also that the surfaces which are characterized by constant $r$ define hyperbolas in the $\boldsymbol{r}$-$\boldsymbol{t}$ plane, since
eq.\,\eqref{intro-eq: KS-t-r--r} takes the form $\boldsymbol{r}^2-\boldsymbol{t}^2=\text{constant}$.
In the special case where $r=2M$ the previous relation simplifies to $\boldsymbol{t}=\pm \boldsymbol{r}$. 
In addition to the above, eq.\,\eqref{intro-eq: KS-t-r--t} indicates that the surfaces which are characterized by constant $t$ appear as straight lines 
in the $\boldsymbol{r}$-$\boldsymbol{t}$ plane. 
The slope of these lines is determined by the function $\tanh(t/4M)$. 
Notice that in cases where $t\ra \pm \infty$ we obtain $\boldsymbol{t}=\pm \boldsymbol{r}$, respectively.
Therefore, the surfaces which are defined by $r=2M$ coincide with the surfaces of $t\ra \pm \infty$.
Everything we described so far is depicted plainly in \fref{intro-fig: KS-spacetime}. 
The spacetime geometry in the Kruskal-Szekeres coordinates has many more interesting properties that one can discover. 
However, for the needs of this dissertation the preceding analysis will suffice.

\vspace*{-0.5em}

\mysubsection{Additional black-hole solutions}

The Schwarzschild solution that we examined previously in great detail is not the only black-hole solution that General Relativity predicts, but is
the simplest one.
If one allows the existence of matter and/or energy in the vicinity of a black hole, new solutions will emerge with a number of extra properties and 
spacetime related phenomena.
In what follows, we present very briefly some more general black-hole solutions. 
These solutions will appear in subsequent chapters in the context of higher-dimensional gravitational models. 
Hence, it will prove beneficial to get a first taste of these solutions here.

\begin{itemize}

\item[\chapc{$\blacktriangleright$}] \textbf{Schwarzschild (anti-)de Sitter solution:}

In an attempt to obtain a more realistic black hole compared to the Schwarz\-schild solution, one may consider a black hole which does not reside in an 
empty universe, but instead, in a universe with a cosmological constant $\Lambda$. 
In case where the cosmological constant is positive, we have the so called Schwarzschild de Sitter\,\footnote{Willem de Sitter (1872,\,Sneek--1934,\,Leiden)
was a Dutch physicist, and astronomer.} spacetime, while the Schwarzschild anti-de Sitter spacetime 
emerges when the cosmological constant is negative. The gravitational field equations in both scenarios are of the form 
\eq$\label{intro-eq: schw-a-dS-gr-eqs}
G_{\mu\nu}-\Lambda\, g_{\mu\nu}=0\,.$
Assuming a static and spherically symmetric spacetime,---see eq.\,\eqref{intro-eq: stat-spher3}---the above equations lead to the Schwarzschild 
(anti-)de Sitter line-element, which is given below
\eq$\label{intro-eq: schw-a-dS-line-el}
ds^2=-\left(1-\frac{2M}{r}-\frac{\Lambda}{3}\,r^2\right)dt^2+\left(1-\frac{2M}{r}-\frac{\Lambda}{3}\,r^2\right)^{-1}dr^2+
r^2\, d\Omega_2^2\,.$
Notice that the above line-element reduces to the Schwarzschild line-element when $\Lambda=0$, as expected. 
However, for $\Lambda\neq 0$, the curvature invariant quantities are evaluated to be
\eq$\label{intro-eq: schw-a-dS-curvs}
R_{\text{S(a-)dS}}=4\Lambda\,,\hspace{1.5em}\mathcal{R}_\text{S(a-)dS}=4\Lambda^2\,,\hspace{1.5em}
\mathcal{K}_\text{S(a-)dS}=\frac{48M^2}{r^6}+\frac{8\Lambda^2}{3}\,,$

while the radial null trajectories are characterized by the relation

\eq$\label{intro-eq: schw-dS}
\frac{dt}{dr}=\pm\left(1-\frac{2M}{r}-\frac{\Lambda}{3}\,r^2\right)^{-1}\,.$

The roots of the depressed cubic polynomial $r^3-(3/\Lambda)\, r+6M/\Lambda=0$ in the denominator, correspond to the horizons in the spacetime. 
Therefore, the number of the horizons may be specified by the value of the discriminant, which for the aforementioned polynomial is evaluated to be 
$\Delta=108(1/\Lambda-9M^2)/\Lambda^2$.

\newpage
Consequently, we may distinguish the following cases:

\begin{itemize}

\item[\chapc{$\bullet$}] For $1/\Lambda>9M^2$, there are three distinct real roots to the polynomial, two of which are positive and one negative.
Therefore, in this case the spacetime has \textit{two horizons}. The inner horizon belongs to the black hole, while the second one is the cosmological horizon.

\item[\chapc{$\bullet$}] For $1/\Lambda=9M^2$, the polynomial has one negative real root and one positive double root.
In this case, the cosmological horizon overlaps with the horizon of the black hole. 
This case is also known as \textit{Nariai limit}\,\footnote{Hidekazu Nariai (1924,\,Izumo-Taisha--1990,\,Takehara) was a Japanese astrophysicist and 
cosmologist.} \cite{Nariai}. 

\item[\chapc{$\bullet$}] Finally, for $1/\Lambda<9M^2$, the polynomial has one real root and two complex conjugate ones.
However, in this case, two distinct subcases emerge. Thus, for $\Lambda<0$, we have a Schwarzschild anti-de Sitter spacetime, and the real root of the aforementioned 
polynomial is positive. 
Consequently, there is only \textit{one horizon} in the spacetime, the black-hole one. 
On the other hand, if $\Lambda>0$, then the real root is negative. 
Hence, although we have a Schwarzschild de Sitter spacetime, the resulting object is a \textit{naked singularity}.

\end{itemize}

\item[\chapc{$\blacktriangleright$}] \textbf{Reissner-Nordstr\"{o}m solution:}

The Reissner-Nordstr\"{o}m\,\footnote{Hans Jacob Reissner (1874,\,Berlin--1967,\,Mt.\,Angel,\,Oregon) was a German aeronautical engineer whose avocation
was mathematical physics.}\footnote{Gunnar Nordstr\"{o}m (1881,\,Helsinki--1923,\,Helsinki) was a Finnish theoretical physicist.}\cite{Reissner,Nordstrom}
solution describes the geometry of an electrically charged black hole. 
Although it is highly unlikely for a black hole to maintain its charge for a long period of time, it is still worth examining this solution since it exhibits some very 
interesting features. 
As with the previous solutions, the spacetime here is also considered static and spherically symmetric.
The electromagnetic field of the black hole is described by the Lagrangian density $\lagr_{em}=-\frac{1}{4}F^{\mu\nu}F_{\mu\nu}$, while the covariant
components of the electromagnetic tensor are defined via the relation $F_{\mu\nu}=\pa_\mu A_\nu-\pa_\nu A_\mu$. 
The gravitational field equations are of the form
\eq$\label{intro-eq: grav-eq-RN}
G_{\mu\nu}=\kappa^2\, T^{(em)}_{\mu\nu}\,,$
while the energy-momentum tensor $T^{(em)}_{\mu\nu}$ is sourced by the electromagnetic field in the vicinity of the black hole. 
Using the preceding expression for the Lagrangian density $\lagr_{em}$ together with eqs. \eqref{intro-eq: var-metr}, \eqref{intro-eq: var-ene}, one can readily 
evaluate that
\eq$\label{intro-eq: RN-en-mom}
T^{(em)}_{\mu\nu}=F_{\mu\lam}F_{\nu}{}^\lam-\frac{1}{4}g_{\mu\nu}F_{\alp\beta}F^{\alp\beta}\,.$
Assuming now that $A_{\mu}=(q/r,0,0,0)$, and solving the field equations for a static and spherically symmetric line-element, one obtains the
Reissner-Nordstr\"{o}m line-element
\eq$\label{intro-eq: RN-line-el}
ds^2=-\left(1-\frac{2M}{r}+\frac{Q^2}{r^2}\right)dt^2+\left(1-\frac{2M}{r}+\frac{Q^2}{r^2}\right)^{-1}dr^2+r^2\, d\Omega_2^2\,.$
In the above, the parameter $Q$ is directly related with the charge $q$, while the curvature invariant quantities in this case are of the form
\gat$\label{intro-eq: RN-curvs}
R_\text{RN}=0\,,\hspace{1.5em}\mathcal{R}_\text{RN}=\frac{4Q^4}{r^8}\,,\hspace{1.5em}
\mathcal{K}_\text{RN}=\frac{56Q^4}{r^8}-\frac{96MQ^2}{r^7}+\frac{48M^2}{r^6}\,.$
Studying the radial null trajectories of the above line-element, one finds that the roots of the quadratic polynomial $r^2-2M r+Q^2=0$ specify the number of
horizons in the spacetime. The roots can be easily determined and they are expressed as
\eq$\label{intro-eq: RN-horzs}
r_{\pm}=M\pm \sqrt{M^2-Q^2}\,.$
Hence, depending on the value of the quantity $M^2-Q^2$, we can distinguish three cases.

\begin{itemize}

\item[\chapc{$\bullet$}] In case where $M^2>Q^2$ the spacetime has \textit{two horizons}. 
The outer horizon, $r_+$, is the black-hole horizon, while the inner horizon, $r_-$, is the Cauchy horizon (or Cauchy surface).

\item[\chapc{$\bullet$}] For $M^2=Q^2$, we have an \textit{extremal black hole}, since now the Cauchy and the black-hole horizons coincide. 

\item[\chapc{$\bullet$}] Finally, for $M^2<Q^2$, there is no horizon and we are left with a \textit{naked singularity}.

\end{itemize}

For more details about the geometry of the Reissner-Nordstr\"{o}m spacetime, and the consequences of the Cauchy horizon to the trajectories of infalling objects 
the interested reader may take a look at the sixth chapter of the book \cite{carroll:2019}.

\item[\chapc{$\blacktriangleright$}] \textbf{Reissner-Nordstr\"{o}m (anti-)de Sitter solution:}

As it can be easily deduced by the preceding discussion, the spacetime geometry of a Reissner-Nordstr\"{o}m (anti-)de Sitter solution describes an electrically 
charged black hole which lives in a universe with non-zero cosmological constant.
The line-element of the aforementioned black-hole solution is given by
{\fontsize{10.5}{10.5}{\eq$\label{intro-eq: RN-adS-line}
ds^2=-\left(1-\frac{2M}{r}+\frac{Q^2}{r^2}-\frac{\Lambda}{3}\, r^2\right)dt^2+\left(1-\frac{2M}{r}+\frac{Q^2}{r^2}-\frac{\Lambda}{3}\, r^2\right)^{-1}
dr^2+r^2\, d\Omega_2^2\,,$}}

\vspace*{-1em}
while the curvature quantities can be directly derived by the previous solutions. Thus, we have
\eq$\label{intro-eq: RN-a-dS-curvs}
R=R_\text{S(a-)dS}+R_\text{RN}\,,\hspace{1.5em}\mathcal{R}=\mathcal{R}_\text{S(a-)dS}+\mathcal{R}_\text{RN}\,,
\hspace{1.5em}\mathcal{K}=\mathcal{K}_\text{S(a-)dS}+\mathcal{K}_\text{RN}\,.$
An extensive study of the roots of the quartic polynomial $\Lambda\, r^4/3-r^2+2Mr-Q^2=0$, which defines the number of horizons in the spacetime, is
performed in \chapref{Chap: P5} and more specifically in \fref{P5fig: Hor-chart}.

\end{itemize}

\begin{figure}[t]
\centering
\includegraphics[width=1\textwidth]{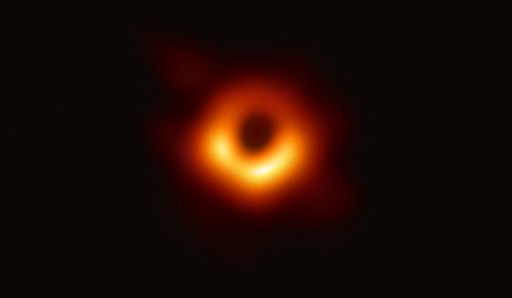}
\caption[]{Image from the Event Horizon Telescope (EHT) depicting the supermassive black hole at the center of the galaxy Messier 87 (M87).}
\label{intro-fig: M87}
\end{figure}

Black holes are indisputably the most astonishing and mind-bending solutions that General Relativity admits.
In April of 2019, the international collaboration Event Horizon Telescope (EHT) published the first image of a supermassive black hole which resides in 
the center of our neighbour galaxy, Messier 87 (M87) \cite{Akiyama:2019cqa}. 
The black hole is $6.5$ billion times more massive than the Sun, and is about $55$\,Mly ($55\times 10^6$ light-years) away from the Earth.
Using the formula $r_s=2G_N M/c^2$ one may calculate that the Schwarzschild radius of the black hole is approximately $1.92\times 10^{10}$\,km, 
a distance which is equivalent to $3.2$ times the distance between Pluto and Sun. 
Although enormous in size, the black hole is so far away from us that a decade ago it was simply impossible to observe such an object.
Given the size of the black hole and its distance from us, its observation is analogous to observing a golf ball from a $576600$\,km distance, that is
$1.5$ times the distance between Earth and Moon.
The fact that the international collaboration EHT managed to <<take a photo>> of such a distant black hole is really a stupendous achievement.
The image of M87 black hole depicted in \fref{intro-fig: M87}, together with the dozens of recent gravitational-wave observations from the 
LIGO-Virgo experiment \cite{Abbott:2016blz,LIGOScientific:2018mvr,Abbott:2020uma} have irrefutably proved that black holes are real astrophysical objects.
Having developed the necessary apparatus to observe and study them, they surely constitute the leading candidates for providing us a better understanding
of gravity at high energies.
Hopefully, in some decades from now, we might even be able to test quantum gravity effects using black-hole observations.
One thing is sure though, the aforementioned experiments have provided us with a new way of gazing the universe, and this can only lead to progress.

\mysection{Gravitational theories with extra dimensions}

Soon after Einstein formulated General Relativity, a quest for a unified theory incorporating both gravity and electromagnetism commenced.
The first endeavor to this direction came from Theodor Franz Eduard Kaluza (1885,\,Wilhelmsthal--1954,\,G\"{o}ttingen) who in 1921 presented
a classical extension of GR to five-dimensions \cite{Kaluza:1921tu}.
The idea of Kaluza was to unify General Relativity and electromagnetism in a purely geometrical way.
To this end, he constructed the five-dimensional metric tensor in such a way that the vacuum solution of the five-dimensional gravitational field 
equations would lead to Einstein's equations on the one hand, and to Maxwell's equations on the other hand.
The metric tensor of a five-dimensional spacetime has $15$ independent components, $10$ of which were identified with the four-dimensional metric components,
$4$ with each component of the electromagnetic potential $A_\mu$, and $1$ with a scalar field $\phi$. 
The aforementioned scalar field, $\phi$, is usually called \textit{\textbf{dilaton}} (or \textit{radion} in the case of Kaluza-Klein model) and it appears in 
higher-dimensional theories when the volume of the compactified extra-dimensional space is a dynamical quantity. 
Kaluza also imposed the ``\textit{\textbf{cylinder condition}}'', namely that no metric component depends on the extra dimension. 
This assumption was made for both physical and practical reasons.
Without this condition, the five-dimensional field equations present an increased complexity.

In 1926, Oscar Benjamin Klein (1894,\,M\"{o}rdby,\,Sweden--1977,\,Stockholm) tried to give a quantum interpretation to Kaluza's five-dimensional theory,
and a physical explanation to the aforementioned cylinder condition \cite{Klein1, Klein2}.
He suggested that the extra dimension was microscopic in size and curled up in a circular way with its radius being of the order of $10^{-30}$\,cm.

\mysubsection{Kaluza-Klein model in a nutshell}

The five-dimensional line-element of the Kaluza-Klein (KK) model is of the form
\eq$\label{intro-eq: KK-line}
ds^2=g_{MN}\, dx^M dx^N=g_{\mu\nu}\, dx^\mu dx^\nu + \phi^2 \left(\xi^2 A_\mu A_\nu dx^\mu dx^\nu + 2 \xi A_\mu dx^\mu dy + dy^2 \right)\,,$
In the above, $y$ denotes the extra spatial dimension, $\xi^2 = 16\pi G_N=2\kappa^2$, while the capital Latin indices $M, N, L, \ldots$ denote the coordinates of 
the higher-dimensional spacetime. 
From now on the Greek indices $\mu,\nu,\ldots$ will be used to denote only the four-dimensional spacetime coordinates.
The cylinder condition also implies that $\pa_y g_{MN}=0,\, \forall M,N$, which by its turn leads to $\pa_y \phi =0$ and $\pa_y A_\mu=0,\, \forall \mu$.

The action functional of the KK model is given by
\eq$\label{intro-eq: KK-act}
S_{KK} = \int d^4x dy \sqrt{-g_{(5)}}\, \frac{R_{(5)}}{2\kappa_{(5)}^2}\,,$
where $R_{(5)}$ is the five-dimensional Ricci scalar, $g_{(5)}$ is the determinant of the five-di\-men\-sion\-al metric, while $\kappa_{(5)}^2$ is defined in terms 
of the five-dimensional Newton's constant $G_{N(5)}$. 
Using the metric defined by the line-element \eqref{intro-eq: KK-line} together with the fact that $\kappa^2 \equiv \kappa_{(5)}^2 /\int dy$, one may calculate that
\eq$\label{intro-eq: KK-act-decom}
S_{KK}=\int d^4x \sqrt{-g}\, \phi \left( \frac{R}{2\kappa^2} + \frac{1}{4} \phi^2 F^{\alp\beta}F_{\alp \beta} +
\frac{2}{3\xi^2} \frac{\pa^\alp\phi\, \pa_\alp \phi}{\phi^2} \right)\,,$
where $F_{\mu\nu}=\pa_\mu A_\nu-\pa_\nu A_\mu$. 
By imposing the principle of least action to the previous action functional one can derive the field equations of the model, which are presented below:
\gat$\label{intro-eq: KK-f-eqs-1}
G_{\mu\nu} = \kappa^2\, \phi^2 T^{(em)}_{\mu\nu} - \frac{1}{\phi} \left[ \nabla_\mu (\pa_\nu \phi) - g_{\mu\nu} \square \phi \right]\,,\\[2mm]
\label{intro-eq: KK-f-eqs-2}
\nabla^\lam F_{\lam\mu} = -3\,\frac{\pa^\rho \phi}{\phi}\, F_{\rho\mu}\,, \\[2mm]
\label{intro-eq: KK-f-eqs-3}
\square \phi = \frac{\kappa^2 \phi^3}{2}\, F^{\alp \beta} F_{\alp\beta}\,.$
In eq.\,\eqref{intro-eq: KK-f-eqs-1}, $T^{(em)}_{\mu\nu}$ are the covariant components of the electromagnetic stress-energy tensor which are evaluated
according to the relation:
\eq$
T^{(em)}_{\mu\nu}=F_{\mu\lam}F_{\nu}{}^\lam-\frac{1}{4}g_{\mu\nu}F_{\alp\beta}F^{\alp\beta}\,.$

Due to the fact that physicists were not familiar with scalar fields back in 1920s, they set $\phi=1$. 
As a result the above equations reduce to
\eq$
G_{\mu\nu} = \kappa^2\, T^{(em)}_{\mu\nu}\,, \hspace{1.5em} \nabla^\lam F_{\lam\mu}=0\,,$
which are no other than the Einstein and Maxwell equations. 
However, the choice $\phi=\text{constant}$ is only consistent with eq.\,\eqref{intro-eq: KK-f-eqs-3} if and only if $ F^{\alp \beta} F_{\alp\beta}=0$, 
something which first noted by Jordan \cite{P.Jordan} and Thiry \cite{Thiry} in 1940s.

Another interesting case of the Kaluza-Klein model is when $\phi\neq \text{constant}$ and $A_\mu=0$. 
In this particular case the line-element \eqref{intro-eq: KK-line} simplifies to
\eq$\label{intro-eq: KK-line-A-0}
ds^2=g_{\mu\nu}\, dx^\mu dx^\nu + \phi^2 dy^2 \,,$
while the action functional \eqref{intro-eq: KK-act} reduces to
\eq$
S=\int d^4x\sqrt{-g}\, \frac{R}{2\kappa^2}\,\phi\,.$
This is the special case $\omega=0$ of the Brans-Dicke action \cite{Brans-Dicke}:
\eq$
S_{BD}=\int d^4x \sqrt{-g}\left(\frac{R}{2\kappa^2}\,\phi +\omega\, \frac{\pa^\lam \phi\, \pa_\lam \phi}{\phi}\right)\,.$
For more details about the Kaluza-Klein model see \cite{Overduin}.

The fact that both General Relativity and electromagnetic radiation emerge purely geometrically from an empty five-dimensional spacetime is simply
remarkable. 
Although we now know that there are much more in nature than gravity and electromagnetism, the idea that reality could be higher-dimensional in a more
fundamental level is clearly fascinating.
It should not be surprising that this simple and elegant idea led physicists in 1970s and 1980s to develop multi-dimensional theories, such as String Theory
\cite{Green,Polchinski} and supergravity (SUGRA), in an attempt to unify all known fundamental interactions and particles.

\begin{figure}[H]
\centering
\includegraphics[width=1\textwidth]{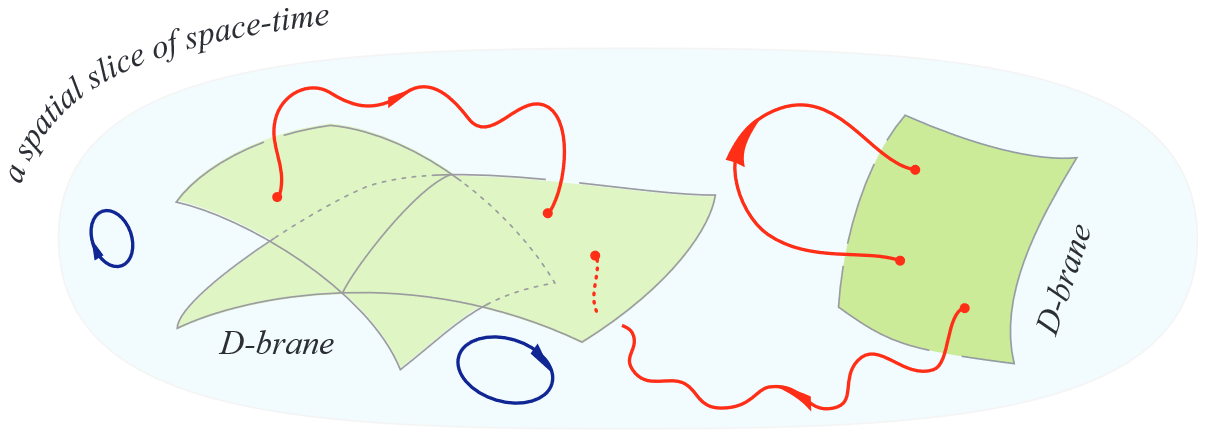}
\caption[]{D-branes, open and closed strings. The above illustration was created by Chien-Hao Liu and Shing-Tung Yau. It was originally presented in
their paper \cite{Liu-Yao}.}
\label{intro-fig: D-branes}
\end{figure}

In 1981, Edward Witten (born 1951,\,Baltimore) showed in his paper ``\textit{Search for a realistic Kaluza-Klein theory}'' \cite{Witten81} that 
eleven-dimensional supergravity with seven compactified dimensions is the smallest multi-dimensional theory which is big enough to incorporate the gauge group 
of the Standard Model, namely $SU(3)\times SU(2)\times U(1)$. 
Witten also conjectured the existence of a theory, later named M-theory, which unifies all five distinct ten-dimensional superstring theories.\,\footnote{Type I, 
$SO(32)$ heterotic, $E_8\times E_8$ heterotic,  Type IIA, and Type IIB.} 
He observed that each one of them constitutes a different limiting case of an eleven-dimensional supergravity theory.
Although the complete formulation of M-theory is not yet known, we know for sure that $p$-branes play an important role in the theory.
In the context of extra-dimensional models, $p$-branes are $(1+p)$-dimensional objects which generalize the notion of point-like particles. 
For instance, a $0$-brane corresponds to a particle, while a $1$-brane to a string. 
A special category of $p$-branes are the so-called D-branes,\,\footnote{The letter ``D'' in D-branes refers to the Dirichlet boundary condition.} which 
become necessary when one considers open strings.
Open strings describe both fermions and bosons, and their endpoints are required to be attached on D-branes.
On the contrary, closed strings describe exclusively gravitational degrees of freedom and they are free to propagate in the bulk.\,\footnote{In higher-dimensional
models the notion of bulk is used to denote the whole (hyper-)volume of the spacetime.} 
The intimate relationship between open strings and D-branes is depicted in \fref{intro-fig: D-branes}.

Besides String Theory and SUGRA, in late 1990s two alternative higher-dimensional models were added to the literature.
The first one in chronological order is known as the ADD model (1998) \cite{ADD1,ADD2,ADD3}, while the second one is the so-called RS model (1999)
\cite{RS1,RS2}.
Utilizing the concept of D-branes, both ADD and RS models assume that our observable universe is a $3$-brane (referred to henceforth simply as the brane) 
embedded in a $(4+n)$-dimensional spacetime (the bulk), with $n$ indicating the number of the extra spatial dimensions.
In such \textit{\textbf{braneworld}} scenarios the Standard Model particles and fields  are confined on the brane, while gravity can freely propagate in the bulk.
Braneworld models were originally motivated by the Hierarchy Problem. 
That is the huge discrepancy between the electroweak scale $m_{EW}\approx 250\,\text{GeV}$ and the Planck scale 
$M_{Pl}\approx  2\times 10^{18	}\,\text{GeV}$, where gravity becomes as strong as the gauge interactions.
In these types of models, the distinction between the fundamental Planck scale---related to the existence of extra dimensions---and
the four-dimensional one, allows one to resolve the	hierarchy problem in a purely geometrical way.
In general, the fundamental Planck scale is assumed to be of the same order as the electroweak scale, while the observed Planck scale on the brane appears to
be only an effective one.
Let us now explore in more detail the aforementioned models.

\mysubsection{ADD model}

The first braneworld model that introduced the idea of large extra dimensions was proposed by Nima Arkani-Hamed, Savas Dimopoulos, and Gia Dvali 
(hence the name ADD) back in 1998 \cite{ADD1,ADD2,ADD3}. 
In their model, they allowed the existence of $n$ extra compact spatial dimensions of the same radius $R_{d}$.
Consequently, gravity experiences a total of $(4+n)$ dimensions, while the action functional regarding the gravitational sector is of the form
\eq$
S_{ADD}=\int d^4x\, d^n y \sqrt{-g_{(4+n)}}\frac{R_{(4+n)}}{2\kappa^2_{(4+n)}}\,.$ 
In the above action the coordinates $\{y^1,\ldots, y^n\}$ denote the extra spatial dimensions, $g_{(4+n)}$ is the determinant of the $(4+n)$-dimensional
metric, $R_{(4+n)}$ is the $(4+n)$-dimensional Ricci scalar, while $\kappa^2_{(4+n)}$ is the gravitational coupling constant which is defined
in terms of the fundamental Planck scale $M_{Pl(4+n)}$ or in terms of the higher-dimensional Newton's constant $G_{N(4+n)}$.

As it is discussed in \cite{ADD1}, the gravitational potential between two test masses $m_1$ and $m_2$ which are separated by a distance $r\ll R_d$
is given by the expression
\eq$\label{intro-eq: ADD-V1}
V(r\ll R_d) = G_{N(4+n)} \frac{m_1\, m_2}{r^{n+1}}\sim \frac{m_1\,m_2}{M_{Pl(4+n)}^{n+2}} \frac{1}{r^{n+1}}\,. $
The above relation can be derived by Gauss' law in $(4+n)$ dimensions.
Assuming now that the masses are placed at a distance $r\gg R_d$, then their gravitational flux lines do not pervade in the extra dimensions, and
thus the classical $1/r$ behavior of the potential is obtained
\eq$\label{intro-eq: ADD-V2}
V(r\gg R_d)\sim \frac{m_1\,m_2}{M_{Pl(4+n)}^{n+2}\, R_d^n} \frac{1}{r}\,.$
Comparing the above relation with the Newtonian potential
\eq$\label{intro-eq: Newton-V}
V_N(r)=G_N\frac{m_1\,m_2}{r}\sim \frac{m_1\,m_2}{M_{Pl}^2} \frac{1}{r}\,,$
it is straightforward to deduce that the fundamental Planck scale, $M_{Pl(4+n)}$, and the four-dimensional one, $M_{Pl}$, are related via
\eq$\label{intro-eq: MPlf-MPl}
M_{Pl}^2\sim M_{Pl(4+n)}^{n+2}\, R_d^n\,.
$
Solving the preceding equation with respect to $R_d$ and setting $M_{Pl(4+n)}=m_{EW}$, we obtain
\gat$
R_d\sim \left(\frac{M_{Pl}^2}{m_{EW}^{n+2}}\right)^{\frac{1}{n}}=\left(\frac{M_{Pl}^2}{m_{EW}^2}\right)^{\frac{1}{n}}\frac{\hbar c}{m_{EW}}
\xRightarrow[m_{EW}\,\approx\, 250\,\text{GeV}]{M_{Pl}\,\approx\, 2\times 10^{18}\,
\text{GeV}}\nonum\\[3mm]
\label{intro-eq: ADD-R}
R_d\sim 2^{3+\frac{6}{n}}\times 10^{\frac{30}{n}-19}\, \text{m}\,. $
For an explicit calculation of the Newtonian potential in $(4+n)$ dimensions one is suggested to take a look at \cite{ADD2,Kehagias:1999my,Floratos:1999bv}.

Equation \eqref{intro-eq: ADD-R} associates the number of the extra spatial dimensions $n$ with the size of their radius $R_d$.
For $n=1$ one can evaluate that $R_d\sim 10^{13}\, \text{m}$, while for $n=2$ one finds that $R_d\sim 1\,\text {mm}$.\,\footnote{
One might notice that the values of $R_d$ differ from the ones that appear in the original paper \cite{ADD1}.
This happens because in \cite{ADD1} the authors have taken the electroweak scale to be $1\,\text{TeV}$ while here it
was assumed to be $250\,\text{GeV}$.}
It is obvious that for $n=1$ and $R_d=10^{13}\,\text{m}$ Newton's gravitational law would differ significantly from the one that we experience in our
everyday lives. 
Therefore, the value $n=1$ can be immediately excluded.
Contemporary Cavendish-type experiments \cite{Kapner,Lee:2020zjt} have verified the conventional Newton's law at distances as small as $60\,\mu\text{m}$.
This means that for $M_{Pl(4+n)}\approx 250\,\text{GeV}$ the case $n=2$ is excluded as well.
Although it is difficult to measure gravitational deviations of the Newton's law in sub-millimeter distances, the fact that the fundamental gravitational scale
$M_{Pl(4+n)}$ can be equal to $250\, \text{GeV}$ (or even $1\, \text{TeV}$) gives us the opportunity to detect indirectly the existence of extra dimensions
in collider experiments through the formation of tiny black holes from highly energetic particles \cite{Franc,Argyres:1998qn,Banks:1999gd,Dimopoulos:2001hw,Giddings:2001bu,Kanti:2004nr,Kanti:2008eq,ATLAS:2015yln}. 
Collider experiments put strict bounds on the number of allowed extra dimensions and the type of compactification topology that the extra dimensions exhibit.

Despite the fact that the ADD model suffers from some serious conceptual problems, it is important to acknowledge that it was the first model that proposed the
existence of large extra dimensions. It was definitely a bold attempt towards the resolution of the Hierarchy Problem, and also provided an alternative way
of thinking on matters concerning higher-dimensional spacetimes.

\newpage
\vspace*{-4em}

\mysubsection{Randall-Sundrum (RS) models}

A year after the formulation of the ADD model Lisa Randall and Raman Sundrum presented their own braneworld scenario \cite{RS1,RS2}.
In fact, they built two separate braneworld models, which although they share a lot of similarities, each one has its own merits and pitfalls.
The compelling characteristic of the RS models is that now the $3$-brane possesses tension itself, and therefore is allowed to interact gravitationally with the bulk.
In particular, the first RS model (RS$1$) \cite{RS1} assumes the existence of two $3$-branes in the bulk and attains to explain the observed hierarchy between electroweak
and Planck scales through an exponentially warped extra dimension.
The aforementioned warping of the extra dimension arises from the presence of a negative five-dimensional cosmological constant $\Lambda_5$ in the bulk.
As a result, the bulk spacetime is AdS$_5$.
What is even more remarkable is that Randall and Sundrum managed to show that even in the case of a single brane and an infinite extra dimension, an observer on 
the brane would still perceive gravity as four-dimensional.
In the latter case where the extra dimension is infinite we have the so-called RS$2$ model \cite{RS2}.
Let us now proceed to a more detailed analysis of both RS models.\\[1mm]

\begin{flushleft}
\chapc{\large{\textbf{RS$\bold{1}$ model}}}
\end{flushleft}
\phantomsection

The RS$1$ model postulates the existence of one extra dimension, $y$, which is compactified on a circle $S^1$ (one-dimensional sphere) of radius $r_c$ and 
also possesses a ${\bf Z}_2$ symmetry. 
Hence the extra dimension is an $S^1/{\bf Z}_2$ orbifold, which by its turn implies that points $(x^\lam,y)$ and $(x^\lam,-y)$ are identical.
As it is illustrated in \fref{intro-fig: orbifold}, this type of compactification contains two fixed points, one at $y=0$ and another at $y=\pi r_c\equiv L$.

\begin{figure}[H]

\centering

\begin{tikzpicture}[scale=1]

\draw[very thick] (0,0) circle (2.5cm);

\draw [thick, dashed] (-2.5,0) -- (2.5,0);
\filldraw[black] (-2.5,0) node[left]{$0$} circle (2.5pt);
\filldraw[black] (2.5,0) circle (2.5pt) node[anchor=south west]{$\hspace{0.8em}\pi r_c$} node[anchor=north west]{$-\pi r_c$};

\def\ya{-2}
\draw[<->,very thick] ({\ya},{-sqrt((2.5)^2-(\ya)^2)}) -- ({\ya},{sqrt((2.5)^2-(\ya)^2)});

\def\ya{-1}
\draw[<->,very thick] ({\ya},{-sqrt((2.5)^2-(\ya)^2)}) -- ({\ya},{sqrt((2.5)^2-(\ya)^2)});

\def\ya{0}
\draw[<->,very thick] ({\ya},{-sqrt((2.5)^2-(\ya)^2)}) -- ({\ya},{sqrt((2.5)^2-(\ya)^2)});

\def\ya{1}
\draw[<->,very thick] ({\ya},{-sqrt((2.5)^2-(\ya)^2)}) -- ({\ya},{sqrt((2.5)^2-(\ya)^2)});

\def\ya{2}
\draw[<->,very thick] ({\ya},{-sqrt((2.5)^2-(\ya)^2)}) -- ({\ya},{sqrt((2.5)^2-(\ya)^2)});

\draw[->,very thick] (-2.8cm,0.7cm) arc (180:140:1.5cm)
node [midway,left]{$y$};

\draw[->,very thick] (-2.8cm,-0.7cm) arc (180:220:1.5cm)
node[midway,left]{$-y$};

\end{tikzpicture}

\caption{$S^1/{\bf Z}_2$ orbifold.}
\label{intro-fig: orbifold}

\end{figure}

\begin{figure}[t]
\centering
\includegraphics[width=0.9\textwidth]{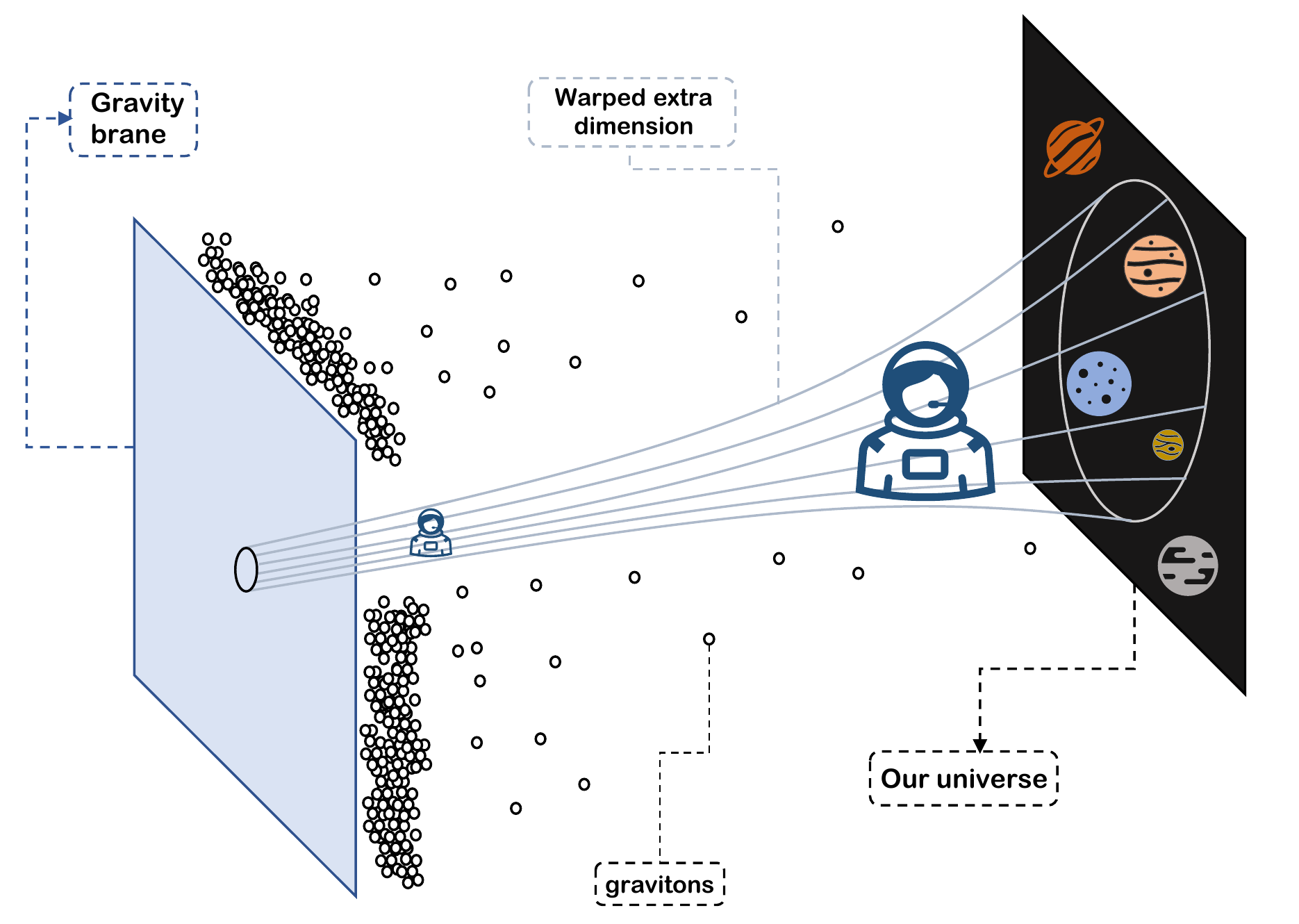}
\vspace*{-0em}
\caption[]{The RS$1$ model.}
\label{intro-fig: RS1}
\end{figure}

Although $y$ ranges from $-L$ to $L$, the metric tensor of the five-dimensional spacetime is entirely specified by the values in the range $0\leq y \leq L$.
The extra dimension is also confined by two $3$-branes, the \textit{hidden} and the \textit{visible} one. 
The former resides at $y=0$, while the latter at $y=L$.
The action describing the previous set-up is given by
\eq$\label{intro-eq: RS1-act}
S_{RS1} = S_{grav} + S_{vis} + S_{hid}\,,$
where
\gat$\label{intro-eq: RS1-act-grav}
S_{grav} = \int d^4x \int_{-L}^L dy \sqrt{-g} \left( \frac{R}{2\kappa^2_{(5)}} - \Lambda_5 \right) \,,\\[2mm]
\label{intro-eq: RS1-act-vis}
S_{vis} = \int d^4x \sqrt{-g_{vis}} \left( \lagr_{vis} - \sigma_{vis} \right) \,,\\[2mm]
\label{intro-eq: RS1-act-hid}
S_{hid} = \int d^4x \sqrt{-g_{hid}} \left( \lagr_{hid} - \sigma_{hid} \right) \,.$
Note that $g_{MN}\equiv g^{(5)}_{MN}=g^{(5)}_{MN}(x^\lam,y)$ denotes the components of the five-dimensional metric tensor, 
$g^{vis}_{\mu\nu}(x^\lam)\equiv g^{(5)}_{\mu\nu}(x^\lam, y=L)$ is the metric defined on the four-dimensional visible brane (our universe),
while $g^{hid}_{\mu\nu}(x^\lam)\equiv g^{(5)}_{\mu\nu}(x^\lam, y=0)$ is the metric defined on the four-dimensional hidden brane (or gravity brane).
In addition to the above, $R_{(5)}$ is the five-dimensional Ricci scalar, $\Lambda_5$ is the higher-dimensional cosmological constant, and
$\kappa_{(5)}$ is the gravitational coupling associated with the fundamental Planck scale $M_{Pl(5)}$ via the relation $\kappa^2_{(5)} \propto M_{Pl(5)}^{-3}$. 
The quantities $\sigma_{vis}$ and $\sigma_{hid}$ represent the tensions of the visible and hidden branes, respectively, whilst the Lagrangian densities
$\lagr_{vis}$ and $\lagr_{hid}$ incorporate the matter and the fields on each brane. 
A perspicuous illustration of the RS$1$ model is depicted in \fref{intro-fig: RS1}.

Employing the principle of least action at the action $S_{RS1}$, we are provided with the gravitational field equations.
The field equations presented below were derived under the assumption that $\lagr_{vis}=\lagr_{hid}=0$.
This particular assumption was made in order to determine the geometrical background of the model.
Thus, the five-dimensional Einstein's equations ensuing from the action $S_{RS1}$ are of the form
\bal$
\sqrt{-g}\, G_{MN} = - \kappa^2_{(5)} \left[\sqrt{-g}\, g_{MN}\, \Lambda_5 \right. +&\, \sqrt{-g_{vis}}\, \sigma_{vis}\, g^{vis}_{\mu\nu}\, \del^{\mu}_M\,
\del^{\nu}_N\, \del(y-L) \nonum\\[2mm]
\label{intro-eq: RS1-eqs}
+& \left. \sqrt{-g_{hid}}\, \sigma_{hid}\, g^{hid}_{\mu\nu}\, \del^{\mu}_M\, \del^{\nu}_N\, \del(y) \right]\,.$
In order to make progress from this point on, it is necessary to introduce an appropriate five-dimensional spacetime geometry. 
A property that is important to be imposed on the metric is to respect the Poincar\'{e} symmetry on both branes.
This property comes naturally from the fact that the induced metric on our brane ought to describe the physical world, hence, it should be Poincar\'{e} invariant.
A five-dimensional metric satisfying the previous requirement is given by the following line-element:
\eq$\label{intro-eq: RS-line-el}
ds^2=e^{2A(y)}\eta_{\mu\nu}dx^\mu dx^\nu+dy^2\,.$
The function $e^{2A(y)}$ is known as \textit{\textbf{warp factor}} and it will be determined by the gravitational field equations.
From eq.\,\eqref{intro-eq: RS-line-el} it is straightforward to deduce that $g^{vis}_{\mu\nu}=e^{2A(L)}\eta_{\mu\nu}$, while
$g^{hid}_{\mu\nu}=e^{2A(0)}\eta_{\mu\nu}$.
Using the metric ansatz of eq.\,\eqref{intro-eq: RS-line-el}, the non-zero gravitational field equations \eqref{intro-eq: RS1-eqs} are given below.
For $M=\mu$ and $N=\mu$ one obtains
\eq$\label{intro-eq: RS1-eq1}
6(\pa_y A)^2+3\pa_y^2 A=-\kappa_{(5)}^2\left[\Lambda_5 + \sigma_{vis}\, \del(y-L) + \sigma_{hid}\, \del(y)\right]\,,$
while for $M=N=y$ one finds
\eq$\label{intro-eq: RS1-eq2}
6(\pa_y A)^2=-\kappa_{(5)}^2 \Lambda_5\,.$
As it is dictated by eq.\,\eqref{intro-eq: RS1-eq2} the five-dimensional cosmological constant $\Lambda_5$ is mandatory to be negative-definite in order
the model to be self-consistent. 
Hence, by setting $k^2\equiv -\kappa_{(5)}^2 \Lambda_5/6$, the previous equation can be solved for $A(y)=\pm ky$.
The integration constant has been ignored since it can be absorbed in an overall constant rescaling of the brane coordinates $\{x^\mu\}$.
Moreover, since we need to maintain the orbifold symmetry of the extra dimension, we are obliged to express the warp factor in terms of $|y|$.
Consequently, the solution of eq.\,\eqref{intro-eq: RS1-eq2} is
\eq$\label{intro-eq: warp}
A(y)=-k|y|\,\hspace{1em}k>0\,.$
The reason that we chose to keep the minus sign will be understood shortly.
Before we proceed to the solution of eq.\,\eqref{intro-eq: RS1-eq1} one first needs to evaluate the derivatives $\pa_yA$ and $\pa_y^2A$.
So, for $y\neq 0$ we have
\eq$\label{intro-eq: warp-diff1}
\pa_yA=-k\pa_y|y|=-k\,sgn(y)=-k\left[H(y)-H(-y)\right]\,,$
where $H(y)\equiv \left\{
\begin{array}{lr}
0\,, & -L < y < 0 \\[1mm]
1\,, & 0 < y \leq L
\end{array}
\right\}\,.$
With the use of \eqref{intro-eq: warp-diff1} and the definition of the Heaviside step function $H(y)$, the evaluation of the second derivative $\pa_y^2A$
becomes a walk in the park. Hence, we find that
\eq$\label{intro-eq: warp-diff2}
\pa_y^2 A=-2k\left[\del(y)-\del(y-L)\right]\,.$
Using eqs. \eqref{intro-eq: RS1-eq2} and \eqref{intro-eq: warp-diff2} in \eqref{intro-eq: RS1-eq1} it is straightforward to derive that
\eq$\label{intr-eq: brane-ten}
\sigma_{hid}=-\sigma_{vis}=\frac{6k}{\kappa_{(5)}^2}\,.$

Let us now examine what does the RS model have to say about the hierarchy problem. 
To this end, we focus on the visible brane (our universe) and we consider the Lagrangian density of a fundamental Higgs field.
We will be evaluating the vacuum expectation value (vev) of the Higgs field on our brane in terms of the vev of the bare Higgs field.
The corresponding action is the form
\eq$\label{intro-eq: RS-higgs1}
S_H = \int d^4x\, \sqrt{-g_{vis}} \left[ g_{vis}^{\mu\nu}\, D_\mu H^\dagger D_\nu H - \lam (|H|^2 - v_0^2)^2\right]\,.$
Taking into account that $g^{\mu\nu}_{vis}=e^{2kL}\eta^{\mu\nu}$ and $\sqrt{-g_{vis}}=e^{-4kL}$, the above action functional can be written as follows
\eq$\label{intro-eq: RS-higgs2}
S_H = \int d^4x \left[ \eta^{\mu\nu} D_\mu \wtild{H}^\dagger D_\nu \wtild{H} - \lam (|\wtild{H}|^2 - v^2)^2 \right]\,,$
where
\eq$\label{intro-eq: higgs-vev}
\wtild{H} \equiv e^{-kL}H\,, \hspace{1.5em} v \equiv e^{-kL} v_0\,.$
Notice that the action \eqref{intro-eq: RS-higgs2} depicts the ordinary action of the Higgs field.
Obviously, in the preceding relations, $v$ is the vev of the normalized Higgs field $\wtild{H}$, while $v_0$ is the vev of the bare Higgs field $H$.
The fact that the vev of the Higgs field determines the mass parameters in the context of the Standard Model, allow us to write
\eq$\label{intro-eq: mass-hier}
m = e^{-kL} m_0\,,$
where $m$ is the physical mass as it would be measured by an observer on the brane.
Equation \eqref{intro-eq: mass-hier} is a simple and powerful result.
Due to the fact that there is an exponential factor in eq.\,\eqref{intro-eq: mass-hier}, one does not require very large hierarchies amongst the fundamental
parameters of the model, i.e. $\kappa_{(5)}$, $L$, $k$, $v_0$.
As a matter of fact, for $m_0=M_{Pl}\approx 2\times 10^{18}\,\text{GeV}$ and $m=m_{EW}\approx 250\,\text{GeV}$, one only needs
\eq$\label{intro-eq: RS-hier}
kL\approx 35\,.$

However, in order to confidently state that the RS model successfully addresses the hierarchy problem, it is also important to examine the dependence of the
effective four-dimensional gravitational scale on the size of the extra dimension.
To achieve this, we perturb the four-dimensional part of the spacetime, and we obtain
\eq$\label{intro-eq: RS-pert}
ds^2=e^{-2k|y|}g^{per}_{\mu\nu}dx^\mu dx^\nu+dy^2\,,$
where $g^{per}_{\mu\nu}(x^\lam)=\eta_{\mu\nu}+h_{\mu\nu}(x^\lam)$ with $|h_{\mu\nu}|\ll 1$.
Using the line-element \eqref{intro-eq: RS-pert} into the original gravitational action \eqref{intro-eq: RS1-act-grav}, one can derive the scale of the
gravitational interactions, namely it is
\eq$\label{intro-eq: RS-eff}
S_{eff}=M_{Pl(5)}^3\int d^4x\int_{-L}^L dy\, e^{-2kL}\sqrt{-g_{per}}\ R_{(4)}\,.$
The effective action $S_{eff}$ should also be equal to
\eq$\label{intro-eq: eff-act}
S_{eff}=M_{Pl}^2\int d^4x\sqrt{-g_{per}}\, R_{(4)}\,.$
Equating the last two relations one can readily derive that
\eq$\label{intro-eq: RS-grav-sc1}
M_{Pl}^2=\frac{M_{Pl(5)}^3}{k}\left( 1-e^{-2kL} \right)\,.$
Substituting now the value of the product $kL$ from eq.\,\eqref{intro-eq: RS-hier}, we find that
\eq$\label{intro-eq: RS-grav-sc2}
M_{Pl}^2=\frac{M_{Pl(5)}^3}{k}\left( 1-e^{-70} \right)\simeq \frac{M_{Pl(5)}^3}{k}\,.$
It is clear from eq.\,\eqref{intro-eq: RS-grav-sc2} that gravity is essentially independent of the size of the extra
dimension. 
Surprisingly, even if one infinitely extends the length $L$ of the extra dimension in eq.\,\eqref{intro-eq: RS-grav-sc1},
the four-dimensional Planck scale $M_{Pl}$ remains finite. 
This particular observation was the central point of the RS$2$ model which is going to be discussed afterwards.

Summing up, it was shown that in the context of the RS$1$ model the hierarchy problem has an extremely simple and clear solution. 
Simultaneously, the RS$1$ model does not introduce new huge hierarchies (in contrast with the ADD model) between its fundamental parameters
($k$, $L$, $M_{Pl(5)}$, and $v_0$). The only constraint that is required by the model is $kL\approx 35$. 
For this purpose, a stabilizing mechanism (\textit{Goldberger-Wise mechanism} \cite{Goldberger-Wise1,Goldberger-Wise2}) must be incorporated in the model, 
but this is beyond our scope at this point.

\begin{flushleft}
\chapc{\large{\textbf{RS$\bold{2}$ model}}}
\end{flushleft}
\phantomsection

In their follow-up paper, Lisa Randall and Raman Sundrum explored the repercussions of a five-dimensional model with a single brane (our universe) and an 
infinite non-compact extra dimension.
By utilizing the warping of the extra dimension, they managed to show that even if the length of the extra dimension becomes infinite, it is nevertheless possible 
to obtain an effectively four-dimensional gravity. 
The justification of the previous assertion was provided by the fact that a curved bulk spacetime can support a bound state of the 
higher-dimensional graviton, which is localized close to our $3$-brane.
Hence, although space is indeed infinite, the graviton is confined to a small region in the vicinity of our brane.
The set-up of the RS$2$ model is similar to the one of RS$1$, with the difference being that the  negative-tension brane is now removed from the
model, whilst the positive-tension brane is the one representing our universe.

In accordance with the above, the RS$2$ action is given by
\gat$\label{intro-eq: RS2-act}
S_{RS2}=S_{grav}+S_{br}\,,\\[2mm]
\label{intro-eq: RS2-grav}
S_{grav}=\int d^4x\int dy \sqrt{-g}\left(\frac{R}{2\kappa_{(5)}^2}-\Lambda_5\right)\,,\\[2mm]
\label{intro-eq: RS2-br}
S_{br}=\int d^4x \sqrt{-g_{br}}\, (\lagr_{br}-\sigma)\,.$
The line-element of the RS$2$ model is given by eq.\,\eqref{intro-eq: RS-line-el}. 
Gravitons correspond to small fluctuations in the background spacetime ``fabric'', hence, we have
\eq$\label{intro-eq: RS2-line-el-flac}
ds^2 = e^{-2k|y|} \left[ \eta_{\mu\nu} + h_{\mu\nu}(x^\lam,y) \right] dx^\mu dx^\nu + dy^2\,,$
where we have chosen $h_{M4}=0,\,\forall M$.
Note that it is always possible to find an appropriate set of coordinates that possess the aforementioned characteristic.
To bring the above line-element in a conformal form we introduce a new coordinate $z$ for the extra dimension, which is defined via 
$z\equiv sgn(y)(e^{k|y|}-1)/k$.
In the coordinate system $\{x^\lam,z\}$ the line-element \eqref{intro-eq: RS2-line-el-flac} takes the form
\eq$\label{intro-eq: RS2-line-el-flac-z}
ds^2=e^{2A(z)}\left[ \eta_{MN} + h_{MN}(x^\lam,z) \right] dx^M dx^N\,,$
where $A(z)=-\ln (k|z|+1)$. 
In the forthcoming calculations we will need the derivatives of $A(z)$, thus, we present them below
\gat$\label{intro-eq: RS-dA-z}
\pa_z A = - \frac{k\, sgn(z)}{k|z|+1} = -k\, \frac{H(z)-H(-z)}{k|z|+1}\,,\\[3mm]
\label{intro-eq: RS-ddA-z}
\pa_z^2 A = - \frac{2k\del(z)}{k|z|+1} + \frac{k^2}{(k|z|+1)^2}=\frac{k^2}{(k|z|+1)^2}-2k\del(z)\,.$

We have now everything we need to derive the equations for the graviton ($h_{\mu\nu}$).
Using the gauge\,\footnote{A complete analysis regarding the derivation and the legitimacy of the aforementioned gauge can be found in 
\cite{Ivanov:1999mt,Myung:2000hu}.}
$h_{M4}=h^{\mu}{}_\mu=\pa_\mu h^{\mu}{}_\nu=0$, and performing the rescaling $h_{\mu\nu}(x^\lam,z)\ra e^{-3A(z)/2}h_{\mu\nu}(x^\lam,z)$, one can 
readily calculate that
\eq$\label{intro-eq: RS2-h-eq}
\square h_{\mu\nu}-\frac{3}{2}\left[\pa_z^2 A+\frac{3}{2}\,(\pa_zA)^2 \right] h_{\mu\nu}=0\,.$
In the above $\square \equiv \eta^{AB}\pa_A\pa_B$.
In order to continue we need to perform a Kaluza-Klein decomposition on the fluctuations $h_{\mu\nu}(x^\lam,z)$, i.e.
\eq$\label{intro-eq: KK-decom}
h_{\mu\nu}(x^\lam,z)=\sum_{n=0}^\infty  e^{i p^n_\mu x^\mu} \psi_n(z)\,.$
Notice that according to the preceding decomposition it holds that $\eta^{\alp\beta}\pa_\alp \pa_\beta h_{\mu\nu}=m_n^2 h_{\mu\nu}$, 
where $\eta^{\alp\beta}p^n_\alp\, p^n_\beta=-m_n^2$.
Substituting the expression of $h_{\mu\nu}(x^\lam,z)$ in eq.\,\eqref{intro-eq: RS2-h-eq} we obtain the Schr\"{o}dinger-like equation for 
the function $\psi_n(z)$. Consequently, we find that
\eq$\label{intro-eq: RS2-psi}
\pa_z^2 \psi_n(z)+\left[m_n^2-V(z)\right]\,\psi_n(z)=0\,,$
with the potential $V(z)$ being
\eq$\label{intro-eq: RS2-V}
V(z)=\frac{3}{2}\,\pa_z^2A+\frac{9}{4}(\pa_zA)^2=\frac{15k^2}{4(k|z|+1)^2}-3k\del(z)\,.$
%

\begin{figure}[t]
\centering
\includegraphics[width=0.65\textwidth]{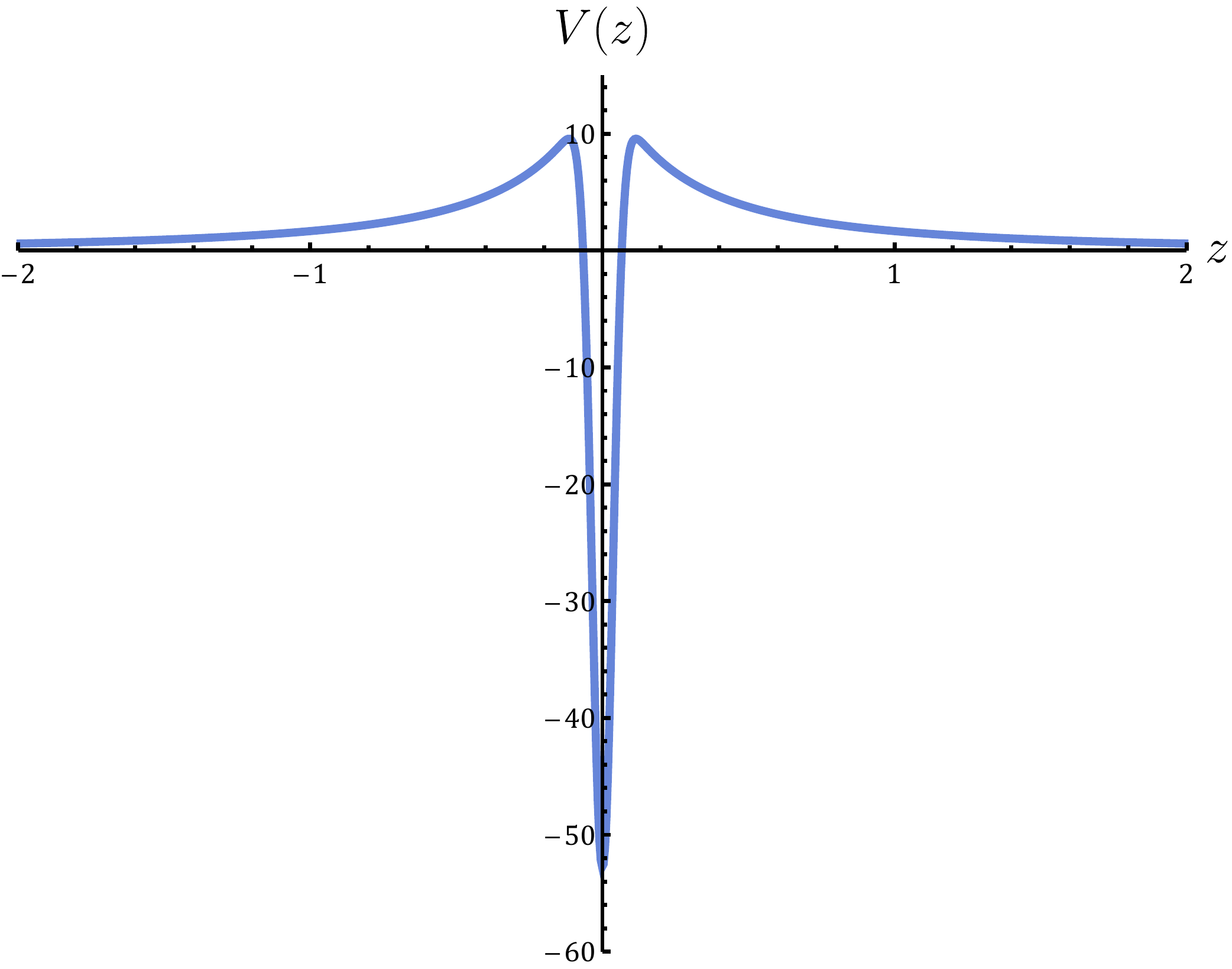}
\vspace*{-0em}
\caption[]{The ``volcano'' potential $V(z)$ for $k=2$ (in units of $M_{Pl(5)}$). The delta function was approximated by $\del(z)\sim e^{-z^2/a^2}/(a\sqrt{\pi})$ with $a=0.05$.}
\label{intro-fig: RS2-pot}
\end{figure}

\begin{figure}[t]
\centering
\includegraphics[width=0.65\textwidth]{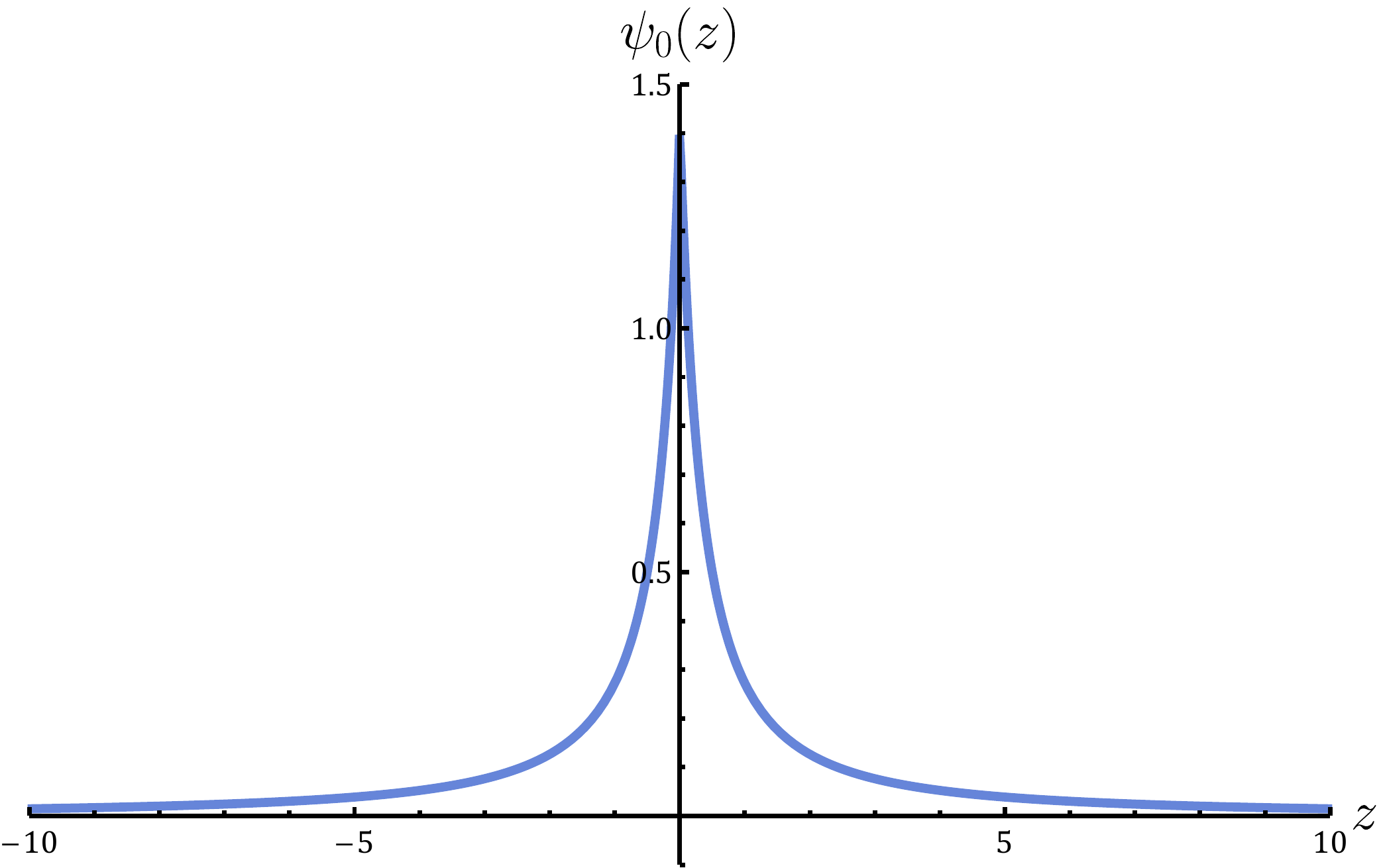}
\vspace*{-0em}
\caption[]{The zero mode $\psi_0(z)$ for $k=2$ (in units of $M_{Pl(5)}$).}
\label{intro-fig: RS2-zero-mode}
\end{figure}

\noindent The qualitative behavior of the potential $V(z)$ is depicted in \fref{intro-fig: RS2-pot}.
The boundary condition at $z=0$ can be found using eq.\,\eqref{intro-eq: RS2-psi}, namely it is
\gat$
\int_{0^-}^{0^+}dz \left[-\pa_z^2\psi_n(z) +V(z) \psi_n(z)\right]=m_n^2 \int_{0^-}^{0^+}dz\, \psi_n(z)\xRightarrow[\pa_z\psi_n(z)=
-\pa_z\psi_n(-z)]{\psi_n(z)=\psi_n(-z)}\,\\[2mm]
\label{intro-eq: bound-cond-modes}
\pa_z \psi_n(0^+)=-\pa_z \psi_n(0^-)=-\frac{3}{2}k\, \psi_n(0)\,.$
\textit{The \textbf{zero-mode} $\psi_0(z)$, for $m_0=0$, corresponds to the massless graviton of the four-dimensional theory}, and it can be easily evaluated
from eq.\,\eqref{intro-eq: RS2-psi}. Its expression is
\eq$
\psi_0(z)=N_0\, e^{3A(z)/2}=N_0\, (k|z|+1)^{-3/2}\,,$
with $N_0$ being a normalization constant. 
Demanding $\int_{-\infty}^{\infty}|\psi_0(z)|^2=1$ one can readily evaluate that $N_0=\sqrt{k}$.
The profile of the zero mode $\psi_0(z)$ for $k=2$ is depicted in \fref{intro-fig: RS2-zero-mode}, and it is straightforward to deduce that is localized close to 
the $3$-brane at $z=0$. 
Note also that this particular massless mode is the one that produces the effective $1/r$ gravitational potential on the brane, while the modes $\psi_n(z)$ for $n>0$
are responsible for the corrections to the well-known Newtonian potential.
The effective gravitational potential on the brane will be determined shortly.
The \textit{\textbf{Kaluza-Klein (KK) modes}} for $n>0$ can be provided from the general differential equation \eqref{intro-eq: RS2-psi}, which for $z\neq 0$
has the solution
\eq$
\psi_n(z)=\left(|z|+\frac{1}{2}\right)^{1/2}\left\{a_n\, J_2\left[m_n\left(|z|+\frac{1}{2}\right)\right]
+b_n\, Y_2\left[m_n\left(|z|+\frac{1}{2}\right)\right] \right\}\,.$
In the above, $a_n, b_n$ are constant coefficients and $J_2(x), Y_2(x)$ are the Bessel functions of the first and second kind, respectively.
The constant coefficients $a_n$ and $b_n$ are related with each other. 
Applying the boundary condition at $z=0$, eq.\,\eqref{intro-eq: bound-cond-modes}, one finds that $a_n=-b_n\,Y_1(m_n/k)/J_1(m_n/k)$.
Hence, by defining the normalization constant $N_n\equiv b_n/J_1(m_n/k)$, the function $\psi_n(z)$ takes the form
{\fontsize{11}{11}\eq$\label{intro-eq: KKmodes}
\psi_n(z)=N_n\left(|z|+\frac{1}{2}\right)^{\frac{1}{2}}\left\{J_1\left(\frac{m_n}{k}\right) Y_2\left[m_n\left(|z|+\frac{1}{2}\right)\right]
-Y_1\left(\frac{m_n}{k}\right) J_2\left[m_n\left(|z|+\frac{1}{2}\right)\right]\right\}\,.$}
\hspace*{-0.5em}For a detailed analysis of the KK modes in the RS models the interested reader may have a look at the master thesis \cite{Nakas}.

Having determined the KK spectrum of the effective four-dimensional theory, we can now compute the non-relativistic gravitational potential
created by an object of mass $M$ on our brane at $z=0$. 
This potential is generated by the exchange of the zero-mode and all KK mode propagators.
The explicit computation can be found in \cite{Garriga:1999yh,Giddings:2000mu}. 
Here we will just present the result without going into details.
So, the energy-momentum tensor for a point-particle of mass $M$ on the brane at $\vec{r}=0$ is given by
\eq$
T_{\mu\nu}=M\, \del^{(3)}(\vec{r})\,\del^{0}{}_\mu\,\del^{0}{}_\nu\,,$
while the spacetime fluctuations $h_{\mu\nu}$ are evaluated to be
\eq$
h_{\mu\nu}\approx \frac{2kM}{M_{Pl(5)}^2}\frac{1}{r}\left[\left(1+\frac{1}{3k^2 r^2}\right)\eta_{\mu\nu}+\left(2+\frac{1}{k^2r^2}\right)
\del^{0}{}_\mu\,\del^{0}{}_\nu \right]\,.$
As we have already shown, for $L\ra+\infty$ it holds that $M_{Pl(5)}^3\simeq M_{Pl}^2\, k\simeq k/G_N$.
Thus, from the previous relation we find that
\eq$
h_{00}\approx\frac{2G_N M}{r}\left(1+\frac{2}{3k^2r^2}\right)\,,\hspace{1.5em}
h_{ij}\approx\frac{2G_N M}{r}\left(1+\frac{1}{3k^2r^2}\right)\del_{ij}\,.$
The Newtonian gravitational potential $V(r)$ is determined by $h_{00}$ as follows
\eq$
V(r)=\frac{1}{2}\, h_{00}\approx \frac{G_N M}{r}\left(1+\frac{2}{3k^2r^2}\right)\,.$
It is obvious that for $kr \gg 1$ the potential $V(r)$ has the well-known $1/r$ behavior of Newtonian gravity.
However, if one considers the case $kr \ll 1$ then the gravitational potential becomes proportional to $1/r^3$.
This term reflects the impact of all the KK modes with $n>0$.
The possibility of observing deviations from the Newton's law of gravitation depends entirely on the value of $k$,
and on the energy levels that we are able to achieve in our collider experiments.

\mysection{Higher-dimensional black objects}

Black holes are among the most fundamental and, at the same time, most fascinating
solutions of the General theory of Relativity. The different types of black holes
predicted by GR have all been determined and classified according to their physical
properties (mass, charge, angular momentum), and uniqueness theorems have been
formulated (see, for example, \cite{MTW, Carter}). The emergence of higher-dimensional theories
\cite{ADD1, ADD2, ADD3, RS1, RS2} based on the early concept of brane \cite{Misha, Akama} 
and postulating
the existence of extra spacelike dimensions in nature has completely changed the 
landscape. Now, the higher-dimensional analogues of black holes cannot easily
be classified or proven to be unique---moreover, they are supplemented by a large number
of black objects such as black strings 
\cite{KNP1,KNP2,KNP3,Charmousis1,Charmousis2,Cisterna1,Cisterna2,CFLO,Rezvanjou,
Estrada,Cisterna2021, Wiseman, Kudoh2, Sorkin1, Sorkin2, Kleihaus2, Headrick, Figueras2, 
Kalisch, Emparan2,Bazeia:2014tua}, black rings
\cite{Emparan:2001wn,Elvang:2003mj,Elvang:2004rt,Bena:2004de,Emparan:2004wy,Gauntlett:2004wh,
Arcioni:2004ww,Bena:2005ni,Elvang:2004ds,Elvang:2005sa,Gaiotto:2005xt,Bena:2005va,Emparan:2006mm,
Emparan:2007wm,Elvang:2007hs,Caldarelli:2008pz,Emparan-review,Jiang:2008gq,Bena:2009ev,Mandlik:2020mxe,
Ruiperez:2020qda,Ali:2020bor,Guo:2021ikr}
or black branes
\cite{Gubser,Kovtun:2003wp,Banerjee:2008th,Berti:2009kk,Goldstein:2010aw,Tarrio:2011de,
Alishahiha:2012qu,Hollands:2012sf,Cisterna3}.

In the limit where the self-energy of the brane is much smaller than the black-hole
mass and the symmetries of the four-dimensional solutions may be extended to
the higher-dimensional spacetime \cite{ADD1, ADD2, ADD3}, analytical forms of higher-dimensional
black holes, either spherically symmetric or rotating, are easy to derive---in fact,
they were derived long ago in \cite{Tangherlini, MP}. If, however, the brane self-energy
is not negligible and contributes toward a particular profile of gravity along
the extra dimensions \cite{RS1, RS2}, the analytic derivation of higher-dimensional black
holes is extremely difficult. A first attempt \cite{CHR}, employing a straightforward
ansatz of a Schwarzschild line element embedded into a warped extra dimension,
has failed to lead to a five-dimensional black hole and has instead led to a black
string, a five-dimensional solution with a horizon and a singularity at every point
of the extra dimension. It was subsequently shown that these black strings are
unstable under linear gravitational perturbations \cite{GL, RuthGL}; therefore these
unphysical objects are unlikely to survive in the context of a fundamental
gravitational theory. 

\mysubsection{Braneworld black-string solutions \label{IntroSec: CHR}}

Let us now investigate the emergence of black-string solutions in the context of RS models. 
To this end, we focus on the first ever documented such solution, namely \cite{CHR}.
In an attempt to find a brane-world black hole on an RS brane, Chamblin, Hawking and Reall (CHR) replaced the flat four-dimensional part of the metric 
\eqref{intro-eq: RS-line-el} with the Schwarzschild geometry. Consequently, the resulted line-element is of the form
\eq$\label{intro-eq: CHR}
ds^2=e^{2A(y)} \left[ -\left(1-\frac{2M}{r}\right) dt^2 + \left(1-\frac{2M}{r}\right)^{-1} dr^2 + r^2 d\Omega_2^2 \right] + dy^2\,, $ 
where $d\Omega_2^2=d\theta^2 + \sin^2\theta\, d\varphi^2$, while $M$ is a constant quantity which is directly related to the black-hole mass.
For $A(y)=-k|y|$ we have the warp factor of the RS model.
Notice now that on the brane, at $y=0$, the induced four-dimensional line-element is identified with the well-known Schwarzschild solution.
Unfortunately though, the above line-element does not correspond to a regular brane-world black hole as one would expect.
The problem with the above geometry is that the singularity on the brane extends into the bulk all the way to the AdS horizon.
To understand why this happens one needs to calculate the Kretschmann scalar that emanates from the preceding line-element.
By doing so, one finds that
\eq$\label{intro-eq: CHR-K}
\mathcal{K}\equiv R^{MNKL}R_{MNKL}=16(\pa_y^2A)^2+40(\pa_yA)^4+32(\pa_yA)^2\pa_y^2A+\frac{48M^2e^{-4A(y)}}{r^6}\,.$ 
The last term of the above relation reveals the existence of a singularity at $r=0$ which extends along the extra dimension $y$.
This means that the line-element \eqref{intro-eq: CHR} generates the topology of a \textit{black-string} rather than the topology of a 
black-hole solution.
In case of the RS2 model, where the extra dimension is infinitely extended, the last term becomes even more problematic, since when $y\ra+\infty$
the quantity $e^{-4A(y)}$ diverges as well.
Obviously, the above behavior is in complete contradiction with the idea of the RS2 model, which intends to keep gravity localized close to the brane at $y=0$.

Despite decades of efforts, the quest for the more physically acceptable solutions
of regular, localized-close-to-our-brane black holes has failed to lead to an
analytical, non-approximate form (see Refs. 
\cite{review1, review2, review3, review4, review5, review6, review7, tidal, Papanto, KT,KOT,
CasadioNew, Frolov, Karasik,Kofinas,GGI, CGKM, Ovalle1, Ovalle2, Ovalle3, Ovalle4, Harko, 
daRocha1, daRocha2, Nakas,Dadhich,Charmousis,Shanka,Andrianov2,Banerjee,Chakraborty1,
Chakraborty2,Chakraborty3,Fitzpatrick, Zegers, Heydari, Dai, Bruni, Tanaka, EFK, EGK, Yoshino,KPZ, KPP}
for an impartial list of works)---however, such solutions were successfully derived in 
lower-di\-men\-sion\-al gravitational models \cite{EHM1, EHM2, AS, Cuadros}. Numerical solutions also
emerged that described either small \cite{Kudoh,Kudoh1} or large black holes
\cite{Tanahashi,Kleihaus,Figueras1, Page1, Page2} in braneworld models. In an effort to derive the
long-sought analytical black-hole solutions, in \cite{KPZ, KPP} the previously proposed
idea \cite{KT,KOT}, of adding a non-trivial profile along the extra dimension to the black-hole
mass function in the  original line element used in \cite{CHR}, was extended to include
also a dependence on the time and radial coordinate; in this way, the rather restricted
Schwarzschild-type of brane background was extended to include additional terms
[of an (anti)--de Sitter or Reissner-Nordstr\"{o}m type] and to allow also for non-static
configurations. A large number of bulk scalar field theories were then investigated;
however, no viable solutions that could sustain the line element of a five-dimensional,
regular, localized-close-to-our-brane black hole was found.\,\footnote{Brane-world models based 
on a five-dimensional bulk scalar field
leading to a cosmologically-evolving brane were also studied in \cite{BDL,Bhatta1, Bhatta2}.}
Following a different approach to the problem, the authors in \cite{NK1} constructed from first 
principles the geometry of an analytic and exponentially localized five-dimensional brane-world black hole.
The singularity of the black hole resides strictly on the brane, thus, the emergence of bulk singularities is avoided altogether. 
The induced metric on the brane is described by the Schwarzschild geometry, while the five-dimensional background
quickly reduces to a pure AdS$_5$ spacetime away from the brane.
A thorough presentation and discussion regarding the derivation of the localized brane-world black hole in \cite{NK1} takes place 
in \chapref{Chap: P4}.
Subsequent works based on the aforementioned brane-world black hole can be found in \cite{NK2,Neves:2021dqx}.

\afterpage{\blankpage}



\mychapter{Antigravitating brane-world solutions for a de Sitter brane \label{Chap: P1}}
\phantomsection
\setcounter{chapter}{2}

\epigraph{\textit{``Be less curious about people and more curious about ideas.''}}{Marie Curie}

\thispagestyle{empty}

{\lettrine[lines=3, lhang=0.03]{\color{chapter(color)}T}{\hspace*{5.5px}he} analyses performed in \cite{KPZ, KPP} 
have hinted toward the existence
of solutions that were not characterized by the desired nontrivial profile of the mass function
in terms of the extra coordinate---these solutions could not therefore be localized black holes
but rather novel black-string solutions. As a result, in this Chapter, which is based on \cite{KNP1}, we focus on the careful
investigation of the existence of these latter types of solutions in the context of a theory
with a scalar field non-minimally coupled to gravity, and on the study of their physical
properties. We demonstrate that, for very natural, simple choices of the coupling function
between the scalar field and the five-dimensional scalar curvature, novel black-string
solutions may indeed be found with rather interesting and provocative characteristics.
Given the fact that the same theory has resisted in giving legitimate black-hole solutions,
even for a wider number of choices of the coupling function \cite{KPZ, KPP}, our present
results add new ``fuel'' to the long dispute around the question of why braneworld models
lead quite easily to black-string solutions but not to localized black holes. 
Indeed, higher-dimensional gravitational theories
often allow for the emergence of uniform or nonuniform black-string solutions
\cite{KNP1,KNP2,KNP3,Charmousis1,Charmousis2,Cisterna1,Cisterna2,CFLO,Rezvanjou,
Estrada,Cisterna2021, Wiseman, Kudoh2, Sorkin1, Sorkin2, Kleihaus2, Headrick, Figueras2, 
Kalisch, Emparan2,Bazeia:2014tua}.

}

In our analysis, we will retain the ``Vaidya form'' of the brane line element used
also in \cite{KOT, KPZ, KPP}, since this form was shown not to lead to additional
spacetime singularities in the bulk. As we are interested in finding static
black-string solutions, here we abandon the dependence of the mass function on
the time and extra-dimension coordinates, and allow for a general, radially dependent
form $m(r)$. Our field equations will be straightforwardly integrated to determine
the form of the mass function that is found to correspond to a Schwarzschild anti-de
Sitter background. We will consider two simple forms of the coupling function, namely a
linear form and a quadratic one in terms of the scalar field. In  both cases, we solve the
set of field equations and derive the scalar-field configurations and the physical properties
of the model. 

A common characteristic of the solutions derived in the two cases is the negative sign
of the coupling function in front of the five-dimensional scalar curvature, either over
the entire bulk (for a quadratic dependence) or at distances larger than a specific value
(for a linear dependence). This clearly leads to the ``wrong sign'' for gravity; however,
as we will see, it is this negative sign that effectively creates an anti--de Sitter spacetime 
and supports a Randall-Sundrum warp factor even in the absence of a negative bulk
cosmological constant. In the case of the linear coupling function, the antigravitating
regime arises away from our brane; this regime is pushed farther away the larger the
warping coefficient and the smaller the cosmological constant on our brane are. In fact,
for particular values of the parameters of the model, the theory resembles an ordinary,
minimally coupled scalar-tensor theory with normal gravity and a Randall-Sundrum
warp factor. 

Although the original objective of our analysis was to investigate the existence of novel
black-string solutions in the context of a non-minimally coupled scalar-tensor theory,
our solutions, in the limit of vanishing black-hole mass on the brane, reduce to
maximally symmetric braneworld solutions that are regular over the entire bulk
apart from the (anti)--de Sitter (AdS) boundary.\,\footnote{Braneworld solutions with a Minkowski spacetime
on our brane were also studied in the context of a non-minimally coupled scalar-tensor
theory in \cite{Farakos1, Bogdanos1, Farakos2, Farakos3}.} In this limit,
the gravitational background on our
brane is a pure AdS spacetime. In fact, we demonstrate that for the
physically motivated case of a positive-cosmological constant on our brane, the
emergence of an antigravitating regime in the bulk is unavoidable.

The current Chapter has the following outline: in \secref{P1sec: Th-Frame}, we present the field equations and
spacetime background. In \secref{P1sec: Linear}, we study in detail the case of a linear coupling
function and determine the complete bulk solution, its physical properties, as well
as the effective theory on the brane. A similar analysis is performed in \secref{P1sec: Quadratic}
for the case of a quadratic coupling function. In \secref{P1sec: Theoretical}, we present a mathematical
argument that underlines the connection between the emergence of an antigravitating
regime in the bulk and the positive sign of the cosmological constant on our brane.
We finally discuss our conclusions in \secref{P1sec: Disc}.

\vspace{-1.5em}

\mysection{The theoretical framework \label{P1sec: Th-Frame}}

\vspace{-0.5em}

In this Chapter, we focus on the following class of five-dimensional
gravitational theories with action functional
\vspace{-0.3em}
\beq
\label{P1eq: action}
S_B=\int d^4x\int dy \,\sqrt{-g^{(5)}}\left[\frac{f(\Phi)}{2\kappa_5^2}R
-\Lambda_5-\frac{1}{2}\,\pa_L\Phi\,\pa^L\Phi-V_B(\Phi)\right].
\eeq
Note that bulk quantities will be denoted with capital $B$, while brane quantities will be denoted with lower-case $b$ (or $br$).
The above theory contains the five-dimensional scalar curvature $R$, a bulk
cosmological constant $\Lambda_5$, and a five-dimensional scalar field $\Phi$
with a self-interacting potential $V_B(\Phi)$ and a
non-minimal coupling to $R$ via the general coupling function $f(\Phi)$.
Also, $g^{(5)}_{MN}$ is the metric tensor of the
five-dimensional spacetime, and $\kappa_5^2=8\pi G_5$ is defined in terms of
the five-dimensional gravitational constant $G_5$. At a particular point along
the fifth spatial dimension, whose coordinate we denote by $y$, a 3-brane
is introduced---without loss of generality, we assume that this takes place
at $y=0$. Then, the above bulk action must be supplemented by the brane one
\beq
\label{P1eq: br-action}
S_{br}=\int d^4x\sqrt{-g^{(br)}}(\lagr_{br}-\sigma)=
-\int d^4x\int dy\sqrt{-g^{(br)}}\,[V_b(\Phi)+\sigma]\,\delta(y)\,.
\eeq
Here, $\lagr_{br}$ is related to the matter/field content of the brane and has been
chosen to consist of an interaction term $V_b(\Phi)$ of the bulk scalar field with
the brane. Also, $\sigma$
is the brane self-energy, and $g^{(br)}_{\mu\nu}=g_{\mu\nu}^{(5)}(x^\lam,y=0)$ is
the induced-on-the-brane metric tensor. Note that, throughout this and forthcoming Chapters, capital
letters $M,N,L,...$ will denote five-dimensional indices while lower-case letters
$\mu,\nu,\lambda,...$ will be used for four-dimensional indices.

\par The variation of the complete action $S=S_B+S_{br}$ with respect to the
metric-tensor components $g^{(5)}_{MN}$ yields the gravitational field equations
that have the form
{\fontsize{11}{11}{\eq$\label{P1eq: Einst-eqs}
f(\Phi)\,G_{MN}\sqrt{-g^{(5)}}=\kappa_5^2\left[(T^{(\Phi)}_{MN}-g_{MN}\Lambda_5)
\sqrt{-g^{(5)}}-[V_b(\Phi)+\sigma]\,g^{(br)}_{\mu\nu} \delta^\mu_M\delta^\nu_N\delta(y)\sqrt{-g^{(br)}}\right],$}}
\hspace{-0.5em}where
\eq$
T^{(\Phi)}_{MN}=\pa_M\Phi\,\pa_N\Phi+g_{MN}\left[-\frac{\pa_L\Phi\pa^L\Phi}{2}-V_B(\Phi)\right]
+\frac{1}{\kappa_5^2}\left[\nabla_M\nabla_Nf(\Phi)-g_{MN}\Box f(\Phi)\right].$
On the other hand, the variation of the action with respect to $\Phi$ leads to the
scalar-field equation
\beq
-\frac{1}{\sqrt{-g^{(5)}}}\,\pa_M\left(\sqrt{-\gfv}g^{MN}\pa_N\Phi\right)=
\frac{\pa_\Phi f}{2 \kappa^2_5} R-\pa_\Phi V_B 
-\frac{\sqrt{-g^{(br)}}}{\sqrt{-\gfv}}\,\partial_\Phi V_b\,\delta(y)\,\,.
\label{P1eq: phi-eq-0}
\eeq

We will also assume that the five-dimensional gravitational background is given by
the expression
\eq$\label{P1eq: metric}
ds^2=e^{2A(y)}\left\{-\left[1-\frac{2m(r)}{r}\right]dv^2+2dvdr+r^2(d\theta^2+\sin^2\theta d\varphi^2)\right\}+dy^2\,.$
The above line element is characterized by the presence of the warp factor
$e^{A(y)}$ that multiplies a four-dimensional background. For $m(r)=M$, this
four-dimensional line element is just the Vaidya transformation of the Schwarzschild
solution describing a black hole with mass $M$, and it leads to the same black-string
solutions found in \cite{CHR}. A generalized Vaidya form, where $m$ is not a
constant but a function of the coordinates, was used in a number
of works \cite{KOT, KPZ, KPP} in an effort to find regular, localized black-hole
solutions. The motivation for the use of the Vaidya form of the four-dimensional
line element, instead of the usual Schwarzschild one, was provided
in \cite{KT, KOT}; in these, it was demonstrated that four-dimensional line elements
with horizons, such as the Schwarzschild one, when embedded in five-dimensional
spacetimes, transform their coordinate singularities at the horizons to true
spacetime ones \cite{KT}. In order to avoid this, in \cite{KOT}
the four-dimensional Schwarzschild line element was first transformed to its Vaidya
form and then embedded in the warped fifth dimension; in that case, no new
bulk singularities emerged. 

Although the desired black-hole solutions have not yet been analytically found in
brane\-world models, the emergence of black-string solutions is more easily realized. 
Indeed, in the context of the theory
(\ref{P1eq: action}), hints for the
existence of novel black-string solutions described by the line element (\ref{P1eq: metric})
were given in \cite{KPP}. Therefore,  here we turn our attention to this question;
we will keep the general $r$-dependence of the mass function, i.e. $m=m(r)$, as
shown in Eq. (\ref{P1eq: metric}), in order to
allow our brane metric background to deviate from the Schwarzschild form. Such a
modification may allow for terms proportional to an effective cosmological constant
or for terms of various forms associated with tidal charges to emerge. As the
explicit forms of the curvature invariant quantities for the line element (\ref{P1eq: metric})
(given in Appendix \ref{P1app: Curv-Inv}) show, such a solution, if indeed supported by
the theory (\ref{P1eq: action}), would describe a black-string solution with only the
black-hole singularity extended over the fifth dimension and no other singularity
present.

For the line element (\ref{P1eq: metric}), one may easily see that the relation
$\sqrt{-g^{(5)}}=\sqrt{-g^{(4)}}$ holds, and the gravitational equations
are then simplified to 
\eq$\label{P1eq: grav-eqs}
{f}(\Phi)\,G^M{}_N=T^{(\Phi)M}{}_N-\del^M{}_N\Lambda_5-[V_b(\Phi)+\sigma]\, g_{\mu\nu} g^{ML}\del^\mu_L\del^\nu_N\del(y),$
with
\beq\label{P1eq: TMN-mix}
T^{(\Phi)M}{}_N=\pa^M\Phi\pa_N\Phi+\nabla^M\nabla_N {f}+\del^M{}_N(\lagr_\Phi-\Box {f}).
\eeq
In the above, we have defined
\beq\label{P1eq: Lagr}
\lagr_{\Phi}=-\frac{1}{2}\,\pa_L\Phi\,\pa^L\Phi-V_B(\Phi).
\eeq
In addition, for simplicity, we have absorbed the gravitational constant
$\kappa_5^2$ in the expression of the general coupling function $f(\Phi)$,
and omitted the superscripts
$^{(5)}$ and $^{(4)}$ from the bulk and brane metric tensors $g_{MN}$ and
$g_{\mu\nu}$, respectively. In fact, we will now focus on the gravitational
equations in the bulk and thus altogether remove the brane term proportional
to $\delta(y)$ from Eq. (\ref{P1eq: grav-eqs})---when the junction conditions are
studied, this term will be reinstated.

The non-vanishing components of the Einstein tensor $G^M{}_N$ for the
background (\ref{P1eq: metric}) are listed below:
\bea
&~&G^0{}_0=G^1{}_1=6A'^2+3A''-\frac{2e^{-2A}\pa_rm}{r^2},\nonumber \\[1mm] 
&~&G^2{}_2=G^3{}_3=6A'^2+3A''-\frac{e^{-2A}\pa_r^2m}{r}, \label{P1eq: GMN-mix-com} \\[2mm] 
&~&G^4{}_4=6A'^2-\frac{e^{-2A}\left(2\pa_rm+r\pa_r^2m\right)}{r^2}, \nonumber
\eea
where a prime ($'$) denotes the derivative with respect to the $y$-coordinate.
We will also assume that the bulk scalar field depends only on the coordinate
along the fifth dimension, i.e. $\Phi=\Phi(y)$. Then, the non-vanishing mixed
components of the energy-momentum tensor $T^{(\Phi)M}{}_N$ take in turn the form 
\begin{gather}
T^{(\Phi)0}{}_0=T^{(\Phi)1}{}_1=T^{(\Phi)2}{}_2=T^{(\Phi)3}{}_3=
A' \Phi'\,\pa_\Phi f+\lagr_\Phi-\Box f, \nonumber \\[2mm]
T^{(\Phi)4}{}_4=(1+\pa_\Phi^2 f)\Phi'^2+\Phi''\,\pa_\Phi f+\lagr_\Phi-\Box f,
\label{P1eq: TMN-mix-com}
\end{gather}
where, under the aforementioned assumptions, the quantities $\lagr_\Phi$ and $\Box f$ have
the explicit forms
\beq
\label{P1eq: Lagr-new}
\lagr_\Phi=-\frac{1}{2}\,\Phi'^2-V_B(\Phi),
\eeq
and
\beq
\label{P1eq: box-f}
\Box f=4A' \Phi'\,\pa_\Phi f+\Phi'^2\,\pa_\Phi^2 f+\Phi''\,\pa_\Phi f.
\eeq

The gravitational field equations may now easily follow by substituting the components of
$G^M{}_N$ and $T^M{}_N$, listed in Eqs. (\ref{P1eq: GMN-mix-com}) and (\ref{P1eq: TMN-mix-com}), respectively,
in Eq. (\ref{P1eq: grav-eqs}) evaluated in the bulk. We thus obtain three equations from the
$(^0{}_0)$, $(^2{}_2)$, and $(^4{}_4)$ components. Subtracting the $(^0{}_0)$ and $(^2{}_2)$
equations as well as the $(^0{}_0)$ and $(^4{}_4)$ equations, we arrive at two simpler 
ones that, together with the $(^0{}_0)$ component, form the following system
\beq
\label{P1eq: mass-eq}
r\,\pa_r^2m-2\pa_rm=0\,,
\eeq
\beq \label{P1eq: grav-eq1}
f\left(3A''+e^{-2A}\frac{\pa_r^2m}{r}\right)=\pa_\Phi f \left(A'\Phi'-\Phi''\right)
-(1+\pa_\Phi^2 f)\Phi'^2\,,
\eeq
\beq
\label{P1eq: grav-eq2}
f\left(6A'^2+3A''-\frac{2e^{-2A}\pa_rm}{r^2}\right)=A'\Phi'\,\pa_\Phi f+
\lagr_\Phi -\Box f-\Lambda_5\,.
\eeq
The above gravitational equations are supplemented by the scalar-field equation of motion
(\ref{P1eq: phi-eq-0}) that has the explicit form
\beq \label{P1eq: phi-eq}
\Phi'' + 4A' \Phi' =\pa_\Phi f \left(10A'^2+4A''-e^{-2A}\frac{2\pa_rm+
r\,\pa_r^2m}{r^2}\right) +\pa_\Phi V_B\,.
\eeq

Equation (\ref{P1eq: mass-eq}) can easily be integrated to yield the general form of the allowed
mass function, and this is
\beq
m(r)=M+ \Lambda r^3/6\,, \label{P1eq: mass-sol}
\eeq
where $M$ and $\Lambda$ are arbitrary integration constants the physical interpretation
of which will be studied later (the coefficient 6 has been introduced for later convenience).
The above solution may now be used in order to simplify the form of the remaining
three equations (\ref{P1eq: grav-eq1})--(\ref{P1eq: phi-eq}). However, not all of them are independent.
As we explicitly demonstrate in Appendix \ref{P1app: Indep-F-Eq}, an appropriate manipulation
and rearrangement of the gravitational equations (\ref{P1eq: grav-eq1}) and (\ref{P1eq: grav-eq2})
lead to the same result following also from a similar manipulation of the scalar-field
equation (\ref{P1eq: phi-eq}). Indeed, in a fully determined theory, i.e. with given $f(\Phi)$
and $V_B(\Phi)$, we would only need three independent equations out of the existing four
to find the two unknown metric functions $m(r)$ and $A(y)$ and the scalar field $\Phi(y)$. 
Therefore, henceforth, we will altogether ignore Eq. (\ref{P1eq: phi-eq}) in our analysis and
retain Eqs. (\ref{P1eq: grav-eq1}) and (\ref{P1eq: grav-eq2}). We will then adopt the following approach:
we will assume the well-known form \cite{RS1, RS2} $A(y)=-k |y|$, with $k$ a positive constant,
for the warp factor of the
five-dimensional line element in order to ensure the localization of gravity near the brane; for
a chosen coupling function $f(\Phi)$, we will then determine the scalar-field configuration
by solving Eq. (\ref{P1eq: grav-eq1}); finally, Eq. (\ref{P1eq: grav-eq2}) will determine the form of the
potential $V_B(\Phi)$ that needs to be introduced to support the solution.

In the following sections, we present two simple choices for the coupling
function $f(\Phi)$, a linear one and a quadratic one; for each one, we determine the
corresponding solution for the scalar field and form of the potential and discuss
their physical characteristics.


\mysection{The case of linear coupling function \label{P1sec: Linear}}

\vspace{-1.5em}

\mysubsection{The bulk solution \label{P1subsec: bulk-linear}}

We will first consider the case where the coupling function is of the general linear
form, $f(\Phi)=a \Phi +b$, where $a$ and $b$ are constants. Employing this
together with the form of the mass function (\ref{P1eq: mass-sol}) and the
exponentially decreasing warp factor $e^{A(y)}=e^{-ky}$ (assuming the usual ${\bf Z}_2$
symmetry in the bulk under the change $y \rightarrow -y$, we henceforth focus on the
positive $y$-regime), Eq. (\ref{P1eq: grav-eq1}) takes the form
\beq
(a \Phi +b)\,\Lambda e^{2ky}=-a\,(k\,\Phi' + \Phi'') -\Phi'^2\,.
\label{P1eq: grav-eq1-linear}
\eeq
In order to solve the above differential equation, we set $\Phi(y)=\Phi_0\,e^{g(y)}$. Substituting in 
Eq. (\ref{P1eq: grav-eq1-linear}) and rearranging, we obtain
\beq
a\Lambda\Phi_{0}\,e^{2ky + g(y)} + b\Lambda\,e^{2ky} =
- a (kg'+ g'' + g'^{2})\Phi_{0}\,e^{g(y)}  -g'^{2}\Phi_{0}^{2}\,e^{2g(y)}\,,
\label{P1eq: main-eq-linear}
\eeq
where a prime in $g$ denotes, as before, the derivative with respect to $y$. The above
leads to a nontrivial solution only if $g(y)=2ky$. In that case, the following
constraints should also hold:
\bea
a = -\frac{4k^2}{\Lambda}\,\Phi_{0}\,, \qquad 
b = \frac{24k^4}{\Lambda^2}\,\Phi_{0}^2\,. \label{P1eq: consts-linear}
\eea
The coefficient $b$ is clearly positive definite; however the sign of the
coefficient $a$ depends on those of $\Phi_0$ and $\Lambda$. 

Let us examine the type of solution that we have derived. Employing the
form of the mass function (\ref{P1eq: mass-sol}) and the general expressions for
the five-dimensional curvature invariants given in Appendix \ref{P1app: Curv-Inv}, the
latter quantities are found to have the form
\bea R &=& -20k^{2} + 4 \Lambda e^{2ky}\,, \nonumber\\[2mm]
R_{MN}R^{MN} &=& 80 k^{4} - 32k^{2}\Lambda e^{2ky} + 4\Lambda^{2} e^{4ky}\,,
\label{P1eq: invar}\\[1mm]
R_{MNKL}R^{MNKL} &=& 40k^{4} - 16k^{2}\Lambda e^{2ky} + \frac{8\Lambda^{2}e^{4ky}}{3}
+\frac{48 M^2 e^{4ky}}{r^6}\,. \nonumber 
\eea
For $M=\Lambda=0$, we recover the curvature invariants of the five-dimensional
AdS spacetime. For $\Lambda=0$ but $M \neq 0$, we obtain the black-string solution
of \cite{CHR}, with the black-hole singularity at $r=0$ extending over the entire
fifth dimension up to the AdS boundary at $ y \rightarrow \infty$. For $M=0$ but
$\Lambda \neq 0$, we find a solution that is everywhere regular apart from the
AdS boundary. Finally, for $M \neq 0$ and $\Lambda \neq 0$, we obtain again a
black-string solution with singular terms from both the black-hole and AdS boundary
appearing in the expressions of the curvature invariants. 

At this point, we should investigate the physical interpretation of the integration
constants $M$ and $\Lambda$ appearing in the expression (\ref{P1eq: mass-sol}) of
the mass function $m(r)$. To this end, we set $y=0$ in the higher-dimensional
line element (\ref{P1eq: metric}), and the projected-on-the-brane four-dimensional
background then reads
\eq$\label{P1eq: metric-4D}
ds^2=-\left(1-\frac{2M}{r} -\frac{\Lambda r^2}{3}\right) dv^2+2dvdr+r^2(d\theta^2+
\sin^2\theta d\varphi^2)\,.$
The above looks like a generalization of the Vaidya form of the Schwarzschild line element
in the presence of a cosmological constant. In order to convince ourselves, we apply a
general coordinate transformation $v = h(t,r)$, where $h(t,r)$ will be defined shortly.
Then, Eq. (\ref{P1eq: metric-4D}) assumes the standard, diagonal form
\eq$\label{P1eq: metric-SdS}
ds^2=-\left(1-\frac{2M}{r} -\frac{\Lambda r^2}{3}\right) dt^2+
\left(1-\frac{2M}{r} -\frac{\Lambda r^2}{3}\right)^{-1} dr^2+r^2(d\theta^2+
\sin^2\theta d\varphi^2)$
provided that $h(t,r)=t +g(r)$ and $g(r)$ satisfies the following condition:
\beq
\frac{dg(r)}{dr}= \left(1-\frac{2M}{r} -\frac{\Lambda r^2}{3}\right)^{-1}.
\eeq
The details of the transformation as well as the explicit form of the function $g(r)$,
which is not of importance in the present analysis, can be found in Appendix
\ref{P1app: Inv-Vaidya}. According to Eq. (\ref{P1eq: metric-SdS}),
the gravitational background on the brane is Schwarzschild (anti-)de Sitter with $M$ being
the mass of the black hole and $\Lambda=\kappa_4^2 \Lambda_4$, where $\Lambda_4$
is the cosmological constant on the brane. 

It is of particular interest to study the profile of the non-minimal coupling
function $f(\Phi)$ along the extra dimension. Using the solution for the scalar
field found above, we obtain
\beq
f(y) = a\,\Phi(y) +b= \frac{4k^2\Phi_{0}^2}{\Lambda^2}\,\left(-\Lambda e^{2ky}
+6k^2 \right)\,. \label{P1eq: f-linear}
\eeq
For $\Lambda<0$, i.e. for a negative cosmological constant on the brane, the above expression
is everywhere positive definite and gravity remains attractive over the whole bulk. However, for
$\Lambda>0$, we find that
\beq
\left\{ \begin{tabular}{ll} $f(y)>0$\,, & $y<\frac{\ln(6k^2/\Lambda)}{2k}$ \\[2mm]
$f(y)=0$\,, & $y=y_0 \equiv \frac{\ln(6k^2/\Lambda)}{2k}$ \\[2mm]
$f(y)<0$\,, & $y>\frac{\ln(6k^2/\Lambda)}{2k}$ \end{tabular}  \right \}.
\label{P1eq: f-lin-Lpos}
\eeq
That is, close to the brane and up to a maximum distance of $y=y_0$ the function
$f(y)$ is positive and gravity acts as normal. However, at $y=y_0$, $f(y)$
vanishes, and gravity locally disappears, whereas,  for $y>y_0$, $f(y)$ turns
negative, and gravity acquires the wrong sign. We may therefore conclude that,
for a positive cosmological constant on the brane, gravity becomes repulsive in the
bulk at some finite distance away from the brane.  

\begin{figure}[t!]
    \centering
    \begin{subfigure}[b]{0.455\textwidth}
        \includegraphics[width=\textwidth]{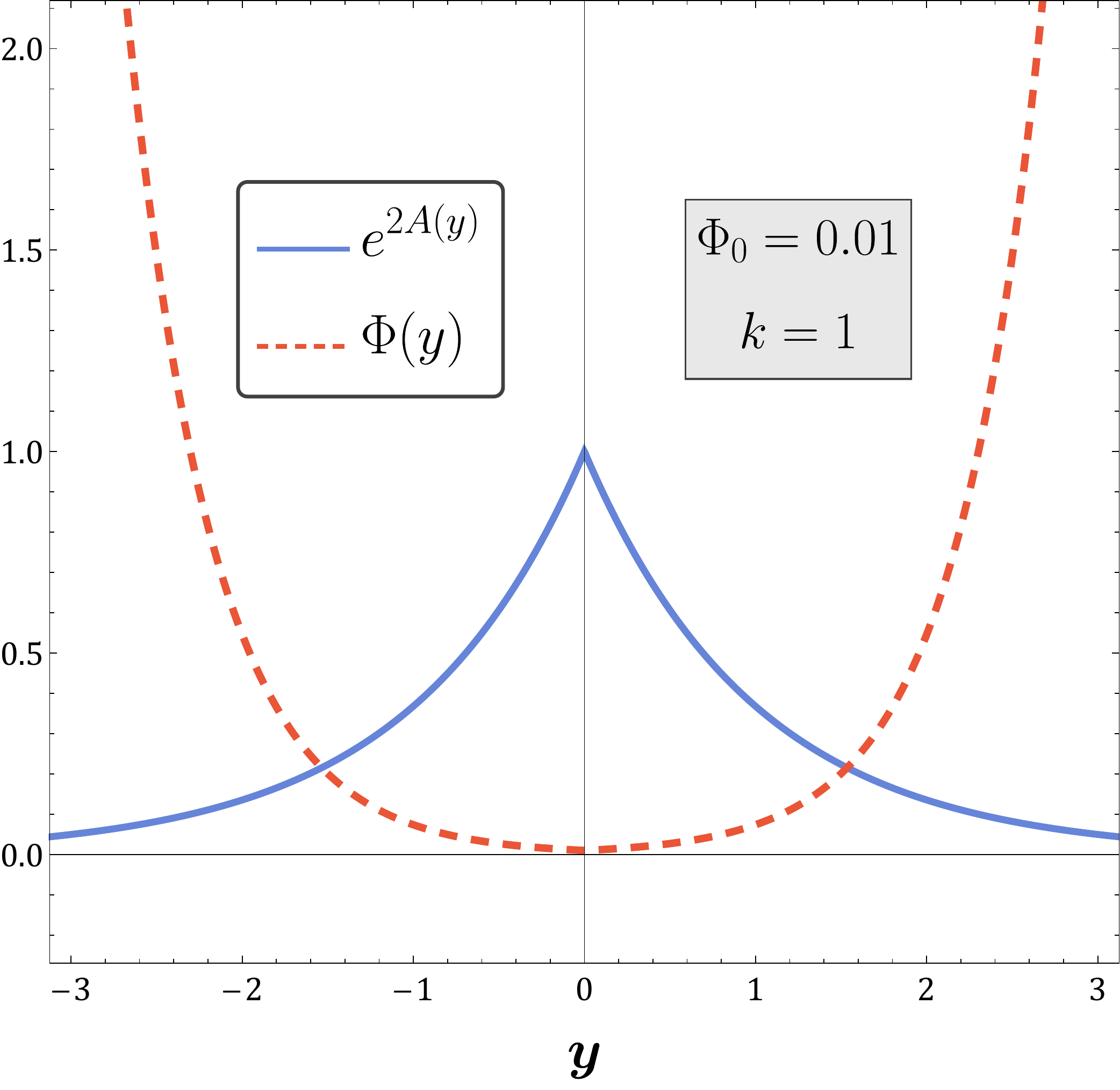}
        \caption{\hspace*{-1.2em}}
        \label{P1subf: warp-phi-linear}
    \end{subfigure}
    ~ 
    \begin{subfigure}[b]{0.5\textwidth}
        \includegraphics[width=\textwidth]{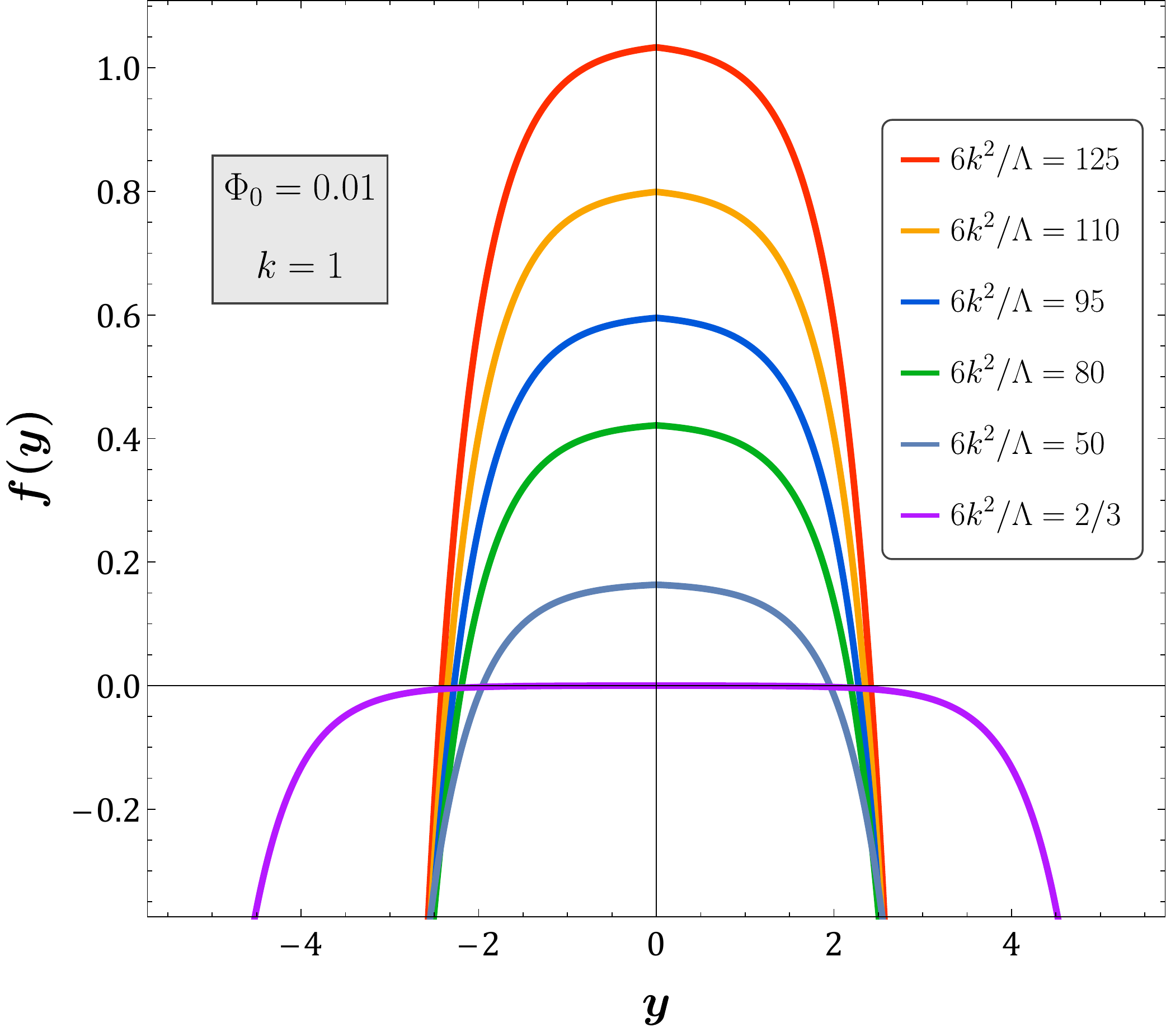}
        \caption{\hspace*{-3em}}
        \label{P1subf: f-y-linear}
    \end{subfigure}
    \vspace*{-1em}
    ~ 
    \caption{(a) The warp factor $e^{-2k |y|}$ and scalar field $\Phi(y)$, and (b)
     the coupling function $f(y)=a\, \Phi(y) +b$, in terms of the coordinate $y$,
  for $k=1$, $\Phi_0=0.01$, and $6k^2/\Lambda=2/3,50,80,95,110,125$ (from bottom to top).\vspace*{-1em}}
   \label{P1fig: warp-phi-f-linear}
\end{figure}

In \myref{P1fig: warp-phi-f-linear}{P1subf: warp-phi-linear} (next page), we depict the form of the warp factor $e^{-2k |y|}$
and the scalar field $\Phi(y)$ in terms of the coordinate $y$ along the fifth dimension,
for $k=1$ and $\Phi_0=0.01$. 
Although the former quantity exhibits the anticipated localization close to our brane,
the latter quantity increases away from the brane diverging at the boundary of
spacetime. The displayed, qualitative behavior of these two quantities is independent
of the particular values of the parameters. In contrast to this, the profile of the
coupling function  $f(y)$, given in Eq. (\ref{P1eq: f-linear}), depends strongly on the value
of the dimensionless parameter $6k^2/\Lambda$. Assuming that $\Lambda>0$ on our brane,
in \myref{P1fig: warp-phi-f-linear}{P1subf: f-y-linear} we display the form of $f(y)$, 
for $k=1$, $\Phi_0=0.01$, and
the values $6k^2/\Lambda=2/3,50,80,95,110,125$. For $6k^2/\Lambda <1$, the
function $f(y)$ does not have
a vanishing point and is always negative; for $6k^2/\Lambda >1$, a regime of positive
values for $f(y)$ appears close to our brane that tends to become larger as $k^2/\Lambda$
gradually increases. In other words, the smaller the cosmological constant
is on our brane, the farther away from our brane the antigravitating regime is
located. It is also interesting to note that the regime of positive values for the
function $f(y)$ around our brane is always characterized by a plateau, an area where the
value of the coupling function is almost constant. Therefore, close to our brane,
gravity would not only act as normal but it would look as if the scalar curvature
$R$ does not have a coupling to the scalar field. In fact, for the particular value
of $6k^2/\Lambda=125$, the coupling function $f(y)$ around the brane is constant
and approximately equal to unity. Thus, the model mimics ordinary, five-dimensional
gravity with the difference that the bulk energy, which as we will see supports
the complete bulk-brane solution, originates, in fact, from the scalar field.

In order to complete the analysis, we need to determine the potential of the scalar
field $V_B(\Phi)$. Substituting the forms of the functions $m(r)$, $A(y)$, and $\Phi(y)$
in Eq. (\ref{P1eq: grav-eq2}), we readily obtain
\beq
V_B(\Phi)=-\Lambda_5 -2 k^2 \left(\frac{72 k^4 \Phi_0^2}{\Lambda^2} -
\frac{20 k^2 \Phi_0}{\Lambda}\,\Phi +3 \Phi^2\right)\,.
\label{P1eq: V-linear}
\eeq
Combining the above expression with the profile of the scalar field along the extra dimension,
$\Phi(y)=\Phi_0\,e^{2ky}$, we notice the following: at the location of the brane, at $y=0$,
the scalar potential reduces to a constant value, namely
\beq
V_B(y=0)=-\Lambda_5 -2 k^2 \Phi_0^2 \left(\frac{72 k^4}{\Lambda^2} -
\frac{20 k^2}{\Lambda} +3 \right).
\label{P1eq: V0-linear}
\eeq
The quantity inside the brackets has no real roots and is thus always positive definite;
that makes the second term a negative-definite quantity for all values of the parameters
of the model. This means that, close to the brane, the scalar potential can mimic the
role of the negative cosmological constant---thus making $\Lambda_5$ redundant---and
support by itself an AdS spacetime in the bulk regime close to the brane. 

\begin{figure}[t!]
    \centering
    \begin{subfigure}[b]{0.47\textwidth}
        \includegraphics[width=\textwidth]{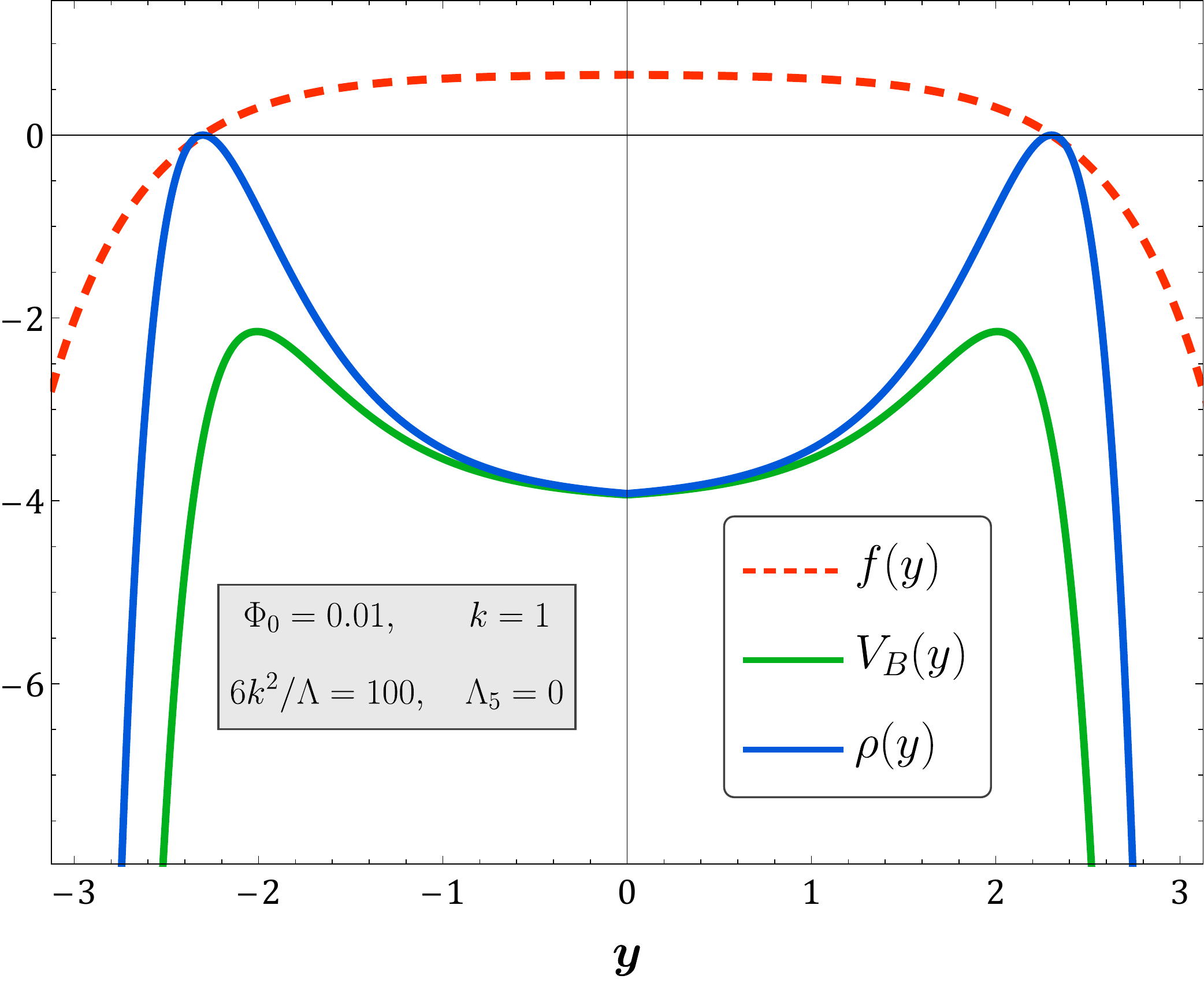}
        \caption{\hspace*{-1.2em}}
        \label{P1subf: f-V-rho-linear}
    \end{subfigure}
    \quad
    ~ 
    \begin{subfigure}[b]{0.47\textwidth}
        \includegraphics[width=\textwidth]{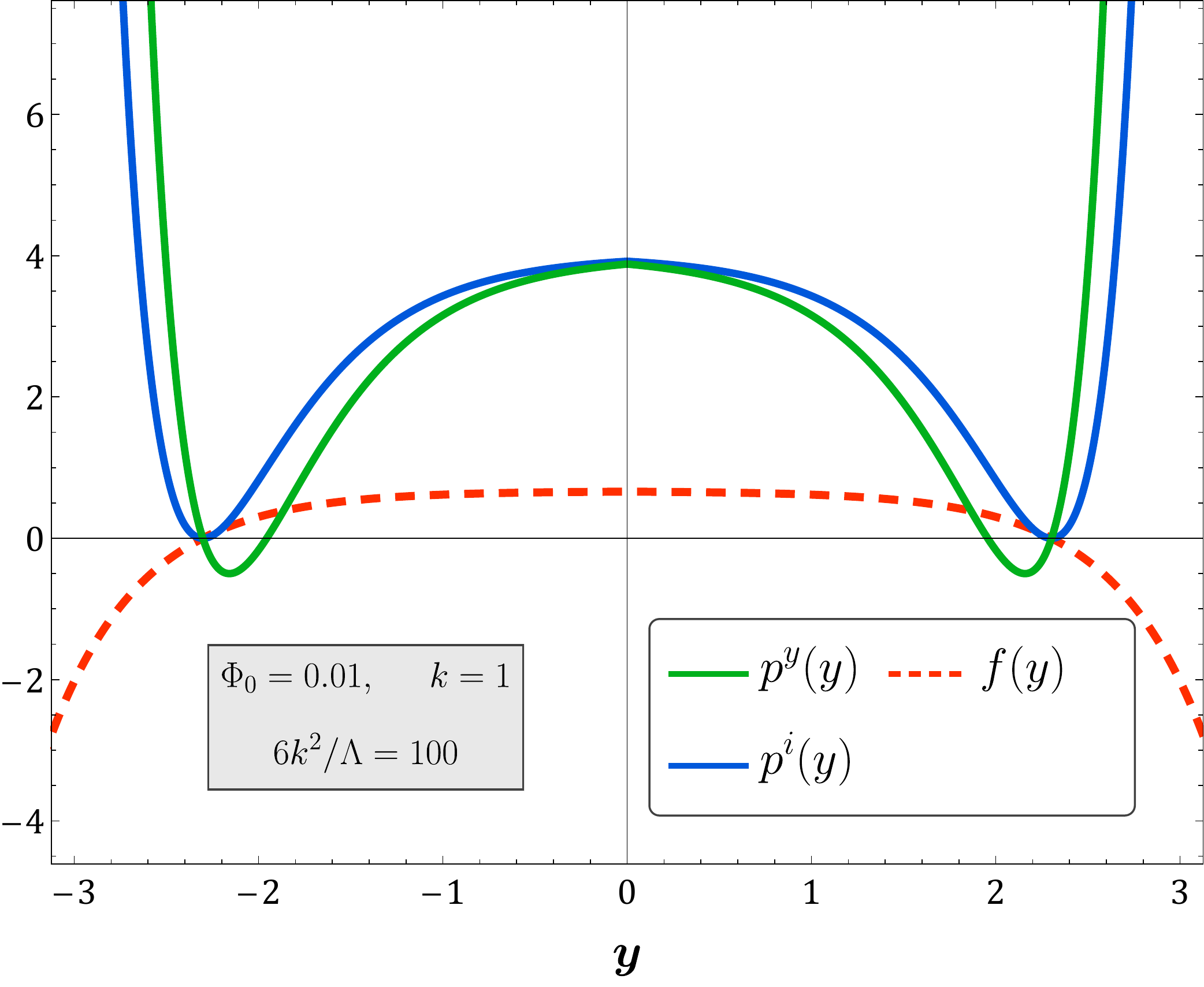}
        \caption{\hspace*{-1.2em}}
        \label{P1subf: f-py-pi-linear}
    \end{subfigure}
    \vspace*{-1em}
    ~ 
    \caption{(a) The scalar potential $V_B$ and energy density $\rho$ of the system, and
(b) the pressure components $p^y$ and $p^i$ in terms of the coordinate $y$
(from bottom to top in both plots), for $6k^2/\Lambda=100$, $\Phi_0=0.01$, $k=1$,
and $\Lambda_5=0$. We also display the coupling function $f$ with its characteristic plateau,
for easy comparison.}
   \label{P1fig: f-TMN-linear}
\end{figure}

In \myref{P1fig: f-TMN-linear}{P1subf: f-V-rho-linear}, we depict the form of the scalar potential
found above, for the choice of parameters $6k^2/\Lambda=100$, $\Phi_0=0.01$, $k=1$, and for
$\Lambda_5=0$. The regime close to our brane where $V_B$ mimics the negative cosmological
constant is clearly present. As we move away from the brane, the scalar field starts
increasing. This leads first to the formation of a small barrier (i.e. a local extremum),
as a result of the competing roles of the linear and quadratic in $\Phi$ terms in
Eq. (\ref{P1eq: V-linear}), and eventually to the divergence of $V$ toward minus infinity
at the boundary of spacetime. In the same plot, we depict the form of the coupling
function $f(y)$, for the same parameter values. This ensures us of the fact that the 
regime of the mimicking of ``negative cosmological constant'' and the location of the
barrier lies well inside the normal gravitating regime. At the point where $f(y)$ 
vanishes and gravity disappears, $V(y)$ retains a moderate, finite value; allowing one,
however, to enter the antigravitating regime leads to arbitrary large negative values
of the scalar potential.

The components of the energy-momentum tensor of the theory may be computed employing
Eqs. (\ref{P1eq: grav-eqs}) and (\ref{P1eq: TMN-mix-com}). Using also the relations $\rho=-T^{0}{}_0$,
$p^i=T^{i}{}_i$, and $p^y=T^{y}{}_y$, we find the explicit expressions
\beq
\rho=-p^i=\Lambda_5 -2a k^2\Phi + 2k^2 \Phi^2 + V_B(\Phi)
= - 4 k^2 \Phi_0^2\left(\frac{6k^2}{\Lambda}-e^{2ky}\right)^2\,,
\label{P1eq: rho-linear}
\eeq
\beq
p^y=-\Lambda_5 + 8a k^2 \Phi + 2k^2 \Phi^2 -V_B(\Phi)
= 8 k^2 \Phi_0^2\left(\frac{18 k^4}{\Lambda^2}-\frac{9 k^2}{\Lambda}\,e^{2ky}
+e^{4ky}\right)\,.
\label{P1eq: py-linear}
\eeq
The behavior of the above quantities is also depicted in \fref{P1fig: f-TMN-linear},
for the same values of parameters as in \fref{P1fig: warp-phi-f-linear} to allow for an easy comparison.
As expected, close to the brane the profile of all components resembles that of a
negative cosmological constant. At an intermediate distance, the energy density $\rho$
reaches a local, maximum value, and, far away from the brane---inside the
antigravitating regime---it diverges to negative infinity. The pressure components
$p^y$ and $p^i$ exhibit the exact opposite behavior. Starting from a constant
value near the brane, they dive toward a local minimum and, inside the antigravitating
regime, diverge to positive infinity. We readily observe that the total energy density
$\rho$ of the system remains negative throughout the bulk---this is also obvious from
its expression in Eq. (\ref{P1eq: rho-linear}). However, this is due to a physical, scalar field
with a potential that turns out to be negative in order to create the local AdS spacetime
and support the decreasing warp factor. Close to the brane, that potential is analytic
and finite---should one wish to ban the diverging, antigravitating regime from the
bulk spacetime, a second brane could easily be introduced at a distance $y=L<y_0$.  
The necessity of introducing a second brane in the model will be discussed shortly.

\mysubsection{Junction conditions and effective theory \label{P1subsec: brane-linear}}

Let us now address the issue of the junction conditions introduced in the
model due to the presence of the brane at $y=0$. The energy content of the brane
is given by the combination $\sigma + V_b(\Phi)$, and it creates a discontinuity
in the second derivatives of the warp factor and scalar field at the location of
the brane. We write $A''=\hat A'' + [A']\,\delta(y)$ and $\Phi''=\hat \Phi'' +
 [\Phi']\,\delta(y)$, where the hat quantities denote the distributional (i.e. regular)
parts of the second derivatives and $[\cdots]$ gives the discontinuities of the corresponding
first derivatives across the brane \cite{BDL}. Then, if we reintroduce the delta-function
terms in both the Einstein equation (\ref{P1eq: grav-eq1}) and the scalar-field equation (\ref{P1eq: phi-eq}),
and match the coefficients of the delta-function terms, we obtain the conditions
\eq$\begin{array}{c}
3 f(\Phi)\,[A'] = -[\Phi']\,\partial_\Phi f- (\sigma + V_b)\,, \\[3mm]
[\Phi'] = 4 [A']\,\partial_\Phi f + \partial_\Phi V_b\,,
\end{array}
\label{P1eq: jun-conds}$
where all quantities are evaluated at $y=0$.  Using the expressions for the warp factor
and the scalar field, as well as the symmetry in the bulk under the change
$y \rightarrow -y$, we find their explicit forms
\bea
\frac{8 k^2}{\Lambda}\,k \Phi_0^2 \left(1-\frac{18 k^2}{\Lambda}\right) &=&
-\sigma -V_b(\Phi_0)\,, \label{P1eq: jun1-linear}\\[2mm]
4k\Phi_0\left(1-\frac{8k^2}{\Lambda}\right)&=&\partial_\Phi V_b\bigl|_{y=0}\,.
\label{P1eq: jun2-linear}
\eea
According to the second junction condition, in the absence of an interaction term $V_b$
of the scalar field with the brane, we should have $k^2=\Lambda/8$. This result
determines the sign of the four-di\-men\-sion\-al cosmological constant that must necessarily
be positive and relates its magnitude to the scale of warping in the bulk.
Moreover, the dimensionless quantity $k^2/\Lambda$, which determines the range
of the gravitating regime, should be exactly $1/8$. This value, being smaller than $1/6$,
does not allow for a normal gravity regime anywhere in the bulk, according to the
discussion above. The first of the conditions also leads to the result $\Phi_0^2=4\sigma/5k$;
for the case $k>0$, which ensures the decrease of the warp factor away from our brane,
the brane self-energy $\sigma$ comes out to be positive, too, and thus is physically
acceptable. 

As we showed above, the existence of a normal-gravity regime close to our brane
demands the presence of an interaction term $V_b$ of the scalar field with the brane.
Although the number of choices for $V_b$ is
in this case infinite, one may draw some general conclusions: if we assume again that
$k>0$ and that $k^2/\Lambda>1/6$, so that a positive $f(\Phi)$-regime exists around
our brane, then Eq. (\ref{P1eq: jun1-linear}) still ensures that the total
energy content of our brane $\sigma + V_b(\Phi_0)$ is always positive. Assuming now,
as an indicative case, a linear form for the interaction term, too, i.e.
$V_b(\Phi)=\lambda_0\,\Phi$, where $\lambda_0$ is a coupling constant, we obtain the
conditions
\beq
\frac{8 k^2}{\Lambda}\,k \Phi_0^2 \left(1-\frac{18 k^2}{\Lambda}\right) =
-\sigma -\lambda_0 \Phi_0\,, \qquad
4k\Phi_0\left(1-\frac{8k^2}{\Lambda}\right)=\lambda_0\,.
\label{P1eq: jun-linear-ex}
\eeq
The above two conditions determine two out of the five parameters of the model:
$k$, $\Lambda$, $\Phi_0$, $\lambda_0$, $\sigma$. Considering the bulk scalar field and the
self-energy of the brane as the constituents of the model that support the complete
bulk-brane solution, the parameters related to them, namely the value of the field
on the brane $\Phi_0$, its coupling constant with the brane $\lambda_0$, and the brane
self-energy $\sigma$, may be naturally chosen as the true independent quantities of
the theory. On the other hand, the scale of the warping $k$ and the effective
cosmological constant on the brane $\Lambda$ are determined through the junction
conditions by the aforementioned three fundamental parameters. In this case, one
may easily see that, for $\lambda_0 \Phi_0>0$, we obtain $k^2/\Lambda<1/8$, which
allows for a bulk that is everywhere antigravitating, while, for $\lambda_0 \Phi_0<0$,
solutions with large values of $k^2/\Lambda$ may be obtained that have their
antigravitating regime pushed away from our brane.

We should finally address the issue of the effective theory on the brane. The negative sign
of the coupling function $f(\Phi)$ emerging at some distance from the brane as well
as the diverging behavior of the field $\Phi$ in the same region raise concerns
about the type of the effective theory that a four-dimensional observer would witness.
In order to answer this question, we need to derive the four-dimensional effective
action by integrating the five-dimensional one, given in Eq. (\ref{P1eq: action}), over the
fifth coordinate $y$. Employing the first of Eqs. (\ref{P1eq: invar}), we write
$R=-20k^2 +R^{(4)} e^{2ky}$, where $R^{(4)}=4 \Lambda$ is the scalar
curvature of the projected-on-the-brane line element (\ref{P1eq: metric-4D}). Then, the
action takes the form
\beq
S=\int d^4 x\,dy\,\sqrt{-g^{(5)}}\left[\frac{f(\Phi)}{2}\,\left(e^{2ky} R^{(4)}-
20k^2\right) -\Lambda_5-\frac{1}{2}\,\Phi'^2 -V_B(\Phi)\right]. 
\label{P1eq: eff-action}
\eeq
Using also that $\sqrt{-g^{(5)}}=e^{-4k|y|} \sqrt{-g^{(4)}}$, the four-di\-men\-sion\-al,
effective gravitational constant would be given by the integral
\beq
\frac{1}{\kappa_4^2}\equiv 2 \int_{0}^{\infty} dy\, e^{-2 k y}\,f(\Phi)
=\frac{8k^2 \Phi_0^2}{\Lambda^2}\,\int_{0}^{\infty} dy\,\left(-\Lambda +
6k^2 e^{-2 k y}\right)\,.
\label{P1eq: effG1-linear}
\eeq
In the above, we have substituted the form of the coupling function $f(\Phi)$ given in
Eq. (\ref{P1eq: f-linear}). We observe that, although the second term inside the brackets will
lead to a finite result even for a noncompact fifth dimension---similar to the
Randall-Sundrum model, the first term will give
a divergent contribution. As a result, the presence of a second brane at a distance
$y=L$ is imperative for a well-defined effective theory. In that case, the upper limit
of the $y$-integral in Eq. (\ref{P1eq: effG1-linear}) is replaced by $L$, and we obtain
\beq
\frac{M_{Pl}^2}{8\pi}=\frac{\Phi_0^2}{k}\,\frac{8 k^2}{\Lambda} \left[
\frac{3k^2}{\Lambda}\,(1-e^{-2 k L})-kL\right]\,.
\label{P1eq: effG2-linear}
\eeq
Compared to the Randall-Sundrum model \cite{RS1, RS2}, the expression for the four-dimensional
gravity scale $M_{Pl}^2$ involves the quantity $\Phi_0^2$---with units $[M]^3$---and
the dimensionless parameter $k^2/\Lambda$ on its right-hand side. This signifies the fact
that, in the context of the theory (\ref{P1eq: action}), the five-dimensional gravity scale $M_5^3$
may altogether be replaced by the coupling function $f(\Phi)$. If one chooses large values
for the $k^2/\Lambda$ parameter, then the value of the effective Planck scale may differ
from that of $\Phi_0$ by orders of magnitude. In fact, the smaller the cosmological constant
is on our brane, the more extended is the positive-value regime for $f(\Phi)$, as we
saw in the previous subsection, and the larger the difference between $M_{Pl}^2$
and $\Phi_0^2$. Equation (\ref{P1eq: effG2-linear}) contains also a term linear in the interbrane
distance $L$, which was absent in the Randall-Sundrum case. Therefore, one should
take care that the inequality $kL <3k^2/\Lambda$ is always satisfied---however,
for small values of the four-dimensional cosmological constant on the brane, as argued
above, this constraint should easily be satisfied. 

The introduction of the second brane in order to ensure a finite effective theory
on our brane is supplemented by a second set of junction conditions at the location
$y=L$. A brane source term of the form $-[\hat \sigma + \hat V_b(\Phi)]\,\delta(y-L)$
should be introduced in the action, where $\hat \sigma$ and $\hat V_b(\Phi)$ are
the self-energy of the second brane and the interaction term of the scalar field
with that brane, respectively. We follow a similar procedure as at $y=0$ and
arrive at a set of junction conditions similar to those in Eq. (\ref{P1eq: jun-conds})
but with all quantities evaluated at $y=L$. Their explicit form reads
\bea
\frac{8 k^2}{\Lambda}\,k \Phi_0^2 \left(\frac{18 k^2}{\Lambda}-e^{2kL}\right) &=&
-\hat\sigma -\hat V_b(\Phi)\bigl|_{y=L}\,, \label{P1eq: jun1-sec-br}\\[2mm]
4k\Phi_0\left(\frac{8k^2}{\Lambda}-e^{2kL}\right)&=&\partial_\Phi \hat V_b\bigl|_{y=L}\,.
\label{P1eq: jun2-sec-br}
\eea
The above set of conditions may be used to determine two more parameters of the
model. One may be the interbrane distance $L$ and the other a parameter 
associated with the interaction term $\hat V_b$. The self-energy of the second
brane $\hat \sigma$ as well as the functional form of $\hat V_b(\Phi)$ remain
completely arbitrary.

To complete the derivation of the effective theory on the brane, we finally
compute the effective cosmological constant---this may be used as a consistency
check of our results. The cosmological constant on the brane is given by the
integral of the remaining terms in Eq. (\ref{P1eq: eff-action})---since $\Phi$ is
only $y$-dependent, no dynamical field will survive in the effective theory.
These terms will be supplemented by the source terms of the two branes as well as
the Gibbons-Hawking terms at the boundaries of spacetime \cite{Gibbons-terms}.
In total, we will have
\bea
-\Lambda_4&=&\int_{-L}^L dy\,e^{-4k|y|}\left[-10 k^2 f(\Phi) -\Lambda_5-
\frac{1}{2}\,\Phi'^2 -V_B(\Phi) +f(\Phi)(-4A'')|_{y=0}\,+\right. \nonumber \\[2mm]
&& \hspace*{1cm}\left.+f(\Phi)(-4A'')|_{y=L} -[\sigma +V_b(\Phi)]\,\delta(y)-
[\hat \sigma + \hat V_b(\Phi)]\,\delta(y-L)\right].
\label{P1eq: L-eff-1}
\eea
Employing the expressions for the coupling function and scalar potential, Eqs. (\ref{P1eq: f-linear})
and (\ref{P1eq: V-linear}), respectively, as well as the junction conditions (\ref{P1eq: jun1-linear}) and
(\ref{P1eq: jun1-sec-br}), and integrating over $y$, we finally obtain the result
\beq
\Lambda_4= 8k \Phi_0^2\left[\frac{3k^2}{\Lambda} \left(1-e^{-2kL}\right) -kL\right] =
\frac{\Lambda}{\kappa_4^2}\,,
\label{P1eq: L-eff-2}
\eeq
where we have used the expression for the effective gravitational scale $M_{Pl}^2/8\pi=1/\kappa_4^2$
given in Eq. (\ref{P1eq: effG2-linear}).
As expected, the derivation of the effective theory has confirmed the interpretation of the
metric parameter $\Lambda$ as the product $\kappa_4^2 \Lambda_4$, that followed
also by comparing the projected-on-the-brane line element (\ref{P1eq: metric-4D}) with
the standard Schwarzschild de Sitter background.

We would also like to note that the presence of the mass parameter $M$ has played no
role either in the profile of the scalar field and the energy-momentum tensor components
or in the derivation of the junction conditions and the effective theory on the brane.
Its presence creates a Schwarzschild de Sitter background on the brane and an extended
singularity into the bulk leading to a five-dimensional black string stretching between the two
branes. If we set this parameter equal to zero, then the four-dimensional background on
the brane reduces to a pure de Sitter spacetime while the five-dimensional background
is free of singularities as long as $L<\infty$. For $L>y_0$, the bulk will also contain
an antigravitating regime (unavoidable for $\Lambda_4>0$, as we will see in the next
section).


\mysection{The quadratic case \label{P1sec: Quadratic}}

We now move to the case where the coupling function has a quadratic form, i.e.
$f(\Phi)=a\,\Phi^2$, where $a$ is again a constant. Employing, as in the previous
subsection, the form of the mass function (\ref{P1eq: mass-sol}) and the warp factor
$e^{A(y)}=e^{-ky}$,  Eq. (\ref{P1eq: grav-eq1}) now takes the form
\beq
a \Lambda e^{2ky} \Phi^2=-2a\,\Phi\,(k\,\Phi' + \Phi'') -(1+2a)\Phi'^2\,.
\label{P1eq: grav-eq1-quad}
\eeq
Again, we set: $\Phi(y)=\Phi_0\,e^{g(y)}$, and the above equation is rewritten as
\beq
a \Lambda e^{2ky} =-2a\,(k g' + g'') -(1+4a)g'^2\,.
\label{P1eq: grav-eq1-2-quad}
\eeq
The above calls for an exponential dependence for the function $g(y)$---we thus set
$g(y)=g_0\,e^{\lambda y}$, with $g_0$ and $\lambda$ constant coefficients, and write
the above equation as
\beq
a\Lambda\,e^{2ky} = - 2a g_0 \lambda\,(k+\lambda)\,e^{\lambda y}  
-(1+4a) g_0^2 \lambda^2\,e^{2\lambda y}\,.
\label{P1eq: main-eq-quad}
\eeq
There is again only one nontrivial solution that satisfies the aforementioned equation,
and this corresponds to the choice $\lambda=2k$. Then, the following constraints
should hold:
\bea
a = -\frac{1}{4}\,, \qquad \quad
g_0 = -\frac{\Lambda}{12 k^2}\,. \label{P1eq: consts-quad}
\eea
The coefficient $a$ is negative definite, and therefore in this case gravity acts as a repulsive
force over the entire bulk. We should note here that an attempt to generalize the form of
the coupling function according to the ansatz $f(\Phi)=a\,\Phi^2 + b\,\Phi +c$, where
($a,b,c$) are constant coefficients, failed to lead to a consistent solution. Had such a
solution been possible, we could perhaps find regimes in the $y$-coordinate where gravity
would act as normal, hopefully close to our brane. Unfortunately such a solution has not
emerged, and therefore for a quadratic coupling function, the theory always leads to an
antigravitating bulk. This feature is strongly connected to the presence of the
cosmological constant on the brane---we will return to this point in the following
section.

\begin{figure}[t!]
    \centering
    \begin{subfigure}[b]{0.44\textwidth}
        \includegraphics[width=\textwidth]{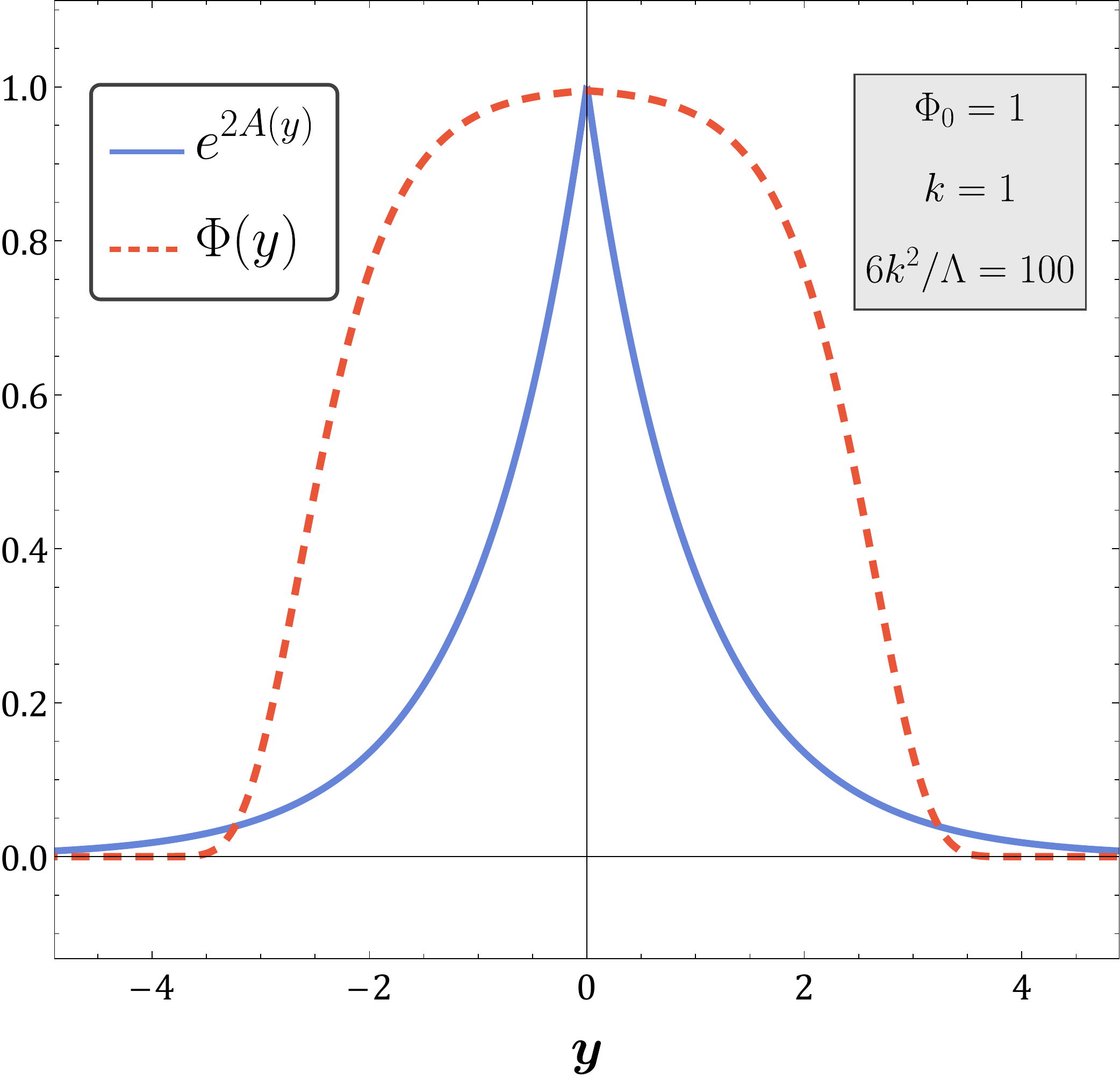}
        \caption{\hspace*{-1.3em}}
        \label{P1subf: warp-phi-quad}
    \end{subfigure}
    \quad
    ~ 
    \begin{subfigure}[b]{0.49\textwidth}
        \includegraphics[width=\textwidth]{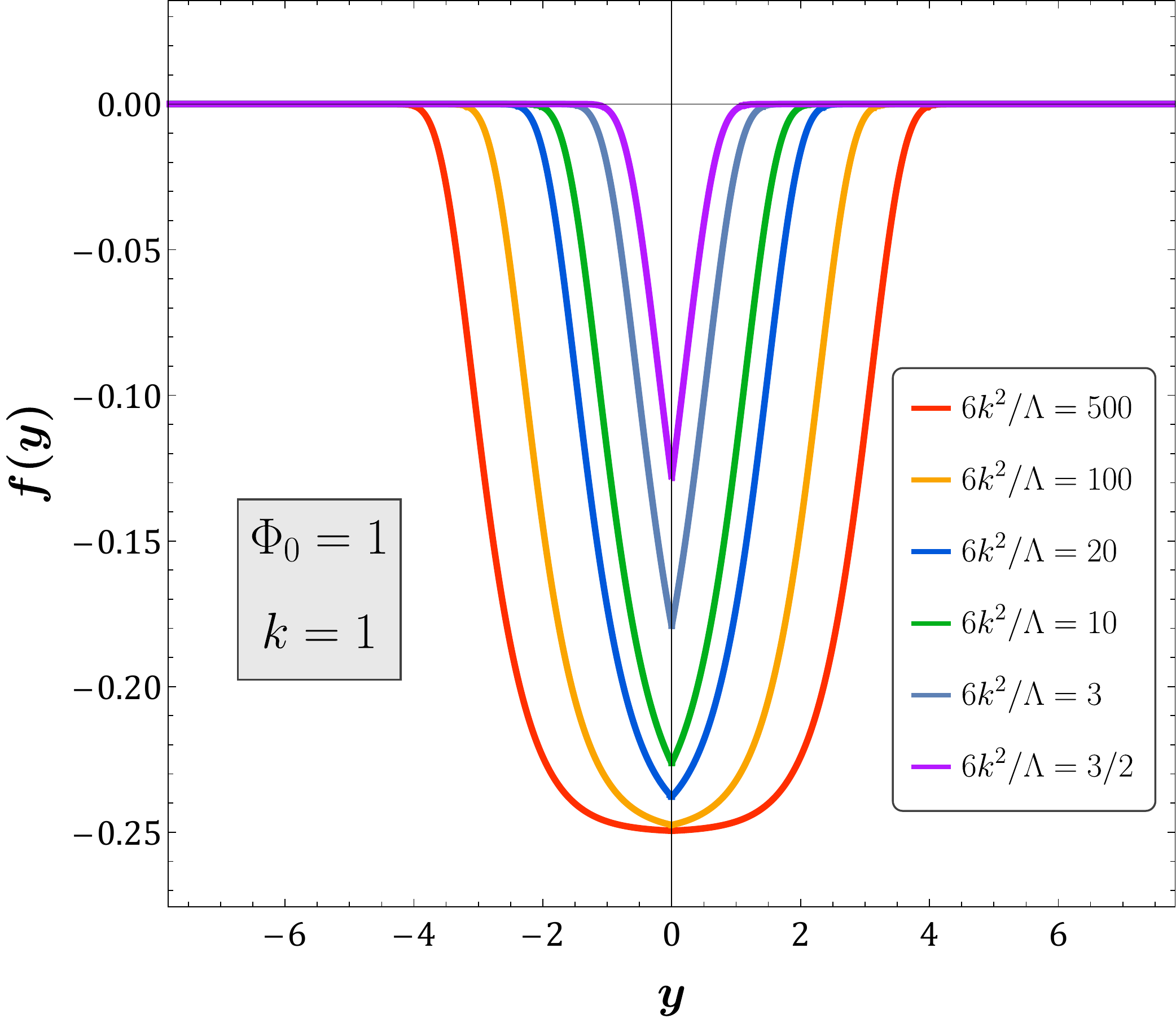}
        \caption{\hspace*{-3.3em}}
        \label{P1subf: f-y-quad}
    \end{subfigure}
    ~ 
    \vspace*{-1em} 
   \caption{(a) The warp factor $e^{-2k |y|}$ and scalar field $\Phi(y)$, and (b)
     the coupling function $f(y)=a\, \Phi^2(y)$, in terms of the coordinate $y$,
   for $k=1$, $\Phi_0=1$, and $6k^2/\Lambda=3/2,3,10,20,100,500$ (from top to bottom).}
   \label{P1fig: warp-phi-f-quad}
\end{figure}

Let us, however, investigate the remaining aspects of the model. The warp factor 
assumes the standard Randall-Sundrum form, i.e. $e^{2A(y)}=e^{-2k |y|}$, and is displayed
in \myref{P1fig: warp-phi-f-quad}{P1subf: warp-phi-quad}. The profile of the scalar field depends 
on the sign of the
parameter $\Lambda$. Using the second of the constraints (\ref{P1eq: consts-quad}), we find that
\beq
\Phi(y)=\Phi_0\,\exp\left(-\frac{\Lambda}{12k^2}\,e^{2ky}\right)\,.
\label{P1eq: Phi-quad}
\eeq
The projected-on-the-brane line element is still given by Eq. (\ref{P1eq: metric-4D});
thus, $\Lambda$ is again proportional to the brane cosmological constant. Then,
Eq. (\ref{P1eq: Phi-quad}) tells us that, for a positive cosmological constant on the brane,
the scalar field takes its maximum value $\Phi=\Phi_0 e^{-\Lambda/12k^2}$ at the
location of our brane, and decreases fast as we move away from the brane.
Therefore, the scalar field exhibits a localization around our brane similar to
that of the warp factor; in particular, for small values of the parameter $k^2/\Lambda$,
the profile of the scalar field exhibits a cusp at the location of the brane ($y=0$)
while, as $k^2/\Lambda$ increases, a plateau appears around the brane.
The coupling function, $f(\Phi)=a \Phi^2$,
assumes a similar profile by decreasing very fast, as $y$ increases; as a result,
the antigravitating regime associated with $f(y)$ is rather small. The 
profiles of the scalar field and coupling function, for $\Lambda>0$, are 
depicted in \myref{P1fig: warp-phi-f-quad}{P1subf: warp-phi-quad} and
\myref{P1fig: warp-phi-f-quad}{P1subf: f-y-quad}, respectively. 
On the other hand, for a negative cosmological constant on the brane, the scalar
field increases very fast away from the brane blowing up at the boundary of the
spacetime, and the same behavior is exhibited by the coupling function $f(\Phi)$. 
In what follows, we ignore this unattractive solution and explore further the more
interesting one with a positive cosmological constant on the brane.

\begin{figure}[t!]
    \centering
    \begin{subfigure}[b]{0.465\textwidth}
        \includegraphics[width=\textwidth]{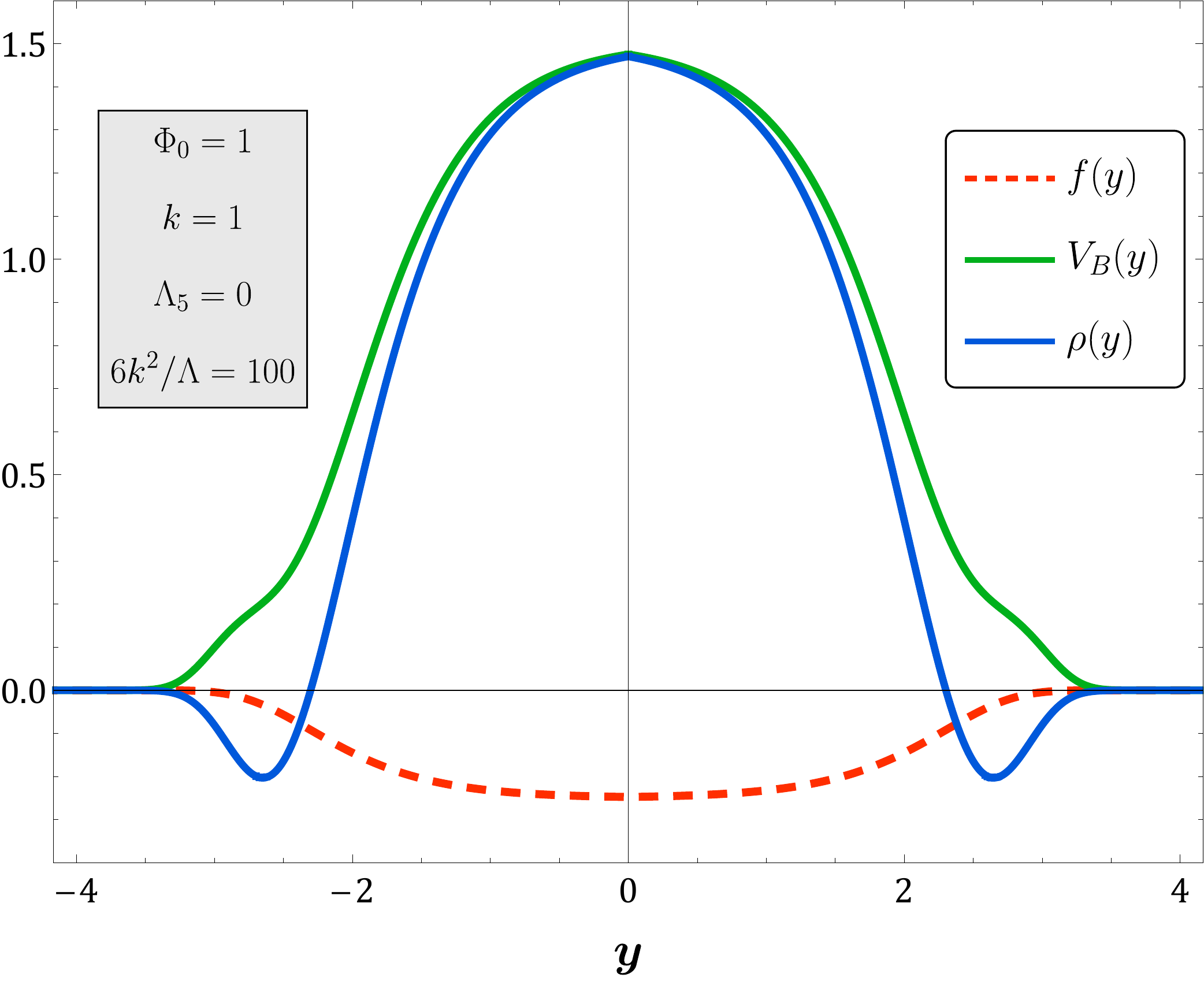}
        \caption{\hspace*{-1.2em}}
        \label{P1subf: f-V-rho-quad}
    \end{subfigure}
    \quad
    ~ 
    \begin{subfigure}[b]{0.47\textwidth}
        \includegraphics[width=\textwidth]{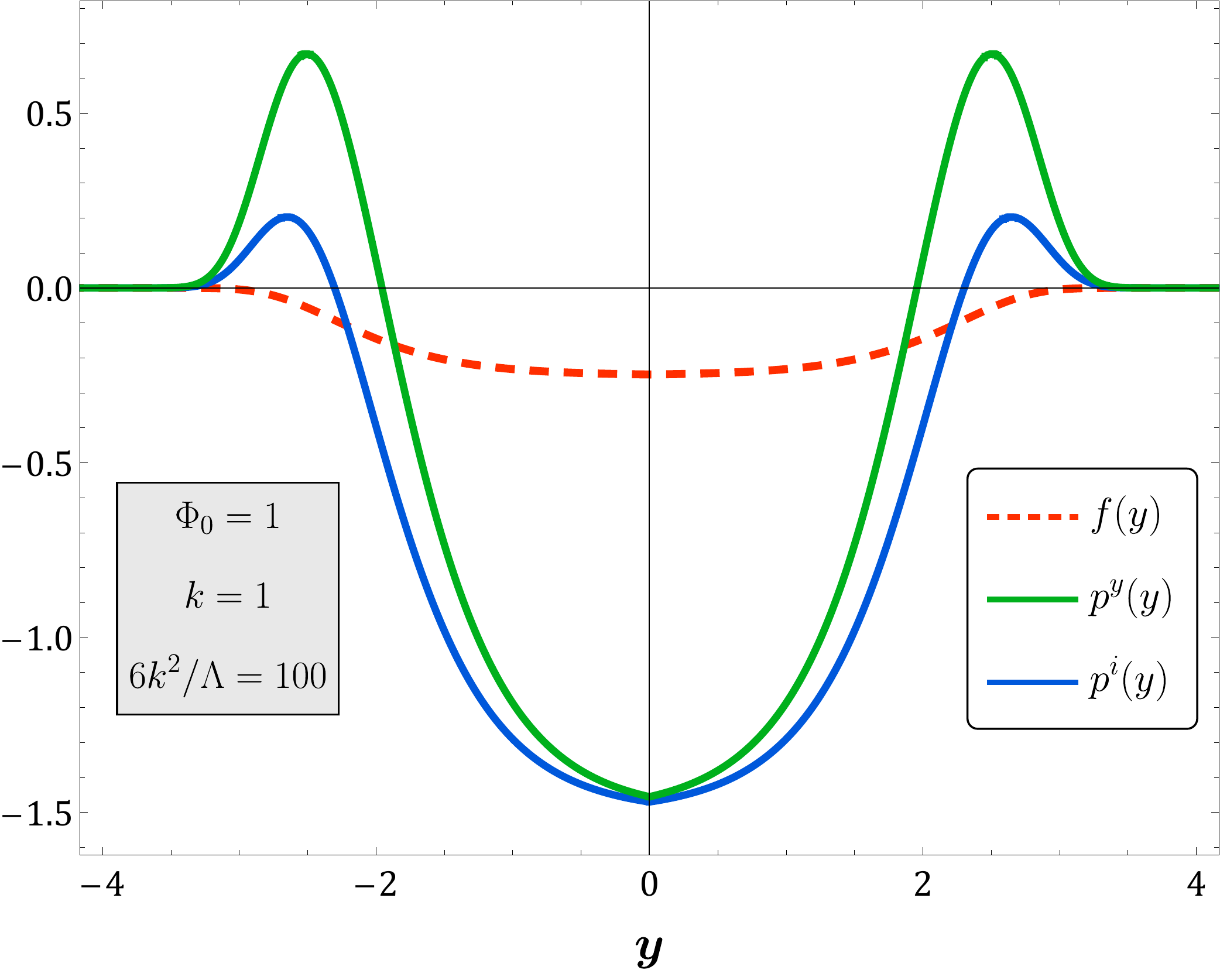}
        \caption{\hspace*{-1.5em}}
        \label{P1subf: f-py-pi-quad}
    \end{subfigure}
    ~ 
    \vspace*{-1em}  
 \caption{(a) The scalar potential $V_B$ and energy density $\rho$ of the system, and
(b) the pressure components $p^y$ and $p^i$ in terms of the coordinate $y$
(from top to bottom in both plots), for $6k^2/\Lambda=100$, $\Phi_0=1$, $k=1$,
and $\Lambda_5=0$. We also display the coupling function $f$,
for comparison.}
\vspace*{-0.5em}
   \label{P1fig: f-TMN-quad}
\end{figure}

We also need to derive the form of the potential $V_B(\Phi)$ of the scalar field in
the bulk. This follows easily from Eq. (\ref{P1eq: grav-eq2}) leading to the expression
\beq
V_B(\Phi)=-\Lambda_5 +k^2 \Phi^2\left[\frac{3}{2} + 2 \ln \left(\frac{\Phi}{\Phi_0}\right)
+2  \ln^2 \left(\frac{\Phi}{\Phi_0}\right)\right],
\label{P1eq: V-quad}
\eeq
or, in terms of the $y$-coordinate,
\beq
V_B(y)=-\Lambda_5 +k^2 \left(\frac{3}{2} -\frac{\Lambda}{6k^2}\,e^{2ky} +
\frac{\Lambda^2}{72k^4}\,e^{4ky} \right) \Phi_0^2 \exp\left(-\frac{\Lambda}{6k^2}\,e^{2ky}\right).
\label{P1eq: V-y-quad}
\eeq
The bulk potential in principle consists of the negative cosmological-constant term and a
term that is related to the scalar field. For $\Lambda>0$, this second term decreases very
fast exhibiting also a localization around our brane---its profile is shown in
\myref{P1fig: f-TMN-quad}{P1subf: f-V-rho-quad} (next page).
Setting $z=\Lambda e^{2ky}/6k^2$,
one may easily see that the second-order polynomial inside the brackets has no real roots,
and is thus always positive definite. As a result, the second term tends to reduce the
negative bulk cosmological constant, if present, with this effect being more important
close to the brane and negligible far away. In fact, the emergence of a decreasing warp
factor has not been related so far to the presence of $\Lambda_5$.

The components of the energy-momentum tensor may also easily be derived from
Eqs. (\ref{P1eq: grav-eqs}) and (\ref{P1eq: TMN-mix-com}) using the solution for the scalar field
and scalar potential. They have the form
\beq
\rho=-p^i=\Lambda_5 +2a \Phi\,(3 A' \Phi' + \Phi'') + V_B(\Phi)
= \frac{3}{2}\,k^2 \Phi^2\left(1-\frac{\Lambda}{6k^2}\,e^{2ky}\right)\,,
\label{P1eq: rho-quad}
\eeq
\beq
p^y=-\Lambda_5 +\frac{1}{2}\,\Phi'^2 -8a \Phi  A' \Phi' -V_B(\Phi)
= -\frac{3}{2}\,k^2 \Phi^2\left(1-\frac{\Lambda}{3k^2}\,e^{2ky}\right)\,.
\label{P1eq: py-quad}
\eeq
The form of the energy-momentum components are depicted in \myref{P1fig: f-TMN-quad}{P1subf: f-V-rho-quad} and
\myref{P1fig: f-TMN-quad}{P1subf: f-py-pi-quad}. The energy density matches the value of the potential at the location
of the brane and decreases slightly faster than the latter away from the brane;
it remains predominantly
positive apart from a small regime at large distances from the brane. The
pressure components exhibit the exact opposite behavior regarding their sign.
Overall, the energy-momentum components resemble those of a {\it positive}
cosmological constant close to our brane, then decrease fast, and finally vanish
at large distances exhibiting a nice localization pattern. We should stress that,
according to our analysis, no negative, bulk cosmological constant needs to be
introduced by hand. It is, in fact, the negative value of the coupling function $f(\Phi)$
that turns the coupling term between the scalar field and the Ricci scalar to a
form of a negative distribution of energy; it is this term then that manages to
support the exponentially falling warp factor even in the absence of a typical
AdS spacetime. 

The presence of the brane, with its nontrivial energy content, introduces once
again discontinuities in the derivatives of the warp factor and scalar field. The
associated junction conditions at $y=0$ have the same form as in 
Eqs. (\ref{P1eq: jun-conds}). Their explicit forms, however,  are bound to be different
and are given by 
\bea
\frac{k \Phi_0^2}{2}\,e^{-\Lambda/6k^2}\left(3+\frac{\Lambda}{3k^2}\right) &=&
-\sigma -V_b\bigl|_{y=0}\,, \label{P1eq: jun1-quad}\\[2mm]
2k\Phi_0\,e^{-\Lambda/12k^2}\left(2+\frac{\Lambda}{6k^2}\right)&=&
-\partial_\Phi V_b\bigl|_{y=0}\,.
\label{P1eq: jun2-quad}
\eea
Since the left-hand sides of the above equations are positive definite, the interaction
term $V_b$ of the scalar field with the brane must be not only nonvanishing but
necessarily negative (with a negative first derivative, too) in order to avoid a negative
brane self-energy $\sigma$. As before, the above conditions may fix the parameters
$k$ and $\Lambda$ while the scalar-field parameters $\Phi_0$ and $V_b$, as well
as $\sigma$, may remain arbitrary.
   
We should, however, stress that this particular solution, being either a black string
or regular in the bulk, cannot constitute a realistic model due to the negative sign
of the coupling function $f(\Phi)$. This sign will be carried over to the four-dimensional
effective theory leading to antigravity on the brane. Indeed, working as in the previous
subsection and isolating the coefficient of $R^{(4)}$ in the action, we arrive at the
relation\vspace*{-0.3em}
\beq
\frac{1}{\kappa_4^2} = -\frac{\Phi_0^2}{2}\,\int_{0}^{\infty} dy\,e^{-2 k y}\,
\exp\left(-\frac{\Lambda}{6k^2}\,e^{2ky}\right) = -\frac{\Phi_0^2}{4k}
\left(e^{-\Lambda/6k^2}-\frac{\Lambda}{3k}\,\mathcal{I}\right),
\label{P1eq: effG1-quad}
\eeq
where
\beq
\mathcal{I} \equiv \int_{0}^{\infty} dy\,\exp\left(-\frac{\Lambda}{12k^2}\,e^{2ky}\right).
\label{P1eq: I-def-quad}
\eeq
The integral $\mathcal{I}$ may be computed numerically and yields a finite result; therefore, there
is no need for the introduction of a second brane in this model.\,\footnote{A similar
analysis to that of \secref{P1subsec: brane-linear}, but simpler due to the absence of the second brane, leads
to the derivation of the effective cosmological constant on the brane, which once again
comes out to be $\Lambda_4=\Lambda/\kappa^2_4$.} Nevertheless, the value
of the effective gravitational constant $\kappa^2_4$ turns out to be negative---this 
becomes clear if one looks at the middle part of Eq. (\ref{P1eq: effG1-quad}), where a 
negative coefficient multiplies a positive-definite integral. This result is catastrophic, and
therefore, the model is not a viable one. Its emergence, however, reveals two facts:
(i) that antigravitating solutions in the context of the theory (\ref{P1eq: action}) are
somehow associated with the positive cosmological constant on the brane since two
such solutions have emerged for two different choices of the coupling function, and
(ii) that, when $M \neq 0$, the theory of a nonminimally coupled scalar field to
gravity gives rise to yet another undesired black-string solution rather than a physically
motivated localized black-hole solution.


\mysection{A theoretical argument \label{P1sec: Theoretical}}

In the previous section, we have constructed explicit solutions that emerge 
from the five-di\-men\-sion\-al field equations, and describe a four-di\-men\-sion\-al
Schwarzschild de Sitter background on the brane. From the bulk point of view,
these solutions describe either black strings, if $M \neq 0$, or regular,
maximally symmetric solutions over the whole bulk apart from its
boundary at $y \rightarrow \infty$, if $M=0$---a second brane could easily shield
the boundary singularity creating two-brane models with a compact fifth
dimension. In both cases, however, the bulk solution is characterized,
either globally or over particular regimes, by a negative coupling function
$f(\Phi)$ that leads to an antigravitating theory. In this section, we
examine from the mathematical point of view why the emergence of such
solutions is possible in the context of the given theory, and why they
do so particularly for the physically motivated case of a positive cosmological
constant on the brane.

The analysis will focus on the gravitational equation \eqref{P1eq: grav-eq1}. By
employing the relations
\beq
\pa_y f= \Phi'\,\pa_\Phi f\,, \qquad 
\pa_y^2 f=\Phi'^2 \,\pa_\Phi^2 f+\Phi''\,\pa_\Phi  f\,,
\eeq
as well as the expressions $A(y)=-k|y|$ and $m(r)=M+\Lambda r^3/6$, Eq. \eqref{P1eq: grav-eq1}
is written as
\eq$
\Lambda\ e^{2k|y|} f=-\Phi'^2-\pa_y^2 f-k\, sgn(y)\, \pa_y f\,.$
We assume once again the existence of the ${\bf Z}_2$-symmetry in the bulk and restrict
our analysis to the positive $y$-regime for simplicity. Consequently, we write
\eq$\label{P1eq: grav1-v1}
\Phi'^2=-\pa_y^2 f-k\pa_y f-\Lambda e^{2ky} f\,.$
The first derivative of the scalar field $\Phi'$ may vanish at particular
values of the coordinate $y$ but is assumed to be in general nonvanishing
to allow for a nontrivial scalar field in the bulk. Also, both functions $f=f(y)$ and
$\Phi=\Phi(y)$ should be real in their whole domain. Therefore, both sides of
Eq. (\ref{P1eq: grav1-v1}) should be non-negative, which finally leads to the constraint
\beq
\label{P1eq: grav1-v2}
\pa_y^2 f+ k\pa_y f + \Lambda e^{2ky} f \leq 0\,.
\eeq
The above constraint should be satisfied for every solution of the field
equations (\ref{P1eq: mass-eq})--(\ref{P1eq: grav-eq2}), including the ones presented in
\chapc{Secs.}\hspace{0.3em}\ref{P1subsec: bulk-linear} and \ref{P1subsec: brane-linear}. These were characterized by $\Lambda>0$, in which case
the combination $\Lambda e^{2ky}$, appearing in the last term of the above
expression, diverges to $+\infty$ at the boundary of spacetime. But there,
the coupling function $f(y)$ is negative for both solutions, and this renders
the last, dominant term smaller than zero as the constraint demands. Also,
for all other values of $y$, one may easily check that the profiles of the
function $f(y)$ found in \chapc{Secs.}\hspace{0.3em}\ref{P1sec: Linear} and \ref{P1sec: Quadratic} always satisfy the constraint
(\ref{P1eq: grav1-v2}).

In what follows, we investigate whether physically acceptable
solutions with $f(y)>0$ may arise in the case where $\Lambda$ is also positive. 
To this, we will add the demand that the components of the energy-momentum
tensor may be localized close to the brane, and are certainly non-diverging
at the boundary of spacetime. These may be written as
\bea
\rho=-p^i &=& \frac{1}{2}\,\Phi'^2 +V_B(\Phi) + \Lambda_5+ 3A' \partial_y f
+ \partial_y^2 f\,, \label{P1eq: rho-general} \\[1mm]
p^y &=&  \frac{1}{2}\,\Phi'^2 -V_B(\Phi) -\Lambda_5- 4A' \partial_y f\,.
\label{P1eq: py-general}
\eea
From Eq. (\ref{P1eq: grav-eq2}), one may solve for the general form of the scalar
potential to find
\beq
V_B(\Phi)=-\Lambda_5 -\frac{1}{2}\,\Phi'^2 - 3A' \partial_y f - \partial_y^2 f
-f\,(6k^2-\Lambda\,e^{2ky})\,. \label{P1eq: potential-general}
\eeq
Employing the above into the expressions (\ref{P1eq: rho-general}) and (\ref{P1eq: py-general}),
together with Eq. (\ref{P1eq: grav1-v1}), the energy-mo\-men\-tum tensor components
simplify to
\bea
\rho=-p^i &=& -6k^2 f(\Phi)+ f(\Phi)\,\Lambda e^{2ky}\,, \label{P1eq: rho-final} \\[2mm]
p^y &=& 6k^2 f(\Phi)-2 f(\Phi)\,\Lambda e^{2ky}\,.
\label{P1eq: py-final}
\eea
We observe that all components contain the diverging combination $\Lambda e^{2ky}$.
Therefore, we should demand the vanishing of the coupling function $f(\Phi)$ at the
boundary of spacetime at least as fast as $e^{-2ky}$. 

We will consider the most general such form, namely $f(y)=A e^{-\sum_{n=1}^N b_n y^n}$,
where $A$ and $b_n$ are arbitrary constants, and $N$ is a positive integer. The
first and second derivatives of $f(y)$ are found to be
\eq$
\pa_y f =-A e^{-\sum_{n=1}^N b_n y^n}\left(\sum_{n=1}^N b_n n y^{n-1}
\right),$
\eq$\label{P1eq: df-exp}
\pa_y^2 f=A e^{-\sum_{n=1}^N b_n y^n} \left[\left(\sum_{n=1}^N b_n n y^{n-1} \right)^2 -\sum_{n=1}^N b_n n(n-1) y^{n-2}\right]. 
$
Both quantities quickly tend to zero which ensures the finiteness of the scalar potential
(\ref{P1eq: potential-general}). Then, the inequality constraint of Eq. (\ref{P1eq: grav1-v2}) reads
\gat$
f(\Phi)\left[\left(\sum_{n=1}^N b_n n y^{n-1} \right)^2 -\sum_{n=1}^N b_n n(n-1) y^{n-2}
-k\sum_{n=1}^N b_n n y^{n-1} +\Lambda e^{2ky}\right] \leq 0\,. \label{P1eq: con-final}$
Since $f(\Phi)$ is demanded to be everywhere positive, it is the expression inside the
square brackets that needs to be negative definite. For $N=1$, the latter reduces to
$b_1 (b_1-k) +\Lambda e^{2ky}$; but this, for $\Lambda>0$, is always positive definite
since $b_1\geq 2k$ according to the argument below Eqs. (\ref{P1eq: rho-final}) and (\ref{P1eq: py-final}).
For $N>1$, as $y$ increases away from the brane, the first and last terms are clearly
the dominant ones in Eq. (\ref{P1eq: con-final}); but these are again positive definite. Therefore,
in all cases the constraint (\ref{P1eq: grav1-v2}) is violated either over the entire $y$-regime
(as in the case studied in \secref{P1sec: Quadratic}) or at a distance from the brane (as in the case
studied in \secref{P1sec: Linear}).

\par To sum up, we have demonstrated that, for a function $f(\Phi)$ positive and decreasing
at large distances from our brane---assumptions that guarantee the correct sign of the
gravitational force and the localization of the energy-momentum tensor in the bulk---no 
viable solutions arise in the context of the theory (\ref{P1eq: action}) when $\Lambda>0$
on our brane. On the other hand, for $\Lambda$ either zero or negative, solutions with
$f>0$ are much easier to arise.\,\footnote{In fact, a static braneworld solution with $M=0$
and $\Lambda=0$ on our brane was presented in \cite{Bogdanos1} where a quadratic coupling
function $f(\Phi)=1-\xi \Phi^2$ between the scalar field and the Ricci scalar was considered.}
For example, for $\Lambda=-|\Lambda|<0$,  Eq. (\ref{P1eq: grav1-v2}) is now written as
\eq$\label{P1eq: con-AdS}
\pa_y^2 f+k \pa_y f - f |\Lambda|\,e^{2ky}\leq 0\,.$
One may readily see that this constraint is much easier to satisfy. For $f(\Phi)$ positive
and decreasing, the second and third terms are already negative definite. For instance,
the choice considered above for $f(\Phi)$, namely $f(y)=A e^{-\sum_{n=1}^N b_n y^n}$,
satisfies the constraint (\ref{P1eq: con-AdS}) over the entire $y$-regime for appropriate choices
of the parameters. A detailed analysis on the emergence of legitimate solutions in the
context of the theory (\ref{P1eq: action}) with an anti-de Sitter or Minkowski background on our
brane is performed in Chapters \ref{Chap: P2} and \ref{Chap: P3}, respectively.


\mysection{Conclusions \label{P1sec: Disc}}

In this Chapter we have focused on the derivation and study of the properties of
black-string solutions that seem to emerge quite naturally in the context of a theory 
with a scalar field non-minimally coupled to gravity. 
To this end, we have retained the Vaidya form of the spacetime line element, which on 
the brane leads to a Schwarzschild black hole while in the bulk produces solutions with 
the minimum number of spacetime singularities. 
We have in addition allowed for an arbitrary mass function $m(r)$ in an effort to accommodate, 
if possible, solutions with a more general profile. 
The brane line element was found to describe a Schwarzschild (anti-)de Sitter spacetime, and 
we chose to study solutions with a positive four-dimensional cosmological constant.
As a result, the brane background assumes the form of a Schwarzschild de
Sitter spacetime. As the expressions of the five-dimensional curvature invariants have
revealed, these solutions may have a dual description from the bulk point of view;
they may describe either black strings, if $M \neq 0$, or braneworld
maximally symmetric solutions, if $M=0$.
The properties of these five-dimensional solutions strongly depend on the 
form of the nonminimal coupling function $f(\Phi)$ between the scalar field and
the five-dimensional scalar curvature. We have considered two simple choices for 
$f(\Phi)$, a linear and a quadratic one in terms of the scalar field.
In the linear case, we found solutions where the theory, close to our brane, mimics an ordinary 
gravitational theory with a minimally coupled scalar field giving rise to an exponentially
decreasing warp factor in the absence of a negative bulk cosmological constant. The solution 
is characterized by the presence of a normal gravity regime around our brane and
an antigravitating regime away from it. In the quadratic case, there is no normal-gravity
regime at all; however, scalar field and energy-momentum tensor components are well defined 
and an exponentially decreasing warp factor emerges again. We also demonstrated that,
in the context of this theory, the emergence of a positive cosmological constant on our
brane is always accompanied by an antigravitating regime in the five-dimensional bulk.

The preceding analysis opens the way for the derivation of solutions
with normal gravity in the case of either a purely Schwarzschild (Minkowski brane) or
Schwarzschild anti-de Sitter spacetime on our brane.  Although less physically motivated, it would still be of interest 
to investigate whether a scalar-tensor theory in the bulk could support a
solution (either a black string or a regular one) with a decaying warp factor
but without the need for a constant distribution of a negative energy density
in the higher-dimensional spacetime. 
The case of an anti-de Sitter brane is studied in \chapref{Chap: P2}, while the
case of a Minkowski brane is examined in great detail in \chapref{Chap: P3}.




\mychapter{Black-string solutions for an anti-de Sitter brane \label{Chap: P2}}
\phantomsection
\setcounter{chapter}{3}

\epigraph{\textit{``We don't see things as they are, we see them as we are.''}}{Ana\"{i}s Nin}

\thispagestyle{empty}

{\lettrine[lines=3, lhang=0.03]{\color{chapter(color)}I}{\hspace*{5.5px}n} this Chapter,
which is based on \cite{KNP2},
we continue the analysis which commenced in \chapref{Chap: P1} and
we focus on the case of a negative cosmological constant on the brane,
which as we will demonstrate removes the condition of the negative sign of the 
non-minimal coupling function in the bulk. With the gravity thus having everywhere
the correct sign, we will look again for analytical solutions describing novel black
strings. We will explicitly solve the coupled system of gravitational and scalar-field
equations to determine both the bulk gravitational background and the scalar field
configuration. Demanding the regularity of the scalar field everywhere in the bulk,
we will reduce the general form of the coupling function to two particular choices.
Both choices lead to analytical black-string solutions that, apart from the infinitely-long
string singularity, are free of any additional bulk singularities associated either with
the scalar field or with the five-dimensional line-element. The solutions that we
found exhibit
also a number of attractive features: the energy-momentum tensor of the theory is
everywhere regular and localized close to our brane leading to a five-dimensional
Minkowski spacetime at large distances away from it. Also, the warping of the fifth
dimension may be supported exclusively by the negative-definite, non-trivial bulk
potential of the scalar field, a result which makes redundant the presence of the
negative bulk cosmological constant. Finally, the five-dimensional theory leads to
a robust four-dimensional effective theory on the brane with the effective gravity
scale being related to the fundamental one by a relation almost identical to the 
one appearing in \cite{RS1, RS2}. It is worth noting that if we set the mass of
the black hole on the brane equal to zero, the black string disappears leaving
behind a regular brane-world model with only a true singularity at the boundary of
the fifth dimension.

}

The outline of the Chapter is as follows: in \secref{P2sec: Th-Frame}, we present our theory, the field
equations and set a number of physical constraints on the scalar field and its
coupling function. In \secref{P2sec: anti1}, we study in detail the case of an exponential coupling
function and determine the complete bulk solution, its physical properties, the 
junction conditions as well as the effective theory on the brane. We repeat the
analysis for another interesting case, that of a double exponential coupling
function, and discuss its properties in \secref{P2sec: anti2}. We finally present our conclusions
in \secref{P2sec: Disc}.

\mysection{The theoretical framework \label{P2sec: Th-Frame}}

As in the previous Chapter the action functional of the bulk spacetime is of the form
\beq
\label{P2eq: action}
S_B=\int d^4x\int dy \,\sqrt{-g^{(5)}}\left[\frac{f(\Phi)}{2\kappa_5^2}R
-\Lambda_5-\frac{1}{2}\,\pa_L\Phi\,\pa^L\Phi-V_B(\Phi)\right],
\eeq
while the four-dimensional action of our world is given by
\beq
\label{P2eq: action_br}
S_{br}=\int d^4x\sqrt{-g^{(br)}}(\lagr_{br}-\sigma)=
-\int d^4x\int dy\sqrt{-g^{(br)}}\,[V_b(\Phi)+\sigma]\,\delta(y)\,.
\eeq
The field equations corresponding to the complete action $S=S_B+S_{br}$ have been already
evaluated in the preceding Chapter. Note that Eqs. \eqref{P1eq: Einst-eqs}-\eqref{P1eq: mass-sol}
continue to hold here as well.
Hence, for the generalized Vaidya line-element
\eq$\label{P2eq: metric}
ds^2=e^{2A(y)}\left\{-\left[1-\frac{2m(r)}{r}\right]dv^2+2dvdr+r^2(d\theta^2+
\sin^2\theta d\varphi^2)\right\}+dy^2\,,$
one should obtain the following independent field equations:
\gat$
\label{P2eq: eq-mass}
r\,\pa_r^2m-2\pa_rm=0\,,\\[2mm]
\label{P2eq: grav-1}
f\left(3A''+e^{-2A}\frac{\pa_r^2m}{r}\right)=\pa_\Phi f \left(A'\Phi'-\Phi''\right)
-(1+\pa_\Phi^2 f)\Phi'^2\,,\\[2mm]
\label{P2eq: grav-2}
f\left(6A'^2+3A''-\frac{2e^{-2A}\pa_rm}{r^2}\right)=A'\Phi'\,\pa_\Phi f+
\lagr_\Phi -\Box f-\Lambda_5\,.$
Note that, for notational simplicity, we have absorbed the gravitational constant
$\kappa_5^2$ in the expression of the general coupling function $f(\Phi)$.
Equation (\ref{P2eq: grav-2}) will serve to determine the scalar potential
in the bulk $V_B(\Phi)$. It is Eq. (\ref{P2eq: grav-1}) that will provide the solution
for the scalar field $\Phi$ once the warp function $A(y)$, the mass function
$m(r)$ and the non-minimal coupling function $f(\Phi)$ are determined. For the
warp factor, we will make the assumption that this is given by the well-known
form $A(y)=-k |y|$ \cite{RS1, RS2}, with $k$ a positive constant, as this ensures
the localization of gravity near the brane. The form of the mass function $m(r)$
readily follows by direct integration of Eq. (\ref{P2eq: eq-mass}) that leads to the
expression
\beq
m(r)=M+ \Lambda r^3/6\,, \label{P2eq: mass-sol}
\eeq
where $M$ and $\Lambda$ are arbitrary integration constants (the numerical
coefficient 1/6 has again been introduced for convenience). Substituting the
above form into the line-element (\ref{P2eq: metric}) and setting $y=0$, we may
easily see that the projected-on-the-brane background is given by the
expression
\eq$\label{P2eq: metric-brane}
ds^2_4=-\left(1-\frac{2M}{r}-\frac{\Lambda r^2}{3}\right)dv^2+2dv dr+
r^2(d\theta^2+\sin^2\theta\ d\varphi^2)\,.$
In Appendix \ref{P1app: Inv-Vaidya}, we explicitly demonstrated that the above Vaidya form of the 
four-di\-men\-sion\-al line-element may be transformed to the usual 
Schwarzschild (anti-)de Sitter solution by an appropriate coordinate
transformation. Therefore, the arbitrary parameter $M$ is the mass of
the black-hole that the four-dimensional observer sees and $\Lambda$
the cosmological constant on the brane. 

The case of a positive cosmological constant on the brane (i.e. $\Lambda>0$)
was studied in the previous Chapter. Therefore, in the context of the present
analysis, we will focus on the case of a negative four-dimensional cosmological
constant ($\Lambda < 0$). Employing the form of the mass function (\ref{P2eq: mass-sol})
and the exponentially decreasing warp factor\,\footnote{We assume a
${\bf Z}_2$-symmetry in the bulk under the change $y \rightarrow -y$ therefore,
henceforth, we focus on the positive $y$-regime.} $e^{2A(y)}=e^{-2ky}$,
Eq. (\ref{P2eq: grav-1}) takes the form
\eq$\label{P2eq: anti-grav-1}
(\Phi')^2=-\pa_y^2 f-k\pa_yf - \Lambda e^{2ky}f\,.$
In the above, we have also used the relations
\eq$\label{P2eq: dif-f}
\pa_yf=\Phi' \,\pa_\Phi f, \hspace{1.5em}\pa_y^2f=
\Phi'^2\,\pa_\Phi^2 f+\Phi''\,\pa_\Phi f\,.$
The l.h.s of Eq. (\ref{P2eq: anti-grav-1}) is positive, therefore the same should hold
for the r.h.s, too. Note that, for $\Lambda>0$, Eq. (\ref{P2eq: anti-grav-1}) demands
that, at least at $y \rightarrow \infty$, the coupling function $f(\Phi)$ should
be negative for the scalar field to have a real first-derivative there. Indeed,
in \chapref{Chap: P1}, we presented two analytic solutions of this theory where $f<0$
either far-away from our brane or in the entire bulk regime. In contrast, in
the present case, where $\Lambda<0$, no such behaviour is necessary, thus in order
to have a normal gravity over the entire five-dimensional spacetime, we will
assume that $f(\Phi)$ is positive everywhere. 

In order to have a physically acceptable behaviour, a few more properties should
be assigned to the functions $\Phi=\Phi(y)$ and $f=f[\Phi(y)]$. Both functions
should, of course, be real and finite in their whole domain and of class $C^{\infty}$.
At $y\ra +\infty$, both functions should satisfy the following relations, otherwise
the finiteness of the theory at infinity cannot be ascertained,
\eq$\label{P2eq: con.1}
\lim_{y\ra+\infty}\frac{d^n[f(y)]}{dy^n}=0,\hspace{1.5em}\forall n\geq 1,$
\eq$\label{P2eq: con.2}
\lim_{y\ra+\infty}\frac{d^n[\Phi(y)]}{dy^n}=0,\hspace{1.5em}\forall n\geq 1.$
These constraints guarantee that all components of the energy-momentum tensor
$T^{(\Phi)M}{}_N$ will be real and finite everywhere, and, in addition, localized
close to our brane. Then, demanding also the finiteness and the vanishing of the
r.h.s. of Eq. (\ref{P2eq: anti-grav-1}) due to the constraint (\ref{P2eq: con.2}), we conclude
that the coupling function $f(y)$ should, at infinity, decrease
faster\,\footnote{Note that allowing the coupling function to vary exactly as
$e^{-2ky}$, i.e. $f(y)=f_0\,e^{-2ky}$, would lead to a finite, constant
value of $\Phi'^2$ at infinity, namely $\Phi'^2_\infty=-f_0\,\Lambda>0$. This would
amount to having a diverging field at the boundary of spacetime but nevertheless
finite, constant values for the components of the energy-momentum tensor. We will
come back to this point later.}
than $e^{-2ky}$, i.e. $f(y)$ should be of the form
\eq$\label{P2eq: anti-f}
f(y)=f_0\ e^{g(y)},\hspace{1.5em} \left\{\begin{array}{c}
f_0>0\\ \\
g(y\ra+\infty)<-2ky 
\end{array}\right\}.$
Consequently, upon integrating Eq. (\ref{P2eq: anti-grav-1}), the following expression
is obtained for the scalar field:
\eq$\label{P2eq: anti-phi}
\Phi(y)=\pm \sqrt{f_0}\int dy\ e^{\frac{g(y)}{2}}\sqrt{\tlam^2 e^{2ky}-g''-g'^2-kg'},$
where, for convenience, we have also set $\Lambda=-\tlam^2$. In order to proceed
further, we need to determine the exact form of the function $g(y)$. As we are
interested in deriving analytical solutions for both functions $f(y)$ and $\Phi(y)$,
the function $g(y)$ should have a specific form in order to result to a solvable
integral on the r.h.s. of equation \eqref{P2eq: anti-phi}. Therefore, we will make the
following two choices:
\gat$\label{P2eq: anti-g1}
g(y)=-\lam ky,\hspace{1.5em}\lam\in(2,+\infty)\,,\\ \nonum\\
\label{P2eq: anti-g2}
g(y)=-\mu^2 e^{\lam y},\hspace{1.5em}\left\{\begin{array}{c}\lam\in(0,+\infty)\\
\\ \mu\in\mathbb{R}\setminus \{0\}\end{array}\right\}.
$
The aforementioned expressions for $g(y)$ ensure that both $f(y)$ and $\Phi(y)$
have the desired properties outlined above and, in addition, lead to analytical
solutions. In the following sections, these two different cases will be studied
separately.

\newpage


\vspace*{-3.5em}

\mysection{The simple exponential case \label{P2sec: anti1}}

We will start with the simple exponential case (\ref{P2eq: anti-g1}), and derive first
the form of the scalar field and its potential in the bulk. We will then study their
main characteristics in terms of the free parameters of the model, and finally
address the effect of the junction conditions and the form of the effective 
theory on the brane. 

\mysubsection{The bulk solution}

\par In this case, we have $f(y)=f_0\ e^{-\lam k y}$, with $f_0>0$ and $\lam>2$.
Then, from Eq. (\ref{P2eq: anti-grav-1}), we obtain
\beq
\Phi'^{\,2}(y)=f(y)\,(\tlam^2 e^{2ky}-\lam^2k^2+\lam k^2) \geq 0\,.
\label{P2eq: anti1-Phi/f}
\eeq
For a non-zero and positive $f(y)$, the above inequality demands that the combination
inside the brackets should be positive. As this is an increasing function of $y$, it
suffices to demand that this holds at the location of the brane, at $y=0$. 
Then, we obtain the following constraint on the parameters of the theory:
\beq
\label{P2eq: anti1-con1}
\frac{\tlam^2}{\lam(\lam-1)k^2}> 1\,.
\eeq
The function $\Phi'^{\,2}(y)$ could, in principle, be zero at the point where
$\Phi(y)$ has an extremum. However, from Eq. \eqref{P2eq: anti1-Phi/f}, we may easily
see that this may happen only at $y_0=\frac{1}{2k}\ln\left(\frac{\lam(\lam-1)k^2}
{\tlam^2}\right)$, which, upon using Eq. (\ref{P2eq: anti1-con1}), turns out to be
negative. Therefore, the scalar field does not have any extremum in the whole
domain $0 \leq y < \infty$, which in turn means that $\Phi(y)$ is an one-to-one
function in the same region. The $\mathbf{Z}_2$ symmetry of the extra dimension
ensures that this result holds in the region $y<0$ as well. We note this
property for later use. 
Equation \eqref{P2eq: anti-grav-1} can be re-written as
\bea(\Phi')^2 &=&f_0\lam(\lam-1)k^2\left[\frac{\tlam^2}{\lam(\lam-1)k^2}e^{2ky}
-1\right]e^{-\lam ky} \nonum\\[3mm]
&=& f_0\lam(\lam-1)k^2\left[\frac{\tlam^2}{\lam(\lam-1)k^2}\right]^{\lam/2}(w-1)\,w^{-\lam/2}\,,
\label{P2eq: phi'-w}
\eea
where we have introduced the new variable $w$ via the definition
\eq$\label{P2eq: anti1-w}
w(y)\equiv \frac{\tlam^2 e^{2 k y}}{\lam(\lam-1)k^2}\,.$
Due to the constraint \eqref{P2eq: anti1-con1}, it is obvious that $w(y)$ is greater than
unity for all values of the extra coordinate $y$. Then, applying the chain rule to
the l.h.s. of Eq. (\ref{P2eq: phi'-w}) and integrating, we obtain for the scalar field
the integral expression 
\gat$
\Phi(w)=\pm\frac{\sqrt{f_0\lam(\lam-1)}}{2}\left[\frac{\tlam^2}{\lam(\lam-1)k^2}
\right]^{\lam/4}\int dw\ (w-1)^{\frac{1}{2}}\,w^{-\frac{\lam}{4}-1}\,.$

In order to evaluate the above integral, we perform a second change of variable,
namely we set $w=1/(1-z)$. Then,
\gat$\int dw\ (w-1)^{\frac{1}{2}}w^{-\frac{\lam}{4}-1}=
\int dz\ z^{\frac{1}{2}}(1-z)^{\frac{\lam}{4}-\frac{3}{2}}=\int_0^z dt\,
t^{\frac{1}{2}}(1-t)^{\frac{\lam}{4}-\frac{3}{2}}+C_1\,,$
where an arbitrary constant $C_1$ has been introduced in order to set the lower
boundary value of the integral equal to zero. Finally, by employing the rescaled variable 
$t'=t/z$, the above integral takes its final form
\gat$
\label{P2eq: integral-zt}
z^{\frac{3}{2}}\int_0^1 dt'\ t'^{\frac{1}{2}}(1-zt')^{\frac{\lam}{4}-\frac{3}{2}}+C_1=
\frac{2}{3}\left(\frac{w-1}{w}\right)^{3/2}\,_2F_1\left(\frac{3}{2}-
\frac{\lam}{4},\frac{3}{2};\frac{5}{2};\frac{w-1}{w}\right)+C_1\,,$
where we used the integral representation of the hypergeometric function \cite{Abramowitz}
\beq
\,_2F_1\left(a,b;c;z\right)=\frac{\Gamma(c)}{\Gamma(b)\Gamma(c-b)}\int_0^1 dt'\
t'^{b-1}(1-t')^{c-b-1}(1-zt')^{-a}\,.
\eeq
where $Re(c)>Re(b)>0$. We now observe that $\Phi$ appears in the field equations \eqref{P2eq: grav-1}
and \eqref{P2eq: grav-2} only through the coupling function $f(\Phi)$ and the bulk potential
$V_B(\Phi)$. Therefore, any shift in the value of the scalar field by an arbitrary
constant would result into a change in the value of $f$ by a constant amount that could
nevertheless be re-absorbed in the redefinition of the value of the arbitrary coefficient
$f_0$; the value of the bulk potential $V_B$ would also change by a constant amount but this
could again be re-absorbed in the value of the arbitrary bulk cosmological constant
$\Lambda_5$. Due to this translation symmetry with respect to the value of the scalar
field $\Phi(y)$, we may set the arbitrary constant $C_1$ in Eq. (\ref{P2eq: integral-zt})
equal to zero. This brings the solution for the scalar field into its final form
{\fontsize{11}{11}\eq$\label{P2eq: anti1-phi}
\Phi_{\pm}(y)=\pm\frac{\sqrt{f_0\lam(\lam-1)}}{3}\left[\frac{\tlam^2}{\lam(\lam-1)k^2}\right]^{\lam/4}\left[\frac{w(y)-1}{w(y)}\right]^{3/2}
\,_2F_1  \left(\frac{3}{2}-\frac{\lam}{4},\frac{3}{2};\frac{5}{2};\frac{w(y)-1}{w(y)}\right).$ }
\hspace{-0.5em}Although the function $w(y)$ is greater than unity for all values of the extra dimension $y$,
the argument $z=\frac{w-1}{w}$ of the hypergeometric function in the previous relation is
always positive and smaller than unity. Hence, we may use the well-known expansion for the hypergeometric function in power series
\beq
\,_2F_1(a,b;c;z)=\,_2F_1(b,a;c;z)=\sum_{n=0}^{\infty}
\frac{a^{(n)}b^{(n)}}{c^{(n)}}\frac{z^n}{n!}\,,
\label{P2eq: Expansion-F}
\eeq
where $|z|<1$, and the quantities of the form $q^{(n)}$ denote (rising) Pochhammer symbols,
namely
\beq
q^{(n)}=\frac{\Gamma(q+n)}{\Gamma(q)}=
\left\{\begin{array}{cc}q(q+1)\cdots(q+n-1)\,, & n>0\\ \\ 1\,, & n=0\end{array}\right\}.
\eeq
Thus, we find
\gat$
\label{P2eq: final-F}
\,_2F_1  \left(\frac{3}{2}-\frac{\lam}{4},\frac{3}{2};\frac{5}{2};\frac{w-1}{w}\right)=
\sum_{n=0}^\infty \frac{\Gamma\left(\frac{3}{2}-\frac{\lam}{4}+n\right)}{\Gamma\left(\frac{3}{2}-\frac{\lam}{4}\right)}\,
\frac{3}{(2n+3)n!}\left(\frac{w-1}{w}\right)^n,$
where we have also used the property $\Gamma(1+z)=z\Gamma(z)$.
There are two interesting categories of values for the parameter $\lam$ which lead to simple
and elegant expressions for the hypergeometric function and subsequently for the scalar field.
These are $\lam=2(1+2q)$ and $\lam=4q$, where $q$ is any positive integer.
Let us examine each case separately. 

\begin{itemize}
\item If $\lam=2(1+2q)$ with $q \in {\mathbb{Z}}^{>}$, then, from Eq. (\ref{P2eq: final-F}),
we have:
{\fontsize{10}{10}\bal$
\,_2F_1  \left(\frac{3}{2}-\frac{\lam}{4},\frac{3}{2};\frac{5}{2};\frac{w-1}{w}\right)
&=\sum_{n=0}^{\infty}(1-q)^{(n)}\frac{3}{(2n+3)n!}\left(\frac{w-1}{w}\right)^n\nonum\\[2mm]
&=\left\{\begin{array}{ccr}
1\,, & & q=1\\[4mm]
1+\sum_{n=1}^{q-1}\frac{3(-q+1)(-q+2)\cdots(-q+n)}{(2n+3)n!}\left(\frac{w-1}{w}\right)^n, & & q>1
\end{array}\right\} .
\label{P2eq: F-series2}$}
\hspace{-0.5em}In the second line of the above expression, the upper limit of the sum has been changed
from $\infty$ to $q-1$ since, for $q$ and $n$ positive, the sum will be trivial for any
value of $n$ equal or higher than $q$ due to the factor $(-q+n)$. As
indicative cases, we present below the form of the scalar field
for\,\footnote{For completeness, we present here also the solution for the limiting case
with $q=0$ (i.e. for $\lam=2$); this has the form
$$\Phi_{\pm}(y)=\pm\sqrt{f_0}\,\sqrt{\frac{\tlam^2}{k^2}}
\left[\,{\rm arctanh}\left(\sqrt{\frac{w-1}{w}}\right)-\sqrt{\frac{w-1}{w}}\,\right].$$
Although the field diverges at infinity---see footnote 2---the components of the
energy-momentum tensor exhibit a regular behaviour as we will comment later.}
$q=1$ (i.e. $\lambda=6$) 
\beq
\Phi_{\pm}(y)=\pm\frac{\sqrt{f_0}}{90}\left(\frac{\tlam^2}{k^2}\right)^{3/2}\left(\frac{w-1}{w}\right)^{3/2},
\eeq
and $q=2$ (i.e. $\lambda=10$)
\beq
\Phi_{\pm}(y)=\pm\frac{\sqrt{f_0}}{3\times 90^2}\left(\frac{\tlam^2}{k^2}\right)^{5/2}
\left(\frac{w-1}{w}\right)^{3/2}\left[1-\frac{3}{5}\left(\frac{w-1}{w}\right)\right].
\eeq
The above expressions follow easily by using Eqs. (\ref{P2eq: anti1-phi}) and (\ref{P2eq: F-series2})
and substituting the aforementioned values of the parameter $q$.

\item If $\lam=4q$ with $q \in {\mathbb{Z}}^{>}$, we can always express the hypergeometric function 
in Eq. (\ref{P2eq: anti1-phi}) in terms of elementary functions, namely 
$\arcsin\left(\sqrt{\frac{w-1}{w}}\right)$, square roots and powers of the argument
$\frac{w-1}{w}$. The process that one follows to obtain this expression is
presented in detail in Appendix \ref{P2app: hyper-analysis}. Thus, for $\lam=4$ (i.e. $q=1$),
using Eq. (\ref{P2app-eq: B.3}), we may straightforwardly write
\beq
\Phi_{\pm}(y)=\pm\frac{\sqrt{f_0}}{4\sqrt{3}}\ \frac{\tlam^2}{k^2}
\left(\frac{w-1}{w}\right)^{1/2}\left[\sqrt{\frac{w}{w-1}}
\arcsin\left(\sqrt{\frac{w-1}{w}}\right)-\sqrt{\frac{1}{w}}\ \right].
\label{P2eq: Phi-lam4}
\eeq
For larger values of $\lambda$ (i.e. for $q=1+\ell$, with $\ell \in {\mathbb{Z}}^{>}$),
we  should use instead Eq. (\ref{P2app-eq: hyper-lam-4q-final}) together with the
constraint (\ref{P2app-eq: alpha-beta}). The latter, as outlined in Appendix \ref{P2app: hyper-analysis},
reduces to a set of linear equations that determine the unknown coefficients
$\alpha, \beta_1, \cdots, \beta_\ell$. For example, for $\ell=1$, the set of
equations that follow from Eq. (\ref{P2app-eq: alpha-beta}) is
\beq
\{2\alpha-\beta_1=0\,, \qquad 2\alpha+3\beta_1=3\}\,,
\eeq
leading to the values $\alpha=3/8$ and $\beta=3/4$. Then, after substituting these in
Eq. (\ref{P2app-eq: hyper-lam-4q-final}), the solution for $q=2$, or equivalently for
$\lam=8$, follows from Eq. (\ref{P2eq: anti1-phi}) and has the form
{\fontsize{10}{10}\gat$\Phi_{\pm}(y)=\pm\frac{\sqrt{f_0}}{896\sqrt{14}}\ \frac{\tlam^4}{k^4}\left(\frac{w-1}{w}\right)^{1/2}
\left[\sqrt{\frac{w}{w-1}}\arcsin\left(\sqrt{\frac{w-1}{w}}\right)+\sqrt{\frac{1}{w}}\left(1-\frac{2}{w}\right)\right].$}
\hspace{-0.5em}Solutions for larger values of $q$, and thus of $\lam$, may be derived in the
same way in terms again of analytic, elementary functions.
\end{itemize}
In all the above, particular expressions for the scalar field $\Phi_{\pm}$, that follow for
specific values of the parameter $\lam$, the dependence on the extra coordinate $y$ is easily
made explicit by employing Eq. (\ref{P2eq: anti1-w}). Also, for all other values of 
$\lam \in {\mathbb{R}}^{>2}$, which do not fall in the aforementioned categories, the
scalar field may still be expressed in terms of the hypergeometric function through
Eqs. (\ref{P2eq: anti1-phi}) and (\ref{P2eq: final-F}).

\begin{figure}[t]
    \centering
    \begin{subfigure}[b]{0.46\textwidth}
        \includegraphics[width=\textwidth]{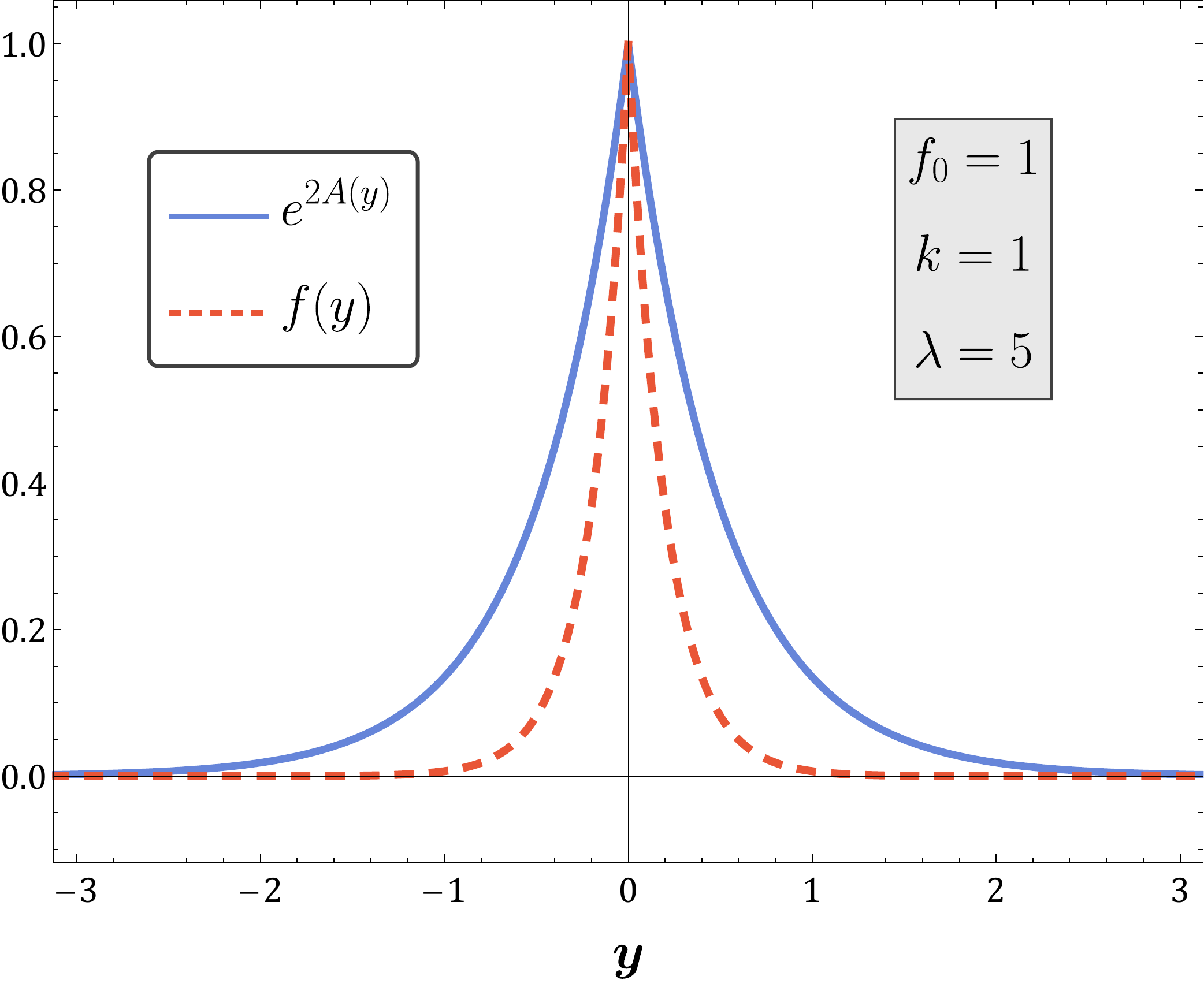}
        \caption{\hspace*{-1.1em}}
        \label{P2subf: fig1a}
    \end{subfigure}
    \quad
    \begin{subfigure}[b]{0.485\textwidth}
        \includegraphics[width=\textwidth]{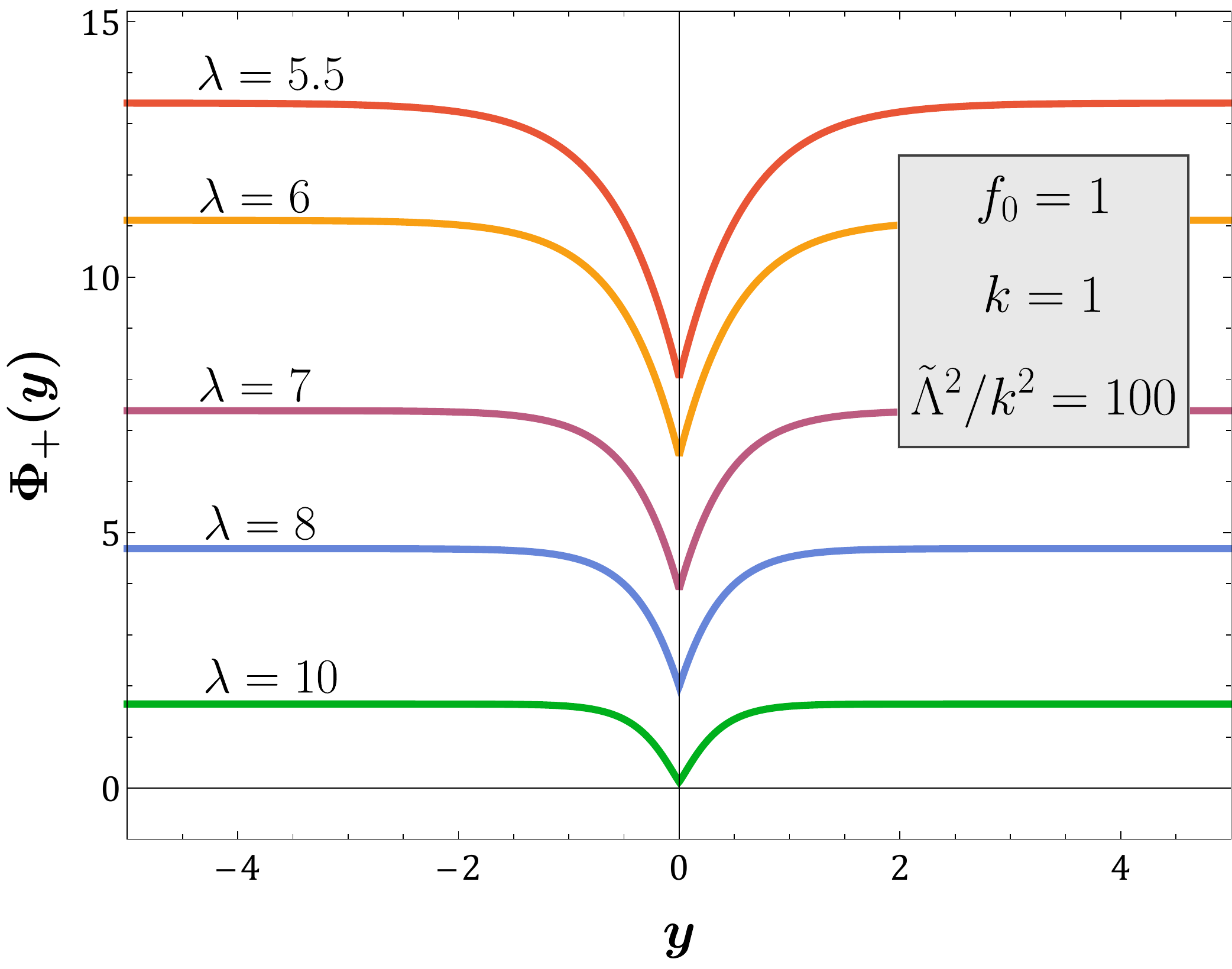}
        \caption{\hspace*{-2.4em}}
        \label{P2subf: fig1b}
    \end{subfigure}
    \caption{(a) The warp factor $e^{2A(y)}=e^{-2k|y|}$ and the coupling function
    $f(y)=f_0\,e^{-\lam k |y|}$ in terms
    of the coordinate $y$, for $f_0=1$, $k=1$, $\lam=5$. (b) The scalar field $\Phi_{+}(y)$ also in
    terms of the coordinate $y$, for $f_0=1$, $k=1$, $\tlam^2/k^2=100$ and $\lam=5.5,6,7,8,10$ (from top to bottom).}
   \label{P2fig: anti1_warp_f_Phi}
\end{figure}

\par Let us now investigate the physical characteristics of the solutions we have derived.
In \myref{P2fig: anti1_warp_f_Phi}{P2subf: fig1a} (next page), we depict the form of the warp factor $e^{-2k|y|}$ and the coupling function
$f(y)=f_0\,e^{-\lam k |y|}$ in terms of the coordinate $y$ along the fifth dimension,
for $f_0=1$, $k=1$ and $\lam=5$. The warp factor exhibits the anticipated localization
close to the brane while the non-minimal coupling function mimics this behaviour by
decreasing exponentially fast away from the brane and reducing to zero at the boundary of
spacetime. In fact, the larger the parameter $\lam$, the faster the decrease rate
of $f$ is; thus by increasing $\lam$, the non-minimal coupling of the scalar field 
to gravity is effectively ``localized'' closer to the brane. 

\myref{P2fig: anti1_warp_f_Phi}{P2subf: fig1b} depicts
the scalar field $\Phi_+(y)$ for different values of the parameter $\lam$ and for $f_0=1$,
$k=1$, $\tlam^2/k^2=100$. It is straightforward to deduce from \myref{P2fig: anti1_warp_f_Phi}{P2subf: fig1b} that the scalar
field $\Phi_+(y)$ exhibits a reverse behaviour, compared to $f(y)$, by increasing away
from the brane and adopting a constant, non-vanishing value at the boundary of spacetime.
Note that, as $\lam$ increases, the scalar field reaches this constant asymptotic value
faster; that is, a more ``localized'' coupling function keeps also the non-trivial profile
of the scalar field closer to the brane. 
Overall, for $\lam > 2$, the scalar field presents a well-defined profile over the entire
extra dimension in accordance to the desired properties set in the previous section.
In \myref{P2fig: anti1_warp_f_Phi}{P2subf: fig1b}, we chose to plot $\Phi_+(y)$, i.e. we chose the positive sign in
Eq. (\ref{P2eq: anti1-phi}) for the expression of the scalar field. A second class of solutions
exists for $\Phi=\Phi_-$, with the only difference being \myref{P2fig: anti1_warp_f_Phi}{P2subf: fig1b} becoming its mirror
image with respect to the horizontal axis. The sign of the scalar field, however, does
not affect either the potential $V_B(y)$ or the components of the energy-momentum tensor,
as we will soon see. Finally, let us emphasize the fact that, as \myref{P2fig: anti1_warp_f_Phi}{P2subf: fig1b} reveals, the 
qualitative behaviour of the scalar field in terms of the parameter $\lam$ remains
unchanged. This holds despite the fact that the value of $\lam$ does affect the exact,
analytic expression of the scalar field, as we have shown in detail above; we may thus
conclude that solutions emerging for non-minimal coupling functions of a simple exponential
form, differing only in the value of the parameter $\lam$, i.e. in the decrease rate of
$f$ with $y$, lead to a class of black-string solutions with the same qualitative characteristics.

\par
\begin{figure}[t!]
    \centering
    \begin{subfigure}[b]{0.45\textwidth}
        \includegraphics[width=\textwidth]{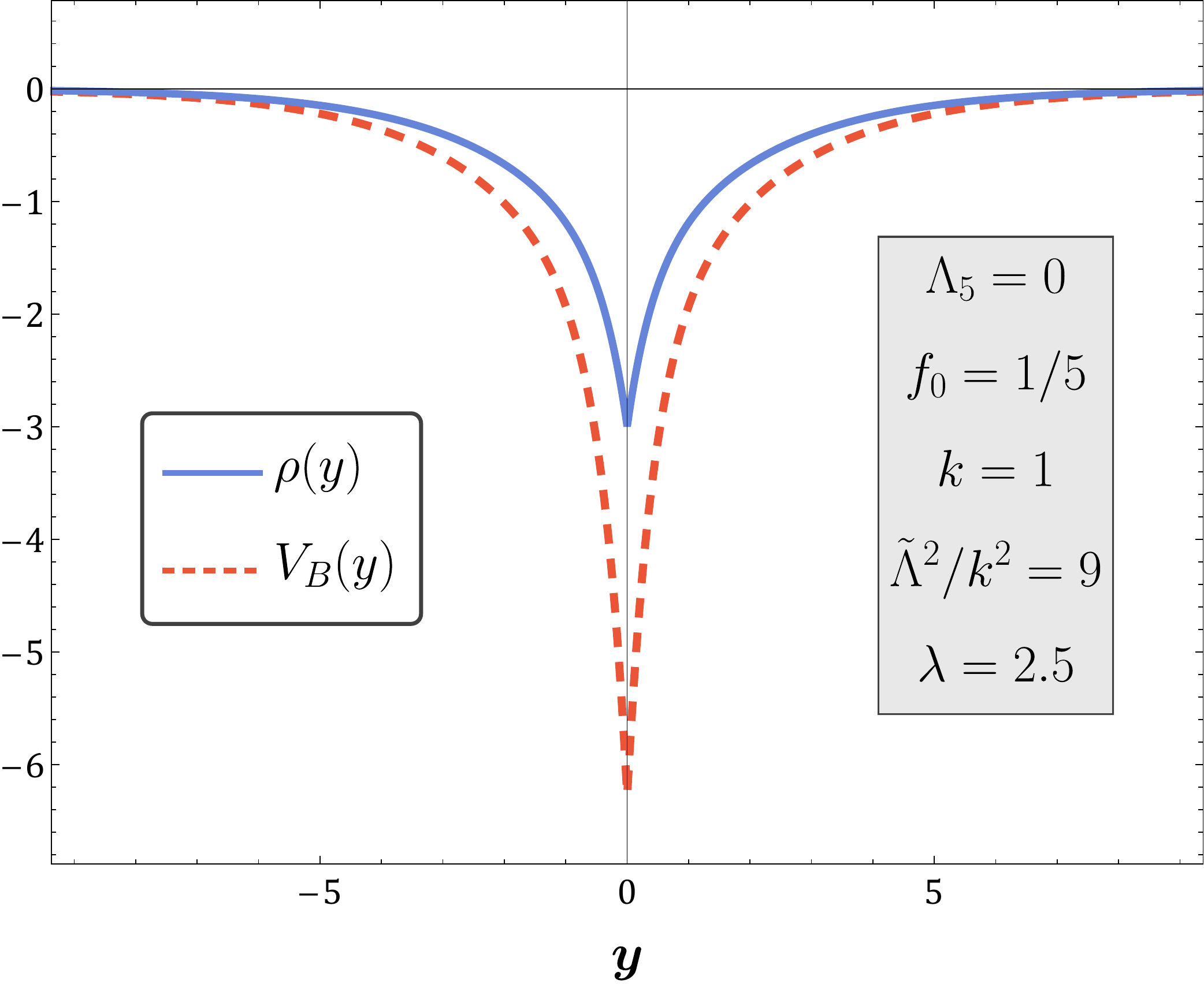}
        \caption{\hspace*{-1.2em}}
        \label{P2subf: fig2a}
    \end{subfigure}
    ~ 
      \hspace{2em}
    \begin{subfigure}[b]{0.44\textwidth}
        \includegraphics[width=\textwidth]{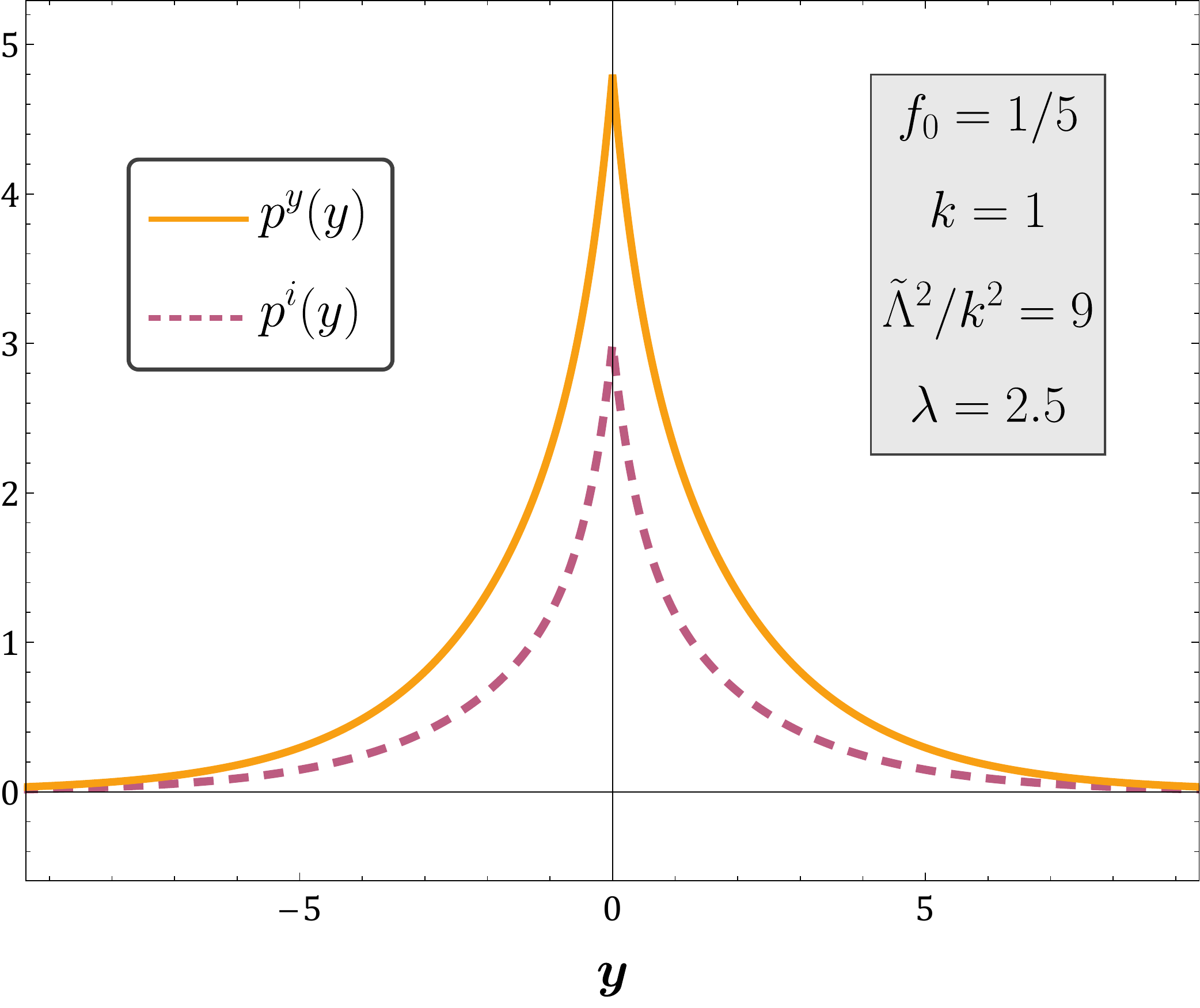}
        \caption{\hspace*{-0.7em}}
        \label{P2subf: fig2b}
    \end{subfigure}
    ~ 
    \vspace*{-1em}
    \caption{(a) The scalar potential $V_B$ and energy density $\rho$ of the system, and
(b) the pressure components $p^y$ and $p^i$ in terms of the coordinate $y$.}
   \label{P2fig: anti1-V-TMN}
\end{figure}

 The potential of the field $V_B$ in the bulk can be determined from Eq. \eqref{P2eq: grav-2}.
Substituting the functions $m(r)$ and $A(y)$, we obtain
\beq
V_B(y)=-\Lambda_5-\frac{1}{2}\,\Phi'^2 +3k \partial_y f -\partial_y^2 f -
f\left(6k^2 -\Lambda e^{2ky}\right),
\label{P2eq: V_B-1}
\eeq
where we have also used the relations \eqref{P2eq: dif-f}. Note that the bulk potential is indeed
insensitive to the sign of $\Phi_{\pm}$ which enters the above expression through $\Phi'^2$.
If we also employ Eq. \eqref{P2eq: anti-grav-1} to substitute $\Phi'^2$, and use the exponential
form for $f(y)$, the following expression readily follows for the potential $V_B$ in terms
of the extra dimension $y$:
\eq$\label{P2eq: anti1-pot}
V_B(y)=-\Lambda_5-\frac{k^2f_0}{2}\ e^{-\lam ky}\left(12+7\lam+\lam^2+\frac{3\tlam^2}{k^2}e^{2ky}\right).$
We observe that the combination $V_B(y)+\Lambda_5$, which appears in the action (\ref{P2eq: action})
as well as in the components of the total energy-momentum tensor as we will shortly see, is
always negative definite. This combination, even for $\Lambda_5=0$, may therefore provide by
itself the negative distribution of energy in the bulk that is
necessary for the support of the AdS spacetime and the localization of gravity. A similar result
was derived in \chapref{Chap: P1} where the case of a positive cosmological constant $\Lambda$ on
the brane was considered. There, the positive $\Lambda$ added a positive contribution to the
value of $V_B$, that was thus decreased in absolute value, while here the negative $\Lambda$
gives an extra boost to the negative value of $V_B$. The profile of the bulk potential of the
scalar field is depicted in \myref{P2fig: anti1-V-TMN}{P2subf: fig2a} for $f_0=1/5$, $k=1$, $\tlam^2/k^2=9$ and $\lam=2.5$. 
The potential is everywhere finite and remains localized close to the brane. For all values of
$\lam>2$, it goes to zero with an exponential decay rate that increases with $\lam$. The
vanishing of $V_B$ at the boundary of spacetime, together with the similar behaviour of
the coupling function $f$ in the same regime and the constant value that the scalar field
assumes there, points to the conclusion that the non-minimally-coupled scalar field, after
serving its purpose of localizing gravity close to the brane, completely disappears leaving
behind a five-dimensional Minkowski spacetime. Only, for $\lam=2$, an asymptotic bulk
cosmological constant equal to $-3f_0 \tilde \Lambda^2/2$ remains, thus leading to an
asymptotically AdS spacetime, but only by paying the price of a diverging scalar field at
infinity.

\par Finally, we may compute the components of the energy-momentum tensor of the theory 
in the bulk. These follow by employing Eqs. \eqref{P1eq: grav-eqs} and \eqref{P1eq: TMN-mix-com}. Using also
the relations $\rho=-T^0{}_0$, $p^i=T^{i}{}_i$ and $p^y=T^y{}_y$, we find the results
\gat$\label{P2eq: anti1-rho-0}
 \rho=-p^i=\frac{1}{2}\,\Phi'^2 + V_B+ \Lambda_5-3k \partial_y f +\partial_y^2 f\,,\\[2mm]
\label{P2eq: anti1-p-0y}
p^y=\frac{1}{2}\,\Phi'^2 - V_B- \Lambda_5 + 4k \partial_y f\,.$
Substituting $V_B$ from Eq. (\ref{P2eq: V_B-1}) and the form of the coupling function, we
finally obtain the following explicit expressions 
\gat$\label{P2eq: anti1-rho}
 \rho=-p^i=-f_0\,e^{-\lam ky}(6k^2+\tlam^2e^{2ky})\,,\\[2mm]
\label{P2eq: anti1-py}
p^y=f_0\,e^{-\lam ky}(6k^2+2\tlam^2e^{2ky})\,.$
We present the behaviour of the energy density $\rho$ in \myref{P2fig: anti1-V-TMN}{P2subf: fig2a} and of the pressure components
$p^i$ and $p^y$ in \myref{P2fig: anti1-V-TMN}{P2subf: fig2b} with respect to the extra dimension $y$. Both figures have the same
values for the parameters of the model to allow for an easy comparison. The energy density is
negative-definite throughout the bulk, due to the negative value of the scalar potential discussed
above, in order to support the pseudo-AdS spacetime and the exponentially falling warp factor.
The spacelike pressure components $p^i$ satisfy the relation $p^i=-\rho$, a remnant of the
equation of state of a true cosmological constant. The fifth pressure component $p^y$ is also
positive but larger than $p^i$ due to the factor of 2 in front of $\tilde \Lambda^2$ 
in Eq. (\ref{P2eq: anti1-py}). All components present a well-defined profile throughout the bulk and 
vanish exponentially fast away from the brane for all $\lambda>2$. The aforementioned behaviour
remains qualitatively the same for all values of the parameters of the model.


\vspace*{-1.5em}

\mysubsection{Junction conditions and effective theory}

We will now turn our attention to the junction conditions that must be incorporated
in the model due to the presence of the brane at $y=0$. We will assume that the
energy content of the brane is given by the combination $\sigma + V_b(\Phi)$, 
where $\sigma$ is the constant self-energy of the brane and $V_b(\Phi)$ an interaction
term of the bulk scalar field with the brane. This energy content is assumed to arise
only at a single point along the extra dimension, i.e. along our brane at $y=0$, and
thus it creates a discontinuity
in the second derivatives of the warp factor, the coupling function and the scalar
field at the location of the brane. Following the same procedure as in \chapref{Chap: P1}, we write
$A''=\hat A'' + [A']\,\delta(y)$,  $f''=\hat f''+[f']\ \del(y)$ and $\Phi''=\hat \Phi''
+ [\Phi']\,\delta(y)$. Then, going back to the field equations
(\ref{P1eq: phi-eq}) and (\ref{P2eq: grav-1}), we reintroduce the delta-function terms, that we
omitted while working in the bulk. If we then match the coefficients of the delta-function
terms appearing in Eqs. (\ref{P1eq: phi-eq}) and (\ref{P2eq: grav-1}), we obtain the following two
conditions
\gat$\label{P2eq: jun_con1}
[\Phi'] = 4 [A']\,\partial_\Phi f + \partial_\Phi V_b\,,\\[3mm]
\label{P2eq: jun_con2}
3 f(y) [A']= -[f']- (\sigma + V_b)\,,$
respectively, where all quantities are evaluated at $y=0^+$. Using the expressions for
the warp function $A(y)=-k |y|$ and the coupling function $f(y)=f_0 e^{-\lam k |y|}$
in Eq. (\ref{P2eq: jun_con2}), and making use of the assumed $\mathbf{Z}_2$ symmetry in the
bulk, we readily obtain the constraint 
\gat$\label{P2eq: anti1-jun1}
\sig+V_b(\Phi)\Big|_{y=0}=2kf_0(\lam+3)\,.$
We note that the combination of parameters on the r.h.s. of the above equation is
positive-definite, therefore, the total energy density of our brane is always positive.
The above constraint may be
used to determine the value of the warp-factor parameter $k$ in terms of the fundamental
quantities of the brane tension $\sigma$ and the scalar-field parameters ($f_0, \lam, V_b$).
We thus observe that, once we decide the form of the non-minimal coupling function, the
warping gets stronger the larger the interaction term of the scalar field with the brane is.

In order to evaluate the first constraint (\ref{P2eq: jun_con1}), we write that:
$\partial_\Phi f= \partial_y f/\Phi'$ and $\partial_\Phi V_b= \partial_y V_b/\Phi'$.
We are allowed to do this since, as we showed previously, the function $\Phi(y)$
does not possess any extrema in the bulk, therefore $\Phi'(y)$ never vanishes. Then, 
multiplying both sides of Eq. (\ref{P2eq: jun_con1}) by $\Phi'$ and using Eq. (\ref{P2eq: anti-grav-1}),
we obtain the condition
\beq
\label{P2eq: anti1-jun2.2}
\partial_y V_b\Bigr|_{y=0}=2f_0\left[\tlam^2-k^2\lam(\lam+3)\right].
\eeq
This second constraint may be used in a two-fold way: for a non-trivial interaction
term $V_b$, it may serve to determine an independent parameter in its expression;
alternatively, under the condition that $V_b=\text{constant}$ and thus $\partial_y V_b=0$,
it may determine the value of the effective cosmological constant on the brane
to be $\Lambda=-\tilde \Lambda^2=-k^2 \lam (\lam+3)$, a value that is absolutely
compatible with the original constraint (\ref{P2eq: anti1-con1}) that should hold on the brane.

Let us finally address the issue of the effective theory on the brane. For this, we need to
derive the four-dimensional effective action by integrating the complete five-dimensional
one $S=S_B+S_{br}$, over the fifth coordinate $y$. Before we proceed though, we present the
explicit forms of the five-dimensional curvature invariants whose general form for the
metric ansatz (\ref{P2eq: metric}) are given in Appendix \ref{P1app: Curv-Inv}. Substituting the
mass function (\ref{P2eq: mass-sol}) and the warp function $A(y)=-k|y|$ in these expressions,
we obtain
\gat$ R = -20k^{2} + 4 \Lambda e^{2k|y|}\,, \nonumber\\[2mm]
R_{MN}R^{MN} = 80 k^{4} - 32k^{2}\Lambda e^{2k|y|} + 4\Lambda^{2} e^{4k|y|}\,,
\label{P2eq: invar}\\[2mm]
R_{MNKL}R^{MNKL} = 40k^{4} - 16k^{2}\Lambda e^{2k|y|} + \frac{8\Lambda^{2}e^{4k|y|}}{3}
+\frac{48 M^2 e^{4k|y|}}{r^6}\,, \nonumber$
where $\Lambda=-\tlam^2$ is the negative constant appearing in the projected-on-the-brane
gravitational background (\ref{P2eq: metric-brane}). The above expressions are valid in the domain
$y\in(-\infty,0)\cup(0,\infty)$, i.e. throughout the bulk where the second derivative of
the warp factor equals zero. Note that, by keeping $M$ in the form of the mass function
(\ref{P2eq: mass-sol}), the obtained solutions describe clearly a black string with its singularity
at $r=0$ extending along the extra dimension. Setting, however, $M=0$, we obtain solutions
that are maximally-symmetric on the brane and possess only a true singularity at the
boundary of the bulk spacetime where $y \rightarrow \infty$. 

This latter singularity was not present in the case of the black-string solution of Ref. \cite{CHR},
when $M=0$. We note that the singular terms in Eq. (\ref{P2eq: invar}) are directly related to the
integration constant $\Lambda$ that, as we will soon see, will be interpreted as the
cosmological constant on the brane. Our scalar-tensor theory allows for solutions with
non-zero cosmological constant on the brane while the model employed in Ref. \cite{CHR}
assumed the Randall-Sundrum fine-tuning between the bulk cosmological constant and the
brane tension to ensure a flat brane. If we also set $\Lambda=0$ in our analysis, this
additional singularity disappears and we recover a regular AdS spacetime as in \cite{CHR}.  
However, we consider the presence of a non-zero cosmological constant on the brane
as an important feature of the solutions, the effect of which has not been adequately
studied in the literature. To this end, our analysis reveals that a non-zero $\Lambda$
on the brane is accompanied by a singularity in the bulk, located at an infinite coordinate
distance from our brane.\,\footnote{The presence of this singularity does not affect the
remaining features of the solution, such as the scalar field configuration, the warping
of spacetime, or the effective theory on the brane---however, if desired, it could easily
be shielded by the introduction of a second brane.}

Returning to the four-dimensional effective theory, we will first calculate the effective
gravitational scale $M_{Pl}^2$ on the brane. To this end,
by employing the first of Eqs. (\ref{P2eq: invar}), we may write $R=-20k^2 +R^{(4)} e^{2k|y|}$,
where $R^{(4)}=4 \Lambda$ is the four-dimensional scalar curvature, that may easily be
computed from the projected-on-the-brane line-element (\ref{P2eq: metric-brane}). Hence, the term
from the complete action $S=S_B+S_{br}$ that is relevant for the evaluation of the effective
gravitational constant is the following:
\gat$ \label{P2eq: action_eff}
S\supset\int d^4 x\,dy\,\sqrt{-g^{(5)}}\ \frac{f(\Phi)}{2}\ e^{2k|y|} R^{(4)}\,.$
Then, using also that $\sqrt{-g^{(5)}}=e^{-4k|y|} \sqrt{-g^{(br)}}$, where
$g^{(br)}_{\mu\nu}$ is the metric tensor of the projected on the brane spacetime,
the four-dimensional, effective gravitational constant is given by the integral
\eq$\frac{1}{\kappa_4^2}\equiv 2\,\int_{0}^{\infty} dy\, e^{-2 k y}\,f(y)
=2f_0\,\int_{0}^{\infty} dy\ e^{-2ky}e^{-\lam ky}=
\frac{2f_0}{k(\lam+2)}\,.
\label{P2eq: anti1_effG}$
Since $1/\kappa_4^2=M_{Pl}^2/(8\pi)$, we obtain
\eq$
M_{Pl}^2=\frac{16 \pi f_0}{k(\lam+2)}=\frac{32 \pi f_0^2}{(\sigma+V_b)|_{y=0}}\,
\frac{(\lam +3)}{(\lam+2)}\,.$
In the last expression above, we have replaced the warp-factor parameter $k$
from Eq. (\ref{P2eq: anti1-jun1}). We first note that the integral in Eq. (\ref{P2eq: anti1_effG})
is finite, therefore there is no need for the introduction of a second brane in the model
(unless one wishes to shield the singularity at the boundary of spacetime by introducing
a second brane). Also, according to the above result, the effective gravity scale
on the brane $M_{Pl}^2$ is determined by the ratio $f_0/k$. Taking into account that $f_0$
has units of $[M]^3$, after the absorption of $1/\kappa^2_5$ in its value, and thus
plays the role of the fundamental energy scale $M_*^3$ (or $M_{pl(5)}^3$), the relation between the
fundamental and the effective gravity scales turns out to be almost the same as the
corresponding one in the Randall-Sundrum II model \cite{RS2}. The difference between
the magnitudes of $M_*$ and $M_{Pl}$ is determined by the combination $f_0/(\sigma+V_b)|_{y=0}$
(note that $\lam$ plays virtually no role). Therefore, if a low-gravity scale is
desired, i.e. a low $f_0$, then the total energy-density of the brane should be
minimized. This may be realised via the presence of a large, negative interaction
term $V_b$ for the scalar field that would result in a small, yet positive as dictated
by Eq. (\ref{P2eq: anti1-jun1}), value for the combination $(\sigma+V_b)|_{y=0}$. 

To complete our study of the effective theory on the brane, we finally compute the
effective cosmological constant. Since the scalar field $\Phi$ is only $y$-dependent,
the effective theory will contain no dynamical degree of freedom. Therefore, the
integral of all the remaining terms of the five-dimensional action $S=S_B+S_{br}$,
apart from the one appearing in Eq. (\ref{P2eq: action_eff}), will yield the effective
cosmological constant on the brane. Due to the existence of the brane, that acts
as a boundary for the five-dimensional spacetime, the bulk integral must be supplemented
by the source term of the brane as well as the Gibbons-Hawking term \cite{Gibbons-terms}.
In total, we have
{\fontsize{11}{11}\bea
\label{P2eq: cosm_eff}
-\Lambda_4&=&\int_{-\infty}^{\infty} dy\,e^{-4k|y|}\Bigl[-10 k^2 f(y)-\Lambda_5-
\frac{1}{2}\,\Phi'^2 -V_B(y)+f(y)(-4A'')|_{y=0}-[\sigma +V_b(\Phi)]\,\delta(y)\Bigr]\nonum\\[1mm]
&=&2\int_0^\infty dy\ e^{-4ky}\left[-10 k^2 f(y)-\Lambda_5-
\frac{1}{2}\,\Phi'^2 -V_B(y)\right]+8kf(0)-[\sigma +V_b(\Phi)]_{y=0}\,.
\eea}
\hspace{-0.5em}Substituting the expressions for the coupling function and the bulk potential of the
scalar field, and employing the junction condition \eqref{P2eq: anti1-jun1}, we finally
obtain the result
\eq$\label{P2eq: anti1_effL}
\Lambda_4=-\frac{2f_0\ \tlam^2}{k(\lam+2)}=\frac{\Lambda}{\kappa_4^2}\,.$
As expected, the constant of integration $\Lambda$ appearing in the form of the mass
function (\ref{P2eq: mass-sol}), and in the projected-on-the-brane line-element (\ref{P2eq: metric-brane})
is indeed the four-dimensional cosmological constant $\Lambda_{4}$, multiplied by $\kappa_4^2$,
as the inverse Vaidya coordinate transformation on the brane had demonstrated (Appendix \ref{P1app: Inv-Vaidya}). 


\mysection{The double exponential case \label{P2sec: anti2}}

We now proceed to the alternative form of the non-minimal coupling function given in 
Eq. (\ref{P2eq: anti-g2}). As in the previous section, the focus will be on the derivation
of analytic solutions of the field equations and the study of the characteristics of
the resulting solutions both in the bulk and on the brane.  



\mysubsection{The bulk solution}

In this case, the coupling function has the form $f(y)=f_0\,e^{-\mu^2 e^{\lambda y}}$,
with $\lam$ any positive real number and $\mu$ a real, non-vanishing number. Substituting
the function $g(y)=-\mu^2 e^{\lambda y}$ in the expression of the scalar field $\Phi(y)$
given by Eq. \eqref{P2eq: anti-phi}, we obtain the integral expression
\eq$\label{P2eq: anti2-phi}
\Phi_{\pm}(y)=\pm \sqrt{f_0}\int dy \exp\left(-\frac{\mu^2}{2}e^{\lam y}\right)
\sqrt{\tlam^2 e^{2ky}+\mu^2\lam(\lam+k)e^{\lam y}-\mu^4\lam^2e^{2\lam y}}\,.$
In general, the above integral does not have an analytic solution. However, if one chooses
appropriate values for the parameters $\mu^2$ and $\lam$, the quantity under the square root
can be expressed as a perfect square and the integral becomes solvable. To this end,
we can rewrite the aforementioned quantity as
\gat$\label{P2eq: per-sqrt}
\tlam^2 e^{2ky}+\mu^2\lam(\lam+k)e^{\lam y}-\mu^4\lam^2e^{2\lam y}=
\left(\sqrt{\tlam^2}\,e^{ky}-\sqrt{\mu^2\lam(\lam+k)}\,e^{\lam y/2}\right)^2\,,$
provided that the following conditions are imposed:
\eq$\label{P2eq: anti2-cons}
k+\frac{\lam}{2}=2\lam\,, \qquad 2\sqrt{\tlam^2\mu^2\lam(\lam+k)}=\mu^4\lam^2\,,$
These lead to the unique values 
\beq
\lam=\frac{2k}{3}\,, \qquad \mu^2=\left(\frac{45}{2}\frac{\tlam^2}{k^2}\right)^{\frac{1}{3}}\,,
\label{P2eq: lam-mu}
\eeq
for the $\lam$ and $\mu$ parameters. Using the form of the coupling function in
Eq. \eqref{P2eq: anti-grav-1}, and substituting the above values for $\lam$ and $\mu^2$ 
in the result, we are led to
{\fontsize{12}{11}\eq$ \label{P2eq: anti2-Phi/f}
\Phi'^2(y)=f(y)\,\Bigl[\tlam^2 e^{2ky}+\mu^2\lam(\lam+k)e^{\lam y}-
\mu^4\lam^2e^{2\lam y}\Bigr]=
\frac{2k^2\mu^2}{45}\,f(y)\,e^{\frac{2ky}{3}}\left(\mu^2e^{\frac{2ky}{3}}-5\right)^2\,.$}
\hspace{-0.5em}The above equation can provide important information on the form of the scalar
field in the bulk even before the explicit integration in Eq. ({\ref{P2eq: anti2-phi}) is
performed. To start with, since $f(y)>0$ for all $y>0$, the r.h.s. of the above relation
is automatically positive-definite, therefore no additional constraint on the parameters
of the theory follow by demanding the positivity of $\Phi'^2$.
The value of the parameter $\mu^2$ though affects significantly the profile of the scalar field
along the extra dimension. In particular, if the value of $\mu^2$ is lower than 5, then the
first derivative of the scalar field will become zero at 
$y_0=\frac{3}{2k}\ln\left(\frac{5}{\mu^2}\right)$. If the value of $\mu^2$ is exactly 5,
then the first derivative of the
scalar field is zero at $y=0$. Finally, if $\mu^2$ is greater than 5, then the first
derivative of the scalar field does not vanish anywhere in the bulk. In summary,
{\fontsize{11}{11}\eq$\label{P2eq: val-for-mu}
\left\{\begin{array}{l}
\mu^2<5: \hspace{1.5em}\Phi'(y_0)=0,\hspace{1.5em}\displaystyle{y_0=\frac{3}{2k}\ln\left(\frac{5}{\mu^2}\right),}\\[2mm]
\mu^2=5: \hspace{1.5em}\Phi'(0)=0,\\[2mm]
\mu^2>5: \hspace{1.5em}\Phi'(y)\neq 0, \hspace{2.3em}\forall y>0.\end{array}\right\}$}
\hspace{-0.5em}Taking however the square-root of Eq. (\ref{P2eq: anti2-Phi/f}), we obtain for $\Phi'(y)$
the expression
\beq
\Phi'(y)=\pm k \sqrt{\frac{2 f_0 \mu^2}{45}}\,\exp\left(-\frac{\mu^2}{2}\,e^{\frac{2ky}{3}}
+\frac{ky}{3}\right)\,\left|\mu^2e^{\frac{2ky}{3}}-5\right|.
\label{P2eq: anti2-phi'}
\eeq
We thus conclude that $\Phi'(y)$ retains a specific sign throughout the bulk, even in
the case where $\mu^2<5$; therefore, the point $y=y_0$ is not an extremum
[where $\Phi'(y)$ owes to change its sign] but rather an inflection point. As a
result, $\Phi(y)$ is a monotonic function throughout the bulk, for all values of $\mu^2$,
and thus it is an one-to-one function in the whole region $y>0$ (as well as in the $y<0$
region due to the ${\mathbf{Z}_2}$ symmetry).

The above behaviour also affects the way that one should proceed in order to find
the solution for the scalar field $\Phi$. For $\mu^2\geq 5$, the quantity inside
the absolute value in Eq. (\ref{P2eq: anti2-phi'}) is positive and non-vanishing for all
values $y>0$; thus, the solution for $\Phi(y)$, at every point in the bulk, follows
by directly integrating Eq. (\ref{P2eq: anti2-phi'}). Then, we obtain
\gat$
\Phi_{\pm}(y)=\pm \left[\mathcal{I}(y)-\mathcal{I}(0)\right],$
where we have defined $\mathcal{I}(y)$ as 
\eq$ \label{P2eq: I-function}
\mathcal{I}(y)\equiv -\sqrt{\frac{f_0}{5}}\left[\sqrt{2\mu^2} \exp\left(\frac{k y}{3}-\frac{1}{2}
\mu^2 e^{\frac{2 k y}{3}}\right)+
4 \sqrt{\pi}\ \text{erf}\left(\sqrt{\frac{\mu^2}{2}}e^{\frac{k y}{3}}\right)\right],$
and $\text{erf}(z)$ is the error function
\beq
\text{erf}(z)=\frac{2}{\sqrt{\pi}}\int_0^z e^{-t^2}dt=\frac{2}{\sqrt{\pi}}
\sum_{n=0}^\infty \frac{(-1)^n\,z^{2n+1}}{n!\,(2n+1)}\,.
\label{P2eq: error}
\eeq
On the other hand, for $\mu^2<5$, we need to address separately the cases where the
solution for $\Phi(y)$ is found at a point in the bulk with $y\leq y_0$ or at a point
beyond the inflection point with $y > y_0$. In the first case, apart from the change
in the order of the terms inside the absolute value in Eq. (\ref{P2eq: anti2-phi'}), no
other action is necessary, and the integration over $y$ is performed as before. In
the second case, however, care must be taken when the inflection point at $y=y_0$
is reached. Then, we write
\bal$
\Phi_{\pm}(y)&=\pm k \sqrt{\frac{2 f_0 \mu^2}{45}}\,\left[\int_0^{y_0} dy' 
\exp\left(-\frac{\mu^2}{2}e^{\frac{2ky'}{3}}+\frac{ky'}{3}\right)
\left(5-\mu^2e^{\frac{2ky'}{3}}\right)+\right.\nonum\\[3mm]
&\hspace{2.5em}\left.+\int_{y_0}^y dy' \exp\left(-\frac{\mu^2}{2}e^{\frac{2ky'}{3}}+
\frac{ky'}{3}\right)\left(\mu^2e^{\frac{2ky'}{3}}-5\right)\right].$
Overall, for $\mu^2<5$, the solution for the scalar field is
\eq$\label{P2eq: anti2-phi-mu<}
\Phi_{\pm}(y)=\left\{\begin{array}{cr}
\mp\left[\mathcal{I}(y)-\mathcal{I}(0)\right],&\hspace{1em} y\leq y_0,\\[4mm]
\pm\left[\mathcal{I}(0)-2\mathcal{I}(y_0)+\mathcal{I}(y)\right],& y > y_0,
\end{array}\right\}$
where $\mathcal{I}(y)$ is still given by Eq. (\ref{P2eq: I-function}).

\begin{figure}[t]
    \centering
    \begin{subfigure}[b]{0.46\textwidth}
        \includegraphics[width=\textwidth]{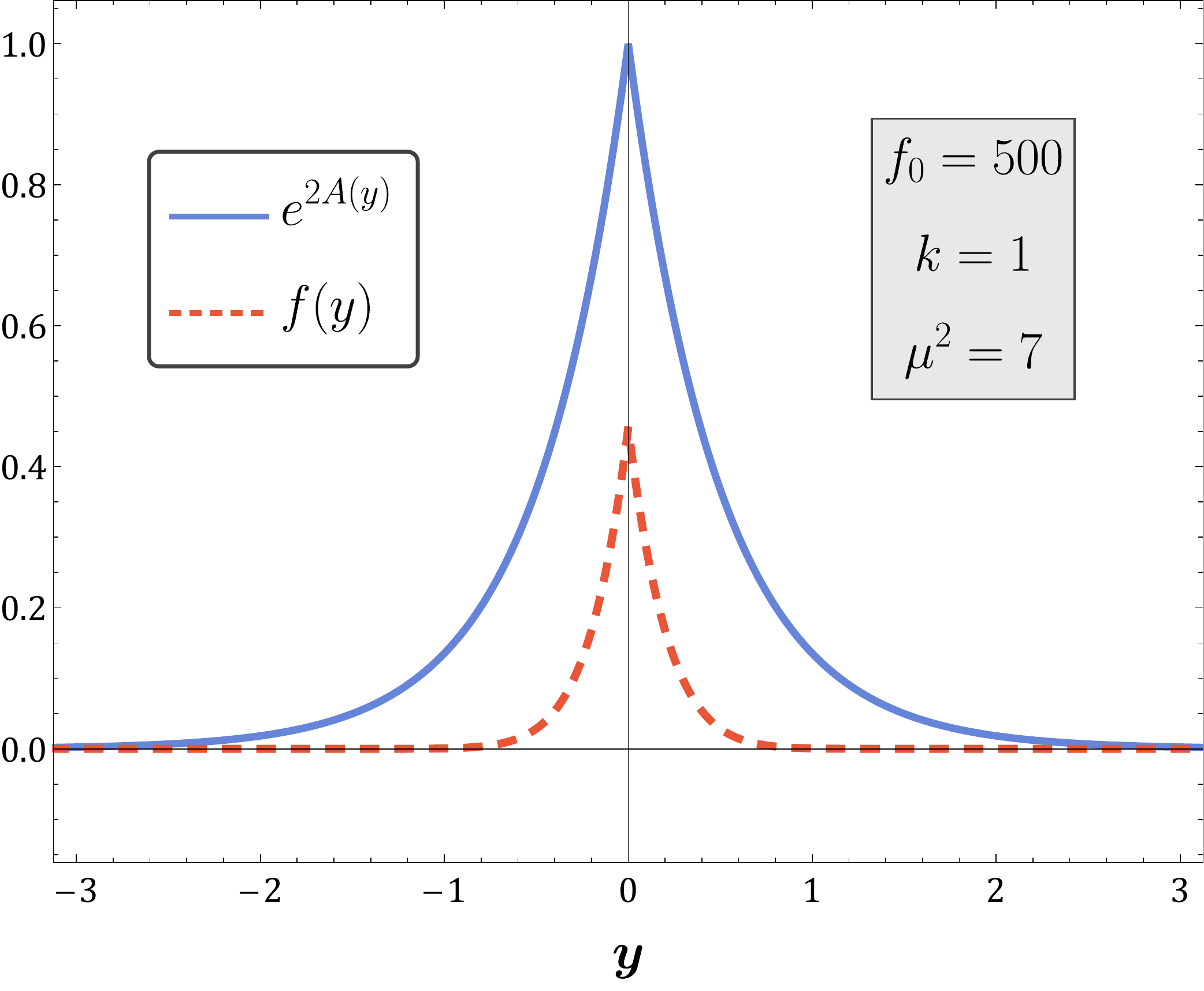}
        \caption{\hspace*{-1.2em}}
        \label{P2subf: fig3a}
    \end{subfigure}
    ~ 
    \begin{subfigure}[b]{0.49\textwidth}
        \includegraphics[width=\textwidth]{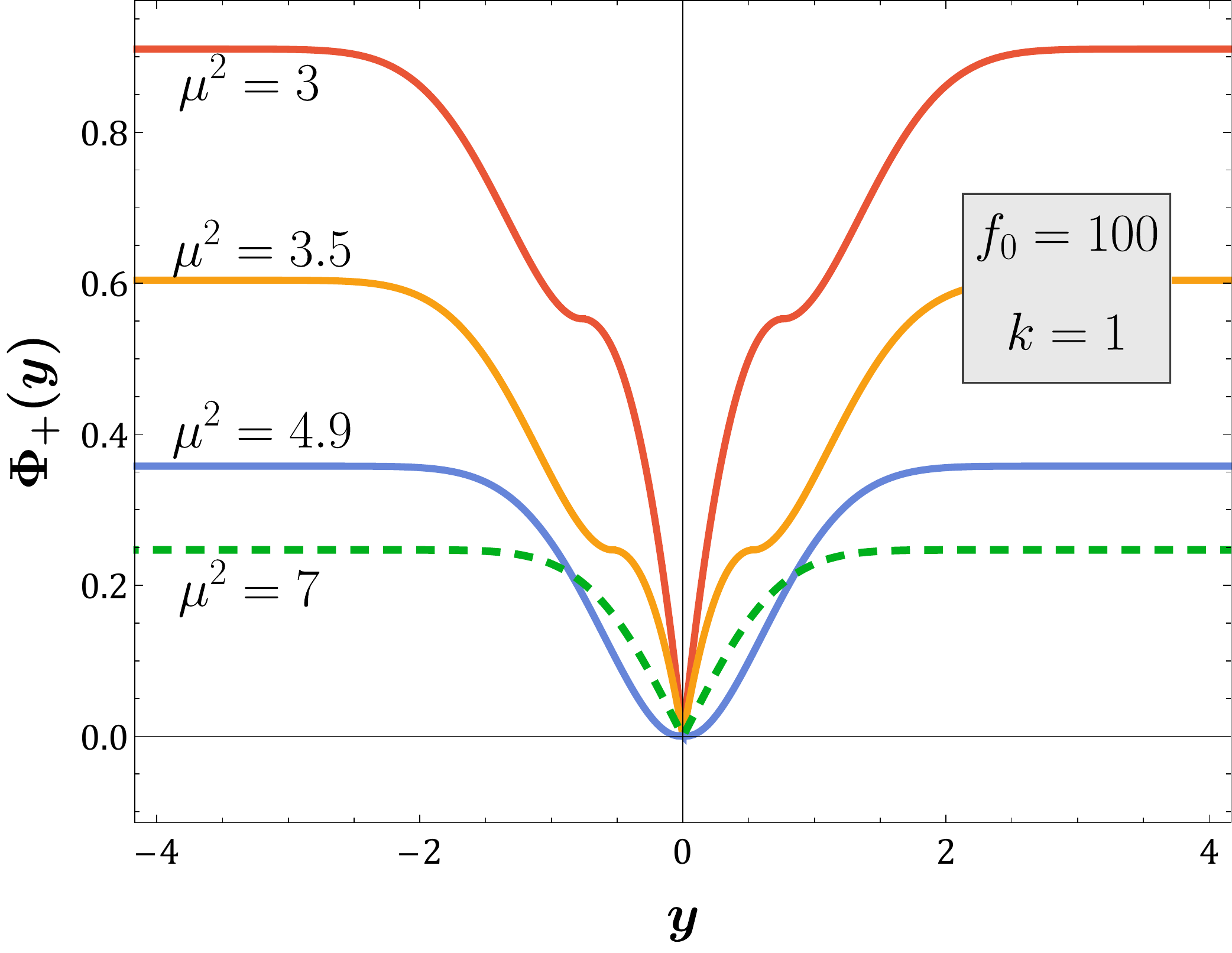}
        \caption{\hspace*{-2.5em}}
        \label{P2subf: fig3b}
    \end{subfigure}
    ~ 
    \caption{(a) The warp factor $e^{2A(y)}=e^{-2k|y|}$ and the coupling function
    $f(y)=f_0\,e^{-\mu^2 e^{2k y/3}}$, and (b) the scalar field $\Phi_{+}(y)$ for
    various values of the parameter $\mu^2$, in terms of the coordinate $y$.}
   \label{P2fig: anti2_warp_f_Phi}
\end{figure}

\par In \myref{P2fig: anti2_warp_f_Phi}{P2subf: fig3a}, we depict the form of the warp factor $e^{-2k|y|}$
and the coupling func\-tion $f(y)=f_0\,e^{-\mu^2 e^{2k y/3}}$ in terms of the coordinate
$y$ along the fifth dimension, for $f_0=500$, $k=1$ and $\mu^2=7$. Both functions
exhibit a localization close to the brane with the coupling function $f(y)$ decreasing,
in fact, much faster due to its double exponential dependence on $y$. At the boundary of
spacetime, both functions go smoothly to zero. The displayed, qualitative behaviour of
these two quantities is independent of the particular values of the parameters.
In contrast, the profile of the scalar field $\Phi(y)$ with respect to the extra dimension
$y$ depends strongly on the value of the parameter $\mu^2$, as one may clearly see
in \myref{P2fig: anti2_warp_f_Phi}{P2subf: fig3b}. We observe that the behaviour of the scalar field
changes significantly as the parameter $\mu^2$ approaches and then surpasses the value 5.
Indeed, for $\mu^2 <5$, the emergence of the  inflection point at $y=y_0>0$ is clearly
visible. As $\mu^2$ approaches the value 5, the inflection point moves towards the brane.
For $\mu^2 \geq 5$, though, this feature completely disappears in accordance to the
analytical study presented above. Overall, the scalar field exhibits a monotonic behaviour
over the entire bulk---for the $\Phi_+(y)$ solution that we have chosen here to plot, the scalar
field presents an increasing profile in the bulk reaching a constant, asymptotic value at
the boundary of spacetime.  In \myref{P2fig: anti2_Phi_2_3}{P2subf: fig4a} and \myref{P2fig: anti2_Phi_2_3}{P2subf: fig4b},
we present the dependence
of $\Phi_+(y)$ on the second parameter $k$ - since the parameter $\lam$ is now fixed to
the value of the warping parameter $k$ through the first of Eqs. (\ref{P2eq: lam-mu}), henceforth
we drop any reference to $\lam$. We observe again the emergence of the inflection point
when $\mu^2 <5$, in \myref{P2fig: anti2_Phi_2_3}{P2subf: fig4a}, and the smooth behaviour when $\mu^2 > 5$,
in \myref{P2fig: anti2_Phi_2_3}{P2subf: fig4b}. The value of the warping parameter $k$ causes only a
rise in the slope of the curve, as $k$ increases, leaving all the other features invariant.

\begin{figure}[t]
    \centering
    \begin{subfigure}[b]{0.475\textwidth}
        \includegraphics[width=\textwidth]{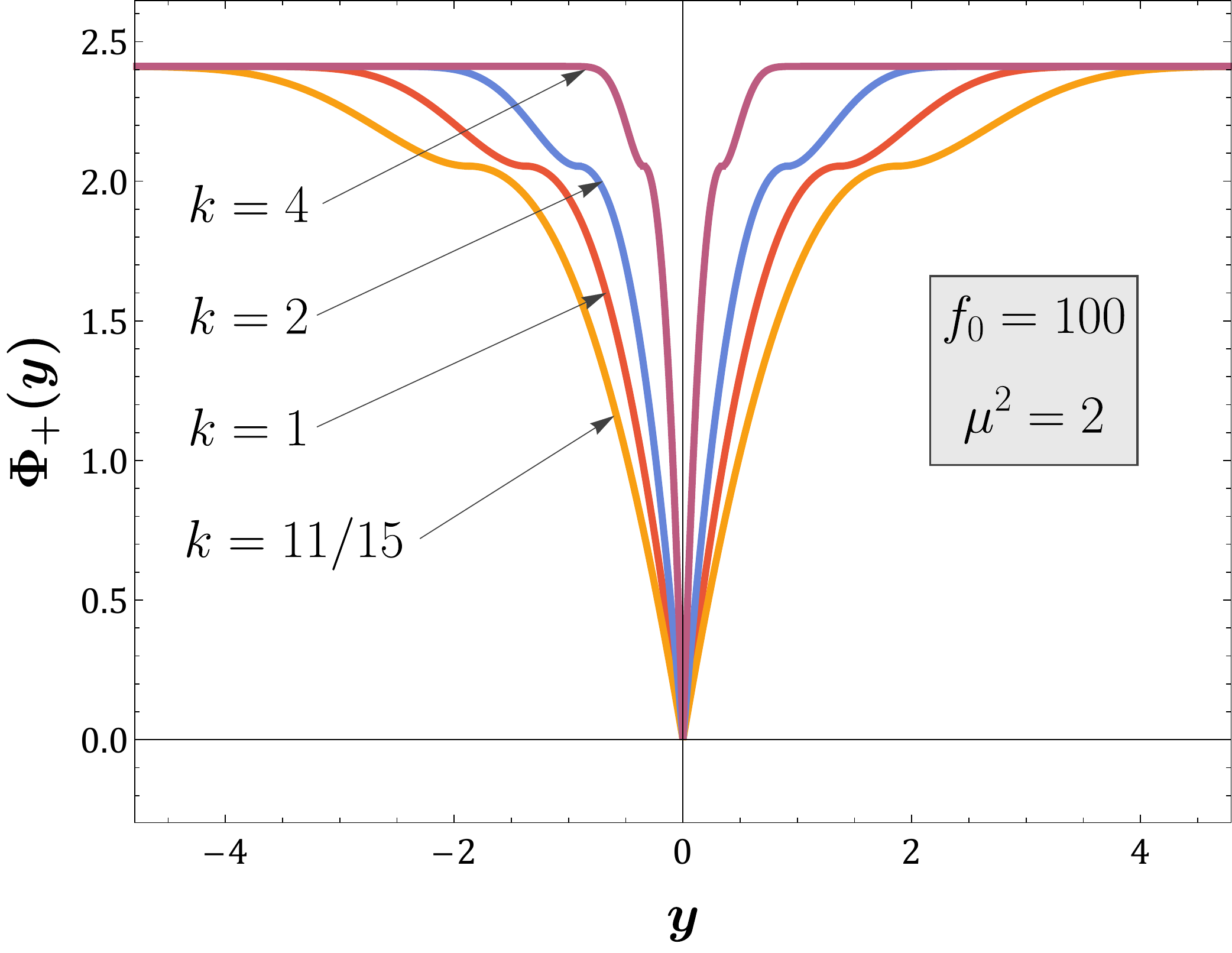}
        \caption{\hspace*{-2.5em}}
        \label{P2subf: fig4a}
    \end{subfigure}\quad
    ~ 
    \begin{subfigure}[b]{0.47\textwidth}
        \includegraphics[width=\textwidth]{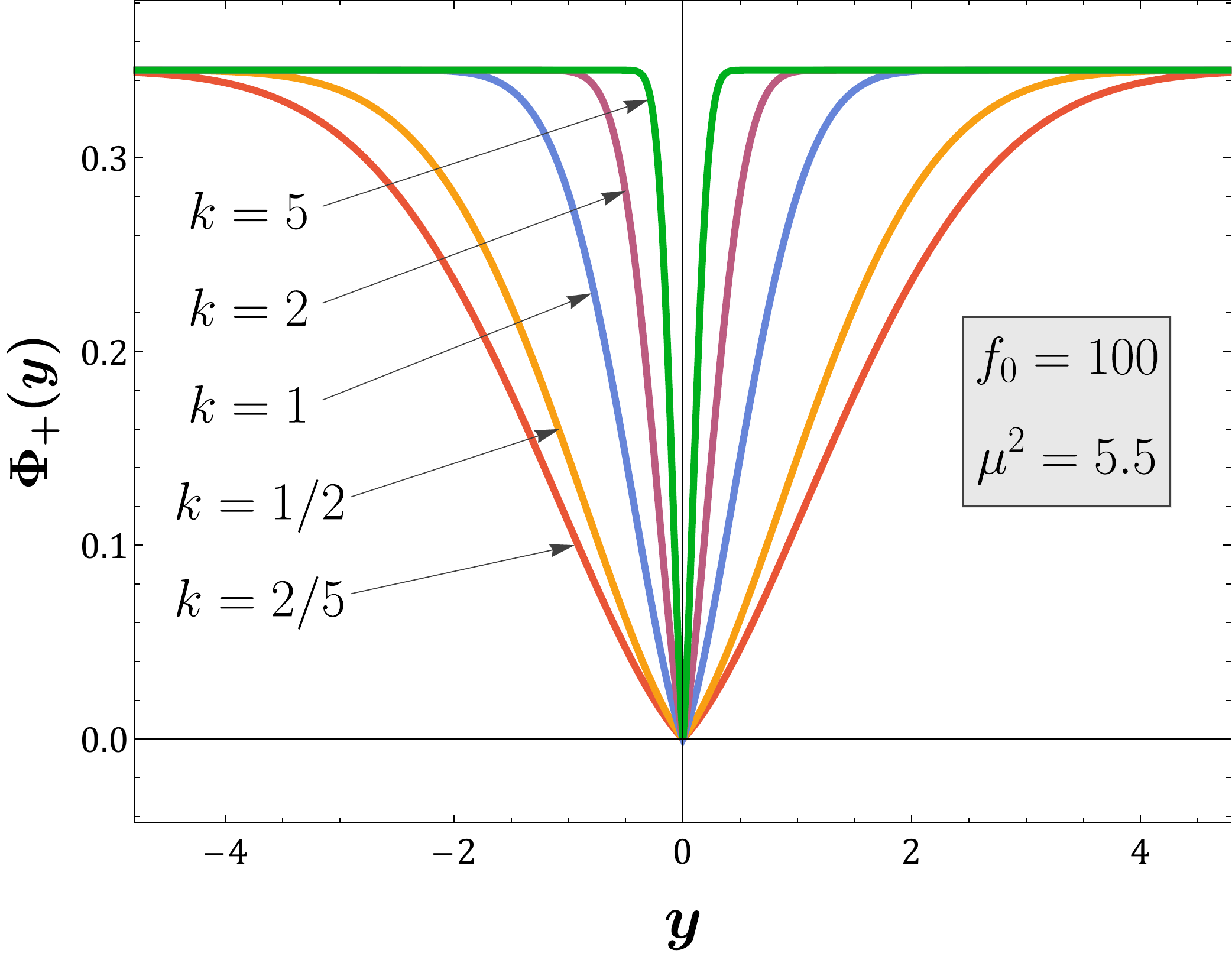}
        \caption{\hspace*{-2.5em}}
        \label{P2subf: fig4b}
    \end{subfigure}
    ~ 
    \vspace*{-1em}
    \caption{The scalar field $\Phi_{+}(y)$, in terms of the coordinate $y$, for $f_0=100$ and
    for (a) $\mu^2=2$, and (b) $\mu^2=5.5$ and various values of $k$. }
   \label{P2fig: anti2_Phi_2_3}
\end{figure}

\par The bulk potential of the field $V_B$ in this case can be determined again by the
general expression (\ref{P2eq: V_B-1}). Substituting the double exponential form of the coupling
function $f(y)$, we now obtain the result
\eq$\label{P2eq: anti2-pot}
V_B(y)=-\Lambda_5-f_0\ e^{-\mu^2e^{\frac{2ky}{3}}}\left[6k^2+\frac{k^2\mu^2}{45}
e^{\frac{2ky}{3}}\left(3\mu^4e^{\frac{4ky}{3}}+10\mu^2e^{\frac{2ky}{3}}+95\right)\right].$
As in the simple exponential case, the combination inside the square brackets
in the above expression is positive-definite, for all values of the parameters of the model,
thus rendering the second term of the bulk potential negative-definite. Therefore, the
presence of a non-minimally-coupled scalar field in the bulk leads to a negative (non-constant) 
potential energy in the bulk that can support again an AdS-type bulk
spacetime with an exponentially decreasing warp factor, even if the quantity $\Lambda_5$ is
set to zero. The potential has a smooth form over the entire bulk, is localized close
to the brane and it goes to zero extremely fast away from it -- all these features are
inherited from the form of the coupling function to which $V_B$ is directly proportional
as  Eq. (\ref{P2eq: anti2-pot}) clearly shows. The aforementioned behaviour of $V_B$ is depicted
in \myref{P2fig: anti2_V-TMN}{P2subf: fig5a} for $f_0=1/5$, $k=1$ and $\mu^2=5$.

\par The components of the energy-momentum tensor of the theory may be computed employing
again Eqs. (\ref{P2eq: anti1-rho-0}) and (\ref{P2eq: anti1-p-0y}). Substituting again the form of
the coupling function together with the expression for the bulk potential (\ref{P2eq: anti2-pot})
presented above, we find the explicit expressions
\eq$\label{P2eq: anti1-rho-2}
\rho=-p^i=-2f_0k^2\left(3+\frac{\mu^6}{45}e^{2ky}\right)e^{-\mu^2e^{\frac{2ky}{3}}}\,,$
\eq$\label{P2eq: anti1-py-2}
p^y=2f_0k^2\left(3+\frac{2\mu^6}{45}e^{2ky}\right)e^{-\mu^2e^{\frac{2ky}{3}}}\,.$
In \myref{P2fig: anti2_V-TMN}{P2subf: fig5a} and \myref{P2fig: anti2_V-TMN}{P2subf: fig5b} we present the behaviour of the energy
density $\rho$ and the pressure components $p^i$ and $p^y$, respectively, in terms of the
extra dimension $y$. These quantities, too, present a smooth profile over the entire bulk,
remain localized
close to our brane, and vanish asymptotically leaving behind a 5-dimensional, flat spacetime
(if $\Lambda_5$ is assumed zero). The energy density $\rho$ is again negative throughout the
bulk (as it should be in order to support by itself a pseudo-AdS spacetime) but this is due to 
the presence of a physical, scalar degree of freedom coupled non-minimally to gravity with
a physically-acceptable positive-definite, and localized close to our brane, coupling
function.  

\begin{figure}[t]
    \centering
    \begin{subfigure}[b]{0.475\textwidth}
        \includegraphics[width=\textwidth]{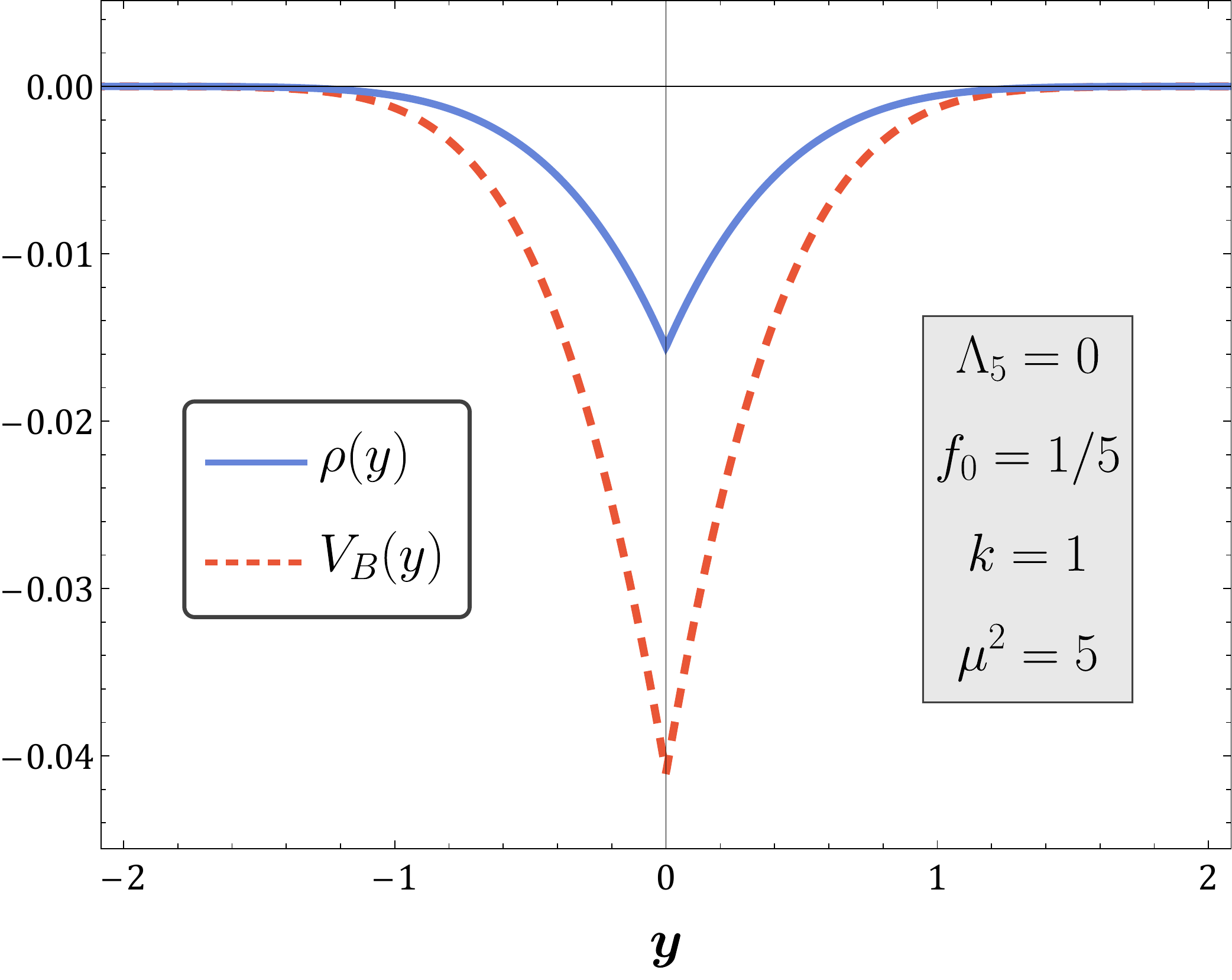}
        \caption{\hspace*{-1.9em}}
        \label{P2subf: fig5a}
    \end{subfigure}\quad
    ~ 
    \begin{subfigure}[b]{0.475\textwidth}
        \includegraphics[width=\textwidth]{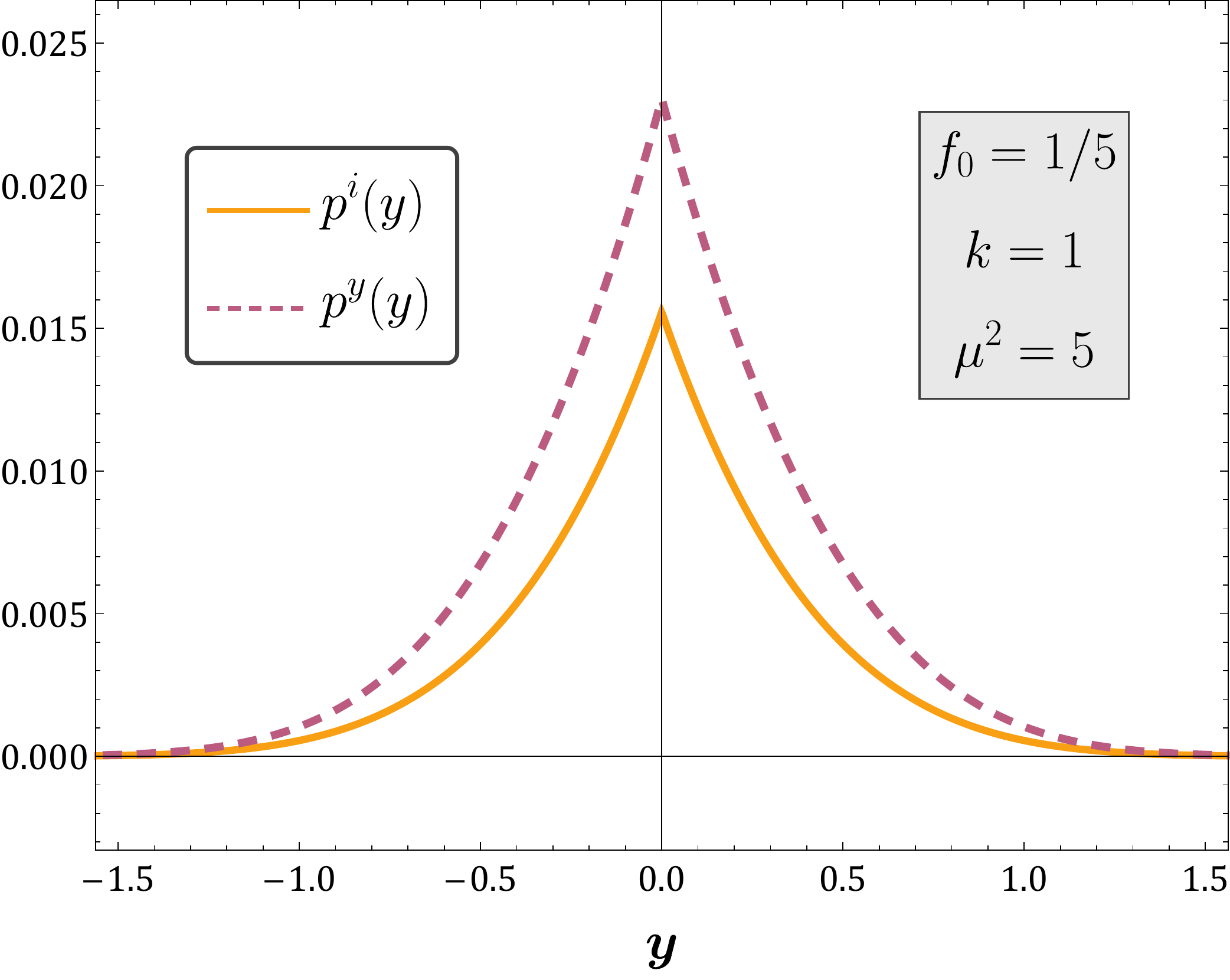}
        \caption{\hspace*{-1.8em}}
        \label{P2subf: fig5b}
    \end{subfigure}
    ~ 
    \vspace*{-1em}
    \caption{(a) The scalar potential $V_B$ and energy density $\rho$ of the system, and
(b) the pressure components $p^y$ and $p^i$ in terms of the coordinate $y$. }
   \label{P2fig: anti2_V-TMN}
\end{figure}


\mysubsection{Junction conditions and effective theory}

For the junction conditions, we use again the general expressions \eqref{P2eq: jun_con1} and
\eqref{P2eq: jun_con2}, in which we substitute the form of the coupling function and employ
also Eq. (\ref{P2eq: anti2-phi'}) to replace $\Phi'(y)$. Then, Eq. \eqref{P2eq: jun_con2}
straightforwardly leads to the constraint
\gat$\label{P2eq: anti2-1-jun1}
2kf_0e^{-\mu^2}\left(3+\frac{2\mu^2}{3}\right)=\sig+V_b(\Phi)|_{y=0}\,,$
that may be used again to fix the warping parameter $k$ in terms of the 
parameters $f_0$ and $\mu$ of the coupling function and the energy-content
$(\sigma+V_b)$ of the brane. On the other hand, Eq. \eqref{P2eq: jun_con1} 
results\,\footnote{Note, that care should be taken in the
evaluation of \eqref{P2eq: jun_con1} due to the different behaviour of the scalar field
$\Phi(y)$ in terms of $\mu^2$. At the end of the evaluation, though, a unique expression
follows from this junction condition for either $\mu^2>5$ or $\mu^2<5$.} in the condition
\gat$
\label{P2eq: anti2-1-jun2.2}
V'_b(0)=f_0\ e^{-\mu^2}\frac{4\mu^2k^2}{3}\left[\frac{(\mu^2-5)^2}{15}-4\right].$
For a non-trivial $V_b$, the above condition can be used to restrict a parameter that
may appear in its expression; on the other hand, for a trivial $V_b$, setting
$V'_b(0)=0$ in Eq. (\ref{P2eq: anti2-1-jun2.2}}), we may fix also the value of $\mu^2$,
and, through Eq. (\ref{P2eq: lam-mu}), the value of the effective cosmological constant
on the brane $\Lambda=-\tilde \Lambda^2$.

We turn finally to the 4-dimensional, effective theory on the brane. In order to
derive the effective gravitational constant on the brane, we may use again the
relation \eqref{P2eq: action_eff}. Substituting the double exponential form of the
coupling function $f(y)$, we find
\bal$\label{P2eq: anti2-kappa4}
\frac{1}{\kappa_4^2}&=2f_0\int_{0}^{\infty} dy\ e^{-2ky}\exp\left(-\mu^2
e^{\frac{2ky}{3}}\right)=\frac{f_0}{2k}\left[e^{-\mu^2}\left(2-\mu^2+\mu^4\right)+
\mu^6\ \text{Ei}\left(-\mu^2\right)\right].$
Above, we have used the exponential-integral function $\text{Ei}(x)$ defined as
\beq
\text{Ei}(x)\equiv-\int_{-x}^\infty dt\ \frac{e^{-t}}{t}\,,
\label{P2eq: fun-Ei}
\eeq
and its property that $\lim_{x\ra\infty}\text{Ei}(-x)=0$. For $x<0$, it can also be shown that
\cite{Abramowitz}
\beq
\text{Ei}(x)=\text{Ei}\left(-|x|\right)=\gamma+\ln|x|+\sum_{k=1}^\infty 
\frac{(-1)^{k}|x|^k}{k\ k!}\,,
\eeq
where $\gamma$ is the Euler-Mascheroni constant. Then, the effective gravitational
energy scale is given by
\bal$\label{P2eq: anti2-MPl}
M_{Pl}^2 &=\frac{4\pi f_0}{k}\left\{e^{-\mu^2}\left(2-\mu^2+\mu^4\right)+\mu^6\left[\gamma+
2\ln|\mu|+\sum_{q=1}^\infty \frac{(-1)^{q}\mu^{2q}}{q\ q!}\right]\right\}.$
Once again, the gravity scale on the brane $M_{Pl}^2$ is determined primarily by the
ratio $f_0/k \sim M_*^3/k$, that resembles again the corresponding relation of the
Randall-Sundrum model \cite{RS2}---note that the combination inside the curly brackets
in Eq. (\ref{P2eq: anti2-kappa4}) is of ${\cal O}(1)$. 

The effective cosmological constant on the brane can be evaluated by employing
Eq. \eqref{P2eq: cosm_eff}. Using again the form of the coupling function $f(y)$ and the
junction condition (\ref{P2eq: anti2-1-jun1}), we find that
\gat$
\Lambda_4=-\frac{f_0 k\mu^6}{45}\left[e^{-\mu^2}\left(2-\mu^2+\mu^4\right)+
\mu^6\,\text{Ei}\left(-\mu^2\right)\right]=
-\frac{\tlam^2}{\kappa_4^2}=\frac{\Lambda}{\kappa_4^2}\,.$
Above, we have used the second of Eqs. (\ref{P2eq: lam-mu}), that relates the parameter
$\mu^2$ with $\tilde \Lambda^2$, and the expression (\ref{P2eq: anti2-kappa4}) for $\kappa_4^2$.
As it was discussed previously, the integration parameter $\Lambda$ is confirmed
to be the effective cosmological constant $\Lambda_4$ on the brane multiplied by the
effective gravitational constant $\kappa_4^2$.



\mysection{Conclusions \label{P2sec: Disc}}

In this Chapter, we have considered a five-dimensional gravitational theory containing 
a scalar field with a non-minimal coupling to the five-dimensional Ricci scalar. 
The coupling is realized through a smooth, real, positive-definite coupling function
$f(\Phi)$. Demanding that all components of the energy-momentum tensor remain
finite throughout the bulk, and looking for analytic solutions for the scalar
field, we have restricted our choices for the coupling function to two particular
forms: a simple exponential and a double exponential, both decreasing
away from the brane. This results into a scalar-tensor five-dimensional
theory with a non-minimal coupling between the scalar field and gravity that is
effectively localized close to the brane. 
Having studied the case of a positive effective cosmological constant
on the brane in the previous Chapter, here, we focused on the case
of a negative four-dimensional cosmological constant.
Hence, by analytically solving the gravitational and scalar field equations in the bulk we produced
black-string solutions which reduce to the Schwarzschild anti-de Sitter spacetime on the brane.
We examined two complete such solutions that are both characterized by a regular scalar field, a localized-close-to-our
brane energy-momentum tensor and a negative-definite, non-trivial bulk potential that may support by
itself the warping of the spacetime even in the absence of the traditional, negative, bulk cosmological
constant. Despite the infinitely long string singularity in the bulk, the four-dimensional effective theory
on the brane is robust with the effective gravity scale being related to the fundamental one and the
warping scale. It is worth noting that if we set the mass of the black hole on the brane equal to zero,
the black string disappears leaving behind a regular braneworld model with only a true singularity at
the boundary of the fifth dimension. 

\afterpage{\blankpage}



\mychapter{\textbf{Incorporating physical constraints in black-string solutions for a Minkowski brane} \label{Chap: P3}}
\phantomsection
\setcounter{chapter}{4}

\epigraph{\justify\textit{``Every scientific inquiry, which is separated from justice and every other virtue, is wile not wisdom.''}}{Plato}

\thispagestyle{empty}

{\lettrine[lines=3, lhang=0.03]{\color{chapter(color)}H}{\hspace*{5.5px}aving} 
covered the cases of the de Sitter and anti-de Sitter spacetimes on our brane
in \chapref{Chap: P1} and \chapref{Chap: P2}, respectively, in this third instalment, 
which is based on \cite{KNP3}, we turn our attention to the case of a
Minkowski brane, i.e. with a vanishing effective cosmological constant. The objective
would be the same, namely to perform a comprehensive study of the complete set
of field equations and derive analytical solutions for the gravitational background and
scalar field in the bulk. As we will demonstrate, this case is the least restrictive and
most flexible of the three, and allows for a variety of profiles for the coupling function
and scalar field along the extra coordinate. In order to construct physically-acceptable
solutions, we will demand the finiteness of both the coupling function and scalar field
everywhere in the bulk. In fact, we will consider forms of the coupling function that
become trivial at large distances from our brane thus leading to a minimally-coupled
scalar-tensor theory in that limit. Even under the above assumptions, we will present
a large number of solutions; they will all be characterised by a regular scalar field
and a finite energy-momentum tensor localized near our brane. In addition, the
bulk potential of the scalar field may take a variety of forms at our will, while
supporting  in all cases an exponentially decaying warp factor even in the absence of
a negative bulk cosmological constant.  Negative values of the coupling function in the
bulk will not be necessary in our analysis, nevertheless, they will be allowed. The form
of the effective theory on the brane will thus be of primary importance and a necessary
ingredient of our analysis in the study of each solution presented. We will naturally
demand a positive effective gravitational constant on our brane, and investigate whether
this demand may be simultaneously satisfied with the condition of a positive total
energy of our brane and the validity of the weak energy conditions in the bulk. 
The gravitational background on the brane will be described by the Schwarzschild solution
leading to either a non-homogeneous black-string solution in the bulk, when the mass
parameter $M$ is non-zero, or a regular anti-de Sitter spacetime, when $M=0$. 

}

The Chapter has the following outline: in \secref{P3sec: th-frame}, we present our theory, the field
equations and impose a number of physical constraints on the scalar field and its
coupling function. In \chapc{Secs.}\hspace{0.3em}\ref{P3sec: linear} to \ref{P3sec: hyper-tang}, we present a large number of complete brane-world
solutions, and discuss in detail their physical properties in the bulk, the junction conditions,
the effective theory on the brane and the parameter space where the
optimum solutions---from the physical point of view---emerge in each case.
Finally, we present our conclusions in \secref{P3sec: Disc}.

\mysection{The theoretical framework \label{P3sec: th-frame}}

The theory that we consider here is the same as in the two previous Chapters, therefore
Eqs. \eqref{P1eq: action} to \eqref{P1eq: mass-sol} hold here as well.
However, in the present Chapter we are interested in solutions emerging for a Minkowski brane,
hence $\Lambda$ will be set to zero. 
As a result, the mass function $m(r)$, which is given by Eq. \eqref{P1eq: mass-sol}, has the constant value $M$.
In the bulk and for $\Lambda=0$ the field equation (\ref{P1eq: grav-eq1}) takes the form\,\footnote{Due to the
${\bf Z}_2$-symmetry in the bulk, henceforth, we focus on the positive $y$-regime.}
\eq$\label{P3eq: grav-1-1}
(1+\pa_\Phi^2f)\Phi'^2+\pa_\Phi f(\Phi''+k \Phi')=0\,,$
or
\eq$\label{P3eq: grav-1-2}
\Phi'^2+\pa_y^2f+k\,\pa_yf=0\, ,$
while Eq. \eqref{P1eq: grav-eq2}, with the use of Eq. \eqref{P3eq: grav-1-2}, can be solved for $V_B(y)$ resulting to
\eq$\label{P3eq: V-B}
V_B(y)=-\Lambda_5-6k^2 f(y)+\frac{7}{2}\,k\,\pa_y f-\frac{1}{2}\,\pa_y^2f\, .$
In the above, we have also used the relations
\eq$\label{P3eq: dif-f}
\pa_yf=\Phi' \,\pa_\Phi f, \hspace{1.5em}\pa_y^2f=
\Phi'^2\,\pa_\Phi^2 f+\Phi''\,\pa_\Phi f\,.$

The topology of the five-dimensional spacetime in the bulk may be inferred from the
form of the curvature invariant quantities. Using the five-dimensional line-element
(\ref{P1eq: metric}), together
with the relations $m(r)=M$ and $A=-k |y|$, we find the following expressions
\beq
R=-20 k^2\,, \quad R_{MN} R^{MN}=80 k^4\,, \quad 
R_{MNRS} R^{MNRS}= 40 k^4 +\frac{48 M^2\,e^{4k|y|}}{r^6}\,.
\eeq
For $M=0$, the bulk spacetime is characterised by a constant
negative curvature at every point, and is therefore an AdS$_5$ spacetime. This holds
despite the presence of a non-trivial distribution of energy in the bulk, i.e. that
of a non-minimally coupled scalar field with a potential, and is ensured through
the field equations which, like Eqs. (\ref{P3eq: grav-1-2}) and (\ref{P3eq: V-B}), relate the
different bulk quantities among themselves. It is for this reason that, as we will
see, the exponentially decaying warp factor will be supported even in the absence
of the negative bulk cosmological constant $\Lambda_5$. In the case where
$M \neq 0$, the above invariants describe a 5-dimensional black-string solution
with an infinitely-long spacetime singularity extending throughout the extra
dimension. The black-string  singularity reaches the boundary of spacetime
which is by itself a singular hypersurface.

The solution for both the scalar field and the bulk potential depends, through 
Eqs. (\ref{P3eq: grav-1-2})-(\ref{P3eq: V-B}), on the form of the non-minimal coupling function $f(\Phi)$. 
In \chapref{Chap: P2}, we assigned the following constraints to the
scalar field $\Phi(y)$ and its coupling function $f[\Phi(y)]$:
\begin{enumerate}
\item[\bf(i)] Both functions should be real and finite in their whole domain 
and of class $C^{\infty}$.
\item[\bf(ii)] At $y\ra \pm\infty$, both functions should satisfy the following relations, otherwise
the finiteness of the theory at infinity cannot be ascertained,
\eq$\label{P3eq: con.1}
\lim_{y\ra\pm\infty}\frac{d^n[f(y)]}{dy^n}=0,\hspace{1.5em}\forall n\geq 1,$
\eq$\label{P3eq: con.2}
\lim_{y\ra\pm\infty}\frac{d^n[\Phi(y)]}{dy^n}=0,\hspace{1.5em}\forall n\geq 1.$
\end{enumerate} 
The second constraint amounts to considering profiles of the scalar field and
forms of the coupling function that both reduce to a constant value far away
from the brane. Together with the first constraint, they ensure a physically
acceptable behaviour for our scalar-tensor theory. The sign, however, of the
coupling function $f(y)$ will not be fixed. In \chapref{Chap: P1}, where the case of
a positive cosmological constant on the brane was studied, i.e. $\Lambda>0$, 
the coupling
function had to be negative-definite away from our brane for the reality of
the scalar field to be ensured; nevertheless, the effective theory on the brane
could still be well-defined. In the case of $\Lambda <0$ (\chapref{Chap: P2}),
no such requirement was necessary and the coupling 
function was assumed to be everywhere positive-definite in terms of the
$y$-coordinate; then, gravity was normal over the entire five-dimensional
spacetime leading to a well-defined effective field theory on the brane.

In the context of the present analysis, where $\Lambda=0$, we may consider
coupling functions that are either positive or negative-definite for particular
regions of the $y$-coordinate. As we will demonstrate, it is possible to obtain
a positive effective four-dimensional gravitational constant in every case. This
will hold even when five-dimensional gravity behaves in an anti-gravitating way
at particular regimes of spacetime---as it turns out, such a behaviour is not 
physically forbidden as long as the effective theory on our brane is well-defined.
To this end, the derivation of the effective theory on the brane is going to play
an important role in our forthcoming analysis, and will thus supplement every
bulk solution we derive. 

%
%

\vspace*{-1em}

\mysection{Linear coupling function \label{P3sec: linear}}

Choosing $\Lambda=0$ on our brane simplifies the set of field equations of
the theory, but more importantly, relaxes constraints that had to be imposed
on the coupling function. As a result, the latter is now allowed to adopt a
variety of physically-acceptable forms, all obeying criteria (i) and (ii)
of the previous section. These forms lead to viable brane-world models (for
$M=0$) or black-string solutions (for $M \neq 0$). In an effort to construct
the most realistic solutions, we will also study, in every case, the energy
conditions both in the bulk and on the brane.

We start our analysis with the case of the linear coupling function
\eq$\label{P3eq: linear-f}
f(\Phi)=f_0+\Phi_0\Phi\, ,$
where $f_0$ and $\Phi_0$ are arbitrary parameters of the theory. In what
follows, we will first solve the system of field equations 
(\ref{P3eq: grav-1-1}) and (\ref{P3eq: V-B}) in the bulk and then consider the
effective theory on the brane as well as the energy conditions.

\vspace*{-1em}

\mysubsection{The bulk solution}

\par Substituting the aforementioned coupling function in Eq. \eqref{P3eq: grav-1-1}
and solving the resulting second-order differential equation, we obtain the solution
\eq$
\Phi(y)=\Phi_0\left[-ky+\ln(e^{ky}+\xi)\right], \label{P3eq: linear-Phi}$
where $\xi$ is an integration constant.
Note that the gravitational field equation \eqref{P3eq: grav-1-1} possesses a
translational symmetry with respect to the scalar field $\Phi(y)$.
Hence, we are free to fix the value of a second integration constant,
that should in principle appear additively on the right-hand-side of
Eq. (\ref{P3eq: linear-Phi}), to zero without loss of generality. Then,
using Eq. \eqref{P3eq: linear-Phi} in \eqref{P3eq: linear-f}, we find
\eq$\label{P3eq: linear-f-y}
f(y)=f_0+\Phi_0^2\left[-ky+\ln(e^{ky}+\xi)\right] .$
As we mentioned earlier, both functions $f(y)$ and $\Phi(y)$ should be real and finite;
therefore $\xi\in(-1,0)\cup(0,\infty)$, and $\Phi_0\in\mathbb{R}\setminus\{0\}$.
It is clear from Eqs. \eqref{P3eq: linear-Phi} and \eqref{P3eq: linear-f-y} that if we allow
$\xi$ to become equal to zero, then we nullify the scalar field everywhere in the
bulk and reduce the coupling function to a constant, which makes our model trivial.
The allowed range of values for the parameter $f_0$ will be determined shortly.

\begin{figure}[t]
\begin{center}
 \begin{subfigure}[b]{0.47\textwidth}
        \includegraphics[width=\textwidth]{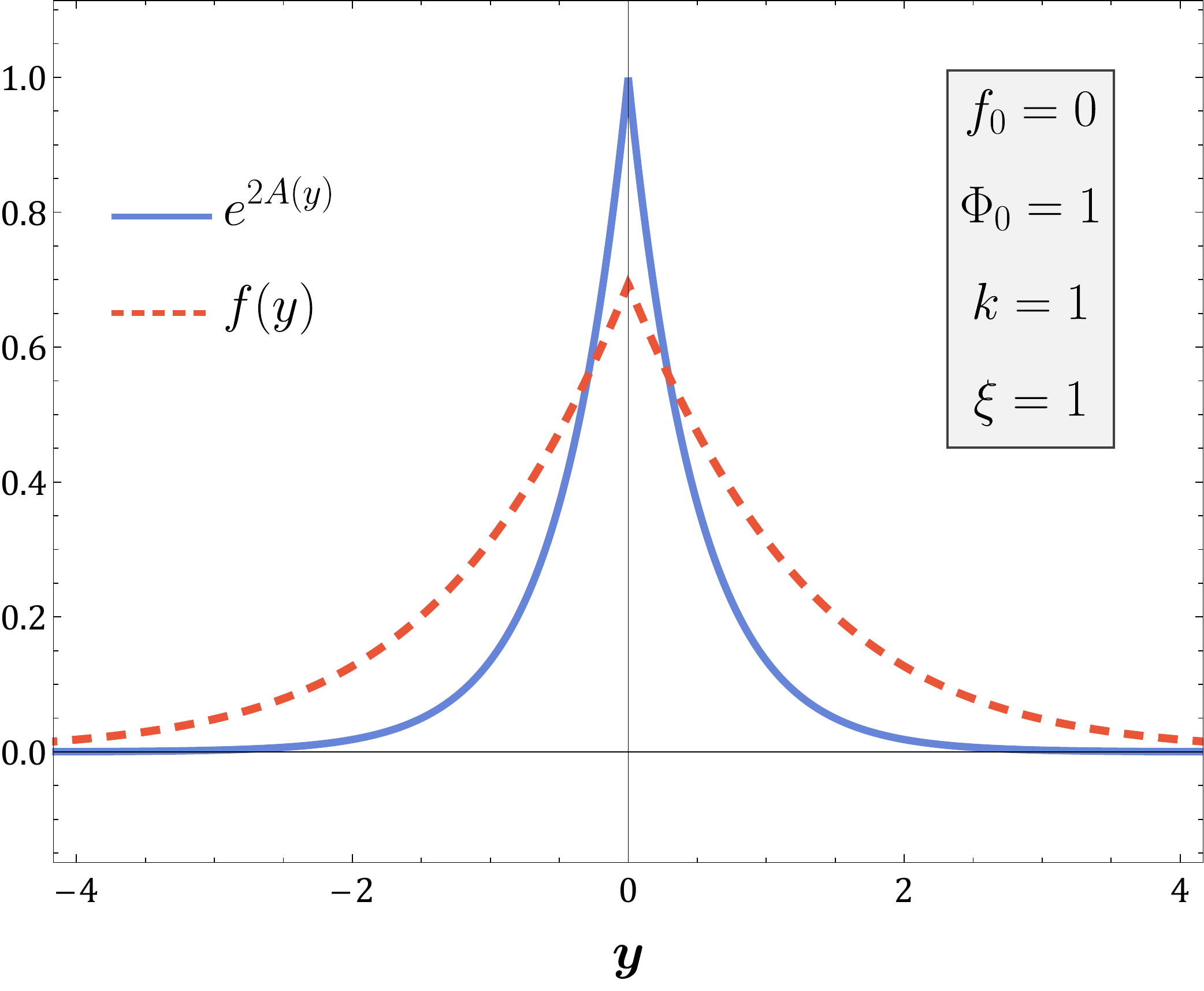}
        \caption{\hspace*{-1.15em}}
        \label{P3subf: linear-plot1}
    \end{subfigure}\quad
    ~ 
 \begin{subfigure}[b]{0.44\textwidth}
        \includegraphics[width=\textwidth]{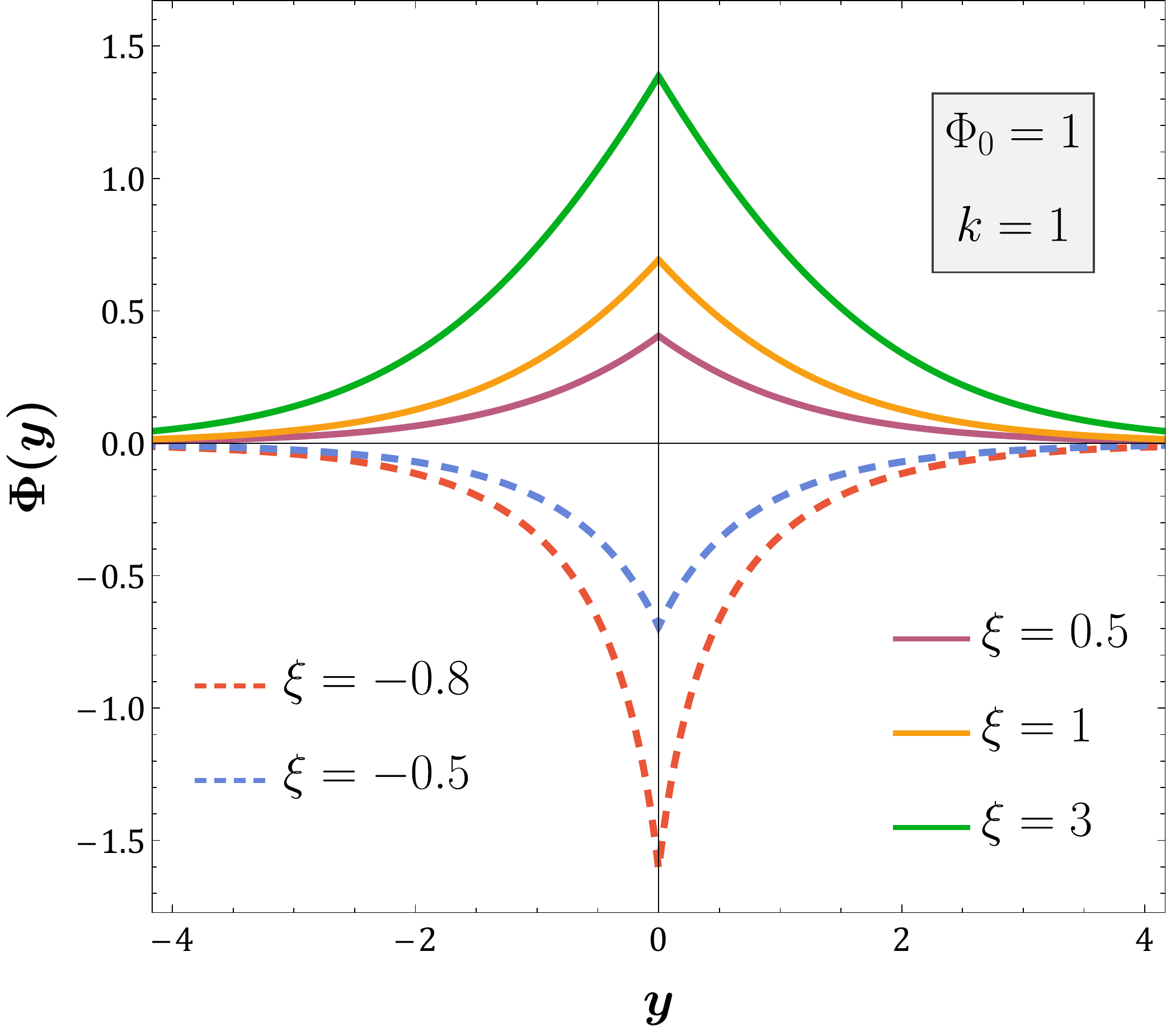}
        \caption{\hspace*{-2.7em}}
        \label{P3subf: linear-plot2}
    \end{subfigure}
    ~ 
    \vspace{-0.3em}
     \caption{(a) The warp factor $e^{2A(y)}=e^{-2k|y|}$ and coupling
    function $f(y)$ in terms of the coordinate $y$ for $f_0=0,\, \Phi_0=1,\,
    k=1,\,\xi=1$, and (b) the scalar field $\Phi(y)$ for
    different values of the parameter $\xi=-0.8,-0.5,0.5,\, 1,\, 3$ (from bottom 
    to top).}
    \vspace{-1em}
   \label{P3fig: linear-plot-1-2}
  \end{center}
\end{figure}

In \myref{P3fig: linear-plot-1-2}{P3subf: linear-plot1}, we depict the warp factor $e^{2A(y)}=e^{-2k|y|}$
and coupling function $f(y)$ in terms of the coordinate $y$ for $f_0=0,\, \Phi_0=1,\,
k=1$, and $\xi=1$. We observe that, similarly to the warp factor, the coupling function
remains localized close to our brane and reduces to zero at large distances although
with a smaller rate. According to this behaviour, the non-minimal coupling of the
scalar field to the five-dimensional Ricci scalar takes its maximum value at the
location of the brane whereas, for large values of $y$, this coupling vanishes leading
to a minimally-coupled scalar-tensor theory of gravity. The profile of the scalar
field $\Phi(y)$ itself is presented in \myref{P3fig: linear-plot-1-2}{P3subf: linear-plot2} for
$\Phi_0=1$ and $k=1$. We also display the dependence of this profile on the
value of the parameter $\xi=-0.8,-0.5,0.5,\, 1,\, 3$ (from bottom to top).
It is clear that also the scalar field exhibits a localized behaviour with
the value of $\xi$ determining the overall sign and maximum value of $\Phi$
on our brane. The dependence of the coupling function $f(y)$ on the value of $\xi$
is similar to that of the scalar field, as one can easily deduce from the relation
\eqref{P3eq: linear-f}. 

\par
\begin{figure}[t]
\begin{center}
\begin{subfigure}[b]{0.47\textwidth}
        \includegraphics[width=\textwidth]{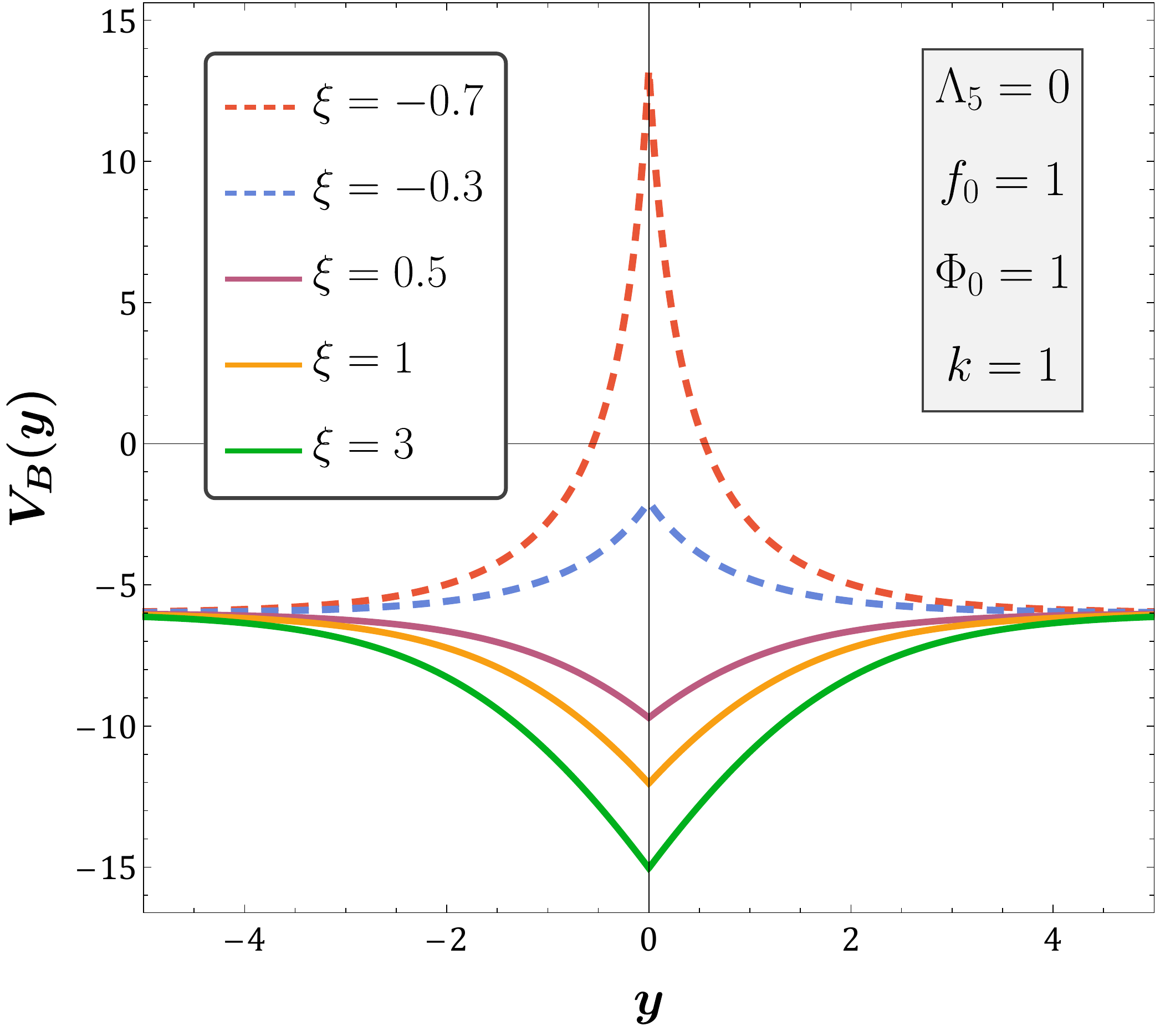}
        \caption{\hspace*{-2.7em}}
        \label{P3subf: linear-plot3}
    \end{subfigure}
    \qquad 
 \begin{subfigure}[b]{0.44\textwidth}
        \includegraphics[width=\textwidth]{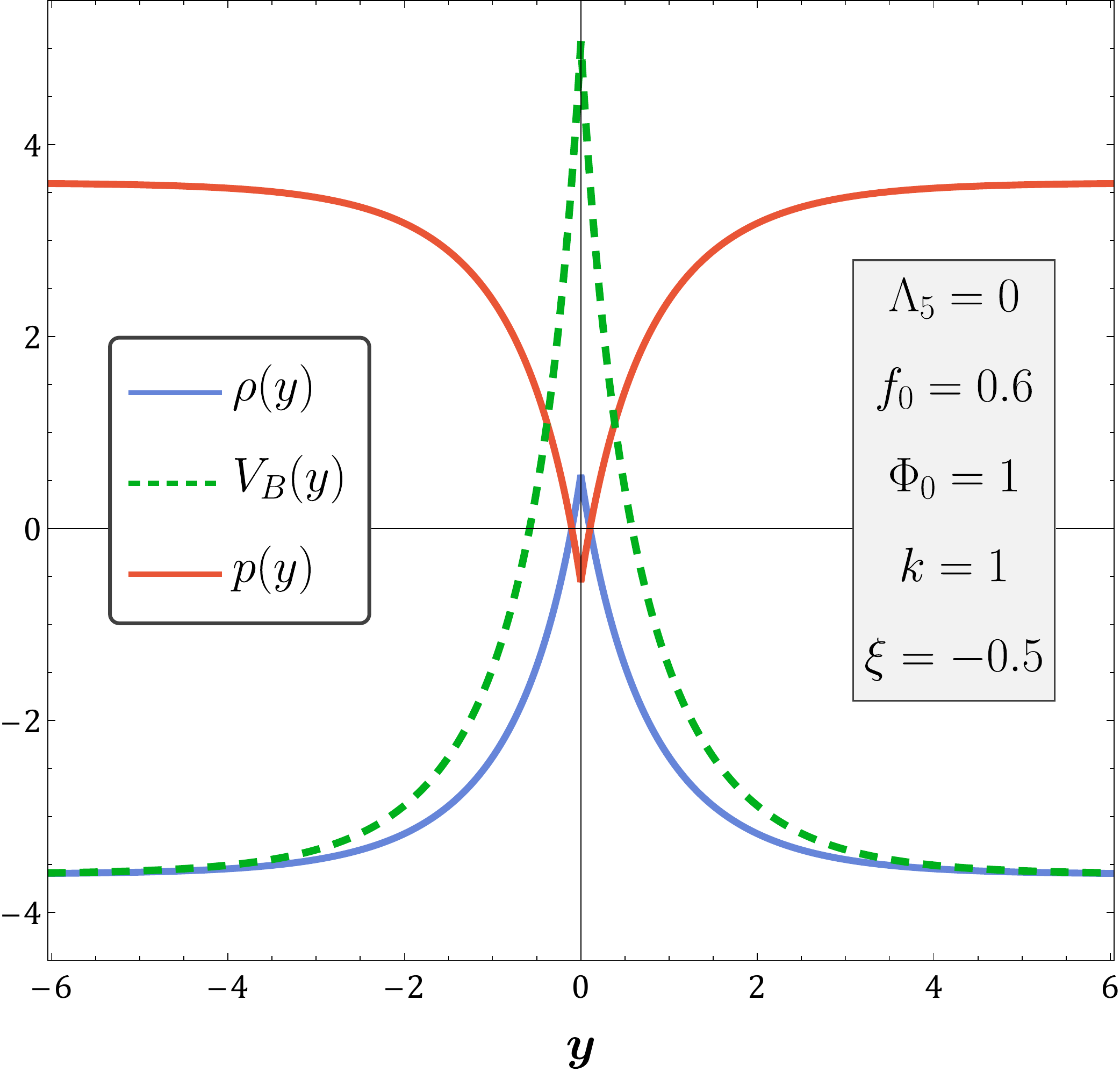}
        \caption{\hspace*{-1.1em}}
        \label{P3subf: linear-plot4}
    \end{subfigure}
    ~ 
    \vspace{-0.5em}
     \caption{(a) The scalar potential $V_B$ in terms of the coordinate $y$ for different
     values of the parameter $\xi=-0.7,\,-0.3,\,0.5,\, 1,\, 3$ (from top to bottom),
     (b) the energy density $\rho$, pressure components 
    $p^i=p^y=p$ and scalar potential $V_B$ in terms of the coordinate $y$ for
    the case $f_0=0.6$ and $\xi=-0.5$.}
    \vspace{-1em}
   \label{P3fig: linear-plot3-4}
  \end{center}
\end{figure}

 The potential of the scalar field $V_B(y)$ in the bulk can be determined from
Eq. \eqref{P3eq: V-B} using the expression of the coupling function $f(y)$ (\ref{P3eq: linear-f-y}).
Thus, we obtain
\eq$\label{P3eq: linear-V-y}
V_B(y)=-\Lambda_5-6k^2f_0+\frac{k^2\Phi_0^2}{2}\left[12ky-\frac{\xi(8e^{ky}+7\xi)}{(\xi+e^{ky})^2}\right]
-6k^2\Phi_0^2\ln(\xi+e^{ky})\, .$
Using Eq. \eqref{P3eq: linear-Phi}, we can express the potential in terms of the scalar
field in a closed form, as follows
\eq$\label{P3eq: linear-V}
V_B(\Phi)=-\Lambda_5-6k^2f_0-6k^2\Phi_0\Phi-4k^2\Phi_0^2\left(1-e^{-\Phi/\Phi_0}\right)
+\frac{k^2\Phi_0^2}{2}\left(1-e^{-\Phi/\Phi_0}\right)^2\,.$
We observe that the parameter $f_0$ appearing in the expression of the coupling
function (\ref{P3eq: linear-f-y}) gives a constant contribution to the scalar bulk
potential. Depending on the value of $f_0$, the asymptotic value of $V_B$ in the
bulk (when $\Phi$ vanishes) can be either positive, zero or negative. In the latter
case, this contribution may be considered to play the role of the negative bulk
cosmological constant $\Lambda_5$, which is usually introduced in an ad hoc way.
Therefore, such a quantity is not necessary any more in order to support
the exponentially decreasing warp factor \'a la Randall-Sundrum.
As mentioned earlier, it is the non-minimal coupling of the scalar field combined
with the form of the bulk potential that supports the AdS bulk spacetime and the
chosen form of the warp factor. To this end, we will henceforth choose a vanishing
value for $\Lambda_5$ in any numerical evaluation, however, for completeness,
we will retain it in our equations. The profile of the bulk potential $V_B$ is presented
in \myref{P3fig: linear-plot3-4}{P3subf: linear-plot3} for $f_0=1$, which leads to a negative
asymptotic value of $V_B$. The figure depicts the dependence of $V_B$ on the
parameter $\xi$. The scalar potential may be negative everywhere in the bulk or
assume a positive value on our brane depending on the value of $\xi$. 

We may also compute the components of the energy-momentum tensor of the theory in the bulk.
Using the relations $\rho=-T^0{}_0$, $p^i=T^i{}_i$, $p^y=T^y{}_y$, we obtain the following
expressions:
\gat$\label{P3eq: linear-rho}
\rho(y)=-\left(T^{(\Phi)0}{}_0-\Lambda_5\right)=-6k^2f(y)\, ,\\[2mm]
\label{P3eq: linear-p}
p^i(y)=T^{(\Phi)i}{}_i-\Lambda_5=6k^2f(y)\, ,\\[3mm]
\label{P3eq: linear-py}
p^y(y)=T^{(\Phi)y}{}_y-\Lambda_5=6k^2f(y)\, .$
The above relations hold in general, for arbitrary form of the coupling function and profile
of the scalar field. From the above expressions, we can immediately observe that the 
energy-momentum tensor in the bulk is isotropic ($p^y=p^i\equiv p$) and satisfies an
equation of state of the form $p=-\rho$. The sign of all energy-momentum tensor
components depends on that of the coupling function. At bulk regimes where $f(y)$
is negative-definite, the energy density $\rho(y)$ will be positive while the pressure
$p(y)$ would have the opposite sign. At these regimes, the weak energy conditions\,\footnote{The
weak energy conditions postulate that $\rho\geq 0,\ \rho+p\geq 0$.} will be satisfied. 
We are primarily interested in satisfying these on and close to our brane.
Thus, if we impose the condition that $f(0)<0$ and combine this inequality with the form of
Eq. \eqref{P3eq: linear-f-y}, we may obtain the range of values for the parameter $f_0$, with respect
to $\xi$  and $\Phi_0$, for which the weak energy conditions on our brane are satisfied.
Hence, we get
\eq$\label{P3eq: linear-weak-con}
\frac{f_0}{\Phi_0^2}<-\ln(1+\xi)\, .$
A particular, indicative case where the weak energy conditions are satisfied on our brane
is depicted in \myref{P3fig: linear-plot3-4}{P3subf: linear-plot4}. It corresponds to the set of values $\Phi_0=1$,
$\xi=-0.5$ and $f_0=0.6$, which satisfy the above inequality. Both the bulk potential and
energy density are positive on our brane while the pressure components assume a negative
value of equal magnitude to that of $\rho$.


\mysubsection{Junction conditions and effective theory}

Let us now address the junction conditions that should be imposed on our bulk solution
due to the presence of the brane at $y=0$. 
Following the same procedure as in the two previous Chapters, and using the complete
field equations (\ref{P1eq: grav-eq1}) and (\ref{P1eq: phi-eq}), we obtain the conditions
\gat$\label{P3eq: jun_con1}
3f(y)[A']=-[\Phi']\,\pa_\Phi f-(\sig+V_b)\, , \\[4mm]
\label{P3eq: jun_con2}
[\Phi']=4[A']\,\pa_\Phi f+\pa_\Phi V_b\, ,$
respectively, where all quantities are evaluated at $y=0$. The above expressions also
hold in general for arbitrary forms of the coupling function $f(\Phi)$. In the case
of a linear $f(\Phi)$, employing the form of the warp function $A(y)=-k |y|$ and the
solution (\ref{P3eq: linear-Phi}) for the scalar field $\Phi(y)$, we obtain the constraints
\gat$\label{P3eq: linear-jc1}
\sig+V_b(\Phi)\Big|_{y=0}=\frac{2k\xi\Phi_0^2}{1+\xi}+6kf_0+6k\Phi_0^2\ln(1+\xi)\,, \\[4mm] 
\label{P3eq: linear-jc2}
\pa_\Phi V_b\Big|_{y=0}=\frac{2k\Phi_0(4+3\xi)}{1+\xi}\,.$
In the above relations, we have used the assumed $\mathbf{Z}_2$ symmetry in the bulk.

The first constraint (\ref{P3eq: linear-jc1}) relates the total energy density of the brane
with bulk parameters. It may be used to fix one of the bulk parameters of our solution,
for example, the warping constant $k$; then, the warping of spacetime is naturally
determined by the distribution of energy in the bulk and on the brane. The second constraint
(\ref{P3eq: linear-jc2}) may in turn be used to fix one parameter of the brane interaction term
$V_b$ of the scalar field. Going further, we may demand that, for physically interesting
situations, the total energy density of the brane should be positive; then, the
r.h.s. of Eq. (\ref{P3eq: linear-jc1})
leads to 
\eq$\label{P3eq: linear-brane-ene-con}
\frac{f_0}{\Phi_0^2}>-\ln(1+\xi)-\frac{\xi}{3(1+\xi)}\, .$
The above is therefore an additional constraint that the bulk parameters ($f_0, \Phi_0, \xi$)
should satisfy which, as the one of Eq. (\ref{P3eq: linear-weak-con}), follows not from the
mathematical consistency of the solution but from strictly physical arguments.

We now turn to the effective theory on the brane that follows by integrating the complete
five-dimensional theory, given by $S=S_B+S_{br}$, over the fifth coordinate $y$. We would
like to derive first the effective four-dimensional gravitational constant that governs
all gravitational interactions on our brane. For this, it is of key importance to express
the five-dimensional Ricci scalar $R$ in terms of the four-dimensional projected-on-the-brane
Ricci scalar $R^{(4)}$. One can easily prove that the five-dimensional Ricci scalar $R$ of
the following line-element
\eq$\label{P3eq: gen-ds2}
ds^2=e^{-2k|y|}g^{(br)}_{\mu\nu}(x)\, dx^\mu dx^\nu+dy^2\,$
can be written in the form
\eq$\label{P3eq: Ricci-5}
R=-20k^2+8k\frac{d^2|y|}{dy^2}+e^{2k|y|}R^{(4)}\, .$
Equation \eqref{P3eq: Ricci-5} holds even if the projected-on-the-brane four-dimensional metric
$g^{(br)}_{\mu\nu}$ leads to a zero four-dimensional Ricci scalar $R^{(4)}$ when the latter
is evaluated for particular solutions (as is the case for our Vaidya induced metric).
The part of the complete action $S=S_B+S_{br}$ that is relevant for the evaluation of the
effective gravitational constant is the following:
\gat$ \label{P3eq: action_eff}
S\supset\int d^4 x\,dy\,\sqrt{-g^{(5)}}\ \frac{f(\Phi)}{2}e^{2k|y|}R^{(4)}\,.$
Then, using also that $\sqrt{-g^{(5)}}=e^{-4k|y|} \sqrt{-g^{(br)}}$, where
$g^{(br)}_{\mu\nu}$ is the metric tensor of the projected on the brane spacetime,
the four-dimensional, effective gravitational constant is given by the integral
\eq$\frac{1}{\kappa_4^2}\equiv 2\,\int_{0}^{\infty} dy\, e^{-2 k y}\,f(y)
=2\,\int_{0}^{\infty} dy\ e^{-2ky}\left[f_0-\Phi_0^2\, ky+\Phi_0^2\ln(e^{ky}+\xi)\right]\,.
\label{P3eq: linear_effG}$
Using the relation $1/\kappa_4^2=M_{Pl}^2/8\pi$ and calculating the above integral,
we obtain the following expression for the effective Planck scale:
\eq$
M_{Pl}^2=\frac{8\pi\Phi_0^2}{k}\left\{\frac{f_0}{\Phi_0^2}-\frac{1}{2}+\frac{1}{\xi^2}\left[\xi+
\left(\xi^2-1\right)\ln(1+\xi)\right]\right\}\,.$
Note, that, due to the localization of both the coupling function and scalar
field close to our brane, no need arises for the introduction of a second brane in
the model. The above value for $M_{Pl}^2$ is therefore finite as demanded, however,
it is not sign-definite. We should therefore demand that the aforementioned expression
is positive-definite which leads to the third, and most, important constraint
on the values of ($f_0, \Phi_0, \xi$), namely 
\eq$\label{P3eq: linear-eff-con}
\frac{f_0}{\Phi_0^2}>\frac{\xi-2}{2\xi}+\frac{1-\xi^2}{\xi^2}\ln(1+\xi)\, .$

The integral of all the remaining terms of the five-dimensional action $S=S_B+S_{br}$,
apart from the one appearing in Eq. (\ref{P3eq: action_eff}), will yield the effective
cosmological constant on the brane. This is due to the fact that the scalar field
$\Phi$ is only $y$-dependent; therefore, when the integration over the extra 
coordinate $y$ is performed, no dynamical degree of freedom remains in the
four-dimensional effective theory. The effective cosmological constant is
thus given by the expression
{\fontsize{11}{11}\bal$\label{P3eq: cosm_eff}
-\Lambda_4&=\int_{-\infty}^{\infty} dy\,e^{-4k|y|}\Bigl[-10 k^2 f(y)-\Lambda_5-
\frac{1}{2}\,\Phi'^2 -V_B(y)+f(y)(-4A'')|_{y=0}-[\sigma +V_b(\Phi)]\,\delta(y)\Bigr]
\nonum\\[1mm]
&=2\int_0^\infty dy\ e^{-4ky}\left[-10 k^2 f(y)-\Lambda_5-
\frac{1}{2}\,\Phi'^2 -V_B(y)\right]+8kf(0)-[\sigma +V_b(\Phi)]_{y=0}\,.$}
\hspace{-0.5em}In the above, we have also added the Gibbons-Hawking term \cite{Gibbons-terms}
due to the presence of the brane, that acts as a boundary for the five-dimensional
spacetime.
Substituting the expressions for the coupling function and the bulk potential of the
scalar field, and employing the junction condition \eqref{P3eq: linear-jc1}, we finally
obtain the result 
\eq$\Lambda_4=0\, .$
As in the previous Chapters for positive (\chapref{Chap: P1}) and negative cosmological constant
(\chapref{Chap: P2}) on the brane, it is clear that the parameter $\Lambda$ appearing in the expression
of the mass function \eqref{P1eq: mass-sol} is indeed related to the four-dimensional cosmological constant
$\Lambda_4$. Therefore, in the context of the present analysis where we have set
$\Lambda=0$, we derived a vanishing $\Lambda_4$ as anticipated.



\mysubsection{The energy conditions in the parameter space}

\par
\begin{SCfigure}[][h]
    \centering
    \includegraphics[width=0.6\textwidth]{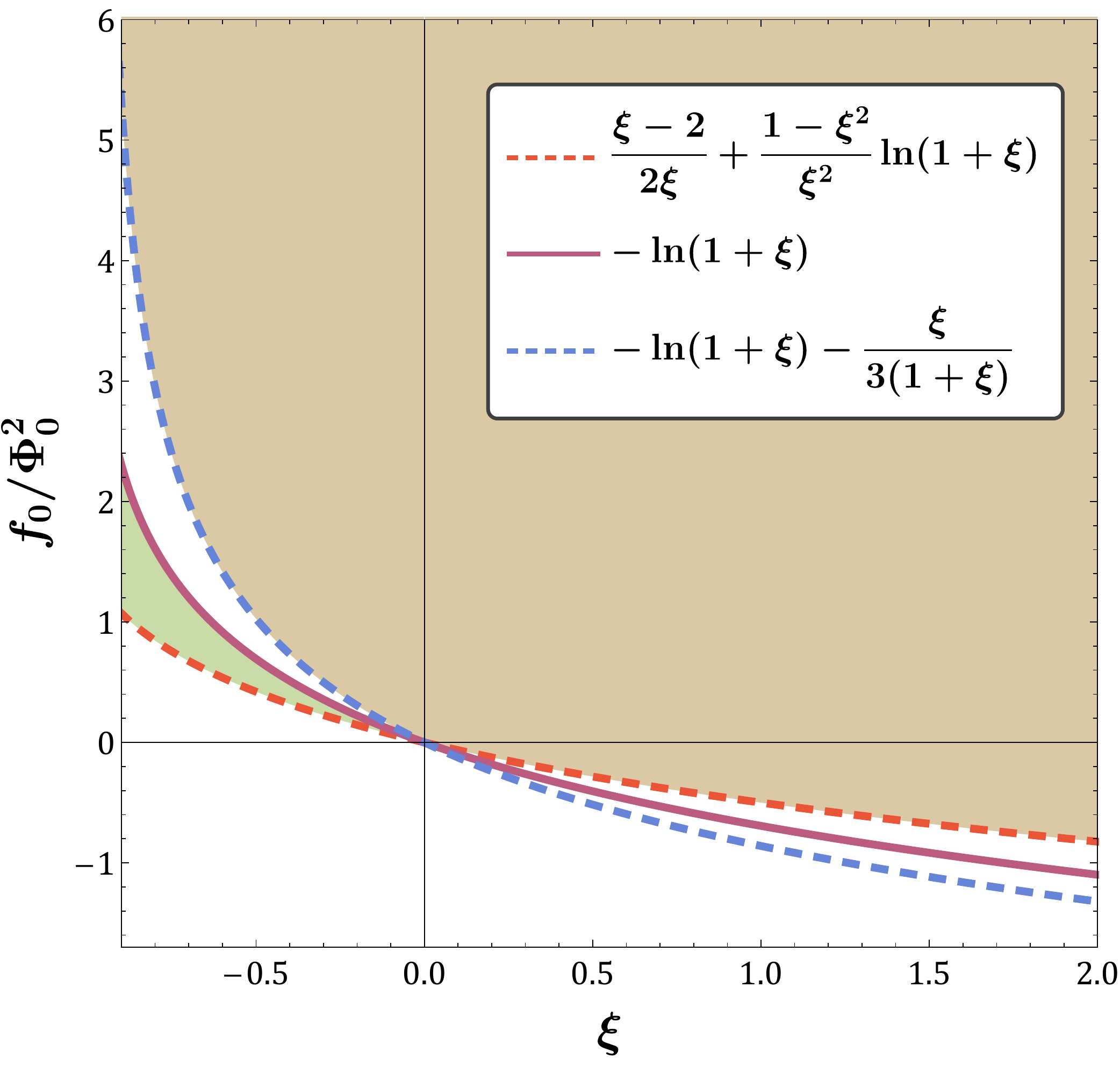}
    \vspace{-0em}
    \caption{The parameter space between the ratio $f_0/\Phi_0^2$ 
    and the parameter $\xi$. The graphs of the expressions of the r.h.s.
    of the inequalities \eqref{P3eq: linear-weak-con}, \eqref{P3eq: linear-brane-ene-con}, 
    \eqref{P3eq: linear-eff-con} are depicted as well.\\
    \vspace{4em}}
    \label{P3fig: linear-plot-par}
\end{SCfigure}

 We will now focus on the inequalities \eqref{P3eq: linear-weak-con}, 
\eqref{P3eq: linear-brane-ene-con}, and \eqref{P3eq: linear-eff-con}, and in particular 
investigate whether it is possible to simultaneously satisfy all three of them.
To this end, we study the parameter space defined by the ratio $f_0/\Phi_0^2$ 
and the parameter $\xi$. This is depicted in \fref{P3fig: linear-plot-par}, where we
have plotted the expressions of the r.h.s.'s of the inequalities \eqref{P3eq: linear-weak-con},
\eqref{P3eq: linear-brane-ene-con}, \eqref{P3eq: linear-eff-con} with respect to the parameter $\xi$.
From a physical point of view, the most important inequality to satisfy is \eqref{P3eq: linear-eff-con},
which ensures that the four-dimensional effective gravitational constant on our brane
is positive: this demands that $f_0/\Phi_0^2$ should be always greater than
$\frac{\xi-2}{2\xi}+\frac{1-\xi^2}{\xi^2}\ln(1+\xi)$ and corresponds to
the area above the red dashed curve in \fref{P3fig: linear-plot-par}.
The inequality \eqref{P3eq: linear-weak-con} ensures that the bulk energy-momentum
tensor satisfies the weak energy conditions at the location of our brane, and demands
that $f_0/\Phi_0^2$ should be smaller than $-\ln(1+\xi)$, this corresponds to the
area below the purple continuous line in \fref{P3fig: linear-plot-par}. Finally,  inequality
\eqref{P3eq: linear-brane-ene-con} expresses the demand that the total energy-density of our brane
is positive; this is satisfied if $f_0/\Phi_0^2$ is greater than $-\ln(1+\xi)-\frac{\xi}{3(1+\xi)}$, this
is the area above the blue dashed curve in \fref{P3fig: linear-plot-par}.

It is straightforward to see that it is impossible to satisfy all three inequalities simultaneously.
However, it is always possible to satisfy two out of these three at a time---in
\fref{P3fig: linear-plot-par}, we have highlighted the regions where the most important
inequality  \eqref{P3eq: linear-eff-con} is one of the two satisfied conditions---we observe that
this area covers a very large part of the parameter space. Which one of the two remaining
inequalities is the second satisfied condition depends on the value of the parameter $\xi$;
therefore, we distinguish the following cases:
\begin{enumerate}
\item[\bf(i)] For $\xi\in(-1,0)$, it is easy to see that the following sequence of
inequalities holds
\eq$\frac{\xi-2}{2\xi}+\frac{1-\xi^2}{\xi^2}\ln(1+\xi)<-\ln(1+\xi)<-\ln(1+\xi)-\frac{\xi}
{3(1+\xi)}\,.$
Thus, we can simultaneously satisfy either the inequalities 
\eqref{P3eq: linear-eff-con} and \eqref{P3eq: linear-weak-con} (green region in 
\fref{P3fig: linear-plot-par}) or \eqref{P3eq: linear-eff-con} and \eqref{P3eq: linear-brane-ene-con} (brown region
in \fref{P3fig: linear-plot-par}). In the former case, we have a physically acceptable four-dimensional
effective theory on the brane and the weak energy-conditions are satisfied on and close to our 
brane; the total energy density of the brane $\sig+V_b(\Phi)|_{y=0}$, however, is negative.
In the latter case, we still have a physically acceptable effective theory and
the total energy density of our brane is now positive; the weak energy conditions though
are not satisfied  by the bulk matter close to our brane. 

\item[\bf(ii)] For $\xi>0$, it now holds
\eq$-\ln(1+\xi)-\frac{\xi}{3(1+\xi)}<-\ln(1+\xi)<\frac{\xi-2}{2\xi}+\frac{1-\xi^2}{\xi^2}
\ln(1+\xi)\, ,$
In this case, we are able to simultaneously satisfy only the inequalities
\eqref{P3eq: linear-eff-con} and \eqref{P3eq: linear-brane-ene-con} (brown region in \fref{P3fig: linear-plot-par}).
Then, we can have a regular four-dimensional effective theory and a positive total 
energy density on our brane. However, in this range of values for the parameter
$\xi$, it is impossible to satisfy the weak energy condition close to our brane
and have a well-behaved effective theory. 

\end{enumerate}

Going back to \myref{P3fig: linear-plot3-4}{P3subf: linear-plot3} and \myref{P3fig: linear-plot3-4}{P3subf: linear-plot4}, we 
observe that the solution depicted in \myref{P3fig: linear-plot3-4}{P3subf: linear-plot4} as well as the solution 
for $\xi=-0.7$ in \myref{P3fig: linear-plot3-4}{P3subf: linear-plot3} fall in the green area of \fref{P3fig: linear-plot-par} 
and thus respect the energy conditions---indeed $V_B$ and $\rho$ are positive on and close to the brane. In contrast, the
remaining solutions of \myref{P3fig: linear-plot3-4}{P3subf: linear-plot3} belong to the brown area of
\fref{P3fig: linear-plot-par}, and thus violate the weak energy conditions; they have,
however, a positive total energy-density through the junction condition \eqref{P3eq: linear-jc1}.
We stress that all depicted solutions have a well-defined four-dimensional effective theory,
i.e. a positive effective gravitational constant. 

%
%

\mysection{Quadratic coupling function \label{P3sec: quad}}

In this section, we proceed to consider the case of the quadratic coupling function, and
thus we write
\eq$\label{P3eq: quad-f}
f(\Phi)=f_0+\Phi_0 \Phi+\lam \Phi^2\,,$
where again ($f_0$, $\Phi_0$, $\lambda$) are arbitrary parameters. Throughout this section,
it will be assumed that $\lambda \neq 0$
otherwise the analysis reduces to the one of the linear case studied in the previous section.
As before, we start with the derivation of the bulk solution and then turn to the effective
theory on the brane. 



\mysubsection{The bulk solution and the effective theory on the brane}

\par Substituting the aforementioned form of the coupling function in Eq. \eqref{P3eq: grav-1-1} we obtain
the equation:
\eq$(1+2\lam)\Phi'^2+(2\lam\Phi+\Phi_0)(\Phi''+k\Phi')=0\, .$
Integrating, we find the following solution for the scalar field
\eq$\label{P3eq: quad-Phi-gen}
\Phi(y)=\left\{\begin{array}{ll}
\displaystyle{\frac{1}{2\lam}\left[\Phi_1(\mu+e^{-ky})^\frac{2\lam}{1+4\lam}-\Phi_0\right]}, & 
\lam\in\mathbb{R}\setminus\{-\frac{1}{4},0\}\\[4mm]
2\Phi_0+\Phi_1\,e^{\,\mu\, e^{-ky}},& \lam=-\frac{1}{4}\end{array}\right\}\,,$
where $\mu$ and $\Phi_1$ are integration constants. We note that the case with $\Phi_0=0$
was studied in \cite{Bogdanos1}; here, we generalize the aforementioned analysis by assuming
that $\Phi_0 \neq 0$. We also perform a more comprehensive analysis of the ensuing solutions
by studying the different profiles of the coupling function, scalar field and bulk potential, which
emerge as the values of the parameters of the model vary. In addition, we supplement our
analysis with the study of the effective theory on the brane and of the physical constraints
imposed on the solutions. In order to simplify our notation, we set $\Phi_1=\xi\Phi_0$, where
$\xi$ is a new integration constant. Then, Eq. \eqref{P3eq: quad-Phi-gen} is written as
\eq$\label{P3eq: quad-Phi}
\Phi(y)=\left\{\begin{array}{ll}
\displaystyle{\frac{\Phi_0}{2\lam}\left[\xi(\mu+e^{-ky})^\frac{2\lam}{1+4\lam}-1\right]}, &
\lam\in\mathbb{R}\setminus\{-\frac{1}{4},0\}\\[4mm]
\Phi_0\left(2+\xi\,e^{\,\mu\, e^{-ky}}\right),& \lam=-\frac{1}{4}\end{array}\right\}\,.$
Substituting the above expression in Eq. \eqref{P3eq: quad-f}, we obtain the following profile for
the coupling function in terms of the extra coordinate
\gat$\label{P3eq: quad-f-y}
f(y)=\left\{\begin{array}{ll}
\displaystyle{f_0+\frac{\Phi_0^2}{4\lam}\left[\xi^2(\mu+e^{-ky})^\frac{4\lam}{1+4\lam}-1\right]}, &
\lam\in\mathbb{R}\setminus\{-\frac{1}{4},0\}\\[4mm]
\displaystyle{f_0+\Phi_0^2\left(1-\frac{\xi^2}{4}\,e^{2\mu\, e^{-ky}}\right)},& \lam=-\frac{1}{4}
\end{array}\right\}\,.$

\par The theory seems to contain five independent parameters: $f_0$, $\Phi_0$, $\lambda$,
$\mu$ and $\xi$. However, the range of values for two of these will be constrained by the
physical demands imposed on the model. To start with, both the scalar field $\Phi(y)$ and the coupling
function $f(y)$ must be real and finite in their whole domain, according to the discussion in \secref{P3sec: th-frame}.
From Eq. \eqref{P3eq: quad-Phi}, we observe that the allowed range of values of the parameter
$\mu$ depends on the values that the parameter $\lambda$ assumes. In Appendix
\ref{P3app: mu}, we consider in detail all possible values for $\lambda$ and the ensuing
allowed ranges of values for $\mu$---the different cases and corresponding results are
summarised in Table \ref{P3tab: quad-par-val} (next page).\,\footnote{The symbol $\land$ that was used in Table \ref{P3tab: quad-par-val}
simply denotes the \textit{logical and}. For example, the statement $A\land B$ is true if $A$ and $B$ are both
true; else it is false.}
\newpage
\begin{center} 
\begin{table}[t]
\centering
\scalebox{0.938}{\begin{tabular}{ |c|c|c| } 
\hline
\multicolumn{3}{|c|}{\textbf{Range of values for all parameters}}\\
\hline
\multirow{20}{10em}{\begin{center}\vspace{-3.5em}$\xi\in\mathbb{R}\setminus\{0\}$,\\ \vspace{1em}$\Phi_0
\in\mathbb{R}\setminus\{0\}$,\\ \vspace{1em}$f_0$ is given by Eq. \eqref{P3eq: quad-eff-con}.
\end{center}}
& \multirow{2}{7em}{$\hspace{2.5em}\lam>0$} & \multirow{2}{5em}{\hspace{1em}$\mu\geq 0$}\\
&  & \\ \cline{2-3}
& \multirow{4}{15em}{\hspace{0.5em}\small{$\displaystyle{\lam\in\left(-\frac{1}{4},0
\right)\ \land\ \frac{2\lam}{1+4\lam}\neq n,\  n\in\mathbb{Z}^<}$}}
& \multirow{4}{5em}{\hspace{1em}$\mu> 0$} \\
& & \\
& & \\
& & \\ \cline{2-3}
& \multirow{4}{15em}{\hspace*{0.5em}\small{$\displaystyle{\lam\in\left(-\frac{1}{4},0\right)\ \land\ 
\frac{2\lam}{1+4\lam}=n,\  n\in\mathbb{Z}^<}$}}
& \multirow{4}{11em}{\hspace{0.2em}$\mu\in(-\infty,-1)\cup(0,+\infty)$} \\
& &\\
& & \\
& & \\ \cline{2-3}
& \multirow{2}{7em}{$\hspace{1.7em}\displaystyle{\lam=-1/4}$} 
& \multirow{2}{10em}{\vspace{0em}$\mu\in(-\infty,0)\cup(0,+\infty)$}\\
&  & \\ \cline{2-3}
& \multirow{3}{17em}{\hspace{2.9em}\small{$\displaystyle{\lam<-\frac{1}{4}\ \land\ \frac{2\lam}
{1+4\lam}\neq n,\  n\in\mathbb{Z}^>}$}}
& \multirow{3}{3em}{\hspace{0.2em}$\mu\geq 0$} \\
& &\\
& & \\ \cline{2-3}
& \multirow{3}{17em}{\hspace{2.9em}\small{$\displaystyle{\lam<-\frac{1}{4}\ \land\ \frac{2\lam}
{1+4\lam}= n,\  n\in\mathbb{Z}^>}$}}
& \multirow{3}{3em}{\hspace{0.2em}$\mu\in\mathbb{R}$} \\
& &\\
& & \\ \cline{2-3}
\hline
\end{tabular}}
\caption{Range of values for all parameters of the model.}
\label{P3tab: quad-par-val}
\end{table}
\end{center}


In addition, from the analysis of the previous section, it became clear that the theory is not
robust unless a positive effective gravitational constant is obtained on the brane. This
demand will impose a constraint on one of the remaining parameters of the theory, we
choose this parameter to be $f_0$. Thus, in order to appropriately choose the values
of $f_0$ to study the profile of the scalar field and coupling function, at this point we
turn to the effective theory and compute the effective gravitational constant. We will
employ Eq. \eqref{P3eq: action_eff}, and consider separately the cases with $\lam\neq-1/4$
and $\lam=-1/4$. In the first case, using also Eq. \eqref{P3eq: quad-f-y}, we obtain
\bal$\frac{1}{\kappa_4^2}=2 \int_{0}^{\infty} dy\, e^{-2 k y}\,f(y)=
\frac{1}{k}\left(f_0-\frac{\Phi_0^2}{4\lam}\right)+\frac{\Phi_0^2\xi^2}{2\lam}
\int_{0}^{\infty} dy\ e^{-2ky}\left(\mu+e^{-ky}\right)^\frac{4\lam}{1+4\lam}\, .$
In order to evaluate the integral on the r.h.s. of the above equation, we perform the change
of variable $t=e^{-ky}$. 
If we also use the integral representation of the hypergeometric function \cite{Abramowitz}
{\fontsize{11}{11}\eq$\label{P3eq: int-rep-hyper}
\,_2F_1\left(a,b;c;z\right)=\frac{\Gamma(c)}{\Gamma(b)\Gamma(c-b)}\int_0^1
dt\ t^{b-1}(1-t)^{c-b-1}(1-zt)^{-a},\hspace{1.5em}Re(c)>Re(b)>0\, ,$
we finally obtain the result
\eq$\frac{1}{\kappa_4^2}=\frac{M_{Pl}^2}{8\pi}=\frac{\Phi_0^2}{k}
\left[\frac{f_0}{\Phi_0^2}-\frac{1}{4\lam}+\frac{\xi^2\,\mu^{
\frac{4\lam}{1+4\lam}}}{4\lam}\,_2F_1\left(-\frac{4\lam}{1+4\lam},2;3;-\frac{1}{\mu}
\right)\right]\, .$}

We can further simplify the above expression using the following relations
\eq$\,_2F_1\left(a,2;3;z\right)=\left\{\begin{array}{ll}
\displaystyle{\frac{2(1-z)^{-a}\left[z(a+z-az)+(1-z)^a-1\right]}{(a-2)(a-1)z^2}}\, , & a\in
\mathbb{R}\setminus\{1,2\}\\[3mm]
\displaystyle{\frac{2}{z^2}\left[-z-\ln(1-z)\right]}\, , & a=1\\[3mm]
\displaystyle{\frac{2}{z^2(1-z)}\left[z+\ln(1-z)-z\ln(1-z)\right]}\, , &a=2
\end{array}\right\}\, .$
Then, for $\lam\neq-1/4$, the four-dimensional effective Planck scale may be written in
terms of elementary functions as follows
{\fontsize{10}{10}\eq$
M_{Pl}^2=\left\{\begin{array}{ll}
\frac{8\pi\Phi_0^2}{k}\left\{\frac{f_0}{\Phi_0^2}-\frac{1}{4\lam}+\frac{(1+
4\lam)^2\xi^2}{4\lam(1+6\lam)(1+8\lam)}\left[\mu^{\frac{2+12\lam}{1+4\lam}}-
(1+\mu)^{\frac{1+8\lam}{1+4\lam}}\left(\mu-\frac{1+8\lam}{1+4\lam}\right)
\right]\right\}, &\lam\in\mathbb{R}\setminus \mathcal{S}\\[4mm]
\frac{8\pi\Phi_0^2}{k}\left\{\frac{f_0}{\Phi_0^2}+2-4\xi^2
\left[1-\mu\ln\left(\frac{1+\mu}{\mu}\right)
\right]\right\}, & \lam=-\frac{1}{8}\\[4mm]
\frac{8\pi\Phi_0^2}{k}\left\{\frac{f_0}{\Phi_0^2}+\frac{3}{2}-\frac{3\xi^2
}{1+\mu}\left[-1+(1+\mu)\ln\left(\frac{1+\mu}
{\mu}\right)\right]\right\}, & \lam=-\frac{1}{6}\end{array}\right\}. \label{P3eq: MPl-quad}$}
\hspace{-0.5em}where the set $\mathcal{S}=\{-\frac{1}{4},-\frac{1}{6},-\frac{1}{8},0\}$. 
On the other hand, for $\lam=-1/4$, we readily obtain
\gat$M_{Pl}^2=\frac{8\pi}{\kappa_4^2}=\frac{8\pi\Phi_0^2}{k}\left\{\frac{f_0}{\Phi_0^2}+1-
\frac{\xi^2}{8\mu^2}\left[1+e^{2\mu}(2\mu-1)
\right]\right\}\,. \label{P3eq: Mpl-1/4}$

Since the effective four-dimensional gravitational scale $M_{Pl}^2$ should be
a positive number, Eqs. \eqref{P3eq: MPl-quad} and \eqref{P3eq: Mpl-1/4} impose the following constraints
on the values of the ratio $f_0/\Phi_0^2$:
\eq$\label{P3eq: quad-eff-con}
\left\{\begin{array}{ll}
\frac{f_0}{\Phi_0^2}>\frac{1}{4\lam}\left\{1-\frac{(1+
4\lam)^2\xi^2}{(1+6\lam)(1+8\lam)}\left[\mu^{\frac{2+12\lam}{1+4\lam}}-
(1+\mu)^{\frac{1+8\lam}{1+4\lam}}\left(\mu-\frac{1+8\lam}{1+4\lam}\right)
\right]\right\}, &\lam\in\mathbb{R}\setminus\mathcal{S}\\[5mm]
\frac{f_0}{\Phi_0^2}>-2\left\{1-2\xi^2
\left[1-\mu\ln\left(\frac{1+\mu}{\mu}\right)
\right]\right\}, & \lam=-\frac{1}{8}\\[5mm]
\frac{f_0}{\Phi_0^2}>-\frac{3}{2}\left\{1-\frac{2\xi^2
}{1+\mu}\left[-1+(1+\mu)\ln\left(\frac{1+\mu}
{\mu}\right)\right]\right\}, & \lam=-\frac{1}{6}\\[5mm]
\frac{f_0}{\Phi_0^2}>-1+\frac{\xi^2}{8\mu^2}\left[
1+e^{2\mu}(2\mu-1)\right], & \lam=-\frac{1}{4}\end{array}\right\}\,.$
We choose to use the above constraints in order to limit the range of values of the
parameter $f_0$. The remaining parameters $\Phi_0$, $\lam$ and $\xi$ may then take
values in almost the entire set of real numbers, specifically $\Phi_0\in\mathbb{R}\setminus\{0\}$, 
$\xi\in\mathbb{R}\setminus\{0\}$ and $\lam\in\mathbb{R}\setminus\{0\}$.  These ranges of
values are also summarised in Table \ref{P3tab: quad-par-val}. We finally note that the above constraints
for the positivity of the effective four-dimensional gravitational constant allow for both positive
and negative values of the parameter $f_0$.

\begin{figure}[t!]
    \centering
    \begin{subfigure}[b]{0.47\textwidth}
        \includegraphics[width=\textwidth]{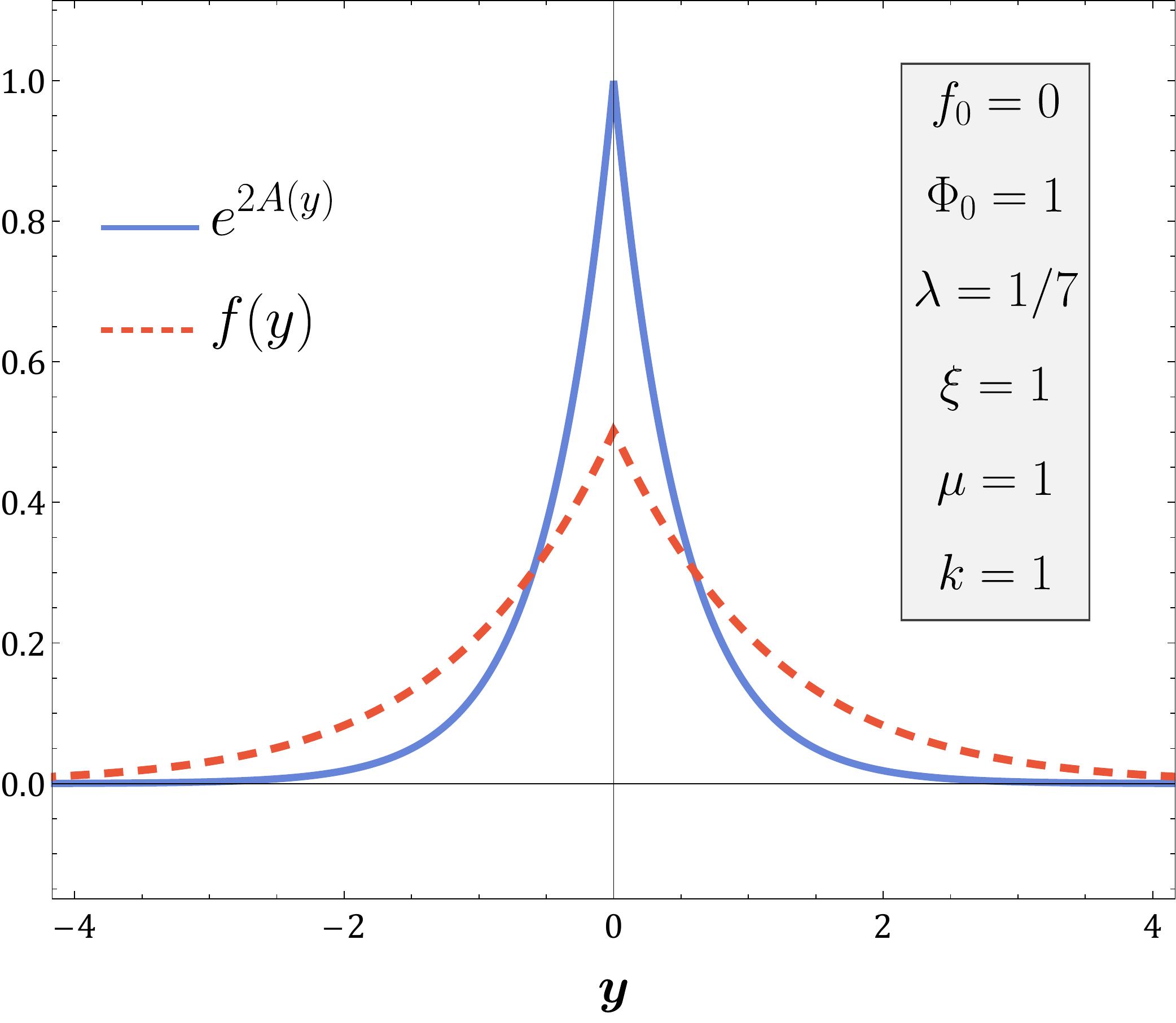}
        \caption{\hspace*{-1.2em}}
        \label{P3subf: quad-plot1}
    \end{subfigure}\quad
    ~ 
    \begin{subfigure}[b]{0.44\textwidth}
        \includegraphics[width=\textwidth]{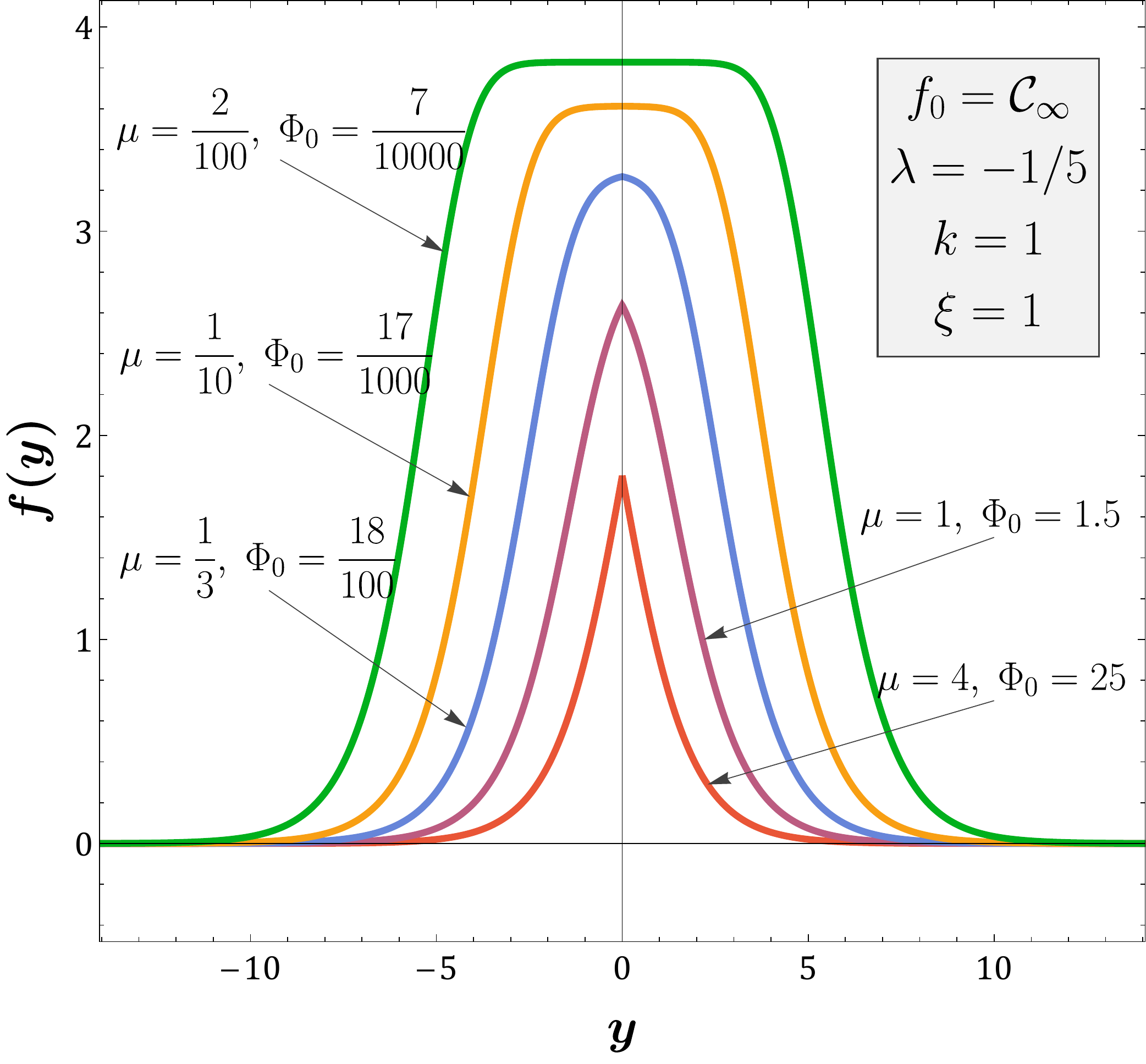}
        \caption{\hspace*{-2em}}
        \label{P3subf: quad-plot2}
    \end{subfigure}
    ~ 
    \vspace{-0.5em}
    \caption{(a) The warp factor $e^{2A(y)}=e^{-2k|y|}$ and coupling function $f(y)$ in terms 
    of the coordinate $y$ for $\lam=1/7$, and (b) the coupling function for $\lam=-1/5$ 
    in the regime $(-1/4,0)$ which satisfies $\frac{2\lam}{1+4\lam}=-2$ and different values 
    of parameters $\mu$ and $\Phi_0$. In
    figure (b), $\cinf$ indicates the value that the parameter $f_0$ should have in
    order to get a vanishing coupling function $f(y)$ at $y\ra\infty$.}
   \label{P3fig: quad-plot-1-2}
\end{figure}

\par We now proceed to study the profile of our solution.
In \myref{P3fig: quad-plot-1-2}{P3subf: quad-plot1}, we depict the form of the warp factor 
$e^{-2k|y|}$ and the coupling function $f(y)$ in terms of the coordinate $y$ along 
the fifth dimension, for $\Phi_0=1$, $\lam=1/7$, $\xi=1$, $\mu=1$, $k=1$ and $f_0=0$, 
which, as one can verify, is allowed by Eq. \eqref{P3eq: quad-eff-con}. The warp factor is
always localized close to the brane and vanishes at the boundary of spacetime 
independently of the values of the parameters. The behaviour of the coupling function 
though depends strongly on the values of the parameters of the model. For
$\lam>0$ and $\mu\geq 0$, the qualitative behaviour of the coupling function is the
same as the one that is  illustrated in \myref{P3fig: quad-plot-1-2}{P3subf: quad-plot1}. In 
\myref{P3fig: quad-plot-1-2}{P3subf: quad-plot2}, we present the behaviour of the coupling function for 
various values of the parameters $\mu$ and $\Phi_0$ while $\lam$ is now in the regime
$(-1/4,0)$. We note that, for generic values of the parameters, the
asymptotic value of $f(y)$, as $y \rightarrow \infty$, is not zero. if so desired, one may
choose $f_0$ to be equal to $\cinf$, which indicates the value that $f_0$ should have
in order to get a vanishing coupling function at infinity; from Eq. \eqref{P3eq: quad-f-y}, we can
immediately calculate that $\mathcal{C}_\infty=\frac{\Phi_0^2}{4\lam}\left(1-\xi^2\,
\mu^{\frac{4\lam}{1+4\lam}}\right)$. In \myref{P3fig: quad-plot-1-2}{P3subf: quad-plot2},
one can clearly see the strong dependence of the profile of the coupling function also
on the value of the parameter $\mu$. As $\mu$ approaches zero, the coupling function is 
characterized by a plateau around our brane; the closer the value of $\mu$ is to zero, the 
wider the plateau. On the contrary, $\Phi_0$ does not significantly affect the behaviour of
$f(y)$; it just scales the function as a whole.  The behaviour depicted in
\myref{P3fig: quad-plot-1-2}{P3subf: quad-plot2} holds for all values of $\lam$ in the regime 
$(-\frac{1}{4},0)$ as long as $\mu>0$. A different behaviour appears in the case where
$\frac{2 \lam}{1+4\lam}=-2n$, $n\in\mathbb{Z}^>$ and $\mu<-1$; in this case the behaviour 
of the coupling function is exactly the same as the one for $\lam<-\frac{1}{4}$ and $\frac{2\lam}
{1+4\lam}\neq n$, with $ n\in\mathbb{Z}^>$, which will be discussed next.

\par
\begin{figure}[t]
    \centering
    \begin{subfigure}[b]{0.477\textwidth}
        \centering
        \includegraphics[height=0.289\textheight]{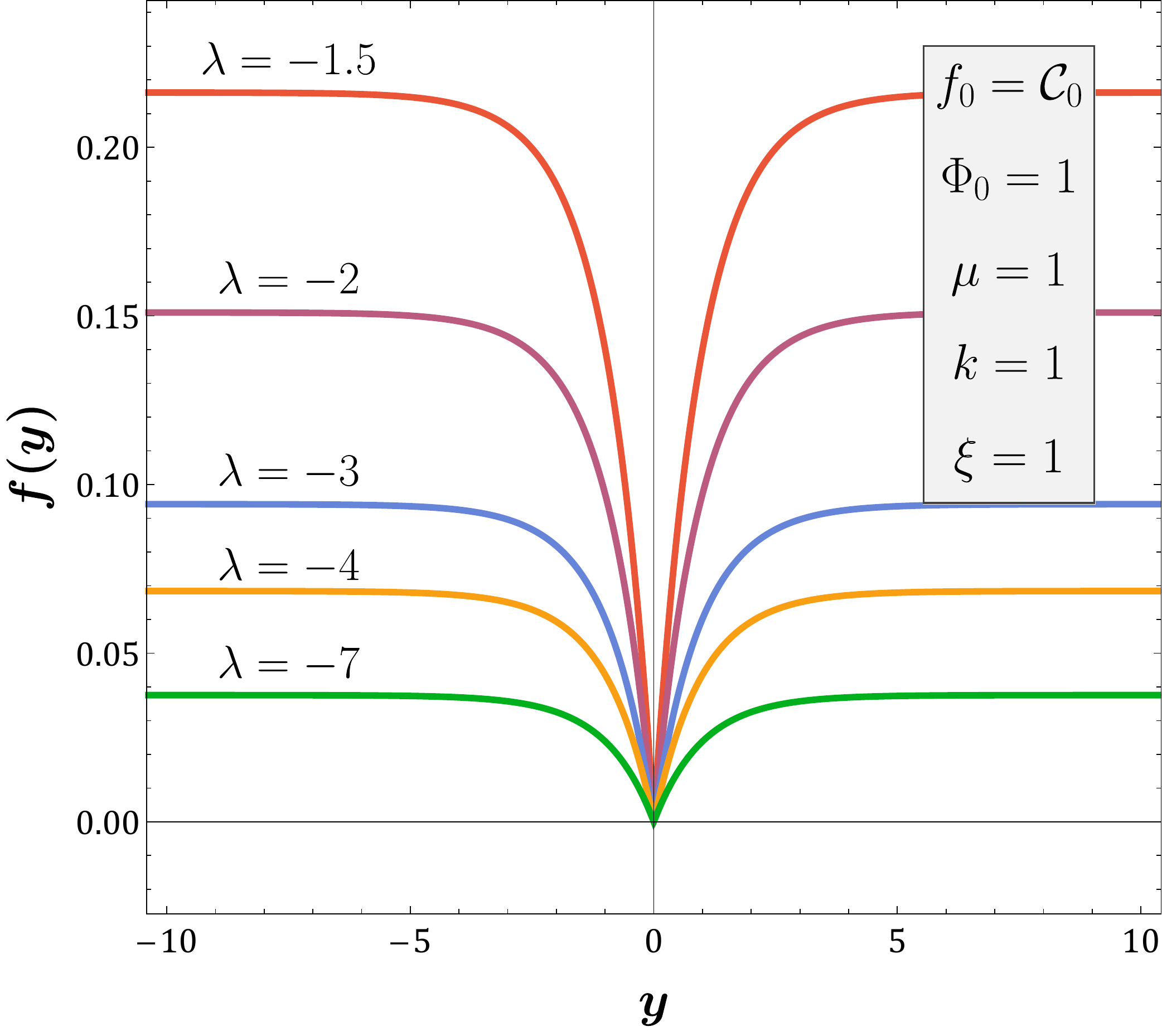}
        \caption{\hspace*{-2.7em}}
        \label{P3subf: quad-plot3}
    \end{subfigure}\hfill
    ~ 
    \begin{subfigure}[b]{0.487\textwidth}
        \includegraphics[height=0.289\textheight]{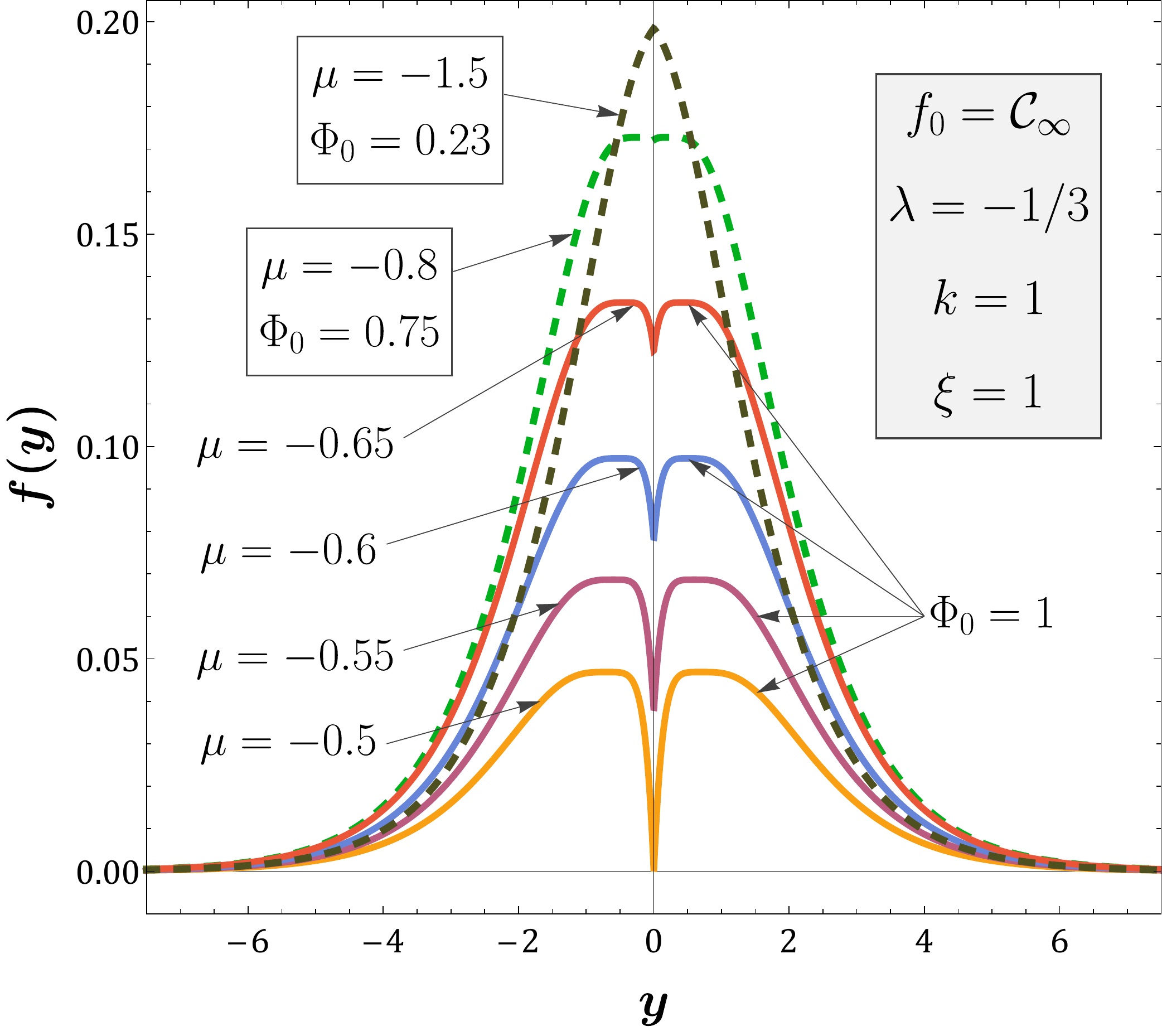}
        \caption{\hspace*{-1.1em}}
        \label{P3subf: quad-plot5}
    \end{subfigure}
    ~ 
    \vspace{-1.8em}
    \caption{(a) The coupling function $f(y)$ in terms of the coordinate $y$ for $\Phi_0=1$, 
    $\mu=1$, $k=1$, $\xi=1$, and values of $\lam$ smaller than $-\frac{1}{4}$ with
    $\frac{2\lam}{1+4\lam}\neq n$, $n\in\mathbb{Z}^>$. (b) The coupling function $f(y)$
    for $\lam=-1/3$, which satisfies  $\frac{2\lam}{1+4\lam}=2$, $k=1$, $\xi=1$ and
    $f_0=\cinf$, while $\mu$ takes values equal or lower than  $-1/2$.}
     \label{P3fig: quad-plot-3-5}
\end{figure}

 In \myref{P3fig: quad-plot-3-5}{P3subf: quad-plot3}, we display the behaviour of the coupling function
$f(y)$ for $\Phi_0=1$, $\mu=1$, $k=1$, $\xi=1$ and values of $\lam$ in the regime $\lam < -1/4$. 
Since it holds that $\frac{2\lam}{1+4\lam} \neq n$, with $n\in\mathbb{Z}^>$, the parameter $\mu$
is constrained to values greater than or equal to zero. For easy comparison, the parameter $f_0$ has
been taken to be equal to $\mathcal{C}_0$, which is the value that leads to $f(0)=0$; again, from
Eq. \eqref{P3eq: quad-f-y}, we find that 
$\mathcal{C}_0=\frac{\Phi_0^2}{4\lam}\left[1-\xi^2(\mu+1)^{\frac{4\lam}{1+4\lam}}\right]$.
In such a model, the non-minimal coupling of the scalar field to the five-dimensional scalar
curvature is non-vanishing in the bulk but disappears at the location of the brane.
In this range of values for the parameter $\lam$, the behaviour of the coupling function, as
depicted in \myref{P3fig: quad-plot-3-5}{P3subf: quad-plot3}, does not change regardless of the values
of all the other parameters. In contrast, when $\lam$ satisfies the condition $\frac{2\lam}{1+4\lam}=n$,
the profile of the coupling function is extremely sensitive to changes in the parameter $\mu$. 
Indeed, \myref{P3fig: quad-plot-3-5}{P3subf: quad-plot5} shows the behaviour of the coupling function $f(y)$
for $k=1$, $\xi=1$ and $f_0=\mathcal{C}_\infty$, while $\lam=-1/3$ or $\frac{2\lam}{1+
4\lam}=2$. In this figure, we focus on values of $\mu$ that are smaller than or equal to $-1/2$. 
We observe that, as $\mu$ approaches and exceeds $-1$, the behaviour of the coupling function
becomes similar to the one in \myref{P3fig: quad-plot-1-2}{P3subf: quad-plot1} 
and \myref{P3fig: quad-plot-1-2}{P3subf: quad-plot2}.  Here, we have chosen again $f_0=\mathcal{C}_\infty$,
therefore, the non-minimal coupling takes its maximum value on or close to our brane while it
vanishes at infinity. On the other hand, as $\mu$ approaches zero and takes on positive values,
the profile of $f(y)$ resembles more the one depicted in \myref{P3fig: quad-plot-3-5}{P3subf: quad-plot3}.

\par Finally, in the case where $\lam=-1/4$, the coupling function, as presented in 
Eq. \eqref{P3eq: quad-f-y}, is given by a double exponential expression. It is not hard 
to realize that the qualitative behaviour of $f(y)$ in this case is similar to 
the one in \myref{P3fig: quad-plot-1-2}{P3subf: quad-plot2} when $\mu<0$ and similar to
\myref{P3fig: quad-plot-3-5}{P3subf: quad-plot3} when $\mu>0$. It is also necessary to stress that the 
behaviour of the scalar field $\Phi(y)$ is similar to that of the coupling function $f(y)$, 
as one may easily conclude by observing Eqs. \eqref{P3eq: quad-Phi} and \eqref{P3eq: quad-f-y}.
Therefore, it is redundant to present any graphs of the scalar field as a function of the 
$y$-coordinate. 

\par The scalar potential $V_B$ can be determined in terms of the extra dimension $y$ from Eq.
\eqref{P3eq: V-B} by substituting the function $f(y)$ given in Eq. \eqref{P3eq: quad-f-y}. Consequently, we obtain
{\fontsize{10}{10}\eq$\label{P3eq: quad-V-y}
V_B(y)=\left\{\begin{array}{ll}
\displaystyle{-\Lambda_5-6k^2f_0+\frac{3k^2\Phi_0^2}{2\lam}-\frac{\xi^2 k^2\Phi_0^2\left(
\mu+e^{-ky}\right)^{-\frac{2+4\lam}{1+4\lam}}}{6\lam(1+4\lam)^2}\times} & \\[4mm]
\hspace{1cm}\times\left\{\left[e^{-ky}(3+16\lam)+3(1+4\lam)\mu\right]^2-\lam(3+16\lam)
e^{-2ky}\right\}\, , & \lam\in\mathbb{R}\setminus\{-\frac{1}{4},0\} \\[2em]
\displaystyle{-\Lambda_5-6k^2(f_0+\Phi_0^2)+\frac{\xi^2\Phi_0^2k^2}{2}\, e^{2\mu\,e^{-ky}}
\left(3+4\mu e^{-ky}+\mu^2e^{-2ky}\right)},&\lam=-\frac{1}{4}\end{array}\right\}.$}
\hspace{-0.5em}As in the linear case, it is possible to express the potential in terms of the scalar field $\Phi$
in closed form and obtain:
{\fontsize{9}{9}\eq$\label{P3eq: quad-V-Phi}
V_B(\Phi)=\left\{\begin{array}{ll}
-\Lambda_5-6k^2f_0+\frac{3k^2\Phi_0^2}{2\lam}+\frac{\xi^2 k^2\Phi_0^2
}{6\lam(1+4\lam)^2}\left(\frac{2\lam\Phi}{\xi\Phi_0}+\frac{1}{\xi}\right)^{-\frac{1+2
\lam}{\lam}}\left\{3\mu^2\lam\right.  & \\[3mm]
\hspace{0em}\left.+6\lam(3+16\lam)\mu\left(\frac{2\lam\Phi}{\xi\Phi_0}+\frac{1}{\xi}
\right)^{\frac{1+4\lam}{2\lam}}-(3+16\lam)(3+15\lam)\left(\frac{2\lam\Phi}{\xi\Phi_0}+
\frac{1}{\xi}\right)^{\frac{1+4\lam}{\lam}}\right\}, 
&\lam\in\mathbb{R}\setminus\{-\frac{1}{4},0\} \\[2em]
\displaystyle{-\Lambda_5-6k^2(f_0+\Phi_0^2)+\frac{\Phi_0^2k^2}{2}\left(\frac{\Phi}{\Phi_0}-2
\right)^2\times} & \\[4mm]
\hspace{4em}\displaystyle{\times\left\{3+4\ln\left[\frac{1}{\xi}\left(\frac{\Phi}{\Phi_0}-2
\right)\right]+\ln^2\left[\frac{1}{\xi}\left(\frac{\Phi}{\Phi_0}-2\right)\right]\right\}}, 
& \lam=-1/4\end{array}\right\}$}

\begin{figure}[t]
    \centering
    \begin{subfigure}[b]{0.48\textwidth}
        \includegraphics[width=\textwidth]{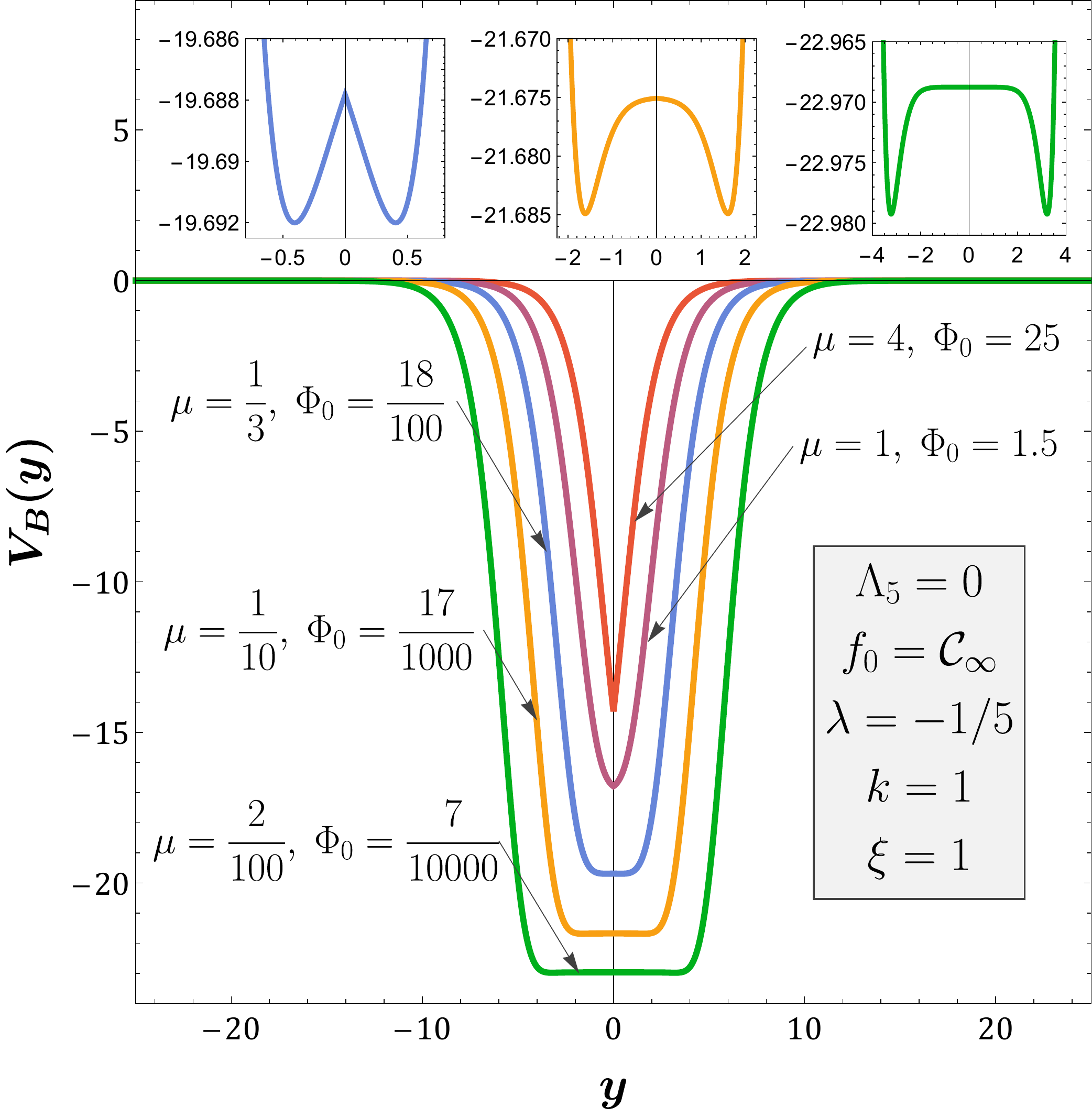}
        \caption{\hspace*{-2.9em}}
        \label{P3subf: quad-plot6}
    \end{subfigure}
    \hfill
    ~ 
    \begin{subfigure}[b]{0.48\textwidth}
        \includegraphics[width=\textwidth]{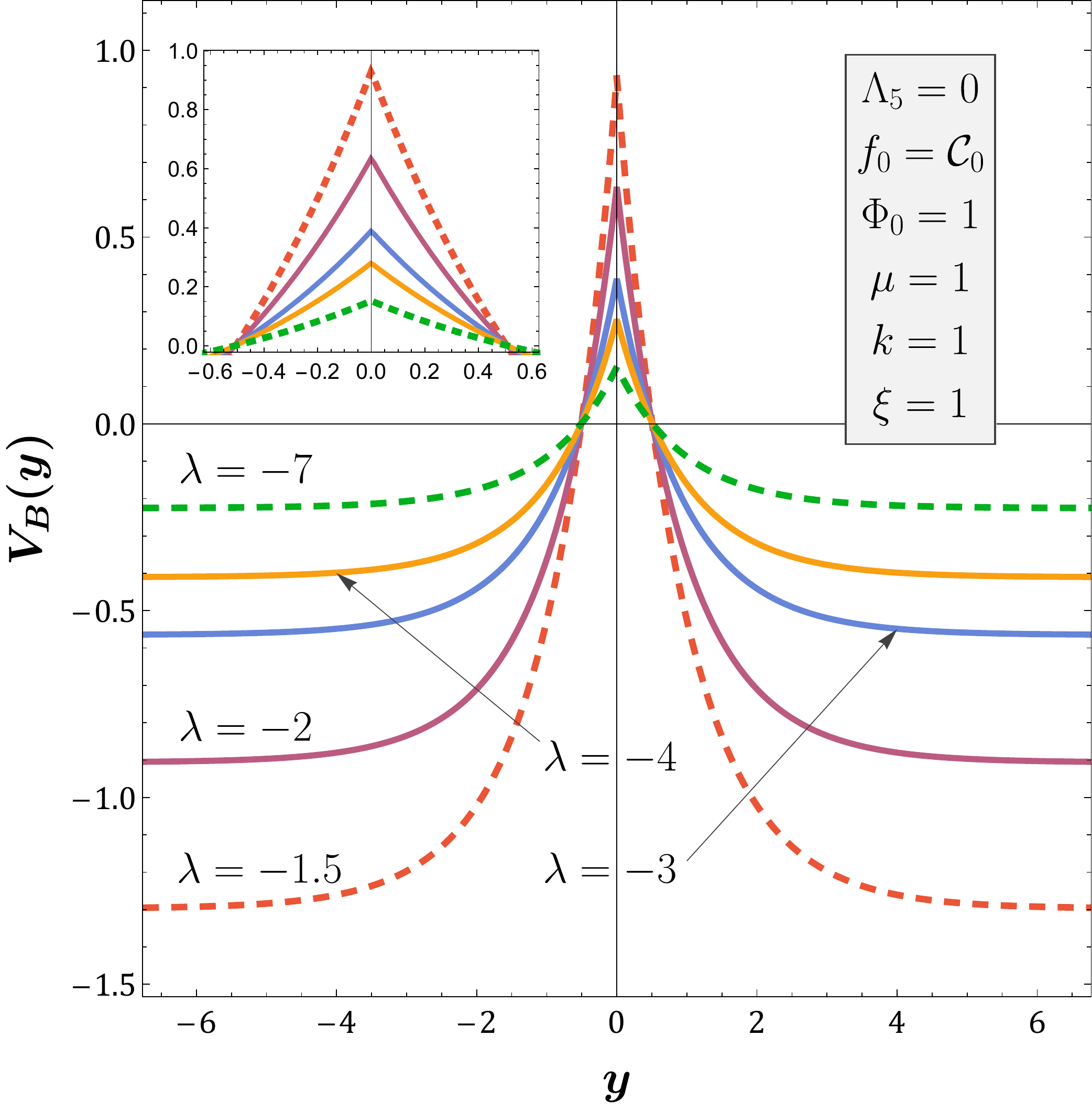}
        \caption{\hspace*{-3em}}
        \label{P3subf: quad-plot7}
    \end{subfigure}
    
    \vspace{-0.5em}  
    \caption{The scalar potential $V_B$ in terms of the extra dimension $y$ for 
    $\Lambda_5=0$, $k=1$, $\xi=1$, and: (a) $\lambda=-1/5$, $f_0=\cinf$, and
    variable $\mu$ and $\Phi_0$, while, in (b) $\Phi_0=1$, $\mu=1$, $f_0=\mathcal{C}_0$
    and $\lam=-1.5, -2, -3, -4, -7$. In 
    each case, $\cinf$ and $\mathcal{C}_0$ should be evaluated separately.}
    \label{P3fig: quad-plot-6-7}
\end{figure}

\begin{figure}[t]
    \centering
    \begin{subfigure}[b]{0.47\textwidth}
        \includegraphics[width=\textwidth]{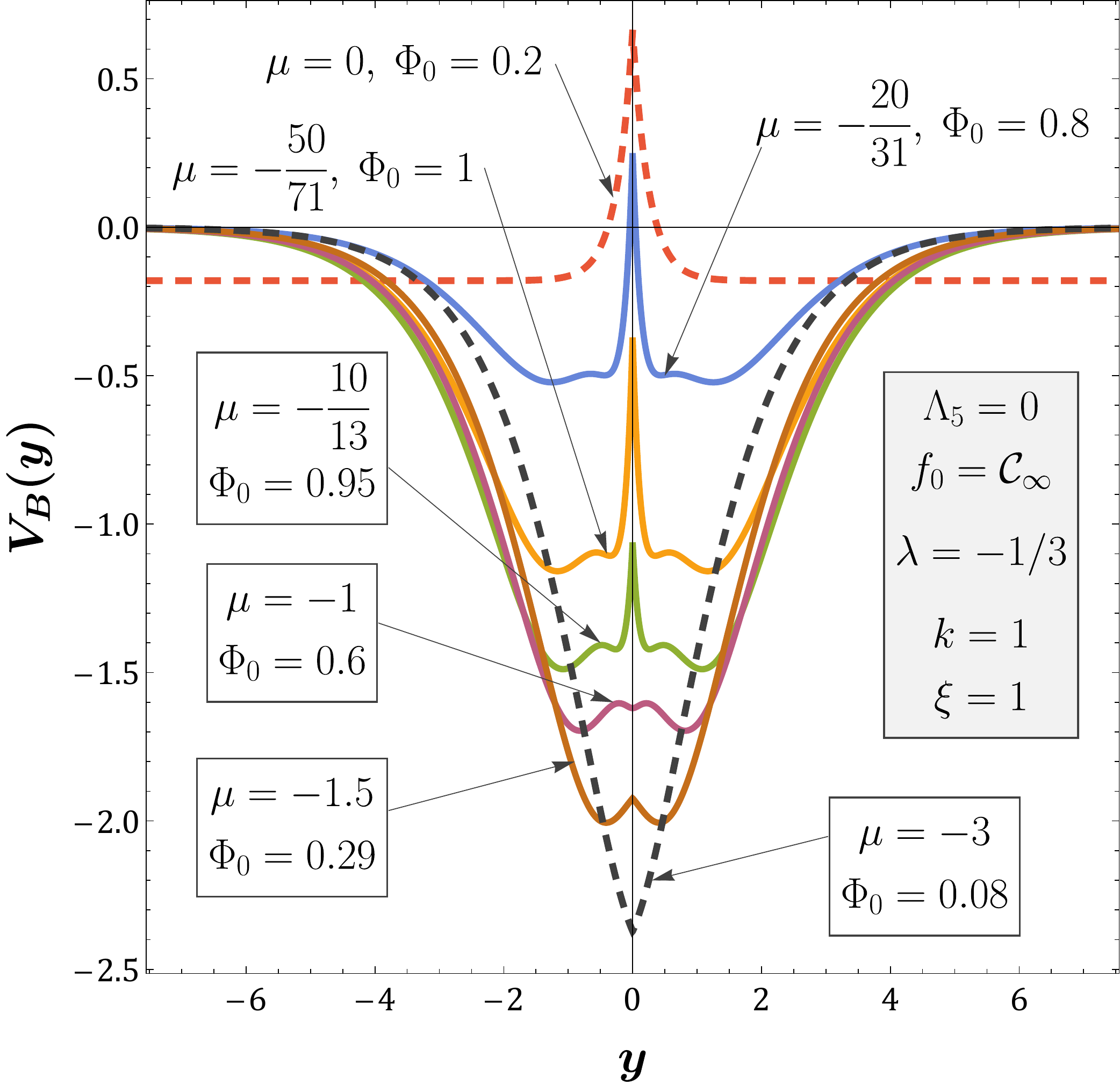}
        \caption{\hspace*{-3em}}
        \label{P3subf: quad-plot8a}
    \end{subfigure}
    \hfill
    ~ 
    \begin{subfigure}[b]{0.475\textwidth}
        \includegraphics[width=\textwidth]{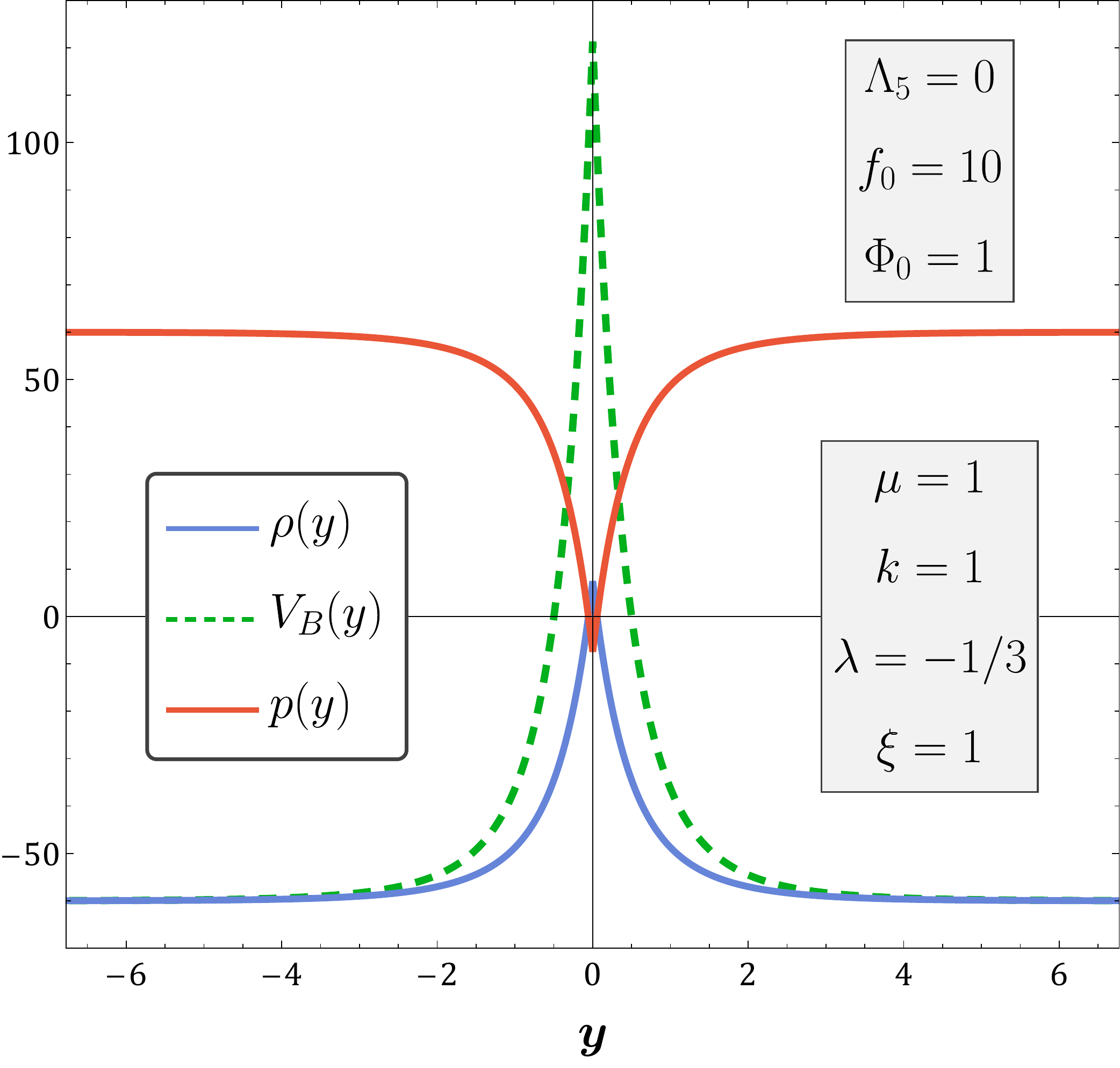}
        \caption{\hspace*{-1.5em}}
        \label{P3subf: quad-plot8b}
    \end{subfigure}  
    \vspace{-0.5em}  
    \caption{(a) The scalar potential $V_B$ in terms of the extra dimension $y$ for 
    $\Lambda_5=0$, $\lam=-1/3$, $k=1$, $\xi=1$ and $f_0=\cinf$. 
    The varying parameters are $\Phi_0$ and $\mu$.
    (b) The energy density $\rho$ and pressure $p$ of the system together with
    the scalar potential $V_B$ in terms of the coordinate $y$ for $\Lambda_5=0$, $f_0=10$,
    $\Phi_0=1$, $\mu=1$, $k=1$, $\xi=1$, and $\lam=-1/3$.}
    \label{P3fig: quad-plot-8}
\end{figure}

In \myref{P3fig: quad-plot-6-7}{P3subf: quad-plot6}, \myref{P3fig: quad-plot-6-7}{P3subf: quad-plot7} and
\myref{P3fig: quad-plot-8}{P3subf: quad-plot8a}, we display the behaviour of the scalar potential $V_B$ as a function
of the extra coordinate $y$ using the same values for the parameters as in
\myref{P3fig: quad-plot-1-2}{P3subf: quad-plot2}, \myref{P3fig: quad-plot-3-5}{P3subf: quad-plot3} and
 \myref{P3fig: quad-plot-3-5}{P3subf: quad-plot5}, respectively. It is worth observing the variety of
forms that one may achieve for $V_B$ by varying the values of the parameters of the model. 
In \myref{P3fig: quad-plot-6-7}{P3subf: quad-plot6}, constructed for $\lam \in (-1/4, 0)$, the
scalar potential adopts a negative value around the location of our brane, thus mimicking
locally a negative bulk cosmological constant $\Lambda_5$, while it vanishes away 
from our brane. In  \myref{P3fig: quad-plot-6-7}{P3subf: quad-plot7}, constructed for values
of $\lam$ in the regime $(-\infty, -1/4)$, the scalar potential has a positive value on
and close to our brane and then decreases rapidly to a constant negative value; this
asymptotic value depends on the values of the parameters and is preserved until the
spacetime boundaries. Finally, \myref{P3fig: quad-plot-8}{P3subf: quad-plot8a} (next page), constructed
for $\lam=-1/3$ and $\frac{2\lam}{1+4\lam}=2$, shows the sensitivity of the scalar
potential to the value of parameter $\mu$ with local
minima and maxima appearing in its profile. It should be however stressed that the warp
factor adopts its exponentially decaying form for all aforementioned profiles of the bulk
potential and independently of whether $\Lambda_5=0$ or not.
 
The components of the energy-momentum tensor of the theory may be finally computed by 
employing Eqs. \eqref{P3eq: linear-rho}-\eqref{P3eq: linear-p}. As in the linear case, we obtain
\eq$\label{P3eq: quad-rho-pi-py}
\begin{gathered}
\rho(y)=-p(y)=-6k^2f(y)\,,\\[1mm]
p(y)=p^i(y)=p^y(y)\,.
\end{gathered}$
We discussed thoroughly in the previous section, that in order to satisfy the 
weak energy conditions on and close to the brane, we should allow the coupling 
function $f(y)$ to take negative values at these regimes. Thus, demanding that $f(0)<0$ 
and using Eq. \eqref{P3eq: quad-f-y}, we obtain the constraints:
\eq$\label{P3eq: quad-weak-con}
\left\{\begin{array}{ll}
\displaystyle{\frac{f_0}{\Phi_0^2}<\frac{1}{4\lam}\left[1-\xi^2(\mu+1)^{\frac{4\lam}
{1+4\lam}}\right]}, &\lam\in\mathbb{R}\setminus\{-\frac{1}{4},0\}\\[4mm]
\displaystyle{\frac{f_0}{\Phi_0^2}<\frac{\xi^2}{4}e^{2\mu}-1}, & \lam=-\frac{1}{4}
\end{array}\right\}\,.$
In \myref{P3fig: quad-plot-8}{P3subf: quad-plot8b}, we present the energy density $\rho(y)$, the pressure
$p(y)=p^i(y)=p^y(y)$ and the scalar potential $V_B(y)$ in terms of the coordinate $y$.
It is obvious that, for this particular set of parameters, chosen to satisfy the above
constraints, the weak energy conditions are satisfied by the bulk matter on and close
to the brane. 


We should complete our bulk solution with the junction conditions introduced in the
model due to the presence of the brane at $y=0$. As discussed in the previous section,
the energy content of the brane is given by the combination $\sigma + V_b(\Phi)$,
and it creates a discontinuity in the second derivatives of the warp factor, coupling
function and scalar field  at the location of the brane. Using Eqs. \eqref{P3eq: jun_con1}
and \eqref{P3eq: jun_con2}, for $\lam\neq\{-\frac{1}{4},0\}$ and $\lam=-1/4$, we obtain
\gat$\label{P3eq: quad-jun1-1}
\sigma+V_b(\Phi)|_{y=0}=6k\left(f_0-\frac{\Phi_0^2}{4\lam}\right)+
\frac{\Phi_0^2\xi^2k}{2\lam(1+4\lam)}\left[3(1+\mu)(1+4\lam)+4\lam\right]
(1+\mu)^{-\frac{1}{1+4\lam}}\, ,\\[3mm]
\label{P3eq: quad-jun1-2}
\pa_\Phi V_b |_{y=0}=\frac{2k\xi\Phi_0}{1+4\lam}(1+\mu)^{-\frac{1+2\lam}{1+4\lam}}
\left[4(1+\mu)(1+4\lam)-1\right]\, ,$
and 
\gat$\label{P3eq: quad-jun2-1}
\sigma+V_b(\Phi)|_{y=0}=6k(f_0+\Phi_0^2)-\frac{\Phi_0^2\xi^2k\,e^{2\mu}}{2}(3+2\mu)\, 
,\\[3mm]
\label{P3eq: quad-jun2-2}
\pa_\Phi V_b |_{y=0}=-2\Phi_0\,k\,\xi\,\mu\,e^{\mu}(\mu+2)\, ,$
respectively. Using the constraints \eqref{P3eq: quad-jun1-1} and \eqref{P3eq: quad-jun2-1}, it is easy
to deduce that in order to have a positive total energy  density on the brane, namely
$\sig+V_b(\Phi)|_{y=0}>0$, we should have, respectively
{\fontsize{11}{11}\eq$\label{P3eq: quad-brane-ene-con}
\left\{\begin{array}{ll}
\displaystyle{\frac{f_0}{\Phi_0^2}>\frac{1}{4\lam}\left\{1-\xi^2\left[(1+\mu)^{\frac{4\lam}{1+4
\lam}}+\frac{4\lam}{3(1+4\lam)}(1+\mu)^{-\frac{1}{1+4\lam}}\right]\right\}}\,,
& \lam\in\mathbb{R}\setminus\{-\frac{1}{4},0\}\\[5mm]
\displaystyle{\frac{f_0}{\Phi_0^2}>-1+\frac{\xi^2}{12}\,e^{2\mu}(3+2\mu)}\,, & \lam=-1/4
\end{array}\right\}\,.$}
Let us also note that, from the constraint \eqref{P3eq: quad-jun2-2}, we see that the brane
interaction term $V_b$ can be a constant, and thus absorbed into the brane tension
$\sigma$, under the condition $\mu=-2$. A similar fixing of the parameter $\mu$ follows
from Eq. \eqref{P3eq: quad-jun1-2}, which leads to the result $\mu=-\frac{3+4\lam}{4(1+4\lam)}$.
However, in this case, care should be taken so that the resulting values of $\mu$, in
terms of $\lambda$, are allowed by Table \ref{P3tab: quad-par-val}.

\par The effective four-dimensional gravitational scale on the brane has already been calculated
and is given in Eqs. \eqref{P3eq: MPl-quad} and \eqref{P3eq: Mpl-1/4}. The effective cosmological constant
on the brane $\Lambda_4$ can be calculated from Eq. \eqref{P3eq: cosm_eff}, and is found to be
zero also in this case, as anticipated. 


\vspace*{-1.5em}

\mysubsection{The energy conditions in the parameter space}

\par We will now study the inequalities \eqref{P3eq: quad-eff-con}, \eqref{P3eq: quad-weak-con}
and \eqref{P3eq: quad-brane-ene-con} and investigate again whether these may be simultaneously
satisfied. In particular, we will study the parameter space between the ratio $f_0/\Phi_0^2$
and the parameters $\lam$, $\mu$, and $\xi$. Given the large number of parameters,
we will present three-dimensional graphs of the parameter space of the ratio $f_0/\Phi_0^2$
with two of the three parameters $\lam,\, \mu,\, \xi$, while keeping the remaining one
fixed. Before we continue, we elucidate that, in the forthcoming analysis, we will denote
the r.h.s. of inequality \eqref{P3eq: quad-eff-con} with $F_{eff}(\lam,\mu,\xi)$, since it is associated
with the effective gravitational constant, the r.h.s. of inequality \eqref{P3eq: quad-weak-con},
which refers to the energy conditions in the bulk, with $F_{B}(\lam,\mu,\xi)$, and
finally, the r.h.s. of inequality \eqref{P3eq: quad-brane-ene-con}, which involves the total
energy density on the brane, with $F_{br}(\lam,\mu,\xi)$. 

While pursuing to satisfy simultaneously all the aforementioned inequalities, we have performed
a comprehensive study of the parameter space of the quantities $f_0/\Phi_0^2$, $\lam$, $\mu$
and $\xi$ following the classification of cases, regarding the values of the free parameters,
presented in Table \ref{P3tab: quad-par-val}. We present the corresponding results below:

\begin{enumerate}
\item[\bf(i)] For $\lam>0$, we have $\mu\geq 0$, while the parameter $\xi$ can take 
values in the whole set of real numbers except zero. In \myref{P3fig: quad-lam-pos-neg}{P3subf: quad-lam-pos},
we depict the parameter space of the quantities $f_0/\Phi_0^2$, $\lam$ and $\mu$, for
$\xi=1$ and $\lam>0$. Although the surfaces representing the functions $F_{eff}(\lam,\mu,\xi)$, 
$F_{br}(\lam,\mu,\xi)$ and $F_{B}(\lam,\mu,\xi)$ change significantly for different values
of the parameter $\xi$, their relative positions remain the same satisfying always the relation
$$F_{eff}(\lam,\mu,\xi)>F_{B}(\lam,\mu,\xi)>F_{br}(\lam,\mu,\xi)\,.$$
This means that there is no point in the parameter space for $\lam>0$ at which all three
inequalities are satisfied simultaneously. It is possible though to satisfy simultaneously the
inequalities \eqref{P3eq: quad-eff-con} and \eqref{P3eq: quad-brane-ene-con}. Particularly, for every 
value of the ratio $f_0/\Phi_0^2$ which is greater than the value of the function $F_{eff}
(\lam,\mu,\xi)$ at any given point in the parameter space the aforementioned two inequalities
will be satisfied. This means that the positivity of both the effective four-dimensional
gravitational constant and the total energy-density on the brane is ensured. In contrast,
there is no point in the parameter space at which we can satisfy the inequality \eqref{P3eq: quad-weak-con}
because the surface of the function $F_{B}(\lam,\mu,\xi)$ lies always below the surface of the
function $F_{eff}(\lam,\mu,\xi)$; as a result, the weak energy conditions are always violated
by the bulk matter close to the brane. 


\begin{figure}[t]
    \centering
    \begin{subfigure}[b]{0.48\textwidth}
        \includegraphics[width=\textwidth]{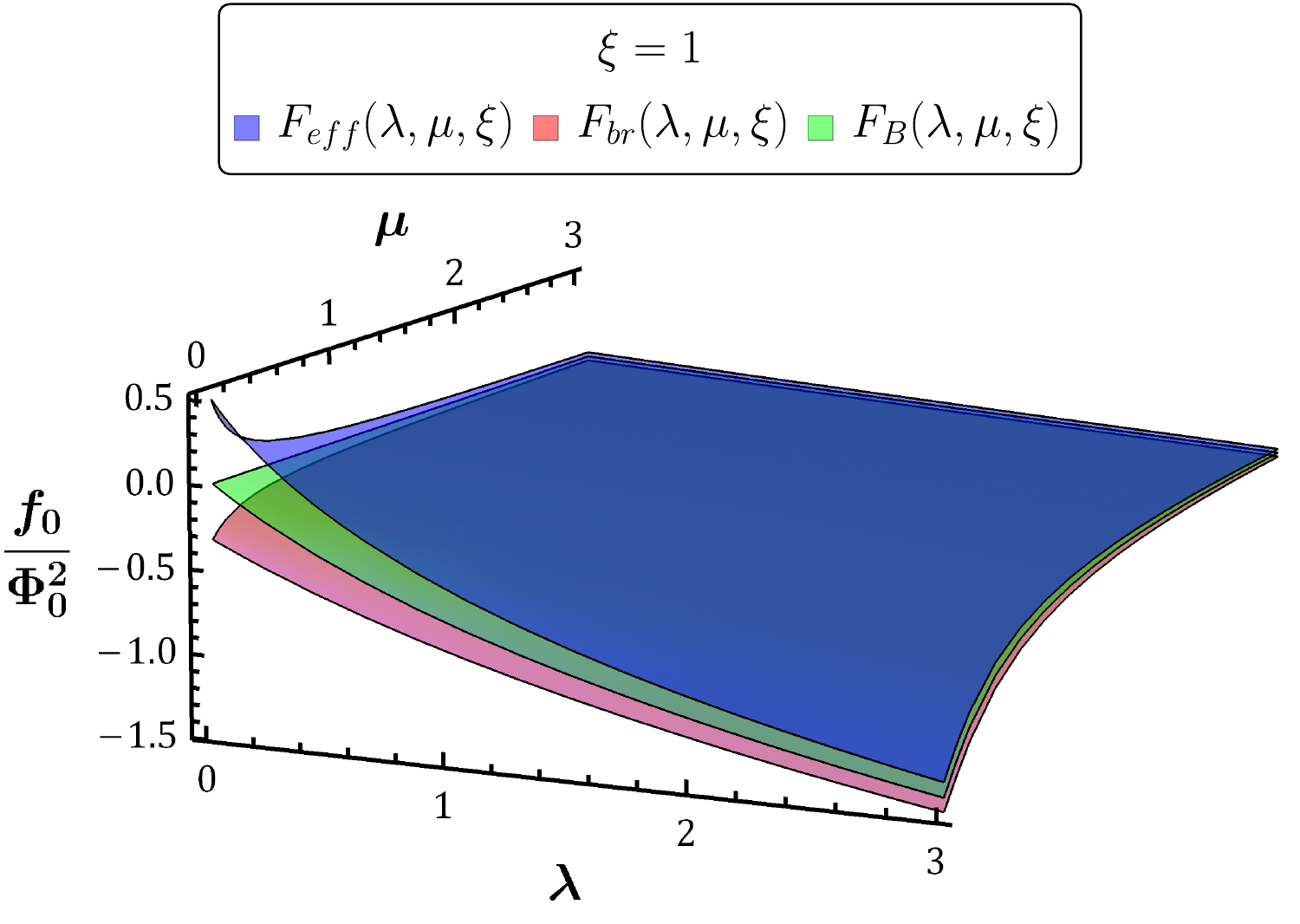}
        \caption{\hspace*{2.5em}}
        \label{P3subf: quad-lam-pos}
    \end{subfigure}
    ~ 
    \begin{subfigure}[b]{0.485\textwidth}
        \includegraphics[width=\textwidth]{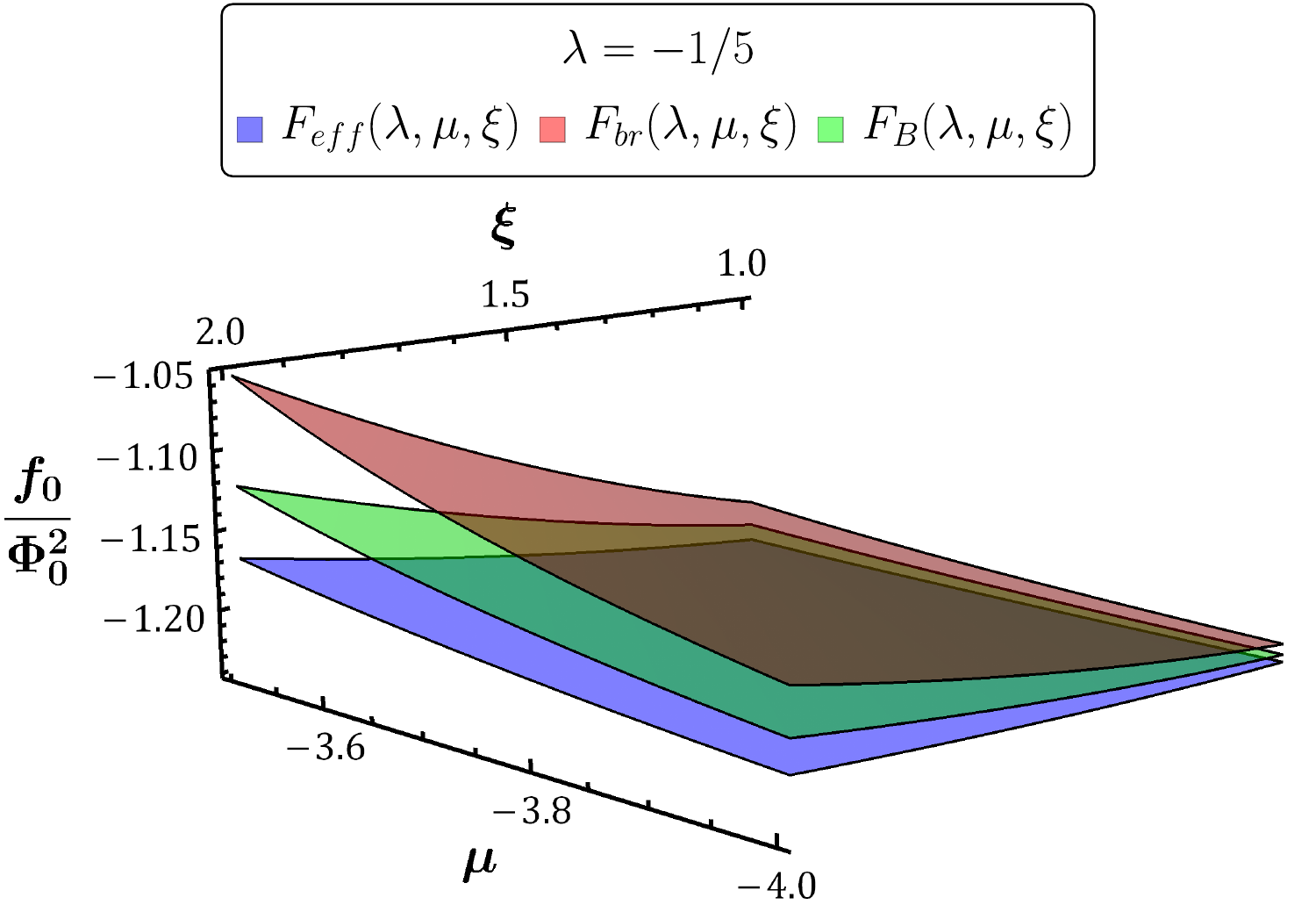}
        \caption{\hspace*{5.5em}}
        \label{P3subf: quad-lam-neg-even1}
    \end{subfigure}
    ~ 
    \vspace{-1.8em}    
    \caption{(a) The parameter space of the quantities $f_0/\Phi_0^2$, $\lam$ and $\mu$, for
    $\xi=1$ and $\lam>0$.
(b) The parameter space of the quantities $f_0/\Phi_0^2$, $\mu$ and $\xi$, for
    $\lam=-1/5$ or $\frac{2\lam}{1+4\lam}=-2$ and $\mu<-1$. The graphs depict the functions
    $F_{eff}(\lam, \mu,\xi)$, $F_{br}(\lam,\mu,\xi)$ and $F_{B}(\lam,\mu,\xi)$.}
   \label{P3fig: quad-lam-pos-neg}
\end{figure}

\item[\bf(ii)] For $\lam\in\left(-\frac{1}{4},0\right)$,  $\frac{2\lam}{1+4\lam}\neq n,\,n\in\mathbb{Z}^<$,
we have $\mu > 0$, and we obtain the same qualitative behaviour as in the previous case.
However, when $\frac{2\lam}{1+4\lam}=n$, we have $\mu\in(-\infty,-1)\cup(0,\infty)$.
In this case, the position of the surfaces $F_{eff}(\lam,\mu,\xi)$, $F_{br}(\lam,\mu,\xi)$ and $F_{B}(\lam,\mu,\xi)$
are different in the region of the parameter space where $\mu<-1$ and in the region where
$\mu>0$.  Specifically, in this case we find that 
\eq$\begin{array}{ll}
\displaystyle{F_{br}(\lam,\mu,\xi)>F_{B}(\lam,\mu,\xi)>F_{eff}(\lam,\mu,\xi)}, & \mu<-1\,,
\\[3mm]
\displaystyle{F_{eff}(\lam,\mu,\xi)>F_{B}(\lam,\mu,\xi)>F_{br}(\lam,\mu,\xi)}, & \mu>0\,.
\end{array}\nonum$
Again, there is no point in the parameter space at which we can satisfy simultaneously all inequalities.
For $\mu>0$, the situation is similar to the one of case (i) depicted in \myref{P3fig: quad-lam-pos-neg}{P3subf: quad-lam-pos}.
In this case, we may easily obtain a positive effective gravitational constant and a positive total
energy-density on the brane. For $\mu<-1$, though, as \myref{P3fig: quad-lam-pos-neg}{P3subf: quad-lam-neg-even1} also reveals,
we have the choice of supplementing the positivity of the effective gravitational constant by either a positive
total energy-density on the brane or by a bulk matter that satisfies the energy conditions close to
our brane.

\item[\bf(iii)] For $\lam=-1/4$, due to the different form of the solution, the functions
$F_{eff}(-1/4,\mu,\xi)$, $F_{br}(-1/4,\mu,\xi)$ and $F_{B}(-1/4,\mu,\xi)$ are given by different
expressions. Now, these are found to satisfy the relations 
\eq$\begin{array}{ll}
\displaystyle{F_{eff}(-1/4,\mu,\xi)>F_{B}(-1/4,\mu,\xi)>F_{br}(-1/4,\mu,\xi)}, & \mu<0\,,
\\[3mm]
\displaystyle{F_{br}(-1/4,\mu,\xi)>F_{B}(-1/4,\mu,\xi)>F_{eff}(-1/4,\mu,\xi)}, & \mu>0\,.
\end{array}\nonum$
In this case, for $\mu<0$, we may obtain only the combination of a positive
effective gravitational scale and a positive total energy-density on the brane, in the region
of the parameter space in which the value of the ratio $f_0/\Phi_0^2$ is greater than
the value of the function $F_{eff}(-1/4,\mu,\xi)$; the relative positions of the different
surfaces are the same as in \myref{P3fig: quad-lam-pos-neg}{P3subf: quad-lam-pos}. On the other hand, for
$\mu >0$, we again have the choice of satisfying either Eqs. \eqref{P3eq: quad-eff-con} and 
\eqref{P3eq: quad-weak-con}, in the region where $F_{eff}(-1/4,\mu,\xi)<f_0/\Phi_0^2<F_{B}(-1/4,\mu,\xi)$,
or Eqs. \eqref{P3eq: quad-eff-con} and \eqref{P3eq: quad-brane-ene-con}, in the region where
$f_0/\Phi_0^2>F_{br}(-1/4,\mu,\xi)$. This situation is in turn similar to the one depicted
in \myref{P3fig: quad-lam-pos-neg}{P3subf: quad-lam-neg-even1}.  

\begin{figure}[t]
    \centering
    \begin{subfigure}[b]{0.47\textwidth}
        \includegraphics[width=\textwidth]{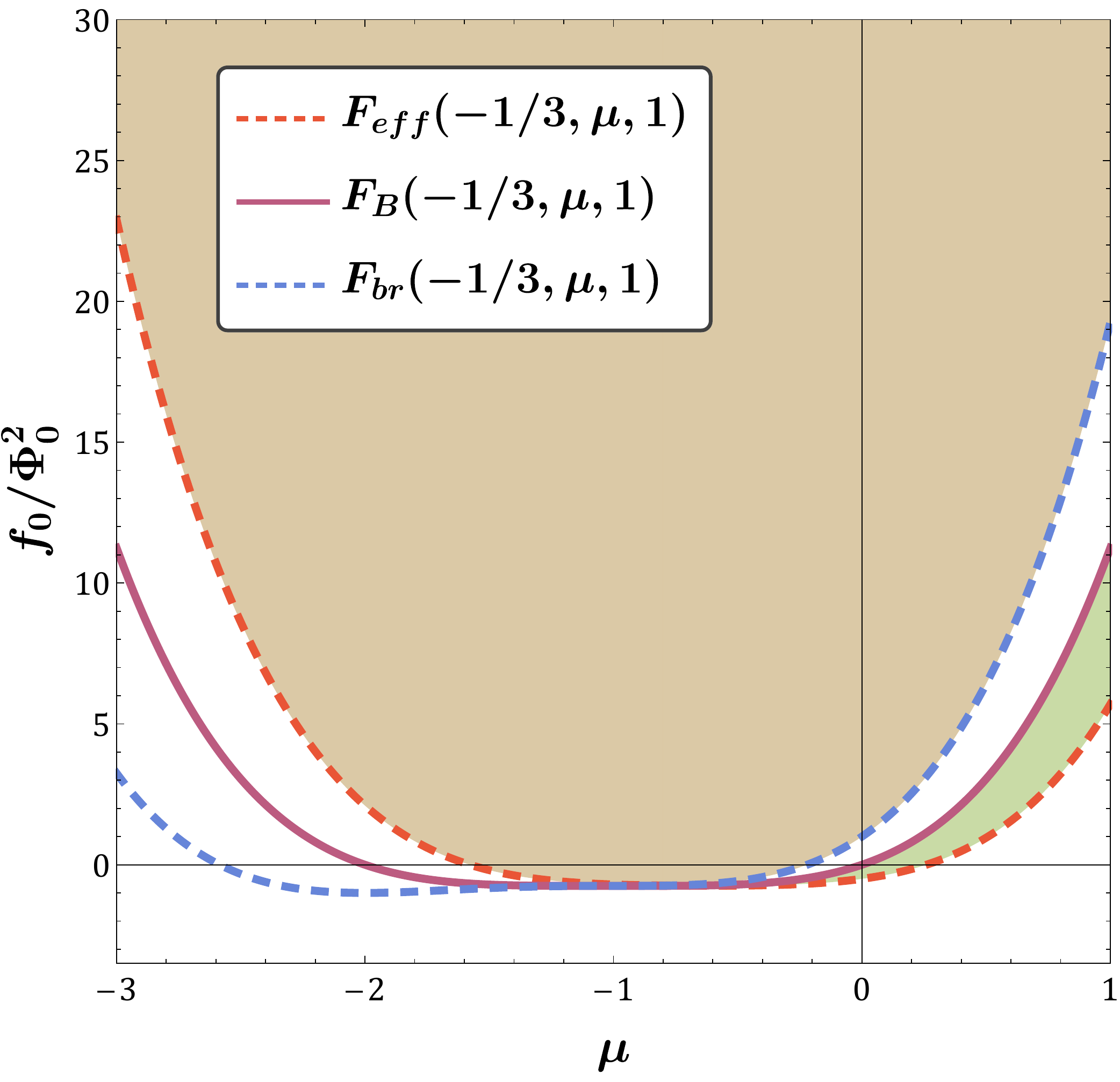}
        \caption{\hspace*{-2.2em}}
        \label{P3subf: quad-lam-neglowereven}
    \end{subfigure}
    \hfill
    ~ 
    \begin{subfigure}[b]{0.49\textwidth}   
        \includegraphics[width=\textwidth]{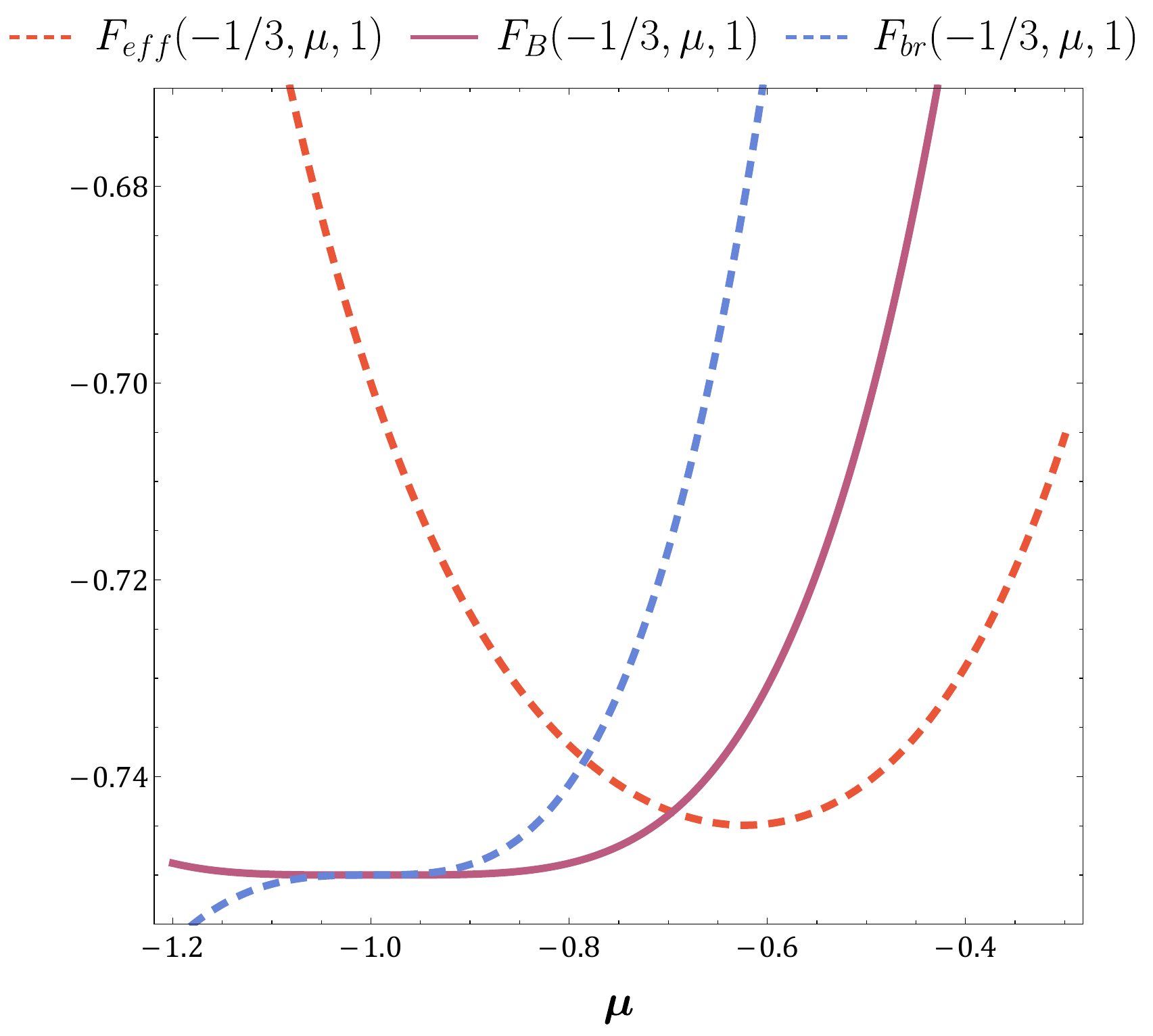}
        \caption{\hspace*{-1.8em}}
        \label{P3subf: quad-lam-neglowerevenzoom}
    \end{subfigure}
    ~ 
    \vspace{-1em}    
    \caption{(a) The parameter space of the quantities $f_0/\Phi_0^2$ and $\mu$, for $\xi=1$,
    $\lam=-1/3$, and (b) a magnification of a particular region of the 
    previous figure in order to get a clear picture of the behaviour of the functions $F_{eff}(
    -1/3,\mu,1)$, $F_{br}(-1/3,\mu,1)$ and $F_{B}(-1/3,\mu,1)$ close to $\mu=-0.8$.}
   \label{P3fig: quad-lam-neglowereven-1-2}
\end{figure}

\item[\bf(iv)] For $\lam<-1/4$, $\frac{2\lam}{1+4\lam}\neq n,\,n\in\mathbb{Z}^>$, and
for every allowed value of the parameters $\mu \geq 0$ and $\xi\in\mathbb{R}\setminus\{0\}$,
we always have
$$F_{br}(\lam,\mu,\xi)>F_{B}(\lam,\mu,\xi)>F_{eff}(\lam,\mu,\xi)\,.$$
In this case, the situation is similar to the one depicted in \myref{P3fig: quad-lam-pos-neg}{P3subf: quad-lam-neg-even1}, 
and we have again the choice of combining a positive effective gravitational constant with
either a positive energy-density on the brane or a bulk matter that satisfies the weak
energy conditions close to and on our brane. 

\item[\bf(v)] For $\lam<-1/4$ and $\frac{2\lam}{1+4\lam}= n,\,n\in\mathbb{Z}^>$, the
parameter $\mu$ is free to take values in the whole set of real numbers. In
\myref{P3fig: quad-lam-neglowereven-1-2}{P3subf: quad-lam-neglowereven} and 
\myref{P3fig: quad-lam-neglowereven-1-2}{P3subf: quad-lam-neglowerevenzoom}, we depict
the parameter space of the ratio $f_0/\Phi_0^2$ and $\mu$ together with the curves of
the functions $F_{eff}(\lam,\mu,\xi)$, $F_{br}(\lam,\mu,\xi)$ and $F_{B}(\lam,\mu,\xi)$. 
Note that, for clarity of the graph, we have fixed the values of two parameters, i.e.
$\lam=-1/3$ and $\xi=1$, and thus present a two-dimensional graph. However, 
the situation remains the same for every other allowed value of the parameters 
$\lam$ and $\xi$. We observe that there always exist a region in the parameter space
in which we can have a positive value for the effective four-dimensional gravitation scale
and satisfy the weak energy conditions close to the brane (green region) and a region in
which both the four-dimensional gravitational constant and the total energy density on
the brane are positive (brown region). Since there is no overlapping between the green and
brown regions, as \myref{P3fig: quad-lam-neglowereven-1-2}{P3subf: quad-lam-neglowerevenzoom} reveals, there is no point
in the parameter space where all three conditions are satisfied. For comparison, we note
that the parameters in \myref{P3fig: quad-plot-8}{P3subf: quad-plot8b} have been chosen so that the depicted
solution falls into the green area of \myref{P3fig: quad-lam-neglowereven-1-2}{P3subf: quad-lam-neglowereven}.

\end{enumerate}

%
%

\vspace*{-3em}

\mysection{Inverse-power coupling function in terms of y}

\par In this and the following two sections, we will consider explicit forms of the coupling function
$f(y)$ in terms of the coordinate $y$. These forms cannot be easily expressed in terms of the scalar
field $\Phi$ in a closed form, they are however legitimate choices that satisfy the reality and
finiteness conditions imposed in \secref{P3sec: th-frame}. We start with the following expression
\eq$\label{P3eq: power-f}
f(y)=f_0+\frac{\Phi_0^2}{k^\lam(y+y_0)^\lam}\, ,$
where $(f_0, \Phi_0) \in \mathbb{R}\setminus\{0\}$ while ($\lam, y_0) \in(0,+\infty)$.
The factor $k^\lam$ in the denominator was introduced to make the product $k(y+y_0)$
dimensionless. 


\mysubsection{The bulk solution}

\par Substituting the aforementioned coupling function in Eq. \eqref{P3eq: grav-1-2} we obtain the
differential equation
\eq$\label{P3eq: power-der-Phi}
[\Phi'(y)]^2=\frac{\lam\,\Phi_0^2}{k^\lam(y+y_0)^{\lam+2}}\left[k(y+y_0)-\lam-1\right]\, .$
The r.h.s. of the above equation should be always positive; evaluating at $y=0$, the above yields 
the following constraint on the parameters of the model
\eq$\label{P3eq: power-con1}
\frac{ky_0}{\lam+1}>1.$
The function $[\Phi'(y)]^2$ could, in principle, be zero at the point where $\Phi(y)$ has
an extremum. However, from Eq. \eqref{P3eq: power-der-Phi} this may happen only at $y=y_0\left(\frac{
\lam+1}{ky_0}-1\right)$ which, upon using the constraint \eqref{P3eq: power-con1}, turns out to be
negative. Therefore, the scalar field does not have an extremum in the whole domain $y\in[0,+
\infty)$, which also means that $\Phi(y)$ is an one-to-one function in the same regime. In 
addition, from Eq. \eqref{P3eq: power-der-Phi} it is straightforward to deduce that, as $y\ra+\infty$, 
the physical constraint \eqref{P3eq: con.2} is satisfied, thus, the scalar field does not diverge at 
infinity.
\par Let us now determine the explicit expression of the scalar field $\Phi(y)$ from Eq.
\eqref{P3eq: power-der-Phi}. For simplicity and without loss of generality, we will assume that
$\Phi_0\in(0,+\infty)$. Then, after taking the square root of Eq. \eqref{P3eq: power-der-Phi},
we have:
\eq$\label{P3eq: power-phi1}
\Phi_{\pm}(y)=\pm \frac{\Phi_0\sqrt{\lam(\lam+1)}}{k^{\lam/2}}\int dy\ (y+y_0)^{-\frac{\lam}{2}-1}
\left[\frac{k(y+y_0)}{\lam+1}-1\right]^{\frac{1}{2}}\, .$
Setting $u=\frac{k(y+y_0)}{\lam+1}$ and then $w=1-\frac{1}{u}$, the above integral takes the form
\bal$\int dy\ (y+y_0)^{-\frac{\lam}{2}-1}\left[\frac{k(y+y_0)}{\lam+1}-1\right]^{\frac{1}{2}}&
=\left(\frac{k}{\lam+1}\right)^{\frac{\lam}{2}}\int dw\ (1-w)^{\frac{\lam}{2}-\frac{3}{2}}\
w^{\frac{1}{2}}\nonum\\[2mm]
&=\left(\frac{k}{\lam+1}\right)^{\frac{\lam}{2}}\int_0^w dt\ t^{\frac{1}{2}}(1-t)^{\frac{\lam}{2}
-\frac{3}{2}}+C_1\nonum\\[2mm]
\label{P3eq: power-int-sol}
&=\left(\frac{k}{\lam+1}\right)^{\frac{\lam}{2}}w^{\frac{3}{2}}\int_0^1 dt'\ {t'}^{\frac{1}{2}}
(1-wt')^{\frac{\lam}{2}-\frac{3}{2}}+C_1\,,$ 
where in the last line we have made the change of variable $t'=\frac{t}{w}$. Using the integral
representation of the hypergeometric function \eqref{P3eq: int-rep-hyper}, 
Eq. \eqref{P3eq: power-phi1} leads to the following expression for the scalar field $\Phi(y)$. 
\eq$\label{P3eq: power-Phi}
\Phi_{\pm}(y)=\pm \frac{2\Phi_0}{3}\sqrt{\frac{\lam}{(\lam+1)^{\lam-1}}}\left[1-\frac{\lam+1}{k(y+y_0)}
\right]^{\frac{3}{2}}\,_2F_1\left(\frac{3}{2}-\frac{\lam}{2},\frac{3}{2};\frac{5}{2};1-\frac{\lam+1}
{k(y+y_0)}\right)\, .$
In the above, we have also used the translational symmetry of the gravitational field equations
with respect to the scalar field to set $C_1=0$.

A solution for the scalar field similar to Eq. (\ref{P3eq: power-Phi}) was derived in the context of
\chapref{Chap: P2} for an exponential coupling function $f(y)$ and an
anti-de Sitter brane ($\Lambda<0$). The mathematical properties of the solution were studied there
in detail, therefore, here, we adapt those results in the present case and present our solutions 
for the scalar field without repeating the analysis. 

Trying to simplify Eq. (\ref{P3eq: power-Phi}), we first note that, for every value of the coordinate $y$,
the argument $1-\frac{\lam+1}{k(y+y_0)}$ of the hypergeometric function is positive and smaller
than unity. Therefore, one can expand the hypergeometric function in power series as
{\fontsize{11}{11}
\eq$\label{P3eq: hyper-expr}
\hspace{2em}\,_2F_1\left(\frac{3}{2}-\frac{\lam}{2},\frac{3}{2};\frac{5}{2};1-\frac{\lam+1}
{k(y+y_0)}\right)=\sum_{n=0}^\infty \frac{\Gamma\left(\frac{3}{2}-\frac{\lam}{2}+n\right)}
{\Gamma\left(\frac{3}{2}-\frac{\lam}{2}\right)}\frac{3}{(2n+3)n!}\left[1-\frac{\lam+1}
{k(y+y_0)}\right]^n\,.$}
\hspace{-0.5em}There are two interesting categories of values for the parameter $\lam$ which lead to even simpler
and more elegant expressions for the hypergeometric function and subsequently for the scalar field.
These are:
\begin{enumerate}
\item[\bf{(i)}] If $\lam=1+2q$ with $q \in {\mathbb{Z}}^{>}$, then, from Eq. \eqref{P3eq: hyper-expr},
we have
{\fontsize{11}{11}\bal$
_2F_1  \left(\frac{3-\lam}{2},\frac{3}{2};\frac{5}{2};1-\frac{\lam+1}
{k(y+y_0)}\right)
&=\left\{\begin{array}{cr}
1\,, &  q=1\\[3mm]
1+\sum_{n=1}^{q-1}\frac{3(-q+1)(-q+2)\cdots(-q+n)}{(2n+3)n!}\times & \\[2mm]
\hspace{2cm}\times\left[1-\frac{\lam+1}{k(y+y_0)}\right]^n, &  q>1
\end{array}\right\} .
\label{P3eq: F-series2}$}

The solution for the scalar field then easily follows by using Eqs. (\ref{P3eq: power-Phi}) and (\ref{P3eq: F-series2})
and substituting the selected values for the parameter $\lam$ (or $q$).
As indicative cases, we present below the form of the scalar field for~\footnote{For completeness,
we present here also the solution for the limiting case
with $\lam=1$ (i.e. for $q=0$); this has the form
$$\Phi_{\pm}(y)=\pm\frac{2\Phi_0}{3}
\left[\,{\rm arctanh}\left(\sqrt{1-\frac{2}{k(y+y_0)}}\,\right)-\sqrt{1-\frac{2}{k(y+y_0)}}
\,\right].$$}
$\lambda=3$ (i.e. $q=1$) 
$$\Phi_{\pm}(y)=\pm\frac{\Phi_0}{2\sqrt{3}}\left[1-\frac{4}{k(y+y_0)}\right]^{3/2},$$
and $\lambda=5$ (i.e. $q=2$)
$$\Phi_{\pm}(y)=\pm\frac{\Phi_0\sqrt{5}}{54}\left[1-\frac{6}{k(y+y_0)}\right]^{3/2}
\left\{1-\frac{3}{5}\left[1-\frac{6}{k(y+y_0)}\right]\right\}.$$

\item[\bf{(ii)}] If $\lam=2q$ with $q \in {\mathbb{Z}}^{>}$, we can always express the 
hypergeometric function in Eq. (\ref{P3eq: power-Phi}) in terms of elementary functions, namely 
$\arcsin$, square roots and powers of its argument. For $\lam=2$ (i.e. $q=1$), it is
\eq$\label{P3eq: hyper-lam-2}
\,_2F_1\left(\frac{1}{2},\frac{3}{2};\frac{5}{2};u^2\right)=\frac{3}{2}\frac{1}{u^2}\left(
\frac{\arcsin\, u}{u}-\sqrt{1-u^2}\right)\,.$
Therefore, from Eq. (\ref{P3eq: power-Phi}), the scalar field for $\lam=2$ can be written in the form
{\fontsize{10}{10}$$\Phi_{\pm}(y)=\pm\frac{\Phi_0\sqrt{2}}{3}\left[1-\frac{3}{k(y+y_0)}\right]^{\frac{1}{2}}
\left[\left(1-\frac{3}{k(y+y_0)}\right)^{-\frac{1}{2}}\arcsin\left(\sqrt{1-\frac{3}{k(y+y_0)}}
\right)-\sqrt{\frac{3}{k(y+y_0)}}\ \right].$$}
For larger values of $\lambda$ (i.e. for $q=1+\ell$, with $\ell \in {\mathbb{Z}}^{>}$),
the following relation holds 
{\fontsize{11}{11}\bal$\,_2F_1\left(\frac{1}{2}-\ell,\frac{3}{2};\frac{5}{2};u^2\right)u^2=&
\alpha\left(\frac{\arcsin u}{u}-\sqrt{1-u^2} \right)\nonum\\[2mm]
&+\sqrt{1-u^2}\left(\beta_1\,u^2+\beta_2\,u^4+\dots+\beta_{\ell-1}\,u^{2(\ell-1)}+
\beta_\ell\,u^{2\ell}\right),
\label{P3eq: hyper-lam>2}$}
\hspace{-0.5em}where $\alpha, \beta_1, \dots, \beta_\ell$ are constant coefficients, which satisfy 
a system of $\ell+1$ linear algebraic equations (Appendix \ref{P2app: hyper-analysis})---the solution
of this system readily determines the unknown coefficients $\alpha, \beta_1, \dots, \beta_\ell$.
For example, for $\ell=1$ (i.e. for $q=2$, or equivalently $\lam=4$), this set of equations
gives $\alpha=3/8$ and $\beta_1=3/4$. Upon substituting these in \eqref{P3eq: hyper-lam>2}, the
solution for the scalar field follows from Eq. \eqref{P3eq: power-Phi} and has the form
\gat$\Phi_{\pm}(y)=\pm\frac{\Phi_0}{5\sqrt{5}}\left[1-\frac{5}{k(y+y_0)}\right]^{\frac{1}{2}}
\left[\frac{1}{2}\left(1-\frac{5}{k(y+y_0)}\right)^{-\frac{1}{2}}\arcsin\left(\sqrt{1-\frac{5}
{k(y+y_0)}}\right)+\right.\nonum\\[3mm]
\hspace{15em}+\left.\sqrt{\frac{5}{k(y+y_0)}}\left(\frac{1}{2}-\frac{5}{k(y+y_0)}\right)\right].$
\end{enumerate}

\begin{figure}[t]
    \centering
    \begin{subfigure}[b]{0.46\textwidth}
        \includegraphics[width=\textwidth]{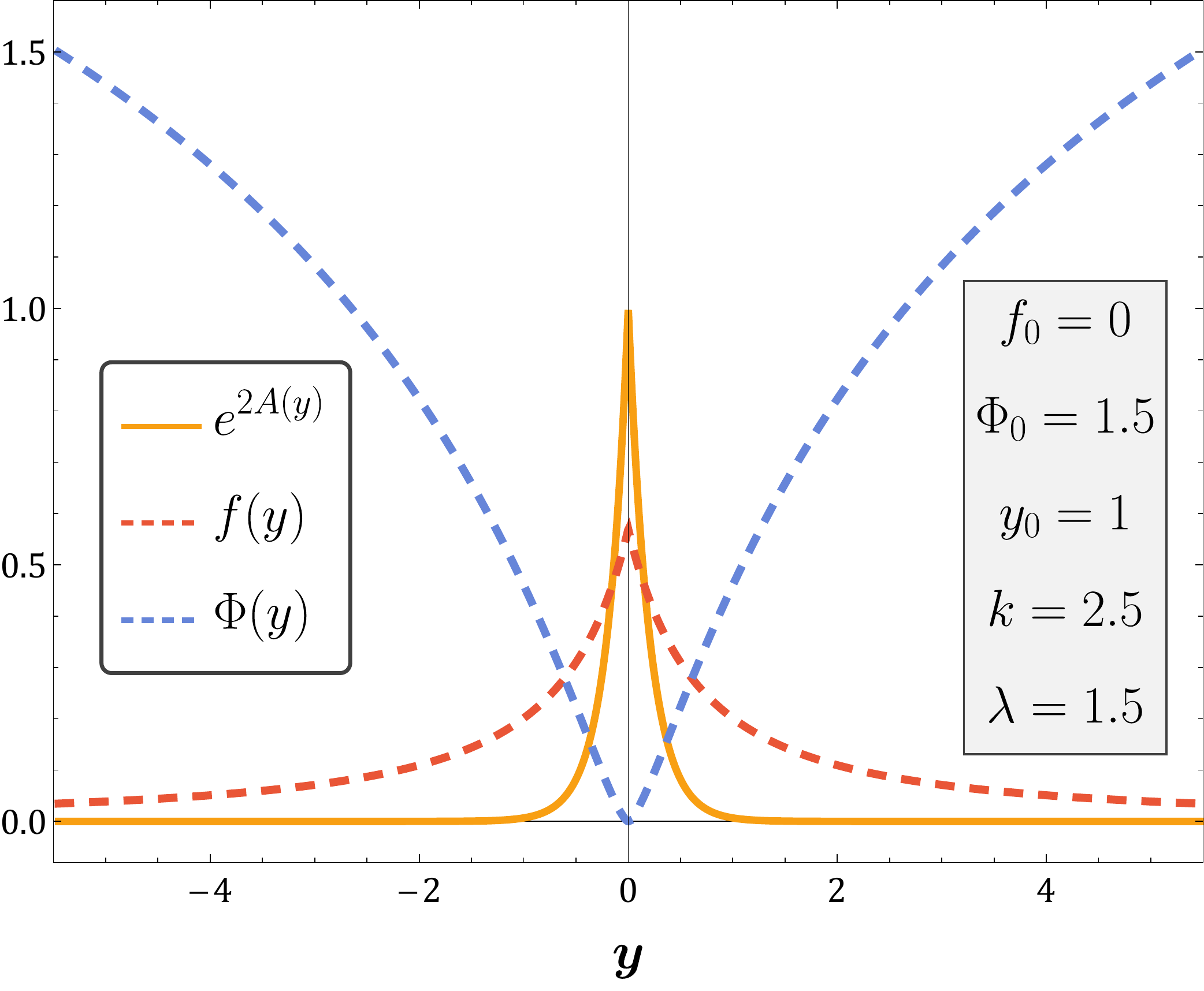}
        \caption{\hspace*{-1.2em}}
        \label{P3subf: hyper-plot1}
    \end{subfigure}
    \hfill
    ~ 
    \begin{subfigure}[b]{0.50\textwidth}   
        \includegraphics[width=\textwidth]{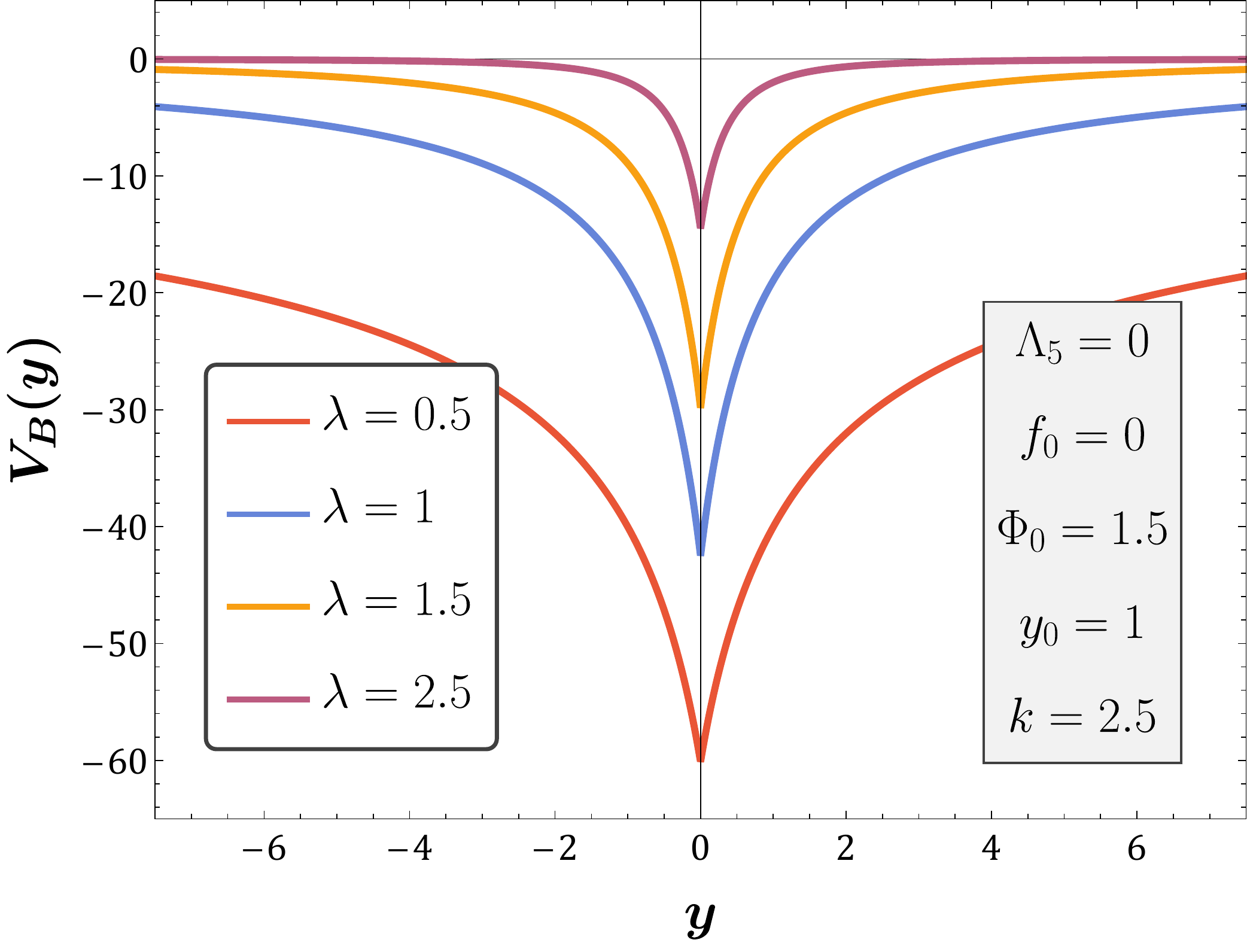}
        \caption{\hspace*{-2.9em}}
        \label{P3subf: hyper-plot-pot}
    \end{subfigure}
    \vspace{-1em}   
    ~ 
    \caption{(a) The warp factor $e^{2A(y)}=e^{-2k|y|}$, the coupling function $f(y)$ and the
    scalar field $\Phi(y)$ in terms of the coordinate $y$ for $f_0=0$, $\Phi_0=1.5$, $y_0=1$,
    $k=2.5$ and $\lam=1.5$. (b) The scalar potential $V_B$ in terms of the extra dimension $y$ for 
    $\Lambda_5=0$, $f_0=0$, $\Phi_0=1.5$, $y_0=1$, $k=2.5$ and $\lam=0.5,\, 1,\, 1.5,\, 2.5$ 
    (from bottom to top).}
   \label{P3fig: hyper-plot-f-pot}
\end{figure}

\vspace*{-1em}

In \myref{P3fig: hyper-plot-f-pot}{P3subf: hyper-plot1}, we depict the coupling function $f(y)$ and the scalar field $\Phi(y)$
for the indicative set of parameters $f_0=0$, $\Phi_0=1.5$, $y_0=1$, $k=2.5$ and $\lam=1.5$. For
comparison, we also display the exponentially decreasing warp factor. The coupling function remains
localized near the brane and asymptotically decreases to the constant value $f_0$, which here has
been taken to be zero. The scalar field starts from a constant value at the location of the brane,
which for this set of parameters turns out to be zero, and goes asymptotically to a constant value
that depends on the values of $\Phi_0$ and $\lambda$. Although this is not very clear from 
\myref{P3fig: hyper-plot-f-pot}{P3subf: hyper-plot1}, it easily follows from Eq. (\ref{P3eq: power-Phi}) with the asymptotic value of
the scalar field, as $y \rightarrow \pm\infty$, coming out to be
\eq$\lim_{y\ra\pm\infty}\Phi_{\pm}(y)=\pm \frac{\sqrt{\pi}\Phi_0}{2}\sqrt{\frac{\lam}{(\lam+1)^{\lam-1}}}
\ \frac{\Gamma\left(\frac{\lam}{2}-\frac{1}{2}\right)}{\Gamma\left(\frac{\lam}{2}+1\right)}\,.$
It is worth noting that the profiles of both $f(y)$ and $\Phi(y)$ do not change with the variation
of the values of the parameters. 
The potential of the scalar field $V_B(y)$ in the bulk can be determined from Eq. \eqref{P3eq: V-B} 
using the expression of the coupling function $f(y)$. Thus, we find
\eq$\label{P3eq: hyper-V-y}
V_B(y)=-\Lambda_5-6k^2f_0-\frac{\Phi_0^2}{2k^\lam (y+y_0)^{\lam+2}}\left[12k^2(y+y_0)^2+7\lam k(y+
y_0)+\lam(\lam+1)\right]\, .$
Since $\lam>0$, the last term in the above expression is negative-definite; it also vanishes as
$y \rightarrow +\infty$ leaving the parameters $\Lambda_5$ and $f_0$ to determine its asymptotic
value. Depending on the values of the parameters, the potential may be either positive or negative
at the location of the brane or asymptotically far away. In \myref{P3fig: hyper-plot-f-pot}{P3subf: hyper-plot-pot}, one
can observe the aforementioned behaviour of the scalar potential $V_B(y)$. The values of the fixed
parameters $f_0,\, \Phi_0,\, y_0,\, k$ are the same as in 
\myref{P3fig: hyper-plot-f-pot}{P3subf: hyper-plot1}, while the parameter $\lam$ varies.

\par The energy density $\rho(y)$ and pressure $p(y)=p^i(y)=p^y(y)$ may be finally computed by
employing  Eqs. \eqref{P3eq: linear-rho} and \eqref{P3eq: linear-p}. Then, we are led to the result
\eq$\label{P3eq: power-rho}
\rho(y)=-p(y)=-6k^2f(y)=-6k^2\left[f_0+\frac{\Phi_0^2}{k^\lam(y+y_0)^\lam}\right].$
If we wish to satisfy the weak energy conditions close and on the brane, we should have $\rho(0)>0$,
which in turn means $f(0)<0$; in that case, the parameters of the model should satisfy the
following inequality: 
\eq$\label{P3eq: power-weak-con}
\frac{f_0}{\Phi_0^2}<-\frac{1}{(ky_0)^\lam}\,.$

\vspace*{-1em}

\mysubsection{Junction conditions and the effective theory}

\par From the relations \eqref{P3eq: jun_con1} and \eqref{P3eq: jun_con2}, we obtain the following
junction conditions for the matter on the brane:
\gat$\label{P3eq: jun-con1}
3f(y)[A']=-[f']-(\sig+V_b)\,, \\[3mm]
\label{P3eq: jun-con2}
[\Phi']=4[A']\pa_\Phi f+\pa_\Phi V_b\,,$
where all quantities are again evaluated at $y=0$. The only difference in this case is that we
have used the derivatives of the coupling function with respect to the coordinate $y$ rather
than the one with respect to the scalar field. This is due to the fact that the explicit expression
of the function $f(\Phi)$ is not know---although $\Phi(y)$ is a one-to-one function, it 
cannot in general be inverted. Taking advantage of the $\mathbf{Z}_2$ symmetry in the bulk,
we can easily evaluate the total energy density on the brane by Eq. \eqref{P3eq: jun-con2}, which
is given by
\eq$\label{P3eq: power-jc1}
\sig+V_b(\Phi)\Big|_{y=0}=6kf_0+\frac{2\Phi_0^2}{k^\lam\,y_0^{\lam+1}}(3ky_0+\lam)=6k\Phi_0^2
\left[\frac{f_0}{\Phi_0^2}+\frac{3ky_0+\lam}{3(ky_0)^{\lam+1}}\right]\,.$
If we demand the total energy density on the brane to be positive, namely $\sig+V_b(\Phi)|_{y=0}>0$,
then we straightforwardly deduce the constraint
\eq$\label{P3eq: power-brane-ene-con}
\frac{f_0}{\Phi_0^2}>-\frac{3ky_0+\lam}{3(ky_0)^{\lam+1}}\,.$
In order to evaluate the first jump condition (\ref{P3eq: jun-con1}), we write:
$\partial_\Phi f= \partial_y f/\Phi'$ and $\partial_\Phi V_b= \partial_y V_b/\Phi'$.
We are allowed to do this since, as we mentioned previously, the function $\Phi(y)$
does not possess any extrema in the bulk, therefore $\Phi'(y)$ never vanishes. Then, 
multiplying both sides of Eq. (\ref{P3eq: jun-con1}) by $\Phi'$ and using Eq. 
(\ref{P3eq: power-der-Phi}), we obtain the condition
\eq$\label{P3eq: power-jc2}
\pa_y V_b\Big|_{y=0}=-\frac{2\lam\Phi_0^2}{k^\lam\, y_0^{\lam+2}}(3ky_0+\lam+1)\,.$
Due to the fact that $\lam>0$, $k>0$ and $y_0>0$, the r.h.s. of the above equation
never vanishes, which means that $V_b\neq \text{constant}$.

\par Let us now focus on the effective four-dimensional theory on the brane. Using
Eq. \eqref{P3eq: action_eff} and the  expression for the coupling function, from
Eq. \eqref{P3eq: power-f} we obtain
\bal$
\frac{1}{\kappa_4^2}&=
\frac{f_0}{k}+\frac{2\Phi_0^2}{k^\lam}\int_0^\infty dy \frac{e^{-2ky}}{(y+y_0)^\lam}
=\frac{f_0}{k}+\frac{2\Phi_0^2\,e^{2ky_0}}{k^\lam}\int_0^\infty dy\frac{e^{-2k(y+y_0)}}
{(y+y_0)^\lam}\, .$
Setting $t=2k(y+y_0)$, the above relation takes the form
\eq$\label{P3eq: power-kappa1}
\frac{1}{\kappa_4^2}=\frac{f_0}{k}+\frac{2\Phi_0^2\, e^{2ky_0}}{k^\lam}(2k)^{\lam-1}
\int_{2ky_0}^{\infty}dt\ t^{-\lam} \,e^{-t}=\frac{f_0}{k}+\frac{2^\lam\Phi_0^2\, e^{2ky_0}}
{k}\,\Gamma(1-\lam,2ky_0)\,.$
Above, we have used the upper incomplete gamma function $\Gamma(s,x)$, defined
as follows
\eq$\label{P3eq: upper-gamma}
\Gamma(s,x)\equiv\int_x^\infty dt\ t^{s-1}\, e^{-t}\,.$
The properties of the incomplete gamma function as well as the expressions giving its
numerical values are discussed in Appendix \ref{P3app: incom-gamma}. With the use of Eq.
\eqref{P3eq: power-kappa1} and the relation $1/\kappa_4^2=M_{Pl}^2/(8\pi)$, we 
finally obtain
\eq$\label{P3eq: power-planck}
M_{Pl}^2=\frac{8\pi\Phi_0^2}{k}\left\{\frac{f_0}{\Phi_0^2}+2^\lam\,e^{2ky_0}\,\Gamma(1-\lam,2ky_0)
\right\}\, .$
Demanding the positivity of the effective four-dimensional gravitational scale $M_{Pl}^2$, we
are led to the additional constraint
\eq$\label{P3eq: power-eff-con}
\frac{f_0}{\Phi_0^2}>-2^\lam\,e^{2ky_0}\,\Gamma(1-\lam,2ky_0)\,.$
\par Finally, substituting the total energy density on the brane from Eq. \eqref{P3eq: power-jc1} and the
expressions of the functions $f(y)$, $\Phi(y)$\,\footnote{For the calculation of the effective
four-dimensional cosmological constant on the brane $\Lambda_4$, it is more convenient to
use the relation \eqref{P3eq: power-der-Phi} instead of the explicit form of the scalar field $\Phi(y)$
as given by Eq. \eqref{P3eq: power-Phi}.} and $V_B(y)$ in Eq. \eqref{P3eq: cosm_eff}, we can verify 
that the effective four-dimensional cosmological constant on the brane is zero, as expected.

\newpage


\vspace*{-4em}

\mysubsection{Energy conditions and the parameter space}

In this subsection, we will study the parameter space of the ratio $f_0/\Phi_0^2\,$
and the dimensionless parameter $ky_0$. The value of the parameter $\lam$ may be also varied,
however, once fixed, it determines the allowed values of the parameter $ky_0$ through the
constraint \eqref{P3eq: power-con1}. As usually, we will investigate the parameter regimes where
the inequalities \eqref{P3eq: power-weak-con}, \eqref{P3eq: power-brane-ene-con} and \eqref{P3eq: power-eff-con} 
are satisfied.

\begin{SCfigure}[][h]
    \centering
    \includegraphics[width=0.6\textwidth]{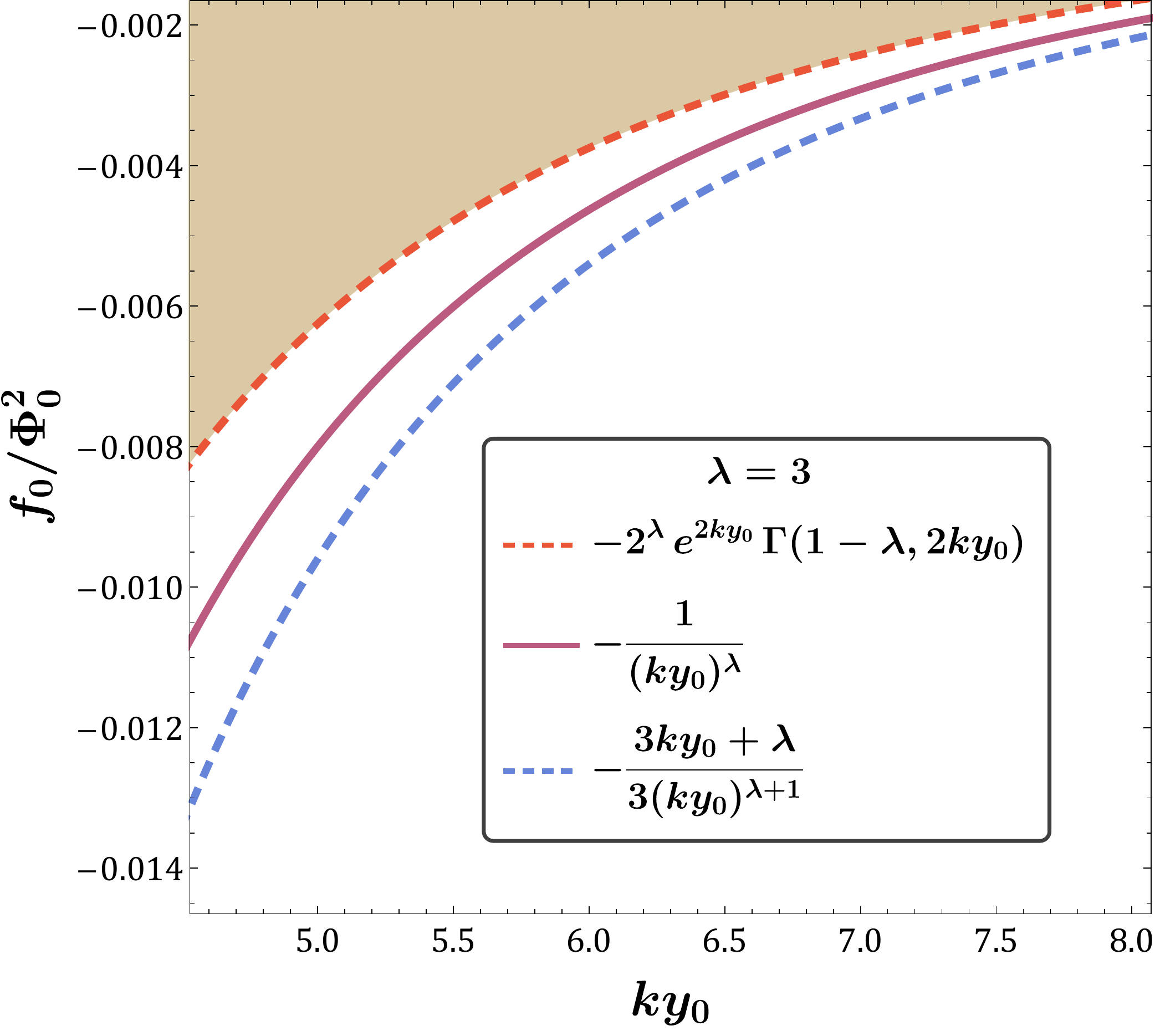}
    \vspace{0em}
    \caption{The parameter space between the ratio $f_0/\Phi_0^2$
    and the parameter $ky_0$, for $\lam=3$. The figure depicts also the plots of the 
    expressions appearing on the r.h.s.'s of the inequalities \eqref{P3eq: power-weak-con},
    \eqref{P3eq: power-brane-ene-con} and \eqref{P3eq: power-eff-con}.\\
    \vspace{3em}}
    \label{P3fig: hyper-plot-par}  
\end{SCfigure}

In \fref{P3fig: hyper-plot-par}, we depict the aforementioned parameter space for the 
value $\lam=3$. We also depict the curves of the expressions on the r.h.s.'s of the inequalities
\eqref{P3eq: power-weak-con}, \eqref{P3eq: power-brane-ene-con} and \eqref{P3eq: power-eff-con}. 
Although the corresponding  curves have been drawn for a particular value of $\lam$, it turns
out that their relative position remains the same for any allowed value of the parameters
$\lam$ and $ky_0$, namely it always holds that
$$-2^\lam\,e^{2ky_0}\,\Gamma(1-\lam,2ky_0)>-\frac{1}{(ky_0)^\lam}>-\frac{3ky_0+\lam}{3(k
y_0)^{\lam+1}}\,.$$
Clearly, this means that only the inequalities \eqref{P3eq: power-brane-ene-con} and \eqref{P3eq: power-eff-con}
can be simultaneously satisfied. Therefore, we may easily obtain a model with a positive four-dimensional 
gravitational constant and a positive total energy density on the brane. However, in that case, 
we will not be able to satisfy the weak energy conditions by the bulk matter close and on the brane. 
This means that the energy density $\rho$ will be negative at the location of the brane with
the pressure having the exact opposite value.

%
%

\mysection{Linear-exponential coupling function in terms of y \label{P3sec: lin-exp}}

In this case, we consider the following coupling function $f(y)$ in terms
of the coordinate $y$:
\eq$\label{P3eq: exp-f}  
f(y)=f_0+f_1\, ky\,e^{-\lam ky}\,.$
We also assume that $f_1\in \mathbb{R}\setminus\{0\}$ and $\lam\in(0,+\infty)$ in order
for $f(y)$ to  satisfy the physical constraints discussed at the end of \secref{P3sec: th-frame}. 

Let us start by deriving first the bulk solution. Substituting the aforementioned coupling
function in Eq. \eqref{P3eq: grav-1-2}, we obtain:
\eq$\label{P3eq: exp-der-Phi}
[\Phi'(y)]^2=f_1\,k^2\,e^{-\lam ky}\left[2\lam-1-\lam(\lam-1)ky\right]\,.$
Since the scalar field $\Phi(y)$ should be a real-valued function, it is obvious
that $[\Phi'(y)]^2\geq 0$ for all values of $y$ which are greater or equal to zero.
Let us first assume that $f_1<0$; then, demanding that $[\Phi'(0)]^2 \geq 0$, we obtain
the constraint $\lam\leq 1/2$. On the other hand, for large values of the $y$-coordinate
(i.e. at $y=y_0\gg 1$), demanding that $[\Phi'(y_0)]^2 \geq 0$ leads to $\lam\geq 1$.\,\footnote{
Here, we have used the fact that, for large values of $y$, only the term proportional to $ky$ mainly
contributes to the value of $[\Phi'(y)]^2$.} However, these two constraints are incompatible,
which leads us to deduce that the parameter $f_1$ should be strictly positive. In that
case, a similar argument as above leads to the allowed regime $\lam\in\left[\frac{1}{2},1\right]$. 
Moreover, since $f_1$ is positive, we may set $f_1=\Phi_0^2$, and assume for simplicity
that $\Phi_0\in(0,+\infty)$.

For $\lam=1$, we can easily integrate Eq. \eqref{P3eq: exp-der-Phi} with respect to $y$, and
determine the expression of the function of the scalar field $\Phi(y)$. Then, we obtain
\eq$\label{P3eq: exp-phi1}
\Phi_{\pm}(y)=\pm\, 2\Phi_0\, e^{-ky/2}\,.$
Above, we have used again the translational symmetry of the gravitational field equations 
with respect to the value of the scalar field in order to eliminate an additive integration
constant. By inverting the above function, we can express the coupling function in terms
of the scalar field $\Phi$, namely
\eq$\label{P3eq: exp-f-phi}
f(\Phi)=f_0-\frac{\ \Phi^2}{2}\ln\left(\frac{\Phi}{2\Phi_0}\right)\,.$

Equation \eqref{P3eq: exp-der-Phi} is more difficult to solve in the remaining $\lam$-parameter
regime, i.e. for $\lam\in\left[\frac{1}{2},1\right)$. In that case, Eq. \eqref{P3eq: exp-der-Phi} 
leads to 
{\fontsize{10}{10}\bal$\Phi_{\pm}(y)&=\pm\,\Phi_0\,k\int dy\, e^{-\lam ky/2}\sqrt{2\lam-1-\lam(\lam-1)
ky}\nonum\\[3mm]
&=\pm\,\frac{2\Phi_0}{\lam}\left[-e^{-\lam ky/2}\sqrt{2\lam-1-\lam(\lam-1)ky}
+\int dy\, e^{-\lam ky/2}\,\frac{d}{dy}\left(\sqrt{2\lam-1-\lam(\lam-1)ky}\right)
\right].
\label{P3eq: exp-phi2-1}$}
\hspace{-0.5em}Focusing on the second term of the r.h.s. of the above relation, and due to the
fact that $\lam\in\left[\frac{1}{2},1\right)$, we can write
\bal$\int dy\,e^{-\frac{\lam ky}{2}}\,\frac{d}{dy}\left(\text{\small{$\sqrt{2\lam
-1-\lam(\lam-1)ky}$}}\right)&=\sqrt{\frac{\pi(1-\lam)}{2}}\,e^{\frac{2\lam-1}{2(1-\lam)}}\times\nonum\\[1mm]
&\hspace{1.5em}\times\,\text{erf}\left(
\text{\small{$\sqrt{\frac{2\lam-1-\lam(\lam-1)ky}{2(1-\lam)}}$}}\,\right)\,,
\label{P3eq: exp-phi2-2}$
where we have used the error function, defined as
$$\text{erf}(x)=\frac{2}{\sqrt\pi}\int_0^x dt\, e^{-t^2}\,,$$
and its property
$$\frac{d}{dx}\, \text{erf}(g(x))=\frac{2}{\sqrt\pi}e^{-g(x)^2}\, \frac{dg(x)}{dx}\,.$$
Combining Eqs. \eqref{P3eq: exp-phi2-1} and \eqref{P3eq: exp-phi2-2} we obtain
\bal$
\Phi_{\pm}(y)=\pm\,\frac{2\Phi_0}{\lam}\Bigg[&-e^{-\lam ky/2}\text{\footnotesize{$\sqrt{
2\lam-1-\lam(\lam-1)ky}$}}\nonum\\[0mm]
&+\sqrt{\frac{\pi(1-\lam)}{2}}\,e^{\frac{2\lam-1}{2(1-
\lam)}}\,\text{erf}\left(
\text{\footnotesize{$\sqrt{\frac{2\lam-1-\lam(\lam-1)ky}{2(1-\lam)}}$}}\right)\Bigg].
\label{P3eq: exp-phi2}$
In this case, it is not possible to invert the function $\Phi(y)$ in order to find the
form of the coupling function $f(\Phi)$. However, from Eq. \eqref{P3eq: exp-der-Phi} and for
$\lam\in\left[\frac{1}{2},1\right)$, it is straightforward to deduce that $\Phi'(y)\neq 0$
for all $y>0$; this, again, means that $\Phi(y)$ does not have any extremum, 
and is therefore a one-to-one function. This property will be of use in the evaluation
of the junction conditions on the brane.

\begin{figure}[t]
    \centering
    \begin{subfigure}[b]{0.495\textwidth}
        \includegraphics[width=\textwidth]{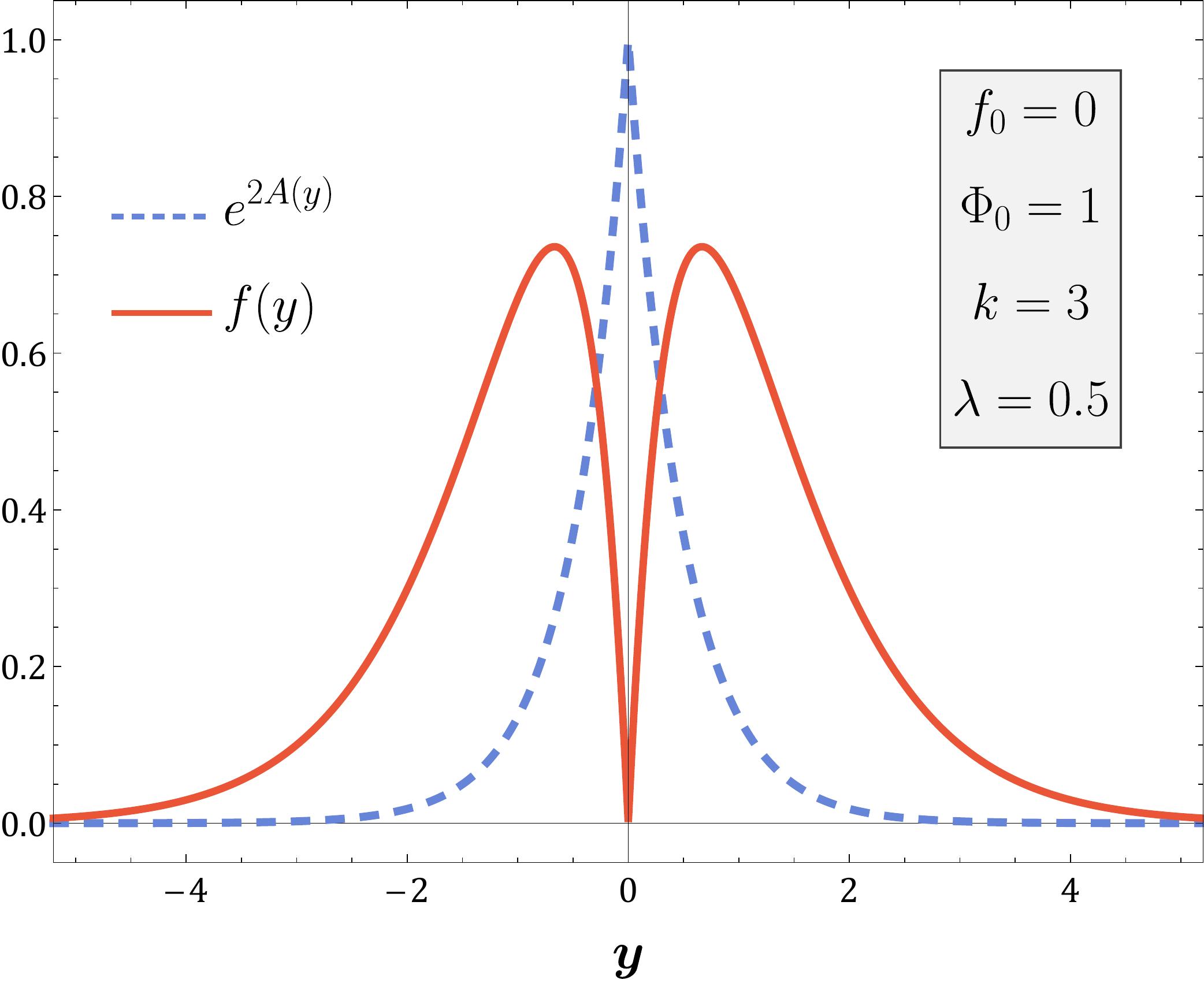}
        \caption{\hspace*{-1.2em}}
        \label{P3subf: erf-plot1}
    \end{subfigure}
    ~ 
    \begin{subfigure}[b]{0.44\textwidth}
        \includegraphics[width=\textwidth]{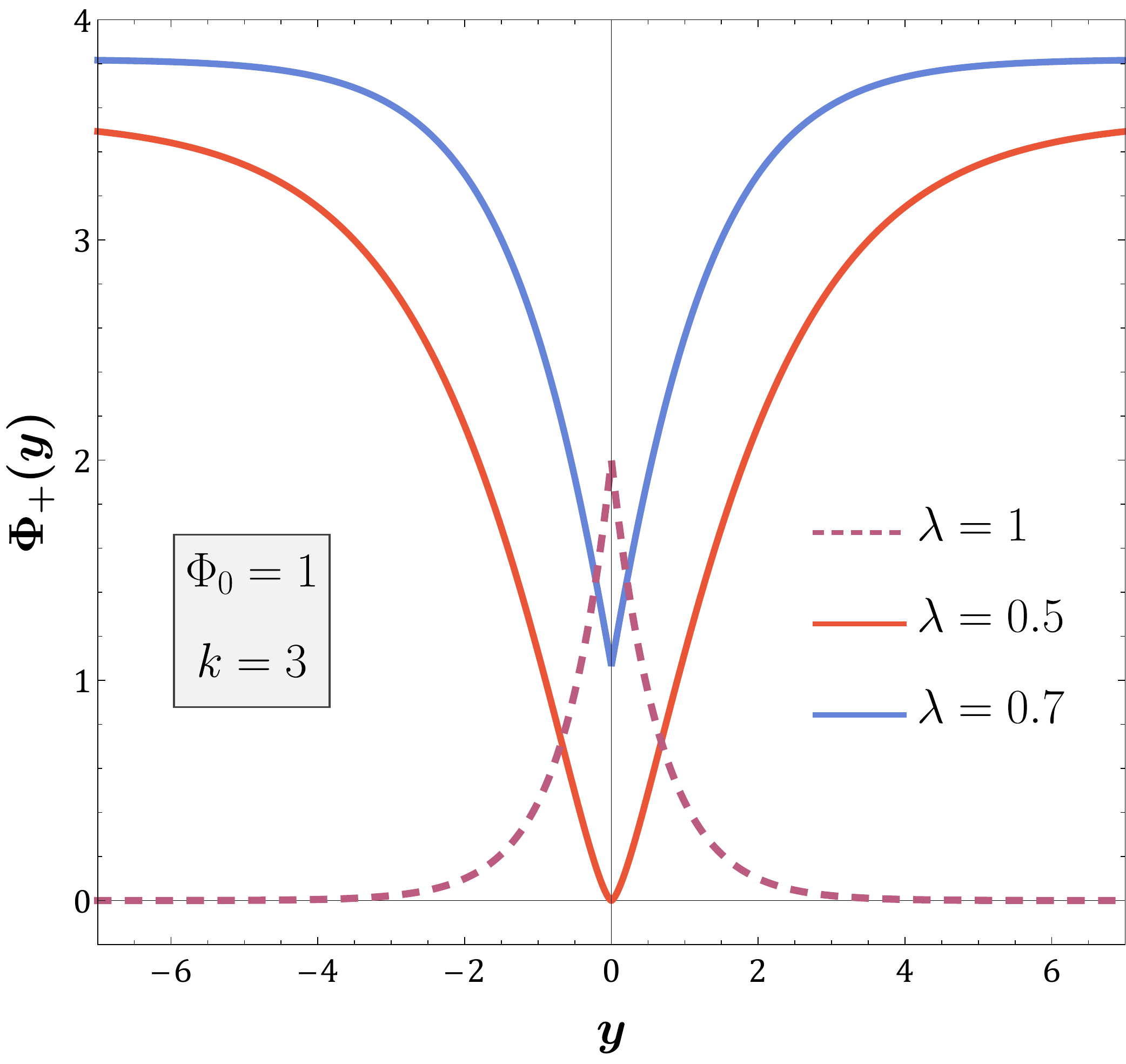}
        \caption{\hspace*{-1.9em}}
        \label{P3subf: erf-plot2}
    \end{subfigure}
    
    \vspace{-0.5em}
    \caption{(a) The coupling function $f(y)$ and the warp factor $e^{-2A(y)}$ in terms
    of the $y$-coordinate for $f_0=0$, $\Phi_0=1$, $k=3$ and $\lam=0.5$, and (b)
    the scalar field $\Phi_+(y)$ for $\Phi_0=1$, $k=3$ and $\lam=1,\,0.5,\,0.7$.}
    \label{P3fig: erf-plot-1-2}
\end{figure}

In \myref{P3fig: erf-plot-1-2}{P3subf: erf-plot1} and \myref{P3fig: erf-plot-1-2}{P3subf: erf-plot2}, we present
the warp factor, the coupling function and the scalar field for particular choices of values
for the parameters of the model. The coupling function $f(y)$ adopts the same constant
value $f_0$ at the location of our brane and at asymptotic infinity while reaching a
maximum value at some intermediate distance off our brane, as depicted in
\myref{P3fig: erf-plot-1-2}{P3subf: erf-plot1}. In \myref{P3fig: erf-plot-1-2}{P3subf: erf-plot2}, the
scalar field presents two distinct profiles, for $\lam=1$ and $\lam \in\left[\frac{1}{2}
,1\right)$ due to the two different solutions given by Eqs. \eqref{P3eq: exp-phi1} and
\eqref{P3eq: exp-phi2}, respectively. In all cases, though, $\Phi_\pm(y)$ remains everywhere finite
approaching a constant value at asymptotic infinity. For $\lam=1$ this constant is zero,
while for $\lam \in\left[\frac{1}{2},1\right)$ this is given by the expression
\eq$\label{P3eq: exp-phi-inf}
\lim_{y\ra\pm\infty}\Phi_{\pm}(y)=\pm\frac{2\Phi_0}{\lam}\sqrt{\frac{\pi(1-\lam)}{2}}\,
e^{\frac{2\lam-1}{2(1-\lam)}},\hspace{1em}\lam\in\left[\frac{1}{2},1\right)\,.$
In the above, we have used the fact that the limit of the error function appearing in
Eq. \eqref{P3eq: exp-phi2}, as $y \rightarrow +\infty$, is unity. Due to the $\mathbf{Z}_2$
symmetry imposed on our model, the same limit will hold for the scalar field also
for $y\ra-\infty$.

\par
\begin{figure}
    \centering
    \includegraphics[width=0.52\textwidth]{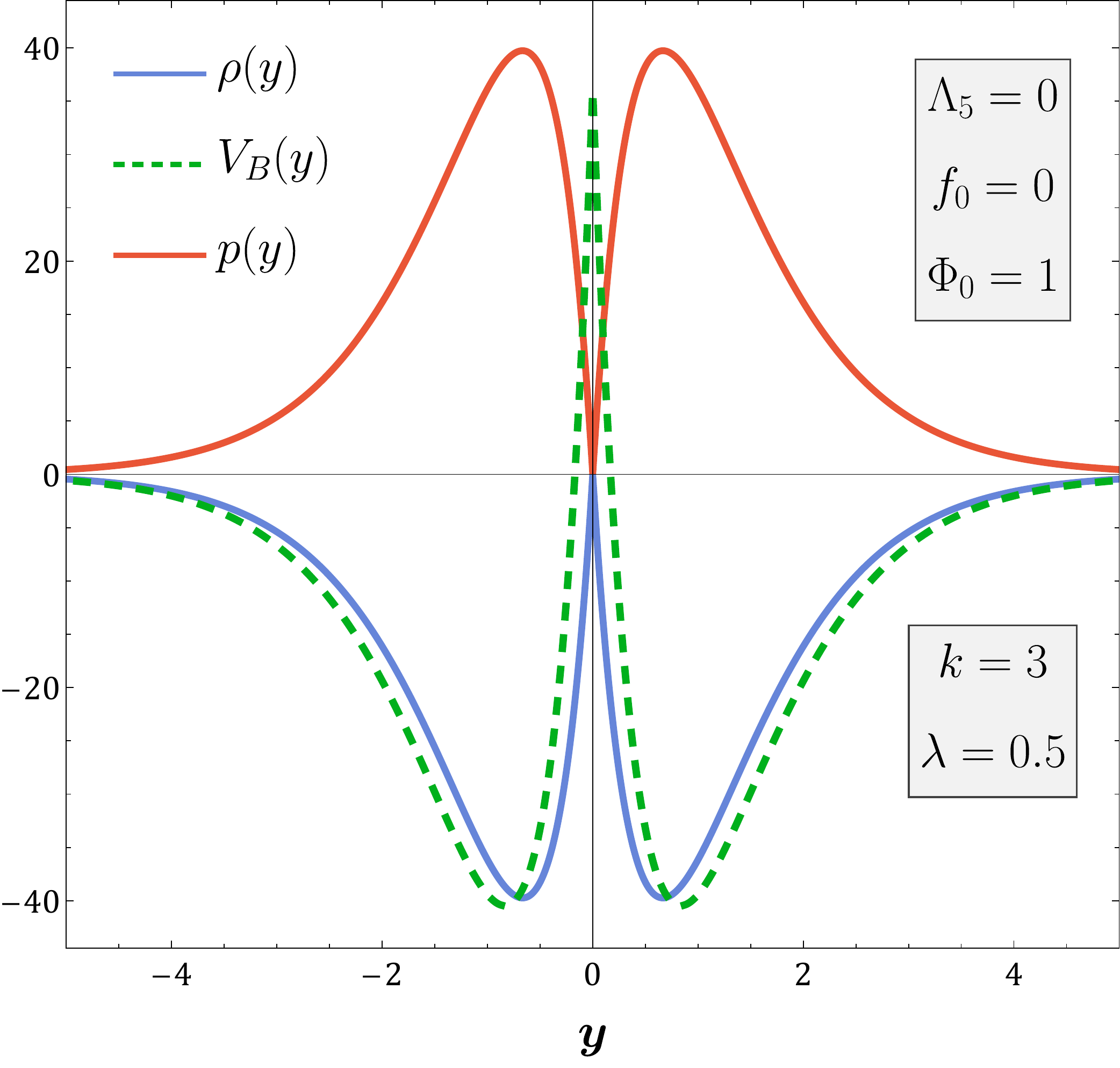}
    \vspace{0em}
    \caption{The energy density $\rho$ and pressure $p$ of the system together with
    the scalar potential $V_B$ in terms of the coordinate $y$ for $\Lambda_5=0$, $f_0=0$,
    $\Phi_0=1$, $k=3$, and $\lam=0.5$.}
    \label{P3fig: erf-plot3}  
\end{figure}

 From Eq. \eqref{P3eq: V-B} , we may now determine the potential of the scalar field $V_B(y)$
in the bulk by using the expression of the coupling function $f(y)$. Then, we find
\eq$\label{P3eq: exp-V-y}
V_B(y)=-\Lambda_5-6k^2f_0+\Phi_0^2\,k^2\,e^{-\lam ky}\left[\frac{7}{2}+\lam-ky\left(\frac{
\ \lam^2}{2}+\frac{7\lam}{2}+6\right)\right],\hspace{1em}\lam\in\left[\frac{1}{2},1\right]\, .$
On the other hand, the energy density $\rho(y)$ and pressure $p(y)=p^i(y)=p^y(y)$ may be
computed by employing Eqs. \eqref{P3eq: linear-rho} and \eqref{P3eq: linear-p}; thus, we obtain
\eq$\label{P3eq: exp-rho}
\rho(y)=-p(y)=-6k^2f(y)=-6k^2\left(f_0+\Phi_0^2\, ky\,e^{-\lam ky}\right),$
In order to satisfy the weak energy conditions close and on the brane, we should have again
$\rho(0) \geq 0$, or equivalently $f(0) \leq 0$; hence, we are led to the following inequality: 
\eq$\label{P3eq: exp-weak-con}
\frac{f_0}{\Phi_0^2} \leq 0\,.$
In \fref{P3fig: erf-plot3}, we present the energy-density and pressure as well as the profile
of the bulk potential for the same values of parameters as in \fref{P3fig: erf-plot-1-2}
for easy comparison. We observe that both components and the bulk potential are
everywhere finite, reach their maximum values at a finite distance from our brane and
reduce to a constant value (which here is taken to be zero) at large distances.

Let us now turn to the junction conditions introduced in the theory at the location of the brane.
From Eqs. \eqref{P3eq: jun-con1} and \eqref{P3eq: jun-con2}, we obtain in a similar way the conditions:
\gat$\label{P3eq: exp-jc1}
\sig+V_b(\Phi)\Big|_{y=0}=2k(3f_0-\Phi_0^2),\\[2mm]
\label{P3eq: exp-jc2}
\pa_yV_b\Big|_{y=0}=8k^2\,\Phi_0^2\left(\lam^2-\lam+\frac{5}{4}\right),$
for $\lam\in\left[\frac{1}{2},1\right]$. The total energy density on the brane will be 
positive if and only if $\sig+V_b(\Phi)|_{y=0}>0$, which results to
\eq$\label{P3eq: exp-brane-ene-con}
\frac{f_0}{\Phi_0^2}>\frac{1}{3}\, .$

Next, we are going to evaluate the effective four-dimensional gravitational constant on the brane.
Using Eq. \eqref{P3eq: action_eff}, we obtain
\gat$\frac{1}{\kappa_4^2}=\frac{M_{Pl}^2}{8\pi}=\frac{\Phi_0^2}{k}\left[\frac{f_0}{\Phi_0^2}+
\frac{2}{(2+\lam)^2}\right].$
For a robust effective theory on the brane, it is imperative to have a positive four-dimensional
gravitational constant, thus, we must satisfy the following constraint:
\eq$\label{P3eq: exp-eff-con}
\frac{f_0}{\Phi_0^2}>-\frac{2}{(2+\lam)^2}\,.$
Once again, as expected, the effective four-dimensional cosmological constant on the brane
may be found to be zero with the use of Eq. \eqref{P3eq: cosm_eff}.
 
It is straightforward to study whether the inequalities \eqref{P3eq: exp-weak-con}, \eqref{P3eq: exp-brane-ene-con}
and \eqref{P3eq: exp-eff-con} can be simultaneously satisfied. By merely observing the first two of
them, it is easy to deduce that they are incompatible since the value of $f_0/\Phi_0^2$ can be
either positive or negative. Additionally, as we already mentioned, the parameter $\lam$
takes values in the range $\left[\frac{1}{2},1\right]$. In this case, it holds that 
\eq$\label{P3eq: exp-eff-val}
-\frac{8}{25}\leq -\frac{2}{(2+\lam)^2}\leq -\frac{2}{9}\,.$
Hence, we can simultaneously satisfy either the inequalities \eqref{P3eq: exp-weak-con} and
\eqref{P3eq: exp-eff-con}, or \eqref{P3eq: exp-brane-ene-con} and \eqref{P3eq: exp-eff-con}. In particular,
a positive four-dimensional gravitational scale $M_{Pl}^2$ can be combined
with the bulk matter satisfying the weak energy conditions close to the brane, for
$$-\frac{2}{9}\leq \frac{f_0}{\Phi_0^2} \leq 0\,,$$
or with a positive total energy-density on the brane, for
$$\frac{f_0}{\Phi_0^2}>\frac{1}{3}\,.$$
The particular solution depicted in \fref{P3fig: erf-plot3} corresponds to the value $f_0=0$;
therefore, it is characterised by a negative energy density inside the bulk, which
violates the energy conditions. Note, however, that at the location of our brane,
both the energy density and pressure are zero while the bulk potential is positive.

%
%

\mysection{Double-exponential scalar field in terms of y \label{P3sec: double}}

In this section, we follow an alternative approach and consider the following expression for the scalar
field in terms of the coordinate $y$:
\eq$\label{P3eq: double-phi}
\Phi(y)=\Phi_0\, e^{-\mu^2 e^{ky}}\,.$
Although this expression seems similar to the sub-case of the quadratic coupling function
with $\lam=-1/4$, it differs significantly as it will become clear from the
expressions of the coupling function $f(\Phi)$ and the scalar potential $V_B(\Phi)$.
Moreover, it is obvious that both parameters $\Phi_0$ and $\mu$ can now take values in the 
entire set of real numbers except zero. 
With the form of the scalar field already known, it is straightforward to derive the corresponding
forms of the coupling function, bulk potential and components of the energy-momentum tensor.
Starting with the coupling function, upon substituting the aforementioned expression of the scalar
field in Eq. \eqref{P3eq: grav-1-2}, we readily obtain
\eq$\label{P3eq: double-f-y}
f(y)=f_0-f_1\,e^{-ky}-\frac{\Phi_0^2}{4\mu^2}\,e^{-2\mu^2e^{ky}}(\mu^2+e^{-ky})\,.$
In the above result, the parameter $f_1$ is allowed to take values in the whole set of real
numbers, while the allowed values for the parameter $f_0$ will be examined shortly. Inverting 
the function $\Phi(y)$, the expression of the coupling function in terms of the scalar field reads
\eq$\label{P3eq: double-f-phi}
f(\Phi)=f_0+\frac{f_1\, \mu^2}{\ln(\Phi/\Phi_0)}-\frac{\ \Phi^2}{4}\left(1-\frac{1}{\ln(
\Phi/\Phi_0)}\right)\,.$

\begin{figure}[t]
    \centering
    \begin{subfigure}[b]{0.47\textwidth}
        \includegraphics[width=\textwidth]{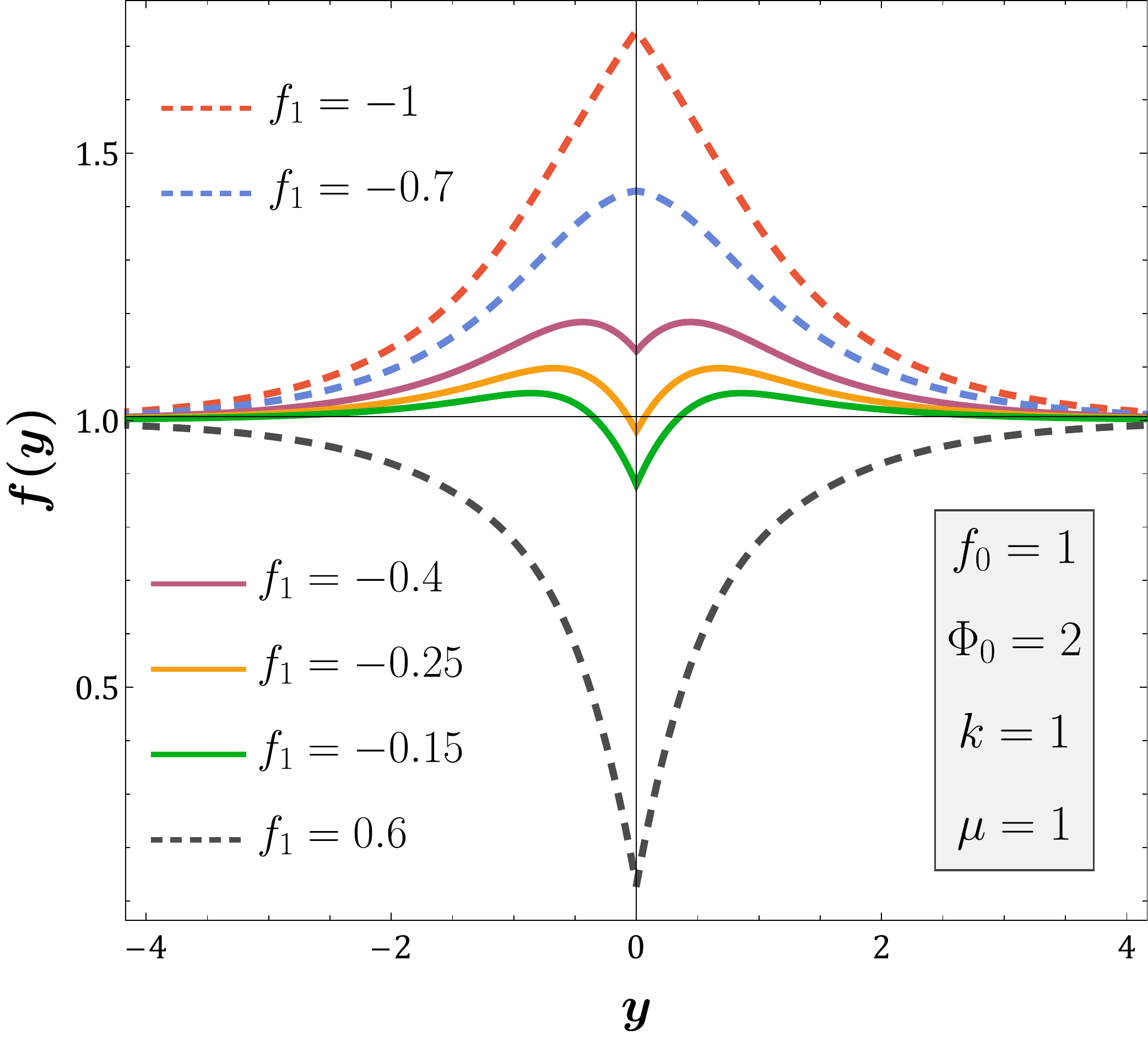}
        \caption{\hspace*{-2.5em}}
        \label{P3subf: double-plot1}
    \end{subfigure}
    \hspace*{-0.1cm}
    ~ 
    \begin{subfigure}[b]{0.47\textwidth}
        \includegraphics[width=\textwidth]{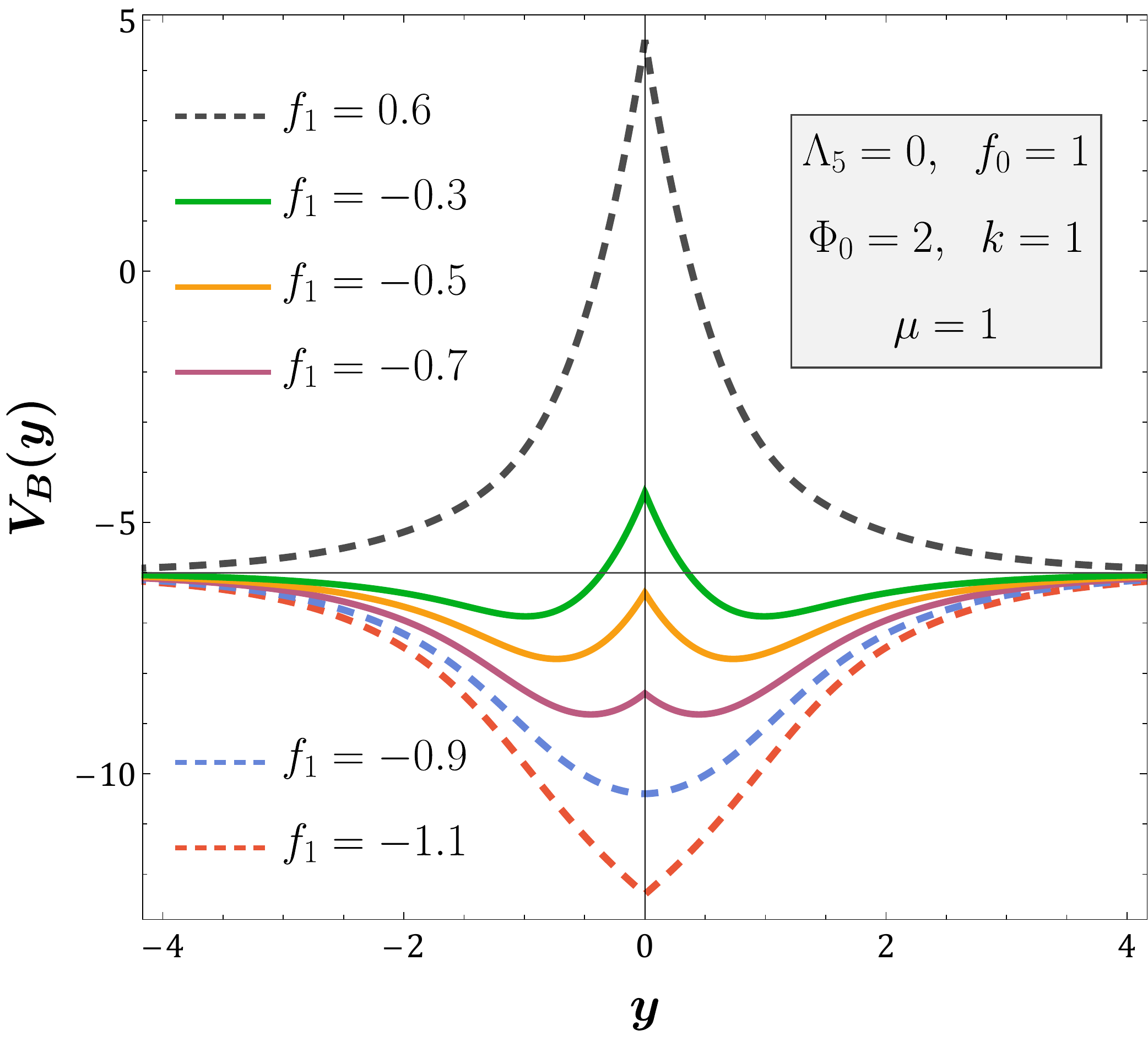}
        \caption{\hspace*{-2.7em}}
        \label{P3subf: double-plot2}
    \end{subfigure}
    \vspace{-0.5em}
    \caption{(a) The coupling function in terms of the $y$-coordinate for $f_0=1$, $\Phi_0=2$,
    $k=1$, $\mu=1$ and $f_1=-1,\,-0.7,\,-0.4,\,-0.25,\,-0.15,\,0.6$ (from top to bottom), and
    (b) the scalar potential $V_B$ in terms of the coordinate $y$ for $\Lambda_5=0$,
    $f_0=1$, $\Phi_0=2$, $k=1$, $\mu=1$ and $f_1=-1.1,\,-0.9,\,-0.7,\,-0.5,\,-0.3,\,0.6$ (from
    bottom to top).}
    \label{P3fig: double-plot-1-2}
\end{figure}

The scalar potential $V_B(y)$ can then be evaluated employing Eqs. \eqref{P3eq: V-B} and
\eqref{P3eq: double-f-y}. Then, we find
\eq$\label{P3eq: double-V-y}
V_B(y)=-\Lambda_5-6k^2f_0+10k^2f_1\,e^{-ky}+\frac{\Phi_0^2\,k^2}{2\mu^2}\,e^{-2\mu^2e^{ky}}
\left(5e^{-ky}+7\mu^2+4\mu^4\,e^{ky}+\mu^6\,e^{2ky}\right)\,,$
in terms of the $y$-coordinate, or
{\fontsize{11}{11}\eq$\label{P3eq: double-V-phi}
V_B(\Phi)=-\Lambda_5-6k^2f_0-\frac{10k^2f_1\,\mu^2}{\ln(\Phi/\Phi_0)}+\frac{\Phi_0^2\,k^2}{
2}\left\{7-\frac{5}{\ln(\Phi/\Phi_0)}-4\ln\left(\frac{\Phi}{\Phi_0}\right)+\left[\ln\left(
\frac{\Phi}{\Phi_0}\right)\right]^2\right\},$}
\hspace{-0.5em}in terms of the scalar field. In \myref{P3fig: double-plot-1-2}{P3subf: double-plot1} and
\myref{P3fig: double-plot-1-2}{P3subf: double-plot2}, we display the profiles of the coupling function and
scalar potential in terms of the $y$ coordinate, for particular values of the parameters of
the model. The varying parameter here is $f_1$, which is clearly the decisive one for the
form of both functions. 
We observe that for positive $f_1$, the coupling function takes its lowest value at the
location of the brane while, as $f_1$ gradually takes larger negative values, the coupling function
eventually exhibits a peak at the location of the brane. The bulk potential has almost the exact
opposite profile of the coupling function: it acquires a maximum, positive value at the location
of our brane for $f_1>0$ while it turns to globally negative values for $f_1<0$. For every set of
values of the parameters, though, both functions are everywhere finite and reduce to a constant
value at large distances---this value, when $\Lambda_5=0$, is determined by $f_0$.

Finally, the energy density $\rho(y)$ and pressure $p(y)=p^i(y)=p^y(y)$ components
may be computed as usually by employing Eqs. \eqref{P3eq: linear-rho} and \eqref{P3eq: linear-p}, in which case
we are led to the results 
\eq$\label{P3eq: double-rho}
\rho(y)=-p(y)=-6k^2f(y)=-6k^2\left[f_0-f_1\,e^{-ky}-\frac{\Phi_0^2}{4\mu^2}\,
e^{-2\mu^2e^{ky}}(\mu^2+e^{-ky})\right].$
In order to satisfy the weak energy conditions close and on the brane, we demand again that
$f(0)<0$; hence, we obtain the following inequality
\gat$\label{P3eq: double-weak-con}
\frac{f_0}{\Phi_0^2}<\frac{f_1}{\Phi_0^2}+\frac{\mu^2+1}{4\mu^2}\,e^{-2\mu^2}\,.$

Turning now to the junction conditions, from Eqs. \eqref{P3eq: jun-con1}, \eqref{P3eq: jun-con2} and using
also the relations \eqref{P3eq: double-phi}-\eqref{P3eq: double-f-phi}, we obtain:
\gat$\label{P3eq: double-jc1}
\sig+V_b(\Phi)\Big|_{y=0}=6kf_0-8kf_1-\frac{\Phi_0^2\,k\,e^{-2\mu^2}}{2\mu^2}(4+5\mu^2+2\mu^4)
\,,\\[3mm]
\label{P3eq: double-jc2}
\pa_yV_b\Big|_{y=0}=8k^2f_1+\frac{2\Phi_0^2\,	k^2}{\mu^2}\,e^{-2\mu^2}(1+2\mu^2+2\mu^4+\mu^6)\,.$
In the second of the above equations, we have used the relation $\pa_y V_b=\Phi'\,\pa_\Phi V_b$.
In order to have a positive total energy-density on the brane, we should have
\eq$\label{P3eq: double-brane-ene-con}
\frac{f_0}{\Phi_0^2}>\frac{4}{3}\frac{f_1}{\Phi_0^2}+\frac{e^{-2\mu^2}}{12\mu^2}(4+5\mu^2+2\mu^4)\,.$
\begin{figure}[t]
    \centering
    \begin{subfigure}[b]{0.51\textwidth}
        \includegraphics[width=\textwidth]{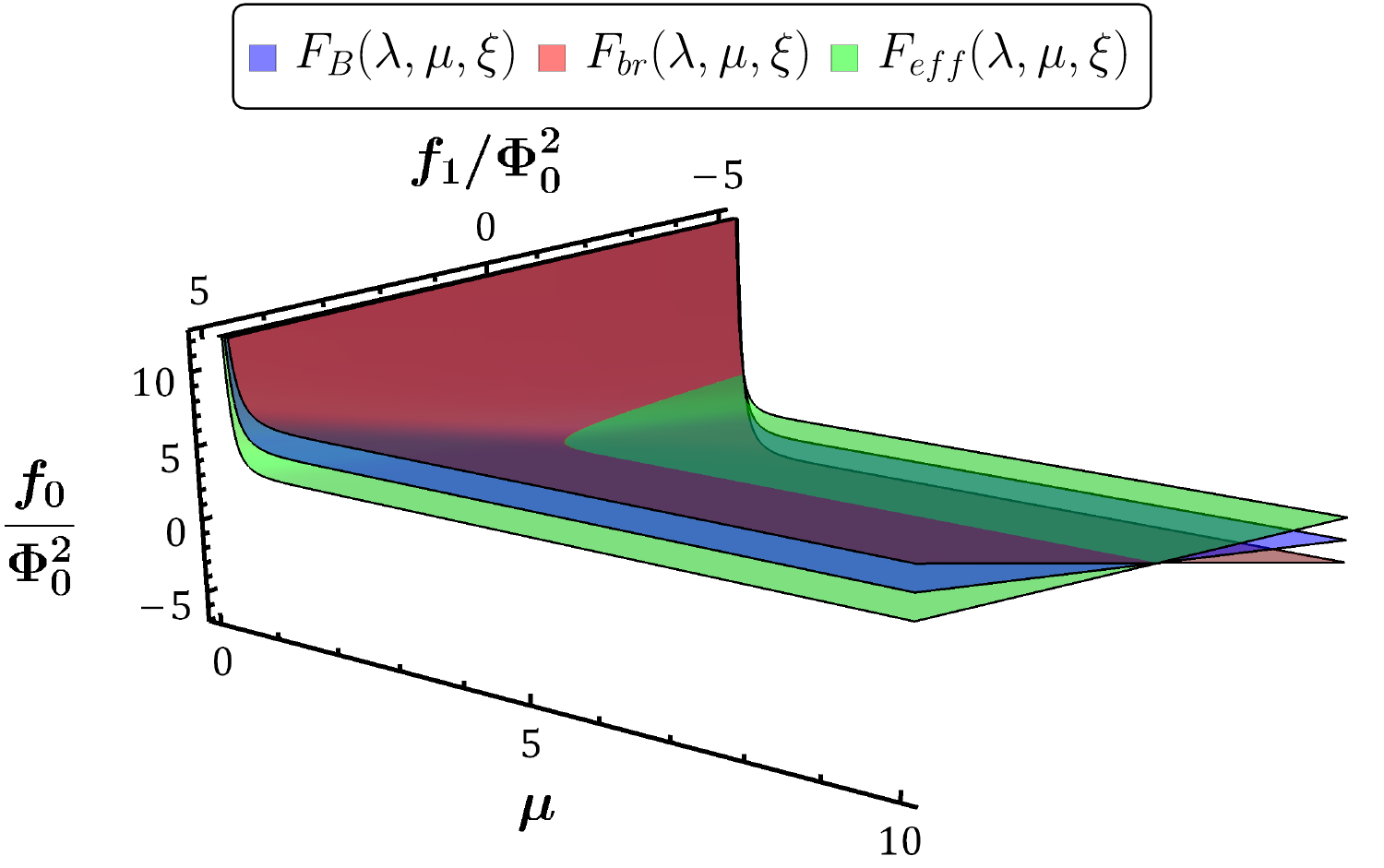}
        \caption{\hspace*{4.7em}}
        \label{P3subf: double-plot-par1}
    \end{subfigure}
    ~ 
    \begin{subfigure}[b]{0.42\textwidth}
        \includegraphics[width=\textwidth]{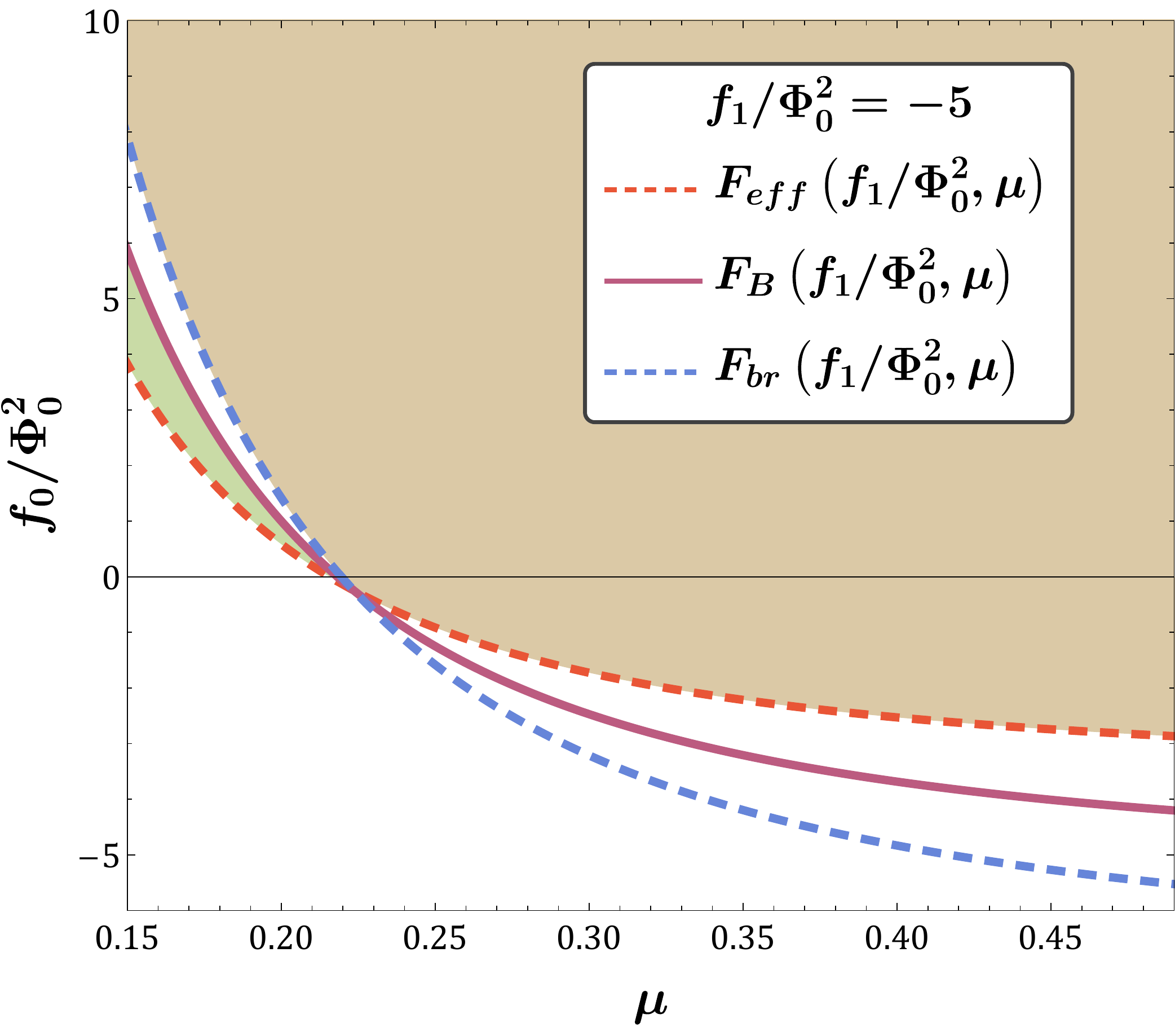}
        \caption{\hspace*{-2.2em}}
        \label{P3subf: double-plot-par2}
    \end{subfigure}\vspace*{1em}
    \vspace{-0.5em}
    \caption{(a) The 3D parameter space of the quantities $f_0/\Phi_0^2$, $f_1/\Phi_0^2$ and $\mu$. 
   (b) The 2D parameter space of the quantities $f_0/\Phi_0^2$ and $\mu$ for $f_1/\Phi_0^2=-5$.
    The figures also present the corresponding surfaces or curves of $F_{B}\left(f_1/\Phi_0^2,\mu\right)$, 
    $F_{br}\left(f_1/\Phi_0^2,\mu\right)$ and $F_{eff}\left(f_1/\Phi_0^2,\mu\right)$.}
    \label{P3fig: double-plot-par}
\end{figure}
\hspace{-0.5em}In the context of the effective theory on the brane, we may evaluate the four-dimensional gravitational
scale using Eqs. \eqref{P3eq: action_eff} and \eqref{P3eq: double-f-y}. Then, 
\bal$ \frac{1}{\kappa_4^2}&=
\frac{f_0}{k}-\frac{2f_1}{3k}-\frac{2\Phi_0^2}{4\mu^2}\int_{0}^{\infty} dy\, 
e^{-2\mu^2e^{ky}-2ky}(\mu^2+e^{-ky}) \nonumber \\[2mm]
& = \frac{f_0}{k}-\frac{2f_1}{3k}-\frac{2\Phi_0^2\, \mu^2}{k}
\left(\frac{e^{-2\mu^2}}{8\mu^4}-\mu^2\int_{2\mu^2}^\infty dt\ t^{-4}\,e^{-t}
\right)\,,
\label{P3eq: double-effG}$
where, in the second line, we first set $e^{-ky}=w$ and then $\frac{2\mu^2}{w}=t$.  The integral in
the above expression is the upper incomplete gamma function $\Gamma\left(-3,2\mu^2\right)$ as
one may easily conclude from Eq. \eqref{P3eq: upper-gamma}. The latter quantity, through 
Eq. \eqref{P3app-eq: upper-gamma-explicit}, may be written as
\gat$\label{P3eq: double-upper-gamma1}
\Gamma\left(-3,2\mu^2\right)=\frac{1}{6}\left[\frac{e^{-2\mu^2}}{4\mu^6}\left(
1-\mu^2+2\mu^4\right)-\Gamma\left(0,2\mu^2\right)\right],$
where
\gat$\label{P3eq: double-upper-gamma2}
\Gamma\left(0,2\mu^2\right)=-\gamma-\ln\left(2\mu^2\right)-\sum_{m=1}^\infty \frac{
(-1)^m\,2^m\,\mu^{2m}}{m(m!)}\,.$
Hence, we finally obtain
\eq$\label{P3eq: double-eff2}
\frac{1}{\kappa_4^2}=\frac{M_{Pl}^2}{8\pi}=\frac{f_0}{k}-\frac{2f_1}{3k}-\frac{\Phi_0^2
\,e^{-2\mu^2}}{12k\,\mu^2}(2+\mu^2-2\mu^4)-\frac{\Phi_0^2\,\mu^4}{3k}\,\Gamma\left(0,2
\mu^2\right).$
Demanding as usually a positive four-dimensional gravitational constant,  we
find that the following inequality must be satisfied:
\eq$\label{P3eq: double-eff-con}
\frac{f_0}{\Phi_0^2}>\frac{2}{3}\frac{f_1}{\Phi_0^2}+\frac{e^{-2\mu^2}}{12\mu^2}(2+\mu^2
-2\mu^4)+\frac{\ \mu^4}{3}\,\Gamma\left(0,2\mu^2\right).$
As before, the evaluation of the effective four-dimensional cosmological constant gives $\Lambda_4=0$.

\begin{figure}[t]
    \centering
    \includegraphics[width=0.56\textwidth]{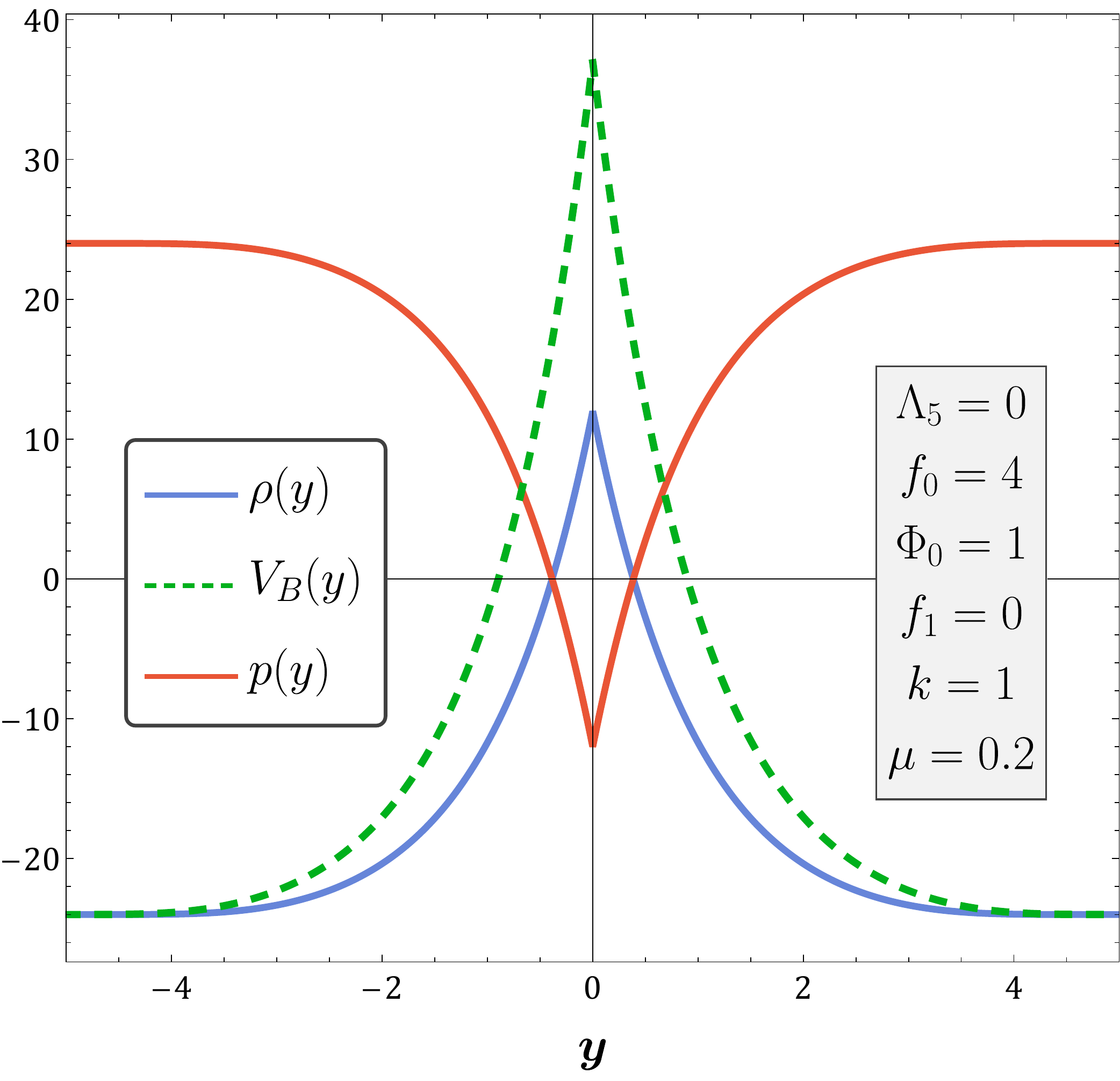}
    \vspace{0em}
    \caption{The energy density $\rho$ and pressure $p$ of the system together with
    the scalar potential $V_B$ in terms of the coordinate $y$ for $\Lambda_5=0$, $f_0=4$,
    $\Phi_0=1$, $f_1=0$, $k=1$, and $\mu=0.2$.}
    \label{P3fig: double-plot3}  
\end{figure}

Finally, we investigate the parameter space of the quantities
$f_0/\Phi_0^2$, $f_1/\Phi_0^2$ and $\mu$, in an attempt to simultaneously satisfy
the inequalities \eqref{P3eq: double-weak-con}, \eqref{P3eq: double-brane-ene-con} and
\eqref{P3eq: double-eff-con}. In \myref{P3fig: double-plot-par}{P3subf: double-plot-par1}, we depict 
the aforementioned parameter
space as well as the surfaces which correspond to the r.h.s.'s of these inequalities.
We observe that there is no point in the parameter space at which all three inequalities can
be satisfied. It is possible though to satisfy two out of these three inequalities simultaneously;
which two are satisfied depends on the values of the parameters. For $f_1/\Phi_0^2=5$, 
for example, the situation is simple as the relative position of the three surfaces remains
the same independently of the value of $\mu$. Thus, we may have a positive effective
cosmological constant and a positive total energy-density on the brane for low values of
$f_0/\Phi_0^2$ whereas for large values of $f_0/\Phi_0^2$ we have a positive $M_{Pl}^2$
and the weak energy conditions are satisfied close to our brane. For $f_1/\Phi_0^2=-5$,
the situation changes and the pair of conditions satisfied depends on the values of all
three parameters---the exact situation is depicted in \myref{P3fig: double-plot-par}{P3subf: double-plot-par2}
where the green region corresponds to the area where the inequalities \eqref{P3eq: double-weak-con}
and \eqref{P3eq: double-eff-con} are satisfied, and the brown region to the area
where \eqref{P3eq: double-brane-ene-con} and \eqref{P3eq: double-eff-con} are satisfied.

In \fref{P3fig: double-plot3}, we present the graphs of the energy density $\rho(y)$ and pressure 
$p(y)$ as well as the potential of the scalar field $V_B(y)$ in terms of the $y$-coordinate for
$\Lambda_5=0$, $f_0=4$, $\Phi_0=1$, $f_1=0$, $k=1$, and $\mu=0.2$. As we can see in the
figure, the values of the parameters are appropriately chosen to satisfy the weak 
energy conditions close to the brane.

%
%

\mysection{Hyperbolic-tangent scalar field in terms of y \label{P3sec: hyper-tang}}

Following the same line of thinking as in the previous section, we now consider the following
expression for the scalar field in terms of the coordinate $y$:
\eq$\label{P3eq: tanh-phi}  
\Phi(y)=\Phi_0\, \tanh(ky)\,,$
where $\Phi_0\in\mathbb{R}\setminus\{0\}$. Substituting the above expression of the scalar 
field in Eq. \eqref{P3eq: grav-1-2}, we obtain the form of the coupling function
\eq$\label{P3eq: tanh-f-y}
f(y)=f_0-f_1\,e^{-ky}+\Phi_0^2\ e^{-ky}\,\arctan(e^{ky})-\frac{\Phi_0^2\ e^{2ky}(e^{2ky}-1)}{
3(e^{2ky}+1)^2}\,.$
Again, the parameter $f_1$ is allowed to take values in the whole set of real numbers, while the
allowed values for the parameter $f_0$ will be examined shortly. By inverting the function
$\Phi(y)$, we may express the coupling function in terms of the scalar field to get
\eq$\label{P3eq: tanh-f-phi}
f(\Phi)=f_0-f_1\sqrt{\frac{\Phi_0-\Phi}{\Phi_0+\Phi}}+\Phi_0^2\sqrt{\frac{\Phi_0-\Phi}{\Phi_0+
\Phi}}\,\arctan\left(\sqrt{\frac{\Phi_0+\Phi}{\Phi_0-\Phi}}\right)-\frac{\Phi(\Phi+\Phi_0)}{6}\,.$ 

Similarly, the scalar potential $V_B(y)$ can be evaluated from Eq. \eqref{P3eq: V-B} with the use of 
\eqref{P3eq: tanh-f-y}; then, we find
\bal$\label{P3eq: tanh-V-y}
V_B(y)=&-\Lambda_5-6k^2f_0+10k^2\,e^{-ky}\left[f_1-\Phi_0^2\,\arctan(e^{ky})\right]\nonum\\[2mm]
&+\frac{2k^2\Phi_0^2}{3}\frac{6+19e^{2ky}+19e^{4ky}-3e^{6ky}+3e^{8ky}}{(e^{2ky}+1)^4}\,.$
In terms of the scalar field, the scalar potential is alternatively written as
\bal$\label{P3eq: tanh-V-phi}
V_B(\Phi)=&-\Lambda_5-6k^2f_0+10k^2\sqrt{\frac{\Phi_0-\Phi}{\Phi_0+\Phi}}\left[f_1-\Phi_0^2\,
\arctan\left(\sqrt{\frac{\Phi_0+\Phi}{\Phi_0-\Phi}}\right)\right]\nonum\\[2mm]
&+\frac{k^2}{6\Phi_0^2}\left(3\Phi^4+8\Phi^3\Phi_0+4\Phi^2\Phi_0^2-14\Phi\Phi_0^3+11\Phi_0^4\right).$
The profiles of the coupling function and scalar potential in this case are qualitatively the
same as the ones in the double-exponential case of the previous section depicted in
\myref{P3fig: double-plot-1-2}{P3subf: double-plot1} and \myref{P3fig: double-plot-1-2}{P3subf: double-plot2}. Again, as the 
parameter $f_1$ changes from positive to negative
values, the coupling function acquires an increasingly larger positive value at the location of our
brane; with the same variation, the scalar potential changes from globally positive-definite to
globally negative-definite values. As before, both functions remain finite everywhere in the bulk
and adopt constant values at large distances. 

\par The energy density $\rho(y)$ and pressure $p(y)=p^i(y)=p^y(y)$ may be computed by 
employing Eqs. \eqref{P3eq: linear-rho} and \eqref{P3eq: linear-p}, and we are led to the result
{\fontsize{10}{10}\eq$\label{P3eq: tanh-rho}
\rho(y)=-p(y)=-6k^2f(y)=-6k^2\left[f_0-f_1\,e^{-ky}+\Phi_0^2\ e^{-ky}\,\arctan(e^{ky})-
\frac{\Phi_0^2\ e^{2ky}(e^{2ky}-1)}{3(e^{2ky}+1)^2}\right].$}
\hspace{-0.5em}In order to satisfy the weak energy conditions close and on the brane, we impose the condition
that $\rho(0)>0$, or $f(0)<0$; hence, we obtain the following inequality: 
\eq$\label{P3eq: tanh-weak-con}
\frac{f_0}{\Phi_0^2}<\frac{f_1}{\Phi_0^2}-\frac{\pi}{4}\,.$
\begin{SCfigure}[][t!]
    \centering
    \includegraphics[width=0.6\textwidth]{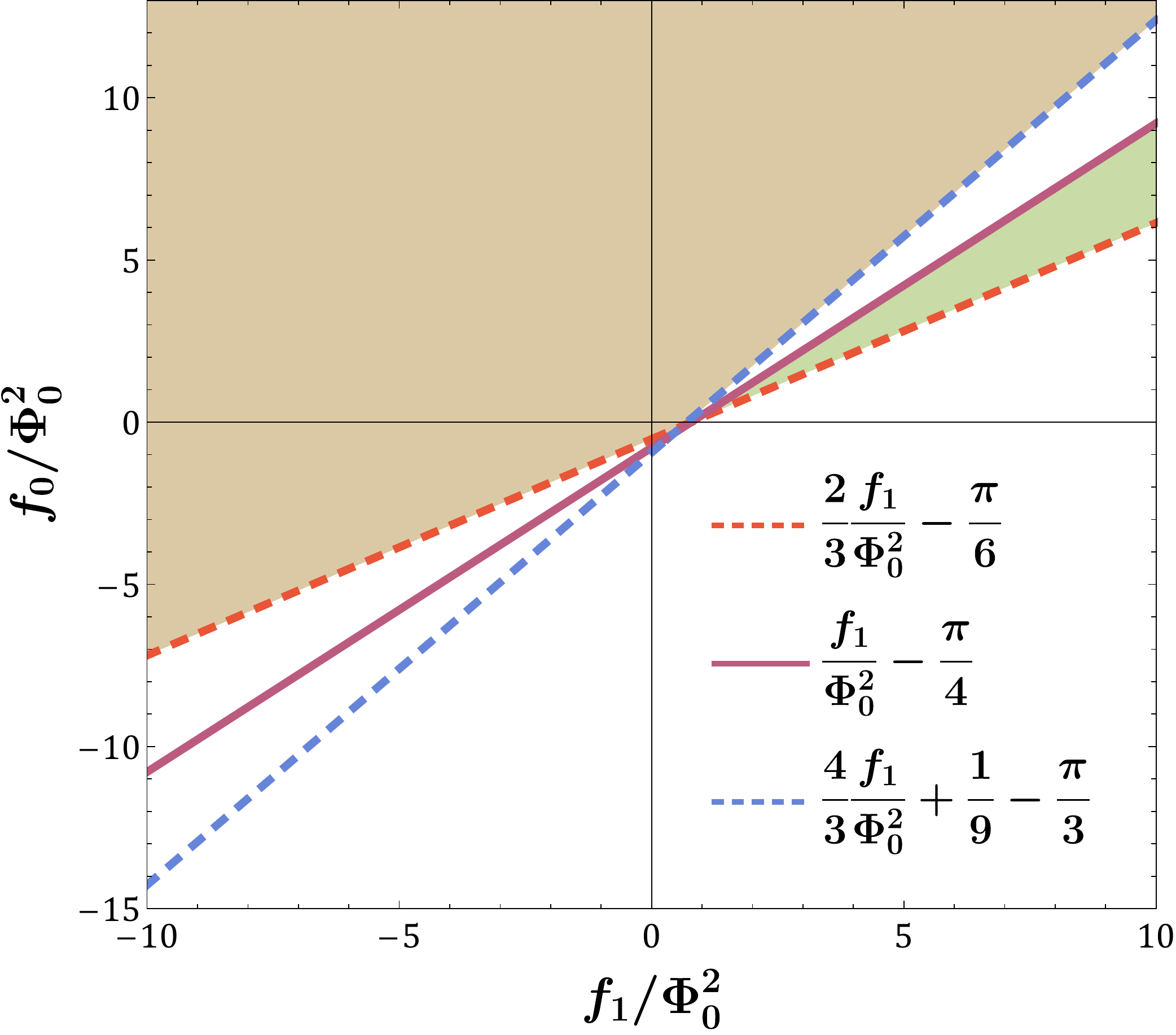}
    \vspace{0em}
    \caption{The parameter space of the quantities $f_0/\Phi_0^2$ and $f_1/\Phi_0^2$.
    The curves corresponding to the expressions on the r.h.s.'s of the inequalities
    \eqref{P3eq: tanh-weak-con}, \eqref{P3eq: tanh-brane-ene-con} and \eqref{P3eq: tanh-eff-con}
    are depicted as well.\\ 
    \vspace{3em}}
    \label{P3fig: tanh-plot-par}
\end{SCfigure}
The junction conditions \eqref{P3eq: jun-con1}, \eqref{P3eq: jun-con2}, employing the relations 
\eqref{P3eq: tanh-phi}--\eqref{P3eq: tanh-f-phi}, now yield:
\gat$\label{P3eq: tanh-jc1}
\sig+V_b(\Phi)\Big|_{y=0}=6kf_0-8kf_1+2k\Phi_0^2\left(\pi-\frac{1}{3}\right)\,,\\[2mm]
\label{P3eq: tanh-jc2}
\pa_yV_b\Big|_{y=0}=8k^2f_1+2k^2\Phi_0^2\left(\frac{7}{3}-\pi\right)\,.$
Therefore, in order to have a positive total energy density on the brane, we demand the condition
\eq$\label{P3eq: tanh-brane-ene-con}
\frac{f_0}{\Phi_0^2}>\frac{4}{3}\frac{f_1}{\Phi_0^2}+\frac{1}{9}-\frac{\pi}{3}\,.$
Let us also evaluate the effective four-dimensional gravitational constant on the brane.
Using Eq. \eqref{P3eq: action_eff}, we obtain
\bal$\frac{1}{\kappa_4^2}&=2\int_0^\infty dy\, e^{-2ky}\left[
f_0-f_1\,e^{-ky}+\Phi_0^2\ e^{-ky}\,\arctan(e^{ky})-\frac{\Phi_0^2\ e^{2ky}(e^{2ky}-
1)}{3(e^{2ky}+1)^2}\right].$
Evaluating the above integral, we obtain the result
\eq$\label{P3eq: tanh-eff}
\frac{1}{\kappa_4^2}=\frac{M_{Pl}^2}{8\pi}=\frac{f_0}{k}-\frac{2}{3}\frac{f_1}{k}+\frac{\pi}{6}
\frac{\Phi_0^2}{k}\,.$
Since it is imperative to have a positive four-dimensional gravitational constant, 
we must satisfy also the following constraint:
\eq$\label{P3eq: tanh-eff-con}
\frac{f_0}{\Phi_0^2}>\frac{2}{3}\frac{f_1}{\Phi_0^2}-\frac{\pi}{6}\,.$

In \fref{P3fig: tanh-plot-par}, we present the parameter space of the quantities
$f_0/\Phi_0^2$ and $f_1/\Phi_0^2$, in an attempt to satisfy simultaneously
the inequalities \eqref{P3eq: tanh-weak-con}, \eqref{P3eq: tanh-brane-ene-con} and
\eqref{P3eq: tanh-eff-con}. As it is clear, there is again no point where all three
inequalities can be satisfied. The green area defines the part of the parameter
space where $M_{Pl}^2$ is positive and the weak energy conditions are satisfied
by the bulk matter close to our brane, while the brown area defines the part
of the parameter space where both the effective gravitational constant and the
total energy-density of the brane are positive.


\mysection{Conclusions \label{P3sec: Disc}}

The present Chapter completes our previous two analyses presented in \chapref{Chap: P1} and \chapref{Chap: P2}, 
where the cases of a de Sitter and an anti-de Sitter brane were considered, respectively. 
Here we focused on the case of a flat, Minkowski brane with $\Lambda=0$.
By assuming a variety of forms for the coupling function, we solved the field equations in the
bulk, and determined in an analytic way the form of the gravitational background and scalar field in
each case. The solutions are always characterized by a regular scalar field, a finite energy-momentum
tensor, and an exponentially decaying warp factor even in the absence of a negative bulk cosmological
constant. The spacetime on the brane is described by the Schwarzschild solution leading to either a
non-homogeneous black-string solution in the bulk, when the mass parameter M is non-zero, or a regular
anti-de Sitter spacetime, when $M = 0$. We constructed physically-acceptable solutions by demanding in
addition a positive effective gravitational constant on our brane, a positive total energy-density for our
brane and the validity of the weak energy condition in the bulk. We found that, although the theory does
not allow for all three conditions to be simultaneously satisfied, a plethora of solutions emerge which
satisfy the first two, and most fundamental, conditions.

\afterpage{\blankpage}



\mychapter{\textbf{Localized brane-world black hole analytically connected to an AdS$\bold{_5}$ boundary} \label{Chap: P4}}
\phantomsection
\setcounter{chapter}{5}

\epigraph{\textit{``Pick a flower on earth and you move the farthest star.''}}{Paul Dirac}

\thispagestyle{empty}

{\lettrine[lines=3, lhang=0.03]{\color{chapter(color)}A}{\hspace*{5.5px}s} we discussed in \chapref{Chap: intro}, the RS
braneworld models \cite{RS1, RS2} made their appearance more than twenty years ago 
and radically changed the way we thought about theories with extra spatial dimensions.
The question of whether the flat background on the brane could be replaced by a curved one
soon emerged. In \cite{CHR}, the four-dimensional Minkowski line-element was replaced by the
Schwarzschild one in an attempt to construct a black hole localized on the brane. 
However, this led instead to an infinitely-long, unstable \cite{GL, RuthGL}
black-string solution plagued also by a curvature singularity at the AdS infinity (see \secref{IntroSec: CHR}). Adopting
a different perspective where the bulk geometry played the leading role, numerical solutions
describing small \cite{Kudoh, Kudoh1} and large \cite{Tanahashi,Kleihaus,Figueras1, Page1, Page2} black holes
were constructed. However, there are countless attempts in the literature (see Refs. \cite{review1, review2, review3, review4, 
review5, review6, review7, tidal, Papanto, KT,KOT, CasadioNew, Frolov, Karasik,Kofinas,GGI, CGKM, Ovalle1, 
Ovalle2, Ovalle3, Ovalle4, Harko, daRocha1, daRocha2, Nakas,Dadhich,Charmousis,Shanka,Andrianov2,Banerjee,Chakraborty1,
Chakraborty2,Chakraborty3,Fitzpatrick, Zegers, Heydari, Dai, Bruni, Tanaka, EFK, EGK, Yoshino,KPZ, KPP} for an impartial 
list of works), adopting either the brane or the bulk perspective, which
fail to describe an analytic five-dimensional black-hole line-element localized close to our brane that also reduces to a
Schwarzschild black hole on the brane.

}

This Chapter, which is based on \cite{NK1}, is occupied with the derivation of an ``algorithm'' that one may use to construct five-dimensional 
analytic braneworld black holes which are localized close to our brane.
In \cite{NK1}, the singularity of the black hole resides strictly on the brane, thus, the emergence of bulk singularities is avoided altogether. 
The induced metric on the brane is described by the Schwarzschild geometry, while the five-dimensional background
quickly reduces to a pure AdS$_5$ spacetime away from the brane.

The structure of the current Chapter is the following: in \secref{P4sec: geo}, we present the general method for
constructing the five-dimensional geometry and study its geometrical properties. In
\secref{P4sec: grav-th}, we turn to the gravitational theory, study the profile of the bulk matter and 
present the field-theory toy model. In \secref{P4sec: junc}, we investigate the junction conditions
and the effective gravitational theory on the brane. We summarize our analysis and
discuss our results in \secref{P4sec: Disc}. 

\vspace*{-1em}

\mysection{The geometrical setup \label{P4sec: geo}}

As we have already seen, the line-element of the Randall-Sundrum model has the following well-known form 
\eq$\label{P4eq: rs-metric1}
ds^2=e^{-2k|y|}\left(-dt^2+d\vec x ^{\,2}\right)+dy^2\,.$
It describes a five-dimensional spacetime comprised by four-dimensional flat slices
stacked together along a fifth dimension denoted by the coordinate $y$. Each slice has
a warp factor $e^{-2k|y|}$, where $k$ is the curvature of the five-dimensional
Anti-de Sitter (AdS) spacetime supported by a negative bulk cosmological constant.  
A $3$-brane must be introduced at the location $y=0$: the cusp in the first derivative of
the warp factor, and thus of the metric tensor, demands the presence of some distribution
of matter at this point, which makes this slice of the AdS spacetime a physical boundary---our 
four-dimensional world. 

The line-element \eqref{P4eq: rs-metric1} may be alternatively written in conformally-flat coordinates.
Introducing the new coordinate $z$ via the relation $z=sgn(y)\,(e^{k |y|}-1)/k$, this takes the form
\eq$\label{P4eq: rs-metr-z}
ds^2=\frac{1}{(k|z|+1)^2}\left(-dt^2+dr^2+r^2\,d\Omega_2^2+dz^2\right)\,.$
In the above expression, we have also employed spherical coordinates for the spatial directions
on the brane with $d\Omega_2^2=d\theta^2+ \sin^2\theta\,d\varphi^2$. We note that, in terms of the new
bulk coordinate, the location of the brane is also at
$z=0$; there, the value of the warp factor is equal to unity. At the AdS asymptotic boundary, i.e. at
$|y| \rightarrow \infty$ or $|z| \rightarrow \infty$, the warp factor vanishes. In the case of
the regular spacetime \eqref{P4eq: rs-metric1} this signifies merely a coordinate singularity,
however, when combined with a black-hole line-element on the brane, it leads to a true
spacetime singularity at the AdS infinity \cite{CHR}, as demonstrated in \secref{IntroSec: CHR}.

\begin{figure}[H]
\centering
\includegraphics[width=0.83\textwidth]{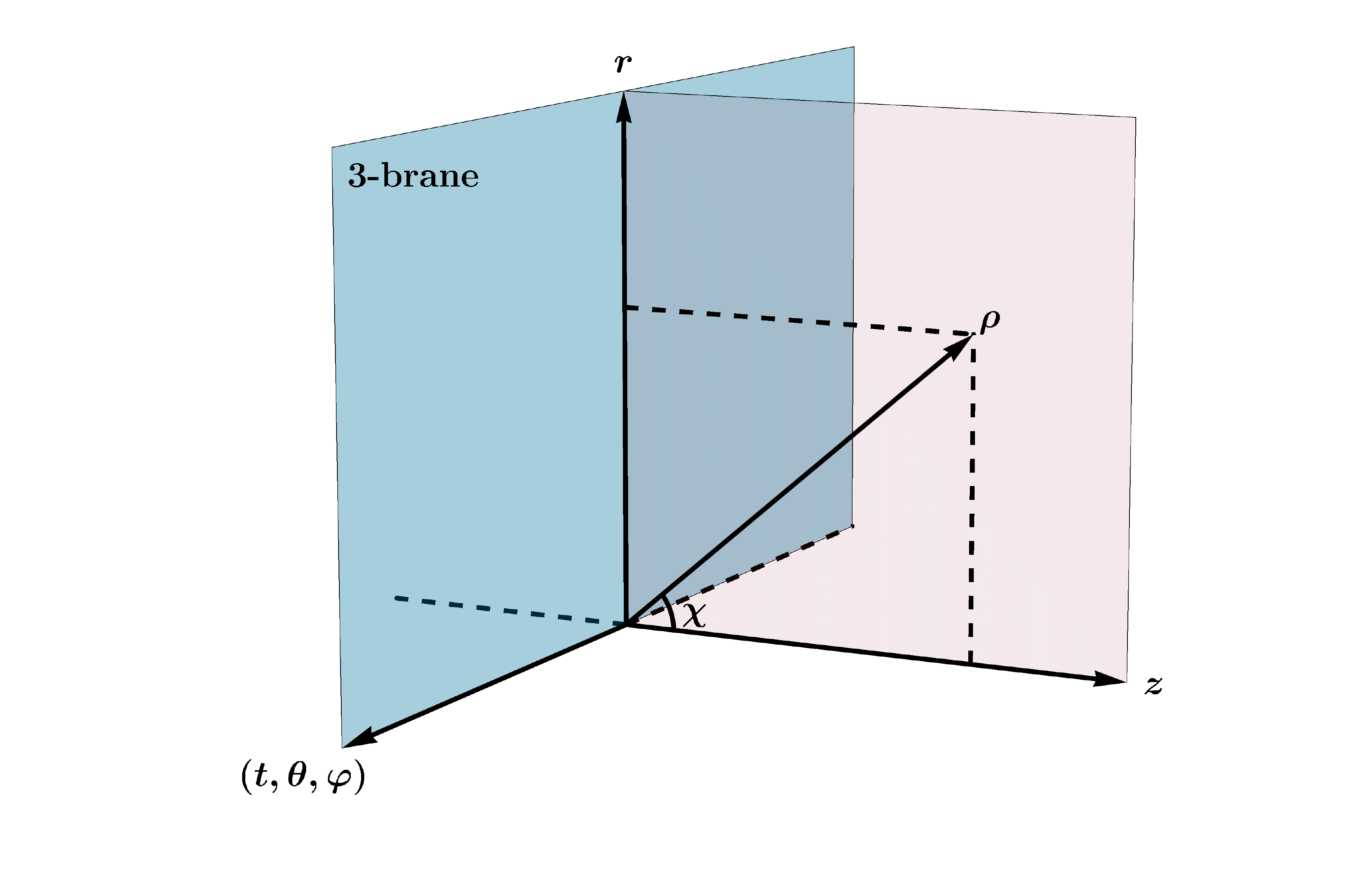}
\vspace*{-1.5em}
\caption{The geometrical set-up of the five-dimensional spacetime and the set of coordinates.}
\label{P4-5fig: coords}
\end{figure}

We will now introduce five-dimensional spherical symmetry. To this end, we perform the
following change of variables:
\eq$\label{P4eq: new-coords}
\left\{\begin{array}{l} r=\rho\,\sin\chi\\[2mm]
z=\rho\,\cos\chi\end{array}\right\}\,, \hspace{1em}{\rm where} \hspace{1em}
\chi\in\left[0,\pi\right]\,.$
Employing these in Eq. \eqref{P4eq: rs-metr-z}, we obtain
\eq$\label{P4eq: rs-metr-sph}
ds^2=\frac{1}{(1+k\rho|\cos\chi|)^2}\left(-dt^2+d\rho^2+\rho^2\,d\Omega_3^2\right)\,,$
where $d\Omega_3^2$ is now the line-element of a unit three-dimensional sphere, namely
\eq$
d\Omega_3^2=d\chi^2+\sin^2\chi\,d\theta^2+\sin^2\chi\,\sin^2\theta\,d\varphi^2\,.$
The inverse transformation reads
\eq$\label{P4eq: new-coords-inv}
\Bigl\{\rho= \sqrt{r^2+z^2}\,, \quad
\tan \chi=r/z\Bigr\}\,.$
As Eq. (\ref{P4eq: new-coords}) dictates, the new radial coordinate $\rho$ is always positive definite
since $\sin\chi \geq 0$ for $\chi\in\left[0,\pi\right]$. On the other hand, $\cos\chi>0$ for $\chi\in\left[0,\pi/2\right)$
and $\cos\chi<0$ for $\chi\in\left(\pi/2,\pi\right]$; thus, the first regime corresponds to positive $z$ and 
describes the bulk spacetime on the right-hand-side of the 3-brane, while the second regime corresponds
to negative $z$ and describes the bulk spacetime on the left-hand-side of the brane.  However, the
corresponding line-elements are related by the coordinate transformation $\chi \rightarrow \pi - \chi$ and thus
describe the same spacetime. The brane itself is located at $\cos \chi=0$, i.e. at $\chi =\pi/2$. 
\fref{P4-5fig: coords} depicts the  geometrical setup of the five-dimensional spacetime. 

The radial coordinate $\rho$ ranges over the interval $[0, \infty)$ since, according to Eq. (\ref{P4eq: new-coords-inv}),
receives contributions both from the (brane) $r$ and (bulk) $z$ coordinates. Therefore, the
5-dimensional radial infinity, $\rho \rightarrow \infty$, may describe both the asymptotic AdS boundary
($|z| \rightarrow \infty$) and the radial infinity on the brane ($r \rightarrow \infty$). 
We note that, on the brane where $z=0$, $\rho$ reduces to the brane radial coordinate $r$. Due to
the aforementioned symmetry of the line-element under $\chi \rightarrow \pi - \chi$, it is adequate
to consider only one of the two $\chi$-regimes; thus, we henceforth focus on the regime
$\chi\in\left[0,\pi/2\right]$ for which $\cos\chi \geq 0$. 

Inspired by \cite{CHR}, we now replace the two-dimensional flat part $(-dt^2 +d\rho^2)$
of the line-element in Eq. \eqref{P4eq: rs-metr-sph} with the corresponding part of
the four-dimensional Schwarzschild solution. Thus, we obtain the following five-dimensional
spacetime 
\eq$\label{P4eq: 5d-schw-metr}
ds^2=\frac{1}{(1+k\rho \cos\chi)^2}\bigg[-f(\rho)\,dt^2+\frac{d\rho^2}{f(\rho)}+
\rho^2\,d\Omega_3^2\bigg],$
where $f(\rho)=1-\frac{2M}{\rho}$. At the location of the brane ($\chi=\pi/2$), $\cos \chi=0$ 
and the warp factor reduces to unity. 
Since there it also holds that $\rho=r$, the line-element on the brane reduces to the usual
four-dimensional Schwarzschild solution with the horizon located at $r=r_h=2M$.\,\footnote{A similar 
construction of the bulk geometry was followed in \cite{Dai},
however, a different form was used for the function $f(\rho)$. As a result, no known black-hole
solution was recovered on the brane. In addition, their choice did not support either an AdS$_5$
spacetime asymptotically in the bulk, in contrast with our choice as we will shortly demonstrate.} But what kind
of five-dimensional gravitational background does the line-element \eqref{P4eq: 5d-schw-metr} describe?
To answer this question, we need to evaluate the five-dimensional curvature invariant quantities.
For instance, the five-dimensional Ricci scalar is found to have the form
\eq$\label{P4eq: ricci-1}
R=-20 k^2 + \frac{12 k^2 M \cos^2 \chi}{\rho }-\frac{24 k M \cos \chi }{\rho ^2}
+\frac{4 M}{\rho ^3}\,.$
The above expression contains a constant contribution $-20 k^2$, attributed to the negative
cosmological constant in the bulk, plus additional terms sourced by the mass $M$ located on
the brane. These terms are inversely proportional to powers of the bulk radial coordinate $\rho$
and  therefore diverge when $\rho \rightarrow 0$. However, since $\rho = \sqrt{r^2 + z^2}$, this 
is realised only when $z=0$ (i.e. on the brane) {\it and} $r=0$ (i.e. at the location of the 
mass $M)$. At any other point in the bulk, characterised by definition by a {\it non-vanishing} value
of $z$, the limit $r \rightarrow 0$ does not lead to a singularity. As a result, the singularity at
$\rho=0$ remains restricted on the brane at the single point $r=0$, and the
line-element \eqref{P4eq: 5d-schw-metr} describes a regular spacetime, which
is analytically connected to a black-hole spacetime on the 3-brane.

In addition, an asymptotically AdS$_5$ spacetime readily emerges when $\rho \rightarrow \infty$:
when we move either far away from the brane, i.e. $z \rightarrow \infty$, or at large distances
along the brane, i.e. $r \rightarrow \infty$, all $\rho$-dependent terms in Eq. (\ref{P4eq: ricci-1})
vanish. A similar behaviour is found for the other two five-dimensional curvature invariants 
$R_{MN} R^{MN}$ and $R_{MNKL} R^{MNKL}$, the  explicit expressions of which can be found
in Appendix \ref{P4app: Curv-Inv}. Taking  the limit $\rho \rightarrow \infty$, we find the following
asymptotic values for the three invariant quantities:
\begin{eqnarray}
&R=-20 k^2\,,&\\[2mm]
&R_{MN}\,R^{MN}=80 k^4\,,&\\[2mm]
&R_{MNKL}\,R^{MNKL}=40 k^4\,,&
\end{eqnarray}
which match the ones for a pure five-dimensional AdS spacetime. Therefore, the line-element
\eqref{P4eq: 5d-schw-metr} describes in particular a regular asymptotically AdS$_5$ spacetime, which
is analytically connected to a black-hole spacetime on the 3-brane.

\begin{figure}[t!]
    \centering
\includegraphics[width=0.7 \textwidth]{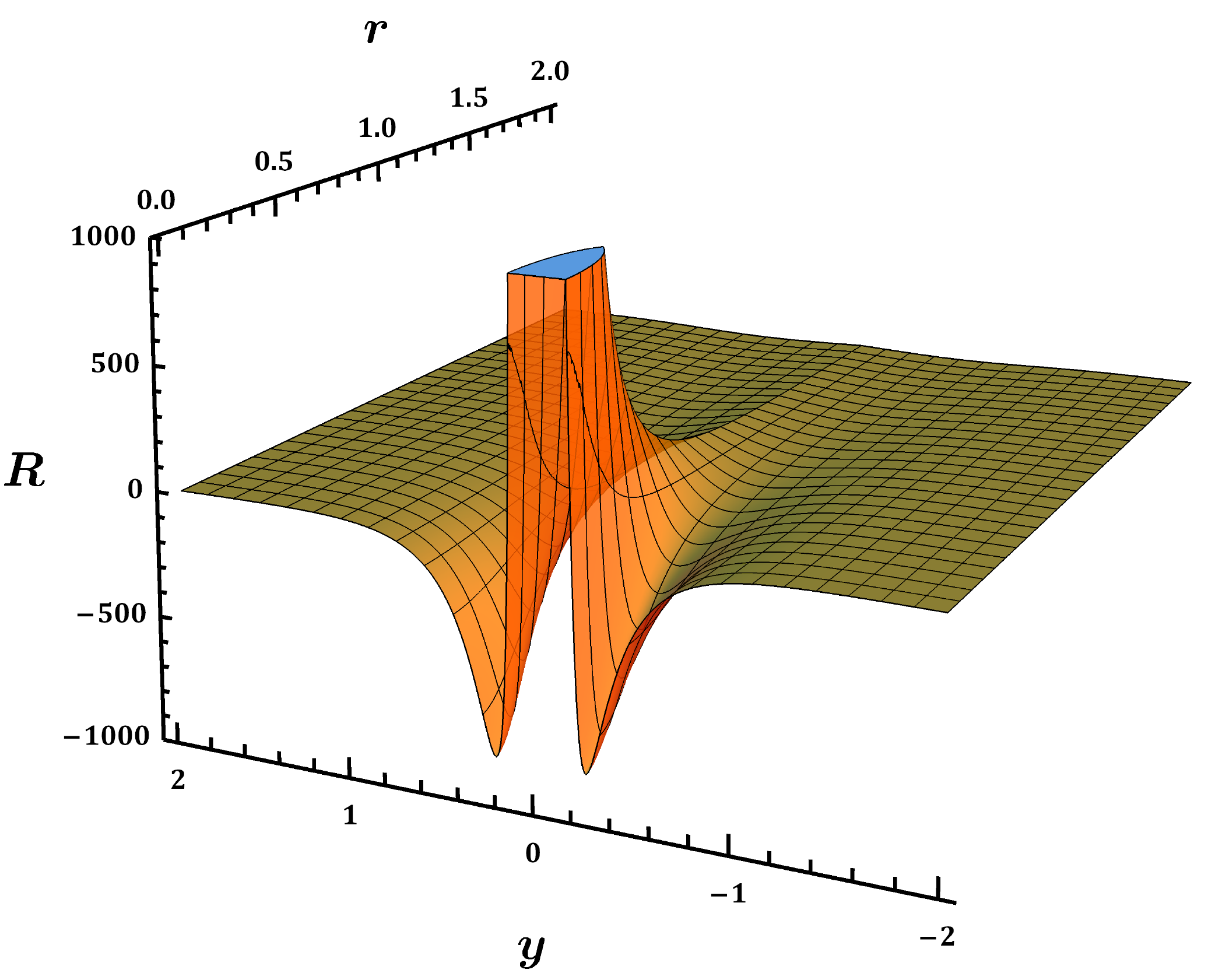}
\vspace*{-0em}
 \caption{The scalar curvature $R$ in terms of the coordinates $\{r,y\}$ for 
    $k=1$ and $M=10$.}
    \label{P4fig: Rscalar}
\end{figure}

The complete regularity of the five-dimensional spacetime and the localization of the black-hole
singularity on the brane are clearly depicted in the profile of the scalar curvature $R$ presented in
\fref{P4fig: Rscalar}. In order to obtain a better understanding of the spacetime
geometry, we have retorted to the original $(r,y)$ coordinates making use of Eq. \eqref{P4eq: new-coords-inv} 
and the relation $z=sgn(y)\,(e^{k |y|}-1)/k$. In terms of these, the Ricci scalar $R$ reads
\eq$\label{P4eq: ricci1-ry}
R=-20k^2+\frac{4k^3M\,\left(10- 12\,e^{k \left| y\right| }+3\,e^{2 k \left| y\right| }\right)}
{\left[\left(e^{k \left| y\right| }-1\right)^2+ k^2 r^2\right]^{3/2}}\,.$
In \fref{P4fig: Rscalar}, we present the overall behaviour of the Ricci scalar
in terms of both $(r,y)$-coordinates; clearly, the singularity arises at $r=0$ if and only if 
$y=0$, too, i.e. at the location of the brane. At every other point of either the bulk or the
brane, the spacetime remains regular. We should note here that the singularity at the AdS
horizon, i.e. as $|y| \rightarrow +\infty$, that plagued the black-string solution of \cite{CHR},
is absent here since the second term in Eq. \eqref{P4eq: ricci1-ry} vanishes altogether in that
limit. The profiles of the other two invariants, namely $R_{MN} R^{MN}$ and
$R_{MNKL} R^{MNKL}$, are similar to that of $R$ with the only difference being their
monotonic rise close to the singularity on the brane and the absence of the double well
observed in \fref{P4fig: Rscalar}.

Let us also re-write the five-dimensional line-element \eqref{P4eq: 5d-schw-metr} in terms
of the original non-spherical coordinates $\{r,y\}$. Employing the inverse transformations
(\ref{P4eq: new-coords-inv}), the line-element takes the form
\bal$\label{P4eq: metr-r-y}
ds^2=e^{-2k|y|}&\Bigg\{-f(r,y) dt^2+\frac{dr^2}{r^2+z^2(y)}\biggl[\frac{r^2}{f(r,y)}+z^2(y)\biggr]+r^2d\Omega_2^2\nonum\\[1mm]
&\hspace{0.7em}+\frac{2r z(y)\,e^{k|y|}}{r^2+z^2(y)}\biggl[\frac{1}{f(r,y)}-1\biggr]drdy \Bigg\}
+\frac{dy^2}{r^2+z^2(y)}\left[r^2+\frac{z^2(y)}{f(r,y)}\right]\,,$
where $z=sgn(y)\,(e^{k |y|}-1)/k$ and 
\eq$\label{P4eq: f-ry}
f(r,y)=1-\frac{2M}{\sqrt{r^2+z^2(y)}}\,.$
We observe that the aforementioned line-element differs significantly from the factorized
line-element employed in \cite{CHR}, or from non-factorized ones which appeared in a number
of subsequent works \cite{KT, KOT, KPZ, KPP}. It is the gradual construction---via the employment
of the spherically symmetric coordinates of the line-element (\ref{P4eq: 5d-schw-metr})---that has
resulted in the expression \eqref{P4eq: metr-r-y}.

\begin{figure}[t!]
    \centering
  \hspace*{1cm}  \begin{subfigure}[b]{0.5\textwidth}
        \includegraphics[width=\textwidth]{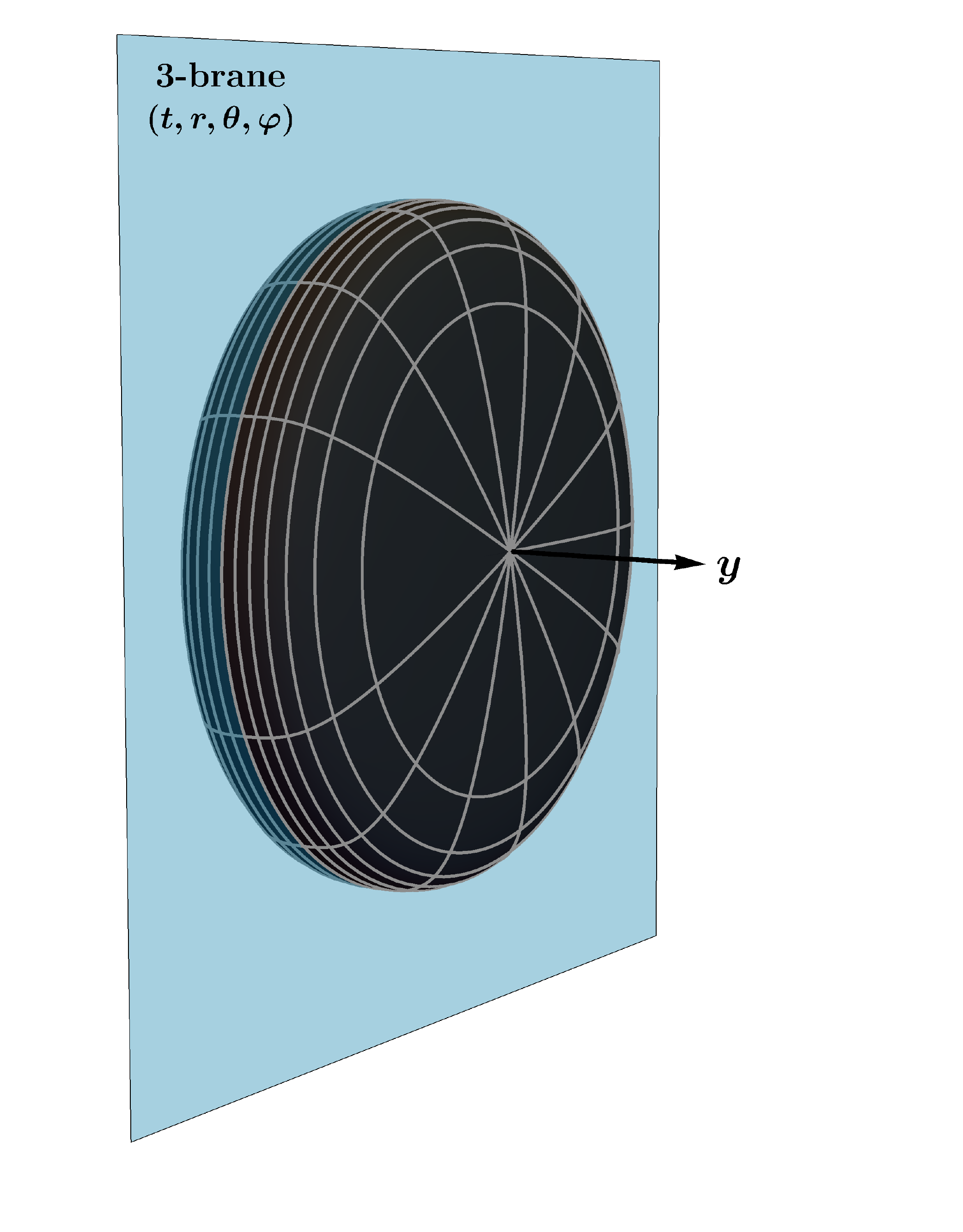}
        \caption{\hspace*{0em}}
        \label{P4subf: loc-BH1a}
    \end{subfigure}
    \hfill
    ~ 
  		  \begin{subfigure}[b]{0.38\textwidth}
        \includegraphics[width=\textwidth]{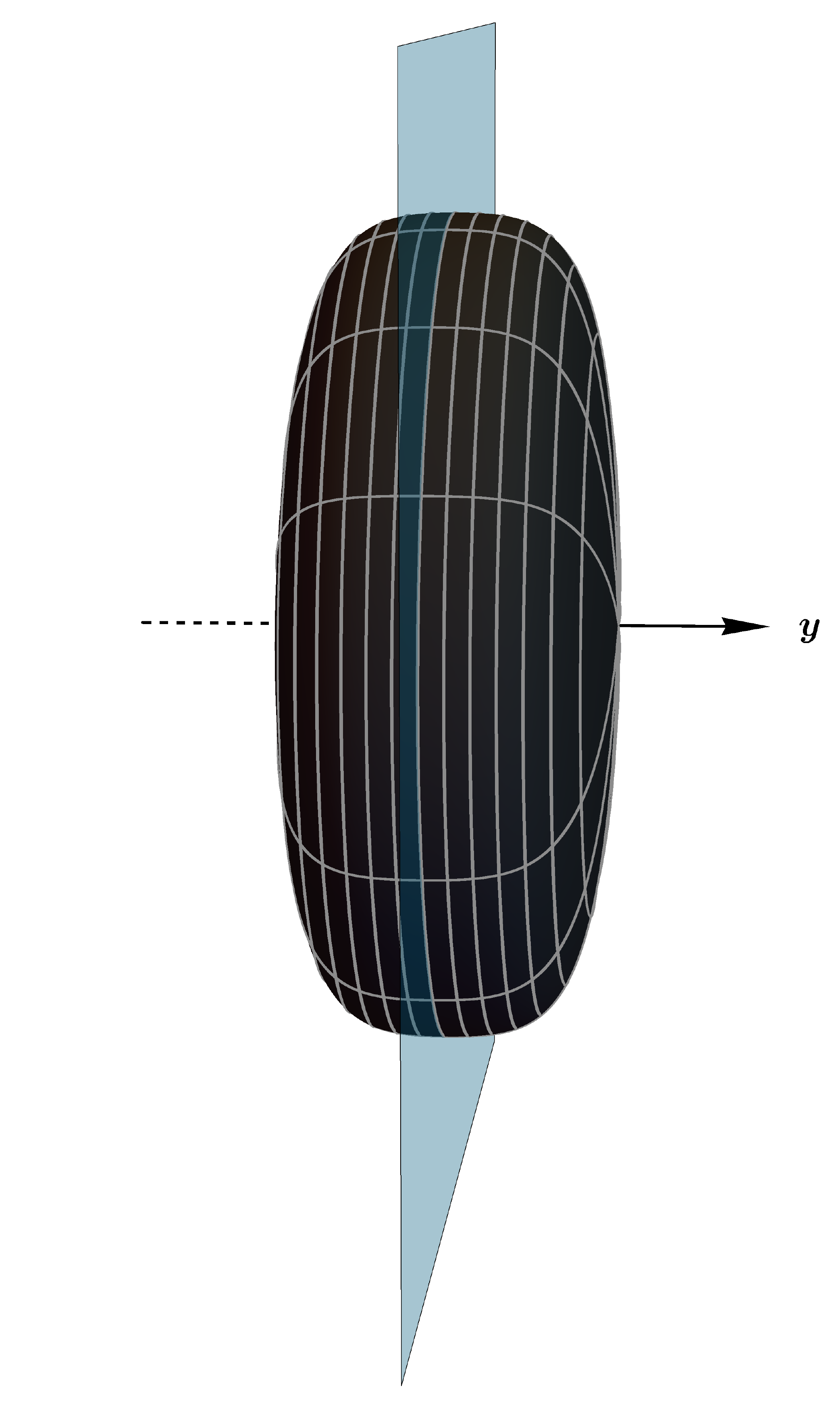}
        \caption{\hspace*{0em}}
        \label{P4subf: loc-BH1b}
    \end{subfigure}  
    \vspace{-0em}  
    \caption{The horizon of the localized five-dimensional Schwarzschild black-hole from the bulk point of view for
    $M=7$, $k=0.5$. Both figures (a) and (b) depict the same image from different 
    angles. The depicted brane coordinates are the radial coordinate $r$ and the angular coordinate $\varphi$.}
    \label{P4fig: loc-BH-plot}
\end{figure}

Although the singularity of the black hole remains localized on the brane as demonstrated
above, the horizon of the black hole does not need to do so; in fact, we expect it to extend
into the bulk. Let us therefore study the causal structure of the bulk spacetime as this is defined
by the light cone. We therefore consider radial null trajectories in the five-dimensional background
\eqref{P4eq: metr-r-y}. For a fixed value $y=y_0$ of the fifth coordinate, the condition $ds^2=0$, with
 $\theta$ and $\varphi$ kept constant,  leads to the result
\eq$ 
\frac{dt}{dr}=\pm\frac{1}{f(r,y_0)}\left[\frac{r^2k^2+f(r,y_0)\left(e^{k|y_0|}-1\right)^2}{r^2k^2+
\left(e^{k|y_0|}-1\right)^2}\right]^{1/2},$
where
\eq$\label{P4eq: horizon-function}
f(r,y_0)=1-2M\left[r^2+\frac{\left(e^{k|y_0|}-1\right)^2}{k^2}\right]^{-1/2}.$
At large distances along the brane, i.e. when $r\ra \infty$, one gets 
$\lim_{r\ra \infty}f(r,y_0)=1$, and the slope ${dt}/{dr}$ goes to $\pm 1$,
as expected. However, at a finite distance $r=r_h$, which is defined through the relation
$f(r_h,y_0)=0$, we obtain $dt/dr=\pm\infty$. Therefore, at $y=y_0$ and $r=r_h$ we
encounter the horizon of the black hole as it extends into the bulk. Its exact location follows
from Eq. \eqref{P4eq: horizon-function},  and is given by
\eq$\label{P4eq: r-hor}
r_h^2=4M^2-\frac{\left(e^{k|y_0|}-1\right)^2}{k^2}\,.$
We note that $r_h$ depends on the parameters $M$, $k$ and $y_0$.
On the 3-brane, where $y_0=0$, $r_h$ equals the Schwarzschild value $2M$ independently
of the value of $k$, in agreement with the discussion above Eq. \eqref{P4eq: ricci-1}. 
But, as we move along the extra dimension, $r_h$ shrinks exponentially fast and
becomes zero at a distance 
\eq$|y_0|=\frac{1}{k}\ln\left(2Mk+1\right)$ 
away from the brane. As a result, the black-hole horizon has the shape
of a ``pancake'' with its long side lying along the brane and its short side extending in the
bulk over an exponentially small distance.  In conclusion, the line-element  \eqref{P4eq: metr-r-y}---or 
its spherically-symmetric analog (\ref{P4eq: 5d-schw-metr})---describes a five-dimensional
black-hole solution which exhibits a localization of its singularity {\it  strictly on our brane}
and a localization of its horizon  {\it exponentially close} to our brane. 
In \fref{P4fig: loc-BH-plot} we give the geometrical representation of the event horizon of the five-dimensional
Schwarzschild spacetime from the bulk point of view for $M=7$ and $k=0.5$.
Note also that in both illustrations, the depicted brane coordinates are the radial coordinate $r$ and the angular 
coordinate $\varphi$.

\newpage

\vspace*{-3.5em}

\mysection{The gravitational theory \label{P4sec: grav-th}}

We now turn to the gravitational theory and consider the following five-dimensional action functional
\eq$\label{P4eq: bulk-action}
S_{B}=\int d^5x\, \sqrt{-g} \left(\frac{R}{2\kappa_5^2}+\lagr^{(B)}_{m}\right).$
In the above, $g_{MN}$ is the metric tensor of the five-dimensional spacetime, $\kappa_5^2=8\pi G_5$ 
incorporates the five-dimensional gravitational constant $G_5$, while $R$ is the five-dimensional Ricci
scalar. The Lagrangian  density $\lagr^{(B)}_m$ describes the matter that exists in the bulk. 

\par The gravitational field equations in the bulk can be obtained by the variation of the action  $S_B$
with respect to the metric tensor $g_{MN}$. These read
\eq$\label{P4eq: field-eqs}
G_{MN}=\kappa_5^2\, T^{(B)}_{MN}\,,$
where $G_{MN}=R_{MN}-\frac{1}{2}\,g_{MN}R$ is the Einstein-tensor, while $T^{(B)}_{MN}$ is the 
energy-momentum tensor associated with the Lagrangian density $\lagr^{(B)}_m$ and is defined as follows
\eq$T^{(B)}_{MN}=-\frac{2}{\sqrt{-g}}\frac{\delta\left(\lagr^{(B)}_{m}\sqrt{-g}\right)}{\delta g^{MN}}\,.$
\par We will now employ the form of the  gravitational background described by the line-element
(\ref{P4eq: 5d-schw-metr}). Substituting on the left-hand-side of Eq. (\ref{P4eq: field-eqs}) and solving for the
components of the energy-momentum tensor, we find:
\gat$\label{P4eq: rho-p1}
T^{(B)t}{}_{t}=T^{(B)\rho}{}_{\rho}=\frac{1}{\kappa_5^2}\biggl(6 k^2+\frac{9 k M \cos \chi}{\rho ^2}
-\frac{3 M}{\rho ^3}\biggr),\\[2mm]
T^{(B)\chi}{}_{\chi}=T^{(B)\theta}{}_{\theta}=T^{(B)\varphi}{}_{\varphi}
=\frac{1}{\kappa_5^2}\left(6 k^2-\frac{6 k^2 M \cos ^2 \chi }{\rho }+\frac{6 k M \cos \chi}{\rho ^2}\right).
\label{P4eq: p2}$
Therefore, the gravitational background (\ref{P4eq: 5d-schw-metr}) may be supported by a bulk energy-momentum tensor
the only non-vanishing components of which are the energy-density $\rho_E \equiv -T^{(B)t}{}_{t}$, the radial pressure
$p_r \equiv T^{(B)\rho}{}_{\rho}$,  and a common tangential pressure $p_\theta \equiv T^{(B)\chi}{}_{\chi}=T^{(B)\theta}{}_{\theta}=
T^{(B)\varphi}{}_{\varphi}$. The necessary matter content of the bulk is thus an anisotropic fluid described by a {\it diagonal}
energy-momentum tensor which in a covariant notation may be written as 
\eq$\label{P4eq: en-mom}
T^{(B)MN}=(\rho_E+p_\theta)U^M U^N+(p_r-p_\theta)X^M X^N+p_\theta\, g^{MN}\,.$
In the above expression,  $U^M$ is the fluid's timelike five-velocity, and $X^M$ is a 
spacelike unit vector in the direction of $\rho$-coordinate satisfying the relations
\gat$
U^M=\{U^t,0,0,0,0\},\hspace{1em} U^M U^N g_{MN}=-1\,,\\[2mm]
X^M=\{0,X^\rho,0,0,0\},\hspace{1em} X^M X^N g_{MN}=1\,.$

We observe that the energy density and radial pressure satisfy the equation of state $p_r=w_r\, \rho_E$ with $w_r=-1$
everywhere in the bulk, whereas $w_\theta$ defined via $p_\theta=w_\theta\, \rho_E$, is $(\rho,\chi)$-dependent. However, we
note that as $\rho\ra+\infty$,  all components of the energy-momentum
tensor reduce to a constant which can be identified as the five-dimensional cosmological constant $\Lambda_5$
\gat$
\lim_{\rho\ra+\infty}\rho_{E}(\rho,\chi)=-\frac{6k^2}{\kappa_5^2} \equiv \Lambda_5\,, \\[1mm]
\lim_{\rho\ra+\infty}p_r(\rho,\chi)=\lim_{\rho\ra+\infty}p_\theta(\rho,\chi)=\frac{6k^2}{\kappa_5^2} \equiv
-\Lambda_5\,.$
This negative cosmological constant supports the AdS$_5$ gravitational background away from the brane,
in agreement with the derived asymptotic behaviour of the line-element (\ref{P4eq: 5d-schw-metr}). It also leads to
the exponentially decaying warp factor $e^{-k |y|}$ of the bulk spacetime and thus incorporates the
Randall-Sundrum model \cite{RS1, RS2}. 

\par Let us now study the profiles of the energy density $\rho_E$ and tangential pressure $p_\theta$ of the bulk
anisotropic fluid. To this end, we employ the coordinates $\{r, y\}$ in terms of which their characteristics are
more transparent.  Using Eqs. (\ref{P4eq: new-coords-inv}), (\ref{P4eq: rho-p1}) and (\ref{P4eq: p2}), we find
{\fontsize{10}{10}\gat$\label{P4eq: rho-p1-ry} 
\rho_E=\frac{3k^2}{\kappa_5^2}\left\{-2+\frac{kM \left(4-3\, e^{k|y|}\right)}{\left[k^2r^2+\left(e^{k|y|}-1\right)^2
\right]^{3/2}}\right\},\hspace{1em}
p_\theta=\frac{6k^2}{\kappa_5^2}\left\{1+\frac{kM\left(e^{k|y|}-1\right)\left(2-e^{k|y|}\right)}
{\left[k^2r^2+\left(e^{k|y|}-1\right)^2\right]^{3/2}}\right\}.$}
\begin{figure}[H]
    \centering
    \includegraphics[width=0.61\textwidth]{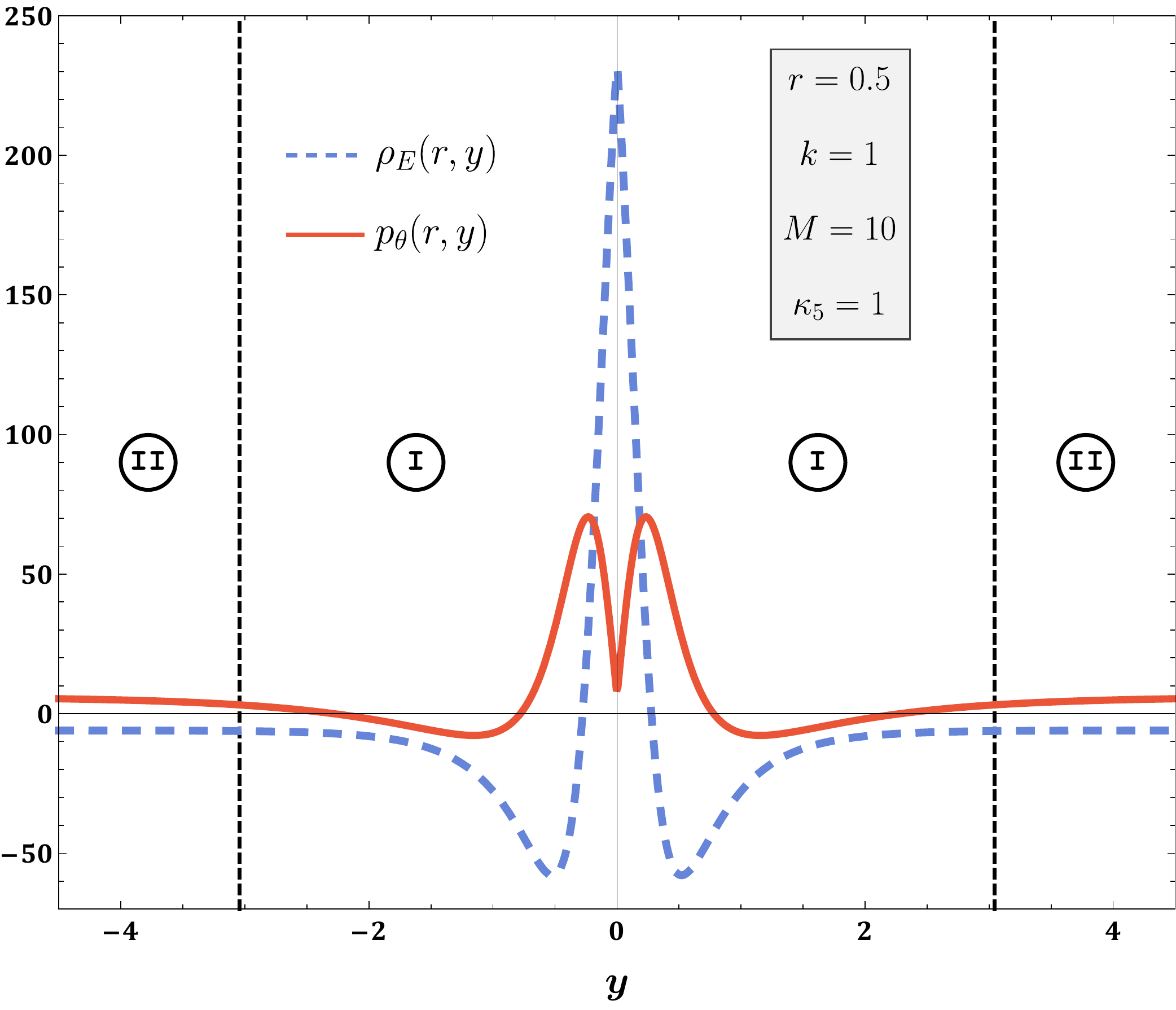}
    \vspace{-0em}
    \caption{The $\rho_E(r,y)$ and $p_\theta(r,y)$ in terms of the $y$-coordinate for $r=0.5$, $k=1$, $M=10$ and $\kappa_5=1$. Region I
     lies inside the black-hole horizon, while region II lies outside the black-hole horizon.}
    \label{P4fig: rho-p2}  
\end{figure}
In \fref{P4fig: rho-p2}, we depict the above quantities in terms of the extra dimension $y$, for $r=0.5$,  $k=1$,
$M=10$ and $\kappa_5=1$. We have divided the bulk spacetime in regions I and II; the former lies inside
the black-hole horizon---which for the selected values is located at the bulk coordinate $|y|\simeq 3$---while 
the latter lies outside of it.  It is worth noticing that outside the black-hole horizon both $\rho_E$ and
$p_\theta$ quickly approach their asymptotic values, which means that the spacetime outside the horizon is
effectively AdS$_5$. On the other hand, on the brane, located at $y=0$, the energy density and tangential
pressure adopt finite values which satisfy all energy conditions since $\rho_E>0$, $\rho_E+p_r=0$
and $\rho_E \gg p_\theta$. It is only off the brane and within the black-hole horizon that the bulk fluid exhibits
a non-conventional behaviour as revealed by the violation of the energy conditions. This is a necessary
feature for the localization of the black hole  near the brane, which would otherwise `leak' into the bulk
resulting in a black string \cite{KNP1,KNP2,KNP3}. In fact, one could compare this requirement with the violation 
of the energy conditions around the throat of a wormhole \cite{MT}---there, as well as in our case, the
violation is only  {\it local} and {\it necessary} for the support of the desired geometry.

\mysection{Junction conditions and effective theory on the brane \label{P4sec: junc}}

In this final section, we turn our attention from the structure and content of the five-dimensional spacetime to issues
related to the presence of the brane itself, namely its consistent embedding in the bulk and the effective four-dimensional
gravitational equations. A detailed derivation of the effective theory on the 3-brane in brane-world models was presented
in \cite{SMS}, however, in order to keep our analysis self-contained, we will reproduce here the main results and equations.
It is also important to note that in \cite{SMS} the bulk matter of the brane-world model was described only by a negative
cosmological constant, whereas in our case the bulk spacetime contains an anisotropic fluid, a feature that slightly
modifies some parts of the analysis.

In the standard brane-world scenario, our 3-brane ($\Sigma$, $h_{MN}$) is embedded in the five-dimensional spacetime
($\mathcal{M}$, $g_{MN}$) at $y=0$. The induced metric on the brane is defined via the relation
$h_{MN}\equiv\left(g_{MN}\right)_{y=0}-n_{M}n_{N}$, where $n^M$ is the unit normal vector to the 3-brane. From
Eq. \eqref{P4eq: metr-r-y}, we may deduce that $n^M=\del^M{}_y$. In what follows, we will denote tensors on $\Sigma$ with
a bar to be distinguished from the corresponding five-dimensional tensors.
Using the \textit{Gauss's Theorema Egregium}\,\footnote{We note that the square brackets $[\ldots]$ in a tensor's indices denote 
anti-symmetrization, namely $A_{[MN]}\equiv\frac{1}{2}\left(A_{MN}-A_{NM}\right)$.}
\eq$\label{P4eq: Gauss-eq}
\bar{R}^A{}_{BCD}=h^{A}{}_M\, h^{N}{}_B\,h^{K}{}_C\, h^{L}{}_D R^M{}_{NKL}+2K^{A}{}_{[C}K_{D]B}\,,$
and the \textit{Codazzi's equation}
\eq$\label{P4eq: Codazzi-eq}
R_{AB}\,h^A{}_M\,n^B=D_L K^{L}{}_M-D_M\, K\,,$
we obtain the following relation for the Einstein tensor on the 3-brane:
\bal$
\bar{G}_{MN}&=h^A{}_Mh^{B}{}_N\,G_{AB}+R_{AB}\,n^An^Bh_{MN}+KK_{MN}-K_{M}{}^LK_{LN}\nonum\\[2mm]
&\hsp-\frac{1}{2}h_{MN}\left(K^2-K^{AB}K_{AB}\right)-\wtild{E}_{MN}\,.
\label{P4eq: einstein-on-brane}$
In the above,\,\footnote{A tensor at a point $P\in\Sigma$ is invariant under the projection $h^{M}{}_N$ if
\eq$\label{P4eq: tensor-on-brane}
T^{M_1M_2\cdots M_p}{}_{N_1N_2\cdots N_q}=h^{M_1}{}_{A_1}\,h^{M_2}{}_{A_1}\cdots h^{M_p}{}_{A_p}\,h^{B_1}{}_{N_1}
h^{B_2}{}_{N_2}\cdots h^{B_q}{}_{N_q}\,T^{A_1A_2\cdots A_p}{}_{B_1B_2\cdots B_q}\,.$
The covariant derivative $D_L$ on $\Sigma$ can be defined via the projection of the covariant derivative on $\mathcal{M}$; for any tensor
obeying \eqref{P4eq: tensor-on-brane} we define 
\eq$\label{P4eq: cov-der-on-brane}
D_{L}T^{M_1\cdots M_p}{}_{N_1\cdots N_q}=h^{K}{}_L\,h^{M_1}{}_{A_1}\cdots h^{M_p}{}_{A_p}\,h^{B_1}{}_{N_1}\cdots h^{B_q}{}_{N_q}\,
\nabla_KT^{A_1\cdots A_p}{}_{B_1\cdots B_q}\,.$} $K_{MN}$ is the extrinsic curvature of the brane defined as
\eq$\label{P4eq: extr-curv}
K_{MN}\equiv h^{A}{}_M\,h^{B}{}_N\,\nabla_A\,n_{B}=h^{L}{}_{M}\,\nabla_L\,n_{N}\,,$
while
\eq$\label{P4eq: E-riem}
\wtild{E}_{MN}\equiv R^A{}_{BCD}\, n_A\,n^C\,h^B{}_M\,h^{D}{}_N\,.$
Decomposing the Riemann tensor into the Weyl curvature, the Ricci tensor and the Ricci scalar, we obtain
\eq$\label{P4eq: riem-weyl}
R_{ABCD}=\frac{2}{3}\left(g_{A[C}R_{D]B}-g_{B[C}R_{D]A}\right)-\frac{1}{6}g_{A[C}\,g_{D]B}R+C_{ABCD}\,.$
Using the five-dimensional gravitational field equations \eqref{P4eq: field-eqs} together with \eqref{P4eq: riem-weyl} in \eqref{P4eq: einstein-on-brane} we are led to
\gat$
\bar{G}_{MN}=\frac{2\kappa_5^2}{3}\left[h^A{}_Mh^{B}{}_N\,T^{(B)}_{AB}+\left(n^An^{B}\,T^{(B)}_{AB}-\frac{T^{(B)}}{4}\right)h_{MN}
\right]+KK_{MN}-K_{M}{}^LK_{LN}\nonum\\
\label{P4eq: effective-theory}
-\frac{1}{2}h_{MN}\left(K^2-K^{AB}K_{AB}\right)-E_{MN}\,,$
where $T^{(B)}\equiv T^{(B)L}{}_L$ is the trace of the bulk energy-momentum tensor, and
\eq$\label{P4eq: E-weyl}
E_{MN}\equiv C^A{}_{BCD}\, n_A\,n^C\,h^B{}_M\,h^{D}{}_N\,.$
As is usual in all brane-world scenarios, we may write the total energy-momentum tensor as the sum of the
bulk  $T^{(B)}_{MN}$ and brane $T^{(br)}_{\mu\nu}$ energy-momentum tensors, namely
\eq$\label{P4eq: ene-mom-dec}
T_{MN}=T^{(B)}_{MN}+\del^{\mu}_M\del^{\nu}_N\,T^{(br)}_{\mu\nu}\del(y)\,.$
The brane  energy-momentum tensor can be decomposed further as follows
\eq$\label{P4eq: brane-ene-dec}
T^{(br)}_{\mu\nu}=-\sigma\, h_{\mu\nu}+\tau_{\mu\nu}\,,$
where $\sigma$ is the tension of the brane, and $\tau_{\mu\nu}$ encodes all the
other possible sources of energy and/or pressure on the brane. A natural question which arises in the
context of our analysis is whether the consistent embedding of our brane in the five-dimensional
line-element (\ref{P4eq: metr-r-y}) demands the introduction of a non-trivial $\tau_{\mu\nu}$ on the brane. 

In order to investigate this, we will study Israel's junction conditions \cite{Israel} at $y=0$. These require
that
\gat$
\label{P4eq: jc1}
[h_{MN}]=0\,,\\[2mm]
\label{P4eq: jc2}
[K_{\mu\nu}]=-\kappa_5^2\left(T^{(br)}_{\mu\nu}-\frac{1}{3}h_{\mu\nu}\,T^{(br)}\right)\,.$
In the above, the bracket notation for a quantity $X$ simply means 
\eq$\label{P4eq: bracket-def}
[X]=\lim_{y\ra 0^+}X-\lim_{y\ra 0^-}X=X^{(+)}-X^{(-)}\,.$
Let us determine first the components of the induced metric on the brane $h_{MN}$. These are found to be
{\fontsize{11}{11}\eq$\label{P4eq: ind-metr-comp}
(h_{MN})=\left(\begin{array}{ccccc}
-\left(1-\frac{2M}{r}\right) 		& 		0 		& 		0 		& 		0 		&   0		 \\[1mm]
0		 &		 \left(1-\frac{2M}{r}\right)^{-1} 		&		 0		 &		 0		& 0	  \\[1mm]
0		&		0		&		r^2		&		0	&	0	\\
0		&		0		&		0		&		r^2\sin^2\theta	&	0\\
0	&  0	&	0	&	0	&  0		
\end{array}\right)\,.$}
\hspace{-0.5em}We may easily see that they indeed satisfy Israel's first condition. Also, employing these, we may easily determine the
components of the extrinsic  curvature close to the 3-brane which have the form
\eq$\label{P4eq: extr-curv2}
K_{MN}=-k \frac{d|y|}{dy}\,\del^{\mu}_M\del^{\nu}_N\, h_{\mu\nu}.$
The trace of $K_{MN}$ is also found to be $K=-4k\,(d|y|/dy)$. We may alternatively write Eq. \eqref{P4eq: jc2}  as\,\footnote{For an explicit
proof of Eq. \eqref{P4eq: jc2-new} see Appendix \ref{P4app: Br-Ene}.}
\eq$\label{P4eq: jc2-new}
T^{(br)}_{\mu\nu}=-\frac{1}{\kappa_5^2}\left([K_{\mu\nu}]-h_{\mu\nu}[K]\right)\,.$
Using Eq. (\ref{P4eq: bracket-def}), the assumed $\mathbf{Z}_2$-symmetry of the model in the bulk
and the components of $K_{\mu\nu}$, we find 
\eq$\label{P4eq: brane-ene-comp}
T^{(br)}_{\mu\nu}=-\frac{6k}{\kappa_5^2}\,h_{\mu\nu}\,.$
Comparing Eq. \eqref{P4eq: brane-ene-comp} with Eq. \eqref{P4eq: brane-ene-dec}, we easily deduce that
$\sigma=6k/\kappa_5^2>0$, while $\tau_{\mu\nu}=0$. This means that the consistent embedding of our
3-brane in the five-dimensional spacetime constructed in \secref{P4sec: geo}---and described by either the line-element
(\ref{P4eq: metr-r-y}) or (\ref{P4eq: 5d-schw-metr})---does not demand the introduction of any additional matter on the
brane.\,\footnote{The absence of the need for the introduction of any brane matter but the necessity for
the presence of bulk fields in order to localize the black-hole geometry close to the brane
could be related to similar conclusions derived following the effective-field-theory point-of-view in
braneworlds\cite{Fichet}.} In the context of the five-dimensional theory, the brane
contains only  its constant positive self-energy $\sigma$.
In fact, it is this quantity together with the five-dimensional gravitational constant $\kappa_5^2$ that
determine the warp parameter $k$ of the line-element in the bulk. 

We may now proceed to derive the effective theory on the brane. The gravitational equations on the 3-brane
can be determined from Eq. \eqref{P4eq: effective-theory} by setting $y=0$. We note that for either $M$ or $N$
equal to $y$, the r.h.s. of \eqref{P4eq: effective-theory} is trivially zero; this implies that $\bar{G}_{yN}=0\ 
\forall N$, as expected. Due to the $\mathbf{Z}_2$-symmetry, we may perform the calculation either on
the $+$ or $-$ side of the brane, therefore we will omit the $\pm$ signs in what follows. Using the 
results for the induced metric $h_{MN}$, the extrinsic curvature $K_{MN}$ and
the normal vector $n^M$ derived above in \eqref{P4eq: brane-ene-dec}, we obtain
\eq$\label{P4eq: grav-eqs-br-new}
\bar{G}_{\mu\nu}=8\pi G_N\left(T^{(eff)}_{\mu\nu}+ \tau_{\mu\nu}\right)+\kappa_5^4\left(\pi_{\mu\nu}-\frac{\ \sig^2}{12}\,h_{\mu\nu}\right)
-E_{\mu\nu}\Big|_{y\ra 0}\,,$
where
\bal$\label{P4eq: New-const}
&G_N=\frac{\kappa_5^4\,\sigma}{48\pi}\,,\\[3mm]
&T^{(eff)}_{\mu\nu}\equiv\frac{2}{3k}\left[T^{(B)}_{\mu\nu}+\left(T^{(B)}_{yy}-\frac{T^{(B)}}{4}\right)h_{\mu\nu}\right]_{y=0}\,,\\[3mm]
&\pi_{\mu\nu}=-\frac{1}{4}\tau_{\mu}{}^\lam\,\tau_{\lam\nu}+\frac{1}{12}\tau\,\tau_{\mu\nu}+\frac{1}{8}\tau^{\alpha\beta}\tau_{\alpha\beta}
\,h_{\mu\nu}-\frac{1}{24}\tau^2\,h_{\mu\nu}\,.$
In the above, $G_N$ constitutes the effective four-dimensional gravitational constant on the brane; this is also defined
in terms of the fundamental gravitational constant $\kappa_5^2$ and the brane tension $\sigma$. The quantity $\pi_{\mu\nu}$
is the well-known quadratic contribution of $\tau_{\mu\nu}$ \cite{SMS} which here, however, trivially vanishes since 
$\tau_{\mu\nu}=0$. Finally, $T^{(eff)}_{\mu\nu}$ can be interpreted as the effective energy-momentum tensor on the brane.
Together with $E_{\mu\nu}$, they constitute the imprint of the dynamics of the bulk fields---gravitational, and possibly gauge
and scalar fields generating the bulk energy-momentum tensor $T^{(B)}_{MN}$---on the brane.  The components of
$T^{(eff)}_{\mu\nu}$ are given by the following relation
\eq$\label{P4eq: new-brane-ene-comp}
T^{(eff)}_{\mu\nu}=\frac{1}{\kappa_5^2k}\left[3k^2h_{\mu\nu}+\frac{M}{r^3}\left(
\begin{array}{cccc}
 -h_{tt} & 0 & 0 & 0 \\
 0 & - h_{rr} & 0 & 0 \\
 0 & 0 & h_{\theta\theta} & 0 \\
 0 & 0 & 0 & h_{\varphi\varphi} \\
\end{array}
\right)\right]\,,$
while the components of the tensor $E_{\mu\nu}$, defined in \eqref{P4eq: E-weyl}, are evaluated to be
\eq$\label{P4eq: E-weyl-comp}
E_{\mu}{}_{\nu}\Big|_{y\ra 0}=\frac{M}{r^3}\left(
\begin{array}{cccc}
  -h_{tt} & 0 & 0 & 0  \\
 0 & -h_{rr} & 0 & 0  \\
 0 & 0 & h_{\theta\theta} & 0  \\
 0 & 0 & 0 & h_{\varphi\varphi} 
\end{array}
\right)\,.$
Notice that $E_{\mu\nu}$ is evaluated infinitesimally close to the brane but not exactly on it, its source
being the five-dimensional Weyl tensor.  Substituting the relations \eqref{P4eq: New-const}-\eqref{P4eq: E-weyl-comp}
in \eqref{P4eq: grav-eqs-br-new}, we readily obtain
\eq$\label{P4eq: 4dEin}
\bar{G}_{\mu\nu}=0\,.$
This is indeed the anticipated result since the induced line-element on the brane is described by the
Schwarzschild solution which is a vacuum solution.


\mysection{Conclusions \label{P4sec: Disc}}

In this Chapter, we have successfully constructed from first principles the geometry of an analytic
five-dimensional black hole exponentially localized close to our 3-brane. We have demonstrated that
the black-hole singularity lies entirely on the brane, while the event horizon extends into the bulk
but is exponentially suppressed as we move along the extra dimension. This exponential localization
alters the shape of the event horizon, making it  appear as a five-dimensional pancake. The
5-dimensional line-element is effectively AdS$_5$ outside the event horizon and
reduces  to the Schwarzschild solution on the brane. 
The derived geometry is supported by an anisotropic fluid in the bulk described by a diagonal
energy-momentum tensor with only two independent components: the energy density $\rho_E$
and tangential pressure $p_\theta$. All energy conditions are satisfied on the brane whereas a 
local violation takes place in the bulk in the region inside the event horizon. No additional
matter needs to be introduced on the brane for its consistent embedding in the bulk geometry
while the effective field equations are shown to be satisfied by the vacuum Schwarzschild
geometry on the brane.



\mychapter{Analytic and exponentially localized brane-world Reissner-Nordstr\"{o}m-(A)dS\\ solution \label{Chap: P5}}
\phantomsection
\setcounter{chapter}{6}

\epigraph{\justify\textit{``Yesterday I was clever, so I wanted to change the world. Today I am wise, so I am 
changing myself.''}}{Jalāl ad-Dīn Mohammad Rūmī}

\thispagestyle{empty}

{\lettrine[lines=3, lhang=0.03]{\color{chapter(color)}I}{\hspace*{5.5px}n} 
the preceding Chapter, we have constructed from first principles,
the geometry of an analytic, spherically-symmetric five-dimensional black hole. This was done by
combining both bulk and brane perspectives, that is by employing a set of coordinates that ensured 
the isotropy of the five-dimensional spacetime and combining it with an appropriately selected metric
function of the four-dimensional line-element. 
In the present Chapter, which is based on \cite{NK2}, we generalize our previous analysis by retaining 
the basic procedure for the construction
of the five-dimensional, spherically-symmetric black hole but by considering an alternative form of the 
metric function. This form is inspired by the one of the four-dimensional Reissner-Nordstr\"{o}m-(A)dS 
solution. In this way, we allow for the presence of a charge term and of a cosmological constant in the
effective metric, thus generalizing our previous assumption of a neutral, asymptotically-flat brane
black hole. However, being also part of a five-dimensional line-element, the richer topological structure
following from this new metric function is transferred also in the bulk. Thus, we perform a thorough
study of both the horizon structure of the five-dimensional spacetime and of all curvature invariants.
We also present a field-theory model for
the realization of the bulk matter, in the form of a five-dimensional tensor-vector-scalar theory, 
and discuss the conditions under which such a description could be viable. We then focus on the
presence of the brane itself, and we study the junctions conditions which govern its consistent
embedding  in the five-dimensional background.
Finally, we derive the 
gravitational equations of the effective theory and demonstrate that they are indeed satisfied
by the induced solution on the brane, namely the Reissner-Nordstr\"{o}m-(A)dS solution.

}

The structure of the Chapter is as follows: in \secref{P5sec: GS}, we present the five-dimensional geometry 
of the black hole and study its geometrical properties. 
In \secref{P5sec: GT}, we turn to the gravitational theory, we study the profile of the bulk matter and 
present the field-theory toy model. In \secref{P5sec: JC-ET}, we investigate the junction conditions
and the effective gravitational theory on the brane. We summarize our analysis and
discuss our results in \secref{P5sec: Disc}.

\mysection{The geometry \label{P5sec: GS}}

As mentioned previously, we are interested in placing a spherically-symmetric black hole on our brane. 
To this end, and by using the analysis of \secref{P4sec: geo}, we assume the line-element of the form
\eq$\label{P5eq: 5d-metr}
ds^2=\frac{1}{(1+k\rho \cos\chi)^2}\bigg[-f(\rho)\,dt^2+\frac{d\rho^2}{f(\rho)}+
\rho^2\,d\Omega_3^2\bigg]\,, \hspace{1.5em}\chi\in[0,\pi/2]\,,$
where
\eq$d\Omega_3^2=d\chi^2+\sin^2\chi\,d\theta^2+\sin^2\chi\,\sin^2\theta\,d\varphi^2\,.$
Here, $f(\rho)$ is a general spherically-symmetric function. Since we are interested in the study of black holes, 
we will therefore assume that $f(\rho)$
has a form inspired by the more general spherically-symmetric black-hole solution of General
Relativity, namely the Reissner-Nordstr\"{o}m-(Anti-)de Sitter solution:
\eq$\label{P5eq: ansatz}
f(\rho)=1-\frac{2M}{\rho}+\frac{Q^2}{\rho^2}-\frac{\Lambda}{3}\, \rho^2\,.$
Note that, on the brane where $\cos\chi=0$ and $\rho=r$, the line-element \eqref{P5eq: 5d-metr} 
{\it does} reduce to a Reissner-Nordstr\"{o}m-(Anti-)de Sitter black hole, with the parameter $M$
being related to its mass, $Q$ to its charge and $\Lambda$ to the effective cosmological
constant on the brane.

However, its interpretation from the bulk point-of-view needs to be carefully examined. 
Indeed, almost all known analytic black-hole solutions on the brane either lack completely a bulk description,
or correspond to bulk solutions with an undesired topology (i.e. that of a black string) or unattractive characteristics
(i.e. non-asymptotically AdS solutions). We will therefore investigate now the topological characteristics
of our five-dimensional construction. To this end, we compute all scalar gravitational
quantities, namely the Ricci scalar $R$, the Ricci tensor combination $\mathcal{R} \equiv R^{MN}R_{MN}$ 
and the Kretchmann scalar $\mathcal{K} \equiv R^{MNKL}R_{MNKL}$. The expression of the Ricci
scalar is the most elegant one and is given below
{\fontsize{11}{11}\eq$\label{P5eq: ricci-1}
R=-20 \left(k^2-\frac{\Lambda}{3} \right)+\frac{12k^2M\cos^2 \chi}{\rho }-\frac{12k\cos\chi \left(k Q^2 
\cos \chi+2 M\right)}{\rho ^2}+\frac{4\left(2 k Q^2 \cos \chi+ M\right)}{\rho ^3}\,,$}
\hspace{-0.5em}while the more extended  $\mathcal{R}$ and $\mathcal{K}$ quantities are presented in Appendix \ref{P5app: Curv-Inv}. 
The above expression contains a constant term which involves the warping parameter $k$ and the effective
cosmological parameter $\Lambda$. It also contains additional terms sourced by the mass and charge
of the black hole. These terms are singular at the value $\rho=0$
of the bulk radial coordinate. However, this singularity arises only when $r$ and $z$ are {\it simultaneously}
zero, i.e. at the location of the black-hole singularity {\it on the brane}. Any bulk point having by definition a non
zero value of $z$, and thus a non-zero value of $\rho$, is regular. In addition, all singular terms vanish in the
limit $\rho \rightarrow \infty$, i.e. 
when approaching the AdS asymptotic boundary or the radial infinity on the brane. Therefore, the spacetime
\eqref{P5eq: 5d-metr} does describe the gravitational background around a five-dimensional localized black hole with
a spacetime singularity entirely restricted on the brane. We also note that no singularity arises at the AdS
asymptotic boundary, a feature which plagues most non-homogeneous  black-string solutions. In our case,
far away from the brane, the spacetime becomes a maximally-symmetric one with a curvature determined
by the combination $-20\,(k^2-\Lambda/3)$. For $\Lambda=0$, we obtain the exact same  AdS spacetime 
of the Randall-Sundrum model. For positive but small values---compared to the warping effect driven by 
$k$---of the effective cosmological constant on the brane, the AdS character of the asymptotic regime is again
retained\,\footnote{Although mathematically possible, we do not consider here the case where $\Lambda > 3k^2$.
Since $k$ is an energy scale close to the fundamental gravity scale, that would demand an extremely large
$\Lambda$. Such an assumption is not supported by current observational data.} while, for $\Lambda<0$,
it is further enhanced.

\begin{figure}[t!]
    \centering
    \begin{subfigure}[b]{0.48\textwidth}
        \includegraphics[width=\textwidth]{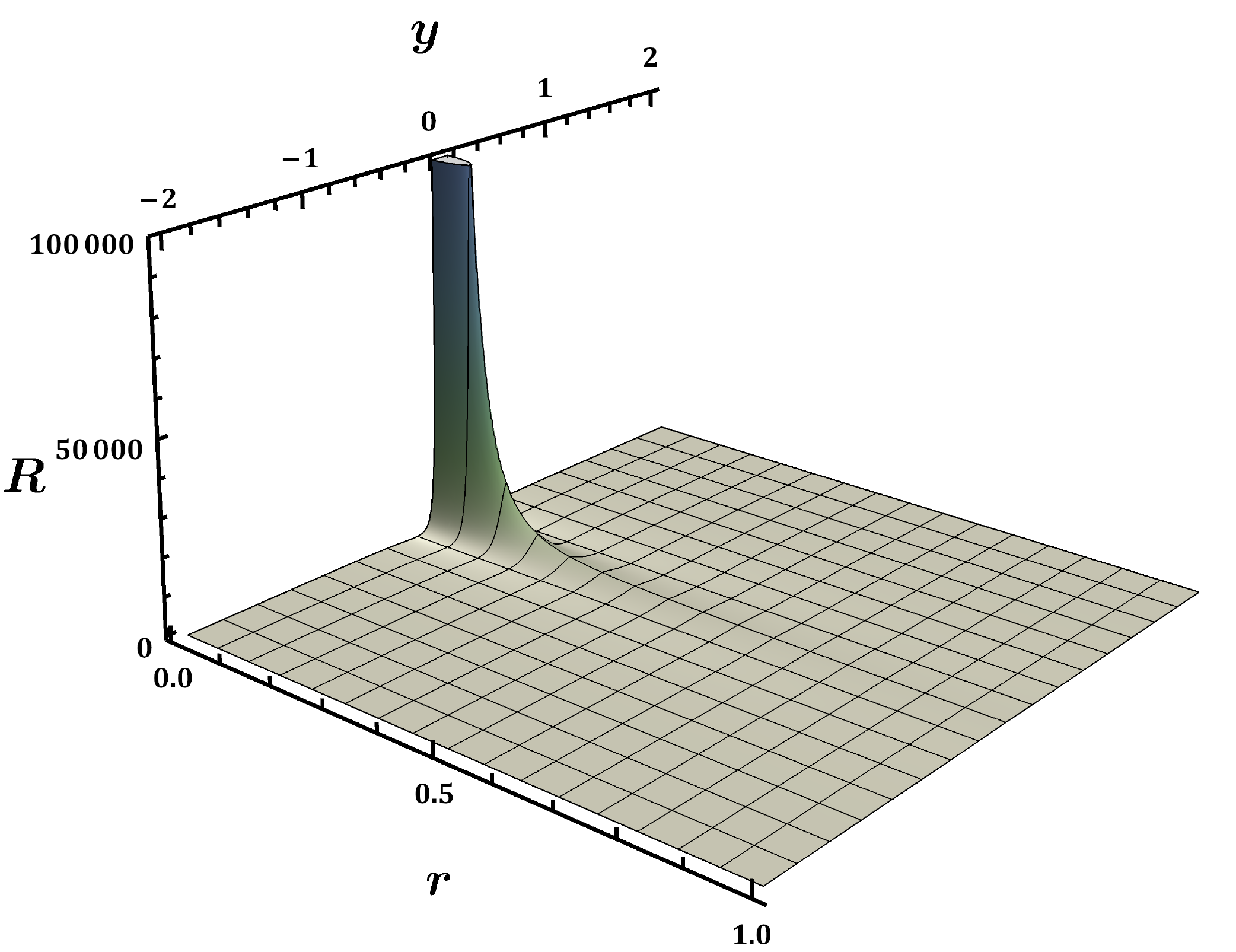}
        \caption{\hspace*{0em}}
        \label{P5subf: Ricci1-1}
    \end{subfigure}
    \hfill
    ~ 
    \begin{subfigure}[b]{0.48\textwidth}
        \includegraphics[width=\textwidth]{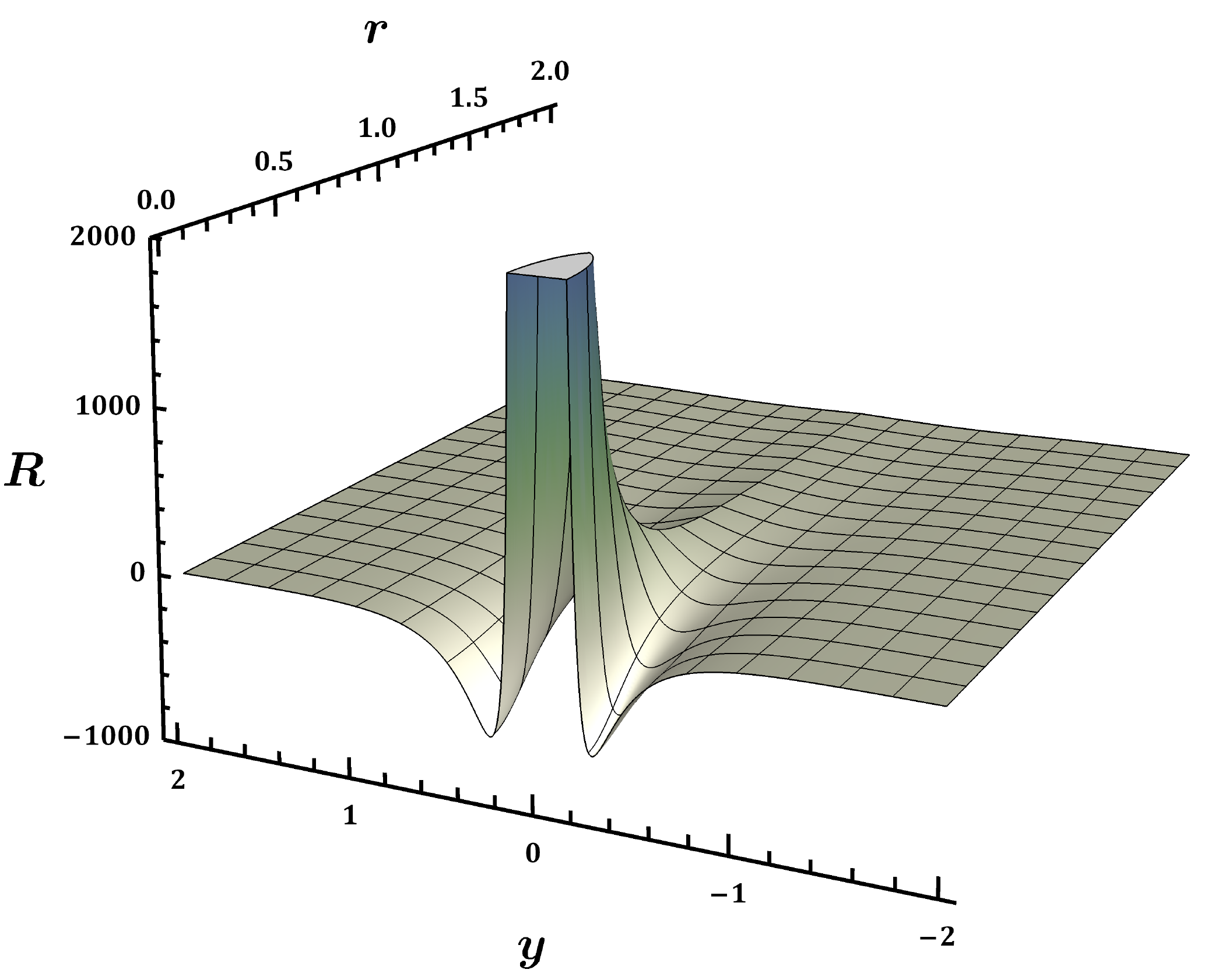}
        \caption{\hspace*{0em}}
        \label{P5subf: Ricci1-2}
    \end{subfigure}  
    \vspace{-0em}  
    \caption{(a) The scalar curvature $R$ in terms of the coordinates $(r,y)$ for 
    $k=1$, $M=10$, $Q=1$, and $\Lambda=5\times 10^{-4}$, while (b) shows a magnification of the geometry near the singularity.    }
    \label{P5fig: Ricci1-plot}
\end{figure}

\begin{figure}[t!]
    \centering
    \begin{subfigure}[b]{0.48\textwidth}
        \includegraphics[width=\textwidth]{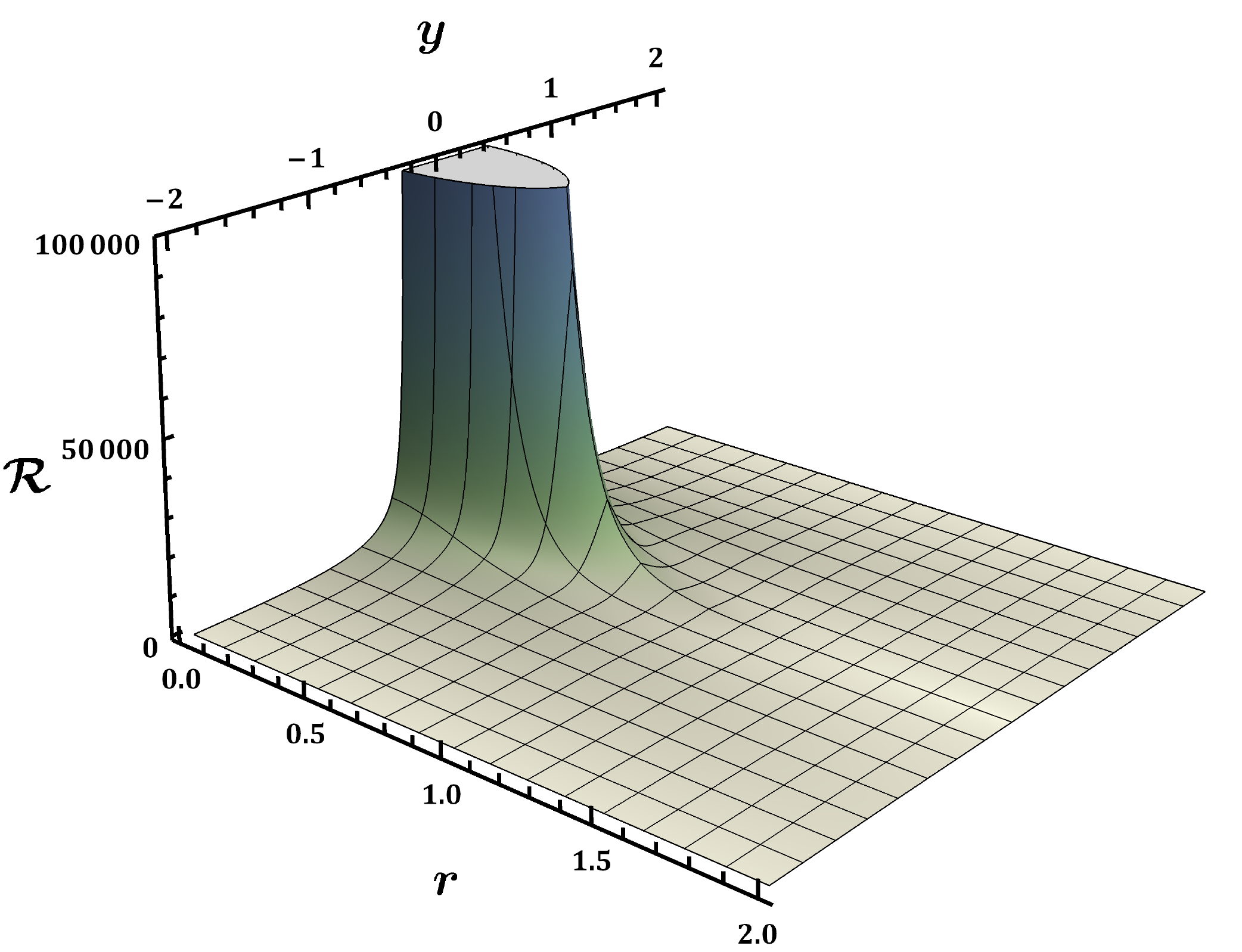}
        \caption{\hspace*{0em}}
        \label{P5subf: Ricci2}
    \end{subfigure}
    \hfill
    ~ 
    \begin{subfigure}[b]{0.48\textwidth}
        \includegraphics[width=\textwidth]{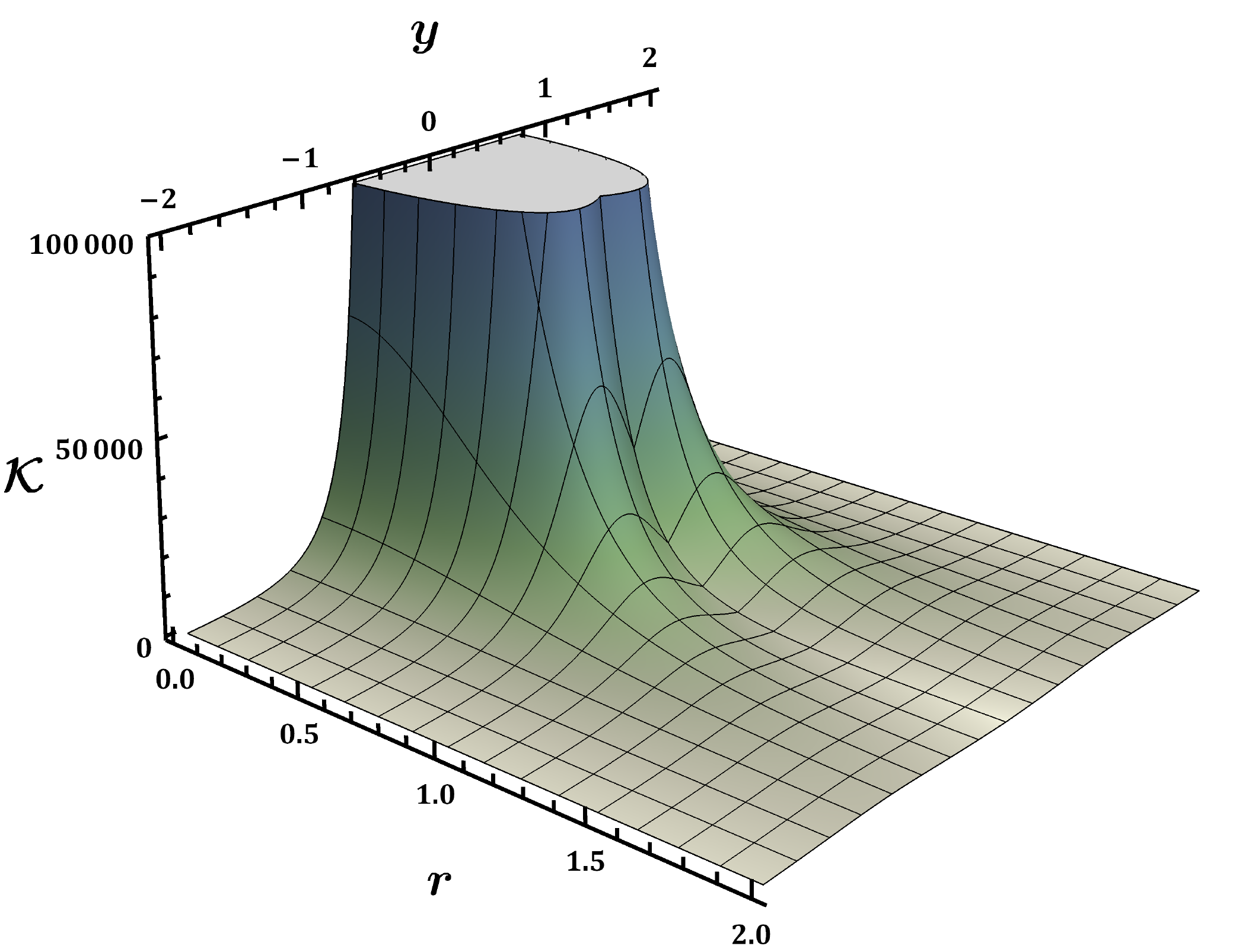}
        \caption{\hspace*{0em}}
        \label{P5subf: RiemSq}
    \end{subfigure}  
    \vspace{-0em}  
    \caption{(a) The invariant quantity $\mathcal{R}\equiv R_{MN}R^{MN}$ in terms of the coordinates $(r,y)$ for 
    $k=1$, $M=10$, $Q=1$, and $\Lambda=5\times 10^{-4}$, and (b) the invariant quantity $\mathcal{K}\equiv R_{MNKL}R^{MNKL}$
    for the same values of the parameters.    }
    \label{P5fig: Ricci2-Riem-plot}
\end{figure}

The expressions of the $\mathcal{R}$ and $\mathcal{K}$ invariant quantities  displayed in Appendix \ref{P5app: Curv-Inv}
also lead to the same conclusions drawn above for the topology of the five-dimensional spacetime.
It is illuminating to plot the behaviour of all curvature quantities. To this end, we use the original
$(r,y)$ brane and bulk coordinates as it is easier to depict the location of the brane. Using \eqref{P4eq: new-coords-inv}
in \eqref{P5eq: ricci-1}, we easily obtain for $R$ the expression
\eq$\label{P5eq: ricci-1-ry}
R=-20 \left(k^2-\frac{\Lambda}{3} \right)+\frac{4k^3M\left(10-12e^{k|y|}+3e^{2k|y|}\right)}{\left[k^2r^2
+\left(e^{k|y|}-1\right)^2\right]^{3/2}}-\frac{4k^4Q^2\left(5-8e^{k|y|}+3e^{2k|y|}\right)}{\left[k^2r^2
+\left(e^{k|y|}-1\right)^2\right]^2}\,.$
Similar expressions may be derived for $\mathcal{R}$ and $\mathcal{K}$, and these are again presented in
Appendix \ref{P5app: Curv-Inv}. In \fref{P5fig: Ricci1-plot}, we depict the Ricci scalar $R$ in terms of both $r$ and 
$y$---we remind the reader that, in this coordinate system, the brane is located at $y=0$. We observe that the
curvature of the 5-dimensional spacetime increases {\it only} when we approach the brane {\it and simultaneously}
take the limit $r \rightarrow 0$. All other bulk or brane points are regular. The curvature quickly decreases
as we move away from the singularity on the brane acquiring its constant, negative, asymptotic value 
corresponding to an AdS spacetime---this value is much smaller than the one adopted in the
vicinity of the singularity and thus is not visible in the plots. In  \myref{P5fig: Ricci1-plot}{P5subf: Ricci1-2}, we present a
magnification of the behaviour of the Ricci scalar close to the singular point. We observe the presence
of an interesting regime in the bulk where the curvature of spacetime dips to a large negative value before 
starting to increase close to the singularity. We will comment on this feature in the following
section. In \myref{P5fig: Ricci2-Riem-plot}{P5subf: Ricci2} and \myref{P5fig: Ricci2-Riem-plot}{P5subf: RiemSq}, we also present the 
behaviour of the $\mathcal{R}$ and $\mathcal{K}$ invariant quantities, respectively. They exhibit the same
asymptotic and near-singularity behaviors as the scalar curvature $R$ with the only difference being the
absence of the negative curvature well. 

\begin{sidewaysfigure}
    \centering
    \includegraphics[width=23.2cm, height=15.3cm]{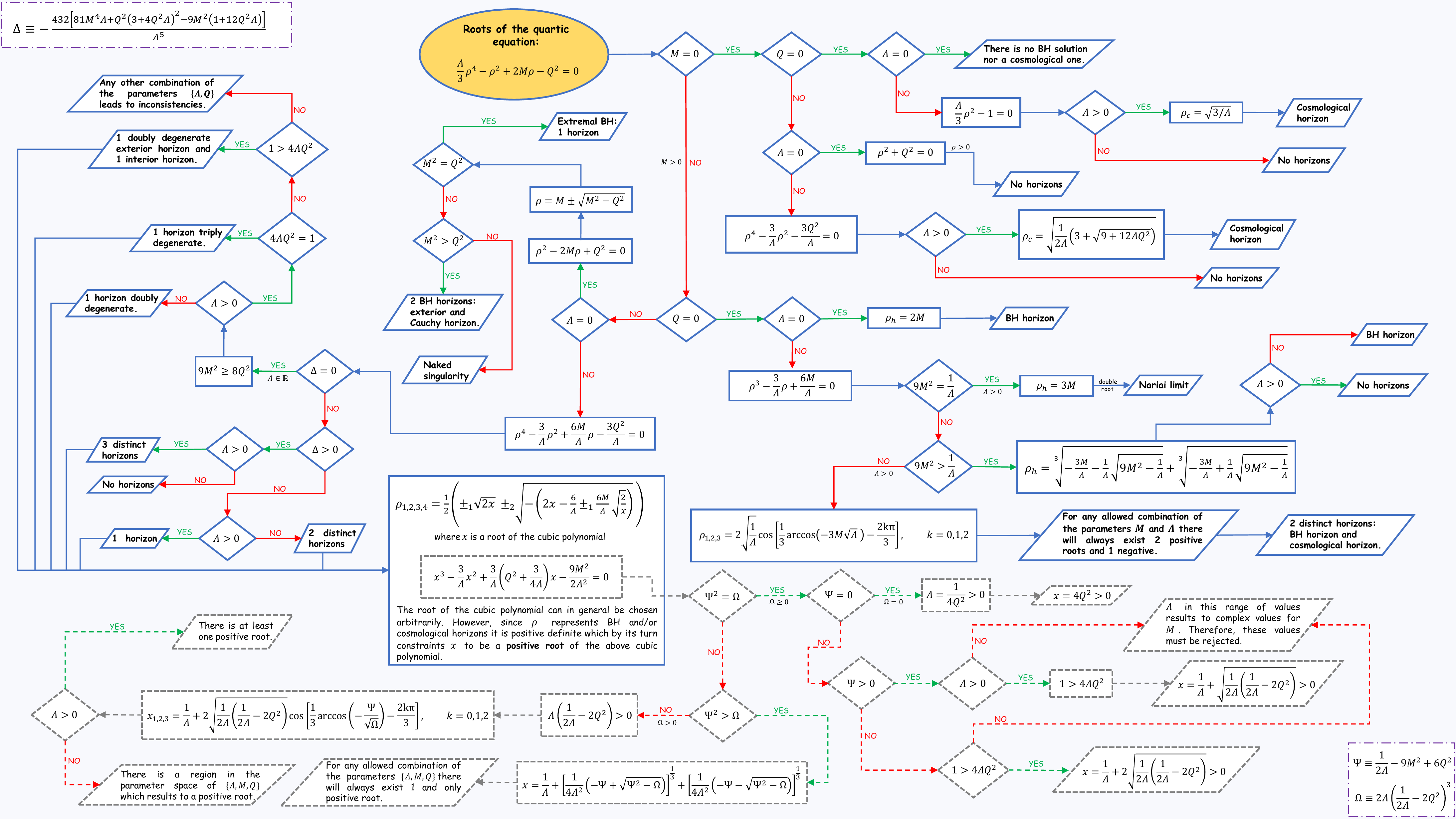}
    \caption{All possible roots of the quartic polynomial $f(\rho)=1-\frac{2M}{\rho}+\frac{Q^2}{\rho^2}-\frac{\Lambda}{3}\rho^2$.
   In the above flowchart, we catalogue the maximum possible number of horizons for each particular case.}
    \label{P5fig: Hor-chart}
\end{sidewaysfigure}

In order to discuss further the topology of the five-dimensional spacetime \eqref{P5eq: 5d-metr}, let us also re-write 
it in terms of the original non-spherical coordinates $(r,y)$. Employing again the inverse transformations
(\ref{P4eq: new-coords-inv}), the line-element takes the form
\bal$\label{P5eq: metr-r-y}
ds^2=e^{-2k|y|}&\Bigg\{-f(r,y) dt^2+\frac{dr^2}{r^2+z^2(y)}\biggl[\frac{r^2}{f(r,y)}+z^2(y)\biggr]+r^2d\Omega_2^2\nonum\\[1mm]
&\hspace{0.7em}+\frac{2r z(y)\,e^{k|y|}}{r^2+z^2(y)}\biggl[\frac{1}{f(r,y)}-1\biggr]drdy \Bigg\}
+\frac{dy^2}{r^2+z^2(y)}\left[r^2+\frac{z^2(y)}{f(r,y)}\right]\,,$
where $z(y)=sgn(y)(e^{k |y|}-1)/k$, and 
\eq$\label{P5eq: f-ry}
f(r,y)=1-\frac{2M}{\sqrt{r^2+z^2(y)}}+\frac{Q^2}{r^2+z^2(y)}-\frac{\Lambda}{3}\big[r^2+z^2(y)\big]\,.$
We are interested in the behaviour of the black-hole horizon(s) in the bulk. If the aforementioned spacetime
describes a regular, localized-on-the-brane black hole, its horizon(s) are expected to extend into the bulk but
stay close to the brane. To investigate this, we will study the causal structure of the bulk spacetime as this
is defined by the light cone. We will consider radial null trajectories in the five-dimensional background
\eqref{P5eq: metr-r-y}, and thus keep $\theta$ and $\varphi$ constant. Then, for a fixed value $y=y_0$ of the fifth
coordinate, the condition $ds^2=0$ leads to the result
\eq$ \label{P5eq: null-traj}
\frac{dt}{dr}=\pm\frac{1}{f(r,y_0)}\left[\frac{r^2k^2+f(r,y_0)\left(e^{k|y_0|}-1\right)^2}{r^2k^2+
\left(e^{k|y_0|}-1\right)^2}\right]^{1/2},$
where
\eq$\label{P5eq: horizon-function}
f(r,y_0)=1-\frac{2M}{\sqrt{r^2+\frac{\left(e^{k|y_0|}-1\right)^2}{k^2}}}+\frac{Q^2}{r^2
+\frac{\left(e^{k|y_0|}-1\right)^2}{k^2}}-\frac{\Lambda}{3}\left[r^2+\frac{\left(e^{k|y_0|}-1\right)^2}{k^2}\right].$
The location and topology of the horizons characterising the line-element \eqref{P5eq: metr-r-y} may be obtained
via Eq.  \eqref{P5eq: null-traj}, by determining the values of $(r,y_0)$ for which $dt/dr=\pm\infty$, or equivalently
$f(r,y_0)=0$. For $y_0=0$, Eq. \eqref{P5eq: horizon-function} reduces to the metric function $f(r)$ of a four-dimensional
Reissner-Nordstr\"{o}m-(anti-)de Sitter spacetime for which the emergence and location of horizons has been
extensively studied (see, for example, \cite{Chambers,Bousso}). A similar analysis may be conducted
also in the context of the five-dimensional spacetime  \eqref{P5eq: metr-r-y}, where the location of horizons is
determined by the equation $f(\rho)=0$, with the bulk radial coordinate being $\rho = \sqrt{r^2 + z^2(y)}$ ---we 
keep the $y$-coordinate fixed in Eqs. \eqref{P5eq: null-traj} and \eqref{P5eq: horizon-function} in order to present
the view of a static ``observer'' located at different slices of the bulk spacetime as we move away from the brane. 

In \fref{P5fig: Hor-chart}, we depict a flowchart\,\footnote{A flowchart is a graphical representation of a process
or a flow of consecutive steps. It was originated from computer science as a tool for representing algorithms and programming logic 
but nowadays plays an extremely useful role in displaying information visually and plainly. It is often the case that different 
flowcharts use different conventions about their symbols, thus, in our case we clarify that:
\begin{itemize}
\item \textbf{Ellipse/Terminator} represents the starting or ending point of the system.
\item \textbf{Rectangle/Process} represents a particular process, or a statement which is \textit{true}.
\item \textbf{Rhombus/Decision} represents a decision or a branching point. Lines coming out from the rhombus 
indicates different possible situations, leading to different sub-processes/sub-cases.
\item \textbf{Parallelogram/Data} represents information entering or leaving the system (input or output). In our
case it has mainly used as the final result/conclusion of each sub-case.
\end{itemize} } which constitutes an attentive 
scrutiny of the roots of the quartic polynomial  $f(\rho)=1-\frac{2M}{\rho}+\frac{Q^2}{\rho^2}-\frac{\Lambda}{3}\rho^2$. Every real, positive root
of this polynomial  corresponds either to a black-hole or a cosmological horizon of the five-dimensional 
spacetime \eqref{P5eq: 5d-metr}.  Let us consider some indicative cases. For $M \neq 0$ but $Q=0$ and $\Lambda=0$, 
we obtain the case of a five-dimensional spacetime with a sole black-hole horizon at $\rho_H=2M$. This
may be written in terms of $(r,y)$ as
\eq$\label{P5eq: hor-loc-Scwarz}
r_H^2=4M^2-\frac{\big(e^{k|y_0|}-1\big)^2}{k^2}\,.$
This case was studied in the previous Chapter where it was shown that the black-hole horizon was exponentially
localized close to the brane. Indeed, the aforementioned equation reveals the exponential decrease of $r_H$ as
$|y_0|$ increases and the existence of a value where the horizon vanishes, namely at $|y_0|=\ln (2M k+1)/k$.
Beyond this point, any $y$-slice of the five-dimensional spacetime is horizon-free and almost pure AdS. 
In addition, the black-hole singularity was strictly localized on the brane as in the present
analysis.

Does the horizon exponential localization persist also in the case of multiple horizons? Let us consider the case
with $M \neq 0$ and $Q \neq 0$ ($M^2>Q^2$), but $\Lambda=0$ for simplicity. In that case, it is easy to see that two horizons
emerge, an internal Cauchy horizon and an external event horizon located at $\rho_\pm= M \pm \sqrt{M^2-Q^2}$. 
Employing again the $(r,y)$ coordinates, these are re-written as
\eq$\label{P5eq: hor-loc-Reissn}
r_\pm^2=\left(M \pm \sqrt{M^2-Q^2}\right)^2-\frac{\big(e^{k|y_0|}-1\big)^2}{k^2}\,.$
We observe that both horizons shrink as we move to $y$-slices of the bulk spacetime located further away from the brane.
Once again both horizons cease to exist beyond a certain value of $y$, namely at the values
\eq$\label{P5eq: hor-loc-Reissn1}
|y_0|_{\pm}=\frac{1}{k} \,\ln\left[1+ kM \left(1 \pm \sqrt{1-\frac{Q^2}{M^2}}\right)\right]\,.$
Note that each horizon will vanish at its own value of the $y$-coordinate and that the horizon corresponding to
the smaller value of the radial coordinate $\rho$, i.e to the smaller root of the equation $f(\rho)=0$, will vanish first.

The most general case arises when $M \neq 0$, $Q \neq 0$ and $\Lambda>0$. Then, we can have at most
three real, positive roots of the equation $f(\rho)=0$, and thus three horizons: an internal Cauchy horizon $\rho_-$,
an external event horizon $\rho_+$ and a cosmological horizon $\rho_{\ssst{C}}$. Their location in terms of the radial coordinate
$\rho$ is determined solely by the parameters $M$, $Q$ and $\Lambda$ and they naturally extend both on the brane
and in the bulk. As above, their profile in the bulk may be studied if we change to the $(r,y)$ coordinates; then,
the following general relation holds
\eq$\label{P5eq: hor-loc}
r_h^2=\rho_h^2-\frac{\big(e^{k|y_0|}-1\big)^2}{k^2}\,,$
where the subscript $h$ has been used to denote the location of all three horizons. Since the value of $\rho_h$
is fixed by $M$, $Q$ and $\Lambda$, it is obvious that as $|y_0|$ increases the value of $r_h$ exponentially
decreases. Thus, as we move along the extra dimension away from the brane, $r_h$ quickly shrinks and becomes
zero at a distance
\eq$
|y_0|_{(h)}=\frac{1}{k}\ln(k\rho_h+1)\,.$
Again, each horizon will vanish at a different point along the extra dimension: the Cauchy horizon will vanish first,
the event horizon will follow next and the cosmological horizon will disappear last.\,\footnote{The vanishing of the
cosmological horizon does not mean that the causal spacetime disappears but rather that a change of coordinates
is necessary (see Appendix \ref{P5app: CHCS}). After this point, a static ``observer'' no longer exists and a set of planar coordinates, such as
the ones used in cosmology to describe a time-depending de Sitter universe, is more appropriate. If one insists
in keeping the static, spherically-symmetric set of coordinates of Eq. \eqref{P5eq: 5d-metr}, and thus the notion of
a static ``observer'', then an interesting bound arises as to how far from the first brane a second one may be
introduced.}
Due to the exponential fall-off of each $r_h$ in terms of the $y$-coordinate, all horizons acquire a ``pancake''
shape with its longer dimension lying along the brane and its shorter one along the bulk. As an indicative case
see \fref{P4fig: loc-BH-plot}, where it is depicted the geometrical representation of the event horizon of the five-dimensional
Schwarzschild spacetime ($Q=0,\Lambda=0$). It is important to stress that by introducing non-vanishing $Q$ or $\Lambda$ 
the depicted general behaviour does not change.

\mysection{The gravitational theory \label{P5sec: GT}}

After constructing the geometrical set-up of our five-dimensional gravitational theory, we now consider its action functional
which is described by the general expression
\eq$\label{P5eq: bulk-action}
S_{B}=\int d^5x\, \sqrt{-g}\left(\frac{R}{2\kappa_5^2}+\lagr^{(B)}_{m}\right).$
By varying the aforementioned action functional $S_B$ with respect to the metric tensor $g_{MN}$, we may
derive the gravitational field equations in the bulk which have the form
\eq$\label{P5eq: field-eqs}
G_{MN}=\kappa_5^2\, T^{(B)}_{MN}\,.$
The quantity  $G_{MN}=R_{MN}-\frac{1}{2}\,g_{MN}R$ denotes the Einstein-tensor while $T^{(B)}_{MN}$
is the bulk energy-momentum tensor associated with the Lagrangian density $\lagr^{(B)}_m$.  If we use the 
gravitational background \eqref{P5eq: 5d-metr} constructed in the previous section, we find that the 
non-zero components of $T^{(B)}_{MN}$ in mixed form are the following:
{\fontsize{11}{11}\gat$\label{P5eq: rho-p1}
T^{(B)t}{}_t=T^{(B)\rho}{}_\rho=\frac{1}{\kappa_5^2}\left[2(3 k^2-\Lambda)+\frac{3 k\cos \chi }{\rho ^2}\left(3M-\frac{2Q^2}{\rho}\right)-\frac{3 M}{\rho ^3}\right],\\[2mm]
\label{P5eq: p2}
T^{(B)\chi}{}_\chi=T^{(B)\theta}{}_\theta=T^{(B)\varphi}{}_\varphi=\frac{1}{\kappa_5^2}\left[2(3 k^2-\Lambda)-\frac{6k^2\cos^2\chi}\rho\left(M-\frac{Q^2}{\rho}\right)+\frac{6 k M \cos \chi }{\rho ^2}\right].$}
\hspace{-0.5em}The bulk energy-momentum tensor is thus characterized solely by three components: the energy-density
$\rho_E \equiv -T^{(B)t}{}_{t}$, the radial pressure $p_r \equiv T^{(B)\rho}{}_{\rho}$,  and a common tangential
pressure $p_\theta \equiv T^{(B)\chi}{}_{\chi}=T^{(B)\theta}{}_{\theta}= T^{(B)\varphi}{}_{\varphi}$. Therefore, the
gravitational background \eqref{P5eq: 5d-metr} of a five-dimensional, localized close to the brane black-hole solution
may be supported by a {\it diagonal} energy-momentum tensor which describes an anisotropic fluid. Employing
the fluid's timelike five-velocity $U^M$ and a spacelike unit vector in the direction of $\rho$-coordinate satisfying
the relations 
\gat$\label{P5eq: U-vec}
U^M=\{U^t,0,0,0,0\},\hspace{1em} U^M U^N g_{MN}=-1\,,\\[2mm]
\label{P5eq: X-vec}
X^M=\{0,X^\rho,0,0,0\},\hspace{1em} X^M X^N g_{MN}=1\,,$
the bulk energy-momentum tensor may be written in a covariant notation as follows
\eq$\label{P5eq: en-mom}
T^{(B)MN}=(\rho_E+p_\theta)U^M U^N+(p_r-p_\theta)X^M X^N+p_\theta\, g^{MN}\,.$
The aforementioned, rather minimal, content of the bulk energy-momentum tensor was first found in the case
where the brane background was assumed to be the Schwarzschild solution \cite{NK1}. As we see, this structure
persists also in the case where the brane background assumes the form of more generalized four-dimensional black-hole
solutions. 

As in \chapref{Chap: P4}, there are only two independent components of the energy-momentum tensor, namely the energy-density
$\rho_E$ and the tangential pressure component $p_\theta$; as Eq. (\ref{P5eq: rho-p1}) reveals, the radial pressure component
$p_r$ is found to satisfy the equation of state $p_r=-\rho_E$ everywhere in the bulk. In addition, at asymptotic
infinity, i.e. as $\rho\ra+\infty$,  all three components of the energy-momentum tensor reduce to a constant
value, which can be identified as the five-dimensional cosmological constant $\Lambda_5$,
\gat$\label{P5eq: as-rho}
\lim_{\rho\ra+\infty}\rho_{E}(\rho,\chi)=-\frac{2(3k^2-\Lambda)}{\kappa_5^2} \equiv \Lambda_5\,, \\[1mm]
\label{P5eq: as-p2}
\lim_{\rho\ra+\infty}p_r(\rho,\chi)=\lim_{\rho\ra+\infty}p_\theta(\rho,\chi)=\frac{2(3k^2-\Lambda)}{\kappa_5^2} \equiv
-\Lambda_5\,.$
As discussed in the previous section and also confirmed here, the asymptotic
form of the bulk spacetime depends on the sign of the quantity $(3k^2-\Lambda)$. If $3k^2>\Lambda$, the asymptotic
form of the energy-momentum tensor reduces to that of a negative bulk cosmological constant and the curvature invariant
quantities $R$, $\mathcal{R}$, $\mathcal{K}$ match the ones of an AdS$_5$ spacetime. Then, the brane parameter
$\Lambda$ is determined through the relation $2\Lambda=6k^2 -\kappa^2_5 |\Lambda_5|$, and its exact value depends
on the inter-balance between the warp parameter $k$ and the bulk cosmological constant $\Lambda_5$. In the special
case where a fine-tuning is imposed so that $\Lambda=0$,  the metric \eqref{P5eq: 5d-metr} incorporates exactly the 
Randall-Sundrum model \cite{RS1, RS2} at the spacetime boundary. Note, however, that the form of the warp factor
remains of an exponential form, i.e. $e^{-k |y|}$,  in our analysis regardless of the value of $\Lambda$. 

\begin{figure}[t!]
    \centering
    \begin{subfigure}[b]{0.485\textwidth}
    	\includegraphics[width=\textwidth]{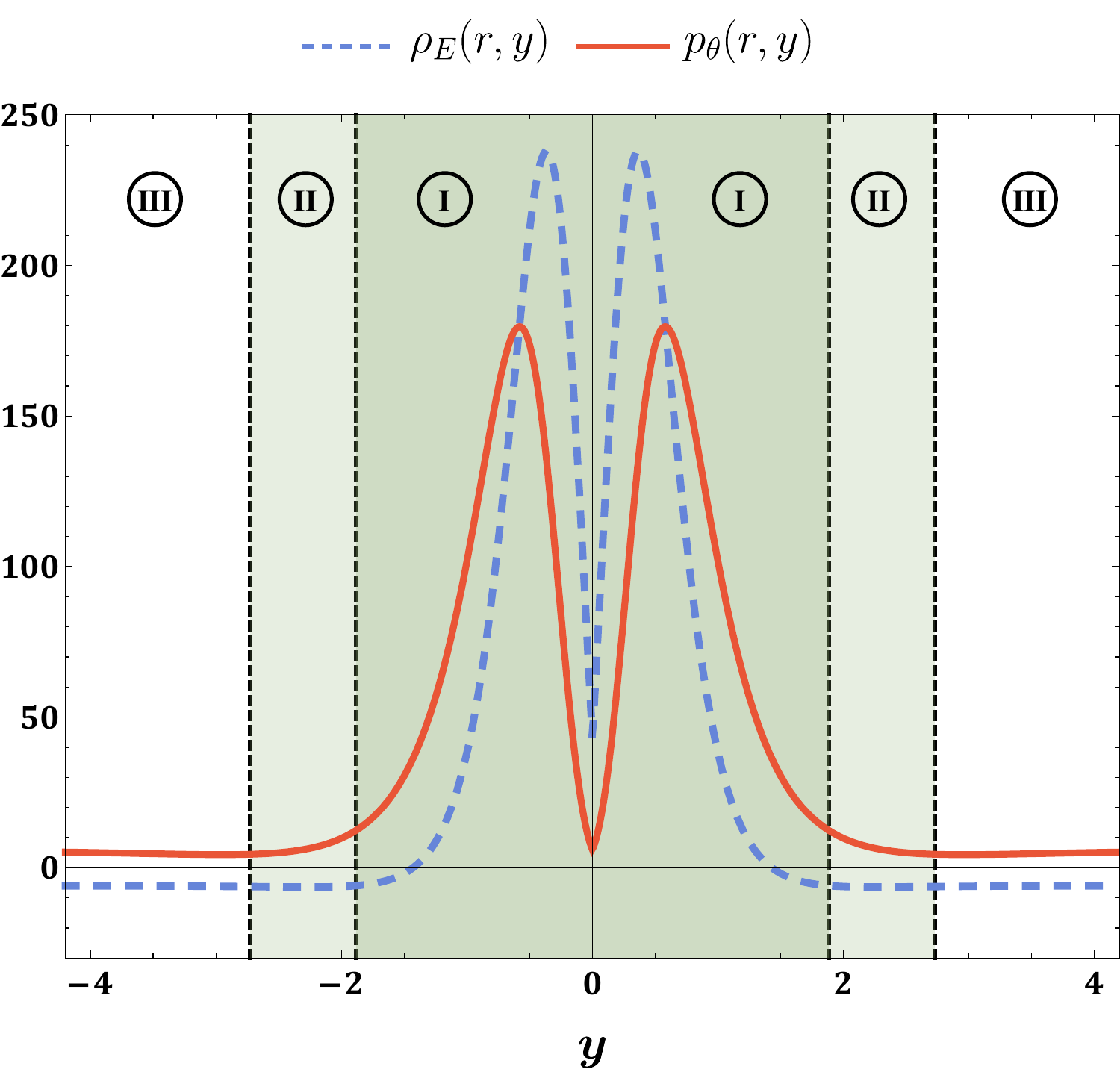}
    	\caption{\hspace*{-1.35em}}
    	\label{P5subf: rho-p2-plot}
    \end{subfigure}
    \hfill
    \begin{subfigure}[b]{0.5\textwidth}
    	\includegraphics[width=\textwidth]{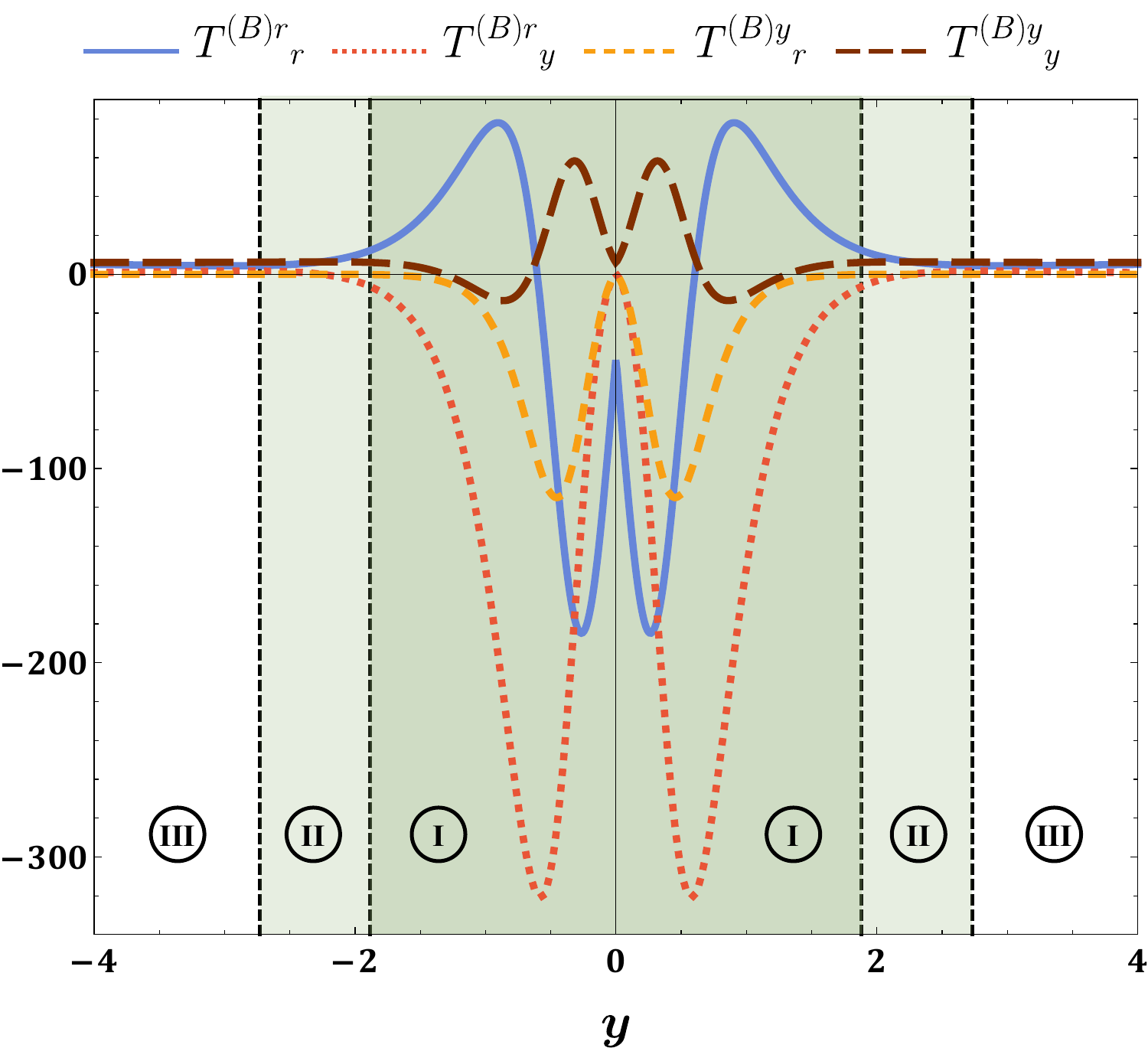}
		\caption{\hspace*{-1.9em}}    	
    	\label{P5subf: Ene-Mom-comp-plot}
    \end{subfigure}
    \caption{(a) The profiles of the energy density $\rho_E$ and tangential pressure $p_\theta$ in terms of the $y$-coordinate
     for $\kappa_5=1$, $k=1$, $M=10$, $Q=9$, $\Lambda=5\times 10^{-21}$ and $r=0.85$. (b) The profiles of $T^{(B)r}{}_r$,
     $T^{(B)r}{}_y$, $T^{(B)y}{}_r$ and $T^{(B)y}{}_y$ depicted for the same values of the parameters.  Region I lies inside
     the Cauchy horizon, region II corresponds to the bulk spacetime between the two black-hole horizons, while
     region III is  located between the exterior black-hole horizon and the cosmological one.}
     \label{P5fig: EneMom-plot}
\end{figure}

In order to study in more detail the profiles of the energy density $\rho_E$ and pressure $p_\theta$ in the bulk, we employ again the 
coordinates $(r,y)$. Using Eqs. \eqref{P4eq: new-coords-inv}, \eqref{P5eq: rho-p1} and \eqref{P5eq: p2}, we find
{\fontsize{11}{11}\gat$\label{P5eq: rho-ry}
\rho_E(r,y)=-\frac{1}{\kappa_5^2}\left\{2(3k^2-\Lambda)-\frac{3Mk^3\left(4-3e^{k|y|}\right)}{\left[k^2r^2+\left(e^{k|y|}-1\right)^2\right]^{3/2}}
-\frac{6Q^2k^4\left(e^{k|y|}-1\right)}{\left[k^2r^2+\left(e^{k|y|}-1\right)^2\right]^{2}} \right\}\,,\\[2mm]
\label{P5eq: p2-ry}
p_\theta(r,y)=\frac{1}{\kappa_5^2}\left\{2(3k^2-\Lambda)+\frac{6Mk^3\left(e^{k|y|}-1\right)
\left(2-e^{k|y|}\right)}{\left[k^2r^2+\left(e^{k|y|}-1\right)^2\right]^{3/2}}+\frac{6Q^2k^4\left(e^{k|y|}-1\right)^2}{\left[k^2r^2
+\left(e^{k|y|}-1\right)^2\right]^{2}} \right\}\,.$}
\hspace{-0.5em}In \myref{P5fig: EneMom-plot}{P5subf: rho-p2-plot}, we present the profiles of the energy density $\rho_E$ 
and tangential pressure $p_\theta$ in terms of the
bulk coordinate $y$. In this indicative case, the values of the parameters were chosen to be $\kappa_5=1$, $k=1$, $M=10$,
$Q=9$, $\Lambda=5\times 10^{-21}$, and we have also fixed the radial coordinate on the brane at the random value $r=0.85$.
Substituting the aforementioned values of $M$, $Q$ and $\Lambda$ in the flowchart of \fref{P5fig: Hor-chart}, one
can evaluate the locations of the three distinct horizons, namely $\rho_{-}=5.64$ (Cauchy horizon), $\rho_{+}=14.36$ (exterior 
black-hole horizon) and $\rho_{\ssst{C}}=2.45 \times 10^{10}$ (cosmological horizon). Given these values and the fixed radial
distance $r=0.85$, it is straightforward to calculate from Eq. \eqref{P5eq: hor-loc} the corresponding values of $y$ at
which we encounter the three horizons in the bulk: the Cauchy horizon lies at $y_{-}=1.88$, the exterior black-hole horizon at 
$y_{+}=2.73$, and the cosmological horizon at $y_{c}=23.92$. We denote the bulk region inside the Cauchy horizon as region I,
the region between the two black-hole horizons as region II, and the region between the exterior black-hole horizon and the
cosmological horizon as region III; we denote these regions also in \myref{P5fig: EneMom-plot}{P5subf: Ene-Mom-comp-plot}. 

We observe that both the energy density $\rho_E$ and tangential pressure $p_\theta$ exhibit a shell-like distribution in region I,
i.e. in the region between the brane, located at $y=0$, and the Cauchy horizon. As the latter is approached, both components
quickly decrease towards their AdS$_5$ asymptotic values given by Eqs. \eqref{P5eq: as-rho} and \eqref{P5eq: as-p2}. These values are
adopted even before the exterior black-hole horizon is reached, therefore, as \myref{P5fig: EneMom-plot}{P5subf: rho-p2-plot} clearly depicts, region III
describes a pure AdS$_5$ spacetime. On the brane, the energy density $\rho_E$ and tangential pressure $p_\theta$ adopt
values which respect all energy conditions since there we have $\rho_E>0$, $\rho_E+p_r=0$ and $\rho_E>p_\theta$. Although the
profiles of $\rho_E$ and $p_\theta$ depend on the chosen values of the parameters of the theory, the behaviour depicted in 
\myref{P5fig: EneMom-plot}{P5subf: rho-p2-plot} is by no means a special one and in fact arises for a large number of sets of parameter values. What we
should also stress is the emergence of a regime close to the Cauchy horizon where the energy conditions are violated
since $\rho_E < p_\theta$. The same behaviour was also observed in the previous Chapter and seems to be a requisite
for the localization of the black-hole topology close to the brane as well as for the transition to a pure AdS$_5$ spacetime which, by
construction, is characterised by the relation $p_i=-\rho_E=|\Lambda_5|$.

 When we perform the coordinate change described via Eq. \eqref{P4eq: new-coords-inv}, the components $T^{(B)M}{}_N$ of the
 energy-momentum tensor are bound to change. The $T^{(B)t}{}_t \equiv -\rho_E$ and $T^{(B)\theta}{}_\theta=T^{(B)\varphi}{}_\varphi \equiv p_\theta$
 components receive no additive corrections and their change amounts to merely substituting $\{\rho, \chi\}$ by $\{r,y\}$ in
 their expressions, thus leading to Eqs. \eqref{P5eq: rho-ry} and \eqref{P5eq: p2-ry}.
 However, the $T^{(B)r}{}_r$, $T^{(B)r}{}_y$, $T^{(B)y}{}_r$ and $T^{(B)y}{}_y$ components receive also additive corrections and their
 expressions are significantly modified. The analysis leading to the new expressions of all the components of the energy-momentum
 tensor is given in Appendix \ref{P5app: En-Mom}.  Therefore, for completeness, in \myref{P5fig: EneMom-plot}{P5subf: Ene-Mom-comp-plot} we 
 depict also the behaviour of
 these four components of the energy-momentum tensor in terms of  the extra dimension $y$ and for the same set of parameter
values as in \myref{P5fig: EneMom-plot}{P5subf: rho-p2-plot}.  As was the case with $\rho_E$ and $p_\theta$, these components remain everywhere regular,
have a shell-like distribution inside region I and quickly adopt their asymptotic values even before the exterior black-hole
horizon is reached: for the $T^{(B)r}{}_r$ and $T^{(B)y}{}_y$ components, this asymptotic value is $|\Lambda_5|$ while,
for the off-diagonal components  $T^{(B)r}{}_y$ and $T^{(B)y}{}_r$, this asymptotic value is zero, as expected.
Note also that the coordinate change \eqref{P4eq: new-coords-inv} destroys the simple relations \eqref{P5eq: rho-p1}, \eqref{P5eq: p2}
between the components of the energy momentum tensor which were
valid in the $\{t,\rho,\chi,\theta,\varphi\}$ set of coordinates. All these reveal that, although the ``axial''  set of coordinates
$\{t,r,\theta,\varphi,y\}$ serve better to illustrate the behaviour of both the curvature and distribution of matter {\it with respect
to the brane observer}, it is the ``spherical'' set of coordinates $\{t,\rho,\chi,\theta,\varphi\}$ which {\it encodes the highest
symmetry} of the five-dimensional theory and leads to the simplest profile of both the spacetime and the energy-momentum tensor. 


\mysubsection{A field-theory toy-model}

\par In this subsection, we will investigate further the nature of the bulk energy-momentum tensor  which is
necessary  to support the geometry of the five-dimensional localized  Reissner-Nordstr\"om-(A)dS black hole
presented in \secref{P5sec: GS}.  Due to the simple structure of $T_{MN}^{(B)}$, given in Eqs. (\ref{P5eq: rho-p1}), (\ref{P5eq: p2}),
the term ``anisotropic fluid'' was used to describe it, and a covariant form for its expression was also found.
However, it would be interesting to see if a field-theory model could be proposed to support it, and under
which conditions on the associated fields this task could be fulfilled. 

In the following analysis, we will use the spherically-symmetric set of coordinates  $\{t, \rho$, $\chi, \theta, \varphi\}$
in which $T^{(B)}_{MN}$ takes its simplest possible form, as argued above. We will employ five-dimensional
fields that are allowed to propagate outside our brane, and thus consider scalar or gauge fields which are
distinct from the corresponding Standard-Model degrees of freedom living on the brane. According to our
results, a theory with only minimally-coupled scalars or with only minimally-coupled vector fields fails to lead
to the desired structure of the bulk energy-momentum tensor.  We therefore consider a tensor-vector-scalar
five-dimensional field theory where the bulk matter Lagrangian density $\lagr^{(B)}_{m}$ appearing in
Eq. (\ref{P5eq: bulk-action}) is given by  
\gat$\label{P5eq: lagr}
\lagr^{(B)}_{m}:= \lagr^{(g)}+\lagr^{(sc)}\,,$
with
\gat$
\label{P5eq: lagr-em}
\lagr^{(g)}:=-\frac{1}{4}\,F^{MN}F_{MN}\,,\\[2mm]
\label{P5eq: lagr-sc}
\lagr^{(sc)}:=-f_1(\xi,\psi)(\pa\xi)^2-f_2(\xi,\psi)(\pa \psi)^2-V(\xi,\psi)\,.$
Above, $F_{MN}=\nabla_M A_N - \nabla_N A_M$ is the field-strength tensor of an Abelian gauge field $A_M$ and 
$\{\xi(\rho,\chi),\psi(\rho,\chi)\}$ are two scalar fields. In addition, we have introduced two arbitrary functions
$f_1(\xi,\psi)$ and $f_2(\xi,\psi)$ in the kinetic terms of the scalar fields as well as an interaction potential $V(\xi,\psi)$.
The variation of $\sqrt{-g}\,\lagr^{(B)}_{m}$ with respect to $g_{MN}$ leads to the result
\eq$
T^{(B)MN} = T^{(g)MN} + T^{(sc)MN}\,,$
where
\eq$\label{P5eq: ene-mom-em}
T^{(g)MN} =F^{MA}F^N{}_A-\frac{1}{4}\,g^{MN}F^{AB}F_{AB}\,,$
and
\eq$\label{P5eq: ene-mom-sc}
T^{(sc)}_{MN}=2f_1(\xi,\psi)\pa_M\xi\,\pa_N\xi+2f_2(\xi,\psi)\pa_M\psi\,\pa_N\psi+g_{MN}\,\lagr^{(sc)}\,.$

In what follows, we will also assume the following configuration for the gauge field-strength tensor
\eq$\label{P5eq: em-tensor}
\left(F^{MN}\right)=\left(\begin{array}{ccccc}
0 & E_1(\rho,\chi) & E_2(\rho,\chi) & 0 & 0\\[1mm]
-E_1(\rho,\chi) & 0 & 0 & 0 & 0\\[1mm]
-E_2(\rho,\chi) & 0 & 0 & 0 & 0\\[1mm]
0 & 0 & 0 & 0 & B(\rho,\chi,\theta)\\[1mm]
0 & 0 & 0 & -B(\rho,\chi,\theta) & 0
\end{array}\right),$
where $E_1$, $E_2$ and $B$ stand for two components of the ``electric'' bulk gauge field and a sole
component of the ``magnetic'' field, respectively. Employing the above in Eq. (\ref{P5eq: ene-mom-em}), one may
easily calculate the components of the gauge-field energy-momentum tensor $T^{(g)MN} $. Using also
the expression (\ref{P5eq: ene-mom-sc}), the components of $T^{(sc)}_{MN}$ for the two scalar fields readily
follow. Taking their sum, we obtain the following results for the non-vanishing mixed components of the bulk
energy-momentum tensor 
{\fontsize{10}{10}\gat$
 T^{(B)t}{}_t=\frac{1}{2}(-b B^2+a_1 E_1^2+a_2 E_2^2)-f_1(\pa\xi)^2-f_2(\pa\psi)^2-V, \nonumber \\[2mm]
  T^{(B)\rho}{}_\rho=\frac{1}{2}(-b B^2+a_1 E_1^2-a_2 E_2^2)+f_1\left(\pa^\rho\xi\pa_\rho\xi-\pa^\chi\xi\pa_\chi\xi\right)
+f_2\left(\pa^\rho\psi\pa_\rho\psi-\pa^\chi\psi\pa_\chi\psi\right)-V, \nonumber\\[2mm]
 T^{(B)\rho}{}_\chi = a_2E_1 E_2+ 2\left(f_1\,\pa^\rho\xi\pa_\chi\xi+f_2\,\pa^\rho\psi\pa_\chi\psi\right), \label{P5eq: total-emt}\\[2mm]
 T^{(B)\chi}{}_\chi=\frac{1}{2}(-b B^2-a_1 E_1^2+a_2 E_2^2)-f_1\left(\pa^\rho\xi\pa_\rho\xi-\pa^\chi\xi\pa_\chi\xi
\right)-f_2\left(\pa^\rho\psi\pa_\rho\psi-\pa^\chi\psi\pa_\chi\psi\right)-V, \nonumber\\[2mm]
 T^{(B)\theta}{}_\theta=T^{(B)\varphi}{}_\varphi=\frac{1}{2}(b B^2-a_1 E_1^2-a_2 E_2^2)-f_1(\pa\xi)^2-f_2(\pa\psi)^2-V. 
\nonumber$}
\hspace{-0.5em}In the above, we have defined the quantities $a_1\equiv g_{tt}\,g_{\rho\rho}$, $a_2\equiv g_{tt}\,g_{\chi\chi}$ and 
$b\equiv g_{\theta\theta}\,g_{\varphi\varphi}$ for simplicity.

Let us now investigate whether the above set of components can be simplified in order to resemble the minimal 
configuration described by Eqs. \eqref{P5eq: rho-p1}, \eqref{P5eq: p2}. We thus first demand that $T^{(B)t}{}_t= T^{(B)\rho}{}_\rho$,
and we obtain the constraint
\eq$\label{P5eq: E2-eq}
E_2^2=\frac{2\left(f_1\,\pa^\rho\xi\pa_\rho\xi+f_2\,\pa^\rho\psi\pa_\rho\psi\right)}{a_2}\,.$
We next observe that the configuration of Eqs. \eqref{P5eq: rho-p1}, \eqref{P5eq: p2} has no off-diagonal component. Thus
demanding that $T^{(B)\rho}{}_\chi =0$ and employing Eq. \eqref{P5eq: E2-eq}, we also obtain 
\eq$\label{P5eq: E1-eq}
E_1^2=\frac{1}{a_2}\frac{2\left(f_1\,\pa^\rho\xi\pa_\chi\xi+f_2\,\pa^\rho\psi\pa_\chi\psi\right)^2}{f_1\,\pa^\rho\xi\pa_\rho\xi
+f_2\,\pa^\rho\psi\pa_\rho\psi}\,.$
Demanding finally that $T^{(B)\chi}{}_\chi= T^{(B)\theta}{}_\theta = T^{(B)\varphi}{}_\varphi$, we are led to a third constraint
\eq$\label{P5eq: B-eq}
B^2=\frac{2\left[f_1(\pa\xi)^2+f_2(\pa\psi)^2\right]}{b}\,.$
Therefore, the co-existence in the bulk of the three components of the field-strength tensor with the two scalar fields,
in a way that they satisfy the above three constraints, ensures that the total energy-momentum tensor in the bulk acquires
the form dictated by Eqs. \eqref{P5eq: rho-p1}, \eqref{P5eq: p2}.  

In addition, setting $\rho_E \equiv -T^{(B)t}{}_t$ and $p_\theta \equiv T^{(B)\chi}{}_\chi$, the remaining two components give 
\eq$\label{P5eq: sc-lagr-eq}
f_1(\pa\xi)^2+f_2(\pa\psi)^2+V=\frac{1}{2}(\rho_E-p_\theta)\,,$
\eq$\label{P5eq: em-eq}
bB^2-a_1 E_1^2-a_2 E_2^2=\rho_E+p_\theta\,.$
Therefore, the two independent components of the energy-momentum tensor in the bulk are determined by the
exact profiles of the gauge and scalar fields. These in turn must satisfy  their own equations of motion. By 
considering the variation of the action $S_B$ with respect to $A_M$, we obtain the five-dimensional equation
for the gauge field in the bulk, namely
\eq$\label{P5eq: maxwell-eqs}
\pa_M\left(\sqrt{-g}\,F^{MN}\right)=0\,.$
Considering the components $N=t$ and $N=\varphi$, we find
\bal$
\label{P5eq: max-eq-1}
& \displaystyle{\pa_\rho E_1+\pa_\chi E_2+E_1\frac{\pa_\rho\sqrt{-g}}{\sqrt{-g}}+E_2\frac{\pa_\chi\sqrt{-g}}{\sqrt{-g}}=0}\,,\\[2mm]
\label{P5eq: max-eq-2}
 & \pa_\theta B+\frac{B\,\cos\theta}{\sin\theta}=0\Ra B(\rho,\chi,\theta)=\frac{B_0(\rho,\chi)}{\sin\theta} \,,$
respectively, while the remaining components are identically zero. Additionally, the variation of $S_B$ with respect to the scalar
fields $\xi$ and $\psi$ results in the equations
\eq$\label{P5eq: xi-eq}
\frac{1}{2}\left[(\pa_\xi f_1)(\pa\xi)^2+(\pa_\xi f_2)(\pa\psi)^2+\pa_\xi V\right]=\frac{\pa_M\left(\sqrt{-g}\,f_1\,g^{MN}\pa_N\xi
\right)}{\sqrt{-g}}\,,$
\eq$\label{P5eq: psi-eq}
\frac{1}{2}\left[(\pa_\psi f_1)(\pa\xi)^2+(\pa_\psi f_2)(\pa\psi)^2+\pa_\psi V\right]=\frac{\pa_M\left(\sqrt{-g}\,f_2\,g^{MN}\pa_N\psi
\right)}{\sqrt{-g}}\,.$

The above set of four differential equations \eqref{P5eq: max-eq-1}-\eqref{P5eq: psi-eq}, together with the constraints \eqref{P5eq: E1-eq}-\eqref{P5eq: em-eq},
may indeed possess a mathematically consistent solution. The complexity of the system would most likely demand numerical
calculation for this solution to be derived. However, instead of attempting to solve this coupled system of equations, we would like to 
examine the ensuing characteristics of the fields. To this end, let us focus on Eq. \eqref{P5eq: em-eq}: employing the exact form of the 
components $\rho_E$ and $p_\theta$ from Eqs. \eqref{P5eq: rho-p1}, \eqref{P5eq: p2}, we may rewrite it as
{\fontsize{11}{11}\eq$ \label{P5eq: rho+p2}
bB^2-a_1 E_1^2-a_2 E_2^2=
\frac{1}{\kappa_5^2}\left[-\frac{6k^2\cos^2\chi}\rho\left(M-\frac{Q^2}{\rho}\right)-\frac{3 k\cos \chi }{\rho ^2}
\left(3M-\frac{2Q^2}{\rho}\right)+\frac{3 M}{\rho ^3}\right].$}
\hspace{-0.5em}The right-hand-side of the above equation is clearly not sign-definite. For small $\rho$, it is positive definite since
in this regime both $\rho_E$ and $p_\theta$ are positive, as \myref{P5fig: EneMom-plot}{P5subf: rho-p2-plot} reveals. However, as $\rho$
increases, negative-valued terms inside the square brackets begin to dominate making this combination  clearly
negative for large values of $\rho$. Since $a_1<0$, $a_2<0$ and $b>0$,  according to their definitions below
Eq. (\ref{P5eq: total-emt}), this means that  at least one of the components of the gauge field strength-tensor
$F_{MN}$ must turn imaginary near the bulk boundary. Due to the constraints \eqref{P5eq: E1-eq}, \eqref{P5eq: B-eq},
this may lead to $\xi$ or $\psi$ also becoming imaginary. 

Simpler variants of the above model may also be built, however, they all suffer from the above problem. For instance,
if we consider the case with $E_2=0$ and $\xi=\xi(\chi)$ together with the condition $f_2=0$, the energy-momentum
tensor comes out to be automatically diagonal and satisfying $T^{(B)t}{}_t=T^{(B)\rho}{}_\rho$. The constraints
\eqref{P5eq: E2-eq}, \eqref{P5eq: E1-eq} now disappear while the one for $B$ still holds. The gauge-field equations 
\eqref{P5eq: max-eq-1} and \eqref{P5eq: max-eq-2} are easily satisfied for a wide range of choices for $E_1$ and $B$.
The second scalar field $\psi(\rho,\chi)$ is now an auxilary field whose equation of motion \eqref{P5eq: psi-eq}
 introduces a constraint between $f_1$ and $V$. Nevertheless, Eq. \eqref{P5eq: rho+p2} still holds with $E_2=0$,
 and thus the necessity for a ``phantom'' gauge field (and a ``phantom'' scalar field) at the bulk boundary still exists. 
 
 Phantom scalar fields are often used in the context of four-dimensional analyses as a mean to create the
 necessary yet peculiar dark energy component with $w < -1$ in our universe. In our analysis, a bulk matter
 with also peculiar characteristics seems to be necessary to localize a five-dimensional black hole on the brane,
 otherwise its singularity would leak in  the bulk.  The desired structure of the bulk energy-momentum tensor
 as well as the introduction of the ``charge'' parameter $Q$ in our metric demand the presence of gauge and
 scalar fields with phantom-like properties at the bulk boundary.  We should stress that all fields are ``ordinary'' 
 close  to our brane and no violation of energy conditions takes place on our brane. Could a gauge field, that
 turns phantom-like at the outskirts of the bulk spacetime, be considered as ``natural'' or at least acceptable?
 Such an analysis, although well-motivated, would take us beyond the scope of  the present study and is thus
 left for a future work.


\mysection{Junction conditions and effective theory \label{P5sec: JC-ET}}

In this final section, we turn our attention from the structure and content of the five-dimensional spacetime to issues
related to the presence of the brane itself, namely its consistent embedding in the bulk and the effective four-dimensional
gravitational equations. A detailed derivation of the effective theory on the 3-brane in brane-world models was presented
in the previous Chapter, thus, we will not repeat the analysis.

Following the same procedure as in \secref{P4sec: junc}, and using eq.\,\eqref{P5eq: metr-r-y}, it is straightforward to
determine the induced metric components $h_{MN}$ on the brane. These are found to be
{\fontsize{11}{11}\eq$\label{P5eq: ind-metr-comp}
(h_{MN})=\left(\begin{array}{ccccc}
-\left(1-\frac{2M}{r}+\frac{Q^2}{r^2}-\frac{\Lambda}{3}r^2\right) 		& 		0 		& 		0 		& 		0 		&   0		 \\[1mm]
0		 &		 \left(1-\frac{2M}{r}+\frac{Q^2}{r^2}-\frac{\Lambda}{3}r^2\right)^{-1} 		&		 0		 &		 0		& 0	  \\[1mm]
0		&		0		&		r^2		&		0	&	0	\\
0		&		0		&		0		&		r^2\sin^2\theta	&	0\\
0	&  0	&	0	&	0	&  0		
\end{array}\right)\,.$}
\hspace{-0.5em}We may now proceed to derive the effective theory on the brane.
Note that eqs.\,\eqref{P4eq: Gauss-eq}-\eqref{P4eq: bracket-def} and \eqref{P4eq: extr-curv2}-\eqref{P4eq: brane-ene-comp}
continue to hold here as well, thus, we are led to
\eq$\label{P5eq: grav-eqs-br-new}
\bar{G}_{\mu\nu}=8\pi G_N\left(T^{(eff)}_{\mu\nu}+ \tau_{\mu\nu}\right)+\kappa_5^4\left(\pi_{\mu\nu}-\frac{\ \sig^2}{12}\,h_{\mu\nu}\right)
-E_{\mu\nu}\Big|_{y\ra 0}\,,$
where
\bal$
&G_N=\frac{\kappa_5^4\,\sigma}{48\pi}\,,\hspace{1.5em}\sigma=\frac{6k}{\kappa_5^2}>0\,,\hspace{1.5em}\tau_{\mu\nu}=0\ (\forall \mu,\nu)\,,\\[3mm]
&T^{(eff)}_{\mu\nu}\equiv\frac{2}{3k}\left[T^{(B)}_{\mu\nu}+\left(T^{(B)}_{yy}-\frac{T^{(B)}}{4}\right)h_{\mu\nu}\right]_{y=0}\,,\\[3mm]
&\pi_{\mu\nu}=-\frac{1}{4}\tau_{\mu}{}^\lam\,\tau_{\lam\nu}+\frac{1}{12}\tau\,\tau_{\mu\nu}+\frac{1}{8}\tau^{\alpha\beta}\tau_{\alpha\beta}
\,h_{\mu\nu}-\frac{1}{24}\tau^2\,h_{\mu\nu}\,.$
In the above, $G_N$ constitutes the effective four-dimensional gravitational constant on the brane; this is also defined
in terms of the fundamental gravitational constant $\kappa_5^2$ and the brane tension $\sigma$. The quantity $\pi_{\mu\nu}$
is the well-known quadratic contribution of $\tau_{\mu\nu}$ \cite{SMS} which here, however, trivially vanishes since 
$\tau_{\mu\nu}=0$. Finally, $T^{(eff)}_{\mu\nu}$ can be interpreted as the effective energy-momentum tensor on the brane.
Together with $E_{\mu\nu}$, they constitute the imprint of the dynamics of the bulk fields---gravitational, and possibly gauge
and scalar fields generating the bulk energy-momentum tensor $T^{(B)}_{MN}$---on the brane.  The components of
$T^{(eff)}_{\mu\nu}$ are given by the following relation
\eq$\label{P5eq: new-brane-ene-comp}
T^{(eff)}_{\mu\nu}=\frac{1}{\kappa_5^2k}\left[3k^2h_{\mu\nu}-{\Lambda}\,h_{\mu\nu}+\frac{M}{r^3}\left(
\begin{array}{cccc}
 -h_{tt} & 0 & 0 & 0 \\
 0 & - h_{rr} & 0 & 0 \\
 0 & 0 & h_{\theta\theta} & 0 \\
 0 & 0 & 0 & h_{\varphi\varphi} \\
\end{array}
\right)\right]\,,$
while the components of the tensor $E_{\mu\nu}$, defined in \eqref{P4eq: E-weyl}, are evaluated to be
\eq$\label{P5eq: E-weyl-comp}
E_{\mu}{}_{\nu}\Big|_{y\ra 0}=\left(-\frac{Q^2}{r^4}+\frac{M}{r^3}\right)\left(
\begin{array}{cccc}
  -h_{tt} & 0 & 0 & 0  \\
 0 & -h_{rr} & 0 & 0  \\
 0 & 0 & h_{\theta\theta} & 0  \\
 0 & 0 & 0 & h_{\varphi\varphi} 
\end{array}
\right)\,.$
We notice that $E_{\mu\nu}$ is evaluated infinitesimally close to the brane but not exactly on it, its source
being the five-dimensional Weyl tensor.  Substituting the above relations in \eqref{P5eq: grav-eqs-br-new}, we obtain
{\fontsize{11}{11}\eq$\label{P5eq: 4dEin}
\bar{G}_{\mu\nu}=\left(\begin{array}{cccc}
  \left(\frac{Q^2}{r^4}+\Lambda_4\right)f(r) & 0 & 0 & 0  \\
 0 & -\left(\frac{Q^2}{r^4}+\Lambda_4\right)\frac{1}{f(r)} & 0 & 0  \\
 0 & 0 & \left(\frac{Q^2}{r^4}-\Lambda_4\right)r^2 & 0  \\
 0 & 0 & 0 & \left(\frac{Q^2}{r^4}-\Lambda_4\right)r^2\sin^2\theta
\end{array}
\right)\,,$}
\hspace{-0.5em}with
\eq$
f(r)\equiv 1-\frac{2M}{r}+\frac{Q^2}{r^2}-\frac{\Lambda_4}{3}r^2\,.$
One can verify that the expression of the Einstein tensor in \eqref{P5eq: 4dEin} matches exactly the Einstein
tensor of the four-dimensional Reissner-Nordstr\"{o}m-(A)dS metric, with $\Lambda_4=\Lambda$ being
the effective cosmological constant on the brane and $Q^2/r^4$ the equivalent of the energy-momentum
tensor component of an electromagnetic field.  Although we have called our five-dimensional black-hole
solution a Reissner-Nordstr\"{o}m-(A)dS one, it is clear that no four-dimensional electromagnetic field has
been---or needed to be---introduced on the brane. The ``charge'' $Q$ is a conserved quantity
carried by the bulk fields and left as an imprint in the four-dimensional spacetime. It is therefore
a tidal charge \cite{tidal} rather than an ordinary electromagnetic one.

\mysection{Conclusions \label{P5sec: Disc}}

In this Chapter, we have generalized the analysis of \chapref{Chap: P4}, where we studied the localization of 
a five-dimensional spherically-symmetric, neutral and asymptotically-flat black hole on our brane,
by considering also a cosmological constant and a charge term in the metric function. We have 
preserved the assumption of spherical symmetry in the five-dimensional bulk and by adopting
an appropriate set of spherical coordinates, we have built a black-hole solution with its singularity
strictly residing on the brane. We have performed a careful classification of the horizons that
this background admits, depending on the values of its parameters, and demonstrated that all of 
them have pancake shapes and one after the other get exponentially localized close to the brane.
The bulk gravitational background is everywhere regular, as the calculation of all scalar gravitational
quantities has shown, and reduces to an AdS$_5$ spacetime right outside the black-hole event horizon. 
To support the geometry of the five-dimensional spacetime, we assumed the presence of a bulk energy-momentum
tensor, which was determined to describe an anisotropic fluid. In addition, we attempted to provide a physical 
interpretation of the nature of the bulk matter by building a
field-theory model involving scalar and gauge fields living in the bulk.
Without determining explicitly
the profiles of these fields---a task that would demand numerical analysis, we obtained the primary
constraints and equations for a viable solution.
Finally, by considering the junction conditions, we studied in detail the consistent
embedding of our 3-brane into the bulk geometry we have constructed.



\mychapter{\textbf{Epilogue} \label{Chap: Concl}}
\phantomsection
\setcounter{chapter}{7}

\epigraph{\justify\textit{``Seek first the good things of the mind, and the rest will either be supplied or its loss will not be felt.''}}{Francis Bacon}

\thispagestyle{empty}

In this Chapter we sum up our studies and ensuing results and discuss possible extensions of the presented work.





In \chapref{Chap: P1}, motivated by the results of previous works \cite{KPZ, KPP}, 
where despite intensive
efforts regular, localized-on-the-brane black hole solutions were not found
in the context of a theory with a scalar field nonminimally coupled to gravity,
we have focused on the derivation and study of the properties of
black-string solutions that, in contrast, seem to emerge quite naturally in the
context of the same theory. To this end, we have retained the Vaidya form
of the spacetime line element, which on the brane leads to a Schwarzschild black
hole while in the bulk produces solutions with the minimum number of
spacetime singularities. We have in addition allowed for an arbitrary mass
function $m(r)$ in an effort to accommodate, if possible, solutions with a more
general profile including an (anti)--de Sitter or Reissner-Nordstr\"{o}m type of 
background.

The integration of an appropriate rearrangement of the equations of motion 
has allowed us to uniquely determine the form of the mass function, namely
$m(r)=M + \Lambda r^3/6$.  Performing an inverse coordinate transformation
on the brane, we readily identified the parameter $M$ with the black-hole
mass and the parameter $\Lambda$ as the product $\kappa_4^2 \Lambda_4$,
where $\Lambda_4$ is the four-dimensional cosmological constant on the brane.
As a result, the brane background assumes the form of a Schwarzschild (anti-)de
Sitter spacetime. As the expressions of the five-dimensional curvature invariants
reveal, these solutions may have a dual description from the bulk point of view:
they may describe either black strings, if $M \neq 0$, or braneworld
maximally symmetric solutions, if $M=0$. 
Note also that in this Chapter we have examined the case of a positive
four-dimensional constant, namely $\Lambda>0$.

The properties of these five-dimensional solutions strongly depend on the 
form of the nonminimal coupling function $f(\Phi)$ between the scalar field and
the five-dimensional scalar curvature. We have considered two simple choices for 
$f(\Phi)$, a linear one and a quadratic one in terms of the scalar field. For a linear
coupling function, the 
scalar field is found to increase exponentially away from the brane and to drive
the coupling function to negative values at a distance from the brane. When
$6k^2/\Lambda>1$, there is always a positive-value regime for $f(\Phi)$ close
to our brane while the antigravitating regime, with $f(\Phi)<0$, is pushed away
from our brane as the value of $6k^2/\Lambda$ gradually increases. For fairly
large values of $6k^2/\Lambda$, i.e. for a large warping factor $k$ or a small
cosmological constant on our brane, the profile of the coupling function exhibits
a wide plateau around our brane. When $6k^2/\Lambda \simeq 125$, this plateau
is centered
around the value of unity, and, therefore, the theory mimics a five-dimensional
scalar-tensor theory with a minimally coupled scalar field and normal gravity
around our brane---the antigravitating regime is, however, still lurking at the
boundaries of the extra dimension. The latter may be cut short or altogether 
removed from the theory by adding a second brane; this is also necessary in
order to obtain a finite four-dimensional gravitational scale, as we have explicitly
demonstrated. The antigravitating regime is also characterized by a diverging
scalar field that results in the divergence of the energy-momentum tensor components,
too. However, after the introduction of the second brane at a finite distance from
the first, all energy-momentum tensor components are well behaved. In fact, the
energy density takes on an almost constant, negative value around our brane,
thus mimicking a bulk cosmological constant (which, in this case, is redundant)
and supporting a Randall-Sundrum warp factor.

For a quadratic coupling function $f(\Phi)$, the scalar field is found to be everywhere
finite and, in fact, to exhibit a localization around our brane---the same behavior
is exhibited by all the energy-momentum tensor components. The four-dimensional
gravitational scale comes out to be finite; therefore, in this case there is no reason
to introduce a second brane. The warp factor takes a form identical to the one in
the Randall-Sundrum model even in the absence of a negative, bulk cosmological
constant and for positive values of the energy-momentum tensor around our brane.
What, in fact, creates the anti--de Sitter spacetime in the bulk and supports the
exponentially decreasing warp factor is the coupling itself between the scalar field
and the bulk scalar curvature, which is everywhere negative. This, of course, leads
to an antigravitating theory over the whole spacetime and eventually to an
unphysical gravitational theory on our brane. This model, being far from a 
realistic theory, is nevertheless a characteristic example of the variety of solutions
that may arise in braneworlds; more specifically, it underlines the easiness with
which unphysical black-string solutions (in the case where $M \neq 0$) emerge
in contrast to the physically motivated localized black-hole solutions. 

The discussion of the second model with the quadratic coupling function also served
another purpose: together with the first one with a linear $f(\Phi)$, they were
both derived under the assumption of a positive cosmological constant on our brane. 
Also, both models were characterized, either globally or over particular regimes, by a
negative coupling function $f(\Phi)$ that led to an antigravitating theory. In order
to investigate the potential connection between a Schwarzschild de Sitter spacetime on our
brane and an antigravitating regime in the bulk, in \secref{P1sec: Theoretical}, we examined from
a mathematical point of view why the field equations in the present theory seem to favor
the emergence of these solutions. By turning a particular combination of the field
equations into a constraint relating solely the coupling function, its derivatives,
and the effective cosmological constant, we demonstrated that, for $\Lambda>0$,
this constraint is impossible to satisfy for $f(\Phi)$ also positive for the entire extra
dimension. 
Therefore, in this class of theories, with a nonminimally coupled scalar field and
a general coupling function, the emergence of an effective four-dimensional theory
on our brane with a positive cosmological constant is always accompanied by
a problematic antigravitating regime in the five-dimensional bulk. 

The aforementioned conclusion opens the way for the derivation of solutions
with normal gravity in the case of either a Minkowski or anti--de Sitter spacetime
on our brane.  Although less physically motivated, it would still be of interest 
to investigate whether a scalar-tensor theory in the bulk could support a
solution (either a black string or a regular one) with a decaying warp factor
but without the need for a constant distribution of a negative energy density
in the higher-dimensional spacetime. 



Thus, in \chapref{Chap: P2}, we continued the study of black-string solutions in the context of the
same theory, but here we considered the case of a negative four-dimensional cosmological constant, i.e. $\Lambda<0$.
Consequently, the induced four-dimensional geometry on the brane is described by the Schwarzschild anti-de Sitter
spacetime.
The coupling is realized through a smooth, real, positive-definite coupling function
$f(\Phi)$. Demanding that all components of the energy-momentum tensor remain
finite throughout the bulk, and looking for analytic solutions for the scalar
field, we have restricted our choices for the coupling function to two particular
forms: a simple exponential and a double exponential, both decreasing
away from the brane. This results into a scalar-tensor five-dimensional
theory with a non-minimal coupling between the scalar field and gravity that is
effectively localized close to the brane. 
In addition to the above, the non-minimal coupling function in both cases is allowed to have everywhere a positive value,
and thus to guarantee a normal gravitational force both in the bulk and on the brane.
However, the calculation of the five-dimensional scalar-invariant
quantities revealed that, for a non-vanishing mass parameter, the five-dimensional
solution is in fact a black-string plagued by a singularity extending over the
entire extra dimension. 

In order to complete the solution, the form of the scalar field in the bulk had
also to be determined. We have managed to analytically attack the problem of
the integration of the scalar field equation and to derive two particular
solutions. For
the exponentially decreasing coupling function, the scalar field was expressed
in terms of the hypergeometric function (that, for particular values of the
parameters, was further reduced to a combination of elementary functions) while
for the double exponential form of $f(y)$, the scalar field was given by a combination
of an exponential and the error function. In both cases, the profile of the
scalar field presented the same set of robust features: $\Phi(y)$ was everywhere a
smooth, regular, monotonic function of the extra coordinate approaching a constant,
finite value at the boundaries of spacetime. 

The same robust behaviour was exhibited also by the bulk potential of the scalar
field and all components of the energy-momentum tensor. All the aforementioned
quantities were finite everywhere in the bulk, remained localized close to our
brane, and vanished asymptotically at large distances. In particular, the scalar
potential was everywhere negative-definite, which led also to a negative-definite
energy density. However, this negative distribution of energy was generated, not
by an exotic form of matter, but by a physical, scalar degree of freedom coupled
non-minimally to gravity with a positive-definite, and localized close to our brane,
coupling function. This energy may therefore support by itself a pseudo-AdS spacetime,
even in the absence of a negative, bulk cosmological constant, and thus to ensure
the localization of gravity close to our brane. 

The presence of the brane in the theory introduces a set of junction conditions
that may serve to fix two of the parameters of the theory, preferably the
warp-factor parameter $k$ and a parameter of the interaction term of the
scalar field with the brane. If the latter term is non-trivial, the effective
cosmological constant remains a free parameter of the model; however, for a
trivial scalar-brane interaction term, the value of the cosmological constant on
the brane may be uniquely determined. The calculation of the effective theory
on the brane led to a finite theory, as expected, without the need to introduce
a second brane in the model. The relation between the fundamental and the
effective gravitational scale had a similar form with the one emerging in the
Randall-Sundrum model although our theory has a dynamical, more realistic
field content.  

In order to produce finite, analytic solutions for the bulk scalar field, we have
made indeed two particular choices for the non-minimal coupling function.
However, despite the apparently different explicit forms of the scalar field,
the main characteristics of the two solutions remained the same, namely, the
smoothness, the regularity, even the constant, finite value at infinity. This
type of ``universality'' is caused by the common characteristics that the
corresponding forms of the coupling function had: they were both smooth,
well-defined throughout the bulk, positive-definite and localized close to
our brane. We may therefore assert that the two particular solutions we have
derived are in fact representative of the behavior that a generic solution for the
scalar field would exhibit if it was sourced by the Ricci scalar through any
coupling function that would respect the same finiteness and positivity-of-value
criteria.



\chapref{Chap: P3} completes our previous two
analyses, where the cases of a de Sitter and an anti-de Sitter brane
were considered, and focuses on the case of a flat, Minkowski brane with $\Lambda=0$.
In this case, the complete five-dimensional solution for the gravitational background in this case may
describe either a non-homogeneous black string, when the metric parameter $M$ is
non-zero, or a regular anti-de Sitter spacetime, when $M=0$.

The above features characterize our solutions irrespectively of the form of the coupling
function between the bulk scalar field and the five-dimensional Ricci scalar. In this Chapter,
we have performed a comprehensive study of the types of brane-world solutions that
emerge in the context of this theory by considering a plethora of forms of the coupling
function, all supported by physical arguments regarding the reality and finiteness of
its value everywhere in the bulk. We have thus considered the cases of a linear and
quadratic coupling function in terms of the scalar field $\Phi$, but also an inverse-power
and a linear-exponential form in terms of the $y$-coordinate. From a different perspective,
we also considered given forms for the scalar field which again satisfied the finiteness condition,
namely a double-exponential and a hyperbolic-tangent form in terms of the $y$-coordinate,
and determined subsequently the form of the coupling function. In all cases, the profile of
the coupling function remains finite along the fifth coordinate as expected, reducing either
to zero or to a constant value far away from our brane---in both cases, the coupling
between the scalar field and the bulk Ricci scalar becomes trivial and as a result the
scalar-tensor theory naturally reduces to a purely gravitational theory at large distances.
Gravity by itself is also localized due to the exponentially decaying warp factor. 

In each case, we have also determined in an analytical way the corresponding solutions
for the profiles of the scalar field and scalar bulk potential. These also remain finite
over the entire bulk for every solution found, and their behaviour resembles the one of
the coupling function reducing either to zero or to a constant value away from our brane.
Depending on the values of the parameters of the solutions, the bulk 
potential in particular could adopt a variety of forms being non trivial close to our brane
and reducing to a constant, positive or negative, value at asymptotic infinity. Also in this case,
the Randall-Sundrum-type, exponentially decaying warp factor is supported independently of the presence of the negative bulk
cosmological constant $\Lambda_5$, which is usually introduced in a ad hoc way in
brane-world models.

The case of a zero effective cosmological constant, studied in the context of this Chapter,
allowed for the maximum flexibility regarding the form and characteristics of the
coupling function, when compared to the cases of a positive or negative effective
cosmological constant \cite{KNP1, KNP2}. For $\Lambda>0$, the coupling function had
to be negative-definite at large distances from our brane, while, for $\Lambda <0$,
a fast localized profile was necessary in order to avoid an ill-defined behaviour for
the scalar field at the bulk boundaries. For $\Lambda=0$, though, no such requirements
are necessary. In order, however, to derive physically-acceptable brane-world solutions,
we have imposed three additional conditions: the positivity of the effective gravitational 
constant $\kappa_4^2$ on our brane defined as
\beq
\frac{1}{\kappa_4^2}\equiv 2\,\int_{0}^{\infty} dy\, e^{-2 k y}\,f(y)\,,
\label{P3eq: Geff_gen}
\eeq
the positivity of the total energy density of our brane, which follows
from the junction condition (\ref{P3eq: jun_con1}) and may be rewritten as
\beq 
\sigma+V_b=6 k f(0) - 2 f'(0)\,,
\label{P3eq: jun_gen}
\eeq
and the validity of the weak energy conditions by the bulk matter in the vicinity of our brane;
the latter, using Eqs. (\ref{P3eq: linear-rho})-(\ref{P3eq: linear-py}), may be expressed as
\beq
f(0)<0\,.
\label{P3eq: weak-gen}
\eeq
In each solution found, we have thus performed a careful study of the effective theory
on the brane, the junction conditions introduced by the presence of the brane and the 
profiles of the energy-density and pressure of the bulk matter. Subsequently, we conducted
a thorough investigation of the corresponding parameter space in order to deduce whether it
is possible to simultaneously satisfy all three aforementioned conditions. 

We have found that, for all solutions presented, this is not possible. The aforementioned
three constraints are not a priori incompatible: Eq. (\ref{P3eq: weak-gen}) constrains the value
of the coupling function $f(y)$ at the location of the brane, Eq. (\ref{P3eq: jun_gen}) dictates
that its first derivative must be also negative and decreasing fast at the same point,
while Eq. (\ref{P3eq: Geff_gen}) imposes a constraint on its integral over the entire bulk.
Note that if we had demanded the validity of the weak energy condition everywhere in the
bulk, i.e. $\rho(y)>0$, that would imply $f(y)<0$, for $\forall y$. This would be in
obvious contradiction with the positivity of the effective gravitational constant through
Eq. (\ref{P3eq: Geff_gen}). Demanding the validity of the weak energy condition only at the
vicinity of our brane, as in Eq. (\ref{P3eq: weak-gen}), allows the coupling function to be
negative close to our brane and become positive at some distance off it, so that the
integral in Eq. (\ref{P3eq: Geff_gen}) turns out to be positive-definite. That was indeed realised
for some of our solutions but the parameter space corresponding to those solutions
was always severely restricted. Imposing the third constraint (\ref{P3eq: jun_gen}) on the
value of $f'(0)$, on top of the previous two constraints, in an attempt to make the
energy-density of the brane also positive, we were led to contradictions for all the
analytical solutions we have found. These contradictions are translated to the absence
of a single point in the parameter space in which all the above constraints can be
simultaneously satisfied. In contrast, relaxing
the weak energy condition, which involves bulk quantities, and demanding instead the 
validity of Eqs. (\ref{P3eq: Geff_gen}) and (\ref{P3eq: jun_gen}), which are relevant for the
4-dimensional observer on the brane, has led to a plethora of analytic solutions with
an extended parameter space. The question of whether a solution satisfying all three
constraints could be constructed, either analytically or numerically, naturally emerges,
and could be pursued in a future work. That solution, however, would have to be not only
a mathematically consistent solution of the set of field equations satisfying the
constraints  (\ref{P3eq: Geff_gen})-(\ref{P3eq: weak-gen}) but to be also characterised by a
physically acceptable behaviour throughout the bulk---the analytical solutions
presented in this Chapter were carefully constructed in order to have a physically
acceptable behaviour regarding the profiles of the scalar field, its coupling
function and potential throughout the bulk.

In conclusion, the well-known generalized gravitational theory of a scalar field 
non-minimally coupled to the Ricci scalar admits, upon embedded in
a five-dimensional brane-world context, a variety of solutions with a number
of attractive features, such as the support of an exponentially decaying warp factor,
and thus of graviton localization, without the need for a negative bulk cosmological
constant. In the particular case of $\Lambda=0$ studied here, this is always supplemented
by a regular scalar field, a finite coupling function, which becomes naturally trivial at
the outskirts of the bulk, a physically-acceptable brane with a positive total energy-density
and a robust effective four-dimensional theory on our brane.

Finally, the stability behaviour of our solutions is an important aspect that needs
to be studied. Compared to the existing stability analyses performed along the lines
of Refs. \cite{GL,RuthGL}, our theory has the additional complexity of the presence of
the scalar field. Our solutions are not purely gravitational, therefore, the perturbation
analysis will involve a coupled system of scalar-field and gravitational equations.
The sign of the cosmological constant $\Lambda$ and the corresponding properties
of the non-minimal coupling function $f(\Phi)$ of the scalar field to the Ricci scalar
are also expected to play a role in this analysis. Such an analysis will reveal whether
the scalar field may stabilise the black string over the bulk regime close to our
brane where it has a non-trivial profile. However, beyond the point where the
scalar-field energy-momentum tensor vanishes, we expect the Gregory-Laflamme
instability to set in, as in all other infinitely extended black-string 
solutions. The stability of the solutions for the
case of $M=0$ should also be carefully examined as the role of the singularity,
arising at the boundary of spacetime when $\Lambda \neq 0$, may be found
to be important.



In \chapref{Chap: P4}, we turned to the question of the existence of localized brane-world black-hole solutions.
By following a different approach this time, and prioritizing the geometry instead of the field-theory, we have successfully 
constructed from first principles the geometry of an analytic
five-dimensional black hole exponentially localized close to our 3-brane. We have demonstrated that
the black-hole singularity lies entirely on the brane, while the event horizon extends into the bulk
but is exponentially suppressed as we move along the extra dimension. This exponential localization
alters the shape of the event horizon, making it  appear as a five-dimensional pancake. The
five-dimensional line-element is effectively AdS$_5$ outside the event horizon and
reduces  to the Schwarzschild solution on the brane. 

The derived geometry is supported by an anisotropic fluid in the bulk described by a diagonal
energy-momentum tensor with only two independent components: the energy density $\rho_E$
and tangential pressure $p_\theta$. All energy conditions are satisfied on the brane whereas a 
local violation takes place in the bulk in the region inside the event horizon. No additional
matter needs to be introduced on the brane for its consistent embedding in the bulk geometry
while the effective field equations are shown to be satisfied by the vacuum Schwarzschild
geometry on the brane.  



In \chapref{Chap: P5}, we have generalized our previous analysis, where we studied the localization of 
a five-dimensional spherically-symmetric, neutral and asymptotically-flat black hole on our brane,
by considering also a cosmological constant and a charge term in the metric function. We have 
preserved the assumption of spherical symmetry in the five-dimensional bulk and by adopting
an appropriate set of spherical coordinates, we have built a black-hole solution with its singularity
strictly residing again on the brane. We have performed a careful classification of the horizons that
this background admits, depending on the values of its parameters, and demonstrated that all of 
them have pancake shapes and one after the other get exponentially localized close to the brane.
The bulk gravitational background is everywhere regular, as the calculation of all scalar gravitational
quantities has shown, and reduces to an AdS$_5$ spacetime right outside the black-hole event horizon. 

We also attempted to provide a physical interpretation of the nature of the bulk matter by building a
field-theory model involving scalar and gauge fields living in the bulk. Without determining explicitly
the profiles of these fields---a task that would demand numerical analysis, we obtained the primary
constraints and equations for a viable solution. Although we demonstrated that this scalar-vector model
could indeed reproduce the general structure of the energy-momentum tensor in the bulk, our analysis
also revealed that the gauge, and inevitably the scalar, fields should become phantom-like at the bulk
boundary. The decision on whether a five-dimensional tensor-scalar-vector theory, whose particle
degrees of freedom are well behaved near and on our brane but they turn phantom-like away from it,
is physically acceptable  is still pending. Alternative field theory constructions could also be considered. 
For instance, the negative sign of the energy density of the bulk matter  points perhaps to a non-minimal
gravitational coupling of the fields that takes over at the outskirts of the bulk---the fact that all terms
proportional to the charge $Q$, and therefore sourced by the bulk gauge field, remain always
positive whereas the gravitational terms proportional to $M$ are the ones that cause the energy
density to turn negative seems to agree with this. 

By considering the junction conditions, we have subsequently studied in detail the consistent
embedding of our 3-brane into the bulk geometry we have constructed. We have demonstrated
that, also in this case, no additional matter needs to be introduced on the brane by hand, and that the only energy
content of our brane in the context of the five-dimensional theory is its constant, and positive
self-energy or tension. In fact it is this quantity together with the five-dimensional gravitational
constant that determine the warp parameter of the bulk metric---we note that the warp factor
of the model has the exact same form as the one of the original Randall-Sundrum model, a feature
which also ensures the localization of gravity close to our brane. These two fundamental quantities
determine also the effective four-dimensional gravitational constant on our brane as the study of
the effective theory on the brane revealed. There, we showed that the combined effect of the
five-dimensional geometry and the bulk matter leaves its imprint on the brane and supports
the Reissner-Nordstr\"{o}m-(A)dS geometry that the four-dimensional observer sees. Let us,
however, stress again that the charge appearing in the metric is a tidal charge rather than an
electromagnetic one as it is sourced by the bulk, gravitational and gauge, fields.

We should note here that a similar perspective for the construction of
the bulk geometry was adopted in \cite{Dai}, however, the form of the 5-dimensional line-element
and bulk energy-momentum tensor did not support either a Schwarzschild solution on the brane or
an AdS$_5$ spacetime right outside the black-hole horizon. Apart from the aforementioned features,
our solution supports an
exponentially decreasing warp factor in the bulk, therefore successfully incorporates
the original Randall-Sundrum brane-world model \cite{RS1,RS2}. Due to this
behavior, our results could be considered also in the context of holography 
\cite{Maldacena,  Gubser2, Witten}. In the asymptotic regime where the spacetime becomes
purely AdS$_5$, a four-dimensional conformal field theory (CFT) can be mapped. As we deviate
from the AdS$_5$ limit, the modification in the 5-dimensional metric can be attributed to matter
added in the boundary CFT and related to interesting field-theory phenomena such as chiral
symmetry breaking \cite{Pomarol, Alho}, confinement/deconfinement \cite{Ballon}, etc.
Future directions of work could also address
the stability behaviour of our solution as the Gregory-Laflamme instability arguments \cite{GL}
do not hold here. In previous studies, a stability analysis led also to observable effects such as
echoes of braneworld compact objects \cite{Chakraborty1,Chakraborty2}, as well as other
exotic compact objects \cite{Cardoso,Zachary,Maggio}. A natural question
emerges of whether gravitational waves from black hole mergers or other astrophysical processes
could provide evidence for extra dimensions and distinguish brane-world solutions of this type from the
corresponding four-dimensional ones \cite{Chakraborty3}. The study of the cosmological aspects
of our construction on the brane is also a future direction of research (see, for example,
 \cite{Antonini1, Antonini2}). Also, could we construct alternative
localized black-hole solutions  by considering different forms of the metric function $f(\rho)$,
such as the Schwarzschild-Rindler-(Anti-)de Sitter solution with an additional  linear term  associated
with dark matter or scalar-hair effects \cite{Alestas}, and what would be in that case the profile
of the bulk matter? Is it finally possible to construct rotating brane-world black holes using a similar
process as the one we developed for static brane-world black holes? We plan to return to, at least, 
some of those questions, in future works.

\afterpage{\blankpage}



\newpage
\thispagestyle{empty}
\vspace*{5cm}
\begin{center}
\chapc{\noindent{\rule{16.5cm}{0.5mm}}\vspace*{1em}
{\fontsize{40}{42} \textbf{Appendices}}\vspace*{1em}
\noindent{\rule{16.5cm}{0.5mm}}}
\end{center}
\afterpage{\blankpage}


\begin{appendices}

\numberwithin{equation}{chapter}


\mychapter{Curvature invariant quantities for the black-string solutions \label{P1app: Curv-Inv}}
\phantomsection

\thispagestyle{empty}

Employing the expression of the line element (\ref{P1eq: metric}), one may compute the
scalar curvature invariant quantities. These have the form
{\fontsize{10}{10}{\eq$
R=-8 A''-20 A'^2+\frac{2 e^{-2 A} \left(r \pa_r^2m+2 \pa_rm\right)}{r^2}\,,$
\begin{align}
R_{MN}R^{MN}=2e^{-4 A} \left[e^{2 A} \left(A''+4 A'^2\right)-\frac{\pa_r^2m}{r}\right]^2&+2
\frac{e^{-4 A} \left[r^2 e^{2 A}\left(A''+4 A'^2\right)-2 \pa_rm\right]^2}{r^4}\nonum\\
&+16 \left(A''+A'^2\right)^2,
\end{align}
\begin{align}
R_{MNKL}R^{MNKL}=&-\frac{8 e^{-2 A} A'^2 \left(r \pa_r^2m+2 \pa_rm\right)}{r^2}+40 A'^4+16 A'' \left(A''+2 A'^2\right)\nonum\\[2mm]
&+4 e^{-4 A}\left[\frac{(\pa_r^2m)^2}{r^2}+\frac{4\left[2 (\pa_rm)^2+ \left(m-r \pa_rm\right) \pa_r^2m\right]}{r^4}
+\frac{4 (3 m^2-4 r m \pa_rm)}{r^6}\right],
\end{align}}}

\noindent{and} may be used for the geometric characterization of the solutions derived from the field
equations. 

\vspace{-10em}
\afterpage{\blankpage}



\mychapter{Independent field equations \label{P1app: Indep-F-Eq}}
\phantomsection

\thispagestyle{empty}

Here, we will demonstrate that the three field equations (\ref{P1eq: grav-eq1})--(\ref{P1eq: phi-eq})
are not all  independent. To this end, we substitute the mass function $m(r)=M+\Lambda r^3/6$
into Eq. (\ref{P1eq: grav-eq1}); as shown in \secref{P1sec: Theoretical}, the latter may then be
brought to the form 
\gat$
\Phi'^2=-f(3A''+\Lambda e^{-2A})+A'\pa_y f-\pa_y^2f\,.$
Taking the derivative of both sides with respect to $y$, we obtain
\gat$ \label{P1app-eq: eq1}
2 \Phi'\,\Phi''=-f(3A'''-2\Lambda A'e^{-2A})-\pa_y f(2A''+\Lambda e^{-2A})
+A'\pa_y^2 f-\pa_y^3 f\,.$

Next, we consider Eq. (\ref{P1eq: grav-eq2}) which we solve for the potential $V$ to find
\gat$
V=-\Lambda_5-\frac{1}{2}\,\Phi'^2-f(6A'^2+3A''-\Lambda e^{-2A})-3A'\pa_y f-\pa_y^2 f\,.$
If we take again the derivative with respect to $y$, we arrive at the result
\gat$
\hspace*{-3cm}\pa_yV=-\Phi'\,\Phi''-f(12A'A''+3A'''+2\Lambda A'e^{-2A})\nonum\\[1mm] 
\hspace{4.8cm} -\pa_yf(6A'^2+6A''-\Lambda e^{-2A}) -3A'\pa_y^2f-\pa_y^3f\,.$
We now use the above expression in the scalar-field equation (\ref{P1eq: phi-eq}) after multiplying
first the latter by $\Phi'$; we eventually obtain
\gat$ \label{P1app-eq: eq2}
2\Phi'\,\Phi''=-f(3A'''-2\Lambda A'e^{-2A})-\pa_yf(2A''+\Lambda e^{-2A})+A'\pa_y^2f-\pa_y^3f\,.$

\par We see that Eqs. \eqref{P1app-eq: eq1} and \eqref{P1app-eq: eq2} are identical, which means that the 
three field equations from which these equations were derived are not independent. We are thus
entitled to keep only two of them in our analysis and to ignore the third one. 




\mychapter{Inverse generalized Vaidya transformation \label{P1app: Inv-Vaidya}}
\phantomsection

\thispagestyle{empty}

Starting from the projected-on-the-brane line element (\ref{P1eq: metric-4D}), which we will write
for simplicity as 
\eq$\label{P1app-eq: vtos1}
ds^2=-\left(1-\frac{2m(r)}{r}\right)dv^2+2dvdr+r^2(d\theta^2+\sin^2\theta\,d\varphi^2)\,,$
where $m(r)=M+\Lambda r^3/6$, we will seek to determine the coordinate transformation of the
Vaidya time-variable $v$, if existent, that will bring the aforementioned line element to a
diagonal, Schwarz\-schild-like form
\beq
ds^2=-f(r)\,dt^2+\frac{dr^2}{f(r)}+r^2(d\theta^2+\sin^2\theta\,d\varphi^2)\,.
\label{P1app-eq: diagonal}
\eeq

We will consider the following general transformation:
\eq$
v=h(t,r) \,\Rightarrow\, dv=\pa_th\ dt+\pa_rh\ dr\,.$
Substituting the above expression of $dv$ into Eq. \eqref{P1app-eq: vtos1}, we obtain
\bal$\label{P1app-eq: vtos2}
ds^2&=-\left(1-\frac{2m(r)}{r}\right)(\pa_th)^2dt^2+\left[-\left(1-\frac{2m(r)}{r}\right)(\pa_rh)^2+2\pa_rh\right]dr^2\nonum\\
&\hspace{12.27em}+2\pa_th\left[-\left(1-\frac{2m(r)}{r}\right)\pa_rh+1\right] dtdr+r^2d\Omega^2\,.$
We now demand the vanishing of the off-diagonal term in Eq. (\ref{P1app-eq: vtos2}): for 
$\partial_t h \neq 0$, this leads to the constraint
\beq
\partial_r h=  \left(1-\frac{2m(r)}{r}\right)^{-1}\,.
\eeq
Provided that the above holds, the coefficient of $dr^2$ in Eq. (\ref{P1app-eq: vtos2}) reduces to
$\partial_r h$, and therefore
\eq$
f(r)=\frac{1}{\partial_r h}=\left(1-\frac{2m(r)}{r}\right).$
Comparing finally the coefficients of $dt^2$ in Eqs. \eqref{P1app-eq: diagonal} and \eqref{P1app-eq: vtos2}, we
conclude that $\pa_th$ must be equal to unity. Therefore, if the coordinate transformation
$v= t+g(r)$ is applied to the line element (\ref{P1app-eq: vtos1}), the latter takes indeed the diagonal form
$$ds^2=-\left(1-\frac{2M}{r} -\frac{\Lambda r^2}{3}\right) dt^2+
\left(1-\frac{2M}{r} -\frac{\Lambda r^2}{3}\right)^{-1} dr^2+r^2(d\theta^2+
\sin^2\theta\,d\varphi^2)$$
that describes a four-dimensional Schwarzschild--(anti)--de Sitter background depending
on the sign of the parameter $\Lambda$, which turns out to be proportional to the
cosmological constant on the brane. 

\par To complete the analysis, we need to determine the value of the function $g(r)$
through the integral
\beq
g(r)= \int \frac{dr}{f(r)}=\int \frac{dr}{1-\frac{2M}{r}-\frac{\Lambda r^2}{3}}\,.
\eeq
Evaluating the above integral amounts to calculating the tortoise coordinate for the
specific black-hole background.
The steps of the evaluation depend on the sign of the parameter $\Lambda$. Let us start with
the case $\Lambda>0$, where the four-dimensional background is a Schwarzschild--de Sitter
one. The function $f(r)$ has, in the most general case, two real, positive roots $r_h$ and
$r_c$ corresponding to the black-hole and cosmological horizon, respectively. Then, the aforementioned integral becomes
\beq
g(r)= \frac{3}{\Lambda}\,\int \frac{r\,dr}{-r^3 +3r/\Lambda -6M/\Lambda}=
\frac{3}{\Lambda}\,\int \frac{r\,dr}{(r-r_h)\,(r_c-r)\,(r+r_c+r_h)}\,,
\label{P1app-eq: integral_g_dS}
\eeq
where the two horizons satisfy the relations
\beq
(r_c+r_h)^2-r_cr_h=\frac{3}{\Lambda}\,, \qquad (r_c+r_h)\,r_c r_h =\frac{6M}{\Lambda}\,.
\eeq
Splitting the fraction in the integral (\ref{P1app-eq: integral_g_dS}) into three separate ones and
performing the corresponding integrations, we arrive at the result \cite{KGB}
\beq
g(r)=\frac{ r_h\,\ln (r-r_h)}{1-\Lambda r_h^2} + \frac{ r_c\,\ln (r_c-r)}{1-\Lambda r_c^2}
-\frac{ (r_c+r_h)\,\ln (r+r_c+r_h)}{1-\Lambda (r_c+r_h)^2} + C_1\,,
\eeq
where $C_1$ is an arbitrary integration constant.

If, on the other hand, $\Lambda=-|\Lambda|<0$, then the background on the brane is of a 
Schwarzschild-anti-de Sitter type. The function $f(r)$ vanishes only at $r=r_h$, i.e. at the
location of the black-hole horizon. We then write:
\beq
g(r)= \int \frac{r\,dr}{\frac{|\Lambda| r^3}{3} +r -2M}=
\frac{3}{|\Lambda|}\,\int \frac{r\,dr}{(r-r_h)\,(r^2+r_h\,r+\beta)}\,,
\label{P1app-eq: integral_g_Anti}
\eeq
where $\beta=6M/|\Lambda| r_h$. Note that the quadratic polynomial $r^2+r_h\,r+\beta$ 
has no real, positive roots.  We then split the fraction inside the integral into two
separate ones of the form
\beq
\frac{1}{(r-r_h)\,(r^2+r_h\,r+\beta)} =\frac{A}{r-r_h} +
\frac{B r + D}{r^2+r_h\,r+\beta}\,,
\label{P1app-eq: split_Anti}
\eeq
where
\beq
A=\frac{1}{2 r_h^2+\beta}\,, \qquad B=-A\,, \qquad
D=-2r_h A\,.
\eeq
Substituting Eq. (\ref{P1app-eq: split_Anti}) into Eq. (\ref{P1app-eq: integral_g_Anti}) and applying standard
integration techniques, we finally arrive at the result
\beq
g(r)=\frac{3}{|\Lambda|(2r_h^2+\beta)}\left[r_h\ln\left(\frac{r-r_h}{\sqrt{r^2+r_h\,r+\beta}}\right)    
+ \frac{r_h^2+2\beta}{\sqrt{4\beta-r_h^2}}
\arctan\left[\frac{2r+r_h}{\sqrt{4\beta-r_h^2}}\right]\right]+C_2\,,
\eeq
where $C_2$ is again an arbitrary integration constant and the horizon radius may be
expressed as 
\beq
r_h=\frac{1}{(-3\Lambda^2 M +\sqrt{9 \Lambda^4 M^2+|\Lambda|^3})^{1/3}}-
\frac{(-3\Lambda^2 M +\sqrt{9 \Lambda^4 M^2+|\Lambda|^3})^{1/3}}{|\Lambda|}\,.
\eeq

\afterpage{\blankpage}



\mychapter{A systematic methodology to express $\scriptstyle{\,_2F_1\left(\frac{3}{2}-q,\frac{3}{2};\frac{5}{2};u^2\right)}$ in terms of elementary\\ functions when $q \in {\mathbb{Z}}^{>}$ \label{P2app: hyper-analysis}}
\phantomsection
\vspace{1em}

\thispagestyle{empty}

\vspace*{-1.5em}

Let us start with the simplest case of $q=1$. For simplicity, we will
use the variable $u^2=\frac{w-1}{w}$. Then, using the expansion of the hypergeometric
function given in Eq. (\ref{P2eq: final-F}) and setting $\lam=4$, we readily obtain
\bal$
\label{P2app-eq: hyper-lam-4}\,_2F_1\left(\frac{1}{2},\frac{3}{2};\frac{5}{2};u^2\right)=
\frac{3}{2\sqrt{\pi}}\sum_{n=0}^\infty \frac{\Gamma\left(n+\frac{1}{2}\right)}
{n+\frac{3}{2}}\frac{u^{2n}}{n!}=\frac{3}{2\sqrt{\pi}}\sum_{n=0}^\infty
\frac{2n-1}{2n+3}\,\Gamma\left(n-\frac{1}{2}\right)\frac{u^{2n}}{n!}$
where we have used the Gamma function property $\Gamma(1+z)=z\Gamma(z)$. In order to express
the above in terms of elementary functions, we observe the following
{\fontsize{10}{10}\bal$
\label{P2app-eq: arcsin-sqrt}
\frac{\arcsin u}{u}-\sqrt{1-u^2}&=\,_2F_1\left(\frac{1}{2},\frac{1}{2};\frac{3}{2};u^2\right)
-\,_2F_1\left(-\frac{1}{2},1;1;u^2\right)\nonum\\[2mm]
&=\sum_{n=0}^\infty \left[\frac{\Gamma\left(n+\frac{1}{2}\right)}{(2n+1)\sqrt{\pi}}+
\frac{\Gamma\left(n-\frac{1}{2}\right)}{2\sqrt{\pi}}\right] \frac{u^{2n}}{n!}=
\sum_{n=1}^\infty \frac{2n}{2n+1}\,\frac{\Gamma\left(n-\frac{1}{2}\right)}{\sqrt{\pi}}\,
\frac{u^{2n}}{n!}\nonum\\[2mm]
&=\sum_{m=0}^\infty \frac{2(m+1)}{2m+3}\,\frac{\Gamma\left(m+\frac{1}{2}\right)}
{\sqrt{\pi}}\,\frac{u^{2(m+1)}}{(m+1)!}=\frac{1}{\sqrt{\pi}}\sum_{m=0}^\infty 
\frac{2m-1}{2m+3}\,\Gamma\left(m-\frac{1}{2}\right)\frac{u^{2(m+1)}}{m!}\,.$}
Note that, in the second sum of the second line of the above equation, we have changed
the lower value of $n$ from $n=0$ to $n=1$ since, due to the ($2n$) factor, this value
has a trivial contribution to the sum. Subsequently, we set $n=m+1$, and by rearranging
we arrived at the final result. Comparing now Eqs. \eqref{P2app-eq: hyper-lam-4} and
\eqref{P2app-eq: arcsin-sqrt}, we find that
\eq$ \label{P2app-eq: B.3}\,_2F_1\left(\frac{1}{2},\frac{3}{2};\frac{5}{2};u^2\right)u^2=
\frac{3}{2}\left(\frac{\arcsin u}{u}-\sqrt{1-u^2}\right).$\vspace{1em}
Let us now address the more general case with $q=1+\ell$, where
$\ell$ a positive number. Then, we may write
\bal$
\label{P2app-eq: hyper-lam-4q}\,_2F_1\left(\frac{1}{2}-\ell,\frac{3}{2};\frac{5}{2};u^2\right)&=
\frac{3}{\sqrt{\pi}}\,\frac{(2\ell)!}{\ell!(-4)^\ell}\sum_{n=0}^\infty \frac{1}{2n+3}\,
\Gamma\left(n-\ell+\frac{1}{2}\right)\frac{u^{2n}}{n!}\,,$
where we have also used the property $\Gamma\left(-\ell+\frac{1}{2}\right)=\frac{\ell!(-4)^\ell\sqrt{\pi}}{(2\ell)!}$.
But, it also holds that
{\fontsize{11}{11}\bal$\label{P2app-eq: gamma-1/2}
\Gamma\left(n-\frac{1}{2}\right)&=\frac{\Gamma\left(n-\ell+\frac{1}{2}+\ell\right)}{n-\frac{1}{2}}
=\frac{2}{2n-1}\left(n-\frac{1}{2}\right)\left(n-\frac{3}{2}\right)\cdots\left(n-\ell+\frac{1}{2}\right)\Gamma
\left(n-\ell+\frac{1}{2}\right)\nonum\\[2mm]
&=2^{1-\ell}\,\frac{(2n-1)(2n-3)\cdots(2n-2\ell+1)}{2n-1}\,\Gamma\left(n-\ell+\frac{1}{2}\right).$}
From Eqs. \eqref{P2app-eq: arcsin-sqrt} and \eqref{P2app-eq: gamma-1/2}, we then obtain
{\fontsize{11}{11}\gat$\label{P2app-eq: arcsin-sqrt-ell}
\frac{\arcsin(u)}{u}-\sqrt{1-u^2}=\frac{2^{1-\ell}}{\sqrt{\pi}}\sum_{n=0}^\infty \frac{(2n-1)(2n-3)
\cdots(2n-2\ell+1)}{2n+3}\Gamma\left(n-\ell+\frac{1}{2}\right)\frac{u^{2(n+1)}}{n!}.$}
In what follows, we are going to discuss also how the multiplication between even powers of $u$
and $\sqrt{1-u^2}$ can result to similar expansions as the one in Eq. \eqref{P2app-eq: arcsin-sqrt-ell}.
The obtained expansions together with Eq. \eqref{P2app-eq: arcsin-sqrt-ell} will help us to express
the r.h.s. of Eq. \eqref{P2app-eq: hyper-lam-4q} in terms of elementary functions. Thus, starting
from the relation
\eq$\sqrt{1-u^2}=-\frac{1}{2\sqrt{\pi}}\sum_{n=0}^\infty
\Gamma\left(n-\frac{1}{2}\right)\frac{u^{2n}}{n!}\,,$
we write, employing also Eq. (\ref{P2app-eq: gamma-1/2}),
\bal$\label{P2app-eq: z2-sqrt}
u^2\sqrt{1-u^2}&=-\frac{2^{1-\ell}}{2\sqrt{\pi}}\sum_{n=0}^\infty
\frac{(2n-1)(2n-3)\cdots(2n-2\ell+1)}{2n-1}\Gamma\left(n-\ell+\frac{1}{2}\right)
\frac{u^{2(n+1)}}{n!}.$
Similarly, we obtain
{\fontsize{11}{11}\bal$\label{P2app-eq: z4-sqrt}
u^4\sqrt{1-u^2}&=-\frac{1}{2\sqrt{\pi}}\sum_{m=0}^\infty m\, \Gamma\left(m-\frac{3}{2}\right)
\frac{u^{2(m+1)}}{m!}=-\frac{1}{2\sqrt{\pi}}\sum_{m=0}^\infty \frac{2m}{2m-3}\, \Gamma\left(m-\frac{1}{2}\right)\frac{u^{2(m+1)}}{m!}\nonum\\[2mm]
u^4\sqrt{1-u^2}&=-\frac{2^{2-\ell}}{2\sqrt{\pi}}\sum_{n=0}^\infty \frac{(2n-1)(2n-3)\cdots(2n-2\ell+1)\ n}{(2n-1)(2n-3)}\,
 \Gamma\left(n-\ell+\frac{1}{2}\right)\frac{u^{2(n+1)}}{n!}.$}
Note that, in the first sum of the above expression, we set $m=n+1$ but retained the lower
value of the sum to be 0 due to the $m$ factor - we have also used, once again,
Eq. (\ref{P2app-eq: gamma-1/2}). Continuing along the same lines, we obtain, for a general $\ell$,
the result 
\bal$\label{P2app-eq: zell-sqrt}
u^{2\ell}\sqrt{1-u^2}&=
-\frac{1}{2\sqrt{\pi}}\sum_{m=\ell-1}^\infty m(m-1)\cdots(m-\ell+2)\ \Gamma\left(m-\ell+\frac{1}{2}\right)\frac{u^{2(m+1)}}{m!}\nonum\\[2mm]
&=-\frac{1}{2\sqrt{\pi}}\sum_{n=0}^\infty n(n-1)\cdots(n-\ell+2)\ \Gamma\left(n-\ell+\frac{1}{2}\right)\frac{u^{2(n+1)}}{n!}\,,$
where, now, we set $m=n+\ell-1$ and again reinstated the lower value of the sum to be 0
due to the multiplying factors that trivialise all terms with $n < \ell-1$. 

Comparing now the r.h.s's of Eqs. \eqref{P2app-eq: arcsin-sqrt-ell}, \eqref{P2app-eq: z2-sqrt} and
\eqref{P2app-eq: zell-sqrt} with the r.h.s of Eq. \eqref{P2app-eq: hyper-lam-4q},
we conclude that we may express the aforementioned hypergeometric function as
\bal$\label{P2app-eq: hyper-lam-4q-final}
\,_2F_1\left(\frac{1}{2}-\ell,\frac{3}{2};\frac{5}{2};u^2\right)u^2=&
\alpha\left(\frac{\arcsin u}{u}-\sqrt{1-u^2} \right)\nonum\\[2mm]
&+\sqrt{1-u^2}\left(\beta_1\,u^2+\beta_2\,u^4+\dots+\beta_{\ell-1}\,u^{2(\ell-1)}+
\beta_\ell\,u^{2\ell}\right),$
where $\alpha, \beta_1, \dots, \beta_\ell$ are constant coefficients. These may be
determined by substituting the explicit expansions (\ref{P2app-eq: hyper-lam-4q}),
\eqref{P2app-eq: arcsin-sqrt-ell}, \eqref{P2app-eq: z2-sqrt} and \eqref{P2app-eq: zell-sqrt} on both sides of
the above equation and demanding its validity. Then, we obtain the following
relation for the coefficients $\alpha, \beta_1, \dots, \beta_\ell$, which must
be true for arbitrary $n\in\mathbb{Z}^{\geq}$,
{\fontsize{11}{11}\bal$
6\frac{(2\ell)!}{ \ell!\,(-4)^\ell}=&\hspace{0.5em}\alpha\,2^{2-\ell}\,(2n-1)(2n-3)\cdots(2n-2\ell+1)\nonum\\[1mm]
&-(2n+3)\bigg[\,\beta_1\,2^{1-\ell}\ \frac{(2n-1)(2n-3)\cdots(2n-2\ell+1)}{2n-1}\nonum\\[1mm]
&\hspace{5em}+\beta_2\,2^{2-\ell}\,\frac{(2n-1)(2n-3)\cdots(2n-2\ell+1)}{(2n-1)(2n-3)}
\,n\nonum\\[1mm]
&\hspace{5em}+\beta_3\,2^{3-\ell}\,\frac{(2n-1)(2n-3)\cdots(2n-2\ell+1)}{(2n-1)(2n-3)(2n-5)}\, n(n-1)\nonum\\[1mm]
&\hspace{5cm}\vdots\nonum\\[1mm]
&\hspace{5em}+\beta_{\ell-1}\,2^{-1}\,(2n-2\ell+1)\ n(n-1)\cdots(n-\ell+3)\nonum\\[1mm]
&\hspace{5em}+\beta_{\ell}\,\frac{(n+1)n(n-1)\cdots(n-\ell+2)}{n+1}\,\bigg]\,.
\label{P2app-eq: alpha-beta}$}
The above equation leads to a system of $\ell+1$ linear equations with $\ell+1$ variables
from which the unknown coefficients $\alpha, \beta_1, \dots, \beta_\ell$ may easily be
derived.

\vspace*{-2em}

\afterpage{\blankpage}



\mychapter{Restrictions on the allowed values of the parameter $\mu$ in quadratic case \label{P3app: mu}}
\phantomsection

\thispagestyle{empty}

We shall now determine the range of values for the parameter $\mu$ in the case
of the quadratic coupling function \eqref{P3eq: quad-f}. The allowed values of $\mu$ will be
obtained by demanding that the scalar field \eqref{P3eq: quad-Phi} remains real and finite,
and depend primarily on the assumed value of the parameter $\lam$. In what
follows, we will consider in detail every possible case:
\begin{enumerate}
\item[\bf(i)] \begin{flushleft}
\underline{$\lam>0$:}
\end{flushleft}
\par Using Eq. \eqref{P3eq: quad-Phi} we get
$$\lim_{y\ra +\infty}\Phi(y)=\frac{\Phi_0}{2\lam}\left(\xi\ \mu^\frac{2\lam}{1+4\lam}-1
\right)\,.$$
Thus, demanding the functions $\Phi(y)$, $f(y)$ to be real-valued in their whole domain,
it is necessary to have $\mu\geq 0$. \\

\item[\bf(ii)] \begin{flushleft}
\underline{$\lam\in\left(-\frac{1}{4},0\right)\ \land\ \frac{2\lam}{1+4\lam}\neq n, \ n\in\mathbb{Z}^<$:}
\end{flushleft}
In this case $\frac{2\lam}{1+4\lam}$ is a negative rational number. Hence, one may write
\eq$\lim_{y\ra +\infty}\Phi(y)=\frac{\Phi_0}{2\lam}\left(\xi\ \mu^\frac{2\lam}{1+4\lam}-1
\right)=-\frac{\Phi_0}{2|\lam|}\left[1-\xi\left(\frac{1}{\mu}\right)^{\left|\frac{2\lam}
{1+4\lam}\right|}\right]\,.\nonum$
Therefore, in order to avoid having a complex scalar field we should demand $\mu>0$.\\

\item[\bf(iii)]\begin{flushleft}
\underline{$\lam\in\left(-\frac{1}{4},0\right)\ \land\ \frac{2\lam}{1+4\lam}=n,\ n\in
\mathbb{Z}^<$:}
\end{flushleft}
In this case we have $\frac{2\lam}{1+4\lam}=n$ or $\lam=\frac{n}{2(1-2n)}$.
Thus, one may write
$$\Phi(y)=\frac{\Phi_0}{2\lam}\left[\xi(\mu+e^{-ky})^{\frac{2\lam}{1+4\lam}}-1\right]=
\frac{\Phi_0(1-2n)}{n}
\left[\xi\left(\frac{1}{\mu+e^{-ky}}\right)^{|n|}-1\right]\,.$$
It is clear that the parameter $\mu$ is allowed to take negative values. However, we 
should not allow values in the range $[-1,0]$, because then at $y_0=-\frac{1}{k}\ln
(-\mu)>0$ we would encounter infinities regarding both the scalar field and the coupling 
function in a finite distance away from the brane. Thus, $\mu\in(-\infty,-1)\cup(0,\infty)$.\\

\item[\bf(iv)]\begin{flushleft}
\underline{$\lam=-\frac{1}{4}$:}
\end{flushleft}
In this particular case it is obvious from Eqs. \eqref{P3eq: quad-Phi} and \eqref{P3eq: quad-f-y} that 
the parameter $\mu$ is allowed to take any value in the set of the real numbers except 
zero. Thus, $\mu\in(-\infty,0)\cup(0,+\infty)$.\\

\item[\bf(v)]\begin{flushleft}
\underline{$\lam<-\frac{1}{4}\ \land\ \frac{2\lam}{1+4\lam}\neq n,\ n\in\mathbb{Z}^>$:}
\end{flushleft}
In this case $\frac{2\lam}{1+4\lam}$ is a positive rational number. Thus, we have
$$\lim_{y\ra + \infty}\Phi(y)=\frac{\Phi_0}{2\lam}\left(\xi\ \mu^\frac{2\lam}{1+4\lam}-1
\right)\,.$$
Therefore, $\mu\geq 0$ to avoid a complex-valued scalar field.\\

\item[\bf(vi)]\begin{flushleft}
\underline{$\lam<-\frac{1}{4}\ \land\ \frac{2\lam}{1+4\lam}=n,\ n\in\mathbb{Z}^>$:}
\end{flushleft}
In this case, it is $\frac{2\lam}{1+4\lam}=n$ and $\lam=\frac{n}{2(1-2n)}$. Thus, from Eq. 
\eqref{P3eq: quad-Phi} we have
$$\Phi(y)=\frac{\Phi_0}{2\lam}\left[\xi(\mu+e^{-ky})^{\frac{2\lam}{1+4\lam}}-1\right]=
\frac{\Phi_0(1-2n)}{n}\left[\xi \left(\mu+e^{-ky}\right)^{n}-1\right],$$
which allows $\mu$ to take values in the whole set of the real number: $\mu\in\mathbb{R}$.\\
\end{enumerate}

The aforementioned results are summarised in Table \ref{P3tab: quad-par-val}.



\mychapter{The upper and lower incomplete gamma functions \label{P3app: incom-gamma}}
\phantomsection

\thispagestyle{empty}

The upper incomplete gamma function $\Gamma(s,x)$ is defined as follows
\eq$\label{P3app-eq: upper-gamma}
\Gamma(s,x)\equiv\int_x^\infty dt\ t^{s-1}\, e^{-t}=\Gamma(s)-\gamma(s,x)\, ,$
where
\eq$\label{P3app-eq: lower-gamma}
\gamma(s,x)\equiv \int_0^x dt\ t^{s-1}\, e^{-t}\, ,$
is the lower incomplete gamma function. Both upper and lower incomplete gamma functions, as
defined above, are valid for real and positive $s$ and $x$. However, both functions can
be extended for almost all combinations of complex $s$ and $x$. One can show that, for
all complex $s$ and $z$, the lower incomplete gamma function can be expanded in the following
power series
\eq$\label{P3app-eq: lower-gamma-exp}
\gamma(s,z)=z^s\,\Gamma(s)\,e^{-z}\sum_{k=0}^\infty \frac{z^k}{\Gamma(s+k+1)}\, .$
Locally, the sum in the r.h.s. of the previous relation converges uniformly for all $s\in\mathbb{C}$ 
and $z\in\mathbb{C}$. Using the relation $\Gamma(s,z)=\Gamma(s)-\gamma(s,z)$ we obtain the
values of the upper incomplete gamma function for complex $s$ and $z$, but only for the
points $(s,z)$ in which the r.h.s. exists. The numerical value of the upper incomplete gamma
function can be given by the following expressions:
\eq$\label{P3app-eq: upper-gamma-explicit}
\Gamma(s,x)=\left\{\begin{array}{ll}
\displaystyle{\Gamma(s)-x^s\,\Gamma(s)\,e^{-x}\sum_{k=0}^\infty \frac{x^k}{\Gamma(s+k+1)}}\, , &
 s\neq-n,\ n\in\mathbb{Z}^>\\[7mm]
\displaystyle{-\gamma-\ln(x)-\sum_{k=1}^\infty \frac{(-x)^k}{k(k!)}}\, , & s=0\\[7mm]
\displaystyle{\frac{1}{n!}\left[\frac{e^{-x}}{x^n}\sum_{k=0}^{n-1}(-x)^k(n-k-1)!+(-1)^n\,\Gamma(0,x)
\right]} , & s=-n,\ n\in\mathbb{Z}^>
\end{array}\right\}\, ,$
where $\gamma$ is the Euler-Mascheroni constant. In our case, for $s=1-\lam$, namely
$s\in(-\infty,1)$ and $x=2ky_0>0$, we obtain the expressions
{\fontsize{10}{10}\eq$\label{P3app-eq: gamma-inc}
\Gamma(1-\lam,2ky_0)=\left\{\begin{array}{ll}
\displaystyle{\Gamma(1-\lam)\left[1-(2ky_0)^{1-\lam}\,e^{2ky_0}\sum_{m=0}^\infty \frac{(2ky_0)^m}{
\Gamma(2+m-\lam)}\right]}, & \lam\neq n+1,\\[-0.8em]
&\ \ n\in\mathbb{Z}^>\\[8mm]
\displaystyle{-\gamma-\ln(2ky_0)-\sum_{m=1}^\infty \frac{(-2ky_0)^m}{m(m!)}}, & \lam=1\\[8mm]
\displaystyle{\frac{1}{n!}\left[\frac{e^{-2ky_0}}{(2ky_0)^n}\sum_{m=0}^{n-1}(-2ky_0)^m(n-m-1)!+(-1)^n
\,\Gamma(0,2ky_0)\right]},& \lam= n+1,\\[-0.8em]
&\ \ n\in\mathbb{Z}^>
\end{array}\right\}.$}




\mychapter{Brane energy-momentum tensor in terms of the extrinsic curvature \label{P4app: Br-Ene}}
\phantomsection

\thispagestyle{empty}

From Eq. \eqref{P4eq: jc2}, we have
\gat$ \label{P4app-eq: app-br-ene-1}
h^{\mu\nu}[K_{\mu\nu}]=
[h^{\mu\nu}K_{\mu\nu}]=-\kappa_5^2 \left(h^{\mu\nu}T^{(br)}_{\mu\nu}-\frac{1}{3}\,h^{\mu\nu}h_{\mu\nu}\,T^{(br)}\right)\Ra
T^{(br)}=\frac{3}{\kappa_5^2}\,[K]\,.$
Using then Eq. \eqref{P4app-eq: app-br-ene-1} in Eq. \eqref{P4eq: jc2}, we obtain
\gat$ \label{P4app-eq: app-br-ene-2}
[K_{\mu\nu}]=-\kappa_5^2\left(T^{(br)}_{\mu\nu}-\frac{1}{3}h_{\mu\nu}\,\frac{3}{\kappa_5^2}[K]\right)\Ra
T^{(br)}_{\mu\nu}=-\frac{1}{\kappa_5^2}\left([K_{\mu\nu}]-h_{\mu\nu}[K]\right)\,.$

\afterpage{\blankpage}



\mychapter{Scalar curvature quantities for the localized black-hole solutions \label{P5app: Curv-Inv}}
\phantomsection

\thispagestyle{empty}

In $(\rho,\chi)$ coordinates, the expressions of the scalar invariants $\mathcal{R}\equiv R_{MN}\,R^{MN}$ and $\mathcal{K}\equiv R_{MNKL}\,R^{MNKL}$ 
are given by
{\fontsize{10}{10}
\bal$\label{P5app-eq: ricci-2}
\mathcal{R}&=\frac{80}{9} \left(3 k^2-\Lambda\right)^2-\frac{32 k^2 M \left(3 k^2-\Lambda \right) \cos ^2\chi }{\rho }
-\frac{8 \left(27k^3 M \cos^3 \chi +12 k^2-4 \Lambda \right) \left(2 k Q^2 \cos \chi+M\right)}{3 \rho ^3}\nonum\\[0.5mm]
&\hsp+\frac{2 k \cos \chi  \left\{k \cos \chi  \left[3 k^2 \left(9 M^2+16 Q^2\right)-16 \Lambda  Q^2\right]+M \left[9 k^3 M \cos (3 \chi )
+96 k^2-32\Lambda \right]\right\}}{\rho ^2}\nonum\\[0.5mm]
&\hsp+\frac{6 k^2 \cos ^2\chi \left[6 k^2 Q^4 \cos (2 \chi )+6 k^2 Q^4+4 k M Q^2 \cos \chi +17 M^2\right]}{\rho ^4}
+\frac{14 \left(2 k Q^2 \cos \chi +M\right)^2}{\rho ^6}\nonum\\[0.5mm]
&\hsp+\frac{12 k \cos \chi  \left[2 k^2 Q^4 \cos (2 \chi )+2 k^2 Q^4-8 k M Q^2 \cos \chi -5 M^2\right]}{\rho ^5}\,,\\[3em]
\label{P5app-eq: riemsq}
\mathcal{K}&=\frac{40}{9} \left(3 k^2-\Lambda\right)^2-\frac{16 k^2 M \left(3 k^2-\Lambda \right) \cos ^2\chi}{\rho }
+\frac{8 k \cos \chi  \left[3 k^2 \left(27 M^2-4 Q^2\right)+4 \Lambda  Q^2\right]}{3\rho^3}\nonum\\[0.5mm]
&\hsp+\frac{4 k \cos \chi \left\{k \cos \chi  \left[3 k^2 \left(9 M^2+4 Q^2\right)-4 \Lambda  Q^2\right]
+M \left[9 k^3 M \cos (3 \chi )+24 k^2-8 \Lambda \right]\right\}}{\rho ^2}\nonum\\[0.5mm]
&\hsp-\frac{4 M \left\{27 k^3 \left[k Q^2 \left(4 \cos (2 \chi )+\cos (4 \chi )+3\right)-2 M \cos (3 \chi )\right]+12 k^2
-4 \Lambda \right\}}{3 \rho ^3}\nonum\\[0.5mm]
&\hsp+\frac{24 k^2 \cos ^2\chi \left[3 k^2 Q^4 \cos (2 \chi )+3 k^2 Q^4-28 k M Q^2 \cos \chi +19 M^2\right]}{\rho ^4}\nonum\\[1mm]
&\hsp+\frac{48 k \cos \chi \left[4 k^2 Q^4 \cos (2 \chi )+4 k^2 Q^4-19 k M Q^2 \cos \chi +5 M^2\right]}{\rho ^5}
+\frac{72Q^4}{\rho^8}\nonum\\[0.5mm]
&\hsp+\frac{8 \left[31 k^2 Q^4 \cos (2 \chi )+31 k^2 Q^4-64 k M Q^2 \cos \chi +11 M^2\right]}{\rho ^6}
-\frac{144 Q^2 \left(M-2 k Q^2 \cos \chi\right)}{\rho ^7}\,,$}
\hspace{-0.5em}while, in $(r,y)$ coordinates, the above expressions take the form
{\fontsize{10}{10}\bal$\label{P5app-eq: ricci-2-ry}
\mathcal{R}&=\frac{80}{9} \left(3 k^2-\Lambda\right)^2+\frac{2 k^6 M^2 \left(160-384 e^{k \left| y\right| }
+375 e^{2 k \left| y\right| }-180 e^{3 k \left| y\right| }+36 e^{4 k \left| y\right| }\right)}{\left[k^2r^2
+\left(e^{k|y|}-1\right)^2\right]^3}\nonum\\[2mm]
&\hsp+\frac{32k^4 Q^2 \left(3 k^2
-\Lambda \right) \left(e^{k \left| y\right| }-1\right) \left(3 e^{k \left| y\right| }-5\right)}{3 \left[k^2r^2
+\left(e^{k|y|}-1\right)^2\right]^2}+\frac{8k^8 Q^4 \left(e^{k \left| y\right| }-1\right)^2 \left(10-12 e^{k \left| y\right| }
+9 e^{2 k \left| y\right| }\right)}{\left[k^2r^2+\left(e^{k|y|}-1\right)^2\right]^4}\nonum\\[2mm]
&\hsp-\frac{8k^3M}{3\left[k^2r^2+\left(e^{k|y|}-1\right)^2\right]^{7/2}}\bigg\{40 \Big[3 k^6 r^4
+k^4 \left(3 Q^2-\Lambda  r^4+6 r^2\right)+k^2 \left(3-2 \Lambda  r^2\right)-\Lambda \Big]\nonum\\[2mm]
&\hsp-16 e^{k \left| y\right| } \Big[9 k^6 r^4+k^4 \left(21 Q^2-3 \Lambda  r^4+48 r^2\right)+k^2 \left(39
-16 \Lambda  r^2\right)-13 \Lambda \Big]\nonum\\[2mm]
&\hsp+e^{2 k \left| y\right| } \Big[36 k^6 r^4+k^4 \left(387 Q^2-12 \Lambda  r^4+888 r^2\right)
-148 k^2 \left(2\Lambda  r^2-9\right)-444 \Lambda \Big]\nonum\\[2mm]
&\hsp+e^{3 k \left| y\right| } \Big[-9 k^4 \left(25 Q^2+48 r^2\right)+48 k^2 \left(3 \Lambda  r^2-31\right)
+496 \Lambda \Big]\nonum\\[2mm]
&\hsp+2 e^{4 k \left| y\right| } \Big[9 k^4 \left(3 Q^2+4 r^2\right)-12 k^2 \left(\Lambda  r^2-38\right)
-152 \Lambda \Big]+12 \left(3 k^2-\Lambda \right) e^{5 k \left| y\right| }\left( e^{k \left| y\right| }-8\right)\bigg\}
\,,\\[3em]
\label{P5app-eq: riemsq-ry}
\mathcal{K}&=-\frac{16k^3 M \left(3 k^2-\Lambda \right) \left(10-12 e^{k \left| y\right| }
+3 e^{2 k \left| y\right| }\right)}{3 \left[k^2r^2+\left(e^{k|y|}-1\right)^2\right]^{3/2}}
-\frac{16k^7 M Q^2 \left(10-28 e^{k \left| y\right| }+39 e^{2 k \left| y\right| }
-30 e^{3 k \left| y\right| }+18 e^{4 k \left| y\right| }\right)}{\left[k^2r^2+\left(e^{k|y|}-1\right)^2\right]^{7/2}}\nonum\\[2mm]
&\hsp+\frac{8 k^6 M^2 \left(20-48 e^{k \left| y\right| }+57 e^{2 k \left| y\right| }
-36 e^{3 k \left| y\right| }+18 e^{4 k \left| y\right| }\right)}{\left[k^2r^2+\left(e^{k|y|}-1\right)^2\right]^{3}}
+\frac{16k^4 Q^2 \left(3 k^2-\Lambda \right) \left(e^{k \left| y\right| }-1\right) \left(3 e^{k \left| y\right| }
-5\right)}{3\left[k^2r^2+\left(e^{k|y|}-1\right)^2\right]^{2}}\nonum\\[2mm]
&\hsp+\frac{8 k^8 Q^4 \left(5-16 e^{k \left| y\right| }+26 e^{2 k \left| y\right| }-24 e^{3 k \left| y\right| }
+18 e^{4 k \left| y\right| }\right)}{\left[k^2r^2+\left(e^{k|y|}-1\right)^2\right]^{4}}+\frac{40}{9} \left(3 k^2-\Lambda\right)^2\,.$}


\mysection{Curvature Invariants for $Q=0$ and $\Lambda=0$ \label{P4app: Curv-Inv}}

In terms of the $(\rho,\chi)$-coordinates, the curvature invariants $\mathcal{R}$ and $\mathcal{K}$ for $Q=0$ and $\Lambda=0$ are simplified as follows
\bal$
\mathcal{R}&=80 k^4-\frac{96 k^4 M \cos^2\chi}{\rho }+\frac{24 k^3 M \cos \chi \left(3 k M \cos ^3\chi+8\right)}{\rho ^2}\nonum\\[1mm]
&\hsp -\frac{8 k^2 M \left(9 k M \cos ^3\chi+4\right)}{\rho ^3}+\frac{102 k^2 M^2 \cos ^2\chi }{\rho ^4}\nonum\\[1mm]
&\hsp-\frac{60 k M^2 \cos \chi }{\rho ^5}+\frac{14 M^2}{\rho ^6}\,,$
\bal$\mathcal{K}&=40 k^4-\frac{48 k^4 M \cos ^2(\chi )}{\rho }+\frac{48 k^3 M \cos \chi  \left(3 k M \cos ^3\chi +2\right)}{\rho ^2}\nonum\\[1mm]
&\hsp+\frac{8 k^2 M \left(36 k M \cos ^3\chi -2\right)}{\rho ^3}+\frac{456 k^2 M^2 \cos ^2\chi }{\rho ^4}\nonum\\[1mm]
&\hsp +\frac{240 k M^2 \cos \chi }{\rho ^5}+\frac{88 M^2}{\rho^6}\,,$

while in terms of $(r,y)$-coordinates, we get

\bal$
\mathcal{R}&=\frac{2k^6M^2 \left(160-384\, e^{k \left|y\right| }+375\, e^{2 k \left| y\right| }-180\, e^{3 k \left| y\right| }+36
\, e^{4 k \left| y\right| }\right)}{\left[k^2r^2+\left(e^{k \left| y\right| }-1\right)^2\right]^3}\nonum\\[2mm]
&\hsp +80k^4-\frac{32 k^5 M \left(10-12\, e^{k \left| y\right| }+3\, e^{2 k \left| y\right| }\right)}{\left[k^2r^2+\left(e^{k \left| y
\right| }-1\right)^2\right]^{3/2}}\,,\\[3em]
\mathcal{K}&=\frac{8k^6M^2 \left(20-48\, e^{k \left|y\right| }+57\, e^{2 k \left| y\right| }-36\, e^{3 k \left| y\right| }+18
\, e^{4 k \left| y\right| }\right)}{\left[k^2r^2+\left(e^{k \left| y\right| }-1\right)^2\right]^3}\nonum\\[2mm]
&\hsp +40k^4-\frac{16 k^5 M \left(10-12\, e^{k \left| y\right| }+3\, e^{2 k \left| y\right| }\right)}{\left[k^2r^2+\left(e^{k \left| y\right| }-1\right)^2\right]^{3/2}}\,.$

\afterpage{\blankpage}



\mychapter{How to remedy the cosmological horizon singularity \label{P5app: CHCS}}
\phantomsection

\thispagestyle{empty}

The line-element \eqref{P5eq: 5d-metr} in terms of the radial, null coordinates $(u,v)$ which are defined by
\eq$\label{P5app-eq: uv-coords}
\left\{\begin{array}{c}
v=t+\rho_*\\[1mm]
u=t-\rho_*
\end{array}\right\}\,,$
takes the form
\eq$\label{P5app-eq: uv-metr}
ds^2=\frac{1}{(1+k\rho\cos\chi)^2}\left[-f(\rho)\,dudv+\rho^2d\Omega_3^2\right]\,.$
In the above, the variable $\rho_*$ is determined by the following relation
\eq$\label{P5app-eq: rho-star}
\rho_*=\int\frac{d\rho}{f(\rho)}=-\frac{1}{2\kappa_{\ssst{C}}}\ln\bigg|\frac{\rho}{\rho_{\ssst{C}}}-1\bigg|+\frac{1}{2
\kappa_{\ssst{+}}}\ln\bigg|\frac{\rho}{\rho_{\ssst{+}}}-1\bigg|-\frac{1}{2\kappa_{\ssst{-}}}\ln\bigg|\frac{\rho}{\rho_{\ssst{-}}}
-1\bigg|+\frac{1}{2\kappa_4}\ln\bigg|\frac{\rho}{\rho_4}-1\bigg|\,,$ where the integration constant has been set to zero. 
The constants $\rho_{\ssst{C}}$, $\rho_{\ssst{+}}$, $\rho_{\ssst{-}}$, $\rho_4$ are the roots~\footnote{It is implied that 
$\rho_1=\rho_{\ssst{C}}$, $\rho_2=\rho_{\ssst{+}}$, $\rho_3=\rho_{\ssst{-}}$.} of the quartic polynomial
$f(\rho)=0$ which for $\Lambda>0$ satisfy the inequality $\rho_{\ssst{C}}>\rho_{\ssst{+}}>\rho_{\ssst{-}}>\rho_4$, 
with $\rho_4<0$. The parameters $\kappa_i$ denote the surface gravity at the corresponding $i$-th horizon located 
at $\rho=\rho_i$ (for more details see \cite{Chambers}). Using the aforementioned roots, the function $f(\rho)$ given
by \eqref{P5eq: ansatz} can be factorised as follows
\eq$\label{P5app-eq: f-rho-fact}
f(\rho)=-\frac{\Lambda}{3}\frac{(\rho-\rho_{\ssst{C}})(\rho-\rho_{\ssst{+}})(\rho-\rho_{\ssst{-}})(\rho-\rho_4)}{\rho^2}\,.$
Combining Eqs. \eqref{P5app-eq: rho-star} and \eqref{P5app-eq: f-rho-fact}, the function $f(\rho)$ near the cosmological horizon reduces
to 
\eq$\label{P5app-eq: f-rho-lim}
\lim_{\rho\ra\rho_{\ssst{C}}^\pm}f(\rho)=\mp2\rho_{\ssst{C}}\kappa_{\ssst{C}}e^{-2\kappa_{C}\rho_*}\,,$
where the minus or plus sign on the right-hand-side depends on the direction from which we approach the cosmological horizon, while
\eq$\label{P5app-eq: kappa-c}
\kappa_{\ssst{C}}=\frac{\Lambda}{6}\frac{(\rho_{\ssst{C}}-\rho_{\ssst{+}})(\rho_{\ssst{C}}-\rho_{\ssst{-}})
(\rho_{\ssst{C}}-\rho_4)}{\rho_{\ssst{C}}^2}\,.$
The future cosmological horizon $\rho_{\ssst{C}}$ lies at $t\ra+\infty$ and $\rho_*\ra+\infty$, i.e. at $v\ra+\infty$.
Consequently, by defining the coordinates
\eq$\label{P5app-eq: UV-coords}
\left\{\begin{array}{l}
V=-e^{-\kappa_{\ssst{C}}v}\\[1mm]
U=e^{\kappa_{\ssst{C}}u}
\end{array}\right\}\,,$
we can readily see that $V\ra0$ as $v\ra+\infty$. Therefore, using the limit \eqref{P5app-eq: f-rho-lim} and the $(U,V)$ coordinates, the
line-element \eqref{P5app-eq: uv-metr} near the future cosmological horizon takes the form
\eq$\label{P5app-eq: UV-metr}
ds^2\simeq\frac{1}{(1+k\rho\cos\chi)^2}\left[\frac{2\rho_{\ssst{C}}}{\kappa_{\ssst{C}}}\,dUdV+\rho^2d\Omega_3^2\right].$
It is easy to see now that in the above coordinate system the geometry close to the cosmological horizon is completely regular.



\mychapter{Bulk energy-momentum tensor components transformed \label{P5app: En-Mom}}
\phantomsection

\thispagestyle{empty}

In this section, we will derive the components of the energy-momentum tensor as we change from the set of coordinates
$x^M=\{t,\rho,\chi,\theta,\varphi\}$ to the set $x'^M=\{t,r,\theta,\varphi,y\}$. We will denote all new quantities with a prime
in order to distinguish them from those in the old coordinates. Thus, using Eq. \eqref{P5eq: en-mom} we have
\eq$\label{P5app-eq: en-mom-ry}
T'^{(B)MN}=(\rho_E+p_\theta)U'^M\, U'^N+(p_r-p_\theta)X'^M\, X'^N+p_\theta\, g'^{MN}\,.$
The quantities $\rho_E$, $p_r$ and $p_\theta$ are scalars and thus they do not change under a coordinate transformation. Their
expressions  in the new coordinates can be easily obtained from Eqs. \eqref{P5eq: rho-p1} and \eqref{P5eq: p2} by using the relations
of Eq. \eqref{P4eq: new-coords-inv}. However, the vectors $U'^M$ and $X'^M$, defined in  \eqref{P5eq: U-vec} and\eqref{P5eq: X-vec},
under the coordinate transformation are transformed as follows
\eq$\label{P5app-eq: U-ry}
U'^M=\frac{dx'^M}{dx^A}\,U^A=\frac{dx'^M}{dt}\,U^t=\frac{e^{k|y|}}{\sqrt{f(r,y)}}\,\del^M{}_t\,,$
{\fontsize{10}{10}\eq$\label{P5app-eq: X-ry}
X'^M=\frac{dx'^M}{dx^A}\,X^A=\frac{dx'^M}{d\rho}\,X^\rho=
\left[r^2+\frac{\left(e^{k|y|}-1\right)^2}{k^2}\right]^{-1/2}e^{k|y|}\sqrt{f(r,y)}\left(r\,\del^{M}{}_r+\frac{1-e^{-k|y|}}{k}\del^{M
}{}_y\right)\,.$}
\hspace{-0.5em}In the above, the function $f(r,y)$ is given in Eq. \eqref{P5eq: f-ry}. One can also verify that $U'^MU'^Ng'_{MN}=-1$ and
$X'^MX'^Ng'_{MN}=1$,  where $g'_{MN}$ is evaluated from the line-element \eqref{P5eq: metr-r-y}. Then, for the mixed components
of the energy-momentum tensor $T'^{(B)M}{}_N$, we obtain
{\fontsize{11}{11}\eq$\label{P5app-eq: en-mom-mix-ry}
T'^{(B)M}{}_N=T'^{(B)ML}g'_{LN}=(\rho_E+p_\theta)U'^M\, U'^t\,g'_{tN}+(p_r-p_\theta)X'^M\left( X'^r\,g'_{rN}+X'^y\,g'_{yN}\right)+p_\theta\, 
\del^{M}{}_{N}\,.$}
\hspace{-0.5em}Using Eqs. \eqref{P5app-eq: U-ry} and \eqref{P5app-eq: X-ry} in Eq. \eqref{P5app-eq: en-mom-mix-ry}, it is straightforward to 
calculate the non-zero mixed components of the energy-momentum tensor in the new coordinate system. These read:
{\fontsize{10}{10}\bal$
T^{(B)t}{}_t&=-\rho_E(r,y)=\frac{1}{\kappa_5^2}\left\{2(3k^2-\Lambda)-\frac{3Mk^3\left(4-3e^{k|y|}\right)}{\left[k^2r^2+\left(e^{k|y|}-1\right)^2\right]^{3/2}}-\frac{6Q^2k^4\left(e^{k|y|}-1\right)}{\left[k^2r^2+\left(e^{k|y|}-1\right)^2\right]^{2}} \right\}\,,$
\bal$T^{(B)r}{}_r&=\frac{1}{\kappa_5^2}\left\{2(3k^2-\Lambda)+\frac{3k^3M \left\{-4\left(1+k^2r^2\right)+e^{k|y|}\left[14-2e^{k|y|}
\left(9-5e^{k|y|}+e^{2k|y|}\right)+3k^2r^2\right]\right\}}{\left[k^2r^2+\left(e^{k|y|}-1\right)^2\right]^{5/2}}\right.\nonum\\[2mm]
&\hspace{2.73cm}\left.+\frac{6 k^4Q^2 \left(e^{k \left| y\right| }-1\right) \left(3 e^{k \left| y\right| }-3 e^{2 k \left| y\right| }
+e^{3 k \left| y\right| }-k^2 r^2-1\right)}{\left[k^2r^2+\left(e^{k|y|}-1\right)^2\right]^{3}}\right\}\,,\\[5mm]
T^{(B)r}{}_y&=e^{2k|y|}\,T^{(B)y}{}_r=\frac{3k^4 r\, e^{2k|y|}}{\kappa_5^2}\left\{ \frac{M \left(e^{k \left| y\right| }-1\right) 
\left(2 e^{k \left| y\right| }-3\right)}{\left[k^2r^2+\left(e^{k|y|}-1\right)^2\right]^{5/2}}-\frac{2 k Q^2 \left(e^{k \left| y\right| }
-1\right)^2}{ \left[k^2r^2+\left(e^{k|y|}-1\right)^2\right]^{3}}\right\}\,,\\[5mm]
T^{(B)\theta}{}_\theta&=T^{(B)\varphi}{}_\varphi=p_\theta(r,y)=\frac{1}{\kappa_5^2}\left\{2(3k^2-\Lambda)+\frac{6Mk^3\left(e^{k|y|}-1\right)
\left(2-e^{k|y|}\right)}{\left[k^2r^2+\left(e^{k|y|}-1\right)^2\right]^{3/2}}+\frac{6Q^2k^4\left(e^{k|y|}-1\right)^2}{\left[k^2r^2
+\left(e^{k|y|}-1\right)^2\right]^{2}} \right\}\,,\\[5mm]
T^{(B)y}{}_y&=\frac{1}{\kappa_5^2}\left\{2(3k^2-\Lambda)+ \frac{3k^3M\left(e^{k \left| y\right| }-1\right) \left[e^{k \left| y\right| } 
\left(3 e^{k \left| y\right| }-2 k^2 r^2-7\right)+4\left(1+ k^2   r^2\right)\right]}{\left[k^2r^2+\left(e^{k|y|}-1\right)^2\right]^{5/2}}\right.\nonum\\[2mm]
&\hspace{2.73cm}\left.+\frac{6 k^4 Q^2 \left(e^{k \left| y\right| }-1\right)^2\left(1+k^2r^2-e^{k \left| y\right| }\right)}{\left[k^2r^2+\left(e^{k|y|}-1\right)^2\right]^{3}}\right\}\,.$}

\end{appendices}



\clearpage
\phantomsection
\addcontentsline{toc}{chapter}{Bibliography}
{\def\chapter*#1{}
\thispagestyle{empty}
\vspace*{4em}
\noindent{\chapc{\rule{16.5cm}{0.5mm}\\[0.5em]
{\fontsize{30}{32} \textbf{Bibliography}}\\[0.25em]
\rule{16.5cm}{0.5mm}\vspace*{2em}}}
\bibliography{Bibliography}{}

\providecommand{\href}[2]{#2}\begingroup\raggedright\begin{thebibliography}{100}

\bibitem{KNP1}
P.~Kanti, T.~Nakas, and N.~Pappas, ``{Antigravitating braneworld solutions for
  a de Sitter brane in scalar-tensor gravity},''
  \href{http://dx.doi.org/10.1103/PhysRevD.98.064025}{{\em Phys. Rev. D}
  {\bfseries 98} no.~6, (2018) 064025},
  \href{http://arxiv.org/abs/1807.06880}{{\ttfamily \footnotesize
  arXiv:1807.06880 [gr-qc]}}.

\bibitem{KNP2}
T.~Nakas, N.~Pappas, and P.~Kanti, ``{New Black-String Solutions for an Anti-de
  Sitter Brane in Scalar-Tensor Gravity},''
  \href{http://dx.doi.org/10.1103/PhysRevD.99.124040}{{\em Phys. Rev. D}
  {\bfseries 99} no.~12, (2019) 124040},
  \href{http://arxiv.org/abs/1904.00216}{{\ttfamily \footnotesize
  arXiv:1904.00216 [hep-th]}}.

\bibitem{KNP3}
T.~Nakas, P.~Kanti, and N.~Pappas, ``{Incorporating Physical Constraints in
  Braneworld Black-String Solutions for a Minkowski Brane in Scalar-Tensor
  Gravity},'' \href{http://dx.doi.org/10.1103/PhysRevD.101.084056}{{\em Phys.
  Rev. D} {\bfseries 101} no.~8, (2020) 084056},
  \href{http://arxiv.org/abs/2001.07226}{{\ttfamily \footnotesize
  arXiv:2001.07226 [hep-th]}}.

\bibitem{NK1}
T.~Nakas and P.~Kanti, ``{Localized brane-world black hole analytically
  connected to an AdS$_5$ boundary},''
  \href{http://dx.doi.org/10.1016/j.physletb.2021.136278}{{\em Phys. Lett. B}
  {\bfseries 816} (2021) 136278},
  \href{http://arxiv.org/abs/2012.09199}{{\ttfamily \footnotesize
  arXiv:2012.09199 [hep-th]}}.

\bibitem{NK2}
T.~Nakas and P.~Kanti, ``{Analytic and exponentially localized braneworld
  Reissner-Nordstr\"om-AdS solution: A top-down approach},''
  \href{http://dx.doi.org/10.1103/PhysRevD.104.104037}{{\em Phys. Rev. D}
  {\bfseries 104} no.~10, (2021) 104037},
  \href{http://arxiv.org/abs/2105.06915}{{\ttfamily \footnotesize
  arXiv:2105.06915 [hep-th]}}.

\bibitem{NR}
T.~Nakas and K.~S. Rigatos, ``{Fermions and baryons as open-string states from
  brane junctions},'' \href{http://dx.doi.org/10.1007/JHEP12(2020)157}{{\em
  JHEP} {\bfseries 12} (2020) 157},
  \href{http://arxiv.org/abs/2010.00025}{{\ttfamily \footnotesize
  arXiv:2010.00025 [hep-th]}}.

\bibitem{BN}
A.~Bakopoulos and T.~Nakas, ``{Analytic and asymptotically flat hairy
  (ultra-compact) black-hole solutions and their axial perturbations},''
  \href{http://dx.doi.org/10.1007/JHEP04(2022)096}{{\em JHEP} {\bfseries 04}
  (2022) 096}, \href{http://arxiv.org/abs/2107.05656}{{\ttfamily \footnotesize
  arXiv:2107.05656 [gr-qc]}}.

\bibitem{einstein1}
A.~{Einstein}, ``{Zur allgemeinen Relativit{\"a}tstheorie},'' {\em
  Sitzungsberichte der K{\"o}niglich Preu{\ss}ischen Akademie der
  Wissenschaften (Berlin)} (Jan, 1915) 778--786.

\bibitem{einstein2}
A.~{Einstein}, ``{Die Feldgleichungen der Gravitation},'' {\em Sitzungsberichte
  der K{\"o}niglich Preu{\ss}ischen Akademie der Wissenschaften (Berlin)} (Jan,
  1915) 844--847.

\bibitem{einstein3}
A.~{Einstein}, ``{Die Grundlage der allgemeinen Relativit{\"a}tstheorie},''
  \href{http://dx.doi.org/10.1002/andp.19163540702}{{\em Annalen der Physik}
  {\bfseries 354} no.~7, (Jan, 1916) 769--822}.

\bibitem{MTW}
C.~W. {Misner}, K.~S. {Thorne}, and J.~A. {Wheeler}, {\em
  \href{https://www.amazon.com/Gravitation-Charles-W-Misner/dp/0691177791}{Gravitation}}.
\newblock 1973.

\bibitem{carroll:2019}
S.~M. Carroll, \href{http://dx.doi.org/10.1017/9781108770385}{{\em Spacetime
  and Geometry: An Introduction to General Relativity}}.
\newblock Cambridge University Press, 2019.

\bibitem{Groen:2007zz}
O.~Groen and S.~Hervik, {\em
  {\href{http://www.springer.com/978-0-387-69199-2}{Einstein's general theory
  of relativity: With modern applications in cosmology}}}.
\newblock
2007.
\newblock

\bibitem{Abbott:2016blz}
{\bfseries LIGO Scientific, Virgo} Collaboration, B.~P. Abbott {\em et~al.},
  ``{Observation of Gravitational Waves from a Binary Black Hole Merger},''
  \href{http://dx.doi.org/10.1103/PhysRevLett.116.061102}{{\em Phys. Rev.
  Lett.} {\bfseries 116} no.~6, (2016) 061102},
  \href{http://arxiv.org/abs/1602.03837}{{\ttfamily \footnotesize
  arXiv:1602.03837 [gr-qc]}}.

\bibitem{LIGOScientific:2018mvr}
{\bfseries LIGO Scientific, Virgo} Collaboration, B.~P. Abbott {\em et~al.},
  ``{GWTC-1: A Gravitational-Wave Transient Catalog of Compact Binary Mergers
  Observed by LIGO and Virgo during the First and Second Observing Runs},''
  \href{http://dx.doi.org/10.1103/PhysRevX.9.031040}{{\em Phys. Rev. X}
  {\bfseries 9} no.~3, (2019) 031040},
  \href{http://arxiv.org/abs/1811.12907}{{\ttfamily \footnotesize
  arXiv:1811.12907 [astro-ph.HE]}}.

\bibitem{Abbott:2020uma}
{\bfseries LIGO Scientific, Virgo} Collaboration, B.~P. Abbott {\em et~al.},
  ``{GW190425: Observation of a Compact Binary Coalescence with Total Mass
  $\sim 3.4 M_{\odot}$},''
  \href{http://dx.doi.org/10.3847/2041-8213/ab75f5}{{\em Astrophys. J. Lett.}
  {\bfseries 892} no.~1, (2020) L3},
  \href{http://arxiv.org/abs/2001.01761}{{\ttfamily \footnotesize
  arXiv:2001.01761 [astro-ph.HE]}}.

\bibitem{Akiyama:2019cqa}
{\bfseries Event Horizon Telescope} Collaboration, K.~Akiyama {\em et~al.},
  ``{First M87 Event Horizon Telescope Results. I. The Shadow of the
  Supermassive Black Hole},''
  \href{http://dx.doi.org/10.3847/2041-8213/ab0ec7}{{\em Astrophys. J. Lett.}
  {\bfseries 875} (2019) L1}, \href{http://arxiv.org/abs/1906.11238}{{\ttfamily
  \footnotesize arXiv:1906.11238 [astro-ph.GA]}}.

\bibitem{RS1}
L.~Randall and R.~Sundrum, ``{A Large mass hierarchy from a small extra
  dimension},'' \href{http://dx.doi.org/10.1103/PhysRevLett.83.3370}{{\em Phys.
  Rev. Lett.} {\bfseries 83} (1999) 3370--3373},
\href{http://arxiv.org/abs/hep-ph/9905221}{{\ttfamily \footnotesize
  arXiv:hep-ph/9905221 [hep-ph]}}.

\bibitem{RS2}
L.~Randall and R.~Sundrum, ``{An Alternative to compactification},''
  \href{http://dx.doi.org/10.1103/PhysRevLett.83.4690}{{\em Phys. Rev. Lett.}
  {\bfseries 83} (1999) 4690--4693},
\href{http://arxiv.org/abs/hep-th/9906064}{{\ttfamily \footnotesize
  arXiv:hep-th/9906064 [hep-th]}}.

\bibitem{EinsteinSR}
A.~Einstein, ``{On the electrodynamics of moving bodies},''
  \href{http://dx.doi.org/10.1002/andp.200590006}{{\em Annalen Phys.}
  {\bfseries 17} (1905) 891--921}.

\bibitem{Lovelock:1972vz}
D.~Lovelock, ``{The four-dimensionality of space and the einstein tensor},''
  \href{http://dx.doi.org/10.1063/1.1666069}{{\em J. Math. Phys.} {\bfseries
  13} (1972) 874--876}.

\bibitem{Nakas}
T.~Nakas, ``{Searching for Localized Black-Hole solutions in Brane-World
  models},'' Master's thesis, Ioannina U., 2017.

\bibitem{Lovelock:1971yv}
D.~Lovelock, ``{The Einstein tensor and its generalizations},''
  \href{http://dx.doi.org/10.1063/1.1665613}{{\em J. Math. Phys.} {\bfseries
  12} (1971) 498--501}.

\bibitem{Padmanabhan:2013xyr}
T.~Padmanabhan and D.~Kothawala, ``{Lanczos-Lovelock models of gravity},''
  \href{http://dx.doi.org/10.1016/j.physrep.2013.05.007}{{\em Phys. Rept.}
  {\bfseries 531} (2013) 115--171},
  \href{http://arxiv.org/abs/1302.2151}{{\ttfamily \footnotesize
  arXiv:1302.2151 [gr-qc]}}.

\bibitem{Buchdahl:1970ynr}
H.~A. Buchdahl, ``{Non-linear Lagrangians and cosmological theory},'' {\em Mon.
  Not. Roy. Astron. Soc.} {\bfseries 150} (1970) 1.

\bibitem{DeFelice:2010aj}
A.~De~Felice and S.~Tsujikawa, ``{f(R) theories},''
  \href{http://dx.doi.org/10.12942/lrr-2010-3}{{\em Living Rev. Rel.}
  {\bfseries 13} (2010) 3}, \href{http://arxiv.org/abs/1002.4928}{{\ttfamily
  \footnotesize arXiv:1002.4928 [gr-qc]}}.

\bibitem{Brans:1961sx}
C.~Brans and R.~H. Dicke, ``{Mach's principle and a relativistic theory of
  gravitation},'' \href{http://dx.doi.org/10.1103/PhysRev.124.925}{{\em Phys.
  Rev.} {\bfseries 124} (1961) 925--935}.

\bibitem{Horndeski:1974wa}
G.~W. Horndeski, ``{Second-order scalar-tensor field equations in a
  four-dimensional space},'' \href{http://dx.doi.org/10.1007/BF01807638}{{\em
  Int. J. Theor. Phys.} {\bfseries 10} (1974) 363--384}.

\bibitem{Kobayashi:2019hrl}
T.~Kobayashi, ``{Horndeski theory and beyond: a review},''
  \href{http://dx.doi.org/10.1088/1361-6633/ab2429}{{\em Rept. Prog. Phys.}
  {\bfseries 82} no.~8, (2019) 086901},
  \href{http://arxiv.org/abs/1901.07183}{{\ttfamily \footnotesize
  arXiv:1901.07183 [gr-qc]}}.

\bibitem{Schwarzschild:1916uq}
K.~Schwarzschild, ``{On the gravitational field of a mass point according to
  Einstein's theory},'' {\em Sitzungsber. Preuss. Akad. Wiss. Berlin (Math.
  Phys. )} {\bfseries 1916} (1916) 189--196,
  \href{http://arxiv.org/abs/physics/9905030}{{\ttfamily \footnotesize
  arXiv:physics/9905030}}.

\bibitem{Eddington}
A.~S.~S. Eddington, ``A {C}omparison of {W}hitehead's and {E}instein's
  {F}ormul{\ae},'' {\em Nature} {\bfseries 113} 192--192.

\bibitem{Finkelstein}
D.~Finkelstein, ``{Past-Future Asymmetry of the Gravitational Field of a Point
  Particle},'' \href{http://dx.doi.org/10.1103/PhysRev.110.965}{{\em Phys.
  Rev.} {\bfseries 110} (1958) 965--967}.

\bibitem{Penrose}
R.~Penrose, ``{Gravitational collapse and space-time singularities},''
  \href{http://dx.doi.org/10.1103/PhysRevLett.14.57}{{\em Phys. Rev. Lett.}
  {\bfseries 14} (1965) 57--59}.

\bibitem{Kruskal}
M.~D. Kruskal, ``{Maximal extension of Schwarzschild metric},''
  \href{http://dx.doi.org/10.1103/PhysRev.119.1743}{{\em Phys. Rev.} {\bfseries
  119} (1960) 1743--1745}.

\bibitem{Szekeres}
G.~Szekeres, ``{On the singularities of a Riemannian manifold},'' {\em Publ.
  Math. Debrecen} {\bfseries 7} (1960) 285--301.

\bibitem{Nariai}
H.~{Nariai}, ``{On some static solutions of Einstein's gravitational field
  equations in a spherically symmetric case},'' {\em Sci. Rep. Tohoku Univ.
  Eighth Ser.} {\bfseries 34} (Jan., 1950) 160.

\bibitem{Reissner}
H.~Reissner, ``Über die eigengravitation des elektrischen feldes nach der
  einsteinschen theorie,''
  \href{http://dx.doi.org/https://doi.org/10.1002/andp.19163550905}{{\em
  Annalen der Physik} {\bfseries 355} no.~9, (1916) 106--120}.

\bibitem{Nordstrom}
G.~{Nordstr\"{o}m}, ``{On the Energy of the Gravitation field in Einstein's
  Theory},'' {\em Koninklijke Nederlandse Akademie van Wetenschappen
  Proceedings Series B Physical Sciences} {\bfseries 20} (Jan., 1918)
  1238--1245.

\bibitem{Kaluza:1921tu}
T.~Kaluza, ``{Zum Unit\"atsproblem der Physik},''
  \href{http://dx.doi.org/10.1142/S0218271818700017}{{\em Sitzungsber. Preuss.
  Akad. Wiss. Berlin (Math. Phys. )} {\bfseries 1921} (1921) 966--972},
  \href{http://arxiv.org/abs/1803.08616}{{\ttfamily \footnotesize
  arXiv:1803.08616 [physics.hist-ph]}}.

\bibitem{Klein1}
O.~Klein, ``Quantentheorie und f{\"u}nfdimensionale relativit{\"a}tstheorie.
  ({German}) [{Quantum} {Theory} and {Five-Dimensional} {Theory} of
  {Relativity},'' \href{http://dx.doi.org/10.1007/BF01397481}{{\em Zeitschrift
  f{\"u}r Physik} {\bfseries 37} no.~12, (1926) 895--906}.

\bibitem{Klein2}
O.~Klein, ``{The Atomicity of Electricity as a Quantum Theory Law},''
\href{http://dx.doi.org/10.1038/118516a0}{{\em Nature} {\bfseries 118} (1926)
  516}.

\bibitem{P.Jordan}
P.~{Jordan}, ``{Erweiterung der projektiven Relativit{\"a}tstheorie},''
  \href{http://dx.doi.org/10.1002/andp.19474360409}{{\em Annalen der Physik}
  {\bfseries 436} no.~4, (Jan., 1947) 219--228}.

\bibitem{Thiry}
Y.~Thiry, ``{The equations of Kaluza's unified theory},'' {\em Compt. Rend.
  Hebd. Seances Acad. Sci.} {\bfseries 226} no.~3, (1948) 216--218.

\bibitem{Brans-Dicke}
C.~Brans and R.~H. Dicke, ``{Mach's principle and a relativistic theory of
  gravitation},'' \href{http://dx.doi.org/10.1103/PhysRev.124.925}{{\em Phys.
  Rev.} {\bfseries 124} (1961) 925--935}.

\bibitem{Overduin}
J.~M. Overduin and P.~S. Wesson, ``{Kaluza-Klein gravity},''
  \href{http://dx.doi.org/10.1016/S0370-1573(96)00046-4}{{\em Phys. Rept.}
  {\bfseries 283} (1997) 303--380},
  \href{http://arxiv.org/abs/gr-qc/9805018}{{\ttfamily \footnotesize
  arXiv:gr-qc/9805018}}.

\bibitem{Green}
M.~B. Green, J.~H. Schwarz, and E.~Witten, {\em
  {\href{http://www.cambridge.org/us/academic/subjects/physics/theoretical-physics-and-mathematical-physics/superstring-theory-volume-1}{SUPERSTRING
  THEORY. VOL. 1: INTRODUCTION}}}.
\newblock Cambridge Monographs on Mathematical Physics. Cambridge University
  Press,
1988.
\newblock

\bibitem{Polchinski}
J.~Polchinski, \href{http://dx.doi.org/10.1017/CBO9780511816079}{{\em {String
  theory. Vol. 1: An introduction to the bosonic string}}}.
\newblock Cambridge Monographs on Mathematical Physics. Cambridge University
  Press,
2007.
\newblock

\bibitem{Liu-Yao}
C.-H. Liu and S.-T. Yau, ``{D-branes and Azumaya/matrix noncommutative
  differential geometry,II: Azumaya/matrix supermanifolds and differentiable
  maps therefrom -- with a view toward dynamical fermionic D-branes in string
  theory},'' \href{http://arxiv.org/abs/1412.0771}{{\ttfamily \footnotesize
  arXiv:1412.0771 [hep-th]}}.

\bibitem{Witten81}
E.~Witten, ``{Search for a Realistic Kaluza-Klein Theory},''
  \href{http://dx.doi.org/10.1016/0550-3213(81)90021-3}{{\em Nucl. Phys. B}
  {\bfseries 186} (1981) 412}.

\bibitem{ADD1}
N.~Arkani-Hamed, S.~Dimopoulos, and G.~R. Dvali, ``{The Hierarchy problem and
  new dimensions at a millimeter},''
  \href{http://dx.doi.org/10.1016/S0370-2693(98)00466-3}{{\em Phys. Lett. B}
  {\bfseries 429} (1998) 263--272},
  \href{http://arxiv.org/abs/hep-ph/9803315}{{\ttfamily \footnotesize
  arXiv:hep-ph/9803315}}.

\bibitem{ADD2}
N.~Arkani-Hamed, S.~Dimopoulos, and G.~R. Dvali, ``{Phenomenology, astrophysics
  and cosmology of theories with submillimeter dimensions and TeV scale quantum
  gravity},'' \href{http://dx.doi.org/10.1103/PhysRevD.59.086004}{{\em Phys.
  Rev.} {\bfseries D59} (1999) 086004},
\href{http://arxiv.org/abs/hep-ph/9807344}{{\ttfamily \footnotesize
  arXiv:hep-ph/9807344 [hep-ph]}}.

\bibitem{ADD3}
I.~Antoniadis, N.~Arkani-Hamed, S.~Dimopoulos, and G.~R. Dvali, ``{New
  dimensions at a millimeter to a Fermi and superstrings at a TeV},''
  \href{http://dx.doi.org/10.1016/S0370-2693(98)00860-0}{{\em Phys. Lett.}
  {\bfseries B436} (1998) 257--263},
\href{http://arxiv.org/abs/hep-ph/9804398}{{\ttfamily \footnotesize
  arXiv:hep-ph/9804398 [hep-ph]}}.

\bibitem{Kehagias:1999my}
A.~Kehagias and K.~Sfetsos, ``{Deviations from the 1/r**2 Newton law due to
  extra dimensions},''
  \href{http://dx.doi.org/10.1016/S0370-2693(99)01421-5}{{\em Phys. Lett.}
  {\bfseries B472} (2000) 39--44},
\href{http://arxiv.org/abs/hep-ph/9905417}{{\ttfamily \footnotesize
  arXiv:hep-ph/9905417 [hep-ph]}}.

\bibitem{Floratos:1999bv}
E.~G. Floratos and G.~K. Leontaris, ``{Low scale unification, Newton's law and
  extra dimensions},''
  \href{http://dx.doi.org/10.1016/S0370-2693(99)01019-9}{{\em Phys. Lett. B}
  {\bfseries 465} (1999) 95--100},
  \href{http://arxiv.org/abs/hep-ph/9906238}{{\ttfamily \footnotesize
  arXiv:hep-ph/9906238}}.

\bibitem{Kapner}
D.~J. Kapner, T.~S. Cook, E.~G. Adelberger, J.~H. Gundlach, B.~R. Heckel, C.~D.
  Hoyle, and H.~E. Swanson, ``{Tests of the gravitational inverse-square law
  below the dark-energy length scale},''
  \href{http://dx.doi.org/10.1103/PhysRevLett.98.021101}{{\em Phys. Rev. Lett.}
  {\bfseries 98} (2007) 021101},
\href{http://arxiv.org/abs/hep-ph/0611184}{{\ttfamily \footnotesize
  arXiv:hep-ph/0611184 [hep-ph]}}.

\bibitem{Lee:2020zjt}
J.~G. Lee, E.~G. Adelberger, T.~S. Cook, S.~M. Fleischer, and B.~R. Heckel,
  ``{New Test of the Gravitational $1/r^2$ Law at Separations down to 52
  $\mu$m},'' \href{http://dx.doi.org/10.1103/PhysRevLett.124.101101}{{\em Phys.
  Rev. Lett.} {\bfseries 124} no.~10, (2020) 101101},
  \href{http://arxiv.org/abs/2002.11761}{{\ttfamily \footnotesize
  arXiv:2002.11761 [hep-ex]}}.

\bibitem{Franc}
R.~Franceschini, P.~P. Giardino, G.~F. Giudice, P.~Lodone, and A.~Strumia,
  ``{LHC bounds on large extra dimensions},''
  \href{http://dx.doi.org/10.1007/JHEP05(2011)092}{{\em JHEP} {\bfseries 05}
  (2011) 092},
\href{http://arxiv.org/abs/1101.4919}{{\ttfamily \footnotesize arXiv:1101.4919
  [hep-ph]}}.

\bibitem{Argyres:1998qn}
P.~C. Argyres, S.~Dimopoulos, and J.~March-Russell, ``{Black holes and
  submillimeter dimensions},''
  \href{http://dx.doi.org/10.1016/S0370-2693(98)01184-8}{{\em Phys. Lett.}
  {\bfseries B441} (1998) 96--104},
\href{http://arxiv.org/abs/hep-th/9808138}{{\ttfamily \footnotesize
  arXiv:hep-th/9808138 [hep-th]}}.

\bibitem{Banks:1999gd}
T.~Banks and W.~Fischler, ``{A Model for high-energy scattering in quantum
  gravity},''
\href{http://arxiv.org/abs/hep-th/9906038}{{\ttfamily \footnotesize
  arXiv:hep-th/9906038 [hep-th]}}.

\bibitem{Dimopoulos:2001hw}
S.~Dimopoulos and G.~L. Landsberg, ``{Black holes at the LHC},''
  \href{http://dx.doi.org/10.1103/PhysRevLett.87.161602}{{\em Phys. Rev. Lett.}
  {\bfseries 87} (2001) 161602},
\href{http://arxiv.org/abs/hep-ph/0106295}{{\ttfamily \footnotesize
  arXiv:hep-ph/0106295 [hep-ph]}}.

\bibitem{Giddings:2001bu}
S.~B. Giddings and S.~D. Thomas, ``{High-energy colliders as black hole
  factories: The End of short distance physics},''
  \href{http://dx.doi.org/10.1103/PhysRevD.65.056010}{{\em Phys. Rev. D}
  {\bfseries 65} (2002) 056010},
  \href{http://arxiv.org/abs/hep-ph/0106219}{{\ttfamily \footnotesize
  arXiv:hep-ph/0106219}}.

\bibitem{Kanti:2004nr}
P.~Kanti, ``{Black holes in theories with large extra dimensions: A Review},''
  \href{http://dx.doi.org/10.1142/S0217751X04018324}{{\em Int. J. Mod. Phys.}
  {\bfseries A19} (2004) 4899--4951},
\href{http://arxiv.org/abs/hep-ph/0402168}{{\ttfamily \footnotesize
  arXiv:hep-ph/0402168 [hep-ph]}}.

\bibitem{Kanti:2008eq}
P.~Kanti, ``{Black Holes at the LHC},''
  \href{http://dx.doi.org/10.1007/978-3-540-88460-6_10}{{\em Lect. Notes Phys.}
  {\bfseries 769} (2009) 387--423},
\href{http://arxiv.org/abs/0802.2218}{{\ttfamily \footnotesize arXiv:0802.2218
  [hep-th]}}.

\bibitem{ATLAS:2015yln}
{\bfseries ATLAS} Collaboration, G.~Aad {\em et~al.}, ``{Search for strong
  gravity in multijet final states produced in pp collisions at $\sqrt{s} =$ 13
  TeV using the ATLAS detector at the LHC},''
  \href{http://dx.doi.org/10.1007/JHEP03(2016)026}{{\em JHEP} {\bfseries 03}
  (2016) 026}, \href{http://arxiv.org/abs/1512.02586}{{\ttfamily \footnotesize
  arXiv:1512.02586 [hep-ex]}}.

\bibitem{Goldberger-Wise1}
W.~D. Goldberger and M.~B. Wise, ``{Modulus stabilization with bulk fields},''
  \href{http://dx.doi.org/10.1103/PhysRevLett.83.4922}{{\em Phys. Rev. Lett.}
  {\bfseries 83} (1999) 4922--4925},
  \href{http://arxiv.org/abs/hep-ph/9907447}{{\ttfamily \footnotesize
  arXiv:hep-ph/9907447}}.

\bibitem{Goldberger-Wise2}
W.~D. Goldberger and M.~B. Wise, ``{Phenomenology of a stabilized modulus},''
  \href{http://dx.doi.org/10.1016/S0370-2693(00)00099-X}{{\em Phys. Lett. B}
  {\bfseries 475} (2000) 275--279},
  \href{http://arxiv.org/abs/hep-ph/9911457}{{\ttfamily \footnotesize
  arXiv:hep-ph/9911457}}.

\bibitem{Ivanov:1999mt}
M.~G. Ivanov and I.~V. Volovich, ``{Metric fluctuations in brane worlds},''
  \href{http://arxiv.org/abs/hep-th/9912242}{{\ttfamily \footnotesize
  arXiv:hep-th/9912242}}.

\bibitem{Myung:2000hu}
Y.~S. Myung, G.~Kang, and H.~W. Lee, ``{Randall-Sundrum gauge in the brane
  world},'' \href{http://dx.doi.org/10.1016/S0370-2693(00)00277-X}{{\em Phys.
  Lett. B} {\bfseries 478} (2000) 294--298},
  \href{http://arxiv.org/abs/hep-th/0001107}{{\ttfamily \footnotesize
  arXiv:hep-th/0001107}}.

\bibitem{Garriga:1999yh}
J.~Garriga and T.~Tanaka, ``{Gravity in the brane world},''
  \href{http://dx.doi.org/10.1103/PhysRevLett.84.2778}{{\em Phys. Rev. Lett.}
  {\bfseries 84} (2000) 2778--2781},
  \href{http://arxiv.org/abs/hep-th/9911055}{{\ttfamily \footnotesize
  arXiv:hep-th/9911055}}.

\bibitem{Giddings:2000mu}
S.~B. Giddings, E.~Katz, and L.~Randall, ``{Linearized gravity in brane
  backgrounds},'' \href{http://dx.doi.org/10.1088/1126-6708/2000/03/023}{{\em
  JHEP} {\bfseries 03} (2000) 023},
  \href{http://arxiv.org/abs/hep-th/0002091}{{\ttfamily \footnotesize
  arXiv:hep-th/0002091}}.

\bibitem{Carter}
B.~Carter, ``{Axisymmetric Black Hole Has Only Two Degrees of Freedom},''
  \href{http://dx.doi.org/10.1103/PhysRevLett.26.331}{{\em Phys. Rev. Lett.}
  {\bfseries 26} (1971) 331--333}.

\bibitem{Misha}
V.~A. Rubakov and M.~E. Shaposhnikov, ``{Do We Live Inside a Domain Wall?},''
\href{http://dx.doi.org/10.1016/0370-2693(83)91253-4}{{\em Phys. Lett.}
  {\bfseries 125B} (1983) 136--138}.

\bibitem{Akama}
K.~Akama, ``{An Early Proposal of 'Brane World'},'' {\em Lect. Notes Phys.}
  {\bfseries 176} (1982) 267--271,
  \href{http://arxiv.org/abs/hep-th/0001113}{{\ttfamily \footnotesize
  arXiv:hep-th/0001113}}.

\bibitem{Charmousis1}
C.~Bogdanos, C.~Charmousis, B.~Gouteraux, and R.~Zegers,
  ``{Einstein-Gauss-Bonnet metrics: Black holes, black strings and a staticity
  theorem},'' \href{http://dx.doi.org/10.1088/1126-6708/2009/10/037}{{\em JHEP}
  {\bfseries 10} (2009) 037},
\href{http://arxiv.org/abs/0906.4953}{{\ttfamily \footnotesize arXiv:0906.4953
  [hep-th]}}.

\bibitem{Charmousis2}
C.~Charmousis, T.~Kolyvaris, and E.~Papantonopoulos, ``{Charged C-metric with
  conformally coupled scalar field},''
  \href{http://dx.doi.org/10.1088/0264-9381/26/17/175012}{{\em Class. Quant.
  Grav.} {\bfseries 26} (2009) 175012},
\href{http://arxiv.org/abs/0906.5568}{{\ttfamily \footnotesize arXiv:0906.5568
  [gr-qc]}}.

\bibitem{Cisterna1}
A.~Cisterna and J.~Oliva, ``{Exact black strings and p-branes in general
  relativity},'' \href{http://dx.doi.org/10.1088/1361-6382/aa9f7b}{{\em Class.
  Quant. Grav.} {\bfseries 35} no.~3, (2018) 035012},
\href{http://arxiv.org/abs/1708.02916}{{\ttfamily \footnotesize
  arXiv:1708.02916 [hep-th]}}.

\bibitem{Cisterna2}
A.~Cisterna, C.~Corral, and S.~del Pino, ``{Static and rotating black strings
  in dynamical Chern–Simons modified gravity},''
  \href{http://dx.doi.org/10.1140/epjc/s10052-019-6910-5}{{\em Eur. Phys. J.}
  {\bfseries C79} no.~5, (2019) 400},
\href{http://arxiv.org/abs/1809.02903}{{\ttfamily \footnotesize
  arXiv:1809.02903 [gr-qc]}}.

\bibitem{CFLO}
A.~Cisterna, S.~Fuenzalida, M.~Lagos, and J.~Oliva, ``{Homogeneous black
  strings in Einstein–Gauss–Bonnet with Horndeski hair and beyond},''
  \href{http://dx.doi.org/10.1140/epjc/s10052-018-6428-2}{{\em Eur. Phys. J.}
  {\bfseries C78} no.~11, (2018) 982},
\href{http://arxiv.org/abs/1810.02798}{{\ttfamily \footnotesize
  arXiv:1810.02798 [hep-th]}}.

\bibitem{Rezvanjou}
S.~Rezvanjou, R.~Saffari, and M.~Masoudi, ``{Particle Dynamics Around the Black
  String},''
\href{http://arxiv.org/abs/1707.02817}{{\ttfamily \footnotesize
  arXiv:1707.02817 [gr-qc]}}.

\bibitem{Estrada}
M.~Estrada, ``{A new exact solution of black-strings-like with a dS core},''
  \href{http://arxiv.org/abs/2102.08222}{{\ttfamily \footnotesize
  arXiv:2102.08222 [gr-qc]}}.

\bibitem{Cisterna2021}
A.~Cisterna, C.~Henríquez-Báez, N.~Mora, and L.~Sanhueza, ``{Quasitopological
  electromagnetism: Reissner-Nordström black strings in Einstein and Lovelock
  gravities},''
\href{http://arxiv.org/abs/2105.04239}{{\ttfamily \footnotesize
  arXiv:2105.04239 [gr-qc]}}.

\bibitem{Wiseman}
T.~Wiseman, ``{Static axisymmetric vacuum solutions and nonuniform black
  strings},'' \href{http://dx.doi.org/10.1088/0264-9381/20/6/308}{{\em Class.
  Quant. Grav.} {\bfseries 20} (2003) 1137--1176},
\href{http://arxiv.org/abs/hep-th/0209051}{{\ttfamily \footnotesize
  arXiv:hep-th/0209051 [hep-th]}}.

\bibitem{Kudoh2}
H.~Kudoh and T.~Wiseman, ``{Connecting black holes and black strings},''
  \href{http://dx.doi.org/10.1103/PhysRevLett.94.161102}{{\em Phys. Rev. Lett.}
  {\bfseries 94} (2005) 161102},
\href{http://arxiv.org/abs/hep-th/0409111}{{\ttfamily \footnotesize
  arXiv:hep-th/0409111 [hep-th]}}.

\bibitem{Sorkin1}
E.~Sorkin, ``{A Critical dimension in the black string phase transition},''
  \href{http://dx.doi.org/10.1103/PhysRevLett.93.031601}{{\em Phys. Rev. Lett.}
  {\bfseries 93} (2004) 031601},
\href{http://arxiv.org/abs/hep-th/0402216}{{\ttfamily \footnotesize
  arXiv:hep-th/0402216 [hep-th]}}.

\bibitem{Sorkin2}
E.~Sorkin, ``{Non-uniform black strings in various dimensions},''
  \href{http://dx.doi.org/10.1103/PhysRevD.74.104027}{{\em Phys. Rev.}
  {\bfseries D74} (2006) 104027},
\href{http://arxiv.org/abs/gr-qc/0608115}{{\ttfamily \footnotesize
  arXiv:gr-qc/0608115 [gr-qc]}}.

\bibitem{Kleihaus2}
B.~Kleihaus, J.~Kunz, and E.~Radu, ``{New nonuniform black string solutions},''
  \href{http://dx.doi.org/10.1088/1126-6708/2006/06/016}{{\em JHEP} {\bfseries
  06} (2006) 016},
\href{http://arxiv.org/abs/hep-th/0603119}{{\ttfamily \footnotesize
  arXiv:hep-th/0603119 [hep-th]}}.

\bibitem{Headrick}
M.~Headrick, S.~Kitchen, and T.~Wiseman, ``{A New approach to static numerical
  relativity, and its application to Kaluza-Klein black holes},''
  \href{http://dx.doi.org/10.1088/0264-9381/27/3/035002}{{\em Class. Quant.
  Grav.} {\bfseries 27} (2010) 035002},
\href{http://arxiv.org/abs/0905.1822}{{\ttfamily \footnotesize arXiv:0905.1822
  [gr-qc]}}.

\bibitem{Figueras2}
P.~Figueras, K.~Murata, and H.~S. Reall, ``{Stable non-uniform black strings
  below the critical dimension},''
  \href{http://dx.doi.org/10.1007/JHEP11(2012)071}{{\em JHEP} {\bfseries 11}
  (2012) 071},
\href{http://arxiv.org/abs/1209.1981}{{\ttfamily \footnotesize arXiv:1209.1981
  [gr-qc]}}.

\bibitem{Kalisch}
M.~Kalisch and M.~Ansorg, ``{Pseudo-spectral construction of non-uniform black
  string solutions in five and six spacetime dimensions},''
  \href{http://dx.doi.org/10.1088/0264-9381/33/21/215005}{{\em Class. Quant.
  Grav.} {\bfseries 33} no.~21, (2016) 215005},
\href{http://arxiv.org/abs/1607.03099}{{\ttfamily \footnotesize
  arXiv:1607.03099 [gr-qc]}}.

\bibitem{Emparan2}
R.~Emparan, R.~Luna, M.~Martínez, R.~Suzuki, and K.~Tanabe, ``{Phases and
  Stability of Non-Uniform Black Strings},''
  \href{http://dx.doi.org/10.1007/JHEP05(2018)104}{{\em JHEP} {\bfseries 05}
  (2018) 104},
\href{http://arxiv.org/abs/1802.08191}{{\ttfamily \footnotesize
  arXiv:1802.08191 [hep-th]}}.

\bibitem{Bazeia:2014tua}
D.~Bazeia, J.~M. Hoff~da Silva, and R.~da~Rocha, ``{Regular Bulk Solutions and
  Black Strings from Dynamical Braneworlds with Variable Tension},''
  \href{http://dx.doi.org/10.1103/PhysRevD.90.047902}{{\em Phys. Rev.}
  {\bfseries D90} no.~4, (2014) 047902},
\href{http://arxiv.org/abs/1401.6985}{{\ttfamily \footnotesize arXiv:1401.6985
  [hep-th]}}.

\bibitem{Emparan:2001wn}
R.~Emparan and H.~S. Reall, ``{A Rotating black ring solution in
  five-dimensions},''
  \href{http://dx.doi.org/10.1103/PhysRevLett.88.101101}{{\em Phys. Rev. Lett.}
  {\bfseries 88} (2002) 101101},
  \href{http://arxiv.org/abs/hep-th/0110260}{{\ttfamily \footnotesize
  arXiv:hep-th/0110260}}.

\bibitem{Elvang:2003mj}
H.~Elvang and R.~Emparan, ``{Black rings, supertubes, and a stringy resolution
  of black hole nonuniqueness},''
  \href{http://dx.doi.org/10.1088/1126-6708/2003/11/035}{{\em JHEP} {\bfseries
  11} (2003) 035}, \href{http://arxiv.org/abs/hep-th/0310008}{{\ttfamily
  \footnotesize arXiv:hep-th/0310008}}.

\bibitem{Elvang:2004rt}
H.~Elvang, R.~Emparan, D.~Mateos, and H.~S. Reall, ``{A Supersymmetric black
  ring},'' \href{http://dx.doi.org/10.1103/PhysRevLett.93.211302}{{\em Phys.
  Rev. Lett.} {\bfseries 93} (2004) 211302},
  \href{http://arxiv.org/abs/hep-th/0407065}{{\ttfamily \footnotesize
  arXiv:hep-th/0407065}}.

\bibitem{Bena:2004de}
I.~Bena and N.~P. Warner, ``{One ring to rule them all ... and in the darkness
  bind them?},'' \href{http://dx.doi.org/10.4310/ATMP.2005.v9.n5.a1}{{\em Adv.
  Theor. Math. Phys.} {\bfseries 9} no.~5, (2005) 667--701},
  \href{http://arxiv.org/abs/hep-th/0408106}{{\ttfamily \footnotesize
  arXiv:hep-th/0408106}}.

\bibitem{Emparan:2004wy}
R.~Emparan, ``{Rotating circular strings, and infinite nonuniqueness of black
  rings},'' \href{http://dx.doi.org/10.1088/1126-6708/2004/03/064}{{\em JHEP}
  {\bfseries 03} (2004) 064},
  \href{http://arxiv.org/abs/hep-th/0402149}{{\ttfamily \footnotesize
  arXiv:hep-th/0402149}}.

\bibitem{Gauntlett:2004wh}
J.~P. Gauntlett and J.~B. Gutowski, ``{Concentric black rings},''
  \href{http://dx.doi.org/10.1103/PhysRevD.71.025013}{{\em Phys. Rev. D}
  {\bfseries 71} (2005) 025013},
  \href{http://arxiv.org/abs/hep-th/0408010}{{\ttfamily \footnotesize
  arXiv:hep-th/0408010}}.

\bibitem{Arcioni:2004ww}
G.~Arcioni and E.~Lozano-Tellechea, ``{Stability and critical phenomena of
  black holes and black rings},''
  \href{http://dx.doi.org/10.1103/PhysRevD.72.104021}{{\em Phys. Rev. D}
  {\bfseries 72} (2005) 104021},
  \href{http://arxiv.org/abs/hep-th/0412118}{{\ttfamily \footnotesize
  arXiv:hep-th/0412118}}.

\bibitem{Bena:2005ni}
I.~Bena, P.~Kraus, and N.~P. Warner, ``{Black rings in Taub-NUT},''
  \href{http://dx.doi.org/10.1103/PhysRevD.72.084019}{{\em Phys. Rev. D}
  {\bfseries 72} (2005) 084019},
  \href{http://arxiv.org/abs/hep-th/0504142}{{\ttfamily \footnotesize
  arXiv:hep-th/0504142}}.

\bibitem{Elvang:2004ds}
H.~Elvang, R.~Emparan, D.~Mateos, and H.~S. Reall, ``{Supersymmetric black
  rings and three-charge supertubes},''
  \href{http://dx.doi.org/10.1103/PhysRevD.71.024033}{{\em Phys. Rev. D}
  {\bfseries 71} (2005) 024033},
  \href{http://arxiv.org/abs/hep-th/0408120}{{\ttfamily \footnotesize
  arXiv:hep-th/0408120}}.

\bibitem{Elvang:2005sa}
H.~Elvang, R.~Emparan, D.~Mateos, and H.~S. Reall, ``{Supersymmetric 4-D
  rotating black holes from 5-D black rings},''
  \href{http://dx.doi.org/10.1088/1126-6708/2005/08/042}{{\em JHEP} {\bfseries
  08} (2005) 042}, \href{http://arxiv.org/abs/hep-th/0504125}{{\ttfamily
  \footnotesize arXiv:hep-th/0504125}}.

\bibitem{Gaiotto:2005xt}
D.~Gaiotto, A.~Strominger, and X.~Yin, ``{5D black rings and 4D black holes},''
  \href{http://dx.doi.org/10.1088/1126-6708/2006/02/023}{{\em JHEP} {\bfseries
  02} (2006) 023}, \href{http://arxiv.org/abs/hep-th/0504126}{{\ttfamily
  \footnotesize arXiv:hep-th/0504126}}.

\bibitem{Bena:2005va}
I.~Bena and N.~P. Warner, ``{Bubbling supertubes and foaming black holes},''
  \href{http://dx.doi.org/10.1103/PhysRevD.74.066001}{{\em Phys. Rev. D}
  {\bfseries 74} (2006) 066001},
  \href{http://arxiv.org/abs/hep-th/0505166}{{\ttfamily \footnotesize
  arXiv:hep-th/0505166}}.

\bibitem{Emparan:2006mm}
R.~Emparan and H.~S. Reall, ``{Black Rings},''
  \href{http://dx.doi.org/10.1088/0264-9381/23/20/R01}{{\em Class. Quant.
  Grav.} {\bfseries 23} (2006) R169},
  \href{http://arxiv.org/abs/hep-th/0608012}{{\ttfamily \footnotesize
  arXiv:hep-th/0608012}}.

\bibitem{Emparan:2007wm}
R.~Emparan, T.~Harmark, V.~Niarchos, N.~A. Obers, and M.~J. Rodriguez, ``{The
  Phase Structure of Higher-Dimensional Black Rings and Black Holes},''
  \href{http://dx.doi.org/10.1088/1126-6708/2007/10/110}{{\em JHEP} {\bfseries
  10} (2007) 110}, \href{http://arxiv.org/abs/0708.2181}{{\ttfamily
  \footnotesize arXiv:0708.2181 [hep-th]}}.

\bibitem{Elvang:2007hs}
H.~Elvang and M.~J. Rodriguez, ``{Bicycling Black Rings},''
  \href{http://dx.doi.org/10.1088/1126-6708/2008/04/045}{{\em JHEP} {\bfseries
  04} (2008) 045}, \href{http://arxiv.org/abs/0712.2425}{{\ttfamily
  \footnotesize arXiv:0712.2425 [hep-th]}}.

\bibitem{Caldarelli:2008pz}
M.~M. Caldarelli, R.~Emparan, and M.~J. Rodriguez, ``{Black Rings in
  (Anti)-deSitter space},''
  \href{http://dx.doi.org/10.1088/1126-6708/2008/11/011}{{\em JHEP} {\bfseries
  11} (2008) 011}, \href{http://arxiv.org/abs/0806.1954}{{\ttfamily
  \footnotesize arXiv:0806.1954 [hep-th]}}.

\bibitem{Emparan-review}
R.~Emparan and H.~S. Reall, ``{Black Holes in Higher Dimensions},''
  \href{http://dx.doi.org/10.12942/lrr-2008-6}{{\em Living Rev. Rel.}
  {\bfseries 11} (2008) 6},
\href{http://arxiv.org/abs/0801.3471}{{\ttfamily \footnotesize arXiv:0801.3471
  [hep-th]}}.

\bibitem{Jiang:2008gq}
Q.-Q. Jiang, ``{Dirac particles' tunnelling from black rings},''
  \href{http://dx.doi.org/10.1103/PhysRevD.78.044009}{{\em Phys. Rev. D}
  {\bfseries 78} (2008) 044009},
  \href{http://arxiv.org/abs/0807.1358}{{\ttfamily \footnotesize
  arXiv:0807.1358 [hep-th]}}.

\bibitem{Bena:2009ev}
I.~Bena, G.~Dall'Agata, S.~Giusto, C.~Ruef, and N.~P. Warner, ``{Non-BPS Black
  Rings and Black Holes in Taub-NUT},''
  \href{http://dx.doi.org/10.1088/1126-6708/2009/06/015}{{\em JHEP} {\bfseries
  06} (2009) 015}, \href{http://arxiv.org/abs/0902.4526}{{\ttfamily
  \footnotesize arXiv:0902.4526 [hep-th]}}.

\bibitem{Mandlik:2020mxe}
M.~Mandlik, ``{de Sitter Static Black Ring in Large $D$ Membrane Paradigm at
  the Second Order},'' \href{http://arxiv.org/abs/2011.13351}{{\ttfamily
  \footnotesize arXiv:2011.13351 [hep-th]}}.

\bibitem{Ruiperez:2020qda}
A.~Ruip\'erez, ``{Higher-derivative corrections to small black rings},''
  \href{http://dx.doi.org/10.1088/1361-6382/abff9b}{{\em Class. Quant. Grav.}
  {\bfseries 38} no.~14, (2021) 145011},
  \href{http://arxiv.org/abs/2003.02269}{{\ttfamily \footnotesize
  arXiv:2003.02269 [hep-th]}}.

\bibitem{Ali:2020bor}
R.~Ali, K.~Bamba, M.~Asgher, and S.~A.~A. Shah, ``{Tunneling under the
  influence of quantum gravity in black rings},''
  \href{http://dx.doi.org/10.1142/S0218271821500024}{{\em Int. J. Mod. Phys. D}
  {\bfseries 30} no.~01, (2021) 2150002},
  \href{http://arxiv.org/abs/2101.07476}{{\ttfamily \footnotesize
  arXiv:2101.07476 [gr-qc]}}.

\bibitem{Guo:2021ikr}
S.-F. Guo, H.~L\"u, and Y.~Pang, ``{Degenerate black rings in D = 5 minimal
  supergravity},'' \href{http://dx.doi.org/10.1007/s11433-021-1761-8}{{\em Sci.
  China Phys. Mech. Astron.} {\bfseries 64} no.~11, (2021) 110411},
  \href{http://arxiv.org/abs/2106.12632}{{\ttfamily \footnotesize
  arXiv:2106.12632 [hep-th]}}.

\bibitem{Gubser}
S.~S. Gubser, ``{On nonuniform black branes},''
  \href{http://dx.doi.org/10.1088/0264-9381/19/19/303}{{\em Class. Quant.
  Grav.} {\bfseries 19} (2002) 4825--4844},
\href{http://arxiv.org/abs/hep-th/0110193}{{\ttfamily \footnotesize
  arXiv:hep-th/0110193 [hep-th]}}.

\bibitem{Kovtun:2003wp}
P.~Kovtun, D.~T. Son, and A.~O. Starinets, ``{Holography and hydrodynamics:
  Diffusion on stretched horizons},''
  \href{http://dx.doi.org/10.1088/1126-6708/2003/10/064}{{\em JHEP} {\bfseries
  10} (2003) 064}, \href{http://arxiv.org/abs/hep-th/0309213}{{\ttfamily
  \footnotesize arXiv:hep-th/0309213}}.

\bibitem{Banerjee:2008th}
N.~Banerjee, J.~Bhattacharya, S.~Bhattacharyya, S.~Dutta, R.~Loganayagam, and
  P.~Surowka, ``{Hydrodynamics from charged black branes},''
  \href{http://dx.doi.org/10.1007/JHEP01(2011)094}{{\em JHEP} {\bfseries 01}
  (2011) 094}, \href{http://arxiv.org/abs/0809.2596}{{\ttfamily \footnotesize
  arXiv:0809.2596 [hep-th]}}.

\bibitem{Berti:2009kk}
E.~Berti, V.~Cardoso, and A.~O. Starinets, ``{Quasinormal modes of black holes
  and black branes},''
  \href{http://dx.doi.org/10.1088/0264-9381/26/16/163001}{{\em Class. Quant.
  Grav.} {\bfseries 26} (2009) 163001},
  \href{http://arxiv.org/abs/0905.2975}{{\ttfamily \footnotesize
  arXiv:0905.2975 [gr-qc]}}.

\bibitem{Goldstein:2010aw}
K.~Goldstein, N.~Iizuka, S.~Kachru, S.~Prakash, S.~P. Trivedi, and A.~Westphal,
  ``{Holography of Dyonic Dilaton Black Branes},''
  \href{http://dx.doi.org/10.1007/JHEP10(2010)027}{{\em JHEP} {\bfseries 10}
  (2010) 027}, \href{http://arxiv.org/abs/1007.2490}{{\ttfamily \footnotesize
  arXiv:1007.2490 [hep-th]}}.

\bibitem{Tarrio:2011de}
J.~Tarrio and S.~Vandoren, ``{Black holes and black branes in Lifshitz
  spacetimes},'' \href{http://dx.doi.org/10.1007/JHEP09(2011)017}{{\em JHEP}
  {\bfseries 09} (2011) 017}, \href{http://arxiv.org/abs/1105.6335}{{\ttfamily
  \footnotesize arXiv:1105.6335 [hep-th]}}.

\bibitem{Alishahiha:2012qu}
M.~Alishahiha, E.~O~Colgain, and H.~Yavartanoo, ``{Charged Black Branes with
  Hyperscaling Violating Factor},''
  \href{http://dx.doi.org/10.1007/JHEP11(2012)137}{{\em JHEP} {\bfseries 11}
  (2012) 137}, \href{http://arxiv.org/abs/1209.3946}{{\ttfamily \footnotesize
  arXiv:1209.3946 [hep-th]}}.

\bibitem{Hollands:2012sf}
S.~Hollands and R.~M. Wald, ``{Stability of Black Holes and Black Branes},''
  \href{http://dx.doi.org/10.1007/s00220-012-1638-1}{{\em Commun. Math. Phys.}
  {\bfseries 321} (2013) 629--680},
  \href{http://arxiv.org/abs/1201.0463}{{\ttfamily \footnotesize
  arXiv:1201.0463 [gr-qc]}}.

\bibitem{Cisterna3}
A.~Cisterna, L.~Guajardo, and M.~Hassaine, ``{Axionic charged black branes with
  arbitrary scalar nonminimal coupling},''
  \href{http://dx.doi.org/10.1140/epjc/s10052-019-6922-1,
  10.1140/epjc/s10052-019-7215-4}{{\em Eur. Phys. J.} {\bfseries C79} no.~5,
  (2019) 418}, \href{http://arxiv.org/abs/1901.00514}{{\ttfamily \footnotesize
  arXiv:1901.00514 [hep-th]}}.
[Erratum: Eur. Phys. J.C79,no.8,710(2019)].

\bibitem{Tangherlini}
F.~R. Tangherlini, ``{Schwarzschild field in n dimensions and the
  dimensionality of space problem},''
\href{http://dx.doi.org/10.1007/BF02784569}{{\em Nuovo Cim.} {\bfseries 27}
  (1963) 636--651}.

\bibitem{MP}
R.~C. Myers and M.~J. Perry, ``{Black Holes in Higher Dimensional
  Space-Times},''
\href{http://dx.doi.org/10.1016/0003-4916(86)90186-7}{{\em Annals Phys.}
  {\bfseries 172} (1986) 304}.

\bibitem{CHR}
A.~Chamblin, S.~W. Hawking, and H.~S. Reall, ``{Brane world black holes},''
  \href{http://dx.doi.org/10.1103/PhysRevD.61.065007}{{\em Phys. Rev.}
  {\bfseries D61} (2000) 065007},
\href{http://arxiv.org/abs/hep-th/9909205}{{\ttfamily \footnotesize
  arXiv:hep-th/9909205 [hep-th]}}.

\bibitem{GL}
R.~Gregory and R.~Laflamme, ``{Black strings and p-branes are unstable},''
  \href{http://dx.doi.org/10.1103/PhysRevLett.70.2837}{{\em Phys. Rev. Lett.}
  {\bfseries 70} (1993) 2837--2840},
\href{http://arxiv.org/abs/hep-th/9301052}{{\ttfamily \footnotesize
  arXiv:hep-th/9301052 [hep-th]}}.

\bibitem{RuthGL}
R.~Gregory, ``{Black string instabilities in Anti-de Sitter space},''
  \href{http://dx.doi.org/10.1088/0264-9381/17/18/103}{{\em Class. Quant.
  Grav.} {\bfseries 17} (2000) L125--L132},
\href{http://arxiv.org/abs/hep-th/0004101}{{\ttfamily \footnotesize
  arXiv:hep-th/0004101 [hep-th]}}.

\bibitem{review1}
R.~Maartens and K.~Koyama, ``{Brane-World Gravity},''
  \href{http://dx.doi.org/10.12942/lrr-2010-5}{{\em Living Rev. Rel.}
  {\bfseries 13} (2010) 5},
\href{http://arxiv.org/abs/1004.3962}{{\ttfamily \footnotesize arXiv:1004.3962
  [hep-th]}}.

\bibitem{review2}
A.~S. Majumdar and N.~Mukherjee, ``{Braneworld black holes in cosmology and
  astrophysics},'' \href{http://dx.doi.org/10.1142/S0218271805006948}{{\em Int.
  J. Mod. Phys.} {\bfseries D14} (2005) 1095},
\href{http://arxiv.org/abs/astro-ph/0503473}{{\ttfamily \footnotesize
  arXiv:astro-ph/0503473 [astro-ph]}}.

\bibitem{review3}
R.~Gregory, ``{Braneworld black holes},''
  \href{http://dx.doi.org/10.1007/978-3-540-88460-6_7}{{\em Lect. Notes Phys.}
  {\bfseries 769} (2009) 259--298},
\href{http://arxiv.org/abs/0804.2595}{{\ttfamily \footnotesize arXiv:0804.2595
  [hep-th]}}.

\bibitem{review4}
P.~Kanti, ``{Brane-World Black Holes},''
  \href{http://dx.doi.org/10.1088/1742-6596/189/1/012020}{{\em J. Phys. Conf.
  Ser.} {\bfseries 189} (2009) 012020},
\href{http://arxiv.org/abs/0903.2147}{{\ttfamily \footnotesize arXiv:0903.2147
  [hep-th]}}.

\bibitem{review5}
P.~Kanti, ``{Footprints of Higher-Dimensional Decaying Black Holes},'' {\em
  Rom. J. Phys.} {\bfseries 57} (2012) 879--893,
  \href{http://arxiv.org/abs/1204.2371}{{\ttfamily \footnotesize
  arXiv:1204.2371 [hep-th]}}.

\bibitem{review6}
P.~Kanti and E.~Winstanley, ``{Hawking Radiation from Higher-Dimensional Black
  Holes},'' \href{http://dx.doi.org/10.1007/978-3-319-10852-0_8}{{\em Fundam.
  Theor. Phys.} {\bfseries 178} (2015) 229--265},
\href{http://arxiv.org/abs/1402.3952}{{\ttfamily \footnotesize arXiv:1402.3952
  [hep-th]}}.

\bibitem{review7}
N.~D. Pappas, ``{The black hole challenge in Randall-Sundrum II model},''
\href{http://arxiv.org/abs/1409.0817}{{\ttfamily \footnotesize arXiv:1409.0817
  [gr-qc]}}.

\bibitem{tidal}
N.~Dadhich, R.~Maartens, P.~Papadopoulos, and V.~Rezania, ``{Black holes on the
  brane},'' \href{http://dx.doi.org/10.1016/S0370-2693(00)00798-X}{{\em Phys.
  Lett.} {\bfseries B487} (2000) 1--6},
\href{http://arxiv.org/abs/hep-th/0003061}{{\ttfamily \footnotesize
  arXiv:hep-th/0003061 [hep-th]}}.

\bibitem{Papanto}
G.~Kofinas, E.~Papantonopoulos, and V.~Zamarias, ``{Black hole solutions in
  brane worlds with induced gravity},''
  \href{http://dx.doi.org/10.1103/PhysRevD.66.104028}{{\em Phys. Rev.}
  {\bfseries D66} (2002) 104028},
\href{http://arxiv.org/abs/hep-th/0208207}{{\ttfamily \footnotesize
  arXiv:hep-th/0208207 [hep-th]}}.

\bibitem{KT}
P.~Kanti and K.~Tamvakis, ``{Quest for localized 4-D black holes in brane
  worlds},'' \href{http://dx.doi.org/10.1103/PhysRevD.65.084010}{{\em Phys.
  Rev.} {\bfseries D65} (2002) 084010},
\href{http://arxiv.org/abs/hep-th/0110298}{{\ttfamily \footnotesize
  arXiv:hep-th/0110298 [hep-th]}}.

\bibitem{KOT}
P.~Kanti, I.~Olasagasti, and K.~Tamvakis, ``{Quest for localized 4-D black
  holes in brane worlds. 2. Removing the bulk singularities},''
  \href{http://dx.doi.org/10.1103/PhysRevD.68.124001}{{\em Phys. Rev.}
  {\bfseries D68} (2003) 124001},
\href{http://arxiv.org/abs/hep-th/0307201}{{\ttfamily \footnotesize
  arXiv:hep-th/0307201 [hep-th]}}.

\bibitem{CasadioNew}
R.~Casadio, A.~Fabbri, and L.~Mazzacurati, ``{New black holes in the brane
  world?},'' \href{http://dx.doi.org/10.1103/PhysRevD.65.084040}{{\em Phys.
  Rev.} {\bfseries D65} (2002) 084040},
\href{http://arxiv.org/abs/gr-qc/0111072}{{\ttfamily \footnotesize
  arXiv:gr-qc/0111072 [gr-qc]}}.

\bibitem{Frolov}
V.~P. Frolov, M.~Snajdr, and D.~Stojkovic, ``{Interaction of a brane with a
  moving bulk black hole},''
  \href{http://dx.doi.org/10.1103/PhysRevD.68.044002}{{\em Phys. Rev.}
  {\bfseries D68} (2003) 044002},
\href{http://arxiv.org/abs/gr-qc/0304083}{{\ttfamily \footnotesize
  arXiv:gr-qc/0304083 [gr-qc]}}.

\bibitem{Karasik}
D.~Karasik, C.~Sahabandu, P.~Suranyi, and L.~C.~R. Wijewardhana, ``{Small black
  holes on branes: Is the horizon regular or singular?},''
  \href{http://dx.doi.org/10.1103/PhysRevD.70.064007}{{\em Phys. Rev.}
  {\bfseries D70} (2004) 064007},
\href{http://arxiv.org/abs/gr-qc/0404015}{{\ttfamily \footnotesize
  arXiv:gr-qc/0404015 [gr-qc]}}.

\bibitem{Kofinas}
G.~Kofinas and E.~Papantonopoulos, ``{Gravitational collapse in brane world
  models with curvature corrections},''
  \href{http://dx.doi.org/10.1088/1475-7516/2004/12/011}{{\em JCAP} {\bfseries
  0412} (2004) 011},
\href{http://arxiv.org/abs/gr-qc/0401047}{{\ttfamily \footnotesize
  arXiv:gr-qc/0401047 [gr-qc]}}.

\bibitem{GGI}
C.~Galfard, C.~Germani, and A.~Ishibashi, ``{Asymptotically AdS brane black
  holes},'' \href{http://dx.doi.org/10.1103/PhysRevD.73.064014}{{\em Phys.
  Rev.} {\bfseries D73} (2006) 064014},
\href{http://arxiv.org/abs/hep-th/0512001}{{\ttfamily \footnotesize
  arXiv:hep-th/0512001 [hep-th]}}.

\bibitem{CGKM}
S.~Creek, R.~Gregory, P.~Kanti, and B.~Mistry, ``{Braneworld stars and black
  holes},'' \href{http://dx.doi.org/10.1088/0264-9381/23/23/004}{{\em Class.
  Quant. Grav.} {\bfseries 23} (2006) 6633--6658},
\href{http://arxiv.org/abs/hep-th/0606006}{{\ttfamily \footnotesize
  arXiv:hep-th/0606006 [hep-th]}}.

\bibitem{Ovalle1}
R.~Casadio and J.~Ovalle, ``{Brane-world stars and (microscopic) black
  holes},'' \href{http://dx.doi.org/10.1016/j.physletb.2012.07.041}{{\em Phys.
  Lett.} {\bfseries B715} (2012) 251--255},
\href{http://arxiv.org/abs/1201.6145}{{\ttfamily \footnotesize arXiv:1201.6145
  [gr-qc]}}.

\bibitem{Ovalle2}
J.~Ovalle and F.~Linares, ``{Tolman IV solution in the Randall-Sundrum
  Braneworld},'' \href{http://dx.doi.org/10.1103/PhysRevD.88.104026}{{\em Phys.
  Rev.} {\bfseries D88} no.~10, (2013) 104026},
\href{http://arxiv.org/abs/1311.1844}{{\ttfamily \footnotesize arXiv:1311.1844
  [gr-qc]}}.

\bibitem{Ovalle3}
J.~Ovalle, L.~A. Gergely, and R.~Casadio, ``{Brane-world stars with a solid
  crust and vacuum exterior},''
  \href{http://dx.doi.org/10.1088/0264-9381/32/4/045015}{{\em Class. Quant.
  Grav.} {\bfseries 32} (2015) 045015},
  \href{http://arxiv.org/abs/1405.0252}{{\ttfamily \footnotesize
  arXiv:1405.0252 [gr-qc]}}.

\bibitem{Ovalle4}
R.~Casadio, J.~Ovalle, and R.~da~Rocha, ``{The Minimal Geometric Deformation
  Approach Extended},''
  \href{http://dx.doi.org/10.1088/0264-9381/32/21/215020}{{\em Class. Quant.
  Grav.} {\bfseries 32} no.~21, (2015) 215020},
\href{http://arxiv.org/abs/1503.02873}{{\ttfamily \footnotesize
  arXiv:1503.02873 [gr-qc]}}.

\bibitem{Harko}
T.~Harko and M.~J. Lake, ``{Null fluid collapse in brane world models},''
  \href{http://dx.doi.org/10.1103/PhysRevD.89.064038}{{\em Phys. Rev.}
  {\bfseries D89} (2014) 064038},
\href{http://arxiv.org/abs/1312.1420}{{\ttfamily \footnotesize arXiv:1312.1420
  [gr-qc]}}.

\bibitem{daRocha1}
A.~M. Kuerten and R.~da~Rocha, ``{Probing topologically charged black holes on
  brane worlds in $f(\mathrm {R})$ bulk},''
  \href{http://dx.doi.org/10.1007/s10714-016-2092-8}{{\em Gen. Rel. Grav.}
  {\bfseries 48} no.~7, (2016) 90},
\href{http://arxiv.org/abs/1407.1483}{{\ttfamily \footnotesize arXiv:1407.1483
  [gr-qc]}}.

\bibitem{daRocha2}
A.~Herrera-Aguilar, A.~M. Kuerten, and R.~da~Rocha, ``{Regular Bulk Solutions
  in Brane-worlds with Inhomogeneous Dust and Generalized Dark Radiation},''
  \href{http://dx.doi.org/10.1155/2015/359268}{{\em Adv. High Energy Phys.}
  {\bfseries 2015} (2015) 359268},
\href{http://arxiv.org/abs/1501.07629}{{\ttfamily \footnotesize
  arXiv:1501.07629 [gr-qc]}}.

\bibitem{Dadhich}
N.~Dadhich, ``{Negative energy condition and black holes on the brane},''
  \href{http://dx.doi.org/10.1016/S0370-2693(00)01101-1}{{\em Phys. Lett. B}
  {\bfseries 492} (2000) 357--360},
  \href{http://arxiv.org/abs/hep-th/0009178}{{\ttfamily \footnotesize
  arXiv:hep-th/0009178}}.

\bibitem{Charmousis}
C.~Charmousis and R.~Gregory, ``{Axisymmetric metrics in arbitrary
  dimensions},'' \href{http://dx.doi.org/10.1088/0264-9381/21/2/016}{{\em
  Class. Quant. Grav.} {\bfseries 21} (2004) 527--554},
  \href{http://arxiv.org/abs/gr-qc/0306069}{{\ttfamily \footnotesize
  arXiv:gr-qc/0306069}}.

\bibitem{Shanka}
S.~Shankaranarayanan and N.~Dadhich, ``{Nonsingular black holes on the
  brane},'' \href{http://dx.doi.org/10.1142/S0218271804005109}{{\em Int. J.
  Mod. Phys. D} {\bfseries 13} (2004) 1095--1104},
  \href{http://arxiv.org/abs/gr-qc/0306111}{{\ttfamily \footnotesize
  arXiv:gr-qc/0306111}}.

\bibitem{Andrianov2}
A.~A. Andrianov and M.~A. Kurkov, ``{Black holes in the brane world: Some exact
  solutions},'' \href{http://dx.doi.org/10.1007/s11232-011-0140-9}{{\em Theor.
  Math. Phys.} {\bfseries 169} (2011) 1629--1642}.

\bibitem{Banerjee}
S.~Banerjee, U.~Danielsson, and S.~Giri, ``{Dark bubbles and black holes},''
  \href{http://arxiv.org/abs/2102.02164}{{\ttfamily \footnotesize
  arXiv:2102.02164 [hep-th]}}.

\bibitem{Chakraborty1}
R.~Dey, S.~Chakraborty, and N.~Afshordi, ``{Echoes from braneworld black
  holes},'' \href{http://dx.doi.org/10.1103/PhysRevD.101.104014}{{\em Phys.
  Rev. D} {\bfseries 101} no.~10, (2020) 104014},
  \href{http://arxiv.org/abs/2001.01301}{{\ttfamily \footnotesize
  arXiv:2001.01301 [gr-qc]}}.

\bibitem{Chakraborty2}
R.~Dey, S.~Biswas, and S.~Chakraborty, ``{Ergoregion instability and echoes for
  braneworld black holes: Scalar, electromagnetic, and gravitational
  perturbations},'' \href{http://dx.doi.org/10.1103/PhysRevD.103.084019}{{\em
  Phys. Rev. D} {\bfseries 103} no.~8, (2021) 084019},
  \href{http://arxiv.org/abs/2010.07966}{{\ttfamily \footnotesize
  arXiv:2010.07966 [gr-qc]}}.

\bibitem{Chakraborty3}
S.~Chakraborty, S.~Datta, and S.~Sau, ``{Tidal heating of black holes and
  exotic compact objects on the brane},''
  \href{http://arxiv.org/abs/2103.12430}{{\ttfamily \footnotesize
  arXiv:2103.12430 [gr-qc]}}.

\bibitem{Fitzpatrick}
A.~L. Fitzpatrick, L.~Randall, and T.~Wiseman, ``{On the existence and dynamics
  of braneworld black holes},''
  \href{http://dx.doi.org/10.1088/1126-6708/2006/11/033}{{\em JHEP} {\bfseries
  11} (2006) 033},
\href{http://arxiv.org/abs/hep-th/0608208}{{\ttfamily \footnotesize
  arXiv:hep-th/0608208 [hep-th]}}.

\bibitem{Zegers}
R.~Gregory, S.~F. Ross, and R.~Zegers, ``{Classical and quantum gravity of
  brane black holes},''
  \href{http://dx.doi.org/10.1088/1126-6708/2008/09/029}{{\em JHEP} {\bfseries
  09} (2008) 029},
\href{http://arxiv.org/abs/0802.2037}{{\ttfamily \footnotesize arXiv:0802.2037
  [hep-th]}}.

\bibitem{Heydari}
M.~Heydari-Fard and H.~R. Sepangi, ``Spherically symmetric solutions and
  gravitational collapse in brane-worlds,'' {\em
  \href{http://stacks.iop.org/1475-7516/2009/i=02/a=029}{Journal of Cosmology
  and Astroparticle Physics}} {\bfseries 2009} no.~02, (2009) 029. arXiv:
  \href{https://arxiv.org/abs/0903.0066}{gr-qc/09030066}.

\bibitem{Dai}
D.-C. Dai and D.~Stojkovic, ``{Analytic solution for a static black hole in
  RSII model},'' \href{http://dx.doi.org/10.1016/j.physletb.2011.09.038}{{\em
  Phys. Lett.} {\bfseries B704} (2011) 354--359},
\href{http://arxiv.org/abs/1004.3291}{{\ttfamily \footnotesize arXiv:1004.3291
  [gr-qc]}}.

\bibitem{Bruni}
M.~Bruni, C.~Germani, and R.~Maartens, ``{Gravitational collapse on the
  brane},'' \href{http://dx.doi.org/10.1103/PhysRevLett.87.231302}{{\em Phys.
  Rev. Lett.} {\bfseries 87} (2001) 231302},
\href{http://arxiv.org/abs/gr-qc/0108013}{{\ttfamily \footnotesize
  arXiv:gr-qc/0108013 [gr-qc]}}.

\bibitem{Tanaka}
T.~Tanaka, ``{Classical black hole evaporation in Randall-Sundrum infinite
  brane world},'' \href{http://dx.doi.org/10.1143/PTPS.148.307}{{\em Prog.
  Theor. Phys. Suppl.} {\bfseries 148} (2003) 307--316},
\href{http://arxiv.org/abs/gr-qc/0203082}{{\ttfamily \footnotesize
  arXiv:gr-qc/0203082 [gr-qc]}}.

\bibitem{EFK}
R.~Emparan, A.~Fabbri, and N.~Kaloper, ``{Quantum black holes as holograms in
  AdS brane worlds},''
  \href{http://dx.doi.org/10.1088/1126-6708/2002/08/043}{{\em JHEP} {\bfseries
  08} (2002) 043},
\href{http://arxiv.org/abs/hep-th/0206155}{{\ttfamily \footnotesize
  arXiv:hep-th/0206155 [hep-th]}}.

\bibitem{EGK}
R.~Emparan, J.~Garcia-Bellido, and N.~Kaloper, ``{Black hole astrophysics in
  AdS brane worlds},''
  \href{http://dx.doi.org/10.1088/1126-6708/2003/01/079}{{\em JHEP} {\bfseries
  01} (2003) 079},
\href{http://arxiv.org/abs/hep-th/0212132}{{\ttfamily \footnotesize
  arXiv:hep-th/0212132 [hep-th]}}.

\bibitem{Yoshino}
H.~Yoshino, ``{On the existence of a static black hole on a brane},''
  \href{http://dx.doi.org/10.1088/1126-6708/2009/01/068}{{\em JHEP} {\bfseries
  01} (2009) 068},
\href{http://arxiv.org/abs/0812.0465}{{\ttfamily \footnotesize arXiv:0812.0465
  [gr-qc]}}.

\bibitem{KPZ}
P.~Kanti, N.~Pappas, and K.~Zuleta, ``{On the localization of four-dimensional
  brane-world black holes},''
  \href{http://dx.doi.org/10.1088/0264-9381/30/23/235017}{{\em Class. Quant.
  Grav.} {\bfseries 30} (2013) 235017},
\href{http://arxiv.org/abs/1309.7642}{{\ttfamily \footnotesize arXiv:1309.7642
  [hep-th]}}.

\bibitem{KPP}
P.~Kanti, N.~Pappas, and T.~Pappas, ``{On the localisation of four-dimensional
  brane-world black holes: II. The general case},''
  \href{http://dx.doi.org/10.1088/0264-9381/33/1/015003}{{\em Class. Quant.
  Grav.} {\bfseries 33} no.~1, (2016) 015003},
\href{http://arxiv.org/abs/1507.02625}{{\ttfamily \footnotesize
  arXiv:1507.02625 [hep-th]}}.

\bibitem{EHM1}
R.~Emparan, G.~T. Horowitz, and R.~C. Myers, ``{Exact description of black
  holes on branes},''
  \href{http://dx.doi.org/10.1088/1126-6708/2000/01/007}{{\em JHEP} {\bfseries
  01} (2000) 007}, \href{http://arxiv.org/abs/hep-th/9911043}{{\ttfamily
  \footnotesize arXiv:hep-th/9911043}}.

\bibitem{EHM2}
R.~Emparan, G.~T. Horowitz, and R.~C. Myers, ``{Exact description of black
  holes on branes. 2. Comparison with BTZ black holes and black strings},''
  \href{http://dx.doi.org/10.1088/1126-6708/2000/01/021}{{\em JHEP} {\bfseries
  01} (2000) 021},
\href{http://arxiv.org/abs/hep-th/9912135}{{\ttfamily \footnotesize
  arXiv:hep-th/9912135 [hep-th]}}.

\bibitem{AS}
M.~Anber and L.~Sorbo, ``{New exact solutions on the Randall-Sundrum 2-brane:
  lumps of dark radiation and accelerated black holes},''
  \href{http://dx.doi.org/10.1088/1126-6708/2008/07/098}{{\em JHEP} {\bfseries
  07} (2008) 098},
\href{http://arxiv.org/abs/0803.2242}{{\ttfamily \footnotesize arXiv:0803.2242
  [hep-th]}}.

\bibitem{Cuadros}
B.~Cuadros-Melgar, E.~Papantonopoulos, M.~Tsoukalas, and V.~Zamarias, ``{BTZ
  Like-String on Codimension-2 Braneworlds in the Thin Brane Limit},''
  \href{http://dx.doi.org/10.1103/PhysRevLett.100.221601}{{\em Phys. Rev.
  Lett.} {\bfseries 100} (2008) 221601},
\href{http://arxiv.org/abs/0712.3232}{{\ttfamily \footnotesize arXiv:0712.3232
  [hep-th]}}.

\bibitem{Kudoh}
H.~Kudoh, ``{Six-dimensional localized black holes: Numerical solutions},''
  \href{http://dx.doi.org/10.1103/PhysRevD.70.029901,
  10.1103/PhysRevD.69.104019}{{\em Phys. Rev.} {\bfseries D69} (2004) 104019},
  \href{http://arxiv.org/abs/hep-th/0401229}{{\ttfamily \footnotesize
  arXiv:hep-th/0401229 [hep-th]}}.
[Erratum: Phys. Rev.D70,029901(2004)].

\bibitem{Kudoh1}
H.~Kudoh, T.~Tanaka, and T.~Nakamura, ``{Small localized black holes in brane
  world: Formulation and numerical method},''
  \href{http://dx.doi.org/10.1103/PhysRevD.68.024035}{{\em Phys. Rev. D}
  {\bfseries 68} (2003) 024035},
  \href{http://arxiv.org/abs/gr-qc/0301089}{{\ttfamily \footnotesize
  arXiv:gr-qc/0301089}}.

\bibitem{Tanahashi}
N.~Tanahashi and T.~Tanaka, ``{Time-symmetric initial data of large
  brane-localized black hole in RS-II model},''
  \href{http://dx.doi.org/10.1088/1126-6708/2008/03/041}{{\em JHEP} {\bfseries
  03} (2008) 041}, \href{http://arxiv.org/abs/0712.3799}{{\ttfamily
  \footnotesize arXiv:0712.3799 [gr-qc]}}.

\bibitem{Kleihaus}
B.~Kleihaus, J.~Kunz, E.~Radu, and D.~Senkbeil, ``{Electric charge on the
  brane?},'' \href{http://dx.doi.org/10.1103/PhysRevD.83.104050}{{\em Phys.
  Rev.} {\bfseries D83} (2011) 104050},
\href{http://arxiv.org/abs/1103.4758}{{\ttfamily \footnotesize arXiv:1103.4758
  [gr-qc]}}.

\bibitem{Figueras1}
P.~Figueras and T.~Wiseman, ``{Gravity and large black holes in Randall-Sundrum
  II braneworlds},''
  \href{http://dx.doi.org/10.1103/PhysRevLett.107.081101}{{\em Phys. Rev.
  Lett.} {\bfseries 107} (2011) 081101},
  \href{http://arxiv.org/abs/1105.2558}{{\ttfamily \footnotesize
  arXiv:1105.2558 [hep-th]}}.

\bibitem{Page1}
S.~Abdolrahimi, C.~Cattoen, D.~N. Page, and S.~Yaghoobpour-Tari, ``{Large
  Randall-Sundrum II Black Holes},''
  \href{http://dx.doi.org/10.1016/j.physletb.2013.02.034}{{\em Phys. Lett.}
  {\bfseries B720} (2013) 405--409},
\href{http://arxiv.org/abs/1206.0708}{{\ttfamily \footnotesize arXiv:1206.0708
  [hep-th]}}.

\bibitem{Page2}
S.~Abdolrahimi, C.~Cattoën, D.~N. Page, and S.~Yaghoobpour-Tari, ``{Spectral
  methods in general relativity and large Randall-Sundrum II black holes},''
  \href{http://dx.doi.org/10.1088/1475-7516/2013/06/039}{{\em JCAP} {\bfseries
  1306} (2013) 039},
\href{http://arxiv.org/abs/1212.5623}{{\ttfamily \footnotesize arXiv:1212.5623
  [hep-th]}}.

\bibitem{BDL}
P.~Binetruy, C.~Deffayet, and D.~Langlois, ``{Nonconventional cosmology from a
  brane universe},''
  \href{http://dx.doi.org/10.1016/S0550-3213(99)00696-3}{{\em Nucl. Phys.}
  {\bfseries B565} (2000) 269--287},
\href{http://arxiv.org/abs/hep-th/9905012}{{\ttfamily \footnotesize
  arXiv:hep-th/9905012 [hep-th]}}.

\bibitem{Bhatta1}
S.~Bhattacharya and S.~R. Kousvos, ``{Constraining the phantom braneworld model
  from cosmic structure sizes},''
  \href{http://dx.doi.org/10.1103/PhysRevD.96.104006}{{\em Phys. Rev.}
  {\bfseries D96} no.~10, (2017) 104006},
\href{http://arxiv.org/abs/1706.06268}{{\ttfamily \footnotesize
  arXiv:1706.06268 [gr-qc]}}.

\bibitem{Bhatta2}
S.~Bhattacharya, S.~R. Kousvos, S.~Romanopoulos, and T.~N. Tomaras,
  ``{Cosmological screening and the phantom braneworld model},''
  \href{http://dx.doi.org/10.1140/epjc/s10052-018-6119-z}{{\em Eur. Phys. J.}
  {\bfseries C78} no.~8, (2018) 637},
\href{http://arxiv.org/abs/1802.07660}{{\ttfamily \footnotesize
  arXiv:1802.07660 [gr-qc]}}.

\bibitem{Neves:2021dqx}
J.~C.~S. Neves, ``{Five-dimensional regular black holes in a brane world},''
  \href{http://dx.doi.org/10.1103/PhysRevD.104.084019}{{\em Phys. Rev. D}
  {\bfseries 104} no.~8, (2021) 084019},
  \href{http://arxiv.org/abs/2107.04072}{{\ttfamily \footnotesize
  arXiv:2107.04072 [hep-th]}}.

\bibitem{Farakos1}
K.~Farakos and P.~Pasipoularides, ``{Second Randall-Sundrum brane world
  scenario with a nonminimally coupled bulk scalar field},''
  \href{http://dx.doi.org/10.1103/PhysRevD.73.084012}{{\em Phys. Rev.}
  {\bfseries D73} (2006) 084012},
\href{http://arxiv.org/abs/hep-th/0602200}{{\ttfamily \footnotesize
  arXiv:hep-th/0602200 [hep-th]}}.

\bibitem{Bogdanos1}
C.~Bogdanos, A.~Dimitriadis, and K.~Tamvakis, ``{Brane models with a
  Ricci-coupled scalar field},''
  \href{http://dx.doi.org/10.1103/PhysRevD.74.045003}{{\em Phys. Rev.}
  {\bfseries D74} (2006) 045003},
\href{http://arxiv.org/abs/hep-th/0604182}{{\ttfamily \footnotesize
  arXiv:hep-th/0604182 [hep-th]}}.

\bibitem{Farakos2}
K.~Farakos and P.~Pasipoularides, ``{Gauss-Bonnet gravity, brane world models,
  and non-minimal coupling},''
  \href{http://dx.doi.org/10.1103/PhysRevD.75.024018}{{\em Phys. Rev.}
  {\bfseries D75} (2007) 024018},
\href{http://arxiv.org/abs/hep-th/0610010}{{\ttfamily \footnotesize
  arXiv:hep-th/0610010 [hep-th]}}.

\bibitem{Farakos3}
K.~Farakos, G.~Koutsoumbas, and P.~Pasipoularides, ``{Graviton localization and
  Newton's law for brane models with a non-minimally coupled bulk scalar
  field},'' \href{http://dx.doi.org/10.1103/PhysRevD.76.064025}{{\em Phys.
  Rev.} {\bfseries D76} (2007) 064025},
\href{http://arxiv.org/abs/0705.2364}{{\ttfamily \footnotesize arXiv:0705.2364
  [hep-th]}}.

\bibitem{Gibbons-terms}
A.~Padilla and V.~Sivanesan, ``{Boundary Terms and Junction Conditions for
  Generalized Scalar-Tensor Theories},''
  \href{http://dx.doi.org/10.1007/JHEP08(2012)122}{{\em JHEP} {\bfseries 08}
  (2012) 122},
\href{http://arxiv.org/abs/1206.1258}{{\ttfamily \footnotesize arXiv:1206.1258
  [gr-qc]}}.

\bibitem{Abramowitz}
M.~Abramowitz and I.~A. Stegun, {\em Handbook of Mathematical Functions with
  Formulas, Graphs, and Mathematical Tables}.
\newblock Dover, New York, ninth {D}over printing, tenth {GPO} printing~ed.,
  1964.

\bibitem{MT}
M.~Morris and K.~Thorne, ``{Wormholes in space-time and their use for
  interstellar travel: A tool for teaching general relativity},''
  \href{http://dx.doi.org/10.1119/1.15620}{{\em Am. J. Phys.} {\bfseries 56}
  (1988) 395--412}.

\bibitem{SMS}
T.~Shiromizu, K.-i. Maeda, and M.~Sasaki, ``{The Einstein equation on the
  3-brane world},'' \href{http://dx.doi.org/10.1103/PhysRevD.62.024012}{{\em
  Phys. Rev. D} {\bfseries 62} (2000) 024012},
  \href{http://arxiv.org/abs/gr-qc/9910076}{{\ttfamily \footnotesize
  arXiv:gr-qc/9910076}}.

\bibitem{Israel}
W.~Israel, ``{Singular hypersurfaces and thin shells in general relativity},''
  \href{http://dx.doi.org/10.1007/BF02710419}{{\em Nuovo Cim. B} {\bfseries
  44S10} (1966) 1}. [Erratum: Nuovo Cim.B 48, 463 (1967)].

\bibitem{Fichet}
S.~Fichet, ``{Braneworld effective field theories — holography, consistency
  and conformal effects},''
  \href{http://dx.doi.org/10.1007/JHEP04(2020)016}{{\em JHEP} {\bfseries 04}
  (2020) 016},
\href{http://arxiv.org/abs/1912.12316}{{\ttfamily \footnotesize
  arXiv:1912.12316 [hep-th]}}.

\bibitem{Chambers}
C.~M. Chambers, ``{The Cauchy horizon in black hole de sitter space-times},''
  {\em Annals Israel Phys. Soc.} {\bfseries 13} (1997) 33,
  \href{http://arxiv.org/abs/gr-qc/9709025}{{\ttfamily \footnotesize
  arXiv:gr-qc/9709025}}.

\bibitem{Bousso}
R.~Bousso and S.~W. Hawking, ``{(Anti)evaporation of Schwarzschild-de Sitter
  black holes},'' \href{http://dx.doi.org/10.1103/PhysRevD.57.2436}{{\em Phys.
  Rev. D} {\bfseries 57} (1998) 2436--2442},
  \href{http://arxiv.org/abs/hep-th/9709224}{{\ttfamily \footnotesize
  arXiv:hep-th/9709224}}.

\bibitem{Maldacena}
J.~M. Maldacena, ``{The Large N limit of superconformal field theories and
  supergravity},'' \href{http://dx.doi.org/10.1023/A:1026654312961}{{\em Int.
  J. Theor. Phys.} {\bfseries 38} (1999) 1113--1133},
  \href{http://arxiv.org/abs/hep-th/9711200}{{\ttfamily \footnotesize
  arXiv:hep-th/9711200}}.

\bibitem{Gubser2}
S.~Gubser, I.~R. Klebanov, and A.~M. Polyakov, ``{Gauge theory correlators from
  noncritical string theory},''
  \href{http://dx.doi.org/10.1016/S0370-2693(98)00377-3}{{\em Phys. Lett. B}
  {\bfseries 428} (1998) 105--114},
  \href{http://arxiv.org/abs/hep-th/9802109}{{\ttfamily \footnotesize
  arXiv:hep-th/9802109}}.

\bibitem{Witten}
E.~Witten, ``{Anti-de Sitter space and holography},''
  \href{http://dx.doi.org/10.4310/ATMP.1998.v2.n2.a2}{{\em Adv. Theor. Math.
  Phys.} {\bfseries 2} (1998) 253--291},
  \href{http://arxiv.org/abs/hep-th/9802150}{{\ttfamily \footnotesize
  arXiv:hep-th/9802150}}.

\bibitem{Pomarol}
L.~Da~Rold and A.~Pomarol, ``{Chiral symmetry breaking from five dimensional
  spaces},'' \href{http://dx.doi.org/10.1016/j.nuclphysb.2005.05.009}{{\em
  Nucl. Phys. B} {\bfseries 721} (2005) 79--97},
  \href{http://arxiv.org/abs/hep-ph/0501218}{{\ttfamily \footnotesize
  arXiv:hep-ph/0501218}}.

\bibitem{Alho}
T.~Alho, N.~Evans, and K.~Tuominen, ``{Dynamic AdS/QCD and the Spectrum of
  Walking Gauge Theories},''
  \href{http://dx.doi.org/10.1103/PhysRevD.88.105016}{{\em Phys. Rev. D}
  {\bfseries 88} (2013) 105016},
  \href{http://arxiv.org/abs/1307.4896}{{\ttfamily \footnotesize
  arXiv:1307.4896 [hep-ph]}}.

\bibitem{Ballon}
C.~Ballon~Bayona, H.~Boschi-Filho, N.~R. Braga, and L.~A. Pando~Zayas, ``{On a
  Holographic Model for Confinement/Deconfinement},''
  \href{http://dx.doi.org/10.1103/PhysRevD.77.046002}{{\em Phys. Rev. D}
  {\bfseries 77} (2008) 046002},
  \href{http://arxiv.org/abs/0705.1529}{{\ttfamily \footnotesize
  arXiv:0705.1529 [hep-th]}}.

\bibitem{Cardoso}
V.~Cardoso, S.~Hopper, C.~F.~B. Macedo, C.~Palenzuela, and P.~Pani,
  ``{Gravitational-wave signatures of exotic compact objects and of quantum
  corrections at the horizon scale},''
  \href{http://dx.doi.org/10.1103/PhysRevD.94.084031}{{\em Phys. Rev. D}
  {\bfseries 94} no.~8, (2016) 084031},
  \href{http://arxiv.org/abs/1608.08637}{{\ttfamily \footnotesize
  arXiv:1608.08637 [gr-qc]}}.

\bibitem{Zachary}
Z.~Mark, A.~Zimmerman, S.~M. Du, and Y.~Chen, ``{A recipe for echoes from
  exotic compact objects},''
  \href{http://dx.doi.org/10.1103/PhysRevD.96.084002}{{\em Phys. Rev. D}
  {\bfseries 96} no.~8, (2017) 084002},
  \href{http://arxiv.org/abs/1706.06155}{{\ttfamily \footnotesize
  arXiv:1706.06155 [gr-qc]}}.

\bibitem{Maggio}
E.~Maggio, V.~Cardoso, S.~R. Dolan, and P.~Pani, ``{Ergoregion instability of
  exotic compact objects: electromagnetic and gravitational perturbations and
  the role of absorption},''
  \href{http://dx.doi.org/10.1103/PhysRevD.99.064007}{{\em Phys. Rev. D}
  {\bfseries 99} no.~6, (2019) 064007},
  \href{http://arxiv.org/abs/1807.08840}{{\ttfamily \footnotesize
  arXiv:1807.08840 [gr-qc]}}.

\bibitem{Antonini1}
S.~Antonini and B.~Swingle, ``{Cosmology at the end of the world},''
  \href{http://dx.doi.org/10.1038/s41567-020-0909-6}{{\em Nature Phys.}
  {\bfseries 16} no.~8, (2020) 881--886},
\href{http://arxiv.org/abs/1907.06667}{{\ttfamily \footnotesize
  arXiv:1907.06667 [hep-th]}}.

\bibitem{Antonini2}
S.~Antonini and B.~Swingle, ``{Holographic boundary states and
  dimensionally-reduced braneworld spacetimes},''
\href{http://arxiv.org/abs/2105.02912}{{\ttfamily \footnotesize
  arXiv:2105.02912 [hep-th]}}.

\bibitem{Alestas}
G.~Alestas, G.~V. Kraniotis, and L.~Perivolaropoulos, ``{Existence and
  stability of static spherical fluid shells in a
  Schwarzschild-Rindler\textendash{}anti\textendash{}de Sitter metric},''
  \href{http://dx.doi.org/10.1103/PhysRevD.102.104015}{{\em Phys. Rev. D}
  {\bfseries 102} no.~10, (2020) 104015},
  \href{http://arxiv.org/abs/2005.11702}{{\ttfamily \footnotesize
  arXiv:2005.11702 [gr-qc]}}.

\bibitem{KGB}
P.~Kanti, J.~Grain, and A.~Barrau, ``{Bulk and brane decay of a
  (4+n)-dimensional Schwarzschild-de-Sitter black hole: Scalar radiation},''
  \href{http://dx.doi.org/10.1103/PhysRevD.71.104002}{{\em Phys. Rev.}
  {\bfseries D71} (2005) 104002},
\href{http://arxiv.org/abs/hep-th/0501148}{{\ttfamily \footnotesize
  arXiv:hep-th/0501148 [hep-th]}}.

\end{thebibliography}\endgroup
\bibliographystyle{utphys}
}

\end{document}